\documentclass{elsart}

\usepackage{epsfig}

 \newcommand{\zr}[1]{\mbox{\hspace*{#1em}}}
 \newcommand{\ID}{\mbox{{\sf 1}\zr{-0.14}\rule{0.04em}{1.55ex}\zr{0.1}}}
 \newcommand{\NN}{\mbox{\zr{0.1}\rule{0.04em}{1.6ex}\zr{-0.05}{\sf N}}}
 \newcommand{\RR}{\mbox{\zr{0.1}\rule{0.04em}{1.6ex}\zr{-0.05}{\sf R}}}
 \newcommand{\CC}{\mbox{\zr{0.1}\rule{0.04em}{1.6ex}\zr{-0.30}{\sf C}}}
 \newcommand{\ZZ}{\mbox{\sf Z\zr{-0.45}Z}}

\begin{document}

    \hspace*{\fill}{\small\sf UNITUE-THEP-00/09,
                    \hskip .5cm FAU-TP3-00/8} \\
    \hspace*{\fill}{\small\sf http://xxx.lanl.gov/abs/hep-ph/0007355} \\ ~\\

\begin{frontmatter}

\title{The Infrared Behaviour of QCD Green's Functions \\ ~\\
{\large Confinement, Dynamical Symmetry Breaking, \\ 
and Hadrons as Relativistic Bound States}\thanksref{DFG}}

\thanks[DFG]{Supported in part by DFG (Al 279/3-3) and COSY (contract
no.\ 41376610).}

\author{Reinhard Alkofer\thanksref{RA}}
\thanks[RA]{E-Mail:     reinhard.alkofer@uni-tuebingen.de}
\address{Institut f\"ur Theoretische Physik,
        Universit\"at T\"{u}bingen, 
        Auf der Morgenstelle 14, 72076 T\"ubingen, Germany}

\and

\author{Lorenz von Smekal\thanksref{LvS}}
\thanks[LvS]{E-Mail:smekal@theorie3.physik.uni-erlangen.de}
\address{Institut f\"ur Theoretische Physik III,
           Universit\"at Erlangen--N\"urnberg, 
           Staudtstr.~7, 91058 Erlangen, Germany}

\end{frontmatter}

\newpage

\begin{frontmatter}
\begin{abstract}
Recent studies of QCD Green's functions and their applications in hadronic
physics are reviewed. We discuss the definition of the generating functional in
gauge theories, in particular, the r\^ole of redundant degrees of freedom,
possibilities of a complete gauge fixing versus gauge fixing in presence of
Gribov copies, BRS invariance and positivity. The apparent contradiction
between positivity and colour antiscreening in combination with BRS invariance
in QCD is considered. Evidence for the violation of positivity by quarks and
transverse gluons in the covariant gauge is collected, and it is argued that
this is one manifestation of confinement.

We summarise the derivation of the Dyson--Schwinger equations (DSEs) of QED
and QCD. For the latter, the implications of BRS invariance on the Green's
functions are explored. The possible influence of instantons on DSEs is
discussed in a two-dimensional model. In QED in 2+1 and 3+1 dimensions, the
solutions for Green's functions provide tests of truncation schemes 
which can under certain circumstances be extended to the DSEs of QCD. We
discuss some limitations of such extensions and assess the validity of
assumptions for QCD as motivated from studies in QED. 
Truncation schemes for DSEs are discussed in axial and
related gauges, as well as in the Landau gauge. Furthermore, we review the
available results from a systematic non-perturbative expansion scheme
established for Landau gauge QCD. 
Comparisons to related lattice results, where available, are presented. 

The applications of QCD Green's functions to hadron physics are summarised.
Properties of ground state mesons are discussed on the basis of the ladder
Bethe--Salpeter equation for quarks and antiquarks. The Goldstone nature of
pseudoscalar mesons and a mechanism for diquark confinement beyond the ladder
approximation are reviewed. We discuss some properties of ground state baryons
based on their description as Bethe--Salpeter/Faddeev bound
states of quark-diquark correlations in the quantum field theory
of confined quarks and gluons.
\end{abstract}

\vskip 2cm

\begin{keyword}
Strong QCD, Green's functions, Confinement, Chiral symmetry breaking,
Dyson--Schwinger equations, Bethe--Salpeter equation.\\[5mm]
  {\small PACS numbers: 02.30.Rz 11.10.Gh 11.10.Hi 11.10.St 11.15.Tk 
  12.20.Ds 12.38.Aw 12.38.Lg 14.20.Dh 14.20.Jn 14.40.Aq 14.65.Bt
  14.70.Dj}

\end{keyword}

\end{frontmatter}

\newpage

\setcounter{tocdepth}{4}

\begin{frontmatter}
\tableofcontents
\end{frontmatter}


\section{Introduction}
\label{chap_intro} 

Looking at the plethora of different hadrons it is evident that baryons and
mesons are not elementary particles in the naive sense of the word
``elementary''. Experimental hadron physics has determined the partonic
substructure of the nucleon to an enormous precision leaving no doubt that the
parton picture emerges from quarks and gluons, the elementary fields of Quantum
Chromodynamics (QCD). On the other hand, it is a well-known fact that these 
quarks and gluons have not been detected outside hadrons. This puzzle was
given a name: confinement. Despite the fact that the confinement hypothesis was
formulated  several decades ago our understanding of the confinement
mechanism(s) is still not satisfactory. And, in contrast to other
non-perturbative phenomena of interest in QCD ({\it e.g.}, dynamical breaking
of chiral symmetry, $U_A(1)$ anomaly, and formation of relativistic bound
states) the phenomenon of confinement might to some extend be in conflict
with nowadays widely accepted foundations of quantum field theory.

Quantum field theory provides the basis for our current understanding of
particle physics. The quantum field theoretical description of elementary 
particles has been impressively successful since it was first developed in the
quantisation of electrodynamics in the late 20ties. After its first
applications to elementary processes like the spontaneous decays of exited
atoms, the photo and the Compton effect, electron-electron scattering, pair
creation and Bremsstrahlung, the next major step was accomplished in the late
40ties. The anomalous magnetic moment of the free electron from the Dirac
theory of relativistic quantum mechanics was observed experimentally. The
development of the covariant perturbation expansion by Tomonaga, Schwinger and
Feynman together with the concept of renormalisation made it possible 
to calculate higher order corrections to the elementary processes of
electrons and photons. 
Its application to the Lamb shift explained the experimental observations, and
higher order corrections subsequently agreed with the results of refined
experiments. 

Since these developments of Quantum Electrodynamics (QED), local quantum
field theory has further been developed and applied to the descriptions of
elementary particles. 
Their processes are accounted for by the collision theory developed by Lehmann,
Symanzik and Zimmermann, the so-called LSZ formalism \cite{Leh55} 
(for a description of its role in modern quantum field theory 
see, {\it e.g.}, Refs.\ \cite{Itz80} or \cite{Haa96}).  
Together with perturbation theory it describes the processes of elementary
particles at high energies based on asymptotically free gauge theories. In
particular, in the weak coupling regime of QCD,
{\it i.e.}, at high energies, the agreement of perturbative calculations with
the huge number of measurements available is impressive (see, {\it e.g.},
Fig.~\ref{Fig:alpha_s} in Sec.~\ref{sec_Trunc}). 

The perturbative description of elementary particles
is essentially based on the field-particle duality which means that each
field in a quantum field theory is associated with a physical particle. A
simple example for this being the Fermi theory of the $\beta$-decay where a
field is associated to all particles involved, the proton, the neutron, as
well as the electron and the neutrino. This is of course not what one has in
mind in describing hadrons as composite states with quark and gluonic
substructure in hadronic processes. On the one hand, scattering theory can be
extended to include processes of composite particles described by ``almost
local'' fields leading to a generalised LSZ formalism for bound states 
\cite{Haa58,Nis58,Zim58} (described also in the book of Ref.~\cite{Haa96}). 
On the other hand, the situation in QCD is more complicated, however. Not
only does the asymptotic state space contain composite states, but the
physical Hilbert space of the asymptotic hadron states does not contain any 
states corresponding to particles associated with the elementary fields in
QCD, the quarks and gluons. For a description of confinement of quarks and
gluons within the framework of local quantum field theory, the elementary
fields have to be divorced completely from a particle interpretation. A
quote from Haag's book~\cite{Haa96} expresses this in a clear way as follows:
 
{\sl ``The r\^ole of fields is to implement the principle of locality. The
number and the nature of different basic fields needed in the theory is
related to the charge structure, not to the empirical spectrum of particles.''}
  
The description of hadronic states and processes based on the dynamics of the
confined correlations of quark and glue is the outstanding challenge in the
formulation of QCD as a local quantum field theory. In particular, assuming
that only hadrons are produced from processes involving hadronic initial
states, one has to explain that the only thresholds in hadronic amplitudes
are due to the productions of other hadronic states, and that possible
structure singularities occur in composite states which are due to their
hadronic substructure only.  

Some theoretical insight into the mechanism(s) for 
confinement into colourless hadrons  could be obtained from
disproving the cluster decomposition property for colour-nonsinglet
gauge-covariant operators. One idea in this direction is based on the
possible existence of severe infrared divergences, {\it i.e.}, divergences
which cannot be removed from physical cross sections by a suitable summation
over degenerate states by virtue of the Kinoshita--Lee--Nauenberg theorem
\cite{Kin62}.\footnote{Also referred to as the non-Abelian Bloch--Nordsieck
prescription, {\it c.f.}, Ref.~\cite{Blo37}.} Such severe infrared divergences
could provide damping factors for the emission 
of coloured states from colour-singlet states (see \cite{Mar78}). However, the
Kinoshita--Lee--Nauenberg theorem applies to non-Abelian gauge theories in
four dimensions order by order in perturbation theory \cite{Kin76,Nel81}. 
Therefore, such a description of confinement in terms of perturbation theory
is impossible. In fact, extended to Green's functions, the absence of
unphysical infrared divergences implies that the spectrum of QCD necessarily
includes coloured quark and gluon states to every order in perturbation theory
\cite{Pog76}.   

An alternative way to understand the insufficiency of perturbation theory
to account for confinement in four-dimensional field theories is that
confinement requires the dynamical generation of a physical mass scale. In
presence of such a mass scale, however, the renormalisation group (RG)
equations imply the existence of essential singularities in physical
quantities, such as the $S$-matrix, as functions of the coupling at $g =
0$. This is because the dependence of the RG invariant confinement scale
on the coupling and the renormalisation scale $\mu$ near the ultraviolet
fixed point is determined by \cite{Gro74}
\begin{equation}
  \Lambda = \mu \exp \left( - \int ^g \frac {dg'}{\beta (g')} \right)
  \stackrel{g\to 0}{\rightarrow } \mu  \exp \left( - \frac 1 {2\beta_0g^2}
\right),    \quad \beta_0>0 .
  \label{Lambda}
\end{equation}
Since all RG invariant masses in massless QCD will exhibit the behaviour
(\ref{Lambda}) up to a multiplicative constant, the ratios of all bound state
masses are, at least in the chiral limit, determined independent of all
parameters. 

Therefore, to study the infrared behaviour of QCD amplitudes non-perturbative
methods are required. In addition, as singularities are anticipated, a
formulation in the continuum is desirable. One promising approach to
non-perturbative phenomena in QCD is provided by studies of truncated systems
of its Dyson--Schwinger equations (DSEs) \cite{Dys49,Sch51}, the equations of
motion for QCD Green's functions. Typical truncation schemes resort to
additional sources of information like the Slavnov--Taylor identities
\cite{Tay71,Sla72}, as entailed by gauge invariance, to express vertex
functions and higher $n$-point functions in terms of the elementary two-point
functions, {\it i.e.}, the quark, ghost and gluon propagators. In principle,
these propagators can then be obtained as self-consistent solutions to the
non-linear integral equations representing the closed set of truncated DSEs. 

The underlying conjecture to justify such a truncation of the originally
infinite set of DSEs is that a successive inclusion of higher $n$-point
functions in self-consistent calculations will not result in  dramatic changes
to previously obtained lower $n$-point functions. To achieve this it is
important to incorporate as much independent information as possible in
constructing those $n$-point functions which close the system.  Such
information, {\it e.g.}, from implications of gauge invariance or symmetry
properties, is sufficiently reliable so that the related properties are
expected to be reproduced by the solutions to
subsequent truncation schemes. 

Until recently, available solutions to truncated DSEs of
QCD did not even fully include all contributions of the propagators
themselves. In particular, even in absence of quarks, solutions for the gluon
propagator in Landau gauge used to rely on neglecting ghost contributions
\cite{Man79,Atk81,Bro89,Hau96} which, though numerically small in
perturbation theory, are unavoidable in this gauge. While this
particular problem can be avoided by ghost free gauges such as the axial
gauge, in studies of the gluon DSE in the axial gauge 
\cite{Bak81a,Bak81b,Ale82,Sch82,Cud91}, 
the possible occurrence of an independent second
term in the tensor structure of the gluon propagator has been disregarded
\cite{Bue95}. In fact, if the complete tensor structure of the gluon propagator
in axial gauge is taken into account properly, one arrives at a coupled system
of equations which is of similar complexity as the ghost-gluon system in the
Landau gauge and which is yet to be solved.

In addition to providing a better understanding of con\-fine\-ment based on
studies of the behaviour of QCD Green's functions in the infrared, DSEs have
proven successful in developing a hadron phenomenology which
interpolates smoothly between the infrared (non-perturbative) and the
ultraviolet (perturbative) regime, for recent reviews see,
{\it e.g.}, \cite{Tan97,Rob94}. In particular, a dynamical description of
the spontaneous breaking of chiral symmetry from studies of the DSE for the
quark propagator is well established in a variety of models for the 
gluonic interactions of quarks~\cite{Mir93}. For a sufficiently large
low-energy quark-quark interaction quark masses are generated dynamically 
in the quark DSE in some analogy to the gap equation in superconductivity. 
This in turn leads naturally to the Goldstone nature of the pion and explains
the smallness of its mass as compared to all other hadrons. 
In this framework a description of the different types of mesons is obtained
from Bethe--Salpeter equations (BSEs) for quark-antiquark bound
states. Recent progress towards a solution of a fully
relativistic three-body equation extends this consistent framework to
baryonic bound states.  

Investigations of QCD Green's functions have been extended successfully to
finite temperatures and densities during the last few years. As this is a
subject of its own, and with regard to the length of the present review we have
refrained from reviewing this topic. Instead we refer the interested reader to
the recent review provided by Ref.\ \cite{Rob00}.

This review is organised as follows: Chapter 2 reviews some basic
concepts of quantum field theory and especially QCD, some derivations and the
notations to  provide the necessary background for the later chapters.  In
Chapter 3 the Dyson--Schwinger formalism is presented. Chapter 4 provides a
short summary of QED Green's functions in 1+1, 2+1 and 3+1 dimensions. These
results are helpful to put the corresponding results for QCD into perspective.
Chapter 5 is the central part of this review: The infrared behaviour of QCD
Green's function and its implications for confinement, for dynamical breaking
of chiral symmetry and the structure of hadrons in general are discussed.
Phenomenological studies of mesons which are based on the results for the
propagators are summarised in Chapter 6.  On the way towards a description of
baryons as bound states in colour singlet 3-quark channels a detailed
understanding of diquark correlations is necessary. In Chapter 7 the
corresponding framework is provided, and baryonic bound states of quarks and
diquarks are described in reduced Bethe--Salpeter/Faddeev equations obtained
for separable diquark correlations. A few concluding remarks are given in the
last chapter. Some more technical issues are provided in several appendices.

We wish to emphasise that this review is a status report on an on-going
effort. The long way from the dynamics of quark and glue to hadrons in one
coherent description is far from paved. Considerable segments are, however,
increasingly well understood, some on a fairly fundamental level, others  from
temporarily used model assumptions. Connections between the various pieces
are made in form of justifications and improvements of the respective model
assumptions. The present review describes some of these segments along the
way.

\section{Basic Concepts in Quantum Field Theory}
\label{chap_Basic} 

In this chapter, some underlying concepts of the subsequent chapters  are
briefly reviewed, mainly to introduce definitions and conventions for later
use. In addition, and maybe more importantly, we discuss some of the
fundamental issues in this chapter which might in future lead to advances in
the understanding of hadronic physics based on the dynamics of quark and glue.

Since a comprehensive and final quantum field theoretic description of a
confining theory is not established yet, it is necessary to cover various
descriptions in this review. However, even the most widely
adopted ones possess some quite complementary aspects. 
A rough classification may be possible in realizing that the least
modifications, necessary to accommodate confinement in quantum field theory,
seem to be given by the choice of either relaxing the principle of locality
or abandoning the positivity of the representation space.\footnote{As a third
alternative both might eventually turn out to be necessary, of course.} We will
discuss some of the implications of both these possibilities in this chapter.
Since abandoning locality has much further reaching consequences,   
the latter choice, the description of QCD based on local quantum
field theory with indefinite metric spaces, might be the more viable
possibility of the two, if the cluster decomposition property of local
fields can be circumvented.

As was originally suggested for QED by Gupta~\cite{Gup50} and
Bleuler~\cite{Ble50}, the starting point for a covariant description of
gauge theories is an indefinite metric space. In particular, this implies
that, apart from positivity, most other properties of local quantum field
theory, and hereby most importantly the analyticity properties of Green's
functions and amplitudes, remain to be valid in such a formulation. 
In QCD, coloured states are supposed to exist in the indefinite metric space
of asymptotic states. A semidefinite subspace is obtained as the kernel of an
operator. Just as in QED, where the Gupta--Bleuler condition is to enforce 
the Lorentz condition on physical states, this subspace, called the
{\em physical subspace}, has a partition in equivalence classes of states which
differ only by their zero norm components.

A first impression of such a description of confinement can be obtained from
the analogy with QED. Quantising the electromagnetic field in a linear
covariant gauge, besides transverse photons one also obtains longitudinal and
scalar (time-like) photons. The latter two are unobservable because one
eliminates indefinite metric states by requiring the Lorentz condition on all
physical states. The $S$-Matrix of QED scatters physical states into physical
ones only, because it commutes with the Lorentz condition.  The scalar
photons are ``eaten up'' by the longitudinal ones with which they form metric
partners. Colour confinement in QCD can be described by an analogous
mechanism:  No coloured states should be present in the positive definite
space of physical states defined by some suitable condition which has to
commute with the $S$-Matrix of QCD to ensure scattering of physical states
into physical ones.  The dynamical aspect of such a formulation resides in
the cluster decomposition property of local field theory. The proof of which,
absolutely general otherwise \cite{Str76}, does not include the indefinite
metric spaces of covariant gauge theories. In fact, there is quite convincing
evidence for the contrary, namely that the cluster decomposition property
does not hold for coloured correlations of QCD in such a description
\cite{Oji80}. This would thus eliminate the possibility of scattering a
physical state into colour singlet states consisting of widely separated
coloured clusters (the ``behind-the-moon'' problem, see also
Ref.~\cite{Nak90} and references therein). We will return in some more detail
in Sec.~\ref{Sec2.4} to the foundations of this description which is based on
the representations of a particular symmetry of covariant gauge theories found
by Becchi, Rouet and Stora, the BRS symmetry \cite{Bec75}.  

The dynamics of the elementary degrees of freedom of QCD is encoded in its
$n$-point correlation functions, {\it i.e}, the hierarchy of the (time-ordered)
Green's functions $G^{(n)}(x_1, \dots x_n)$. Owing to the axioms of local
quantum field theory, in Minkowski space-time these are defined to be
boundary values (by $\eta_k \to 0$ from below) of analytic functions of $n-1$
complex 4-vectors $z_k^\mu = \xi_k^\mu + i \eta_k^\mu$ which are complex
extensions of the relative coordinates  $\xi_k \equiv x_k - x_{k+1}$ (with
$k=1, \ldots n-1$). The complicated domain of holomorphy of the correlation
functions is then established in several steps, see Refs.~\cite{Fer91,Haa96}
for more details. First, one observes that it contains the {\sl primitive
domain} which is defined by the requirement that the negative imaginary parts
of all $z_k$ lie in the forward cone, $-\eta_k \in V_+$. Then, all ``points''
are included which can be reached from the primitive domain by {\em complex}
Lorentz transformations, {\it i.e}, by extending $SL(2,\CC )$, the double cover
of the proper orthochronous  Lorentz group, to $SL(2,\CC )\times SL(2,\CC )$.
Permutations of the $n-1$ variables $z_k$ and the theory of functions of
several complex variables then lead to the so-called envelope of holomorphy of
{\sl permuted extended tubes}. This connects the primitive domain with the
non-coincident Euclidean region, $\{ (x_1,\ldots x_n) \in \RR^{4n} \, : \; x_k
\not= x_l \; \forall_{k\not= l \in \{ 1, \ldots n \} } \}$. Or {\it vice
versa}, with the Euclidean $SU(2) \times  SU(2)$ symmetry as subgroup of
$SL(2,\CC ) \times SL(2,\CC )$, the domain of holomorphy  allows a complex
extension of the Euclidean space. This apparently technical issue is quite
important to realize, however,  since it justifies the incorporation
of time-like vectors ({\it e.g.}, the total momenta of bound states) as complex
4-vectors in an analytically continued Euclidean formulation.

We will adopt such a Euclidean formulation throughout the following chapters 
with few exceptions which will be mentioned explicitly where they occur.

The first of the alternatives mentioned in the beginning of this section
is based on describing confinement by an absence of coloured states from the
asymptotic state-space altogether. In particular, for a formulation in terms
of some elementary quark and gluon fields this requires a relaxation of
the principle of locality in order to admit singularity structures of their
Green's functions that cannot occur in a local quantum field theory. We will
discuss some consequences of this in Sec.~\ref{Sec2.5}.  

One way to implement confinement in such a description might be provided by
assuming that the elementary correlations are given by entire functions in
momentum space, {\it e.g.}, that no singularities are present at all in any
finite region of the complex $p^2$-plane of the 2-point correlations reflecting
their confined character. While (finite) time-like momenta are readily
incorporated in such a description, singularities for $p^2 \to \infty$ are
indispensable for non-trivial entire functions. Asymptotic freedom, however,
entails that analytic 2-point functions need to vanish in this limit for all
directions of the complex $p^2$-plane~\cite{Oeh80,Oeh95}, see also
Sec.~\ref{Sec2.3} below.  
Perturbation theory, yielding the perturbative logarithms for large $p^2$, thus
seems hard to be reconciled with the idea of entire 2-point functions.  A
singularity structure which generates the perturbative logarithms and complies
at the same time with the analyticity considerations above entails that
singularities occur on the time-like real $p^2$-axis only. The 2-point
correlations of quarks and transverse gluons are then analytic functions in
the cut complex $p^2$-plane. 
While the phenomenologically appealing models employing entire 2-point
functions can therefore be motivated only as approximations for not too large
$|p^2|$, the required analyticity structure is evident in the 
local description of covariant gauge theories based on indefinite 
metric spaces. Confinement of quarks and transverse gluons is hereby
attributed to violations of positivity which should result in indefinite
spectral densities for their respective correlations, see, {\it e.g.},
Ref.~\cite{Oeh95}. It has in fact been argued that such a violation of
positivity can already be inferred from asymptotic freedom in combination
with the unbroken BRS invariance in QCD~\cite{Oeh95,Nis96}. The subtleties 
in this argument which might be regarded as not absolutely conclusive 
are discussed further in Sec.~\ref{Sec2.3}. Independent of this perturbative
argument, however, such violations of positivity are observed in 
the presently available solutions to Dyson--Schwinger equations as well as in
lattice results for the transverse gluon propagator, see Chapter 
\ref{chap_QCD}. Since these results
together seem to provide quite convincing evidence for such violations of
positivity of the elementary correlations of QCD in the covariant
formulation, we will return to this issue repeatedly in the following
chapters.

\subsection{Generating Functional of QED and QCD}
\label{Sec2.1}

The Feynman--Schwinger functional integral representation of the generating
functional for a gauge theory coupled to fermions is in the Euclidean domain 
formally given by,
\begin{eqnarray}
Z[j,\bar \eta, \eta] &=& \int {\mathcal D}[A,q,\bar q]  \, \Delta[A] \,
\delta(f^a(A)) \label{Zj} \\
&&  \hskip -2mm \exp \bigg\{ - \int d^4x \left( \frac{1}{4} F^a_{\mu\nu}
F^a_{\mu\nu} \, + \, \bar q (-{D\kern-.6em \slash} + m) q \right) \, + \,
\int d^4x \, \left( A_\mu^a j_\mu^a \, + \, \bar \eta q \, + \, \bar q \eta \right)
\bigg\} \; .  \nonumber
\end{eqnarray}
Hereby sources $j_\mu^a$ for the gauge fields $A_\mu^a$, and Grassmannian
sources $\bar \eta$ and $\eta$ for the fermion fields $q$ and $\bar q$, have
been introduced. We furthermore employ a positive definite Euclidean metric
$g_{\mu\nu} = \delta_{\mu\nu}$ with hermitian $\gamma$-matrices, $\{\gamma_\mu
, \gamma_\nu\} = 2 \delta_{\mu\nu}$.
In the case of QED, an Abelian gauge theory with gauge coupling $e$, 
the field strengths are given by
\begin{equation}
F_{\mu\nu}  \, = \,  \partial_\mu A_\nu \, - \, \partial_\nu A_\mu
\; , \quad \mbox{and} \; \; D_\mu \,
=\,  \partial_\mu \, + \, ie A_\mu 
\end{equation} 
is the covariant derivative. The other case of interest here is QCD, {\it
i.e.}, the gauge group $SU(3)$.
It is often convenient, however, to consider a variable number of colours
$N_c$, and we present the following discussions mostly in a way applicable to
general $SU(N_c$) gauge groups.  The field strengths are in either case 
given by
\begin{equation}
F^a_{\mu\nu}  \, = \,  \partial_\mu A^a_\nu \, - \, \partial_\nu A^a_\mu
\, - \, g f^{abc} \, A^b_\mu A^c_\nu  \; , \quad \mbox{and} \; \; D_\mu^{ab} \,
=\, \delta^{ab} \partial_\mu \, + \, g f^{abc} A^c_\mu 
\end{equation}
is the covariant derivative in the adjoint representation of $SU(N_c$)  with 
$f^{abc} $  being the corresponding structure constants, and  
$g$ is the coupling constant. 
Denoting the generators of $SU(N_c$) in the fundamental representation
as $t^a$ we can rewrite the covariant derivative: 
\begin{equation}
A_\mu \, = \,  t^a  A^a_\mu \; , \quad \mbox{and}  \;\;   D_\mu \, = \,
 \partial_\mu  \, + \, ig A_\mu  \; , \quad \mbox{with} \;\;  [t^a, t^b] \, =
 \,  i f^{abc} t^c \; .
\end{equation}

The functional integration of the gauge fields over the hypersurface $f^a(A)
= 0 $ involves the measure $\Delta[A]$, called the Faddeev--Popov
determinant~\cite{Fad67}. In linear covariant gauges one uses 
$f^a(A) = \partial_\mu A^a_\mu $. In the case of QED 
the corresponding condition $f(A) = \partial_\mu A_\mu $ leads to a 
field-independent Faddeev--Popov determinant, {\it i.e.}, a pure number, and 
Faddeev--Popov ghosts do not couple to physical fields. The situation is 
different in non-Abelian gauge theories despite the fact that the underlying 
idea is quite similar. To obtain the physical configuration space it is
necessary to divide the set of all gauge potentials by the set of all gauge
transformations including the homotopically non-trivial ones
\cite{Baa92a,Baa92b}. A local gauge fixing condition is 
introduced to select a particular gauge field configuration $A^{U_0}$ by
$f^a(A^{U_0}) = 0 $ from the equivalence class of gauge fields belonging to
the same orbit,
\begin{equation}
[ A^U ]  \, := \, \big\{ A^U = U A U^\dagger + U d U^\dagger \, : \; U(x) \in
SU(N_c) \big\} \; . \label{gorbi}
\end{equation}
This procedure is locally unique if for infinitesimally neighbouring
configurations along the orbit,
$A^U \, = \, A^{U_0} \, + \, \delta
A^\theta $ with  $ \delta A^{a\, \theta}_\mu \, = \,  - D_\mu^{ab}
\delta\theta^b $, one has:
\begin{equation}
    \Delta[A]  \; = \; {\rm Det} \left. \left(\frac{\delta f^a(A^U(x))}{
   \delta\theta^b(y) }\right)\right|_{\theta = 0} \not= 0 \quad .
\end{equation}
In linear covariant gauges the Faddeev--Popov determinant reads explicitly:
\begin{equation}
    \Delta[A]  \; = \; {\rm Det} \left( - \partial_\mu D^{ab}_\mu \right)   \quad .
\end{equation}
Perturbatively this Jacobian factor is taken care of by introducing ghost
fields, {\it i.e.}, scalar Grassmann fields $\bar c^a$ and $c^a$ in the
adjoint representation, such that the Faddeev--Popov determinant is written
as a Gaussian integral of these ghost fields.

The scalar ghost fields belong to the trivial representation 
of $SL(2,\CC )$, the cover of the connected part of
the Lorentz group. As local fields with space-like anti-commutativity,
they violate the spin-statistics theorem and are thus necessarily
unphysical. The domain of holomorphy of the vacuum expectation values of any 
product of local fields, and the positive definiteness of the scalar
product between any two states generated from the vacuum by the polynomial
algebra of those fields, together entail that anti-commutativity is normally 
tied to fields belonging to half-odd integer spin representations of
$SL(2,\CC )$, see, {\it e.g.},~\cite{Haa96}. This is not necessarily so in
the indefinite-metric spaces of covariant gauge theories, however, in which
the scalar product is replaced by an indefinite sesquilinear form. It implies
of course that Faddeev--Popov ghosts are unobservable, see also~\cite{Gre97}. 

As we shall describe in a little more detail in the context of BRS
invariance in Sec.~\ref{Sec2.4}, in the covariant operator formulation of
gauge theories ghosts and antighosts together with 
longitudinal and time-like gluons form quartets of metric partners,
see, {\it e.g.},~\cite{Nak90}. With the exception that ghosts decouple in QED
this is no different from the case of longitudinal and time-like 
photons~\cite{Gup50,Ble50}. In contrast to QED, however, in QCD positivity is
violated for transverse gluon states too. This can be
inferred already in perturbation theory from asymptotic freedom (and unbroken
global gauge invariance) for less than 10 quark flavours~\cite{Oeh90,Oeh94},
see the discussion at the end of Sec.~\ref{Sec2.3}. It implies that the
massless transverse asymptotic gluon states of perturbation theory belong to
unphysical quartets also.\footnote{Together with massless
states in certain ghost-gluon composite operators, see Sec. 4.4.3 in
\cite{Nak90}.}   
This alone is {\em not} sufficient for a realization of confinement, however,
for which it is absolutely crucial that non-perturbatively no such {\em
massless} transverse gluon states exist, regardless of the fact that 
possible asymptotic single particle states in transverse gluon correlations
generally form quartets. We will come back to this point in Sec.~\ref{Sec2.4}.
The results presented in Chap.~\ref{chap_QCD} for the Landau gauge gluon
propagator from both, solutions to Dyson--Schwinger
equations~\cite{Sme97b,Sme98} and lattice simulations~\cite{Lei98,Bon00}
demonstrate the violation of positivity for transverse gluons 
non-perturbatively and,\footnote{Positivity violation was already observed in
the lattice studies of Refs.~\cite{Mar95a,Nak95}.} in addition, agree in
confirming the absence of massless asymptotic transverse single gluon states.

The gauge condition $f^a(A) = 0$, formally represented by a delta functional,
is usually relaxed into $f^a(A) = i\xi B^a$ with a Gaussian distribution of
width $\xi$. In the linear covariant gauges this amounts to the replacement,
\begin{equation}
   \delta(f^a(A)) \, \to \, \exp - \frac{1}{2\xi} \int d^4x  \,
   (\partial_\mu A^a_\mu)^2   \, = \, \int \mathcal{D}B \,  \exp -
  \int d^4x \left(  i B^a\partial_\mu A^a_\mu  + \frac{\xi}{2} B^a B^a  
    \right) \; ,
\end{equation}
which may or may not be represented by a Gaussian integration of the
(Euclidean) Nakanishi--Lautrup auxiliary field $B^a$. The Lorentz condition 
$\partial_\mu A_\mu^a = 0 $ is strictly implemented only in the limit 
$\xi \to 0$ which defines the Landau gauge.

Perturbation theory can then be defined by choosing the Gaussian measure of 
{\sl free} quark, gluon and ghost fields as a starting point for a power 
series expansion of the
non-Gaussian interaction terms in the effective Lagrangian ${\mathcal
L}_{\mbox{\tiny eff}} $ of covariant perturbation theory,
\begin{eqnarray}
Z[j,\bar \eta, \eta , \bar\sigma ,\sigma] & = &\\  
&& \hskip -2cm
\int {\mathcal D}[A,q,\bar q,\bar c,c]  \, \exp \bigg\{
- \int d^4x \,  {\mathcal L}_{\mbox{\tiny eff}} \, + \, \int d^4x \,
\left( A^a j^a \, + \, \bar \eta q \, + \, \bar q \eta 
                 + \, \bar \sigma c + \bar c \sigma \right) \bigg\} \; ,
                 \nonumber
\end{eqnarray}
with \vskip -1.5cm
\begin{eqnarray}
\hskip 1cm
 {\mathcal L}_{\mbox{\tiny eff}} \,  &=&  \,  \frac{1}{2} \, A^a_\mu \big( -
 \partial^2 \delta_{\mu\nu} \, - \, \Big( \frac{1}{\xi} - 1 \Big)
 \partial_\mu  \partial_\nu  \big) A^a_\nu \label{Leff}\\
 && + \, \bar c^a \partial^2 c^a \, + \, g f^{abc} \,  \bar c^a \partial_\mu
 (A_\mu^c c^b ) \, - \, g f^{abc} \, (\partial_\mu A_\nu^a) \, A_\mu^b  A_\nu^c
 \nonumber \\
&& + \, \frac{1}{4} g^2 f^{abe} f^{cde} \, A^a_\mu A^b_\nu A^c_\mu A^d_\nu
     \, + \,   \bar q \big( -{\partial \kern-.5em\slash} + m \big) q \, - \, i g \, \bar q
 \gamma_\mu  t^a q \, A^a_\mu   \; . \nonumber
\end{eqnarray}
Here, sources $\sigma$ and $\bar\sigma$ have been introduced for the
(anti)ghost fields ($\bar c$)$c$, in exactly the same way as the
sources for the quark fields. The sign convention adopted for Grassmann
fields is that derivatives generically denote,
\begin{equation}
\frac{\delta}{\delta (\bar\eta ,\bar\sigma)} \, := \, \mbox{left derivative}
\; , \quad   \frac{\delta}{\delta (\eta ,\sigma)} \, := \, \mbox{right
derivative} \; .
\end{equation}

Despite the  gauge invariance of the generating functional the Green's
functions, obtained as the moments of this functional by taking derivatives
with respect to the sources, are of course gauge dependent.  The  underlying
gauge invariance, however, leads to relations between different Green's
functions: the Ward--Takahashi identities in QED  \cite{War50,Tak57}
and the Slavnov--Taylor identities in QCD \cite{Tay71,Sla72}.
The most convenient device to derive the Slavnov--Taylor identities is to
exploit the Becchi--Rouet--Stora (BRS) symmetry~\cite{Bec75}
of Green's functions \cite{Lle80}. This will be discussed in detail in Section
\ref{sec_BRS}. Here, for completeness,  we give the (on-shell) nilpotent BRS 
transformations for linear covariant gauges: 
\begin{equation}
\begin{array}{cc}
\delta A^a_\mu \, =\,  D^{ab}_\mu c^b \, \lambda  \; , \quad &  \delta q
\, = \,-  i g t^a \, c^a \, q \, \lambda \; , \\
\delta c^a \, = \, - \, \frac{g}{2} f^{abc} \, c^b c^c \, \, \lambda \; ,
\quad  & \delta\bar c^a \, = \, \frac{1}{\xi} \partial_\mu A_\mu^a
\, \lambda \; , \end{array}   \label{BRS_2}
\end{equation}
with the global parameter $\lambda $ belonging to the Grassmann
algebra of the ghost fields. This parameter thus (anti)commutes with
monomials in the fields that contain (odd)even powers of ghost or antighost
fields. Here it commutes with the quark fields since we assumed
commutativity of ghosts with quark fields (without loss of generality, because
of ghost number conservation, either commutativity or anti-commutativity can
be assumed, see Ref.~\cite{Nak90}). Therefore, one can assign $\lambda$ 
the ghost number $N_{\mbox{\tiny FP}} = -1$ reflecting the fact that the BRS
charge has ghost number $N_{\mbox{\tiny FP}} =1$.
The BRS invariance of the total Lagrangian in Eq.~(\ref{Leff}) follows from the
gauge invariance of the classical action and the fact that gauge fixing and
ghost terms can be expressed as a BRS variation themselves ({\it i.e.}, they
are BRS-exact).

We use complex ghost fields with $\bar c \equiv c^\dagger $. 
With this hermiticity assignment the Lagrangian of 
Eq.~(\ref{Leff}) is not strictly hermitian and, furthermore, the BRS
transformation given in Eq.~(\ref{BRS_2}) is not compatible with this
assignment, as discussed, {\it e.g.}, in \cite{Nak90}. To avoid this,
independent (here Euclidean) real Grassmann fields $u, \, v$ should be
introduced by substituting $c \to u$ and $\bar c \to i v$ in Eqs.~(\ref{Leff})
and (\ref{BRS_2}) above.  In Landau gauge ($\xi = 0$) we can make use of the
additional ghost-antighost symmetry, however, to maintain the hermiticity of
the Lagrangian and compatibility with the larger {\em double} BRS symmetry
also for the assignment $\bar c = c^\dagger $.  For $\xi \not=0$, in the more
general covariant gauge, this assignment is possible only at the expense of a
quartic ghost interaction (which vanishes for $\xi \to 0$). 
Since we are mainly interested in the Landau gauge, we can disregard this
subtlety and employ the naive BRS transformations together with the
apparently {\it wrong} hermiticity assignment for the ghost fields in the
derivations of Slavnov--Taylor identities. The results obtained in this way
will be correct as long as we let $\xi \to 0$ eventually in these derivations. 
The explicit connection between independent real and complex ghost fields is
provided by realizing that, in Landau gauge, the ghost number $Q_c$ and the
ghost-antighost symmetry of the real formulation are actually both part of a
larger global $SL(2,\RR)$ symmetry (which can be maintained for $\xi \not= 0 $
by introducing the quartic ghost self-interactions). The connection with the
complex formulation is provided by the Cayley map and $SL(2,\RR)
\simeq SU(1,1)$. Some details of this connection are provided in
App.~\ref{App.BRS}. 

Noether's theorem implies that there is a conserved
anti-commuting charge associated with the BRS symmetry. 
The existence of this non-trivial nilpotent and
hermitian charge is only possible because the state space has indefinite
metric. From Noether's theorem one deduces furthermore that the BRS charge is 
the generator of BRS transformations: The BRS transform of an operator is
given  by the (anti-)commutator of it with the BRS charge. An operator which
is the BRS transform of another operator is called exact. From the nilpotency
of the BRS charge one immediately concludes that the BRS transform of an
exact operator vanishes. Taking furthermore into account that the BRS charge
commutes with the $S$-matrix this has lead to the conjecture that physical
states are the ones which are annihilated by the BRS charge
\cite{Kug79,Nis96}. Furthermore, we note here that $\lambda $ need not
be infinitesimal nor need it be field-independent for (\ref{BRS_2}) to be a
symmetry of the Faddeev--Popov gauge fixed action \cite{Jog00}. However, the
use of a field-dependent BRS transformation is aggravated by the fact that
the functional measure is in general not invariant under this non-local
transformation.  

Within linear covariant gauges the Green's functions also depend on the gauge
parameter $\xi$. Using BRS symmetry one can furthermore derive the Nielsen
identities \cite{Nie75} which control the gauge parameter dependence of 
Green's functions. These identities can be used to prove the gauge independence
of particle poles in the standard model to all orders in perturbation theory,
for recent applications see Ref.\ \cite{Gam99} and the references therein.

Concluding this section we would like to add a remark regarding the
non-perturbative use of the generating functional.  
In perturbation theory, the Gaussian
measure over the free fields can formally be defined as a probability measure
with support over the space of tempered distributions, see 
Refs.~\cite{Gli87,Fer91}. For
ghosts and longitudinal gluons this measure is not positive. The products of
free fields occurring in the interactions, the composite fields, may also be
well defined as tempered distributions. Ambiguities arise for products of these
composite fields at coinciding Euclidean points. This is the origin for the
need of renormalisation. In a renormalisable theory there exists a finite set
of composite fields such that the product of any of them at coinciding points
contains composite fields within the same set multiplied by (formally infinite)
renormalisation constants. This is usually proven at all orders in perturbation
theory. Multiplicative renormalisability beyond perturbation theory has the
status of a conjecture.

Beyond perturbation theory, the only safe way to define the measure in the 
Euclidean generating functional, and thus the Euclidean Green's functions as
its moments, is given
by the continuum limit of the lattice formulation of quantum
field theory. The need for gauge fixing, the presence of long-range
correlations such as the infrared divergences caused by the soft photons in
QED,\footnote{For two recent reviews on the treatment of soft and collinear
infrared divergences see, {\it e.g.}, \cite{Pro99,Bag99}.}
the possibility of infrared slavery in QCD, and triviality
are some obstacles in a proper definition of the generating functional beyond
perturbation theory. Some of these can be taken care of, others are less
understood. In order to proceed, the existence of the generating functional has
to some extend still be postulated for many realistic theories.
This will be also the point of view in the following chapters. 
Before we move on, however, we briefly discuss the incompleteness of the
standard gauge fixing procedure and some related issues in the next section.


\subsection{Gribov Copies, Monopoles and Gauge Fixing}
\label{Sec2.2}

It is a well known problem that the Lorentz gauge condition $\partial A^a = 0$
is not complete~\cite{Gri78}. On compact space-time manifolds it has been 
proven that solutions to local gauge conditions of the form $f(A) =0$ are
generally unable to uniquely specify the connection, {\it i.e.}, the gauge
potentials $A$. The problem is generic and due to the topological structure
of the non-Abelian gauge group~\cite{Sin78}, 
for a pedagogical discussion see Chapter 8 of \cite{Nas91}.

Gribov's observation from the Coulomb, or analogously, from the Lorentz gauge
condition $\partial A = 0$ is intuitively easy to understand. Consider the
set of connections $\Gamma := \big\{ A : \partial A = 0  \big\}$
with the further constraint that all $A$
connected by global transformations $SU(N_c)_{\mbox{\tiny global}}$ have to
be identified in addition.
For sufficiently strong $A(x)$ the Faddeev--Popov
operator $-\partial D(A)$ can be shown not to be positive. Very much like
bound states in quantum mechanics arise for sufficiently strong potentials,
there is a critical $A_c$ for which the lowest eigenvalue $\lambda_0$ of the
Faddeev--Popov operator is zero. A normalisable zero-mode always arises from
the analogue of the bound state wave-function in the limit $A \to A_c $ from
that side for which $\lambda_0 \to 0^-$.  

Field configurations for which such zero modes occur in the
Faddeev--Popov operator $-\partial D(A)$ constitute the Gribov
horizons. In particular, the configurations $A_c$ where this happens
for the lowest eigenvalue are said to lie on the first Gribov horizon
$\partial\mathbf\Omega$, {\it i.e.}, the set of field configurations 
for which the lowest eigenvalue of the Faddeev--Popov operator vanishes. 
It can be shown that any point on $\partial\mathbf\Omega$ has a
finite distance to the origin in field space \cite{Baa92b}. 
Furthermore, in Coulomb gauge on any compact three-manifold every Gribov 
copy obtained by a homotopically non-trivial gauge transformation of the 
trivial gauge field $A=0$ has a vanishing Faddeev--Popov
determinant~\cite{Baa92a}.   
An example of this has already been given in the
appendix of Gribov's original paper \cite{Gri78}.
He considered a pure gauge potential $A^\mu = (0,\vec A^{pg}) = 
(0,-iU^\dagger \vec \nabla U)$ in Coulomb
gauge $\vec \nabla \vec A^{pg} =0$. If the gauge condition was unique, the only
solution should be $\vec A^{pg} =0$. However, choosing an hedgehog
configuration in an SU(2) subgroup parametrised by the generators $\vec \tau$
\begin{equation}
U(\vec r) = \cos \frac {\theta (r) }{2} + i \vec \tau \hat r \sin \frac {\theta
(r) }{2} \; , \quad \hat r = \frac {\vec r }{|\vec r|}   \label{hedg}
\end{equation}
the gauge  condition becomes
\begin{equation}
\frac {d^2\tilde\theta}{dt^2} + \frac {d\tilde\theta}{dt} - 2\sin
\tilde\theta =0,  
\quad {\rm with} 
\quad t=\ln\mu r \quad {\rm and} \quad \tilde\theta(t) = \theta(r). 
\label{dampSinG}
\end{equation} 
This is the classical equation of motion of a damped pendulum corresponding
to the motion of a particle in the potential $V(\tilde\theta) = 2\cos
\tilde\theta $ with friction $d\tilde\theta/dt$ of unit strength. 
The static solutions $d\tilde\theta/dt = 0$ are given by  
$\tilde\theta = \theta = l\pi $ which decomposes into 
two sequences, the even and the odd multiples of $\pi$ for 
$l = 2n $ and $l=2n+1$ with $n \in \ZZ$, respectively.
For all pure gauge field configurations that approach these solutions at
spatial infinity,  
\begin{equation}
\vec A^{pg} \stackrel{r\to\infty}{\longrightarrow}  -i U^\dagger \vec \nabla
U =  -i \exp (-i \frac{\theta}{2} \vec \tau \hat r ) \vec \nabla  \exp
 (i\frac{\theta}{2}\vec \tau \hat r) \, ,
\end{equation}
the Pontryagin index (winding number) is found to be half integer,
\begin{equation}
\nu = \frac{-i}{24\pi ^2} \int d^3x  \, \epsilon_{ijk} \, {\rm tr}
(A^{pg}_iA^{pg}_jA^{pg}_k) = \frac{l}{2} \quad {\rm with} \quad l\in \ZZ \, .
\end{equation}
Note that even though 
\begin{equation}
\vec A^{pg}_{(2n)} = 0 \;, \quad \mbox{and} \quad  \vec A^{pg}_{(2n+1)} =
\Big(\frac{\vec\tau}{2} \times \hat r \Big) \, \frac{2}{r} \;  , \quad
\mbox{at} \; r\not= 0  \label{nvacua}
\end{equation}
for the even $\theta = 2n \pi $ and the odd $\theta = (2n+1)\pi $ 
static solutions, respectively, neither of these need to be entirely
trivial. They can carry winding number concentrated at $\vec r = 0$ 
(for $l\not=0$). For regularised $\theta^{l}_\epsilon(r) = l \pi \,
r/\sqrt{r^2 + \epsilon^2} $ one verifies, 
\begin{equation}
 -\frac{1}{24\pi^2}  \epsilon_{ijk} \mbox{tr} \Big[ (U_\epsilon^\dagger
 \nabla_i U_\epsilon)  
 (U_\epsilon^\dagger \nabla_j U_\epsilon)   (U_\epsilon^\dagger \nabla_k
 U_\epsilon )   \Big]  \,\stackrel{\epsilon\to 0}{\longrightarrow}\,
 \frac{l}{2}  \delta^{3}(\vec x) \; , \quad \mbox{for} \; 
U_\epsilon =  \exp( i \frac{\theta_\epsilon^l(r)}{2} \vec\tau \hat r) \; .
\end{equation}
The even sequence $l=2n$ yields infinitesimal and thus singular $n$-vacua
obtained from the regular ones, $\theta^{2n}_\epsilon(r) = 2\pi n \,
r/\sqrt{r^2 + \epsilon^2} \to  2n\pi $, for $\epsilon\to 0$. These, of
course, correspond to the classification of configurations according to
$\pi_3[SU(N)] = \ZZ$, {\it i.e.}, of configurations with the boundary
condition that $U \to U_0 $ for a unique $U_0\in SU(2)$ in all directions at
spatial infinity.  
The odd sequence, on the other hand, corresponds to configurations for which
two group elements at spatial infinity in opposite directions differ by a
non-trivial central element (here by $-\ID \in Z_2 = \{\pm \ID\}$ for
$SU(2)$) which does not affect adjoint fields such as the gauge
potentials.\footnote{For the pure gauge theory these boundary conditions are
equivalent and combined in $U \to Z_N U_0$.}  
This additional classification into even and odd sequences 
generalises for the $SU(N)$ pure gauge theory according to $\pi_1[SU(N)/Z_N]
= Z_N$ corresponding to fractional topological indices $k/N$ with $k = 0,
\dots N-1$, see~\cite{Zhi87}.

In addition to the even and odd static solutions with $\theta = l\pi$
discussed so far, there are of course also solutions to the equation
(\ref{dampSinG}) for $\tilde\theta(t)$ which start at one of the maxima of
the potential $V(\tilde\theta) = 2\cos \tilde\theta $ 
at $\tilde\theta = 2n\pi$ with infinitesimal velocity for $t\to -\infty$, and
which approach one of the two neighbouring minima at $\tilde\theta = (2n\pm 1)
\pi$ for $t\to \infty$ where they eventually 
come to rest due to the friction term. For
$n=0$ these correspond to everywhere regular pure gauge field configurations,
{\it i.e.}, Gribov copies of the vacuum, with $U(\vec r) \to \pm \ID$ for
$r\to 0$ and $U(\vec r) \to \pm  i \vec \tau \hat r$ for $r\to \infty$, and
with topological index $\nu = \pm 1/2$. The property of these regular
solutions to reach the odd sequence asymptotically at $r\to \infty$
suffices to show that any Wilson line-integral along a curve
$\gamma$ starting from spatial infinity in some direction and leading to
spatial infinity in the opposite direction is $-\ID$, see Ref.~\cite{Zhi87}, 
\begin{equation} 
W_{\mbox{\tiny fd}} (\vec A^{pg}) = P \exp\Big\{ i \int_\gamma   \vec A^{pg}
 d\vec s \Big\}  = - \ID \, , \;\; {\rm for } \; \vec A^{pg} = -i \exp (-i
\frac{\theta(r)}{2} \vec \tau \hat r ) \vec \nabla  \exp 
 (i\frac{\theta(r)}{2}\vec \tau \hat r)  
\label{WiL}
\end{equation}
in the fundamental representation for these configurations. Note that the $n$-vacua
yield $+\ID$ just as the trivial vacuum; and adjoint Wilson lines are $+\ID
$ in either case, of course.

Instantons change the index $\nu$ by one
unit. Regular ones of infinitesimal size $\rho = \epsilon \to 0$, or
equivalently, those with $\rho \to \infty$ in a singular gauge,  
connect the vacua from the even sequence with the adjacent even ones and
those of the odd sequence with adjacent odd ones, {\it i.e.}, they correspond
to transitions between the above vacua of constant angles $\theta = l\pi$
with $l \to l \pm 2$. At zero temperature these instantons 
are discontinuous at a space-like surface passing through their
centres. This is a ramification of the general argument for a necessarily
discontinuous time-evolution of such transitions in the Coulomb
gauge~\cite{Jac78}. In the Hamiltonian description this was related to the
observation that, within such a transition, the configurations pass through the
Gribov horizon at which the Coulomb gauge Hamiltonian fails to generate a
continuous time-evolution~\cite{Gri78,Jac78,Chr80}. 
 
Instantons at finite temperature $T=1/\beta$ are the Harrington--Shepard
calorons~\cite{Har78}. Those calorons that connect the $n$-vacua of
Eq.~(\ref{nvacua}), with $l \to  l \pm  2$, at the opposite ends of the
finite time interval can be obtained from (anti-periodic) gauge
transformations of special types of static, (anti)self-dual
Bogomolnyi--Prasad--Sommerfield (BPS) monopoles, namely the ones with
topological charge $Q = \beta\mu/2\pi = 1$, where $\mu$ is the scale of the
BPS monopole, see~\cite{Ros79,Gro81}.\footnote{The Polyakov loop of these
special BPS monopoles passes through the centre of $SU(2)$ at the position of
the monopole, and it approaches the centre at spatial infinity. They thus
correspond to two axial-gauge monopoles, the ramifications of the Gribov
problem in the axial gauge, one at the position of the BPS monopole and the
other one at infinity with zero total magnetic charge. This is a special case
of the general relation between instantons and axial-gauge monopoles which
was clarified in Refs.~\cite{Rei97,Jah98,For98,For99}. 
Instantons corresponding to two axial-gauge monopoles
at a finite distance of each other, called the non-trivial holonomy instantons
because for these the Polyakov-loop does not approach the centre at spatial
infinity,  were found in Refs.~\cite{Kra98a,Kra98b}. For the relation
between instantons and the magnetic monopoles of general Abelian gauges, see
Ref.~\cite{Jah99}. For gauge fixing and instantons in a field strength
formulation, see Ref.~\cite{Rei93}. A relation to monopoles might be provided
by a field strength formulation in the maximal Abelian gauge~\cite{Qua98}.} 

Studying the limit $T\to 0$ of these calorons with first, at finite $T$, {\it
e.g.}, $\rho\to \infty $ for the singular gauge, one can see the
aforementioned discontinuity arise explicitly. In this limit, they reduce to
sequences of the form,
\begin{equation}
\label{eq:sinst}
     \vec A(\vec x,t) \, = \, \left\{ \begin{array}{ll}
                \vec A^{pg}_{(2n+2)}   \;, \quad &  t > t_0 \\
                \vec A^{pg}_{(2n+1)}   \;, \quad &  t = t_0 \\     
                \vec A^{pg}_{(2n)}   \;, \quad &  t < t_0 
                  \end{array}      \right.  
\end{equation} 
where $t = t_0$ defines the central time-slice of the original
caloron. At the expense of the finite action for an instanton,
the vacua of the odd sequence of constant angles $\theta = (2n+1) \pi $ 
can therefore exist for infinitesimally short times only.                     

Cutting the time interval for finite $\beta = 1/T$ at
$t=t_0$,
one obtains self-dual configurations in each of the two parts which
are separately  gauge transforms of BPS monopoles, now with topological charge
$Q = 1/2$.\footnote{Since for these BPS monopoles the Polyakov loop passes
through the centre only once, at their positions and, in particular, does not
approach a central element at spatial infinity, they each correspond to one
single axial-gauge monopole and thus have non-trivial holonomy.} 
These configurations have half the instanton action and   
connect the even with the odd vacua, with integer and half-odd integer winding
numbers $\nu = l/2$, respectively, corresponding to transitions $l \to l\pm
1$ from one end of their finite time intervals to the other. Quite obviously,
however, for temperatures $T\to 0$ these transitions occur at infinitesimally
early or late times leading to the pure gauge configurations $ \vec
A^{pg}_{(2n)} $ with integer winding number $\nu = 2l= n$ for all finite
times. In the Hamiltonian description, it therefore seems rather questionable
whether such transitions can in the end give rise to a ground-state
wave-functional that is centre symmetric in the sense of the Wilson lines of
Eq.~(\ref{WiL}). 
  
An alternative possibility to connect vacua of the even sequence with
neighbouring odd ones is provided by merons. These are classical solutions of
the $SU(N)$ pure gauge theory which are singular, non-selfdual
and thus of infinite action~\cite{Alf76,Alf78}. 
Explicitly, an Euclidean one-meron solution in
Coulomb gauge, and for an arbitrary SU(2) subgroup, see the review in
Ref.~\cite{Act79}, is given by    
\begin{equation} 
 \vec A^{\mbox{\tiny mn}}\!(\vec r,t)  = \Big(\frac{\vec\tau}{2} \times \hat
 r\Big)  \,  \frac{1}{r}  \, \left(1 - \frac{t}{\sqrt{t^2+r^2}}\right) \; .
\end{equation}
Clearly, for $t\to \infty$ (and $r\not= 0$) the meron configuration
vanishes. More carefully, including the singularity at $r=0$, one finds that
it approaches a vacuum of the even sequence, corresponding to the $\epsilon
\to 0$ limit of $\theta^{2n}_\epsilon (r) \to 2n\pi $.
For $t\to -\infty$ one has $\vec A^{\mbox{\tiny mn}} \to \vec A^{pg} =
(\vec\tau \times \hat r)/r$ corresponding to a $\theta_\epsilon^{2n\!+\!1}(r)
\to (2n+1) \pi $ configuration of the odd sequence.\footnote{Its connection
to monopoles is quite obvious,  at $t=0$  the meron passes through a
chromomagnetic monopole with $B_i^a = - r_i r_a /r^4 $ and $E_i^a  =0$ which
was first reported as a static solution to the classical Yang-Mills equations
in Ref.~\cite{Wu68}.}
As might intuitively seem reasonable for configurations that
connect the even with the odd vacua, the gauge potentials of single-meron
configurations are exactly half the gauge potentials of the instantons in the
special limit discussed above, {\it i.e.}, of those with $\rho = \epsilon \to
0$ or $\rho \to \infty $ in the regular or the singular gauge,
respectively. Or, {\it vice versa}, these instantons can in fact be viewed as
the special case of analytically known two-meron solutions for vanishing
distance, see Refs.~\cite{Alf78,Act79}.\footnote{The hypothesis that
exact solutions for two merons at a finite distance~\cite{Alf76}
might provide a more general connection between instantons and monopoles, in
particular also for Coulomb and Landau gauges, is a long-standing
conjecture, see Refs.~\cite{Cal77,Cal78}.} Though single-meron configurations 
are neither self-dual nor have finite action, in contrast to the $Q=1/2$ BPS
monopoles discussed above, they do connect vacua with integer
and half-odd integer winding number without discontinuity in time. At finite
separations in the time-direction 
two-meron configurations might therefore, at least in
principle, be employed to populate the half-odd winding number configurations
$\vec A^{pg}_{(2n+1)}$ for finite time periods. It was furthermore argued
that the logarithmically diverging action of meron pairs, which can made
explicit in a regularisation in terms of so-called ``instanton
caps'', might be compensated in the free energy by their contributions to
the entropy~\cite{Cal77,Cal78}. This observation has induced some renewed
interest in the role of merons with respect to confinement, {\it e.g.}, for a
recent lattice study of meron pairs, see Ref.~\cite{Ste00}. 

To summarise, while a simple picture of confinement, {\it e.g.}, 
as intuitive as the dual Meissner effect by a condensation of
magnetic monopoles proposed for the maximal Abelian gauge \cite{tHo75,Man76}, 
is not yet 
available for Coulomb or Landau gauge, relations between the various 
types of monopoles in the various gauges, gauge singularities and
the role of topologically non-trivial gauge copies are increasingly well
understood. 
Some analogies of these issues can be found also for the Coulomb or Landau
gauge along the directions mentioned above. Whether these subtleties will
be relevant for a description of confinement or not, whether any kind
of semi-classical analysis might in the end be swamped by genuine quantum
effects, the mere existence and the possible couplings of the fractional
$n$-vacua seems to show that the capacity exists, at least in principle, to
introduce a sufficient disorder which might eventually be all that is needed
in the pure gauge theory  to lead to an area law for large Wilson loops
also in the Coulomb or the Landau gauge.

For reasons that will become clear below, consider now the square norm of
the various pure gauge configurations obtained from Eq.~(\ref{hedg}). It is
straightforward to see that 
\begin{equation} 
  \Arrowvert {A^{pg}} \Arrowvert^2  \, = \,     
     \int d^3x \, \mbox{tr} A_i^{pg} A_i^{pg} \, = \, 2\pi \int_0^\infty r^2 dr
\, \left( (\theta'(r))^2 + \frac{4}{r^2} (1-\cos\theta(r)) \right) 
\label{sqnorm}
\end{equation}
which vanishes for $\vec A =0$ and for all static $\theta = 2n\pi$ copies
thereof.\footnote{The fact that these are singular at $\vec r=0$ is irrelevant 
here. One readily verifies that the norm of the regular $n$-vacuum
configurations $\theta_\epsilon(r) = 2\pi n r/\sqrt{r^2 + \epsilon^2}$ is of
order $\epsilon$ and thus vanishes for $\epsilon\to 0$.}   
From Eq.~(\ref{sqnorm}) it is clear that $\theta(r) \to 2n\pi $ for $r\to
\infty$, corresponding to the maxima of the potential $V(\tilde\theta)$ 
in order for the square norm of such a pure gauge configuration to be
finite. This is the case only for those configurations that approach the
integer $n$-vacua $\vec A^{pg}_{(2n)}$ at large $r$. In particular, it is not
the case for the regular Gribov copies discussed in the paragraph above
Eq.~(\ref{WiL}) nor for regularised vacua with half-odd winding numbers
obtained from $\theta_\epsilon(r) = \pi (2n+1) r/\sqrt{r^2 + \epsilon^2}$. 
In addition, the first term in the norm integral, with
\begin{equation} 
     \int_0^\infty r^2 dr  \, (\theta'(r))^2 \, =\, \int_{-\infty}^\infty dt'
      \, \left(\frac{d\tilde\theta(t')}{dt'}\right)^2   \; ,
\end{equation}
then corresponds to the total energy dissipated (for $t\to \infty$) by the
particle $\tilde\theta(t) $ moving in the potential $V(\tilde\theta) =
2\cos\tilde\theta $.  This implies that, due to the friction
$d\tilde\theta(t)/dt $ in its equation of motion~(\ref{dampSinG}), any
particle that comes to rest at one of the maxima $\tilde\theta =  2\pi l$ for
$t\to\infty$ must at $t\to-\infty$ initially have come from (positive or
negative) infinity with infinite initial energy. Therefore, the
dissipated energy of this particle is also infinite except for
$\tilde\theta(t) \equiv 2n\pi  $ at all $t$. The only configurations with
finite square norm are thus the singular $n$-vacua of the even sequence with
integer winding number, for which $ \Arrowvert {A^{pg}} \Arrowvert^2 \, =\,
0$ degenerate with the trivial configuration $\vec A = 0$. The even sequence
$\theta = 2n\pi $ with $\nu = n\in\ZZ$ provides a set of degenerate absolute
minima of the square norm.

\begin{figure}[t]
  \centering\epsfig{file=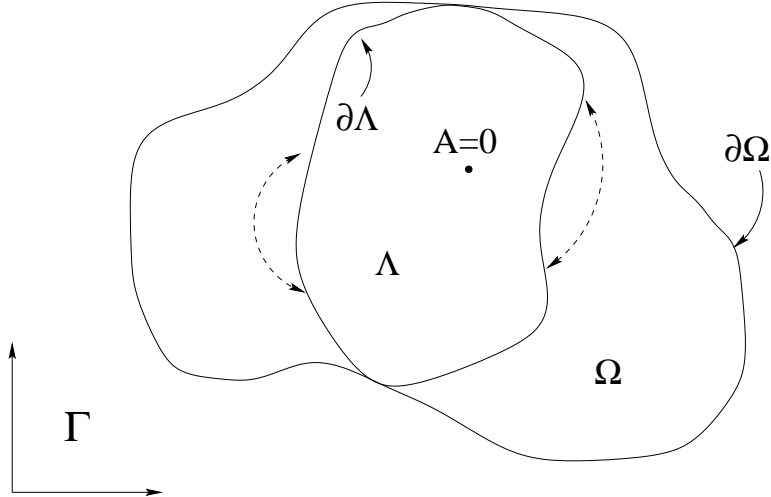,width=.6\linewidth}
  \caption[Sketch of the gauge-fixing hyperplane $\Gamma$, the Gribov and 
           the fundamental modular region.]
    {Sketch of the hypersurface $\Gamma= \big\{ A : \partial A = 0  \big\}$, 
    the Gribov and the fundamental
  modular region, $\Omega$ and $\Lambda$, respectively. The necessity of
  identifications on the boundary of $\Lambda$ is indicated by dashed arrows.}
   \label{fig:Gribov}
\end{figure}

Generally, the Gribov region $\mathbf\Omega$ is defined as the set of
connections within the first Gribov horizon, {\it i.e.}, as the set of gauge
potentials for which the Faddeev--Popov operator is positive. In the linear
covariant gauges it explicitly reads 
$\mathbf\Omega := \big\{ A : \partial A = 0 \; ,-\partial D(A) \ge 0 \big\}$.
This convex Gribov region $\mathbf\Omega $ is determined by 
the set of the {\em local} minima of the functional
\begin{equation}
E_A[U]  \, \equiv \, \Arrowvert A^U \Arrowvert^2  \, := \,
\frac{1}{2} \, \int d^4x \;  A^{a\, U}_\mu (x)  \, A^{a\, U }_\mu (x)
\;  \label{gf_func}
\end{equation}
for the equivalence classes of gauge fields $[A^U]$ given in 
Eq.\ (\ref{gorbi}). 
There are in general many local minima, and the Gribov region still contains
gauge copies. As a further restriction, the fundamental modular region
$\Lambda $ is defined as the set of absolute minima of the
functional~(\ref{gf_func}).  Each orbit~(\ref{gorbi}) intersects $\Lambda$
exactly once~\cite{Del91}. 
The fundamental modular region is contained within the Gribov region. 
In the interior of $\Lambda$ the absolute
minima are non-degenerate. Degenerate minima exist, however,
on the boundary $\partial\Lambda$. These minima have to be 
identified~\cite{Baa92a,Baa92b}. The
Gribov horizon, {\it i.e.}, the boundary of the Gribov region
$\mathbf\Omega$, touches $\partial\Lambda$ at the so-called singular
boundary points. The situation is sketched in Fig.~\ref{fig:Gribov}.
Note that the relevant configuration space is $\Lambda /
SU(N_c)$, since the origin of the fundamental modular region, $A=0$, is
invariant under global gauge transformations \cite{Baa92a,Baa92b}.

In a continuum formulation of QCD
it seems unlikely that a systematic elimination of gauge
copies is possible at all.\footnote{For completeness we mention that 
using Stochastic Quantisation there is no need for a gauge fixing term and 
the Gribov problem is thus avoided, see Ref.\ \cite{Riv87} for a pedagogical
treatment of this topic. A related continuum formulation \cite{Bau00a,Bau00b}
considers QCD from a five-dimensional point of view, the fifth dimension
playing the role of the ``stochastic  time''. This leads to parabolic equation
for the propagators of the various ghost fields in the five-dimensional bulk
and thus yields a trivial Faddeev--Popov determinant.
There are also recent numerical investigations on the lattice
based on Stochastic Quantisation, {\it e.g.}, see
Refs.~\cite{Miz94,Ren99,Sho00}.}  
Their presence may or may not be a serious problem.
On the other hand, there has been recently some progress treating the Gribov 
problem in lattice calculations.
The lattice analogue of restricting to the absolute minima of $E_A $ is called
minimal Landau gauge~\cite{Zwa92,Zwa94}. Various algorithms are used in gauge
fixed lattice calculations to minimise this functional, {\it e.g.}, in
Refs.~\cite{Ber94,Mar95a,Nak95,Sum96}. Methods to find the absolute minima
and the influence of Gribov copies are assessed in Ref.~\cite{Cuc97}.
Therefore, a solution of the Gribov problem might in principle be feasible on
the lattice. However, the question of existence and uniqueness of the
continuum limit for corresponding quantities still remains an open question.

On compact manifolds, gauge fixing without the necessity of 
elimination of Gribov copies can
be formulated systematically in terms of a (Witten type) topological quantum
field theory on the gauge group $\mathcal{G}$ (see Refs.~\cite{Bir91,Nas91}).
In such a formulation, the standard Faddeev--Popov procedure of inserting unity
into the unfixed generating functional which generalises 
to a weighted average over all $U$ with $A^U \in \Gamma = \{ A : \partial A =
0\}$ in presence of Gribov copies~\cite{Fuj79,Hir79} or, equivalently, the
perturbative BRS quantisation, essentially correspond to constructing a
topological quantum field theory whose partition function computes the
generalised Euler characteristic $\chi (\mathcal{G})$ of the gauge
group~\cite{Bau98}.  
This can vanish, however, just as the Witten index vanishes in theories with
spontaneously broken supersymmetry.   
For the $SU(2)$ lattice gauge theory the vanishing of the Euler character 
follows quite trivially from $\chi(\otimes_{\mbox{\tiny sites}} SU(2) ) =
\chi(S^3)^{\#\mbox{\tiny sites}} = 0$, see also Ref.~\cite{Sha84}. 
In the continuum this remains to be the case due to the global gauge
transformations which provide one vanishing factor $\chi(S^3)$ that survives
the continuum limit.  
One way to cure this problem is to remove the global ghost zero modes by 
constructing a topological quantum field theory that computes the Euler
character of the coset space $\otimes _{x} SU(N_c)/SU(N_c)_{\mbox{\tiny
global}}$ which was shown not to vanish for $SU(2)$ in
Ref.~\cite{Bau98}. Within the framework of BRS quantisation this procedure 
has been worked out for QCD in the covariant gauge on the 4-torus in
Refs.~\cite{Bau96,Sch98}.\footnote{Among compact space-time manifolds, 
the special choice of the torus was adopted for simplicity,
to avoid global topological obstructions. The infinite volume limit is
believed to be independent of this particular choice, of course.}  

An alternative way to avoid a vanishing Euler character 
proposed in Ref.~\cite{Sch98b} 
is to fix the $SU(N)$ gauge symmetry only partially to the maximal Abelian
subgroup $U(1)^{N-1}$. For the $SU(2)$ lattice gauge theory the BRS
construction to compute $\chi(\otimes_{\mbox{\tiny sites}} SU(2)/U(1) ) =
\chi(S^2)^{\#\mbox{\tiny sites}} = 2^{\#\mbox{\tiny sites}} $ can then be 
used to obtain a reduced $U(1)$ lattice gauge theory~\cite{Sch98b}.  
Within covariant Abelian gauges in the continuum this kind of BRS
quantisation by ghost-antighost condensation can lead 
to mass generation for off-diagonal gauge bosons and
thus to finite propagators at all Euclidean momenta except for the
the diagonal ``Abelian'' gauge boson which remains massless
\cite{Sch99,Sch00}.  
The most important difference between this scheme and the standard
Faddeev--Popov gauge fixing adopted for the maximal Abelian gauge, {\it e.g.},
in Ref.~\cite{Qua98}, is that the maximal Abelian gauge condition is not
implemented exactly in the former but softened by a Gaussian weight of width
$\xi $ analogous to the Lorentz condition in the linear covariant gauge. 
This leads to the occurrence of
quartic ghost self-interactions (which formally vanish for $\xi = 0$) 
together with a global $SL(2,\RR)$ symmetry in the BRS construction of
Refs.~\cite{Sch99,Sch00}. The global $SL(2,\RR)$ can dynamically break down
to the usual ghost number symmetry in the phase with ghost-anti\-ghost
condensation with the condensate as the order parameter. Besides being
responsible for the ghost-antighost condensation and mass generation in some
analogy to the BCS theory of superconductivity (with Higgs mechanism for the
plasmon excitation), technically, the quartic ghost-selfinteractions
eliminate the global gauge zero modes due to the constant (in this case the 
off-diagonal) ghosts. In this description, screening masses for the
off-diagonal gauge bosons thus emerge naturally which persist in the high
temperature phase~\cite{Sch99}. This last conclusion is due to the relation
of the ghost condensate with the scale anomaly which at the same time 
seems to show that it cannot provide an order parameter for the chiral symmetry
breaking and/or confinement transition. It rather suggests that the
global $SL(2,\RR)$ is broken in both, the high {\em and} the low
temperature phases.  In order to understand the possible origin of
confinement in such a formulation, which in the usual BRS framework 
is related to the realization of the global gauge symmetry on unphysical
states, an application to the Higgs mechanism in the $SU(2)\times
U(1)$ electroweak interactions would seem to be a natural next
step.\footnote{In particular, the question might arise why the massive gauge
boson belongs to an unphysical quartet (see below) in one case while it
definitely is a BRS singlet in any of the known Higgs models.}    

It might be interesting for our present purposes, however, to note that 
a global $SL(2,\RR)$ symmetry containing ghost number
and ghost-anti\-ghost symmetry emerges also in the Landau gauge, {\it i.e.},
in the special
case of the linear covariant gauge with $\xi=0$ in which the Lorentz
condition is implemented ``exactly''. 
Maintaining this symmetry in covariant gauges for gauge parameters
$\xi \not=0$ leads to the (massless) Curci--Ferrari gauges discussed in
Appendix~\ref{App.BRS}. The significance of this symmetry seems not entirely
clear at present. The differences between these Curci--Ferrari gauges
and the standard linear covariant gauge seems, however, quite analogous to the
situation in the maximal Abelian gauge discussed above.

While the presence of the quartic ghost self-interactions might at first 
not seem to be a very appealing feature of the Curci--Ferrari 
generalisation of the Landau gauge apart from maintaining its special
symmetry, they do have one possibly quite interesting effect: 
The  quartic ghost-selfinteractions could be effective to eliminate all
constant ghost and antighost zero modes for the gauge group $SU(3)$ of
QCD.\footnote{For $SU(3)$ there are 16 constant (anti)ghost modes, and the
expansion of the exponential of their quartic interaction to fourth order
yields exactly one term that contains each of them exactly
once, provided the prefactor of this term does not vanish.
The same was not possible for $SU(2)$ with 6 constant modes, since no
6-(anti)ghost term arises in this expansion in the first place.} 
This might therefore provide
for a BRS formulation by a topological quantum field theory with non-vanishing
partition function without need to eliminate the global gauge invariance.  
As we shall discuss in Section~\ref{Sec2.4} the realization of the global gauge
symmetry is of particular importance in the BRS formulation of the 
linear covariant gauge. The Kugo--Ojima criterion is based on
the necessity of this global symmetry to be unbroken for a realization of
confinement. Its breaking, on the other hand, leads via the {\em converse of
the 
Higgs mechanism} to massive physical (BRS singlet) states in transverse gauge
boson channels. In light of this, a formulation that allows both these
possibilities by leaving the global gauge invariance untouched clearly seems
desirable.

Among the $SL(2,\RR)$-symmetric covariant gauges Landau gauge is special in
that the quartic ghost interactions disappear for $\xi = 0$ with the effect 
that constant ghost zero modes arise. These are certainly problematic for a
proper formulation of the gauge fixed theory at a finite volume as discussed
above. It is not inconceivable, however, that it might
suffice to deform the Landau gauge just slightly into an $SL(2,\RR)$-symmetric
covariant gauge without such zero modes at large but finite volume without
modifying the naive Landau gauge results presented in later chapters of this
review in the infinite volume limit (in which $\xi\to 0$ might be retained).
This is certainly a quite optimistic assessment of the current situation
about gauge fixing in presence of Gribov copies, and considerable further
studies will be necessary to clarify this issue. 
Apart from some evidence in favour of the naive procedure, by comparing 
the results from Dyson--Schwinger equations to the conjectures of
Gribov~\cite{Gri78} and Zwanziger~\cite{Zwa92} and to lattice results, we
will not have much more to say about this problem in the following chapters.  
 
\goodbreak

\subsection{Positivity versus Colour Antiscreening}
\label{Sec2.3}

In this section we briefly review and discuss quite a long-known
contradiction between asymptotic freedom, implying antiscreening of the colour
charge in the sense of K\"all\'en, and the positivity of the spectral density
for gluons in the covariant gauge~\cite{Oeh80}. While the apparent paradox
was argued to be resolved  for the (space-like) axial gauge by
West~\cite{Wes83}, as will be discussed in Sec.~\ref{sub_Ax3}, present
knowledge of the axial gauge suggests that this resolution is itself likely
to be an artifact of a violation of positivity introduced by the axial-gauge
singularity of the gluon propagator.\footnote{In fact, recent studies of
the axial gauge and, in particular, of the singularities in the corresponding
tree-level gluon propagator, start from linear covariant gauges in defining
the axial gauge, see Ref.~\cite{Jog00} and the references therein.} 
The root of the contradiction thus seems to be more generic and
not special to the covariant gauge. As will become clear in subsequent
chapters, combined evidence from different non-perturbative calculations
indicates quite convincingly that gluonic correlations do indeed violate
positivity. As mentioned in the beginning of this chapter, and discussed in
more detail in the next section, this can be interpreted as a manifestation of
confinement. In the present section, we describe the original argument that 
this might be inferred already from asymptotic freedom and BRS invariance 
\cite{Oeh90,Nis94,Nis96}.

To understand the origin of the problem some basic properties of interacting
fields are briefly recalled. For the moment a one-to-one correspondence
between basic fields and stable particles is assumed which is of course not
the case in QCD. Assuming field-particle duality and asymptotic
completeness, the Lehmann--Symanzik--Zimmermann asymptotic condition 
for $t\to -\infty$ on an interacting field $\Phi(\varphi ;t)$ states
that it converges weakly on a dense domain ${\mathcal D}$ to the creation
operator $a^\dagger_{in}(\varphi)$ of a single particle state with wave
function $\varphi(x) $:
\begin{equation}
 \langle \alpha | \Phi(\varphi ;t) |\beta \rangle \, \stackrel{t \to -
 \infty}{\longrightarrow}  \,  Z^{1/2} \,  \langle \alpha |
 a^\dagger_{in}(\varphi)  |\beta \rangle  \; ,
\label{LSZ_as}
\end{equation}
{\it i.e.}, the matrix elements of all states $ |\alpha \rangle ,\;
|\beta\rangle $ in ${\mathcal D}$ converge to those of the asymptotic field
involving a normalisation constant $Z$ for the overlap with the corresponding
single particle state of mass $m_\varphi$. This constant 
appears, of course, in the Lehmann representation of the
propagator of the $\Phi$-field ($m > m_\varphi$),\footnote{The difficulties
encountered in presence of massless particles are ignored here. 
Losing the correspondence between single particle states and the
discrete eigenvalues of the mass operator one has to account for the infrared
divergences caused by, {\it e.g.}, the soft photons in QED. These can be dealt
with by introducing coherent states. Thus, this complication is
of no further significance for the arguments sketched in the following.}
\begin{equation}
D_\Phi(k)  \, = \, \frac{Z}{k^2 + m_{\varphi}^2} \, + \, \int_{m^2}^\infty
d\kappa^2  \, \frac{\tilde\rho(\kappa^2)}{k^2 + \kappa^2}  \; .
\label{KL_rep}
\end{equation}
The single particle contribution is explicitly separated here,  {\it i.e.},
a full spectral function can be defined as $\rho(k^2) \, := \, Z \,
\delta(k^2 -m_\varphi^2) \, + \tilde \rho(k^2)$. If one insisted on
equal-time commutation relations for the interacting fields ({\it e.g.}, as
in \cite{Itz80}), the following spectral sum rule would be obtained,
\begin{equation}
1 \, = \, Z \, +\, \int_{m^2}^\infty d\kappa^2 \,  \tilde\rho(\kappa^2) \; ,
\label{naive_sr}
\end{equation}
which would thus imply $ 1 \ge Z \ge 0$ for positive $\tilde\rho$.
In general however, the second term on the r.h.s.\ of Eq.~(\ref{naive_sr}) 
is a divergent quantity associated with the field renormalisation necessary
in a renormalisable theory. This reflects the fact that, in contrast to free
fields which are as operator valued distributions defined at fixed times, the
interacting fields in a renormalisable theory are more singular objects. In
particular, smearing over both the space {\sl and} time variables is
necessary in their definition. Their equal time commutation relations are no
longer well-defined. Heuristically, they involve the divergent field
renormalisation constants. In the case of the gluon field being primarily
under consideration here this constant is usually called $Z_3$, and the 
resulting spectral sum rule reads,
\begin{equation} \label{true_sr}
Z_3^{-1} \, = \, Z \, +\, \int_{m^2}^\infty d\kappa^2 \,
\tilde\rho(\kappa^2) \; .
\end{equation}
Turning the above argument around, the necessity of renormalisation can be
understood as follows: If the canonical equal-time commutation relations of
the free theory ($g = 0$) are retained in the
interacting theory ($g\not= 0$), the constant $Z $ has to acquire a
divergence so as to cancel the one on the r.h.s.\ of
Eq.~(\ref{naive_sr}). This would imply that the asymptotic condition,
Eq.~(\ref{LSZ_as}), is lost. The representation of the interacting fields and
the Fock space representation of the free canonical fields are inequivalent
which is referred to as Haag's theorem. It is an example of the general
representation problem in quantum field theory, the existence of inequivalent
representations of the canonical commutation relations being the rule rather
than the exception. Of course, in constructive field theory the equal-time
canonical commutation relations are replaced by space-like commutativity as
the more general implementation of locality for interacting fields, see
Haag's book for a thorough presentation and an account of the mathematical
background~\cite{Haa96}.

The renormalised version of the spectral sum rule for the interacting theory
given in Eq.~(\ref{true_sr}) is in conflict with positivity of the spectral
density of the (transverse) gluon propagator in Landau gauge QCD
as was first observed in Ref.\ \cite{Oeh80}. 

To see this, we note that Eq.~(\ref{true_sr}) implies 
$Z_3  \le Z^{-1}$
for a positive spectral function, $\tilde\rho(\kappa^2) \ge 0$. 
Near the renormalisation group fixed point of vanishing $g^2$, 
however, one has in linear covariant
gauges, 
\begin{equation}
Z_3 \, = \,   \left( \frac{g^2}{g_0^2} \,
\right)^\gamma  \, =\, \frac{\xi_0}{\xi }
\end{equation}
with the renormalised gauge parameter $\xi $, the bare one $\xi_0$, and $\gamma
$ being the leading coefficient of the anomalous dimension of the gauge
field. The second equality is meaningful only, of course, if one is not
considering the Landau gauge $\xi = \xi_0 = 0$. The first equality, however,
holds for general covariant gauges including $\xi = 0$.

In QED the spectral density $\rho(k^2) $ of the photon propagator in the
covariant gauge is identical to its axial gauge counterpart $\rho_g(k^2)$,
{\it c.f.},  Sec.~\ref{sub_Ax2}, which is a consequence of the gauge
invariance of the Coulomb potential.  In QED one furthermore has $\gamma =
1$, and from the gauge invariance of $\rho$ it was argued that for the bare
gauge parameter only the choices $\xi_0 = 0 $ and $\xi_0 = \infty$ exist,
implying the possible values of the bare coupling to be $e_0^2 \in \{ \infty,
0\}$~\cite{Nis96b}. The second choice corresponding to asymptotic freedom, 
in QED one might thus expect that the bare coupling diverges, {\it i.e.},
that the running coupling behaves as $\bar e^2(\mu) \to 
\infty $ for $\mu \to \infty$ (beyond one-loop). Here, a possible problem
might rather be triviality of QED in the absence of an ultraviolet fixed
point (for a pedagogical discussion see, {\it e.g.}, Huang's text book on 
quantum field theory \cite{Hua98}).
In fact, recent evidence in favour of triviality is obtained for
instance in the lattice simulation of Ref.~\cite{Goe98}. Without a further
fixed point, it follows that $Z_3 \to 0 $. Due to the infrared fixed point
at $e^2 = 0$, however, it is sufficient to note that the positive
$\beta$-function near this fixed point generally implies a renormalised
charge which is smaller than the bare charge. This is referred to as K\"all\'en
screening. Therefore, in QED one has $Z_3 < 1 \le Z^{-1}$ as one should. 

In contrast, asymptotic freedom corresponds to the scaling limit $g_0 \to
0$. Therefore, for $\gamma > 0$ one has $Z_3 \to \infty$, and from
Eq.~(\ref{true_sr}) one thus concludes that the spectral density cannot be
positive. In perturbative QCD in Landau gauge one has $ 1 > \gamma > 0$
for $N_f < 10 $ quark flavours. Then, $Z_3^{-1} \to 0$ leads to the
Oehme--Zimmermann superconvergence relation~\cite{Oeh80}. Its generalisation
to the complete class of linear covariant gauges leads to the following form
of the spectral sum rule for the transverse gluon
propagator~\cite{Oeh90,Oeh94,Nis94},  
\begin{equation} \label{OZ}
Z \, +\, \int_{m^2}^\infty d\kappa^2 \,  \tilde\rho(\kappa^2) \, = \, \left\{
{ 0 \; , \quad \; \; \mbox{for} \; \xi \le 0  \atop \xi/\xi_0 \; , \quad
\mbox{for} \; \xi > 0  } \right. \; .
\end{equation}
Positivity of the gluon spectral density in Landau gauge is thus apparently in
contradiction with anti\-screening. As a result of this, it was concluded that
positivity for gauge-boson fields is indeed violated in gauge theories
with $Z^{-1}_3 \to 0$~\cite{Nis94}.

This has been interpreted as a manifestation of confinement from asymptotic
freedom and unbroken BRS invariance, since the existence of a semi-definite
{\sl physical} space of transverse gluon states obtained after projecting out
longitudinal gluon and ghost degrees of freedom would imply that
$\rho(\kappa^2) \ge 0$ \cite{Oeh90,Oeh95,Nis96}. The significance of the
global BRS charge structure here, to describe confinement in QCD on one hand
versus the Higgs mechanism in the standard model of electroweak interactions
on the other, will be discussed in the next section. 

In order to look into the origin of the superconvergence relation, we 
recall that the renormalised gluon propagator in linear covariant gauges
(explicitly including the dependence on the renormalisation scale $\mu$ for
the moment) has the general structure, 
\begin{equation}
  D_{\mu\nu}(k,\mu) \, =\,  \left(\delta_{\mu\nu} - \frac{k_\mu k_\nu}{k^2}
  \right) \,   \frac{Z(k^2,\mu^2)}{k^2} \, + \, \xi \, \frac{k_\mu
  k_\nu}{k^4}  \;  . \label{glp_cg}
\end{equation}
The subtlety of the argument can be made a little more explicit by considering
the gluon renormalisation function $Z(k^2,\mu^2)$ which depends on the
invariant momentum $k^2$, the scale $\mu$ and the gauge parameter
$\xi$. In a perturbative momentum subtraction scheme in Landau gauge ({\it
i.e.}, $\xi= 0$) one obtains for sufficiently large $\mu$ its leading
logarithmic behaviour to be (also compare Sec.~\ref{sub_Sub}),
\begin{equation} \label{sc_large_k}
  Z(k^2,\mu^2) \, = \,  \biggl( \frac{\bar g^2(t_k, g)}{g^2} \biggr)^\gamma
  \, = \, Z_3^{-1}(\mu^2,k^2) \, \to \, 0 \;, \quad
  \mbox{for} \; k^2 \to \infty \; ,
\end{equation}
with a positive anomalous dimension $\gamma $ (for $N_f < 10 $). Here,
$\bar g^2(t_k, g)$ is the one-loop running coupling with $t_k =
\frac{1}{2}\ln(k^2/\mu^2)$ and $\bar g^2(0,g) = g^2$. $Z_3(\mu^2,{\mu'}^2)$
is the multiplicative constant of finite renormalisation group
transformations. Depending on the details of the regularisation scheme it is
related to the gluon field renormalisation constant essentially by
$Z_3(\mu^2,\Lambda^2) \to Z_3$ for $\Lambda \to \infty$. The spectral
representation of the gluon propagator, on the other hand, leads to
\begin{equation}
  Z(k^2,\mu^2) \, = \,   \int_0^\infty \, dm^2 \, \frac{k^2}{k^2 + m^2} \,
  \rho(m^2,\mu^2,g)  \, 
      \; , \label{spec_Z}
\end{equation}
where the dependence of the spectral function $\rho$ on $\mu^2$ and $g$ was
made explicit again (the pair $(g,\mu)$ really represents only one parameter,
of course). The renormalisation condition of the momentum subtraction scheme
fixes the gluon propagator to the tree-level one at a sufficiently large
space-like renormalisation point $k^2 = \mu^2$, 
\begin{eqnarray} \label{sc_large_mu}
  Z(\mu^2,\mu^2) \, &=& \, 1 \nonumber \\
                    &=& \,  \int_0^\infty \, dm^2 \,
  \frac{\mu^2}{\mu^2 + m^2} \,
  \rho(m^2,\mu^2,g)  \, \to \,  \int_0^\infty \, dm^2 \, \rho(m^2)  \;, \quad
  {\rm for} \quad \mu^2 \to \infty .
 \end{eqnarray}
Comparing Eq.~(\ref{sc_large_mu}) to the limit in~(\ref{sc_large_k}) one thus
realizes that
\begin{equation}
  \int_0^\infty \, dm^2 \, \rho(m^2,\infty,0)  \, = \, 1 \; , \label{mu_lim} 
\end{equation}
as expected for the free theory, whereas from Eq.~(\ref{spec_Z}) for $k^2\to
\infty$ one obtains,
\begin{equation}
  \int_0^\infty \, dm^2 \, \rho(m^2,\mu^2 <\infty, g > 0)  \, = \, 0 \;
  .\label{k_lim}
\end{equation}
Note that this last limit results by choosing a strictly finite
renormalisation point $\mu^2$ and employing the limit $k^2/\mu^2 \to
\infty$. This thus demonstrates explicitly that it is not possible to
renormalise the interacting theory at a strictly finite scale, and a small
but finite coupling $g$, to the free theory (corresponding to $g\equiv
0$). The superconvergence relation might therefore be interpreted as a
reincarnation of Haag's theorem. The free theory and the interacting theory
are inequivalent no matter how small the coupling is. The contradiction with
positivity from the superconvergence relation could be avoided at this
stage by supplying the definition of the asymptotic subtraction
scheme with an implicit limit $\mu \to \infty$, 
\begin{equation}
\lim_{\mu^2 \to \infty} \; \, D(k) \big|_{k^2 = \mu^2} \, = \,
D(k)_{\mbox{\tiny tree-level}}
\end{equation}
In practical applications of the renormalisation group this means that 
momenta larger than the renormalisation point can be considered, their  
rough order of magnitude, however, is bound by that of the renormalisation 
scale (with the momentum dependence for $k^2 \sim \mu^2$ governed by the
scaling fixed point). Ambiguities in the non-commuting limits 
$k^2 \to \infty$ and $\mu^2 \to \infty$ as those leading to
Eqs.~(\ref{mu_lim}) versus (\ref{k_lim}) arise, if momenta are taken to
infinity relative to the subtraction point. 

This same spirit of renormalising the interacting theory to the free one {\sl
asymptotically} was actually adopted previously also for the quark
propagator~\cite{Sme91}. In that context it turned out to be necessary in
order to implement  the quark-confinement mechanism of infrared slavery
into the asymptotically free theory. 

The alert reader will have noticed that the Landau gauge considered so far
is exceptional ($\xi = 0$), and that for other possible choices of the gauge
parameter $\xi $ in Eq.~(\ref{OZ}) the superconvergence relation might not
rule out positivity anyway. The generalisation of the Oehme--Zimmermann
argument to the whole family of linear covariant gauges is less obvious. It
is, in fact, long known that QCD in the covariant gauge has an ultraviolet
fixed point in the $(g^2,\xi)$ plane at a finite positive value of the gauge
parameter $(0,\xi_0)$ with $\xi_0 = 13/3 - 4N_f/9 $, see, {\it e.g.},
Ref.~\cite{Mar78}. Therefore, 
naively one might think that with $\xi \to \xi_0 $ the spectral sum rule of
the free theory~(\ref{naive_sr}) is recovered. A more detailed analysis shows,
however, that the gluon spectral function $\rho(k^2)$ is negative for
sufficiently large $k^2$ also in this case. Assuming that the only
singularities of the gluon propagator lie on the time-like real axis in the
complex $k^2$-plane, one can still show that the discontinuity at the cut
behaves asymptotically~\cite{Oeh90,Oeh94},
\begin{equation} \label{ll_neg_rho}
\rho(k^2) \, \equiv \, \rho(k^2,g^2,\mu^2,\xi) \, \to \, - \, \gamma
C_R(g^2,\xi)  \frac{1}{k^2} \, \left(\ln \frac{k^2}{\mu^2}
\right)^{-\gamma-1} \; , \quad \mbox{for} \quad k^2 \to \infty \; .
\end{equation}
Here, $\gamma$ is the same positive (for $N_f < 10$) anomalous dimension of
the gluon field and $C_R(g^2,\xi)$ some positive constant. Therefore,
$\rho(k^2) $ is shown not to be positive also in the general, linear covariant
gauges. Analogous results from the renormalisation group analysis
employing analyticity in the cut complex $k^2$-plane exist also for the ghost
and the quark propagator for asymptotically large but complex $k^2$, see
Ref.\ \cite{Xu96}. While this demonstrates the violation of positivity of
transverse gluons independent of the spectral sum rule in Eq.~(\ref{OZ}), and
already at the level of perturbation theory, by itself it is of course not
sufficient to yield confinement. It can serve to demonstrate that the
massless transverse gluon states of perturbation theory have to belong to
unphysical BRS quartets (see the next section). For the realization of
confinement it is necessary in addition that there is a mass gap in
the transverse gluon correlations, {\it i.e.}, that the massless one-gluon
pole of perturbation theory is screened non-perturbatively. Only then the
Kugo--Ojima confinement criterion can establish the equivalence of BRS singlets
with colour singlets by requiring the global colour symmetry to be unbroken.  
This will be discussed next. Note, however, that the requirement of an 
unbroken global gauge symmetry, the absence of both, physical as well as
unphysical massless states from the spectrum of the global gauge current, is
a necessary condition in the derivation of the superconvergence relations
discussed in this section~\cite{Nis96}. This condition is violated in models
with Higgs mechanism which prevents one from concluding a positivity violation
of the massive physical vector-states in the transverse gauge-boson
correlations in that case.  

The non-positivity of the gluon spectral density will be discussed in
Sec.~\ref{sub_Pos} again. There we collect the present evidence for
its violation from two sources of non-perturbative results, from lattice
simulations of the gluon propagator in the Landau gauge, and from the 
solutions to truncated Dyson--Schwinger equations. These two kinds
of non-perturbative results furthermore both agree in indicating that no
massless one-particle pole exists in the transverse gluon correlations.

\goodbreak

\subsection{Description of Confinement in the Linear Covariant Gauge}
\label{Sec2.4}

Covariant quantum theories of gauge fields require indefinite metric
spaces. This implies some modifications to the standard
(axiomatic) framework of quantum field theory. Modifications are also
necessary to accommodate confinement in QCD. 
These seem to be given by the choice of either relaxing the principle of
locality or abandoning the positivity of the representation space.
The much stronger of the two principles being locality, non-local
descriptions (see Sec.~\ref{Sec2.5}) have received far less attention than
local ones. Great emphasis has therefore been put on the idea of relating
confinement to the violation of positivity in QCD.  
Just as in QED, where the Gupta--Bleuler prescription is to
enforce the Lorentz condition on physical states, a semi-definite {\em
physical subspace} can be defined as the kernel of an operator.  
The physical states then correspond to equivalence classes of states 
in this subspace differing by zero norm components.  
Besides transverse photons covariance implies the existence of 
longitudinal and scalar photons in QED. The latter two form metric partners
in the indefinite space. The Lorentz condition eliminates
half of these leaving unpaired states of zero norm which do not contribute to
observables. Since the Lorentz condition commutes with the 
$S$-Matrix, physical states scatter into physical ones exclusively. 
Colour confinement in QCD is ascribed to an analogous
mechanism: No coloured states should be present in the positive definite space
of physical states defined by some suitable condition maintaining physical
$S$-matrix unitarity. A comprehensive and detailed account of most of the
material summarised in this section can be found in the textbook by Nakanishi
and Ojima~\cite{Nak90}. Here, we briefly recall those of the
general concepts that relate to some of the results presented in
Chapter~\ref{chap_QCD}. In particular, we would like to emphasise the
following three aspects: positivity violations of transverse gluon and quark
states, the Kugo--Ojima confinement criterion, and the conditions necessary
for a failure of the cluster decomposition. We describe each of these in the
next three subsections, and we will find that the results of
Chapter~\ref{chap_QCD} nicely fit into these general considerations which
thus together lead to a quite coherent, though certainly still somewhat 
incomplete picture.

\subsubsection{Representations of the BRS Algebra and Positivity}
\label{Sec.2.4.1}

Within the framework of BRS algebra, in the simplest version for
the BRS-charge $Q_B$ and the ghost number $Q_c$ (both hermitian with respect
to an indefinite inner product) given by,
\begin{equation} 
           Q_B^2 = 0 \; , \quad \left[ iQ_c , Q_B \right] = Q_B \; ,
\end{equation}
completeness of the nilpotent BRS-charge $Q_B$ in a state space $\mathcal{V}$
of indefinite metric is assumed. This charge generates the BRS 
transformations ($\delta\Phi \equiv \lambda \delta_B\Phi$ with Grassmann
parameter $\lambda $) of a generic field $\Phi $ by the ghost number graded 
commutator,
\begin{equation}
            \delta_B \Phi = \{ i Q_B , \Phi \} \,
\end{equation}
{\it i.e.}, by a commutator or anticommutator for fields $\Phi$ with even
or odd ghost number $Q_c$, respectively. In presence of ghost-antighost
symmetry by Faddeev-Popov conjugation, this structure generalises to that 
for the semi-direct product of the global $SL(2,\RR) $ with the double BRS
invariance,\footnote{Corresponding to a Inonu--Wigner contraction of a
$OSp(1,2)$ superalgebra, see Refs.~\cite{Thi85,Nak90}.} see
App.~\ref{App.BRS}.      
The semi-definite {\em physical} subspace 
$\mathcal{V_{\mbox{\tiny phys}}}  = \mbox{Ker}\, Q_B$ is 
defined on the basis of this algebra by those states which are annihilated by
the BRS charge $Q_B$, 
\begin{equation} 
       \mathcal{V_{\mbox{\tiny phys}}}  = \Big\{ 
            |\psi \rangle \in \mathcal{V}\,  : \;\;  Q_B |\psi\rangle = 0 \,
            \Big\} = \mbox{Ker}\, Q_B  \; .
\end{equation}
Since $Q_B^2 =0 $ this subspace contains the space of so-called daughter
states which are images of others, their parent states in $\mathcal{V}$,
\begin{equation}
                \mbox{Im}\, Q_B  =   \Big\{ 
            |\psi \rangle \in \mathcal{V}\,  :   \;\; |\psi\rangle =   Q_B
            |\phi \rangle \, , \;   |\phi\rangle  \in \mathcal{V}\,     
             \Big\} \subset  \mathcal{V_{\mbox{\tiny phys}}} \; .
\end{equation}
A physical Hilbert space is then obtained as (the completion of) the 
covariant space of equivalence classes, the BRS-cohomology of states in the
kernel modulo those in the image of $Q_B$,
\begin{equation}
         \mathcal{H}(Q_B,\mathcal{V}) = {\mbox{Ker}\, Q_B}/{\mbox{Im} Q_B} 
       \simeq  \mathcal{V}_s \; , 
\end{equation}
which is isomorphic to the space $\mathcal{V}_s$ of BRS singlets. 
It is easy to see that the image is furthermore contained in the orthogonal
complement of the kernel. Given completeness they are identical, $\mbox{Im}\,
Q_B = (\mbox{Ker}\, Q_B )^\perp = \mbox{Ker}\, Q_B \cap (\mbox{Ker}\, Q_B
)^\perp $ which is the isotropic subspace of $\mathcal{V_{\mbox{\tiny phys}}}
$. It follows that states in $\mbox{Im}\, Q_B$, in the language of de Rham
cohomology called BRS-coboundaries, do not contribute to the inner product in
$\mathcal{V_{\mbox{\tiny phys}}}$.   
Completeness is thereby important in the proof of positivity for physical
states \cite{Kug79,Nak79,Nak90}, because it assures the absence of metric
partners of BRS-singlets, so-called ``singlet pairs'' which would otherwise 
jeopardise the proof. 

With completeness all states in $\mathcal{V}$ can be shown to be either BRS
singlets in $\mathcal{V}_s$ or belong to so-called quartets which are 
metric-partner pairs of BRS-doublets (of parent with daughter states), and
that this exhausts all possibilities. The generalisation of the
Gupta--Bleuler condition on physical states, $Q_B |\psi\rangle = 0$ in
$\mathcal{V}_{\mbox{\tiny phys}}$, eliminates half of these metric partners
leaving unpaired states of zero norm (in the isotropic subspace of
$\mathcal{V}_{\mbox{\tiny phys}}$) which do not contribute to any observable.
This essentially is the quartet mechanism: 
\begin{itemize}
\item[] Just as in QED, one such quartet, the elementary quartet, is formed by
the massless asymptotic states of longitudinal and time-like gluons together 
with ghosts and antighosts which are thus all unobservable. 

\item[] In contrast to QED, however, one expects the quartet mechanism also 
to apply to transverse gluon and quark states, as far as they exist
asymptotically. A violation of positivity for such states then entails
that they have to be unobservable also. 
\end{itemize}

The combined evidence for
this, as collected in the present review, provides strong indication 
in favour of such a violation for possible transverse gluon states. 

The members of quartets are frequently said to be confined
kinematically. This is no comprehensive explanation of confinement, of
course, but one aspect (among others as we shall describe below) of its
description within the covariant operator formulation~\cite{Nak90}.   
In particular, asymptotic transverse gluon and quark states
may or may not exist in the indefinite metric space $\mathcal{V}$. If either 
of them do exist and the Kugo--Ojima criterion is realized (see below), they
belong to unobservable quartets. In that case, the BRS-transformations of their
asymptotic fields entail that they form these quartets together with
ghost-gluon and/or ghost-quark bound states, respectively, see Sec. 4.4.3
in~\cite{Nak90}. We reiterate that it is furthermore  crucial for
confinement, however, to have a mass gap in transverse gluon correlations,
{\it i.e.}, the massless transverse gluon states of perturbation theory have to
disappear (even though they should belong to quartets due to
superconvergence in asymptotically free and local theories, see the
discussion at the end of Sec.~\ref{Sec2.3}).        

Before we continue we add two brief remarks. The BRS construction of the
physical state space sketched above is endowed with a quantum mechanical  
interpretation in terms of transition probabilities and measurements as
expectation values of observables. A necessary {\em and} sufficient condition
on a (smeared local) operator\footnote{One in the polynomial algebra 
of fields on a bounded open set of space-time, see, {\it e.g.},
Ref.~\cite{Haa96}.} 
$A$ is that the isotropic subspace of zero norm states does not affect its
expectation values in $\mathcal{V}_{\mbox{\tiny 
phys}}$~\cite{Oji78,Nak90}, {\it i.e.}
\begin{equation} 
        \delta_B A \, = \{ iQ_B , A \} \, = 0 \; .   \label{ObsDef}
\end{equation}
$A$ is then called a (smeared local) observable in the present context
thereby slightly generalising the usual notion of an observable (by
self-adjointness). 
It then follows that for all states generated from the vacuum
$|\Omega\rangle$ by any such observable, {\it i.e.} a BRS-closed operator,   
one has
\begin{equation} 
          \langle \Omega | A^\dagger A |\Omega \rangle \ge 0 \; .
\end{equation}
The interesting ones among the BRS-closed operators here are those, of course,
which are not BRS-exact, {\it i.e.}, which are not BRS variations of others. 
On the other hand, the vacuum should be a BRS-invariant physical state, 
and thus for any BRS-exact operator $A = \delta_B B$,
\begin{equation}
                  \langle \Omega |  \delta_B B |\Omega \rangle =  
         \langle \Omega |   \{ iQ_B , B \} |\Omega \rangle = 0     \; ,
         \label{genSTI} 
\end{equation}
from which all the famous Slavnov-Taylor identities can be derived by 
BRS transformations with choosing some suitable product of fields for $B$.

If this construction is shown to apply to a QCD description of hadrons as the
genuine physical particles of $\mathcal{H}$, it is intuitively quite clear, 
and it can be established rigorously from physical $S$-matrix unitarity (with
respect to the indefinite inner product), that absorptive thresholds in
hadronic amplitudes can only be due to intermediate hadronic states.  
The $S$-matrix commutes with the BRS-charge, it is an observable in the above
sense, and it thus transforms physical sates into physical ones exclusively
and without leading to measurable effects of possible zero norm
components~\cite{Nak90}. Anomalous thresholds, the singularities related to
the substructure of hadrons, can in this description also arise only 
from substructure of a given hadron as a composite state of other hadrons. 
The argument to establish this employs standard analyticity 
properties for hadronic amplitudes and crossing to relate them to
absorptive singularities of other hadronic amplitudes which by the first
argument above can only be due to intermediate hadronic states~\cite{Oeh95}.

\subsubsection{The Kugo--Ojima Confinement Criterion}
\label{Sec.2.4.2}

In the BRS formulation of gauge theories, the realization of confinement by
the quartet mechanism explained above depends on the realization of the
unfixed global gauge symmetries. In particular, the identification of the 
BRS singlet states in the physical Hilbert space $\mathcal{H}$ with
colour singlets is possible only if the charge of global gauge transformations
is BRS-exact {\em and} unbroken, {\it i.e.}, well-defined in the whole of the
indefinite metric space $\mathcal{V}$. The sufficient conditions for this are
provided by the Kugo--Ojima criterion. 

The starting point for this discussion, for the details of which we again 
refer to Ref.~\cite{Nak90} and the references 
therein, is the globally conserved current $J^a_\mu$, {\it i.e.}, with
$\partial_\mu J^a_\mu = 0$, given by     
\begin{equation} 
    J^a_\mu = \partial_\nu F_{\mu\nu}^a  + \{ Q_B , D_{\mu}^{ab} \bar c^b \} 
    \; .
       \label{globG}
\end{equation}
It consists of two terms, the first one corresponding to a coboundary term
with respect to the space-time exterior derivative and the other to a
BRS-coboundary term. The spatial integrations of the zeroth components give 
their corresponding charges $G^a$ and $N^a$, respectively, and the charge of
the global gauge symmetry as the sum of the two,
\begin{equation} 
      Q^a =  \int d^3x \,  \partial_i F_{0 i}^a \,  +\,  \int d^3x \, 
             \{ Q_B , D_{0}^{ab} \bar c^b \} \, = \, G^a \, + \, N^a \; .
        \label{globC}
\end{equation}
For the first term herein there are only two options, it is either ill-defined
due to massless states in the spectrum of $\partial_\nu F_{\mu\nu}^a $, or else
it vanishes because the integrand is a total derivative. 

Without going into detail, due to the masslessness of photons in QED,
massless states contribute to both currents in~(\ref{globG}), and both
charges in (\ref{globC}) are separately ill-defined (the second term 
is given by the Nakanishi--Lautrup $B$-field as $ N = \int d^3x \, \partial_0
B $ in this case). One can, however, employ an arbitrariness in the
definition of the generator of the global gauge transformations (\ref{globC})
to multiply the first term by a constant suitably chosen so that the massless
contributions cancel, and one arrives at a well defined and unbroken global
gauge charge to replace the naive definition in (\ref{globC}) above, see also
\cite{Kug95}. Roughly speaking, 
there are two independent structures in the globally conserved gauge currents
in QED which both contain massless photon contributions. These can be combined
to yield one well-defined charge as the generator of global gauge
transformations leaving any other combination as a spontaneously broken
global symmetry. One such combination generates the displacement symmetry,
the vector symmetry corresponding to gauge transformations  
$ \theta (x) = a_\mu x^\mu $ with global parameters $a_\mu$,
and the photon has conversely been interpreted as the massless
Goldstone boson of its spontaneous breaking~\cite{Fer71,Bra74,Len94a}.

If on the other hand, $\partial_\nu F_{\mu\nu}^a $ contains no massless
discrete spectrum ({\it i.e.}, if there is no massless particle pole in the
Fourier transform of transverse gluon correlations), then $G^a \equiv
0$~\cite{Kug79}.
In particular, this is the case for channels with massive vector fields in
theories with Higgs mechanism, and it is expected to be also the case in any 
colour
channel for QCD with confinement for which it actually represents one of the
two conditions formulated by Kugo and Ojima. 
In both these situations one first has, however, 
\begin{equation}
                       Q^a \, = \, N^a \, = \, \Big\{   Q_B \, , 
       \int d^3x \,   D_{0}^{ab} \bar c^b \Big\} \; ,
\end{equation}
which is BRS-exact. The second of the two conditions for confinement
which together are sufficient to establish that all BRS-singlet physical
states in $\mathcal{H}$ are also colour singlets, and that, the other way
around, all coloured states are subject to the quartet mechanism, is to
guarantee that the BRS-exact charge of global gauge transformations above be
well-defined in the whole of the indefinite metric space $\mathcal{V}$.  

This second condition obviously provides the essential 
difference between Higgs mechanism and confinement at the present stage. 
The operator $D_\mu^{ab}\bar c^b$ determining the charge $N^a$ usually
contains a contribution from the {\em massless} asymptotic field
$\bar\gamma^a(x)$ of the elementary quartet contained in the asymptotic 
antighost field,  $\bar c^a\, \stackrel{x_0 \to \pm\infty}{\longrightarrow}
\, \bar\gamma^a + \cdots $,\footnote{These are understood as weak limits
in the sense of LSZ, see also Sec.~\ref{Sec2.3}.}   
\begin{equation}
          D_\mu^{ab}\bar c^b \; \stackrel{x_0 \to \pm\infty}{\longrightarrow}
              \;   ( \delta^{ab} + u^{ab} )\,   \partial_\mu \bar\gamma^b(x) +
                 \cdots  \;  .
\end{equation}
Here, the $ u^{ab} $ are dynamical parameters which determine the contribution 
of the massless asymptotic state to the composite field, $g f^{abc} A^c_\mu
\bar c^b  \, \stackrel{x_0 \to \pm\infty}{\longrightarrow}  \,
u^{ab} \partial_\mu \bar\gamma^b + \cdots $. These parameters can be obtained
in the limit $p^2\to 0$ from (here Euclidean) correlation functions of this
composite field, {\it e.g.},
\begin{equation}
\int d^4x \; e^{ip(x-y)} \,
\langle  \; D^{ae}_\mu c^e(x) \; gf^{bcd}A_\nu^d(y) \bar c^c (y) \; \rangle
\; =: \; \Big(\delta_{\mu \nu} -{p_\mu p_\nu \over p^2} \Big) \, u^{ab}(p^2)
\; .  \label{Corru}
\end{equation}
The theorem by Kugo and Ojima~\cite{Kug79} asserts that all $Q^a = N^a$ are
well-defined in the whole of  $\mathcal{V}$ (and do not suffer from
spontaneous breakdown), if and only if
\begin{equation}
                 u^{ab} \equiv u^{ab}(0)  \stackrel{!}{=} - \delta^{ab} \; .
\label{KO1}
\end{equation}
Then the massless states from the elementary quartet do not contribute to 
the spectrum of the current in $N^a$, and the equivalence between physical
states as BRS-singlets and colour singlets can be
established, details and proofs are provided in
Refs.~\cite{Kug79,Nak90,Kug95}.      

For a discussion of the fate of the non-Abelian version of the displacement
symmetry, $ \theta^a (x) = a^a_\mu x^\mu $ with global parameters $a^a_\mu$, 
see  Refs.~\cite{Kug95,Hat82}. Essentially, one finds that the Kugo--Ojima
criterion $ \ID + u \equiv 0$, also removes the massless states from the
correlation functions of its corresponding current.

If, however, as the other extreme, no eigenvalue of $ \ID + u $
vanishes, {\it i.e.}, for $\mbox{det}(  \ID + u ) \not=0$, then the 
so-called {\em converse of the Higgs theorem} asserts that the global
gauge symmetry generated by the charges $Q^a$ in Eq.~(\ref{globC}) is
spontaneously broken in each channel in which the gauge potential $A_\mu^a$  
contains an asymptotic massive vector field (leading to a mass gap in
$\partial_\nu F_{\mu\nu}^a $ and thus $G^a = 0$ for that channel), see
Refs.~\cite{Kug79,Nak90}.  
 
Spontaneous breakdown of any (combination) of the charges $Q^a$ really is 
in one-to-one correspondence with the occurrence of a massive gauge boson in
this framework. And as a consequence of this breaking, the massive vector
asymptotic field contained in the gauge potential results to be a BRS-singlet
and thus physical. 
Whereas the massless Goldstone boson states usually occurring in
some components of a Higgs field, belong to the elementary quartet and are
thus unphysical. Therein they take the place of one of the components of the
gauge field which in turn becomes a BRS-singlet to yield the third
component of the massive physical vector field. This is explicitly worked out
in detail for the Abelian Higgs and the Higgs--Kibble models also in
Ref.~\cite{Nak90}.  

Since the Goldstone bosons are all unphysical and the broken charges
BRS-exact in this case, called {\em hidden} spontaneous symmetry breaking,
nothing of this breaking is observed in the Hilbert space of physical states
$\mathcal{H}$, and no conflict with Elitzur's theorem~\cite{Eli75} arises
which asserts that local gauge symmetries should not be spontaneously broken
for reasons similar to ordinary quantum mechanics where spontaneous symmetry
breaking is precluded by von Neumann's uniqueness theorem for the
representations of the canonical commutation relations.\footnote{Some spin
transformation symmetries can also be spontaneously broken even in quantum
mechanics. This is important in systems with spontaneously broken
supersymmetry for instance~\cite{Wit81}.}  
The same can be understood from an alternative description. Eliminating the
redundant variables of gauge theories in a non-covariant canonical
formulation explicitly, the Higgs mechanism can be shown to generate the
masses of physical vector bosons without breaking of the gauge
symmetry~\cite{Fro81,Sto95}. Mass generation without symmetry breaking is in
fact a well known mechanism. It occurs in the large $N$ limit of the
$O(N)$-vector model \cite{Abb76,Lan93a,Sme94,Lan96} of $N$ selfinteracting
scalar fields in four dimensions,\footnote{Scalar $\phi^4$-theory likely is
trivial~\cite{Fer91} but it can be analytically continued in the complex
coupling to yield a stable Euclidean field theory, in general, on the cost of
reflection positivity~\cite{Gaw85}.} as well as in the $O(N)$-symmetric
Gross--Neveu or $SU(N)$ Thirring model \cite{Gro74,Lan95} of selfinteracting
fermions in two dimensions.\footnote{In two dimensions there cannot be a
chiral phase transition due to the Coleman--Mermin--Wagner theorem. The large
$N$ results can, however, be interpreted as a Kosterlitz--Thouless
transition. What occurs is a mass gap for the fermions but no Goldstone
bosons \cite{Wit78}.} 
It usually requires breaking of scale invariance only, which is an anomaly
rather than a broken symmetry \cite{Col77,Lan93b}.    

Nevertheless, it is instructive to classify the different scenarios
by considering the realization of the global gauge symmetry on the whole of
the indefinite metric space $\mathcal{V}$ of covariant gauge theories. 
In this formulation, however, as long as the global gauge symmetry is
unbroken, {\it i.e.}, for QED and QCD, the first of the two conditions
becomes the relevant one. Namely, it is then necessary for confinement to
have a mass gap in transverse gluon correlations ({\it i.e.}, in
$\partial_\nu F_{\mu\nu}^a $), since otherwise one could in principle have
{\em non-local} physical (BRS-singlet and thus gauge invariant) states which
are no colour singlets, however, just as one has non-local gauge invariant
charged states in QED, {\it e.g.}, the state of one electron alone in the
world with its long-range Coulomb field. 
Indeed, with unbroken global gauge invariance QED and QCD have in common
that any gauge invariant localised state must be
chargeless/colourless, see Sec 4.3.3 in~\cite{Nak90}. The
question is the extension to non-local states as approximated by local ones.
In QED this leads to the so-called charge superselection sectors,
see~\cite{Haa96}, and non-local charged physical states arise.   
In QCD, with mass gap {\em and} unbroken global gauge symmetry, the
representations of (the net of) algebras of local observables will be
irreducible in $\mathcal{H}$ and every gauge-invariant state can therefore be
approximated by gauge-invariant localised ones (which are colourless), thus
{\em every} gauge invariant (BRS-singlet) state in $\mathcal{H}$ will be a
colour singlet.

We close this subsection with one more comment relevant to the results
presented in Chap.~\ref{chap_QCD}:
It concerns the observation  pointed out by Kugo in Ref.~\cite{Kug95} that 
the (2nd condition in the) Kugo--Ojima confinement criterion, $u = -\ID$,  
in Landau gauge is equivalent to an infrared enhanced ghost propagator. 
In particular, based on standard arguments employing Dyson--Schwinger
equations and Slavnov-Taylor identities, one can show that 
the non-perturbative ghost propagator of  Landau gauge QCD in momentum space 
is related to the form factor occurring in the correlations of Eq.~(\ref{Corru}) 
as follows,
\begin{equation}
    D_G(p) = \frac{-1}{p^2}      \, \frac{1}{ 1 + u(p^2) } \, , \;\;
                 \mbox{with}  \; \;   
                 u^{ab}(p^2)  = \delta^{ab}  u(p^2) \, .
\end{equation}
The Kugo--Ojima criterion, $u(0) = -1 $, thus entails that the Landau gauge
ghost propagator should be more singular than a massless particle pole in the
infrared.

Indeed, we will present quite compelling evidence from DSEs and lattice
simulations for this exact infrared enhancement of ghosts in Landau gauge
necessary for a realization of the Kugo--Ojima confinement criterion. 
It is furthermore interesting to note that there are recent lattice
simulations also testing this criterion directly~\cite{Nak99,Nak00a}: Instead
of $-\delta^a_b$ they obtain numerical values of around $u = -0.7$  for the
unrenormalised diagonal parts and zero (within  statistical errors) for the
off-diagonal parts. After renormalisation, diagonal parts very close to
-1 result. Taking into account the finite size effects on the
lattices employed in the simulations, these preliminary results
might be perfectly reconciled with the 
Kugo--Ojima confinement criterion (\ref{KO1}).
Certainly, they serve to demonstrate the utility of such an independent test.

\subsubsection{Cluster Decomposition Property and Observables}
\label{Sec.2.4.3}

The remaining dynamical aspect of confinement in this
formulation resides in the cluster decomposition property of local quantum 
field theory. Its proof \cite{Ara62,Rue62}, which is absolutely general
other\-wise, does not include the indefinite metric spaces of covariant gauge
theories \cite{Str76,Str78}. The situation in local quantum field theory 
including indefinite metric spaces can roughly be summarised as
follows, see Refs.~\cite{Nak90,Haa96}: For the vacuum expectation values of
two (smeared local) operators $A$ and $B$, translated to a large space-like
separation $R$ of each other one obtains the following bounds depending on
the existence of a finite gap $M$ in the spectrum of the 
mass operator in $\mathcal{V}$, 
\begin{eqnarray} 
        \Big|  \langle  \Omega | A(x) B(0) |\Omega \rangle  &-&         
 \langle  \Omega | A(x) |\Omega \rangle  \,  \langle  \Omega
             |  B(0) |\Omega \rangle  \Big|   \nonumber\\
   && \quad \le  \;  \Bigg\{ \begin{array}{ll} 
   \mbox{\small Const.} \, \times \, R^{-3/2 + 2N} \, e^{-MR} \!\!, \quad 
                      & \mbox{mass gap } M \; , \\
   \mbox{\small Const.} \, \times \, R^{-2 + 2N} \,, \;\; 
                      & \mbox{no mass gap} \; ,  \end{array}
 \label{cluster_property} 
\end{eqnarray}
for $R^2 = - x^2 \to \infty $. Herein, positivity entails that $N = 0$, but a
positive integer $N$ is possible for the indefinite inner product structure in
$\mathcal{V}$. Therefore, in order to avoid the decomposition property
for products of unobservable $A$, $B$ which 
together with the Kugo--Ojima criterion is equivalent to avoiding the
decomposition property for coloured clusters, there should 
be no mass gap in the indefinite space $\mathcal{V}$. 
Of course, this implies nothing on the physical spectrum of the mass operator
in $\mathcal{H}$ which certainly should have a mass gap. 
In fact, it was shown within the covariant BRS formulation of gauge theories
that if the cluster decomposition property holds for a product $A(x) B(0)$ 
which together forms a (smeared local) observable in the sense of
condition~(\ref{ObsDef}), that both, $A$ and $B$ also satisfy this
condition separately and are thus observables themselves~\cite{Oji80}. 
This would then eliminate the possibility of scattering a physical state into
colour singlet states consisting of widely separated coloured clusters (the
``behind-the-moon'' problem, see also Ref.~\cite{Nak90} and references
therein).\footnote{It is always an alternative possibility, of course, that
space-like commutativity does not hold for unphysical fields, see the
discussion in the next section.}   

The necessity for the absence of the massless particle pole in $\partial_\nu
F^a_{\mu\nu} $ in the Kugo--Ojima criterion shows that the unphysical
massless correlations to avoid the cluster decomposition property are {\em
not} the transverse gluon correlations. In that sense,
an infrared suppressed propagator for the transverse gluons in Landau gauge
confirms this condition. This holds equally well for the infrared vanishing
propagator obtained from DSEs and conjectured in Refs.~\cite{Gri78,Zwa92}
studying the implications of the Gribov horizon, as for the suppressed but
possibly infrared finite ones extracted from improved lattice actions for
quite large volumes~\cite{Bon00}. At the same time, however, it shows that
transverse gluons are not responsible for a failure of the cluster
decomposition property. The infrared enhanced correlations responsible for
this in Landau gauge can be identified with the ghost correlations which at
the same time provide for the realization of the Kugo--Ojima criterion.

Then, the question for massless unphysical single particle states remains. 
Here, for the Landau gauge results the infrared fixed point obtained in the
non-perturbative running coupling as defined in Chapter~\ref{chap_QCD}, if it
will be confirmed, implies the existence of unphysical massless bound states,
{\it i.e.}, massless single particle poles corresponding to asymptotic
states in coloured composite operators which by virtue of the Kugo--Ojima
criterion belong to unobservable quartets. This should be further studied,
in particular, with emphasis on the relation of these massless states to a
hidden spontaneous breakdown of some symmetry. As a candidate for this
symmetry breaking, the global $SL(2,\RR ) \to \RR $ (ghost number) common to
the Landau or more generally the Curci--Ferrari gauge 
and the BRS formulation of covariant Abelian
gauges has been suggested in Refs.~\cite{Sch99,Sch00}. This certainly offers an
interesting possibility for further studies 
extending the Landau gauge results to the more general, $SL(2,\RR )$
symmetric gauges discussed in Appendix~\ref{App.BRS}. 

We close this section with an interesting technical side remark on implications
of the results from the Landau gauge DSEs presented in Chapter~\ref{chap_QCD}.
This DSE solution entails an infrared vanishing gluon propagator  together with
an infrared fixed point for the coupling. The latter arises from to the
infrared finiteness of the product of $G^2(k^2) Z(k^2)$, where $Z(k^2)$ is the
infrared vanishing gluon renormalisation function and $G(k^2) = 1/(1+u(k^2))$
is the infrared singular ghost renormalisation function. This scenario
corresponds to the third of three different possibilities discussed by Kugo in
Ref.~\cite{Kug95} for the realization of the Kugo--Ojima criterion in Landau
gauge. This third possibility, the one favoured by Kugo from
plausibility arguments, in addition to a vanishing gluon propagator 
also leads  to an infrared vanishing renormalisation function for the 3-gluon
vertex at symmetric momenta, see also Secs.~\ref{sub_Ghost} and
\ref{sub_Sub}.

\subsection{Alternative Singularity Structures of Green's
Functions}
\label{Sec2.5}

As stated above, possibilities other than the description of confinement in
QCD in terms of local quark and gluon fields might also be viable. Since
space-like commutativity of local fields is such a strong principle, and
any relaxation of this has quite severe consequences, non-local descriptions
have received far less attention than local ones. The most elementary
examples to discuss differences between local and non-local descriptions are
provided by 2-point correlation functions, {\it i.e.}, the quark and gluon
propagators in QCD.  
In local quantum field theory any 2-point correlation function 
is an analytic function in the cut complex $p^2$-plane with singularities
along the time-like real axis only. The assumptions to establish this, also
with indefinite metric, are exactly the same as those for the cluster
property in Eq.~(\ref{cluster_property}) discussed in the previous
subsection. In particular, these are: covariance under (space-time)
translations, space-like (anti)commutativity, uniqueness and cyclicity 
of the vacuum, and the spectrum condition.\footnote{On indefinite metric
spaces the exact form of the spectrum condition has to be generalised
slightly~\cite{Str77}. Its essence remains the same, {\it i.e.}, that the 
spectrum of the energy-momentum operator is to be contained in the forward cone.} 
For any other singularity structure of 2-point correlations at least
one of these prerequisites is violated. 

Typically, this concerns space-like commutativity. The prime example here is
given the phenomenologically appealing models of confinement based on entire
2-point correlation functions. Asymptotic freedom entails, however, that
these functions would have to vanish for $p^2 \to \infty$ in all directions
of the complex $p^2$-plane~\cite{Oeh80,Oeh95}. Non-trivial entire functions
with that property do not exist.  The use of entire functions with
an essential singularity at infinity might nevertheless be interesting as a
phenomenological concept. Such 2-point correlations on the other hand,
cannot correspond to vacuum expectation values of two local fields, rather of
fields which have (anti)commutators with non-vanishing support into a region
of space-like separations~\cite{Iof69}. This region is determined by
space-like hyperboloids $(x-y)^2 < - l^2$ which might for 
possibly small but finite $l$ be restricted to a small neighbourhood 
of the light-cone. 
It is thus unbounded, however, and the problem arises, how to restrict 
the violations of causality to a finite (microscopic) region in
space-time~\cite{Efi68},  
or how to prevent the ``signals'' resulting from such unphysical correlations
which will grow exponentially in time to lead to measurable effects.

Alternatively, complex singularities with time-like real part 
might be considered as
acceptable for the propagators of unphysical (coloured) fields.  
Such singularities might then conspire to cancel with singularities or 
zeros in other unphysical correlation functions so as to be absent from 
physical amplitudes. This will give rise to an infinite hierarchy of
constraints on such unphysical singularities in arbitrarily high $n$-point
functions. An example of such compensating singularities are those 
inherent in the non-perturbative expansion scheme of Refs. \cite{Hae90,Sti95}
which we describe briefly in Sec.~\ref{CompSing}, and later in
Sec.~\ref{sec_Stingl}.

\subsubsection{Entire correlation functions}

Entire correlation functions occur, {\it e.g.}, in self-dual 
constant background fields~\cite{Leu80}.
Historically the Euler--Heisenberg Lagrangian ({\it e.g.}, see Sec.\ 4.3 of
Ref.\ \cite{Itz80}) provided the first example. The constant electric and
magnetic background fields introduce a non-local interaction. Correspondingly,
the correlation functions for charged particles, usually expressed as proper
time integrals, display a non-standard singularity structure, see, {\it e.g.},
Chapter 2 of Ref.\ \cite{Dit00} and the references therein. Note that a
constant field in an infinite volume is an infinitely large energy reservoir. 
This underlines the impossibility of such a background field in QED: 
Pair creation would immediately set in. Therefore it is safe to conclude 
that the appearance of such effects in certain treatments of QED in strong
fields is due to the employed  approximations.\footnote{This remark does not
apply to the time evolution in a plasma based on a kinetic description, see
Chapter 5 of Ref.\ \cite{Rob00} and the references therein.} 
For QCD one might adopt a different point of view and consider entire
correlation functions possible for unphysical fields hereby accepting 
the existence of non-local interactions.  

Causality might be kept in a non-local quantum field theory if one requires
that only exponentially localised sources are allowed. Of course, as the aim is
to describe confinement such a restriction seems to be quite natural:
Confinement implies that coloured sources are strongly localised. Despite the
non-local interaction, and the resulting non-vanishing commutators of
operators over space-like distances, physical amplitudes then possess a causal
behaviour. For the purpose of this review it is instructive to turn the
argument around. Suppose the propagator of a field is, apart from the tensorial
structure related to the spin of the field, given by 
\begin{equation} 
D(p^2) =  b \int _0^1 ds  e^{-sb(p^2+m^2)} = 
\frac {1-e^{-b(p^2+m^2)}}{p^2+m^2} \; .
\label{entireD}
\end{equation}   
For space-like (Euclidean) momenta $p^2>0$ it differs from the
propagator of a free field with mass $m$ only by an exponentially small number,
for time-like momenta, however, the pole at $p^2=-m^2$ is removed at the
expense of an essential singularity at $-p^2\to \infty$. Note that this
qualitative type of behaviour occurs is some terms of the propagator of
charged particles in  constant background fields. Obviously, with the usual
constraints on the sources, 
such a propagator would violate causality. On the other hand, if the Fourier
transforms of all sources, $\tilde J(p)$, are restricted such that
\begin{equation}
|\tilde J(p)| < e^{-|p^2|/\mu^2+\epsilon} \quad {\rm with} \; \mu^2, 
               \, \epsilon >0 \; ,
\end{equation}   
causality is restored. This implies that
\begin{equation}
|J(x_0,\vec x )| < e^{- \mu' |x_0|  +\delta} \quad {\rm with} \; \mu',\, 
            \delta>0 \; .
\end{equation}
This, in turn, can be interpreted as confinement \cite{Efi93}: With all
charged, {\it i.e.}, coloured, sources exponentially localised in time, 
the problem has been turned into a feature. Assuming the realization of
confinement via entire correlation functions one has to require localised
sources for consistency.

Ans\"atze for propagators which consist of sums and/or products of the form
(\ref{entireD}) have been widely used in models for meson and baryon physics,
see Refs.\ \cite{Rob00,Efi93,Efi95,Efi96,Bur96} and the references therein as
well as Chapters \ref{chap_Meson} and \ref{chap_Bary} of this review. These
propagators are entire functions in the whole complex $p^2$-plane and are
non-trivial only due to their essential singularity  at infinity. Of course,
these propagators contain a number of parameters which are fitted to 
observables, {\it e.g.},  in Ref.\ \cite{Rob96} the quark propagator being an
entire function with six parameters has been fixed in a least-squares fit to
light meson observables. It should be noted, however, that such propagators
do have zeros in the complex $p^2$-plane. As we will see in the course of this 
review Ward identities require that vertex functions are proportional to
inverse propagators. Obviously, these vertex functions will then in general
possess singularities at finite (complex) values of $p^2$. Of course, these
singularities are highly problematic when calculating hadronic reactions with
time-like momentum transfers.


There are other models in which a momentum dependent quark mass is
tuned such that the quark cannot go on-shell, {\it e.g.}, see Refs.\
\cite{Bub92,Hab95}. Nevertheless, the quark propagator possesses singularities
in the complex $p^2$-plane. These singularities are again highly problematic if
such a propagator is used to calculate hadronic processes. 

Summarising this subsection it is probably fair to say that the idea to realize
confinement via the use of entire correlation functions is neither understood
on a fundamental level nor is it without technical problems. 

\subsubsection{Compensating singularities}
\label{CompSing}

Apparently unphysical singularities in proper vertex functions induced by 
zeros in the momen\-tum-space propagators are inherent in the scheme of
Refs.~\cite{Hae90,Sti95}. In this approach rational Ans\"atze are employed 
in DSEs to account for the essential singularities that
occur in the Green's functions as functions of the coupling 
due to the dynamical generation of a non-perturbative mass scale.
At short distances, the results for the gluon and ghost propagators 
in this scheme can readily be related to their 
operator product expansions~\cite{Ahl92}. 

As we discuss in more detail in Sec.~\ref{sec_Stingl}, the hierarchical
coupling of the different vertex functions then implies common poles in the
external momenta.  
In particular, poles in the 2-point vertices, {\it i.e.}, the zeros of the
propagators, reappear in all higher vertex functions. For scattering 
amplitudes, the poles in the external lines are compensated by the
corresponding zeros of the propagators attached to these lines. Thus, these
artifacts of the approximation will not give rise to unphysical asymptotic
particle states.  

The situation is more complicated, however. It can be inferred from the
hierarchy of DSEs, together with the exchange symmetries, 
that the presence of poles in the external momenta of lower $n$-point ($n=2,3$)
vertex functions furthermore implies poles in the Mandelstam variables of
successively higher $n$-point vertices ($s,t,u,$..., for $n \geq 4$). These
pole contributions cannot be represented by one-particle reducible
contributions and thus in no way contradict the 1-PI property of these
vertex functions (just as bound state poles do not). However, analogous poles
occur also in 1-particle reducible contributions to scattering
amplitudes: In such contributions, an internal propagator having 
say one simple zero is connected to two vertex functions each contributing a
simple pole at the momentum of the zero in the propagator. One simple pole
remains in this exchange. The important but non-trivial observation is now
that these poles exactly cancel those appearing in the Mandelstam variables
of the 1-PI contributions to the scattering amplitudes in the 
scheme of Refs.~\cite{Dri98a,Dri98b}. The respective contributions to
scattering amplitudes that remain after this cancellation can be classified
according to an extended notion of irreducibility. It involves subtracted,
{\sl softened} one-particle reducible exchanges and equally subtracted 1-PI
contributions. As a result, no unphysical particle productions according to
Cutkosky's rule will occur in the scattering amplitudes due to these
compensating poles. In the integration kernels of the DSEs
for arbitrary 1-PI $n$-point functions, on the other hand, this
cancellation is incomplete. The necessary pole contributions reproduce
themselves in the hierarchy~\cite{Dri98a,Dri98b}.

The first studies within this approach, which were 
restricted to the lowest non-trivial
order in this rational approximation scheme to pure QCD and which furthermore
assumed a perturbative ghost propagator in the Landau gauge, suggested an
infrared vanishing gluon propagator $\sim k^2$ for small space-like
momenta~\cite{Hae90,Sti95}. Extended to include the DSE
for the 3-gluon vertex it was later concluded that the pure gauge theory 
no-longer allowed for a physically acceptable solution with infrared vanishing
gluon propagator. Solutions were instead obtained for an infrared finite gluon
propagator~\cite{Dri98a}. This statement, of course, is restricted
to the lowest order of the rational approximation employed in these studies
without radiative corrections.  
Furthermore, the inclusion of two flavours of dynamical quarks 
led to a sign change in the gluon propagator and thus to yet another  
conclusion about its possible infrared behaviour. 
This result might seem particularly questionable as
it implied tachyonic poles in the vertex functions. While the mechanism of
pole compensation does not make a formal difference between unphysical
non-tachyonic and tachyonic compensating poles, the presence of the latter is
usually a sign of a vacuum instability. The infrared limits ranging from
small but finite positive to negative values in the various cases,
these studies might nevertheless indicate that the scenario of an
infrared vanishing gluon propagator is not ruled out at
the present stage.

\section{The Dyson--Schwinger Formalism}
\label{chap_Form} 

Being the vacuum expectation values of fields the Green's functions are 
constrained by the classical action principle and the 
equal-time commutation relations. Thus, at least formally, their equations
of motion, the Dyson--Schwinger equations (DSEs)~\cite{Dys49,Sch51}, and
the existence of a generating functional are tied together. Based on purely
combinatorial arguments, DSEs can be established from
the way how particles interact based on canonical quantisation. The
generating functional can in simple cases be constructed from these
equations. On the other hand, a proper definition of the generating
functional for the Green's functions, together with a linked cluster 
theorem to classify the (dis)connected contributions, allows to derive the
corresponding DSEs. 
      
Practically, such definitions are full of mathematical difficulties. In neither
direction, from the combinatorics of interactions to the generating functional
nor the other way round, from the generating functional to DSEs, can the
derivation of both as yet be fully divorced from perturbation theory. Just as
multiplicative renormalisability, proven at every order in perturbation theory,
is assumed to hold non-perturbatively, the DSEs are nevertheless considered to
reflect the full non-perturbative dynamics of the quantum field theory they
describe.

Since lattice formulations seem to provide non-perturbative 
constructions of quantum field theory of unprecedented rigour,
a corresponding formulation of DSEs  might in principle be considered a
promising approach to their non-perturbative definition also
(see, {\it e.g.}, Ref.\ \cite{Gur99} and the references therein). In such an
approach the problem arises, however, that lattice regularised DSEs do in
general not possess unique solutions. This has been traced to ambiguities in
the selection of boundary conditions \cite{Gar96}. The search for the correct
ones is aggravated by the fact that there are so many of them. In the continuum
limit, however, many of the solutions associated with different boundary
conditions coalesce while solutions for others disappear. 
In matrix models countable sets of solutions are obtained 
in some instances, and continua of distinct solutions
somewhat analogous to the occurrence of $\theta$-vacua in others~\cite{Gar96}. 
It was furthermore observed that the set of boundary conditions for which the 
``thermodynamic limit'' exists can vary along certain paths in the space of
coupling constants thus leading to discontinuous changes in the solutions as
such ``phase boundaries'' are crossed. The phase structures of quantum field 
theories might therefore have one explanation in terms of the boundary
conditions on the solutions to their DSEs. These studies provide 
some indication towards the considerable power and flexibility  
of DSEs in describing non-perturbative physics, at least in
principle, which to a large extend is yet to be explored.

\subsection{Dyson--Schwinger Equations for QED and QCD
Propagators}
\label{sec_DSE} 

Assuming the existence of a well-defined measure in a functional
integral representation of the generating functional for QED or QCD, 
here we briefly present the derivation of the corresponding DSEs describing
the non-perturbative dynamics of electrons and photons or quarks and gluons,
respectively. To avoid unnecessary confusion, we will adopt somewhat symbolic
and simplified notations to outline the basic steps in this derivation first.

Corresponding to the classification of full, connected and proper ({\it i.e.},
one-particle irreducible) Green's functions various kinds of different
generating functionals are introduced. First, the generating functional of
Sec.~\ref{Sec2.1},
\begin{equation}
Z[j] = \int {\mathcal D} \phi \, \exp \left\{ -S[\phi ] +  j_i\phi_i \right\}
= \langle \, e^{j_i\phi_i} \, \rangle = \sum_{n=0}^\infty \, G_{i_1\ldots i_n}
\, j_{i_1} \ldots j_{i_n} \; ,      \label{genZj}
\end{equation}
generates the full Green's functions abbreviated here as $G_{i_1\ldots
i_n}$. The symbolic notation adopted herein is such that the indices 
include the (Euclidean) space-time variables the summations of which
represent 4-dimensional integrations, $j_i\phi_i := \int d^4 x
j_a(x) \phi_a(x)$ with $a$ denoting the totality of additionally 
possible discrete (Lorentz, Dirac, colour, flavour ... ) indices. 
The irreducible (connected) contributions to the full $n$-point Green's
functions can be extracted recursively by subtracting all 
possible partitionings of the $n$ points that lead to disconnected
contributions described by products of lower
$m$-point functions ($m<n$). The combinatorics in the relation between the
full Green's functions and their connected contributions is combined in 
the linked cluster theorem: With the generating functional for the full
Green's functions $Z[ j ]$ the connected ones are obtained 
from the Taylor expansion of the functional
\begin{equation}
W[j] \, := \, \ln Z[j] \; ,\quad \mbox{with normalisations} \; \; Z[0] \, =
1\; , \; \; W[0] \, = \, 0 \; ,    
\end{equation}    
which are assumed implicitly from now on. A functional Legendre transform
then leads to the generating functional $\Gamma$ for the one-particle
irreducible (1-PI) vertex functions called the effective action,
\begin{equation}
\Gamma[\phi^c] \, := \, - \ln Z[j] \, +  \phi_i^c j_i 
\; ,\; \; \mbox{with} \; \; \phi_i^c  \, = \, \frac{\delta W[j] }{\delta
j_i } \, = \, \langle \phi_i \rangle_{[j]}  \; .
\end{equation}    
Its leading term in a derivative expansion, formally obtained by inserting 
constant fields $\phi_i^c \equiv \phi^c$, with
$\Gamma[\phi^c=const] = \int d^4x \, V(\phi_c)$, yields the effective
potential  $V(\phi^c) $ which gives the energy density as a function of the 
expectation values of the fields in states $|\psi \rangle $ with 
$ \langle \psi | \phi(x) | \psi \rangle = \phi_c $. The identification
$\Gamma_{\rm tree}[\phi^c ] \, = \, S[\phi^c] $ is useful to derive the explicit 
forms of the tree-level vertices ({\it e.g.}, that of the 
tree-level ghost-gluon vertex in the ghost DSE~(\ref{ghDSE_exp})). 

Given the functional integral representation~(\ref{genZj}) 
of the generating functional is well-defined, DSEs follow from the
observation that the integral of a total derivative vanishes, 
\begin{equation} \label{genDSEzero}
0 \, = \,  \int  {\mathcal D}\phi  \; \exp\Big\{-S[\phi ] +  j_i\phi_i \Big\}
\;  \left( \frac{\delta}{\delta \phi_k} \, S[\phi] \, - j_k \right) 
 =: \, \Big\langle \, \Big(\, \frac{\delta}{\delta \phi_k} \, S[\phi]\,  -\,
 j_k  \Big)   \, \Big\rangle_{[j]}    \;  ,  \label{totDvan}
\end{equation}  
provided the measure herein, ${\mathcal D}\phi \,\exp -S[\phi] $, is invariant
under field translations, $\phi(x) \, \to \, \phi(x) + \Lambda(x) $,
for arbitrary $\Lambda(x)$, see, {\it e.g.}, Refs.~\cite{Fri72,Riv87}.   

In our condensed notation the whole infinite tower of DSEs is generated by
either of the following functional equations:
\begin{eqnarray} 
\left(- \frac{\delta S}{\delta \phi_i} \left[ \frac{\delta}{\delta j} \right] 
+ j_i \right) Z[j] &=& 0 \quad {\rm (for \; full \; Green's \; functions),} 
\label{fGF} \\
- \frac{\delta S}{\delta \phi_i} \left[\frac{\delta W}{\delta j} + 
\frac{\delta}{\delta j} \right] + j_i &=& 0 \quad {\rm (for \; connected  \;
Green's \; functions),} \label{cGF} \\
\frac{\delta \Gamma [\phi ]}{\delta \phi_i} - 
\frac{\delta S}{\delta \phi_i} \left[ \phi + \frac{\delta ^2 W}{\delta j \delta
j } \frac{\delta }{\delta \phi } \right] &=& 0 \quad {\rm (for \; proper \;
Green's \; functions).} \label{pGF}
\end{eqnarray}
This very dense notation\footnote{It can also be applied to 
obtain an overview over generating identities reflecting symmetries other
than field translations.
For a symmetry of the action in the form $\frac{\delta S}{\delta \phi_i}
F_i=0$, one obtains the Ward identities from the functional relation
\begin{equation}
j_iF_i \left[ \frac{\delta}{\delta j} \right] Z[j] = 0 .
\end{equation} 
In presence of an anomaly the anomalous Ward identities can be inferred from
\begin{equation}
\left( j_iF_i \left[ \frac{\delta}{\delta j} \right] 
- \frac{\delta F_i}{\delta \phi_i} \left[ \frac{\delta}{\delta j} \right] 
\right)  Z[j] = 0 .
\end{equation} 
The functional Slavnov--Taylor identities can be derived using non-local
transformations involving the Faddeev-Popov matrix $\mathcal M$,
\begin{equation}
j_iF_i \left[ \frac{\delta}{\delta j} \right] Z[j] =
-j_iD_i \left[ \frac{\delta}{\delta j} \right] {\mathcal M}^{-1} 
\left[ \frac{\delta}{\delta j} \right] Z[j] .
\end{equation} 
} provides the guideline for the following derivations of the DSEs for specific
Green's functions: Upon taking the $n$th derivative of
the appropriate functional, the vacuum expectation values (setting all
sources to zero) then yield the equation for the desired $n$-point function.

As a first example we will briefly review the derivation of  
the DSE for the photon propagator in QED with one species of fermions as given,
 {\it e.g.}, in Ref.~\cite{Itz80}. The DSE for the fermion propagator is
of a structure equivalent to the one for the quark propagator in QCD discussed
below. We will therefore not present its derivation separately here.
In the linear covariant gauge the QED analogue of the
Lagrangian (\ref{Leff}), {\it i.e.}, without ghost fields
and gauge field self-interactions, reads:
\begin{equation} 
 {\mathcal L}_{\mbox{\tiny QED}}   \,  =  \,  \frac{1}{2} \, A_\mu \left( -
 \partial^2 \delta_{\mu\nu} \, - \, \Big( \frac{1}{\xi} - 1 \Big)
 \partial_\mu  \partial_\nu  \right) A_\nu 
  \, + \,  \bar q \big( -{\partial \kern-.5em\slash} +  m \big) q \, - \,
  i e  \, \bar q  \gamma_\mu  q \, A_\mu   \; . 
\label{LQED}
\end{equation}
To any order in perturbation theory one effectively employs a Gaussian
measure in Eq.~(\ref{genDSEzero}) for which there is no need to worry about
possible boundary terms in the derivation of DSEs. 
Quite likely, however, this might have to be reconsidered
beyond perturbation theory~\cite{Gur99}. Ignoring this potential subtlety 
and, in addition, also assuming multiplicative renormalisability,
the renormalisation constants are formally introduced by replacing the
effective Lagrangian of Eq.~(\ref{LQED}) with the renormalised one, 
\begin{eqnarray} 
 {\mathcal L}_{\mbox{\tiny QED}}   \,  &=&  \,   Z_3 \, \frac{1}{2} \,
  A_\mu \left( -
 \partial^2 \delta_{\mu\nu} \, - \, \Big( \frac{1}{Z_3\xi} - 1 \Big)
 \partial_\mu  \partial_\nu  \right) A_\nu  \nonumber \\
 && \hskip 2cm  + \,  Z_2 \,  \bar q \big( -{\partial \kern-.5em\slash} +  
  Z_m  m \big) q \, - \,
   Z_{1F} \,  i e  \, \bar q  \gamma_\mu  q \, A_\mu   \; .
\label{LQEDren} 
\end{eqnarray} 
This defines the multiplicative constants for the photon wave function
renormalisation ($Z_3$), the fermion wave function renormalisation ($Z_2$),
the fermion mass renormalisation ($Z_m$), and for the renormalisation of the 
fermion-photon vertex (here denoted by $Z_{1F}$ to adopt a unified  
naming of constants for QED and QCD). The variation of the action 
$S_{\mbox{\tiny QED}} := \int d^4x \, {\mathcal L}_{\mbox{\tiny QED}}$
with respect to the photon field yields
\begin{equation} 
\frac{\delta S_{\mbox{\tiny QED}}}{\delta A_\mu (x)} \, = \,
Z_3 \, \left( - \partial^2 \delta_{\mu\nu} \, - \, \Big( \frac{1}{Z_3\xi} - 1
\Big) \partial_\mu  \partial_\nu  \right) A_\nu 
\, - \,  Z_{1F} \,  i e  \, \bar q  \gamma_\mu  q    \; . \label{SQEDderA}
\end{equation}
One way to proceed here is to employ Eq.\ (\ref{fGF}):
Replacing the photon field by a derivative with respect to the current
$j_\mu$, the fermion fields by derivatives with respect to their Grassmann
sources (introduced as in Eq.~(\ref{Zj})), and applying these derivatives
together with an appropriate combination of further derivatives with respect
to these sources on the full generating functional $Z$ yields the DSE for the
full Green's function under consideration. For the photon propagator, with
one additional derivative $\delta/\delta j_\nu(y) $ this amounts to
\begin{equation}
       \Big\langle \, \frac{\delta S_{\mbox{\tiny QED}}}{\delta A_\mu (x)} \,
       A_\nu(y) \, \Big\rangle \, = \, \delta_{\mu\nu} \, \delta^4(x-y)
       \label{PhDSE0} 
\end{equation}
which could have been inferred directly from Eq.~(\ref{totDvan}) in this
simple case. Inserting Eq.~(\ref{SQEDderA}) herein, one still has to work out
the linked cluster theorem for the decomposition of the full Green's
functions occurring in the DSE of the form (\ref{PhDSE0}) into 
1-PI ones explicitly by hand, however. In general, this can be a quite 
tedious task, {\it e.g.}, when applied to the gluon propagator DSE in QCD
which formally looks just as simple as Eq.~(\ref{PhDSE0}) at first.   

It might therefore be more convenient to employ the generating identity 
for proper ({\it i.e.}, 1-PI) Green's functions of Eq.\ (\ref{pGF})
directly. In the present case, this leads to:
\begin{equation} 
\left( \frac{\delta \Gamma_{\mbox{\tiny QED}}}{\delta A_\mu (x)} 
\right) _ {\bar q = q = 0} \!\! = \,
Z_3 \, \left( - \partial^2 \delta_{\mu\nu} \, - \, \Big( \frac{1}{Z_3\xi} - 
1 \Big) \partial_\mu  \partial_\nu  \right) A_\nu 
\, - \,  Z_{1F} \,  i e  \, {\rm tr} \Big( \gamma_\mu S(x,x,[A_\mu ]) \Big) .
\label{Gamma1A}
\end{equation}
Here, use has been made of the fact that one may set the fermionic fields to
zero already at this stage in the derivation of the DSE for the
photon propagator.
We furthermore defined the fermionic propagator in an external photon field
$A_\mu (x)$ as
\begin{equation} 
S(x,y,[A_\mu ]) \, = \, \left( \left(
\frac{\delta^2 \Gamma_{\mbox{\tiny QED}}}{\delta \bar q (x) \delta q (y)} 
\right) _ {\bar q = q = 0} \right) ^{-1} \, .
\end{equation}
The fermion propagator in the vacuum is obtained from the one above for 
vanishing external photon field, $A_\mu =0$, of course,
\begin{equation} 
S(x,y):= S(x,y,[A_\mu =0]) \; .
\label{fermionprop}
\end{equation}
Applying a  second derivative with respect to the
photon field  on Eq.\ (\ref{Gamma1A}), and setting $A_\mu =0$ afterwards,
we arrive at the DSE for the photon propagator in the form,
\begin{eqnarray}
D^{-1}_{\mu\nu}(x,y) \,& := & \, 
-\left( \frac{\delta^2 \Gamma_{\mbox{\tiny QED}}}{\delta A_\mu (x) 
\delta A_\nu (y)} \right) _ {A_\mu =\bar q = q = 0} 
\, = \, {D_{(0)}^{-1}}_{\mu\nu}(x,y) + \Pi_{\mu\nu}(x,y)\, ,
\label{photonDSE} \\
{D_{(0)}^{-1}}_{\mu\nu}(x,y) \, &=& \,
Z_3 \, \left( - \partial^2 \delta_{\mu\nu} \, - \, \Big( \frac{1}{Z_3\xi} - 
1 \Big) \partial_\mu  \partial_\nu  \right) \delta (x-y)  \, ,
\label{photonprop0}
\end{eqnarray} 
where we introduced the inverse of the tree-level propagator,
${D_{(0)}^{-1}}_{\mu\nu}(x,y)$, and the photon polarisation tensor
$\Pi_{\mu\nu}(x,y)$. Its explicit form, 
\begin{equation}
\Pi_{\mu\nu}(x,y) =  
e^2 Z_{1F} \int d^4u \int d^4v \, \, {\rm tr} (\gamma_\mu S(x,u) 
\Gamma_\nu (y;u,v) S(v,x) ) \; ,
\label{photonpolarization}
\end{equation}
involves the proper fermion-photon vertex function defined as,
\begin{equation}
e\Gamma_\mu (x;y,z) := \left(
\frac{\delta^3 \Gamma_{\mbox{\tiny QED}}}{\delta A_\mu (x) \delta \bar q (y) 
\delta q (z)} \right) _ {A_\mu =\bar q = q = 0} \, .
\label{photonfermionvertex}
\end{equation}
\begin{figure}[t]
  \centering\epsfig{file=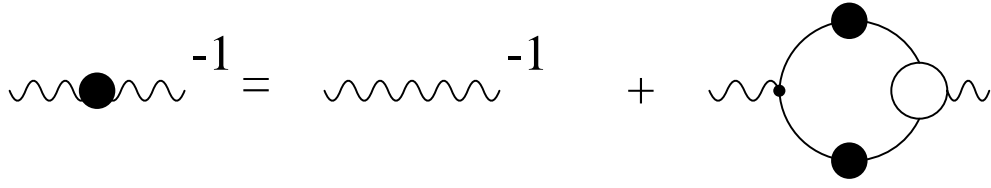,width=0.8\linewidth}
  \caption[Pictorial representation of the photon DSE.]
    {Pictorial representation of the photon DSE (\protect{\ref{photonDSE}}).}
   \label{fig:Photon-DSE}
\end{figure}
In Eqs.\ (\ref{photonDSE}) and (\ref{photonpolarization}) the general
structure of DSEs can be seen quite clearly, see also
Fig.~\ref{fig:Photon-DSE}. Here, the
DSE for the photon propagator $D_{\mu\nu}(x,y)$ contains the inverse of the 
tree-level propagator (\ref{photonprop0}),
the photon-fermion vertex (\ref{photonfermionvertex}), and the fermion  
propagator (\ref{fermionprop}). This structure describes the way the photon
propagates from $y$ to $x$. In doing so, the photon either does not interact
at all, reflected in the tree-level propagator (\ref{photonprop0}), 
or it generates a fermion-antifermion pair in one elementary interaction
which then propagates and recombines in all possible ways, {\it i.e.}, with
all radiative corrections included. This is
reflected by the fully dressed propagators (\ref{fermionprop}) and the 
fully dressed vertex (\ref{photonfermionvertex}).
Equivalently, in momentum space 
(see Appendix \ref{app_FT} for the definition of Green's functions in momentum 
space), the photon DSE reads, 
\begin{eqnarray}
D^{-1}_{\mu\nu}(k) \, &=& \, {D_{(0)}^{-1}}_{\mu\nu} (k) + \Pi_{\mu\nu}(k) \,
,
\nonumber \\
\nonumber \\
\Pi_{\mu\nu}(k) \, &=& \, - e^2Z_{1F}\int \frac{d^{\rm 4}p}{(2\pi)^{\rm 4}}
{\rm tr}\bigl(\gamma_\mu S(p) \Gamma _\nu (p,p-k))S(p-k)\bigr)\; .
\label{photonDSEmom}
\end{eqnarray}  
The gluon DSE in QCD is considerably more complicated due to
the additional elementary (self)interactions of gluons,
the elements therein are, however, all of this same general structure entailed
by the combinatorics of these interactions.   

As in QED, before we derive the DSEs of QCD, we replace 
the Lagrangian of the linear covariant gauge given in Eq.~(\ref{Leff}) by its
renormalised version,
\begin{eqnarray} 
 {\mathcal L}_{\mbox{\tiny QCD}} \,  &=&  \,  Z_3 \, \frac{1}{2} \, A^a_\mu 
 \left( -
 \partial^2 \delta_{\mu\nu} \, - \, \Big( \frac{1}{Z_3\xi} - 1 \Big)
 \partial_\mu  \partial_\nu  \right) A^a_\nu  \nonumber\\
 && + \, \widetilde Z_3 \, \bar c^a \partial^2 c^a \, + \, \widetilde Z_1 \,
 g f^{abc} \,  \bar  c^a \partial_\mu 
 (A_\mu^c c^b ) \, - \, Z_1 \, g f^{abc} \, (\partial_\mu A_\nu^a) \, A_\mu^b
 A_\nu^c  \nonumber \\
&& + \,Z_4 \,  \frac{1}{4} g^2 f^{abe} f^{cde} \, A^a_\mu A^b_\nu A^c_\mu
 A^d_\nu  \, + \, Z_2 \, \bar q \big( -{\partial \kern-.5em\slash} + Z_m m \big) q \, - \,
 Z_{1F} \,  i g  \, \bar q  \gamma_\mu  t^a q \, A^a_\mu   \;
 . \label{renLeff}  
\end{eqnarray}
This defines all multiplicative renormalisation constants. 
In addition to the QCD counterparts of those introduced for QED in
Eq.~(\ref{LQEDren}) above, these are $Z_1, Z_4, \widetilde Z_3$ and
$\widetilde Z_1$ for the gluonic self-interactions and ghost terms. 
With the coupling renormalisation defined by $Z_g g = g_0$
one all together has nine constants which are not independent of each other,
of course. In particular, one has,
\begin{eqnarray}
           Z_{1F} \,=\, Z_g Z_2 Z_3^{\frac{1}{2}} \; , \quad  Z_1 \, = \,
           Z_g Z_3^{\frac{3}{2}} \; , \quad  
           \widetilde Z_{1} \,=\, Z_g \widetilde Z_3 Z_3^{\frac{1}{2}} \; ,
           \quad && Z_4 \, = \, Z_g^2 Z_3^{2} \; . \label{renSTI}
\end{eqnarray}
These relations are maintained in the multiplicative renormalisation as a
result of the Slavnov-Taylor identities.
           
In the renormalisation procedure for $n$-point functions one first considers 
the primitively divergent functions, {\it i.e.}, the ones with a
tree-level counterpart in the Lagrangian. In QED these are the (inverse)
propagators and the fermion-photon vertex, in QCD the primitively divergent
$n$-point functions are the (inverse) quark, gluon and ghost propagators, and
the 3-gluon, 4-gluon, ghost-gluon and quark-gluon vertex functions. 
These are the seven primitively divergent vertex functions of QCD with their 
respective renormalisation constants $Z_2, \, Z_3,\, \widetilde Z_3,\, 
Z_1, Z_4, \widetilde Z_1$ and $Z_{1F}$ defined in 
Eq.\ (\ref{renLeff}) and constrained by the Slavnov-Taylor identities to obey
the relations in (\ref{renSTI}). As we will see in the following
chapters the determination of these renormalisation constants in a
non-perturbative approximation scheme is quite non-trivial. 
For all other $n$-point functions except the primitively divergent ones
renormalisation effectively amounts to replacing bare Green's functions by
renormalised ones in their corresponding skeleton expansions (which contain
only the primitively divergent Green's functions by construction).

The simplest example of a DSE in QCD is the one for the ghost propagator. 
Here we present its derivation based on Eq.\ (\ref{genDSEzero}) directly as
discussed below Eq.~(\ref{SQEDderA}). In this particular example,  
starting from Eq.~(\ref{fGF}) for a (left)
derivative of the action $S_{\mbox{\tiny QCD}} :=  \int d^4x \, {\mathcal
L}_{\mbox{\tiny QCD}}$ with respect to the anti-ghost field $\bar c^a$, and
with $\sigma $ and $\bar\sigma $ denoting the sources for (anti)ghosts as
before, acting on the full generating functional one obtains  
\begin{equation} 
 \Bigg( -\frac{\delta S_{\mbox{\tiny QCD}}}{ \delta \bar c^a(x) }
 \Bigg[\frac{\delta}{\delta j},\frac{\delta}{\delta \bar\sigma}\Bigg]
       \, + \, \sigma^a(x) \Bigg) \,  Z[j,\bar\eta , \eta ,\bar\sigma, \sigma] \Bigg|_{\bar\eta = \eta=0 }  \hskip -.4cm = \,
\Big\langle  - \frac{\delta S_{\mbox{\tiny QCD}}}{\delta \bar c^a(x)}
 \, + \, \sigma^a(x) \,
\Big\rangle_{[j,\bar\sigma,\sigma]} \hskip -.1cm = \, 0 \; ,  
\end{equation} 
where the subscript $[j,\bar\sigma,\sigma]$ indicates which non-zero sources
are retained for further derivatives. Acting with a further (right) derivative 
$\delta/\delta\sigma^b(y)$ on this equation, from (\ref{renLeff}) one
explicitly obtains, 
\begin{equation} 
\Big\langle \,\frac{\delta S_{\mbox{\tiny QCD}}}{\delta \bar c^a(x)} \bar
c^b(y) \,  
\Big\rangle  \, = \, \delta^{ab} \, \delta^4(x-y) \, = \, \widetilde Z_3 \,
\langle \, \Big( \partial_\mu D^{ac}_\mu \, c^c(x) \Big) \, \bar c^b (y) \,
\rangle  \; ,
\label{ghDSE}  
\end{equation} 
where all sources are set to zero now. This is the equation of motion for the
ghost propagator. We note here already that it will be used in the 
derivation of the gluon Slavnov--Taylor identity~(\ref{glSTI_ch2}) in the
next section. With $\langle  \, c^a(x) \, \bar c^b (y) \, \rangle =:
D_G^{ab}(x-y) $ denoting the ghost propagator, this DSE reads more explicitly,
\begin{eqnarray} 
\widetilde Z_3 \, \partial^2 \,  D_G^{ab}(x-y)  &-& \widetilde Z_1  \, g
f^{acd}  \, \int \, d^4z d^4z' \, \Big( \partial^z_\mu \delta^4(z-x) \Big)
\delta^4(z - z')  \, \times \nonumber \\
&&  \hskip 3cm \big\langle \, c^c(z') \, \bar c^b(y) \, A^d_\mu (z) \,
\big\rangle \, = \, \delta^{ab} \, \delta^4(x-y)  \; .  \label{ghDSE_exp}
\end{eqnarray} 
Again the general structure of DSEs is clearly visible. Here, the
DSE for the ghost propagator $D_G$ contains the inverse of the tree-level
propagator, $\partial^2$, the tree-level ghost-gluon vertex in coordinate
space, $ - \widetilde Z_1  \, g f^{abc}  \big( \partial^x_\mu \delta^4(x-y)
\big) \delta^4(x-z)$, and the 3-point correlation function $\big\langle \,
c^c(z) \, \bar c^b(y) \, A^a_\mu (x) \, \big\rangle $. 
As we used the  full generating functional up to now,
the latter correlation function, being a moment of $Z[j,\bar\sigma,\sigma]$,
still is a full Green's function.

\begin{figure}[t]
  \centering\epsfig{file=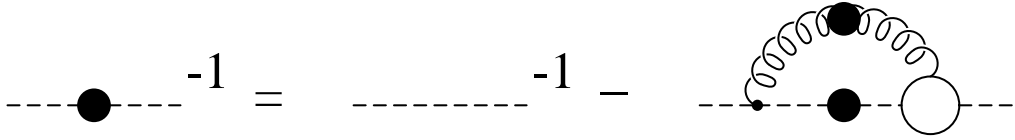,width=.8\linewidth}
  \caption[Pictorial representation of the ghost DSE.]
    {Pictorial representation of the ghost DSE.}
   \label{fig:Ghost-DSE}
\end{figure}

\begin{figure}[t]
  \centering\epsfig{file=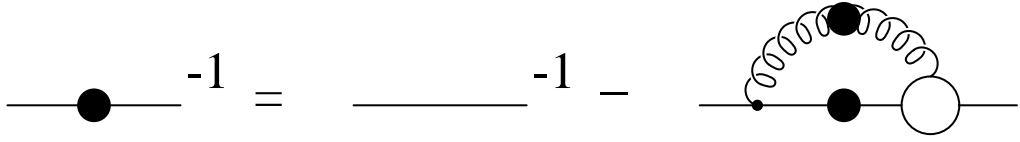,width=.8\linewidth}
  \caption[Pictorial representation of the quark DSE.]
    {Pictorial representation of the quark DSE.}
   \label{fig:Quark-DSE}
\end{figure}

In the notation introduced at the beginning of this section
the connected 3-point ghost-gluon correlation,
\begin{equation}
\big\langle \, c^c(z) \, \bar c^b(y) \, A^a_\mu (x) \, \big\rangle _{\rm conn.}
\, := \, 
\frac{\delta^3 W[j,\bar\sigma,\sigma]}{\delta\sigma^b(y)
\delta\bar\sigma^c(z) \delta j^a_\mu (x) } \Bigg|_{\bar\sigma =\sigma = j=0}\;
, \end{equation} 
can be decomposed into the proper (1-PI) ghost-gluon vertex function,
\begin{equation}
G^{abc}_\mu(x,y,z) \, = \, \frac{\delta^3 \Gamma [A,\bar c , c]}{\delta c^c(z)
\delta\bar c ^b(y) \delta A_\mu^a (x) } \Bigg|_{c = \bar c = A = 0}\; , 
\end{equation}
with ghost and gluon propagators, $D^{ab}_G(x)$ and $D^{ab}_{\mu\nu}(x)$,
respectively, attached to its legs. In the covariant formalism full and
connected 3-point functions are equivalent, and we thus obtain
\begin{eqnarray}
\big\langle \, c^c(z) \, \bar c^b(y) \, A^a_\mu (x) \, \big\rangle
&=& \big\langle \, c^c(z) \, \bar c^b(y) \, A^a_\mu (x) \, \big\rangle _{\rm
conn.} \,  \, = \nonumber \\
&&\hskip -1cm  - \int d^4u \, d^4v\, d^4w 
 D^{ad}_{\mu\nu}(x-u) \, D^{ce}_G(z-v) \, G^{def}_\nu(u,v,w) 
D^{fb}_G(w-y) \; .  \label{irgghc}
\end{eqnarray}
From such decompositions of all connected correlations,
the combinatorics of the possible interactions is reflected in the DSEs
explicitly. For the ghost propagator $D_G(x-y)$, after rearranging
Eq.~(\ref{ghDSE_exp}), this structure describes (similar to the general
structure of the DSE for the photon propagator) the way a ghost field 
propagates from $y$ to $x$ (see also Fig.\ \ref{fig:Ghost-DSE}): 
it does it by either not interacting at all, reflected in the
tree-level propagator $1/\partial^2$, or by generating a ghost-gluon pair
in one elementary interaction which then propagates and recombines in all
possible ways, {\it i.e.}, with all radiative corrections included. This is
again reflected by the fully dressed propagators and the fully dressed vertex
contained in the correlations~(\ref{irgghc}) and thus in the
DSE~(\ref{ghDSE_exp}). In momentum space the ghost propagator DSE reads:
\begin{equation}
(D_G^{-1})^{ab} (k) \, = \,  -\delta ^{ab} \widetilde Z_3 k^2 
+ g^2 f^{acd} \widetilde Z_1 \int \frac{d^{\rm 4}q}{(2\pi)^{\rm 4}}
ik_\mu D_G^{ce} (q) G_\nu^{efb}(q,k)D_{\mu\nu}^{df}(k-q) \, .
\label{ghDSEmom}
\end{equation}
Analogous to the derivation of the ghost DSE~(\ref{ghDSE_exp}), the DSE for
the quark propagator, \\$S(x-y) :=  \langle \,q(x) \bar q(y)  \, \rangle $, is
obtained as follows,   
\begin{eqnarray} 
\Big\langle \,\frac{\delta S_{\mbox{\tiny QCD}}}
{\delta \bar q(x)} \bar q(y) \, 
\Big\rangle  &=&  {\mathbf 1} \, \delta^4(x-y) \, = \, Z_2  \, 
\big( -{\partial \kern-.5em\slash }
\, +\, Z_m m \big) \, S(x-y) \nonumber\\
&& \hskip -1cm  - Z_{1F} \, i g  
\, \int \, d^4z d^4z' \, \delta^4(x-z) \,
\delta^4(x - z') \;   (\gamma_\mu t^a )   \big\langle \, q(z) \, \bar q(y) \,
A^a_\mu (z') \, \big\rangle \; .
\end{eqnarray} 
Again, it contains the inverse tree-level propagator 
$(- {\partial \kern-.5em\slash} +m
)$, the tree-level vertex, \\
$ \Gamma^{a\, \mbox{\tiny tl}}_\mu (x,y,z)  = -
Z_{1F} \, i g  \,  \gamma_\mu t^a \, \delta^4(y-x) \, \delta^4(z-x)$, and the
full correlations $ \big\langle \, q(z) \, \bar q(y) \,
A^a_\mu (x) \, \big\rangle $. These are decomposed into the proper (1-PI)
vertex function, $ \Gamma^{a}_\mu (x,y,z) $, with quark/gluon propagators
attached, in an analogous way as the correlations in Eq.~(\ref{irgghc}),
see Fig.\ \ref{fig:Quark-DSE}. For completeness we give the quark propagator
DSE in momentum space:
\begin{equation}
S^{-1} (k) = Z_2(- i {k \kern-.5em\slash } +Z_m m) + g^2 Z_{1F}
\int \frac{d^{\rm 4}q}{(2\pi)^{\rm 4}}
t^a\gamma_\mu S(q) \Gamma_\nu^b(q,k) D_{\mu\nu} ^{ab}(k-q) \, .
\label{quarkDSEmom}
\end{equation}

A first idea of the complexity to expect in general (see, {\it e.g.}, Sec.
\ref{sec_nPoint} on 3- and higher n-point Green's functions)
can be obtained from the gluon DSE,  
\begin{eqnarray}
&& 
\Big\langle \, \frac{\delta S_{\mbox{\tiny QCD}}}
{\delta A_\mu^a(x)}  A_\nu^b(y) \, \Big\rangle
\, = \, \delta^{ab} \, \delta_{\mu\nu} \, \delta^4(x-y) \\
&& \hskip .5cm =   \, Z_3 \, \Big( - \partial^2 \, \delta_{\mu\rho}\,  - \,
\bigg(\frac{1}{Z_3\xi} - 1 \bigg)\,  \partial_\mu \partial_\rho \, \Big) \,
\big\langle \, A^a_\rho(x) A^b_\nu(y) \, \big\rangle  \nonumber\\
&& \hskip .7cm - \, \widetilde Z_1 \, g f^{ade} \big\langle \, \big(
\partial_\mu\bar c^d(x) 
\big) \, c^e(x) \, A^b_\nu(y) \, \big\rangle \,  
+ \, Z_{1F} \, i g (\gamma_\mu t^a)_{ij} \, \big\langle \, q_j(x) \, \bar
q_i(x) \, A^b_\nu(y) \, \big\rangle  \nonumber\\[8pt]
&& \hskip .7cm
+ \, Z_1 g f^{ade}\,  \big\langle \, \big\{ (\partial A^d(x)) \, A_\mu^e(x)  \,
+\, (\partial_\mu A^d_\rho(x) ) \, A_\rho^e(x)  \,  - \, 2\, (\partial_\rho
A_\mu^d(x) )\, A_\rho^e(x)  \big\}  \, A_\nu^b (y)  \,\big\rangle \nonumber\\[8pt]
&& \hskip 2cm   + \, Z_4 \,  g^2 f^{ack} f^{dek} \, \big\langle \, 
A^c_\rho(x)  \, A^d_\mu(x) \, A^e_\rho(x) \, A^b_\nu(y) \, \big\rangle 
\; .\nonumber
\end{eqnarray}
The full decomposition of all disconnected correlations herein blows up the
notations considerably. It is straightforward but lengthy and is therefore 
given in Appendix \ref{app_gluon-DSE}. 
The probably least suspicious term in the last line generates the majority 
of terms in this
decomposition. It contains the tree-level 4-gluon vertex and gives rise to
a tadpole as well as explicit 2-loop contributions to the gluon DSE. These
will both not be included in the DSE solutions presented in
Sec.\ \ref{sec_Trunc} below. The discussion and the necessary truncation of
the DSEs for the propagators of QCD, derived formally
here, is continued in Chapter~\ref{chap_QCD}. 
Before discussing some issues on DSEs for 3- and higher $n$-point
functions in Sec.\ \ref{sec_nPoint} we will exploit their BRS invariance 
to derive the Slavnov-Taylor identities.

\begin{figure}[t]
  \centering\epsfig{file=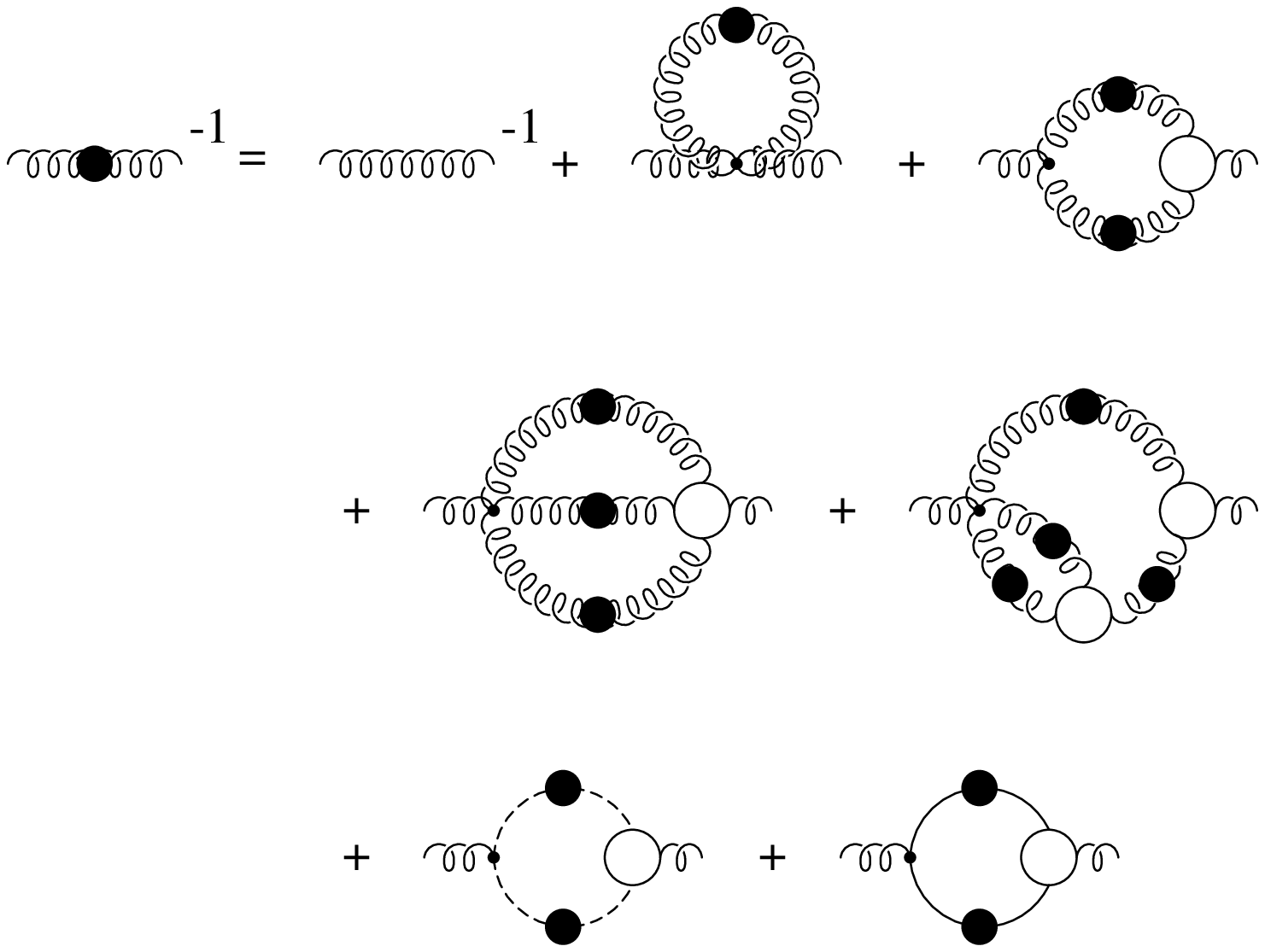,width=.8\linewidth}
  \caption[Pictorial representation of the gluon DSE.]
    {Pictorial representation of the gluon DSE.}
   \label{fig:Gluon-DSE}
\end{figure}

Despite the formal equivalence of the DSEs in different gauges their explicit
form as integral equations are, of course, different. We would therefore like
to add two comments on the gluon propagator in non-covariant gauges at the 
end of this section.   
One might think that the particular problem with ghosts in the Landau gauge can
be avoided using gauges such as the axial gauge, in which there are no ghosts
in the first place.\footnote{A further note of warning seems appropriate here.
In a recent study it has been shown by using ``interpolating'' gauges that
axial and light-like gauges are highly singular in the sense that infinite
contributions to Feynman graphs are generated in the limit of a pure axial or
light-like gauge \cite{Bau99}.} As for studies of the gluon DSE in
the axial gauge, in which the gluon field is taken transverse to a fixed
gauge vector $t_\mu$, $t_\mu A^a_\mu = 0$, it is important to note that,
so far, these rely on a simplifying assumption on the tensor structure of the
gluon propagator \cite{Bak81a,Bak81b,Sch82,Cud91,Gei99}. 
In particular, an independent
additional term in the structure of the gluon propagator has not been
included in presently available studies of the gluon DSE in axial
gauge~\cite{Bue95}. This term vanishes in perturbation theory, see Appendix
\ref{app_gluon-DSE} for an explicit definition of the corresponding 
tensor multiplying this term. If, however,
the complete tensor structure of the gluon propagator in axial gauge
is taken into account properly, one arrives at a coupled system of equations
which is of considerably higher complexity than the ghost-gluon system in the
Landau gauge, see Sec.~\ref{sec_Axial}.
Thus, despite the fact the gluon propagator DSE contains in axial gauge one
term less than in linear covariant gauges (the ghost loop) its explicit form
is much more complicated. 

A final comment concerns the use of the Coulomb gauge in the present context. 
From the fact that the field $gA_0$ is invariant under
renormalisation in Coulomb gauge one concludes that the zero-zero component
of the gluon propagator allows to extract the running coupling, see, {\it
e.g.}, Ref.~\cite{Bau99} and the references therein. However, 
considerable complications due to the additional terms which 
arise in the proper Coulomb gauge Lagrangian as compared to 
a naive canonical quantisation procedure~\cite{Chr80,Bau99}, see also Chapter
11 in \cite{Riv87}, have meant that the non-perturbative gluon propagator has
received very little attention in Coulomb gauge. While comprehensive studies  
of the gluon propagator in the Coulomb gauge are not available as
yet,\footnote{Only very recently a lattice study of the gluon
propagator in the Coulomb gauge has been preformed, see Ref.\ \cite{Cuc00}. }
it has widely been used to
study the quark propagator employing ans\"atze for the gluon propagator and the
quark-gluon vertex, see
Refs.~\cite{Fin82,Adl84,LeY84,Koc86,Alk87,Alk88,Alk89,Kle88} as well as    
\cite{Wil97} for a recent discussion.

\subsection{BRS Invariance of Green's Functions}
\label{sec_BRS} 

A classical gauge theory is by construction invariant under local gauge
transformations. As discussed in the last chapter, quantising a gauge theory 
necessitates gauge fixing and in general the introduction of Faddeev-Popov
ghosts. We noted already that the BRS transformation (\ref{BRS_2})
acting on gluons and quarks is nothing else than a gauge transformation with
the ghost field as ``gauge parameter'', and that
the total Lagrangian is invariant under the global BRS transformations. 

Here, we make use of the observation that the BRS invariance is an (unbroken)
symmetry of the gauge fixed
theory which, from Eq.~(\ref{genSTI}) in Sec. 2.4.1, implies
that all QCD Green's functions are BRS invariant \cite{Lle80}. This provides 
the easiest way to derive the Slavnov--Taylor identities of QCD
~\cite{Tay71,Sla72}. 
First, we introduce the renormalised BRS
transformations for linear covariant gauges ({\it c.f.}, the unrenormalised
ones in Eqs.\ (\ref{BRS_2})) by defining $\delta\lambda _R :=  Z_3^{-1/2}
\widetilde Z_3^{-1/2} \delta\lambda$:
\begin{equation}
\begin{array}{cc}
\delta_{\mbox{\tiny R}} A^a_\mu \, =\,   \widetilde Z_3 D^{ab}_\mu c^b 
\delta\lambda_R  \; , \quad &  \delta_{\mbox{\tiny R}} q
\, = \,-  \widetilde Z_1 i g \, t^a c^a \, q \delta\lambda_R \; , \\
\delta_{\mbox{\tiny R}} c^a \, = \, - \widetilde Z_1{\displaystyle\frac{g}{2}} 
           f^{abc} \, c^b c^c \,  \delta\lambda_R \; , \quad
& \delta_{\mbox{\tiny R}}\bar c^a \, = \, {\displaystyle\frac{1}{\xi}} 
      \partial_\mu A_\mu^a \delta\lambda_R  \; .
\end{array}   \label{BRS_R}
\end{equation}
Note that in non-linear gauges the last relation has to be replaced by
\begin{equation}
\delta_{\mbox{\tiny R}}\bar c^a \, = \,  Z_3^{-1/2} \frac{1}{\xi} F^a [ Z_3^{1/2}
A]  \delta\lambda_R  
\end{equation}
with $F^a$ being the appropriate gauge fixing condition.

As the most elementary example to demonstrate how this symmetry constrains
Green's functions we consider 
\begin{equation}
\langle A^a(x) \bar c^b(y) \rangle \, :=\;  \frac{\delta^2
Z[j,\bar\eta , \eta ,\bar\sigma, \sigma]}{ \delta\sigma (y) \delta j^a(x) }
\Bigg|_{j=\bar\eta = \eta=\bar\sigma=\sigma=0 } \; .
\end{equation}
From the BRS transformations given in Eqs.~(\ref{BRS_R}) one obtains:
\begin{equation}
\delta_{\mbox{\tiny R}} \langle A^a(x) \bar c^b(y) \rangle \, = \,
- \widetilde Z_3 
\langle \Big(D^{ac}_\mu c^c(x) \Big) \bar c^b(y) \rangle \, \delta\lambda_R
\, + \, \frac{1}{\xi} \langle A^a_\mu(x) \partial_\nu A_\nu^b (y) \rangle \,
\delta\lambda _R\, = \, 0 \; .   \label{BRSgluon}
\end{equation}
Using the ghost DSE (\ref{ghDSE}),
$\widetilde Z_3 
\langle\, \big(\partial_\mu D^{ac}_\mu c^c(x)\big) \, \bar c^b(y)\, \rangle \,
\, = \, \delta^{ab} \, \delta^4(x-y)$,
it immediately follows from acting with $\partial_\mu$ on
Eq.~(\ref{BRSgluon}) that the longitudinal part of
the gluon propagator $D_{\mu\nu}^{ab}(x-y)$ is not modified by
interactions,
\begin{equation}
\langle \partial A^a_\mu(x) \, \partial_\nu A_\nu^b (y) \rangle \,  = \,  -
\partial_\mu \partial_\nu \, D^{ab}_{\mu\nu} (x-y) \, = \,
\xi \; \delta^{ab}  \, \delta^4(x-y) \; .   \label{glSTI_ch2}
\end{equation}
This easiest example of a Slavnov--Taylor identity implies that, 
in Euclidean momentum space, the gluon propagator in the
covariant gauge has the general structure,
\begin{equation}
 D^{ab}_{\mu\nu}(k) \, =\,  \delta^{ab} \left(
 \left(\delta_{\mu\nu} - \frac{k_\mu k_\nu}{k^2}
  \right) \,   \frac{Z(k^2)}{k^2} \, + \xi \, \frac{k_\mu
  k_\nu}{k^4} \right)\; ,
 \label{eq:Gluon-Prop}
\end{equation}
involving one unknown invariant function, the gluon renormalisation function
$Z(k^2)$. 

Obviously, an analogous statement is true for the photon propagator. In QED
there is a Ward--Takahashi identity similar to Eq.\ (\ref{glSTI_ch2}). This
means that the Fourier transform of the photon polarisation tensor 
(\ref{photonpolarization}),\footnote{As we will consider
QED for different values of space-time dimensions in the next chapter we
generalise here already from $D=4$ to an arbitrary dimension $D$.}
\begin{equation}
\Pi_{\mu\nu}(k)= - e^2 Z_{1F}\int \frac{d^{\rm D}p}{(2\pi)^{\rm D}}
{\rm tr}\bigl(\gamma_\mu S(p) \Gamma _\nu (p,p-k))S(p-k)\bigr)\;,
\label{PiD}
\end{equation} 
is purely transverse, $k_\mu \Pi_{\mu\nu}(k)=0$. Defining the dimensionless
function $\Pi(k^2)$ via
the relation $\Pi_{\mu\nu}(k)=\left({k^2}\delta_{\mu\nu} - {k_\mu k_\nu}
\right)  \Pi(k^2)$ one finds that the general structures of the photon and the
gluon propagator agree provided one identifies $Z(k^2)= 1/(1+\Pi(k^2))$.
Of course, the corresponding renormalisation functions $Z(k^2)$ are obtained
from their respective DSEs and will thus be completely different.

Similarly, other Slavnov--Taylor identities can be derived, most importantly,
for 3-point correlation functions which typically relate their longitudinal
pieces to the lower 2-point correlations, {\it i.e.}, to the propagators. 
An important example is already provided by the Ward identity for the 
electron-photon vertex function in QED,
\begin{equation}
k_\mu  \Gamma _\mu (p+k,p) = S^{-1} (p+k) - S^{-1} (k).
\label{Ward}
\end{equation}  
For $k\to 0$ the differential Ward identity results, 
\begin{equation}
\Gamma _\mu (p,p) = \frac {\partial }{\partial p_\mu} S^{-1} (p).
\label{dWard} 
\end{equation} 
Note that the differential identity also constrains the transverse
parts of the electron-photon vertex function in this limit.

Another important example is provided by the derivation of a
Slavnov-Taylor identity for the ghost-gluon vertex
function obtained recently in Ref.~\cite{Sme98}.\footnote{This vertex
function will play a quite central role 
throughout the presentations of the solutions for the propagators of Landau
gauge QCD in the following chapters.}
In order to construct a non-perturbatively dressed gluon-ghost vertex, one
can start from the BRS invariance of the following particular correlator,
\begin{equation}
  \delta_{\mbox{\tiny R}} \langle c^a(x) \, \bar{c}^b(y) \, \bar{c}^c(z)
   \rangle    = 0  \, ,
\end{equation}
from which one obtains with Eqs.\ (\ref{BRS_R}),
\begin{eqnarray}
  \widetilde Z_1 \, \frac{1}{2} \, g f^{ade}
    \langle c^d(x) \, c^e(x) \, \bar{c}^b(y) \, \bar{c}^c(z) \rangle  
  = \frac{1}{\xi} \, \langle c^a(x) \bigl( \partial A^b(y) \bigr) \bar{c}^c(z) 
  \rangle
    - \frac{1}{\xi} \, \langle c^a(x) \, \bar{c}^b(y) \bigl( \partial A^c(z) 
    \bigr) \rangle \,  . \nonumber \\
  \label{eq:ghost-STI}
\end{eqnarray}
Note that this symbolic notation refers to full (and thus reducible)
correlation functions. 
The two terms on the r.h.s.\ of Eq.\ (\ref{eq:ghost-STI}) can be
decomposed in the gluon-ghost proper vertex and the respective propagators. The
derivative on the gluon leg thereby projects out the longitudinal part of the
gluon propagator which, by virtue of its own Slavnov-Taylor identity $k_\mu
D_{\mu\nu}(k) = \xi k_\nu/k^2$, remains undressed in the covariant gauge, see
Eq.\ (\ref{eq:Gluon-Prop}). The l.h.s.\ contains a disconnected part plus terms
due to ghost-ghost scattering, 
\begin{equation}
  \langle c^d c^e \bar{c}^b \bar{c}^c \rangle
    = \langle c^e \bar{c}^b \rangle \langle c^d \bar{c}^c \rangle
      - \langle c^e \bar{c}^c \rangle \langle  c^d \bar{c}^b \rangle
      + \langle c^d c^e \bar{c}^b \bar{c}^c \rangle_{\rm conn.} \; .
\end{equation}
If the truncating assumption is made at this
point, and this is done consistently for all irreducible scattering kernels in
the truncation scheme for the DSEs developed in Ref.~\cite{Sme98}, that
ghost-ghost scattering contributions to the vertex are neglected herein,
only the reducible part of the correlation function on the l.h.s.\ of
the Slavnov-Taylor identity in Eq.\ (\ref{eq:ghost-STI}) is maintained which
corresponds to the disconnected ghost propagation. Decomposing the
ghost-gluon 3-point correlation functions on the r.h.s of
Eq.\ (\ref{eq:ghost-STI}) into the irreducible ghost-gluon vertex function
$G^{abc}_\mu(x,y,z)$ and gluon and ghost propagators according to 
Eq.~(\ref{irgghc}) then yields, 
\begin{eqnarray}
 &&\widetilde Z_1  \frac{1}{2} g f^{ade} \left\{ D_G^{eb}(x-y) \, D_G^{dc}(x-z)
         - D_G^{db}(x-y) \, D_G^{ec}(x-z) \right\} \, = \nonumber \\
 && \hskip 1cm   \frac{1}{\xi} \int d^4\!u \, d^4\!v \, d^4\!w \;
    \partial^z_\mu D^{cd}_{\mu\nu}(z-u) \, D_G^{ae}(x-v) \,
      G^{def}_\nu(u,v,w) \, D_G^{fb}(w-y) \nonumber \\
 && \hskip 2cm  - \frac{1}{\xi} \int d^4\!u \, d^4\!v \, d^4\!w \;
    \partial^y_\mu D^{bd}_{\mu\nu}(y-u) \, D_G^{ae}(x-v) \,
      G^{def}_\nu(u,v,w) \, D_G^{fc}(w-z) \, .
\end{eqnarray}
Fourier transforming this expression one obtains,
\begin{eqnarray}
  \widetilde Z_1  \frac{1}{2} g f^{ade} \, & (2\pi)^4 \, \delta^4(p+q+k)
    \biggl\{ D_G^{eb}(-q) \, D_G^{dc}(-k) - D_G^{db}(-q) \, D_G^{ec}(-k)
    \biggr\} = \nonumber\\ 
   &  \frac{1}{\xi} i k_\mu \, D^{cd}_{\mu\nu}(k) \, D_G^{ae}(p)
                     \, G^{def}_\nu(k,p,-q) \, D_G^{fb}(-q) \nonumber\\
    &\hskip 1cm -\frac{1}{\xi} i q_\mu \, D^{bd}_{\mu\nu}(q) \, D_G^{ae}(p)
                     \, G^{def}_\nu(q,p,-k) \, D_G^{fc}(-k)
\end{eqnarray}
Therefore, when irreducible ghost-ghost correlations are neglected,
the Slavnov-Taylor identity (\ref{eq:ghost-STI}) yields an equation for the   
ghost-gluon vertex which relates its longitudinal part to a sum of inverse 
ghost propagators in some sense reminiscent of the Abelian Ward--Takahashi
identities in QED. Separating the momentum conserving $\delta$-function and
the colour structure from the ghost-gluon vertex (see below), and defining the
ghost renormalisation function $G(k^2)$, via  
\begin{equation}
           G^{abc}_\mu(k,q,-p) =  (2\pi)^4 \delta^4(k+q+p)
           gf^{abc} \, G_\mu(q,-p) \; ,  \;\mbox{and} \;\;       
                       D_G^{ab}(p) = -\delta^{ab} G(p^2)/p^2 \, , 
\end{equation}
respectively, one finally arrives at
\begin{equation}
    i k_\mu  G_\mu(p,-q) \, G(q^2)
 \,  +\,  i q_\mu  G_\mu(p,-k) \, G(k^2)
          = p^2  \widetilde Z_1 \, \frac{G(k^2) \, G(q^2)}{G(p^2)}   ,
  \label{eq:ghost-WTI}
\end{equation}
where $p+q+k=0$. Note that this equation is valid for all $\xi$ including
$\xi=0$, {\it i.e.}, for Landau gauge in which one has $\widetilde Z_1 =
1$, see Sec.~\ref{sub_Sub}. A solution to this identity for the
ghost-gluon vertex will be given in Sec.~\ref{sub_Ghost}. 

Interestingly, the simplest solution to Eq.~(\ref{eq:ghost-WTI}) given by,
\begin{equation}
       G_\mu(q,p) = i q_\mu \frac{G(k^2)}{G(q^2)} \; ,
\end{equation}
was employed also in Ref.~\cite{Elw95,Elw96} in which solutions to the flow
equations in Wilson's renormalisation group approach were studied.  
As we will see in Sec.~\ref{sub_Ghost}, this solution admits a particularly
simple solution also to the Slavnov--Taylor identity for the 3-gluon vertex
function. However, it does not reflect the ghost-antighost
symmetry of the Landau gauge, see appendix~\ref{App.BRS}. A ghost-antighost 
symmetric solution to Eq.~(\ref{eq:ghost-WTI})  
which leads to the same 3-gluon vertex function~\cite{Sme98}
will be presented in Sec.~\ref{sub_Ghost}.

\subsection{The Structure of Vertex Functions}
\label{sec_nPoint} 

DSEs for 3- and higher $n$-point
functions become increasingly more complicated in structure, see
Ref.~\cite{Eic74}. 
Before we continue their discussion, we first have to 
introduce some further definitions regarding the 3-point functions of
QCD. The following notations 
are used to separate the structure constants  $f^{abc}$ of the gauge group
$SU(N_c = 3)$ (and the coupling $g$) from the 3-gluon vertex 
function,
\begin{equation}
    \Gamma^{abc}_{\mu\nu\rho}(k,p,q) =
                                   g f^{abc} (2\pi)^4 \delta^4(k+p+q)
\Gamma_{\mu\nu\rho}(k,p,q)     \; .
\quad \hbox{
\begin{minipage}[c]{0.25\linewidth}
  \epsfig{file=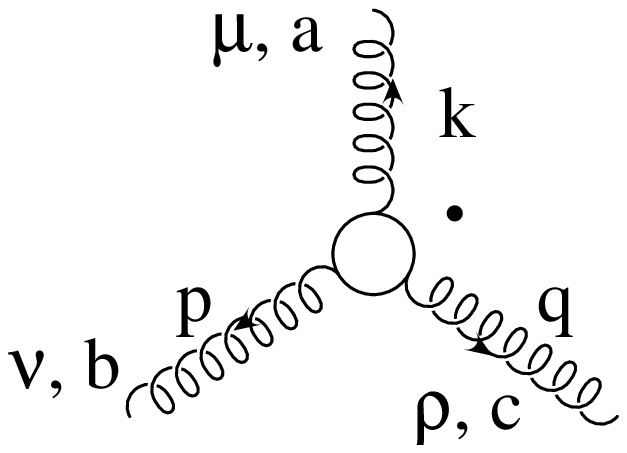,width=0.8\linewidth}
\end{minipage}}            \label{3GVdef}
\end{equation}
The arguments of the 3-gluon vertex denote the three outgoing gluon momenta
according to its Lorentz indices (counter clockwise starting from the dot).
With this definition, the tree-level vertex has the form, 
\begin{equation}
\Gamma^{(0)}_{\mu\nu\rho}(k,p,q) \, = \,-  i(k-p)_\rho
\delta_{\mu\nu} \, - \, i(p-q)_\mu \delta_{\nu\rho} \, - \, i(q-k)_\nu
\delta_{\mu\rho} \; .
\end{equation}
The possibility of colour structures other than the perturbative ones assumed so
far was recently assessed for the pure Landau gauge theory 
by lattice simulations in Ref.~\cite{Bou98}. 
In this study, the (gluon and ghost) propagators were found to be
proportional to $\delta^{ab}$ to an accuracy of the order of 1\%. For the
3-gluon Green's function, a possibly symmetric colour structure $\sim d^{abc}$
was also verified to be negligible except at the very largest lattice momenta
considered.  

The arguments of the ghost-gluon vertex denote in the following order two
outgoing momenta for gluon and ghost, and one incoming ghost momentum,
\begin{eqnarray}
     G^{abc}_\mu(k,q,p) \, = \,  (2\pi)^4 \delta^4(k+q-p) G^{abc}_\mu(q,p) \;
     &,&  \label{ghglvdef} \\
     G^{abc}_\mu(q,p) \, = \,  g f^{abc} G_\mu(q,p)
                      = g f^{abc} iq_\nu \widetilde{G}_{\mu\nu}(q,p) \; &,&
\quad \hbox{\hskip .8cm
\begin{minipage}[c]{0.25\linewidth}
  \epsfig{file=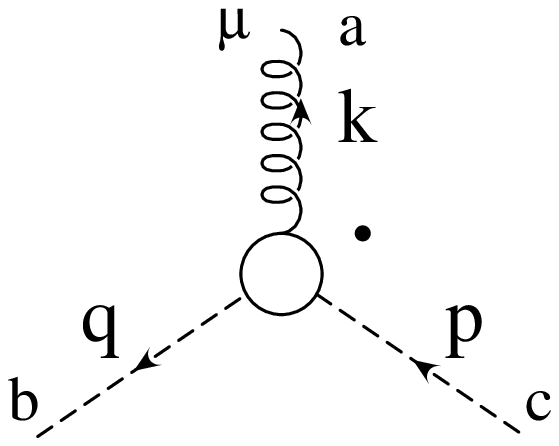,width=0.7\linewidth}
\end{minipage}} \nonumber
\end{eqnarray}
where the tensor $\widetilde{G}_{\mu\nu}(q,p)$ contains the ghost-gluon
scattering kernel (for its definition see, {\it e.g.}, Ref.~\cite{Bar80}). At
tree-level this kernel vanishes which corresponds to
$\widetilde{G}_{\mu\nu}(q,p) = \delta_{\mu\nu}$. Note that the colour
structure of the ghost and the gluon loop diagrams in the gluon DSE
(\ref{glDSE}) as well
as the one in the ghost equation (\ref{ghDSE}) are simply given
by $f^{acd} f^{bdc} = - N_c \delta^{ab}$. The quark loops involve the colour
generators in the fundamental representation $t^a$ normalised to $ {\rm tr} (t^a
t^b ) = \delta^{ab}/2 $ and giving rise to the quadratic Casimir operator
$t^a t^a = C_f  $ (with $C_f = 4/3 $ for $N_c = 3$). The colour
structure of the quark-gluon vertex is made explicit by
\begin{equation}   \label{qkv_def}
     \Gamma^a_\mu(k,q,p) =   - ig t^a \, (2\pi)^4 \delta^4(k+q-p)
     \Gamma_\mu(q,p)  \; ,
\quad \hbox{\hskip .8cm
\begin{minipage}[c]{0.25\linewidth}
  \epsfig{file=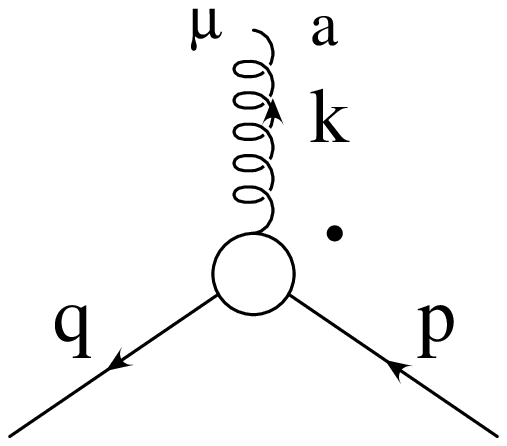,width=0.7\linewidth}
\end{minipage}}
\end{equation}
the convention for momentum arguments is analogous to the one adopted for the
ghost-gluon vertex, and the tree-level vertex is given by $ \Gamma_\mu(q,p) =
\gamma_\mu $. The trivial colour structure of the propagators $\sim
\delta^{ab}$ is thus reproduced by the contractions in the loops.

In linear covariant gauges 
the Slavnov--Taylor identity for the 3-gluon vertex in momentum space is
given by \cite{Bar80,Bal80,Kim80},
\begin{eqnarray}
  i k_\rho \Gamma_{\mu\nu\rho} (p,q,k) &=&  G(k^2) \, \left\{ \,
  \widetilde{G}_{\mu\sigma}(q,-k) \, {\mathcal P}_{\sigma\nu} (q) \,
  \frac{q^2}{Z(q^2)}
  \right. \label{glSTI}\\
  && \hbox{\hskip 3cm}  - \left.
  \widetilde{G}_{\nu\sigma}(p,-k) \, {\mathcal P}_{\sigma\mu}
  (p) \, \frac{p^2}{Z(p^2)} \right\}  \; ,
  \nonumber
\end{eqnarray}
where ${\mathcal P}_{\mu\nu}(k) = \delta_{\mu\nu} - k_\mu k_\nu /k^2$ is the
transverse projector. A simple solution to (\ref{glSTI}) is possible if
ghosts are neglected completely. However, this is not satisfactory
for a complete truncation scheme including ghost contributions in linear
covariant gauge neither does it account for the correct renormalisation
scale dependence of the 3-gluon vertex. We will therefore
postpone a more complete discussion to Sec.\ \ref{sub_Ghost} where the
inclusion of ghosts into the coupled gluon-ghost DSEs is discussed in detail.

One of the appealing aspects of the axial gauge is that the Slavnov--Taylor
identity for the 3-gluon vertex $\Gamma_{\mu\nu\rho}$
in axial gauge has the comparatively simple form,
\begin{equation}  \label{STI_ag}
ik_\rho \Gamma_{\mu\nu\rho} (p,q,k) \, =\, \Pi_{\mu\nu}(q) \, -  \,
\Pi_{\mu\nu} (p) \; .
\end{equation}
In particular, no unknown contributions from higher correlation functions such
as the 4-point ghost-gluon scattering kernel as in linear covariant gauges 
enter in this axial gauge identity for the 3-gluon vertex. Retaining the most
general tensor structure, however, its solution becomes quite involved. With
the additional requirement that it should be free from kinematic
singularities  the solution was obtained in Ref.~\cite{Kim80}. Its rather
complicated form is given in Appendix \ref{app_3-gluon}. It should be
emphasised that the solution \ref{axg_3gv} for the longitudinal part of the
3-gluon vertex is not only free from singularities of the type $1/(p^2 -q^2)$
but also from the typical axial gauge singularities of the form
$1/(pt)$~\cite{Kim80}. 

Of course, Slavnov--Taylor or Ward identities are only helpful in constructing
the longitudinal parts of vertex functions (except at vanishing or infinite
momenta where they also constrain the transverse components). In order to
determine the full vertex functions one would have to solve their corresponding
DSEs. Being the least complicated case we will discuss the fermion-photon 
vertex $\Gamma_\mu (k,q,p) =  -i e (2\pi)^4 \delta^4(k+q-p) \Gamma_\mu (q,p)$. 
Its DSE, 
\begin{eqnarray}  \label{fermion--photon}
\Gamma_\mu (q,p) = Z_2 \gamma_\mu + \int  \frac{d^{\rm D}l}{(2\pi)^{\rm D}}
 S(q+l) \Gamma_\mu (q+l,p+l) S(p+l) K(p+l,q+l,l) \; ,
\end{eqnarray}
is an inhomogeneous Bethe--Salpeter equation \cite{Rob94}. The kernel $K$ is
defined as the sum of all amputated  two-particle irreducible 
contributions,\footnote{More precisely: all contributions which are 
one- and two-particle irreducible in the $s$-channel.}
{\it i.e.}, it is the renormalised amputated fermion-antifermion scattering
kernel. 

It should be noted that the vertex function $\Gamma_\mu (q,p)$ can have 
a pole at $p^2=-m^2$ corresponding to bound states of mass $m$ with these
particular quantum numbers. This is especially
interesting when the fermions are not only interacting electromagnetically. It
has been verified explicitely in the ladder-rainbow approximation using 
a model gluon propagator as interaction between quark and antiquark that the
quark-photon vertex displays such a pole at the vector meson mass
\cite{Mar00,Fra95}. 

The general form of the fermion-photon vertex can be
decomposed into twelve linearly independent Lorentz covariants. Whereas the 
four longitudinal components are highly constrained by the Ward identity
(\ref{Ward}), for all photon momenta $k_\mu = p_\mu - q_\mu$ the eight 
transverse components are only affected by the differential Ward identity
(\ref{dWard}) at $k=0$. The solution for the quark-photon vertex given in
Ref.\ \cite{Mar00} exhibits several interesting features. First, the
longitudinal components agree perfectly with the Ball-Chiu ansatz \cite{Bal80}
(see also the Secs.\ \ref{sec_QED3} and \ref{sec_QED4} for its explicit form).
Secondly, all eight transverse components were demonstrated clearly to contain
the vector meson pole. We will discuss these issues further in Sec.\
\ref{sub_EMFF}. 
Unfortunately, however, already the ladder-rainbow calculation presented 
in Ref.\ \cite{Mar00} turned out to be technically quite involved.
 
\begin{figure}
\centerline{\epsfxsize 15.0cm\epsfbox{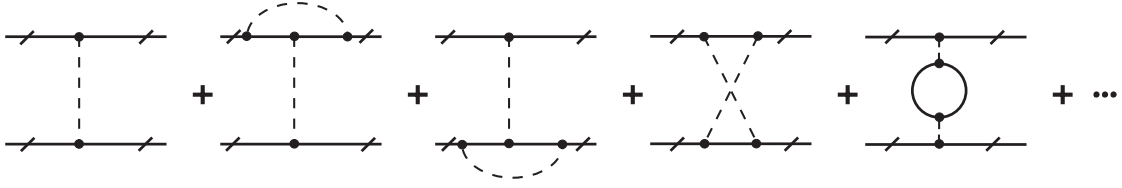}}
\caption[Feynman diagrams of the first few terms in a
perturbative expansion of the kernel $K$.]
{Feynman diagrams of the first few terms in a
perturbative expansion of the kernel $K$. Solid lines represent propagators of
the constituents, dashed lines the propagator of the exchange particle,
{\it i.e.} photons for QED and gluons for QCD.
\label{K1}}
\end{figure}

An obvious complication for considering DSEs of vertex functions is the fact
that the kernel $K$ is defined in a diagrammatic language, a first few terms
in a perturbative (or a skeleton) expansion which occur in QED and QCD are
shown in Fig.\  \ref{K1}.\footnote{As the kernel $K$ is a two-particle
irreducible amplitude it can be obtained by two-field Legendre transforms, see
Refs.\ \cite{Dom64,Vas72}.} Thus, a self-consistent treatment of vertex
functions, especially in QCD, hardly seems feasible. It is therefore quite
remarkable that the DSEs for all primitively divergent functions have been
studied in the approximation scheme of Ref.~\cite{Sti95}, see Refs.\
\cite{Dri98a,Dri98b}. 
As this approach will briefly be presented in Sec.\ \ref{sec_Stingl} below,
we also postpone the corresponding discussion until then.

Some toy models have also been employed to solve DSEs of vertex functions. 
A simplified equation for the 3-gluon vertex has,
{\it e.g.},  been solved in Ref.\ \cite{Cor89} based on the so-called pinch
technique, see Refs.~\cite{Cor82,Pap00,Wat99}. It might be interesting
to note that an infrared singular 3-gluon vertex results
in this study for massless gluons. In Ref.\ \cite{Cor89} an explicit gluon
mass was introduced to avoid this singularity. This ambiguity underlines the
necessity to understand the qualitative behaviour of 2-point functions, {\it
i.e.}, propagators, before a more systematic study of
higher $n$-point functions can be attained.

\section{Green's Functions of QED}
\label{chap_QED} 

\subsection{1+1 Dimensional QED (Schwinger model):
Dyson--Schwinger Equations in a Model with Instantons}
\label{sec_QED2} 

QED with one massless fermion in 1+1 dimensions is called the Schwinger model
\cite{Sch62a}. There is an enormous number of papers on this exactly solvable
model which displays phenomena like confinement and dynamical mass generation
in a Kosterlitz--Thouless phase.
It therefore serves as a ``theoretical laboratory'' (for a recent 
review see \cite{Ada97}).  Here we are particularly interested in the aspect 
that instantons exist in the Schwinger model. In fact, an instanton vacuum
is needed to obtain a consistent quantisation.  In the Schwinger model there is
an infinite number of vacua labelled by the (integer) winding
number $n$. As in QCD, instantons relate different vacua and the instanton
number $k=n_+-n_-$ can be calculated from the corresponding Pontryagin index
density. Furthermore, a superposition of these vacua is introduced as the
physical vacuum,
\begin{equation}
|\Theta \rangle = \sum_{n=-\infty}^{\infty} e^{in\Theta} |n\rangle ,
\end{equation}
which is characterised by a vacuum angle $\Theta$.

It has been shown by explicit calculation \cite{Ada96} that the DSEs for the 
$n$-point functions of the Schwinger model hold separately in every instanton 
sector. This fact is
related to the invariance of the model with respect to the vacuum angle
$\Theta$: For general $\Theta$ a Green's function would pick up a phase $\exp
(ik\Theta )$ from a sector with instanton number $k$. If the DSEs were to
enforce cancellations between different instanton sectors, the related
conditions would acquire a $\Theta$ dependence. On the other hand, as the
Schwinger model is invariant with respect to $\Theta$ the DSEs have to be, too.
And the fact that the DSEs are separately fulfilled in every sector is the
simplest possibility to ensure the required $\Theta$ invariance.  Thus, at least
for the Schwinger model, it has been demonstrated in Ref.~\cite{Ada96}  
that the DSEs can be extended to cases in which the true
vacuum is not the perturbative one but a $\Theta$ vacuum  (or a vacuum
with a definite number of instantons).

Of course, the fact that the DSEs hold independently in each sector does not
imply that the solutions to Green's functions are unaffected by the presence
of instantons. This is demonstrated by an explicit solution for the fermion
propagator and the 4-fermion Green's function in Ref. \cite{Rad99}.
The corrections due to the sectors with non-zero instanton number 
are shown to be just the terms allowed by the general structure of the
DSEs. The $\Theta$
dependence of the Green's functions found in this calculation can be removed by
a suitably chosen chiral transformation and is thus trivial. 
{\it E.g.}, the fermion propagator is of the form ($S_0(x)$ being the tree-level
propagator)
\begin{equation}
S(x) = S_0(x) e^{-ie^2\beta (x)} + ie \displaystyle{\frac {e^{\gamma_E}}
{4\pi ^{3/2}}} e^{-i\gamma_5 \Theta} e^{ie^2\beta (x)}
\end{equation}
where the first term is the one which would result in a calculation 
without instantons.
Further details can be found in Refs.\ \cite{Ada96,Rad99}. Since we will not 
touch upon the issue about the influence of topological field configurations 
on DSEs in the reminder of this review,
we would like to point out that the Schwinger model can be solved exactly by 
calculating its Green's functions using DSEs. Hereby, it is not only possible
but also necessary to take all topological effects into account. Whereas the
DSEs stay formally correct in different instanton vacua, the explicit
expressions for Green's function acquire additional  
terms allowed by the DSEs. 
To our knowledge similar investigations in other models or theories are
unavailable as yet: Possible complications due to a presence of topologically
non-trivial configurations are usually
not addressed in DSE based studies.\footnote{Note, however, 
that effects of Abrikosov--Nielsen--Olesen strings on the gauge boson
propagator in the dual Abelian Higgs model have been considered in terms of an
unknown string-string correlator in Ref.\ \cite{Che99} which nevertheless
allows some general restrictions on the infrared behaviour of
the gauge boson propagator. In particular, it was demonstrated that an infrared
enhanced gauge boson propagator is excluded by confinement in this model.}

\subsection{Confinement in 2+1 Dimensional QED}
\label{sec_QED3} 

There are basically two different objectives in studies of the Green's
functions in 2+1 dimensional QED. One line of research is based on
applications to condensed matter systems. The systems of main interest
are hereby high-$T_c$ superconductors, see, {\it e.g.}, Ref.\ \cite{Cam98} 
for a  recent review in this direction. Other 
applications to condensed matter systems include the fractional quantum Hall
effect by adding a Chern--Simons term to the action, for a summary of studies
of the topological structure due to this term, see Ref.\ \cite{Tre95}. 

On the other hand, and that will be
the main issue discussed in this section, there also is
a considerable number of investigations which
employ 2+1 dimensional QED in order to study general aspects of DSEs in a
confining and chiral symmetry breaking theory. 
Here we restrict to non-compact electrodynamics\footnote{The difference
between a compact and a non-compact U(1) gauge theory is best understood on
the lattice: There it amounts to allow for gauge transformations 
to be valued on a circle or on the real line, respectively \cite{Pol87}. 
In the continuum this translates, roughly speaking, into the question whether
monopole field configurations are in principle allowed or not.} at zero
temperature. In 2+1 dimensional QED, fermions can be introduced by  
two-- or four-component Dirac spinors, for details as well as a discussion of
possible mass terms see below, where it will also become clear why we restrict
to four-component spinors here,\footnote{Studies of DSEs
in two-component 2+1 dimensional QED exist with Chern--Simons term included,
see Ref. \cite{Mat99}, and without, see \cite{Hos89}. The main motivation 
for both these studies are applications to condensed matter systems.} with
vanishing bare mass. 
The possibility to generate a parity invariant and chiral-symmetry
breaking dynamical mass is still included, of course. 
Finally, having outlined the basis for the discussion   
of this subsection, we refer the reader to chapter three of the
review \cite{Rob94} in which corresponding
studies are summarised with status as of the 
early nineties in some detail. We therfore put the emphasis here on the 
developments since then. 

The coupling in 2+1 dimensional QED, $e^2$, has the dimension of mass. The
theory is super-renormalisable, and no new scale needs to be introduced
during renormalisation. The dynamical generation of a fermion mass, {\it i.e.},
the dynamical breaking of chiral symmetry (DB$\chi$S), is thus directly 
related to the scale provided by $e^2$.  An order parameter for DB$\chi$S is
provided by the fermion condensate $\langle \bar \psi \psi \rangle = - {\rm tr}
S(x)$ in the case of vanishing bare mass in the Lagrangian. As mentioned 
already, two-component spinors are sufficient to describe
``Dirac'' fermions in three dimensions. Furthermore, there are two
inequivalent \mbox{$2\times 2$} representations of the Clifford algebra
\mbox{$\{\gamma_{\mu},\gamma_{\nu}\} = 2\delta_{\mu\nu}$}.\footnote{As before,
we adopt a positive definite Euclidean  metric $g_{\mu\nu}=\delta_{\mu\nu}$  
with hermitean Dirac matrices.}
Using two-component spinors, however, any mass term,
whether explicit or dynamically generated, is odd under parity
transformations.  Furthermore, for the two-component fermion which belongs to
the irreducible spinor representation in three dimensions, chiral symmetry
cannot be well-defined. These features are unacceptable in studies of 
three-dimensional QED as a testing ground for (four-dimensional) QCD. 
An alternative is provided by using four-component
spinors or, phrased otherwise, an even number of flavours of two-component
fermions. The Euclidean Clifford algebra \mbox{$\{\gamma_{\mu},\gamma_{\nu}\}
= 2\delta_{\mu\nu}$} can be realized by employing three of the Dirac matrices
conventionally used in four dimensions, {\it e.g.}, \mbox{$\gamma_1$,
$\gamma_2$, $\gamma_4$}. Since both $\gamma_5 :=
-\gamma_1\,\gamma_2\,\gamma_4$ and $\gamma_3$ anticommute with these three
matrices the massless theory is invariant under an U(2$N_f$) group of chiral
transformations where $N_f$ is the number of (four-component) flavours. The
four generators of this group (per flavour) 
are the unit matrix, $\gamma_3$, $\gamma_5$ and $\frac 12
[\gamma_3, \gamma_5]$. 
One way to construct a mass term is given by 
$ \tilde m \overline{\psi} \frac{1}{2}
[\gamma_3,\gamma_5] \psi $ which is invariant under chiral transformations
but not under parity.  This is obviously not the analogue of a
Dirac mass in four dimensions, and we will not consider it here. The
term $ m \overline{\psi}  \psi $ on the other hand does have the desired
transformation properties: It is invariant under
parity but not under chiral transformations.
The generation of a parity-invariant dynamical mass 
breaks the U(2$N_f$) symmetry down to U($N_f$)$\times$U($N_f$). 
This leads to $2N_f^2$ Nambu--Goldstone bosons, 
$N_f^2$ scalars and $N_f^2$ pseudoscalars. 

The fermion DSE is given by ({\it c.f.}, the quark DSE in
Eq.~(\ref{quarkDSEmom})) 
\begin{equation}
S^{-1} (p) = Z_2(- i {p \kern-.5em\slash } +Z_m m) + e^2 Z_{1F}
\int \frac{d^{\rm 3}q}{(2\pi)^{\rm 3}}
\gamma_\mu S(q) \Gamma_\nu (q,p) D_{\mu\nu} (p-q) \, .
\label{fermDSEqed3}
\end{equation}
Besides the full photon propagator $ D_{\mu\nu} (k)$ it contains the 
fermion-photon vertex function $\Gamma_\nu (q,p)$. 
During the past years some effort has been put in finding
suitable Ans\"atze for this fermion-photon vertex with the intention 
to minimise the gauge dependence of the results for observables. To this end, 
the so-called {\em transverse} condition has been suggested and employed in an
explicit construction of the vertex function, {\it e.g.}, see Refs.\ 
\cite{Don94,Bur98}. It was pointed out in Ref.\ \cite{Bas99}, however, that 
this transverse condition is violated at two-loop level in perturbation theory. 
This might be an indication that the vertex does contain transverse parts
which cannot be determined from properties of the single particle propagators
alone. We will return to this issue in the next section where we discuss
four-dimensional QED.

On the other hand, in Ref.\ \cite{Mar96} it was observed that the detailed
structure of the vertex does not play a too significant role 
in the solution of the coupled DSEs for fermion and photon propagator
in three-dimensional QED, both in the massless and in the massive phase. 
Despite the fact that only a quite 
restricted class of Ans\"atze has been used in this study, it nevertheless
shows that maintaining consistency at the level of the propagators does 
somewhat improve the robustness of the solutions with respect to truncations
in higher order $n$-point functions. 
In the following we briefly summarise the work presented in Ref.\ \cite{Mar96}.
In three-dimensional QED in Landau gauge $Z_2=Z_{1F}=1$ is finite,
and starting with bare mass $m=0$ the renormalisation constant $Z_m$ is
irrelevant.  Assuming that parity is not broken dynamically one can decompose
the fermion propagator as usual,
\begin{equation}
S^{-1}(p) =  - i {p \kern-.5em\slash } A(p^2) + B(p^2) \, .
\label{S-1AB}
\end{equation}
With this decomposition one obtains from Eq.\ (\ref{fermDSEqed3}), 
\begin{eqnarray}
 A(p^2) &=& 1 -
 \frac{e^2}{p^2} \int\!\frac{{\rm d^3}q}{(2\pi)^3}\;  {\textstyle\frac14}
    {\rm tr}[{p \kern-.5em\slash }\,\gamma_\mu S(q)\Gamma_\nu(q,p) D_{\mu\nu}(p-q)]
 \,,
\\
 B(p^2) &=& e^2 \int\!\frac{{\rm d^3}q}{(2\pi)^3}
        \; {\textstyle\frac14}
 {\rm tr}[\gamma_\mu S(q)\Gamma_\nu(q,p)D_{\mu\nu}(p-q)]\,.
\end{eqnarray}
The DSE of the photon propagator which occurs 
in these equations  was solved simultaneously in Ref.\ \cite{Mar96}. 
It is fully determined by the tranverse part of the photon polarisation tensor
(see the discussion below Eq.\ (\ref{PiD})),
\begin{equation}
D^{-1}_{\mu\nu}(k) \, = \, {D_{(0)}^{-1}}_{\mu\nu} (k) + \Pi_{\mu\nu}(k) \,
\end{equation}
with
\begin{equation}
\Pi_{\mu\nu}(k) = \left( k^2  \delta_{\mu\nu} -  {k_\mu k_\nu}\right)
\Pi(k^2) = - e^2 Z_{1F}\int\!\frac{{\rm d^3}q}{(2\pi)^3}
       {\rm tr} ( \gamma_\mu S(q) \Gamma_\nu(q,q-k) S(q-k) )\,.
\end{equation}
In principle, a gauge-invariant regularisation scheme is required here,
because of an ultraviolet divergence in the longitudinal part of the photon
polarisation tensor. Note that one possible source of spurious longitudinal 
terms in the photon (or gluon) DSE can be the regularisation by an
$O(d)$-invariant Euclidean cutoff $\Lambda $. 
This regularisation violates the residual invariance
under gauge transformations generated by harmonic gauge functions ($\partial^2
\Lambda(x) =0$) in linear covariant gauges. 
A straightforward elimination of spurious longitudinal
terms by contracting with the transverse projector ${\mathcal
P}_{\mu\nu}(k)$ is known to result in quadratically ultraviolet divergent
contributions which are of course artifacts of the regularisation not being
gauge invariant. As observed in Ref.\ \cite{Bro88a}, in
general, quadratic ultraviolet divergences can occur only in
the term proportional to $\delta_{\mu\nu}$. 
Therefore, this part cannot be determined unambiguously, its finite part 
necessarily also depends on the momentum routing in the loop integration. An
unambiguous procedure is, however, to isolate the part free of quadratic
ultraviolet divergences by contracting with the projector 
\begin{equation}
{\mathcal R}_{\mu\nu} (k) = \delta_{\mu\nu} - d \frac{k_\mu k_\nu}{k^2} =
\delta_{\mu\nu} - 3 \frac{k_\mu k_\nu}{k^2} \, ,
\end{equation}
which is constructed such that 
${\mathcal R}_{\mu\nu} (k) \delta^{\mu\nu} =0$, and therefore
the ambiguous term proportional to $\delta_{\mu\nu}$ is projected out.
Thus, in the numerical solution of the
DSEs, the problem introduced by the gauge non-invariant cutoff regularisation 
can be avoided in contracting $\Pi_{\mu\nu}(k)$ with ${\mathcal R}_{\mu\nu}
(k)$ to project onto the unambiguous and finite part of the polarisation
tensor, even though gauge invariance is in principle broken by the sharp
cutoff.\footnote{In a gauge-invariant regularisation this problem 
is avoided all together, of course. As such, dimensional regularisation was
employed for the fermion propagator DSE in four-dimensional QED in Refs.
\cite{Sch98a,Gus99,Kiz00} as discussed in the next section.
Note also that a gauge-invariant regularisation would allow to check the
transversality of the Landau gauge photon propagator for numerical 
self-consistency. Using a contraction with ${\mathcal R}_{\mu\nu}$ 
as done in most DSE calculations the corresponding Ward identity is
not tested independently.} 
The coupled equations for the photon and fermion propagator form a set 
of three non-linear equations for three
scalar functions, and the only unknown therein is the full vertex
function. Up to this point the DSEs for the fermion propagator
and the photon propagator are, in principle, exact. As discussed in the last 
chapter the vertex satisfies its own DSE which could in principle be solved
in some approximation. In the exploratory study of Ref.\ \cite{Mar96}, on the
other hand, the following Ans\"atze have been investigated:
\begin{equation}
 \Gamma_\mu(p,k) = f(A(p^2),A(k^2),A((p-k)^2))\, \gamma_\mu \, ,
\end{equation}
with 
$$
f(A(p^2),A(k^2),A((p-k)^2)) = \cases{1&\cr 
                                     \frac{1}{2}(A(p^2)+A(k^2)&\cr 
				     A(p^2)A(k^2)/A((p-k)^2&\cr 
				    \frac{1}{4}(A(p^2)+A(k^2))^2&\cr}\,\, . 
$$
The coupled equations for $N_f$ different fermion flavours,
\begin{eqnarray}
\label{DSeqAqed3}
 A(p^2) &=& 1 + \frac{2\,e^2}{p^2} \int\!\frac{{\rm d^3}k}{(2\pi)^3}
        \frac{A(k^2) (p\cdot q)(k\cdot q)/q^2}{A^2(k^2)k^2 + B^2(k^2)} \,\,
        \frac{f(A(p^2),A(k^2),A(q^2))}{q^2(1+\Pi(q^2))} \,, \\
\label{DSeqBqed3}
 B(p^2) &=& 2\,e^2 \int\!\frac{{\rm d^3}k}{(2\pi)^3}
        \frac{B(k^2)}{A^2(k^2)k^2 + B^2(k^2)} \,\,
        \frac{f(A(p^2),A(k^2),A(q^2))}{q^2(1+\Pi(q^2))} \,, \\
\label{DSeqPiqed3}
 \Pi(q^2) &=& 4N_f e^2 \!\int\!\frac{{\rm d^3}k}{(2\pi)^3}
   \left(k^2 + 2 k\cdot q - 3 \frac{(k\cdot q)^2}{q^2}\right)
   \frac{A(k^2)}{A^2(k^2)k^2 + B^2(k^2)} \,
   \nonumber \\
  && \hskip 3cm \times  f(A(p^2),A(k^2),A(q^2))
   \frac{A(p^2)}{A^2(p^2)p^2 + B^2(p^2)}  ,
\end{eqnarray}
(with $q=p-k$) were solved for both, the chirally symmetric and the broken
phase. 

The most important result of this study is a critical number for chiral
symmetry breaking, $N_{crit} \approx 3.3$ such that 
dynamical mass generation occurs for $N_f<N_{crit}$; above this
critical number, only the chirally symmetric solution exists. This is in
qualitative agreement with studies based on the $1/N_f$ expansion of the DSEs
employing a bare vertex and assuming no wavefunction renormalisation ($A(p^2)
\equiv 1$) \cite{App86,App88}.\footnote{This phase transition
is of infinite order and resembles a conformal phase transition, see Ref.\
\cite{Gus98}. Adding a Chern--Simons term, however, the phase
transition becomes one of first order \cite{Kon95}.}
This result is in contradiction to previous claims, however, that the effect
of wavefunction renormalisation would lead to DB$\chi$S for all $N_f$, see the
discussion in section 3.5 of the review \cite{Rob94} as well as references
therein. What the results of Ref.\ \cite{Mar96} thus 
seem to tell us is that the photon polarisation is very important here. Without
it, {\it i.e.}, in the quenched approximation, the occurrence of DB$\chi$S is
sensitive to the wavefunction renormalisation and to the particular Ansatz
employed for the fermion-photon vertex function. Quite in contrast to this
situation, the solutions to the complete system of coupled DSEs for the
propagators do not seem to be overly sensitive at all to the precise
assumptions made to truncate the system at the level of the vertex function.  

In the chirally broken phase one has an infrared vanishing photon polarisation,
and thus logarithmic confinement as might be expected from dimensional
reasons. In the chirally symmetric phase, an infrared divergent  photon
polarisation  function $\Pi(q^2)$ is found \cite{Mar96}.\footnote{Note that a
somewhat unconventional definition of the photon polarisation, by the
denominator $q^2 + \Pi(q)$ in the photon propagator, was
adopted in Ref.\  \cite{Mar96}. In order to compare to the dimensionless
polarisation defined by $q^2 (1 + \Pi(q))$ herein, Eq.\ (28) in Ref.\
\cite{Mar96} has to be replaced by $\Pi_{pert} = \alpha /q$. Some typos in
Ref.\ \cite{Mar96} can easily be identified by comparing to the 
formulas given above.} This leads to screening instead of confinement. 
The infrared behaviour of the fermion propagator in the chirally symmetric
phase is determined by an exponent $\eta$ indicative of a conformal regime:
The wavefunction renormalisation $A(p^2)$ (being approximately equal to one
in the ultraviolet as required by perturbation theory) vanishes with positive
exponent $\eta$ in the infrared. From Fig.\ 3 of Ref.\ \cite{Mar96} one might
estimate,  
\begin{equation}
S^{-1}(p) = A(p^2) {p \kern-.5em\slash }  \, 
\stackrel{p^2 \to 0^+}{\longrightarrow} \, \left( \frac p{N_fe^2/8}
\right) ^\eta {p \kern-.5em\slash } \quad {\rm with} \quad \eta = 
{\frac 8{3N_f\pi^2}}.
\label{S-1A}
\end{equation}
This same exponent was obtained from inverse Landau--Khalatnikov
transformations in Ref. \cite{Ait97}, and it has furthermore been verified in
a recent calculation \cite{Ahl00} to a very high precision. Note that
for all possible values of $N_f$ which lead to the chirally  symmetric phase,
$\eta $ is an irrational number much smaller than one.  This is very likely
to lead to a cut (with infinitely many Riemann sheets, {\it i.e.},
logarithmic-like) in the fermion propagator. This cut on the time-like
$p^2$-axis is, at least in principle, understandable from the existence of
massless photons. The infrared behaviour of the photon propagator is of the
form $const./q = const./\sqrt{q^2}$ and thus indicates that the photon
propagator has a two-fold cut. One might interpret this cut as being generated
by the massless fermions. The chirally symmetric phase is often called the
Coulomb phase for this reason. 
We would also like to emphasise here that the infrared behaviour of the
propagators as well as their analytic structures related to this behaviour 
reflect the physics of a given phase, as this explicit example
demonstrates. 

The analytic structure of the propagators is much more interesting
in the confining phase, of course.
In the chirally broken and confining phase $A(p^2)$ is not of the order
one either, but considerably smaller in this case. As the number of fermion
flavours approaches the critical number $N_{crit}$ (from below), $A(0)$ tends
to go to zero. This behaviour suggests the conjecture that $A(p^2)$ might
vanish by an irrational power at the zeros of the denominator
$A^2(p^2)p^2+B^2(p^2)$ for $N_f < N_{crit}$. This
would imply that the fermion propagator had no isolated poles in the
confining phase. This hypothesis is currently being investigated more closely
\cite{Ahl00}. Results in this direction based on simultaneous solutions of
the coupled photon {\em and} fermion DSEs are not available yet, however.  

Studies of singularities of the fermion propagator in the complex $p^2$ plane 
are available only for solutions to fermion DSEs with assuming different
Ans\"atze for the photon polarisation and the fermion-photon vertex
\cite{Mar93,Mar95}. Not too surprisingly, however, the locations of the poles
then depend strongly on the Ansatz for the vertex function
\cite{Mar93}. Improving the vertex function such that, {\it e.g.}, Ward
identities are satisfied, the poles of the fermion propagator are found to
occur  much closer to the time-like real axis  as compared to
the cases in which bare vertices are employed. This might be taken
as an evidence that the (unknown) full renormalised fermion propagator does 
have singularities on time-like axis despite confinement. 
Independent of this, however, in connection with the results for QCD
presented in Chapter~\ref{chap_QCD} below, it seems interesting  
to note that the fermion propagator does violate reflection positivity 
in all the cases for which the underlying Ans\"atze lead to fermion confinement
\cite{Mar95}. This result was established by two different methods: 
One method is to solve the fermion DSE for complex values of $p^2$ in order
to determine the singularities of the propagator directly. The other method 
is based on studying the Fourier
transform of the fermion propagator at large (Euclidean) times. 
As a central result of Ref.\ \cite{Mar95}, oscillations in the Fourier
transform of the fermion propagator as a function of time $t$ are observed
clearly. This result thus demonstrates the violation of positivity quite
unambiguously. It would therefore seem to be interesting to perform an
analogous investigation of positivity on possible solutions for the 
fermion propagator as obtained from the coupled fermion and photon DSEs.
Such a study was initiated during the write-up of this review
\cite{Ahl00}.

Summarising this subsection we conclude that in non-compact
four-component massless three-dimensional QED with $N_f < N_{crit}$ flavours
D$\chi$SB and  confinement are realized.\footnote{The investigations described 
above start from a vanishing bare mass. In the limit of
infinitely heavy fermion masses, on the other hand, it was proposed from
bosonisation techniques in  Ref.\ \cite{Abd98} that one should always
obtain screening  instead of confinement.
This can be related to the occurrence of a massive pole in the effective
bosonic action found in Ref. \cite{Dal99}. One should keep in mind, however,
that these bosonisation technique are applicable only if the dimensionless
parameter $e^2/m$ is small. This condition is certainly always violated for 
sufficiently small or vanishing bare fermion masses $m$.}
In this phase fermions acquire a dynamical mass and decouple from the infrared
dynamics. The resulting potential thus grows at large separation similar to
the tree-level one leading to logarithmic confinement. The corresponding
analytic structure of the fermion propagator is, however, not known in detail.
Nevertheless, based on the investigations reported in Refs.\ \cite{Mar93,Mar95}
it seems safe to conclude that {\it the fermion propagator of
three-dimensional QED in the confining phase violates reflection positivity}.
For $N_f > N_{crit}$ a Coulomb phase with massless fermions and photons is 
found in a solution of the coupled photon and fermion DSEs. This result agrees
qualitatively with the leading order in the $1/N_f$ expansion. 
This well understood case serves as an explicit example of how
the infrared behaviour of the propagators, as well as 
their analytic structures related to this behaviour, have a direct
interpretation in terms of the physics of this phase 
which is characterised by massless, interacting fermions and photons.

As stated in the beginning of this section, three-dimensional QED was analysed
here as a ``toy model'' for four-dimensional QCD. To this end we have to learn 
the lesson that even the qualitative behaviour of DSE solutions can be 
sensitive to the truncations employed. A nevertheless encouraging lesson is,
however, that those truncations which are complete at the level of the 
propagators lead to fairly stable results. In particular, the
qualitative behaviour of the corresponding solutions is then found to be 
quite insensitive to modifications in the 3-point vertex function. As a
result, a consistent picture of the two-phase structure of this theory
emerges: For a large number of flavours one has a Coulomb phase, whereas for a
small number of flavours D$\chi$SB and confinement occur. 

\subsection{Dynamical Chiral Symmetry Breaking in 3+1 Dimensional QED}
\label{sec_QED4} 

In the sense of Wilson's renormalisation group, full 3+1 dimensional QED is
most likely to be trivial: It has only an infrared fixed point, see, {\it
e.g.},  chapter 16 in 
Ref.\ \cite{Hua98} for a pedagogical discussion and Ref. \cite{Goe98}  for a
non-perturbative verification based on a lattice calculation. As a matter of
fact, the underlying goals when studying the Green's functions of QED are based
on the observation that massless QED  with $N_f$ flavours possesses the same
chiral symmetry as QCD. Therefore  the quark DSE in 3+1 dimensional QED has
been used to study technical questions like: the analytic structure of
propagators, numerical cancellations of divergences, gauge dependencies
introduced by numerical cutoffs, and the question of gauge covariance in
general. Obviously, with this restriction in mind in most cases it is
sufficient to consider the quenched approximation.\footnote{There 
are very few unquenched calculations. The only one to our knowledge in which
the quadratic divergence in the vacuum polarisation was treated correctly 
is that of Ref.\ \cite{Blo95}.}

DB$\chi$S has been discussed in detail in the review \cite{Rob94} and in the
book \cite{Mir93}. We briefly summarise the corresponding knowledge and
describe the progress made since then in some more detail 
(parts of the material presented here follow Ref.\ \cite{Pen98}). 
Of course, the generation of a dynamical
mass for fermions with bare mass zero is a genuinely non-perturbative problem.
Note that it is not possible to put massless fermions on a lattice. Thus every
lattice calculation which wants to include DB$\chi$S has to rely on
extrapolations provided, {\it e.g.}, by ``chiral perturbation theory'' or other
continuum techniques. (Note also that ``chiral perturbation theory'' describes
DB$\chi$S rather than explaining its origin.)
Studying this issue with non-perturbative methods based on a continuum
formulation is therefore required. Already the very simplest truncation of the
fermion propagator DSE does provide for a dynamically generated
mass. Replacing the photon propagator and the fermion-photon vertex by their
tree-level forms,
$D_{\mu\nu} (k) \to {D_{(0)}}_{\mu\nu} (k)$ and $\Gamma_\mu (p,q) \to
\gamma_\mu$, {\it i.e.}, applying the so-called rainbow
approximation together with the quenched
approximation,\footnote{Diagrammatically the use of a bare vertex in the
fermion DSE amounts to a summation of horizontal arrays of ``rainbows'' of 
parallel photon exchanges. In many-body theory this is identical to the
Green's function formulated Hartree--Fock approximation, see {\it e.g.} Ref.\
\cite{Alk88}. Bosonisation techniques as employed, {\it e.g.}, in the
``Global Colour Symmetry Model'' \cite{Rob87} (see also the reviews
\cite{Rob94,Tan97,Cah98} and references therein) naturally lead to this
approximation also.} one obtains
\begin{eqnarray}
 A(p^2) &=& 1 + \xi \frac{e^2}{p^2} \int\!\frac{{\rm d^4}q}{(2\pi)^4}
        \frac{A(q^2) (p\cdot (p-q))(q\cdot (p-q))}{A^2(q^2)q^2 + B^2(q^2)} \,\,
        \frac{1}{(p-q)^4} \,, \nonumber \\
        &=& 1 + \xi \frac{e^2}{(4\pi )^2} \int\! dq^2 
	\frac{A(q^2) }{A^2(q^2)q^2 + B^2(q^2)} \left( \Theta (p^2-q^2) \frac
	{q^2}{p^2} + \Theta (q^2-p^2) \right) \,, 
\label{DSeqAqed4r} \\
 B(p^2) &=& Z_m m + (3+\xi ) e^2 \int\!\frac{{\rm d^4}q}{(2\pi)^4}
        \frac{B(q^2)}{A^2(q^2)q^2 + B^2(q^2)} \,\,
        \frac{1}{(p-q)^2} \, , \nonumber \\
        &=&Z_m m + (3+\xi ) \frac{e^2}{(4\pi )^2} \int\! dq^2 
	\frac{B(q^2)}{A^2(q^2)q^2 + B^2(q^2)} \left( \Theta (p^2-q^2) \frac
	{q^4}{p^4} + \Theta (q^2-p^2) \right) \,. 
\label{DSeqbqed4r} 
\end{eqnarray}
The bare mass $m_0 = Z_m m$ is sent to zero by an appropriately defined limit
in studies of DB$\chi$S. One furthermore observes that in this approximation
there is no wave function renormalisation, {\it i.e.}, $A(p^2)=1$ for the
choice of the Landau gauge $\xi = 0$. 
In this gauge, and with introducing an $O(4)$-invariant Euclidean momentum
cutoff $\Lambda_{UV}$, the equation for the dynamical mass function $M(p^2)
\equiv B(p^2)/A(p^2) = B(p^2)$ reads:
\begin{equation}
M(p^2) = m_0 + \frac{3e^2}{(4\pi )^2} \int_0^{p^2} dq^2 \frac {q^2}{p^2} 
\frac{M(q^2)}{q^2 + M^2(q^2)} + \frac{3e^2}{(4\pi )^2} \int_{p^2}^{\Lambda_{UV}^2}
dq^2 \frac{M(q^2)}{q^2 + M^2(q^2)} \, .
\label{MQED3}
\end{equation}
This equation obviously has the trivial solution $M(p^2) = 0$ in the chiral
limit $m_0=0$. 
The interesting question is, however: Under what circumstances can 
the solution for $M(p^2)$ be non-vanishing in 
the chiral limit? The simplicity of Eq.\ (\ref{MQED3}) allows one to 
convert it into a differential equation,
\begin{equation}
\frac{d\phantom{p^2}}{dp^2} \left( p^4 \frac{d\phantom{p^2}}{dp^2} M(p^2)
\right) = - \frac{3e^2}{(4\pi )^2}  \frac{p^2M(p^2)}{p^2 + M^2(p^2)} \; ,
\end{equation}
together with the {\it boundary conditions} 
\begin{equation}
\lim _{p^2\to 0} \left( \frac{d\phantom{p^2}}{dp^2} \left(p^4 M(p^2) \right)
\right)  = 0 \, \quad {\rm and} \quad \lim _{p^2\to \Lambda_{UV}^2} \left(
\frac{d\phantom{p^2}}{dp^2} \left(p^2 M(p^2) \right) \right)  = m_0 \, .
\label{bcQED} 
\end{equation}
For large $p^2$, {\it i.e.}, $p^2\gg M^2(p^2)$, one has $p^2 + M^2(p^2)
\approx p^2 $ and the differential equation can
be linearised. The Ansatz $M(p^2) \propto (p^2)^a$ then yields  
$a(a+1) = - {3e^2}/{(4\pi )^2}$. Writing the corresponding roots in terms
of the fine structure constant $\alpha =  {e^2}/{(4\pi )}$ and defining
$\alpha_c := \frac{\pi}{3}$,
\begin{equation}
a= -\frac 1 2 \pm \frac 1 2 \sqrt{1-\frac {\alpha }{\alpha_c}} \, ,
\end{equation}
one realizes that solutions of qualitatively different character arise in 
the two distinct cases $\alpha < \alpha_c$ and $\alpha > \alpha_c$.
In the latter case, having a complex power $a$, the solution oscillates
at large $p^2$. This oscillatory behaviour makes it possible to obey the
boundary condition at the ultraviolet cutoff, even in the limit $m_0\to 0$. 
Having the real power for $\alpha < \alpha_c=\pi /3$ this is not the case, and
DB$\chi$S is excluded. This is a nice and quite 
intuitive result: Only if the interaction
is sufficiently strong, and this is the case for a fine structure constant
larger than the critical value of the order one, a mass will be generated
dynamically \cite{Fuk76}.  

\begin{figure}
  \centering\epsfig{file=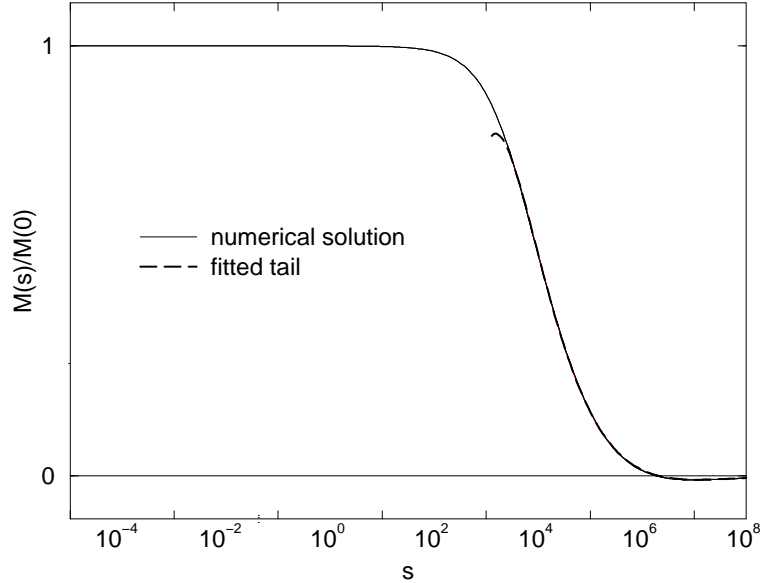,width=0.6\linewidth}
  \caption[The mass function $M(s)$ in QED.]
    {The mass function $M(s)$ ($s=p^2/\mu^2$ with $\mu$ being some arbitrary
    numerical scale) for some coupling ($\alpha =1.6$) above the critical value.
    Also shown is the asymptotic behaviour according to the linearised
    differential equation. }
   \label{fig:MQED}
\end{figure}

It is instructive to plot the numerical result for the dynamical
mass (with $ \alpha > \pi /3$) as a function of $p^2$ on a
logarithmic scale, see Fig.\ \ref{fig:MQED}. At large $p^2$ the 
numerical solution reproduces the (analytically obtained)
asymptotic behaviour quite well. Note that the plot in Fig.\ \ref{fig:MQED} is
extended beyond the ultraviolet cutoff where the mass function is set to
$m_0=0$ in order to demonstrate the oscillatory behaviour.  
At small $p^2$ the mass function is
almost momentum independent and approaches a constant value. This indicates
that an analytic continuation of the mass function (and thus the propagator) 
to time-like momenta $p^2<0$ might be possible.\footnote{For this
analytic continuation the integral equation should be used. Relying only 
on the numerical solution on the Euclidean axis will not be sufficient.
Complex analysis states that as long as an analytic function is only known
at a finite set of points (here, a finite set of Euclidean points)
its analytic continuation away from this set is infinitely ambiguous:
Outside the set the function may take any value. Employing, on the
other hand, the integral equation does provide the possibility of analytic 
continuation.}
It would then furthermore be quite evident that the position of the pole
should approximately be determined by $M(0)$, {\it i.e.}, $p^2_{\rm pole}
\approx - M^2(0)$. 

For a coupling smaller than its critical value, the boundary condition
(\ref{bcQED}) cannot be fulfilled for $m_0=0$, and thus only the trivial
solution is possible. At the critical coupling the solution bifurcates:
One still has the trivial solution which is unstable, however. The chiral
symmetry breaking solution with $M(p^2)>0$ is energetically favoured as can be
seen by analysing the Cornwall--Jackiw--Tomboulis action \cite{Cor74}.
(There exists a third solution with $M(p^2)<0$ which is only a mirror of the
massive solution with positive mass function. This third solution is  
excluded if one considers $m_0\to 0^+$ instead of putting $m_0=0$ immediately.)

As discussed above a non-vanishing mass function implies a non-vanishing pole
mass. As the latter is an observable and thus a gauge-independent quantity,
the critical coupling has to be gauge invariant also. However, in the rainbow
approximation this is not the case within the class of linear covariant gauges.
Using, {\it e.g.}, the Feynman gauge ($\xi=1$) instead of Landau gauge
($\xi=0$), the value of the critical coupling increases by approximately
fifty percent. As the use of a bare vertex violates Ward identities this
should be no big surprise. 
Requiring that (i) the vertex function solves the Ward identity (\ref{Ward}),
that (ii) it is free of kinematical singularities ({\it i.e.}, that $\Gamma
_\mu (p,q)$ has a unique limit for $p^2\to q^2$), that (iii) it has the same
transformation properties as the bare vertex under parity, time reversal and
charge conjugation invariance, and that (iv) it reduces to the bare vertex
$\gamma_\mu$ for a vanishing coupling, the longitudinal part of the vertex
function is completely fixed in terms of the fermion propagator \cite{Bal80}:
\begin{equation}
\Gamma _\mu ^{\rm BC} (p,q) = \frac 1 2 (A(p^2)+A(q^2)) \gamma_\mu
+ \frac 1 2 \Delta A_{pq} (p+q)_\mu (p\kern-.5em\slash + q \kern-.5em\slash ) 
+i \Delta B_{pq} (p+q)_\mu 
\label{BallChiu}
\end{equation}
with 
\begin{equation}
\Delta A_{pq} = \frac {A(p^2)-A(q^2)}{p^2-q^2} \, , \quad
\Delta B_{pq} = \frac {B(p^2)-B(q^2)}{p^2-q^2} \, ,
\end{equation}
and where $p$ and $q$ are the momenta attached to the fermion legs.
As stated in Sect.\  \ref{sec_nPoint} in an explicit solution of the DSE
for the vertex function, the form (\ref{BallChiu}) for its longitudinal
part is {\it exactly} reproduced.  

If one refrains from solving the vertex function DSE, on the other hand, the
transverse part of the vertex function has to be modelled. 
Transverse part here refers to those contributions transverse to
the photon momentum $k=q-p$, {\it i.e.}, $(p-q)_\mu \Gamma _\mu ^{\rm T}
(p,q) = 0$,  which, in addition, satisfy the requirement that
$\Gamma _\mu^{\rm T} (p,p) = 0$. This additional requirement results from the
differential Ward identity (\ref{dWard}).  
The general form of the fermion-photon vertex can be
decomposed into twelve linearly independent Lorentz covariants, four of them
are  longitudinal and eight are transverse. As the Ward identity has no
$\sigma_{\mu \nu}$  component one coefficient is zero, and one obtains the
three terms given in  Eq.\ (\ref{BallChiu}) for the longitudinal part. 
As a further constraint, the condition that the propagators and vertices 
should be gauge covariant can then be imposed on the coefficients
of the transverse parts. Restricting to the purely
longitudinal vertex (\ref{BallChiu}), however, the mass generation mechanism 
is still gauge dependent. This is related to the fact that a consistent
regularisation and renormalisation of the fermion propagator DSE cannot be
achieved unless the fermion propagator is multiplicatively renormalisable. 
Note that proofs of the gauge independence of the pole mass implicitly use
this property, {\it e.g.}, see \cite{Atk79,Gam99}. Since multiplicative
renormalisability and gauge covariance are satisfied order by order in
perturbation theory,\footnote{In perturbation theory this just 
means that violations of gauge covariance are of
higher order than the highest one taken into account in a given
approximation.} the perturbative result is a good 
starting point to resolve this issue \cite{Cur90}. Renormalisability is related
to the ultraviolet behaviour of the loop, and it is thus sufficient to consider
momenta $p^2\gg q^2$. In this limit the transverse part of the vertex function
to ${\mathcal O}(e^2)$ is given by\footnote{The complete one-loop form of the 
fermion-photon vertex is given in Ref.\ \cite{Kiz95}. This reference also
gives a list of minor errors in Ref.\ \cite{Bal80}.}
\begin{equation}
\Gamma _\mu ^{\rm T,1-loop} (p,q) = \frac {e^2 \xi}{32\pi ^2} \left(
\gamma _\mu - \frac{p_\mu p\kern-.5em\slash }{p^2} \right) \log \frac {p^2}{q^2}
\, . 
\end{equation}
On the other hand, taking the limit $p^2\gg q^2$ in the purely longitudinal
vertex (\ref{BallChiu}) one then realizes that a non-perturbative
expression for the transverse part has to include the functions $A(p^2)$ and
$A(q^2)$. These considerations have been the underlying motivation for Curtis 
and Pennington to suggest the following form for the fermion-photon 
vertex \cite{Cur90},
\begin{eqnarray}
\Gamma _\mu ^{\rm CP} (p,q) &=&  \Gamma _\mu ^{\rm BC} (p,q) \label{CP}\\
&-& \frac 1 2 \, \Delta A_{pq}  \,
\frac {(p^2+q^2)(p^2-q^2)}{(p^2-q^2)^2 +(M^2(p^2)+M^2(q^2))^2} 
\, 
\left( \gamma _\mu  (p^2-q^2) - 
(p+q)_\mu (p\kern-.5em\slash - q \kern-.5em\slash ) \right)
\, . \nonumber 
\end{eqnarray}
To understand how multiplicative renormalisability constrains the
transverse part of the vertex it is instructive to study the function $A(p^2)$
in the massless, {\it i.e.}, the chirally symmetric phase. A leading
logarithmic expansion of $A(p^2)$ has the form
\begin{equation}
A(p^2) = \sum _{n=0}^{\infty} c_n (e^2)^n \log^n \left( 
\frac {p^2}{\Lambda_{\rm UV}^2}\right)
\, . \label{Aleadlog}
\end{equation}
In the chirally symmetric phase of the quenched approximation $A(p^2)$ is the
only quantity which has to be renormalised. Therefore, the coefficients have to
be of the form $c_n = c_1^n/n!$ in order to ensure multiplicative
renormalisability. Then the series (\ref{Aleadlog}) can be summed, and using
the 1-loop perturbative result one obtains
\begin{equation}
A(p^2) = \left( \frac {p^2}{\Lambda_{\rm UV}^2} \right) ^\eta \quad {\rm with}
\quad \eta = \frac {e^2 \xi}{16\pi ^2} \, .
\end{equation}
Since there is only one scale in the problem, $\Lambda_{\rm UV}$,
a power-law is indeed inevitable:
\begin{equation}
A(p^2) = \left( \frac {p^2}{\Lambda_{\rm UV}^2} \right) ^\gamma \, .
\label{Apow}
\end{equation}
Using the vertex (\ref{CP},\ref{BallChiu})  $A(p^2)$ turns out to be
of the form (\ref{Apow}) with the difference $\gamma - \eta$
 being of the order $e^4$.
This demonstrates explicitly that the transverse part of the 
vertex function is sufficient to maintain multiplicative renormalisability.

If the vertex function (\ref{CP},\ref{BallChiu}) is employed in the quark
DSE, {\it i.e.}, with this particular transverse part included, the critical
coupling above which D$\chi$SB occurs becomes almost independent of the gauge
parameter $\xi$ \cite{Cur93,Atk94}. For $0\le \xi \le 20$ it varies by less
than ten percent, see Fig.~1 of Ref.\ \cite{Gus99}. From this figure it can 
also be seen that imposing the transverse condition \cite{Bur93} leads to
slightly different values for  the critical coupling. 
This condition (which has been
mentioned already in the last section) is imposed in order to remove
gauge-covariance violating terms. An explicit solution to this condition is
given in Ref.\ \cite{Don94}. It should also be mentioned that some of the 
simplifying assumptions used there have been questioned \cite{Bas94,Bas98}.
The interesting point here is that the values of the critical coupling
obtained when imposing the transverse condition show an even stronger
variation with the gauge parameter, {\it e.g.}, dropping from $\alpha_c =
0.933$  in Landau gauge to $\alpha_c \approx 0.80$ at $\xi = 5$. 
Note that for small
values of the gauge parameter $\xi$ these results have been verified in a
direct numerical calculation in Ref.\cite{Haw96}. Using a bifurcation
analysis \cite{Atk94} 
it is straightforward to show that the critical coupling $\alpha_c^{CP}$ based
on the Curtis--Pennington vertex (\ref{BallChiu},\ref{CP}) is related to the
critical coupling $\alpha_c^{TC}$, if the transverse condition is enforced. 
In particular, one obtains \cite{Gus99}, 
\begin{equation}
\alpha_c^{TC} = \frac {\alpha_c^{CP}}{1+\xi \alpha_c^{CP}/8\pi } \, .
\end{equation}
Note that both values coincide in Landau gauge $\xi =0$. For large positive
values of $\xi$, however, the difference becomes large.

Identifying the regularisation with the help of an Euclidean momentum cutoff as
a source of gauge-invariance violating terms, the fermion propagator DSE of
quenched four-dimensional QED has been studied using dimensional
regularisation in Refs. \cite{Sch98a,Gus99,Kiz00}. Note that the use of
dimensional regularisation increases the numerical effort considerably. 
The values for the critical coupling are within numerical errors equal to the
ones obtained by using cutoff schemes \cite{Gus99}. 

Thus we have to report that the goal to maintain gauge covariance by a
suitable choice of the fermion-photon vertex has not fully been achieved. 
Nevertheless, the qualitative conclusions are unaffected: Above the critical
coupling any vertex which satisfies the Ward identity and leads to
multiplicative renormalisability will lead to a diverging mass function for
quenched QED in four dimensions \cite{Haw97}. This supports the 
conjecture mentioned in the beginning of this subsection that
four-dimensional QED might not have a non-trivial continuum limit. 

Finally, we emphasise the special role that the Landau gauge plays here 
amongst the linear covariant gauges. Even in the most simple truncation, the
rainbow approximation, it provides a reasonably accurate value for the critical
coupling. As we have to acknowledge that even gauge-invariant observables
can be sensitive to the details of the truncation (in this example the Ansatz 
for the fermion-photon vertex), and that they can furthermore depend on
technicalities such as the choice of the regularisation within a certain
renormalisation scheme, it seems worthwhile to remember that most
approximations (and truncations) can best be justified in combination with
the Landau gauge. This remark seems especially appropriate in view of  
the Green's functions in QCD discussed in the next chapter.

\section{The Infrared Behaviour of QCD Green's Functions}
\label{chap_QCD} 

In this chapter the present knowledge on the infrared behaviour of QCD Green's
functions is summarised. As we will see, most of the results are obtained
in the Landau gauge. This might seem somewhat surprising at first, as studies
of Green's functions in Landau gauge should include 
the Faddeev--Popov ghosts. Thus, it could seem more natural to employ 
the axial gauge.
And indeed, especially the gluon propagator has been subject to a
considerable number of investigations in the axial gauge. Before
discussing the corresponding results, however, a few general remarks on the
infrared behaviour of the gluon propagator are in order.

\subsection{``Confined'' or ``Confining'' Gluons?}
\label{sec_IR} 

In the literature one sometimes finds the distinction between {\em confined}
and {\em confining} gluons (see, {\it e.g.}, Ref.\ \cite{Bue96} and references
therein). The first phrase is usually attributed to a gluon propagator which is
infrared suppressed. The underlying idea is simply that under these
circumstances the gluons themselves do not propagate over long distances.  It
is also clear that an infrared suppressed propagator almost necessarily violates
reflection positivity, and we will present the quite
convincing evidence for exactly that picture later in this chapter. 
Infrared enhanced gluonic correlations have, on the other hand, been referred
to as confining gluons. As we describe below, if
the gluon propagator was as singular in the infrared as $1/k^4$
in some gauge, this alone would establish an area law and  
the linearly rising interquark potential. Thus, such gluon (2-point)
correlations alone would suffice to generate quark confinement in a simple
and intuitive picture which was frequently connected with the notion of 
{\em infrared slavery} in the past. As will become clear in the course of
this chapter, there is increasing evidence nowadays, however, that this is
actually not what happens in QCD. And neither would it fit into the formal
development of the covariant description of confinement outlined in
Sec.~\ref{Sec2.4}.  It is important to realize that the more complicated
picture emerging for QCD in the covariant gauge
can certainly accommodate confined (but not confining)
gluons in coexistence with an effective quark interaction which is confining,
however. 

Since the intuitive picture of confining gluons had quite
a long history, we will describe some of its implications and applications
first in the following. Despite the necessity of a reinterpretation about the
origin of effective quark interactions, especially the applications thereby
continue to provide many interest aspects.
 
As stated, by infrared enhanced gluonic correlations one usually refers to 
a gluon two-point function which (in momentum space) is more singular than a
massless free particle pole in the infrared. In particular, for the gluon
propagator in the Landau gauge,
\begin{equation} 
  D_{\mu\nu}(k) \, =\,  \left(\delta_{\mu\nu} - \frac{k_\mu k_\nu}{k^2}
  \right) \,   \frac{Z(k^2)}{k^2} \; , \label{glp_LG}
\end{equation} 
this implies that the invariant function $Z(k^2)$ diverges for $k^2 \to
0$. The most singular behaviour that does not lead too obviously to
contradictions in the theory is given by $Z(k^2) \to \sigma/k^2$. Such a
behaviour can, of course, at best be meaningful when supplemented by a
prescription to define this singularity in the sense of distributions. 
An early Lagrangian model of confinement was given in Ref.~\cite{Bla74}, 
the dipole gluon model which uses the derivative of
the principal value prescription,  
$ \mbox{P}\frac 1{k^4}  :=  - \frac d{dk^2}  \, \mbox{P}\frac 1{k^2} $.
However, this model is in quite obvious conflict with causality.

Other frequently adopted prescriptions designed to subtract the leading
singularity in $Z(k^2)$ at $k^2 = 0$ are the $[\sigma/k^2]_+ $ prescription
used in Ref.~\cite{Bro89} which is defined as acting
on a function $F$ by,
\begin{equation} 
\int_0^\infty dx \, \left[ \frac{1}{x} \right]_+ \, F(x,y) \, = \,
\int_0^\infty dx \,  \frac{1}{x} \, \big( F(x,y) - \theta(y-x) F(0,y)
\big) \; ,
\end{equation}
and a prescription leading to a 3-dimensional Fourier transform corresponding
to a linear rising potential $ V(r) = \sigma r$ (in instantaneous
approximation,  $k_0 \to 0$),
\begin{equation} \label{reg_enh}
   8\pi \left[ \frac{\sigma}{k^4} \right] :=  \lim_{\mu \to 0} \left\{
   \frac{8\pi\sigma}{(k^2 + \mu^2 )^2 } -   C(k_0,\mu) \,
   \delta^3(\vec k) \right\} \; , 
   \;\;  C(k_0,\mu) = \int \frac{d^3k}{(2\pi)^3}
   \frac{8\pi\sigma}{(k^2 + \mu^2)^2} \; ,   
\end{equation} 
with $C = \sigma/\sqrt{k_0^2 + \mu^2} $ introduced to implement
the regularity of the potential at $r \to 0$, {\it i.e.}, $\int d^3q V(q) =
0$~\cite{Gro90}. The need for such a prescription introduces an additional
ambiguity as this can usually not be determined from the
numerical solutions of the gluon propagator DSE. The more dramatic problem with
such an infrared enhanced propagator, however, is that the $\sigma/k^4$
behaviour cannot be well-defined as a distribution over a finite neighbourhood
of the origin in the Euclidean momentum space by simply providing a small
mass $\mu^2 \to 0$. This is most clearly seen from Eq.~(\ref{reg_enh}). The
regularity constraint on its Fourier transform results in the necessity to
subtract a term which is not a well-defined distribution but a more singular
object in the limit $\mu\to 0$ (the contribution one subtracts for $k_0 = 0$
is essentially proportional to $\delta^3(\vec k)/\mu $ while for $k_0 \not=
0$ it can be cast into the form $\propto \delta^4(k) \sqrt{k_0^2 + \mu^2}/\mu
$).          

While this is a serious conceptual problem, infrared enhanced gluon
correlations have been phenomenologically quite successful, {\it e.g.}, in
describing quark confinement by infrared slavery \cite{Pag77,Mar78,Sme91}. 
This term refers to the possible existence of severe infrared singularities, 
{\it i.e.}, infrared singularities which cannot be removed from
the $S$-matrix by virtue of the Kinoshita--Lee--Nauenberg theorem in analogy
to those arising from soft photons in QED.\footnote{As mentioned in the
introduction, such severe infrared singularities do not arise at any finite
order in perturbation theory in QCD.} Occurring in the inverse (ultraviolet
renormalised) propagator of interacting quark fields, these divergences can
then remove asymptotic states with quark quantum numbers from the asymptotic
Hilbert space. 

It will be corroborated repeatedly in the following that
the effective interaction thereby arises from combinations of different
contributions. This holds independently of whether that combination 
in the end gives rise to an interaction as strong as the $\sim
\sigma/k^4$ infrared-slavery model or not. 
In particular, in covariant gauges it already follows from the
necessity of renormalisation group invariance of the colour octet
quark-current interactions that these interactions are mediated by both,
gluon and ghost correlations. An infrared enhancement of these interactions,
{\it e.g.}, the two-current interaction of the ``Global Calor Model''
\cite{Rob87,Rob94,Tan97,Cah98}, thus is 
still a viable possibility even though the elementary gluon and ghost
correlations should separately give rise to tempered distributions and show the
corresponding analyticity properties in the non-coincident Euclidean domain. 

By the same remark, such effective quark-current interactions should replace
the literal meaning of a gluon propagator in the following argument:
Infrared enhanced correlations $\propto \sigma/k^4$ in the purely
gluonic Yang--Mills theory are known to provide sufficient damping in
expectation values of large Wilson loops to give rise to an area law in
analogy to the $1/k^2$ behaviour of photons in the 2-dimensional Schwinger
model. For the non-Abelian Yang--Mills theory in four
dimensions this would follow from an infrared enhanced gluon propagator. 
The gluonic correlations besides being gauge and, in
general, also renormalisation scale dependent give rise to the following
upper bound on the behaviour of large Wilson loops with contour $C$
\cite{Wes82}, 
\begin{eqnarray} 
W(C)  \,=\,  \left\langle  P \exp \left\{ i g \oint_C A_\mu dx_\mu \right\}
\right\rangle 
\, \le \,  \exp \left\{ - g^2 N_c C_f \oint\!\!\oint_C dx_\mu  dy_\nu \, 
D_{\mu\nu}(x-y)  \right\} \; .
\end{eqnarray}
Note that a gluon propagator which is an analytic function for all
non-coincident Euclidean $x\not=y$ cannot provide the sufficient damping to
lead to an area law. Under those circumstances, its upper bound on Wilson
loops is thus likely to be useless. The argument can, however, be
strengthened by the replacement of the gluon propagator with an effective 
current interaction as mentioned above, and a stronger bound might be
obtained from that renormalisation group invariant combination of
contributions here also. This seems plausible, in particular, since 
the expectation values of Wilson loops are physical objects and should as such
have renormalisation scale invariant bounds. Gauge invariance in this context
simply means that such bounds from gauge covariant correlations will in
general depend on the specific gauge employed in their derivation. 
Judging from nowadays' knowledge, the original intention in Ref. \cite{Wes82} 
to obtain the area law from the gauge dependent upper bound by
demonstrating the existence of an infrared enhanced 2-point correlator in just
one specific gauge, seems too optimistic.

An example in which the infrared behaviour of the 2-point ghost correlations
provide the essential infrared strength for the area law was given in
Ref.~\cite{Zwa94}. Assuming that configurations close to the Gribov horizon 
overcompensate the suppression from the Faddeev--Popov determinant and
dominate the Yang--Mills partition function, an infrared enhanced ghost
propagator was conjectured. Together with additional ghost fields which were
introduced into the action via a Boltzmann factor this led to the following 
estimate for the Wilson loop,
\begin{equation}
W(C)  \, \propto \, \exp  - c \sigma^2  \oint\!\!\oint_C
dx_\mu  dy_\nu  \int d^4\!u \, d^4\!v  \,  
g^2D_{\mu\rho}(x-u) \,  g^2D_{\nu\rho}(y-v) \, D_G(u-v) \; . \nonumber
\end{equation}
The dimensionless constant $c$ herein is related to the leading infrared
behaviour of the gluon and ghost propagators, $D_{\mu\nu}(k) \propto 1/\sigma
$ and  $D_G(k)\propto \sigma/k^4 $ for $k^2 \to 0$, respectively, as
suggested by the estimates in this approach. 
The renormalisation group invariance of this bound is
somewhat unclear here again. Naively, one would at least  expect the
constant $c$ to have a suitable scale dependence to compensate  the one of the
other factors in the exponent in order to obtain an invariant
string tension. At least in the Landau gauge  the particular  combination
of ghost and gluon propagators does certainly not compensate the scale
dependence of the explicit couplings (the factor $g^4$). A second question is,
of course, whether it is more acceptable for the ghost correlations to violate
temperedness than it is for the gluons. However, this example once more 
demonstrates that ghosts can provide for the infrared dominant correlations in
large Wilson loops in covariant gauges. As will be discussed in 
Sec.~\ref{sec_lattice}, a behaviour of the gluon and ghost propagators, 
$D(k) \propto 1/\sigma $ and $D_G(k) \propto \sigma/k^4 $ for $k^2 \to 0$, 
is not incompatible with present lattice data. 
 
After these cautioning remarks on a possible $\sigma/k^4$ behaviour for any
two-point correlations in a sensible quantum field theory, in the next
sections early studies of the gluon Dyson--Schwinger equation will be described
in which exactly such an infrared enhancement was obtained. It will become
clear in the sequel that ghosts affect this conclusion. We reiterate that this 
alone would not necessarily exclude a $\sigma/k^4 $ behaviour for the effective 
interactions of quarks in the infrared.

\subsection{The Gluon Propagator in Axial Gauge}
\label{sec_Axial} 

Historically, an infrared enhanced gluon propagator $\propto \sigma/k^4$ 
was first obtained in the covariant gauge much before the importance of ghosts
became evident which had been neglected in the approximation summarised in
Sec. \ref{sub_Mand} below. The same infrared behaviour 
was confirmed in the early studies of the gluon DSE in the axial
gauge~\cite{Bak81a,Bak81b,Ale82}. In this section we will discuss why it now
seems plausible that these axial gauge results are just as inconclusive,
since they were based on a simplifying assumption quite analogous to
disregarding the ghosts of the covariant gauge.

\subsubsection{Dyson--Schwinger Equation Studies}
\label{sub_Ax1} 

In the axial gauge the gluon field is taken transverse to a fixed gauge vector
$t_\mu$, {\it i.e.} $t_\mu A^a_\mu = 0$. The tensorial structure of the gluon
propagator in this gauge is given in Eq.\ (\ref{glp_axg}) in Appendix
\ref{app_gluon-DSE}.  Note that two independent tensor structures and therefore
two scalar functions $f(p^2,(pt)^2)$ and $g(p^2,(pt)^2)$ are involved. The
tree-level propagator is recovered from setting $g(p^2,(pt)^2)\equiv 1$ and
$f(p^2,(pt)^2)\equiv 0$. A corresponding construction \cite{Kim80} of the
longitudinal part of the 3-gluon vertex  with the help of the Slavnov--Taylor
identity relating this vertex to the gluon propagator is summarised in Appendix
\ref{app_3-gluon}. Note that in this gauge the Slavnov--Taylor identity reduces
to a kind of Ward identity. In particular, no unknown contributions from higher
correlation functions such as the 4-point ghost-gluon scattering kernel in
covariant gauges enter in this axial gauge identity for the 3-gluon vertex
given in Eq.~(\ref{STI_ag}).
Nevertheless, keeping the full tensor structure of the gluon propagator leads
to a longitudinal part of the 3-gluon vertex which is of an 
enormous complexity, however.

The DSE for the gluon propagator is given explicitely in Eq.\
(\ref{eq:Gluon-DSE}) and pictorially represented in Fig.\ 
(\ref{fig:Gluon-DSE-A}) in Appendix \ref{app_gluon-DSE}.  Of course, the
ghost loop does not contribute in axial gauge.  As already mentioned in section
\ref{sec_DSE}, studies of the axial-gauge gluon  DSE 
\cite{Bak81a,Bak81b,Ale82,Sch82,Cud91,Gei99} 
rely on a simplifying assumption on the tensor structure of the gluon
propagator. That particular term which strength is parametrised by 
the function $f(p^2,(pt)^2)$, and that can only be non-vanishing 
non-perturbatively, has not been included in presently available studies of 
the gluon DSE in axial gauge~\cite{Bue95}. The original 
studies of the axial-gauge gluon DSE assumed that a possibly
infrared enhanced part of the gluon propagator has the same structure as the
tree-level propagator, {\it i.e.}, $ f = 0$ was set for simplification and
the structure corresponding to the function $f$ had been dismissed. This
also simplifies the longitudinal part of the 3-gluon vertex considerably: 
Only with the approximation $f = 0$, and with assuming, in addition,
$g(p^2,(pt)^2) \to g(p^2) $ in Eq.~(\ref{axg_3gv}), {\it i.e.}, that the function $g$ is independent of $t$,
this solution reduces to the simple form used in Refs.~\cite{Bak81a,Bak81b}. 
It was further noted that upon contracting the axial gauge gluon
Dyson--Schwinger 
equation for the vacuum polarisation tensor $\Pi_{\mu\nu}$ with the tensor
$t_\mu t_\nu$ only the 3-gluon loop 
plus a tadpole term give non-vanishing contributions to a
single integral equation for the one scalar function $g(p^2)$ of this
approximation scheme. And it was found numerically that $g(p^2) \propto
p^2/ \sigma $, {\it i.e.}, the
solution gives rise to an infrared enhanced gluon propagator $\propto
\sigma/p^4$. 

Before discussing some implications of this approximation scheme in the 
following subsection we will comment on a further simplification: In Ref.\
\cite{Sch82} the $t$-dependence (that had anyhow been disregarded in the
solution for the 3-gluon vertex) has  been consistently neglected also in the
Dyson--Schwinger equation. Then  the tadpole contribution disappears in
addition to the other contributions with 4-gluon  vertices~\cite{Bue96}. 
Furthermore, a one-dimensional approximation and subtractions have been
employed to ensure the masslessness condition 
\begin{equation}
  \lim_{k^2 \to 0} D_{\!\mu\nu}^{-1}(k^2) = 0 \, . \label{massless.MA}
\end{equation}
This
masslessness condition follows from the axial gauge Slavnov--Taylor identity
(\ref{STI_ag}) in the limit $k \to 0$  with assuming that the vertex does not
have massless poles in the external momenta.\footnote{This condition was
in some older studies assumed to hold also for the Landau gauge. In
Sec.~\protect\ref{sub_Ghost} it 
will be demonstrated, however, that ghosts change this conclusion and that the
masslessness condition in the present form (\protect\ref{massless.MA}) does not
hold in general in Landau gauge.} The resulting equation, of a structure
similar to the Mandelstam equation to be 
discussed in Sec.\ \ref{sub_Mand}, was found to
confirm the findings of Refs.\ \cite{Bak81a,Bak81b} with respect to the
infrared behaviour of the solution.

Subsequent studies based on the identical approximation scheme for the axial
gauge came to a somewhat dissentive conclusion proposing the
possibility of a gluon propagator less singular than a massless free
particle pole $1/p^2$ in the infrared~\cite{Cud91}. This possibility was,
however, later attributed to a sign error in the resulting
equation~\cite{Bue95,Bue96}. The infrared soft solution to 
the approximate gluon DSE of Ref.~\cite{Sch82}
for this wrong sign also yields the wrong sign in the well-known perturbative
contributions valid at high momenta \cite{Bue96}.  
In Ref.~\cite{Bue96} a detailed comparison of the approximate gluon DSEs of
the axial gauge (in the scheme of Ref.~\cite{Sch82}) with the Mandelstam
approximation to the Landau gauge DSE (see Sec.~\ref{sub_Mand} below) showed 
some analogies in these two schemes. In light of this, it might thus not be too
surprising anymore that previous DSE studies of the gluon propagator in both
these gauges seemed to agree in indicating an infrared enhancement of the gluon
propagator, $D(k) \propto \sigma/k^4 $ for $ k^2 \to 0$.\footnote{It
is nevertheless not entirely trivial that the solutions to different
non-linear integral equations, may they look similar or not, show this same
behaviour.}

Studies of the gluon Dyson--Schwinger equation using the light-cone gauge,
$t^2 = 0$, in which the second tensor structure (as well as the spurious gauge
singularities $\propto 1/(kt)^2$, see the next section) are obviously avoided,
seem to come to similar conclusions, see Refs.~\cite{Nat85,Vac89,Vac94}. The
light-cone gauge has its own problems, however, for an extensive discussion
see,  {\it e.g.}, Ref.~\cite{Gai90}. Here it suffices to mention that ghost
fields are indispensable in light-cone gauges, see Ref.\ \cite{Nak00} and
references therein for the recent developments concerning this issue.  

The gluon propagator in axial gauge has recently also been investigated  
\cite{Lit98} within the renormalisation group approach based on Wilson's flow 
equations \cite{Wil74}. For a finite infrared regulator the typical axial-gauge
singularities of the tree-level gluon propagator (see below) are thereby 
removed. A functional form for modified Ward identities which take 
the non-vanishing infrared regulator into account has been derived.
It remains to be shown, however, whether a possible solution for the gluon 
propagator of the corresponding flow equations, in a reasonable truncation 
scheme,
stays well-behaved and finite for all momenta when these equations are 
integrated down towards a vanishing infrared scale.

\subsubsection{The Spectral Representation}
\label{sub_Ax2} 

On one hand, an infrared enhanced gluon propagator is known to lead to
an area law for expectation values of large Wilson loops~\cite{Wes82},  on
the other hand it was pointed out in Ref.~\cite{Wes83} that a full axial
gauge gluon propagator more singular than $1/k^2$ in the infrared violates
positivity of the norm in the corresponding Hilbert space.
Such an infrared enhanced behaviour is thus generally excluded for physical
correlation functions, see also Sec.~\ref{sub_Pos}. 
We will therefore discuss the possibilities to infer positivity violations of
transverse gluon states from the axial-gauge propagator in the following two
subsections.    

The spectral representation for the gluon propagator in the
(spacelike) axial gauge involving the spectral functions $\rho_g$ and
$\rho_f$ corresponding to the two different tensor structures ({\it c.f.}, Eq.\
(\ref{glp_axg})) is given
by~\cite{Wes83}, 
\begin{equation}  \label{specrep_ag}
D_{\mu \nu}(k) \, = \, \int_0^\infty d\mu^2 \bigg\{  \frac{\rho_g(\mu^2,kt)}{
k^2 +  \mu^2 }\,  {\mathcal M}_{\mu \nu}(k)  \, -  \frac{\rho_f(\mu^2,kt)}{
k^2 +  \mu^2 }\,  {\mathcal P}_{\mu \nu}(t)   \bigg\} \; .
\end{equation}
Imposing equal-time canonical commutation relations in the standard way, 
one obtains their respective spectral sum rules as follows,
\begin{eqnarray}
\int_0^\infty dq^2 \, \rho_g(q^2,qt) &=& 1 \label{specsum_g}\\
\int_0^\infty dq^2 \, \rho_f(q^2,qt) &=&  -\frac{g^2}{3}
\int_{-\infty}^\infty \, dl  \, \epsilon(l) \, e^{-il(qt)} \,
\langle A^a_i( l \hat t) A^a_i(0) \rangle   \; , \quad \hat t_\mu  =
t_\mu /t^2  \; , \label{specsum_f}
\end{eqnarray}
where $\epsilon(l)$ is the signature function.
For the interacting theory modifications are necessary since
equal-time commutation relations are lost in general, see Sec.~\ref{Sec2.3}.
Note though that quite interestingly a finite value for the integral of the
second spectral function, the r.h.s in Eq.~(\ref{specsum_f}), might be used to
obtain a bound on the coupling $g^2$ for finite $\langle A^a_i A^a_i \rangle$.

The positivity of the two spectral functions separately being affected by the
axial gauge singularity (see the next section), for the difference of these
spectral functions it was shown unambiguously in Ref.~\cite{Wes83}
that positive definiteness in a space of physical gluon states in the axial
gauge would imply the positivity condition,
\begin{equation}
\rho_g(k^2,kt) \, -\,  \rho_f(k^2,kt) \ge 0 \; .   \label{pos_ag}
\end{equation}
To see how this generalises the analogous condition for QED, recall that in
QED the field strengths are gauge invariant and thus, in particular,
independent of the gauge vector $t$ in axial gauge from which it can be
inferred that $\rho_f  \equiv 0 $ as well as $\rho_g(k^2,kt) \equiv
\rho_g(k^2) $ which is identical to the spectral density  $ \rho(k^2) $ in
the covariant gauge. Then, $ \rho_g(k^2)\ge 0 $ from Eq.~(\ref{specsum_g})
implies K\"all\'en screening for the renormalised (physical) charge (to be
smaller than the bare charge), and the gauge invariance of $\rho_g(k^2)$
reflects the invariance of the Coulomb potential.

From the discussion above it is clear that the assumptions underlying the
studies of refs~\cite{Bak81a,Bak81b,Ale82,Sch82,Cud91,Bue95}, of the gluon
DSE in axial gauge can be thought of as an Abelian approximation to 
the more general structures that can occur in the non-Abelian theory. 
QED does provide an example to demonstrate, on the other hand,  
that it is the coefficient of the metric tensor (here the Euclidean
$\delta_{\mu\nu}$) in the spectral representation of the photon which is
gauge invariant and which thus has a physical interpretation. 
For the full non-Abelian structure of the axial gauge gluon
propagator and its corresponding spectral condition~(\ref{specrep_ag}) this
coefficient is given by $\rho_g(k^2,kt) \, -\,  \rho_f(k^2,kt)$, and the
positivity condition (\ref{pos_ag}) certainly suggests to study this
difference. Therefore, the possibility of the additional tensor structure
introduced by $f\not=0$ should be considered not only to explore physical
consequences such as long range forces but to first of all assess the
question of positivity. Whether or not the gluonic degrees of
freedom in the axial gauge have a particle interpretation cannot be
decided from the results of~\cite{Bak81a,Bak81b} for $g(k^2)$ alone.

This clearly demonstrates that progress is desirable in axial gauge. 
Another important
pre\-requisite for this will be a proper treatment of the spurious infrared
divergences which are well-known to be present in axial gauge due to the zero
modes of the covariant derivative \cite{Gai90}.
This can be achieved by either introducing redundant  degrees of freedom,
{\it i.e.}, ghosts, also in this gauge \cite{Lav89} (by which it obviously
looses its particular advantage) or by using a modified axial gauge
\cite{Len94a,Len94b}, which is specially designed to account for those zero
modes. Ultimately, progress in more than one gauge will be the only
reliable way to asses the influence of spurious gauge dependencies.

\subsubsection{Positivity in Axial Gauge}
\label{sub_Ax3} 

The same contradiction between positivity and antiscreening as discussed in
Sec.~\ref{Sec2.3} applies to the axial gauge  (here with $\gamma = 1$ and
$Z_3^{-1} \to 0$),  
if it can be shown that positivity of the norm implies not only 
the inequality in (\ref{pos_ag}) but also the stronger
condition $\rho_g(k^2, kt) \ge 0$. It was argued in Ref.~\cite{Wes83} 
that this condition is generally not satisfied in the axial gauge, however. 
In particular, for large $k^2$, {\it i.e.}, in the ultraviolet, which for
dimensional reasons 
corresponds to very small $kt$ in the spectral function, it depends on the
prescription adopted for the gauge singularities $\sim 1/(kt)$. This argument
is based on the observation that positivity of the norm of the states in
axial gauge (spacelike corresponding to Euclidean $t^2 > 0 $) actually 
implies the modified condition,  
\begin{equation}
\frac{t^2}{(kt)^2} \, \rho_g(k^2, kt) \, \ge\,  0 \; . 
\end{equation}
Accordingly, the positivity of $\rho_g$ depends on the positivity
of the prescription adopted for the gauge singularity \cite{Gai90}. 
In particular, for the
principal value prescription (which is {\em not} positive) it follows that 
$\rho_g(k^2, 0) \le 0 $ which in generalisation of the perturbative argument
given in Ref.~\cite{Fre76} would be sufficient to resolve the
contradiction between positivity and asymptotic freedom in the axial
gauge. It is clear that this resolution of the issue of positivity is itself 
based on an artifact of the principal value prescription, however. 
In fact, it is known by now that defining the
square of the principal value prescription (usually denoted by square
brackets) in the gluon propagator through its derivative, 
\begin{equation}
\frac{1}{kt}  \,\to\, \mbox{P}\frac{1}{kt} \,=:\, \frac{1}{[kt]} \; , \quad
\mbox{and} \quad \frac{1}{(kt)^2} \, \to\,  - \frac{d}{d(kt)} \frac{1}{[kt]}
\; , \end{equation}
results in loosing positivity of the polarisation sum at the Feynman poles
of the tree-level gluon propagator. This demonstrates that such a  
prescription in fact introduces unphysical degrees of freedom also in the axial
gauge~\cite{Nak83,Bas89}. Exactly
this prescription was also employed, however, 
for the double pole in the spectral representation of the
full propagator in Ref.~\cite{Wes83}. 
As a result, the resolution of the Oehme--Zimmermann
paradox proposed in Ref.~\cite{Wes83} is most likely 
itself based on the presence of negative norm states. 
It is thus inconclusive. In contrast to this, other prescriptions such as the
positive Mandelstam--Leibbrandt 
prescription or the planar gauge choice naturally lead to redundant degrees of
freedom as a result of the impossibility to fully eliminate $t_\mu
A_\mu^a$. These degrees of freedom give rise to negative norm states. Their
elimination by projection onto a subspace of semi-definite norm suffices to
recover Gauss' law and unitarity and, in addition, canonical quantisation
then results in a free (tree-level) gluon propagator without double pole of
the form $1/(kt)^2$~\cite{Bas89}. While the method of canonical quantisation
may not be too appropriate for interacting theories, this nevertheless
suggests that the stronger positivity condition
\begin{equation}
\rho_g(k^2,kt) \, \ge 0 \;  \label{stposag} 
\end{equation}
should also hold in the axial gauge, implying that the Oehme--Zimmermann
superconvergence relation (\ref{OZ}) does as well.

In either case, the considerations based on the principal value prescription
cannot establish positivity in the axial gauge. Note however, that 
the positivity condition for the difference of the spectral functions
(\ref{pos_ag}) is unaffected by this, {\it i.e.}, by the question whether the
stronger condition (\ref{stposag}) has to hold in general or not. 
The relevance of the second tensor structure in the axial gauge gluon
propagator as emphasised in Refs.\ \cite{Bue95,Bue96} holds independent of
the artifacts of the principal value 
prescription for the gauge singularity. In addition, this singularity 
is less relevant for
the infrared behaviour than it is in the ultraviolet, as the limit $k^2 \to 0$
can be studied for $kt > 0$. On the other hand, the typical zero modes of QCD
in the axial gauge make an important difference as compared to QED: these zero
modes prevent the spectral density  $\rho_g(k^2,kt)$ from being independent of
the gauge vector $t_\mu$~\cite{Wes83}.

As a result, it seems fair to say that available studies of the gluon
propagator from the axial gauge DSE cannot be regarded any more conclusive
than those of the Mandelstam approximation to the Landau gauge DSE, see Sec.\
\ref{sub_Mand} below. As for
the Landau gauge, it will be demonstrated in Sec.~\ref{sub_Ghost} that, instead
of the gluon propagator, the previously neglected ghost propagator assumes an
infrared enhancement similar to what was then obtained for the gluon. It is
certainly not inconceivable that the additionally possible structure in the
axial gauge has some similar effect there, too. To this end it is interesting
to note that a study of the gauge boson propagator of the dual Abelian Higgs 
model in axial gauge does lead to a non-vanishing function $f(p^2,(pt)^2)$
\cite{Che99}. 
Herein, an infrared suppressed gauge boson propagator has been
obtained, and an inspection of the signs involved in the functions
$f(p^2,(pt)^2)$ and $g(p^2,(pt)^2)$ reveals that this gauge boson propagator
does indeed violate positivity also in the axial gauge.
 
\subsection{Truncation Schemes for Propagators in Landau Gauge}
\label{sec_Trunc} 

\begin{figure}
  \leftline{ \hskip 2.5cm
        \epsfig{file=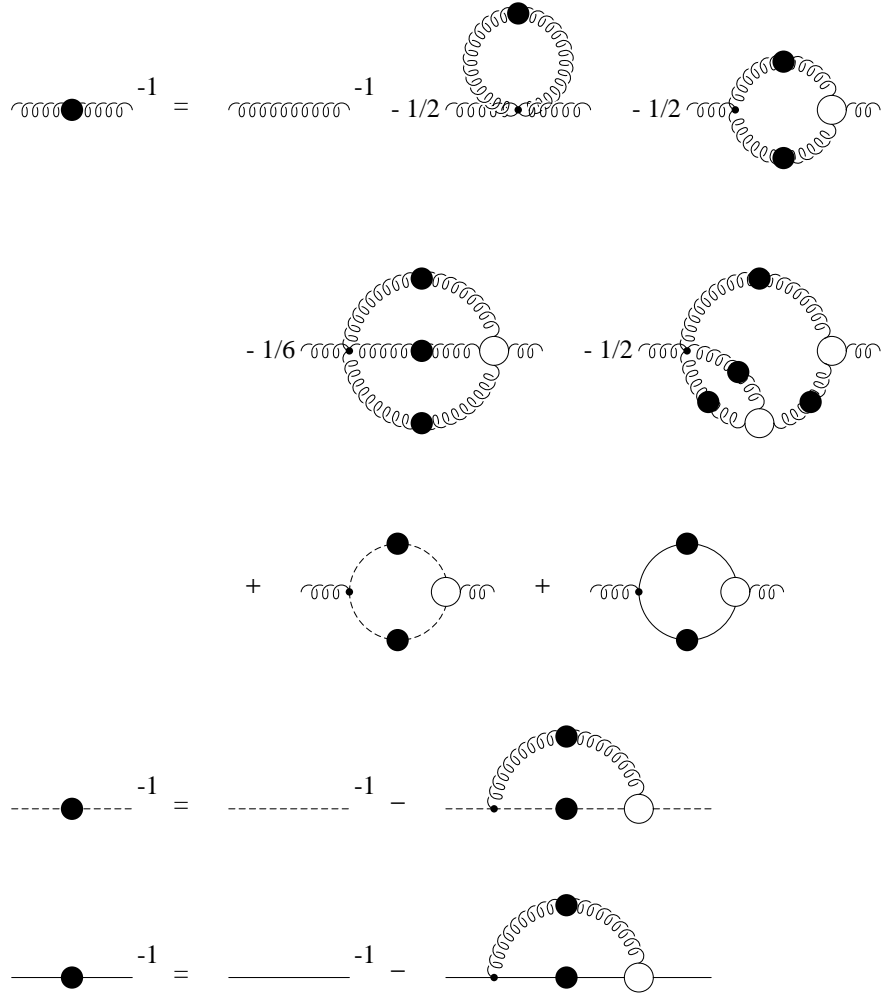,width=0.7\linewidth} }
  \caption[Diagrammatic representation of the gluon, ghost and quark
           DSEs of QCD.]
          {Diagrammatic representation of the gluon, ghost and quark
           DSEs of QCD. The wiggly, dashed and solid
           lines represent glue, ghosts and quark, respectively. Filled blobs
           stand for fully dressed propagators and open circles for
           one-particle irreducible vertices. }
  \label{GlGhQuark}
\end{figure}

In this section we present a truncation scheme for the Landau gauge
QCD Green's functions which is complete at the level of two-point functions,
{\it i.e.}, the propagators. The corresponding system of DSEs whose
derivation has been reviewed in Sec.\ \ref{sec_DSE} is diagrammatically
represented in Fig.\  \ref{GlGhQuark}. In addition to the emphasis on 
maintaining the correct infrared properties as much as possible, 
the known ultraviolet behaviour of the Green's functions provides
another important guideline in the assumptions employed to simplify the
system. Hereby we  start with a discussion of the
gluon DSE. In a first step towards its truncation consider the 
terms containing explicit four-gluon vertices. These are the momentum
independent tadpole term, an irrelevant constant which vanishes
perturbatively in Landau gauge, as well as explicit two-loop contributions
to the gluon DSE. A possibly non-vanishing contribution from the tadpole term
beyond perturbation theory can nevertheless be eliminated from the equation
upon contraction of the free Lorentz indices with 
momentum projector $\mathcal{R}_{\mu\nu}(k) = \delta_{\mu\nu} - 4 k_\mu k_\nu
/k^2$, see Sec. \ref{sec_QED3}.
For the explicit two-loop terms one first notes that these lead to subdominant
contributions to the gluon propagator 
in the ultraviolet. Possible solutions can therefore be expected to resemble
the correct leading perturbative behaviour for asymptotically high momenta
also without those terms. It should be kept in mind though that these terms
will have an effect, {\it e.g.}, on the running QCD coupling, at next to
leading order in perturbation theory. While this can partially be compensated
for by an adjustment of the scale parameter of the subtraction scheme as will
be discussed in Sec.\ \ref{sub_Sub} below, discrepancies with two-loop 
perturbative
results in the energy range where these are phenomenologically well supported
serve as one indicator of the quantitative effects of the truncations.

Secondly, and more importantly, for the infrared behaviour it has been argued
that the singularity structure of the two-loop terms does not interfere with
that of the one-loop terms~\cite{Vac95}. The argument is based on studies of
certain Ans\"atze for the leading infrared behaviour of all correlation
functions and their self-reproduction in the coupled equations. A
dimensional regularisation was adopted and integer powers of the invariant
momenta for these Ans\"atze were studied with the result that the explicit
two-loop contributions to the gluon DSE turn out to be subleading in the
infrared and do not influence the leading contributions from the one-loop
terms. The conditions for self-reproduction of the respective contributions
(with different exponents) in the gluon DSE were found to decouple from each
other. This
argument is unfortunately still incomplete, however, because the possibility
for non-integer exponents in the leading infrared behaviour was not
considered. The solutions to the coupled system of gluon and ghost equations
as presented below in this section provide strong evidence for precisely such
an infrared behaviour determined by a non-integer exponent. For this type of
solutions, however, it has been verified a posteriori by an explicit
calculation that the two-loop terms are subleading in the infrared
\cite{Wat00}. 

In light of this evidence  and of the 
tremendously rich structure of the non-perturbative 4-gluon vertex
function introducing a whole new degree of complexity, it seems reasonable to
study the non-perturbative behaviour of the gluon propagator in the infrared
without explicit contributions from four-gluon vertices, {\it i.e.}, 
without the
two-loop diagrams in Fig.~\ref{GlGhQuark}. The renormalised
Dyson--Schwinger equation for the inverse gluon propagator $D^{-1}_{\mu\nu}$
in Euclidean momentum space with positive definite metric, $g_{\mu\nu} =
\delta_{\mu\nu}$, (colour indices suppressed) then simplifies as follows,
\begin{eqnarray}
  D^{-1}_{\mu\nu}(k)
    &=& Z_3 \, {D^{\hbox{\tiny tl}}}^{-1}_{\mu\nu}(k)\,
 - \, g^2 N_c \, \widetilde Z_1 \int \frac{d^4q}{(2\pi)^4} \; iq_\mu \,
D_G(p)\, D_G(q)\, G_\nu(p,q)
\label{glDSEsim}\\
&& - \, g^2 Z_{1F} N_f \frac{1}{2} \int \frac{d^4q}{(2\pi)^4} \;
   {\rm tr} \left( \gamma_\mu \, S(p) \, \Gamma_\nu (p,q) \, S(q) \right)
   \nonumber \\
&+& g^2 N_c\, Z_1  \frac{1}{2} \int \frac{d^4q}{(2\pi)^4} \;
\Gamma^{\hbox{\tiny tl}}_{\mu\rho\alpha}(k,-p,q)
       \, D_{\alpha\beta}(q) D_{\rho\sigma}(p) \,
                \Gamma_{\beta\sigma\nu}(-q,p,-k)
  \; ,\nonumber
\end{eqnarray}
where $p = k + q$, $D^{\hbox{\tiny tl}}$ and $\Gamma^{\hbox{\tiny tl}}$ are
the tree-level propagator and three-gluon vertex, $D_G$ is the ghost
propagator, $S$ the quark propagator, and $\Gamma$ and $G$ are the respective
fully dressed 3-point vertex functions. The DSEs for the ghost propagator
and the quark propagator, formally without any truncations, will be used
as given in Eqs.\ (\ref{ghDSEmom},\ref{quarkDSEmom}).

In order to arrive at a closed set of equations for these functions,
it is still necessary to specify the form of the three remaining vertex
functions, the ghost-gluon, the  quark-gluon and the 3-gluon vertex 
functions. As discussed in Sec.\ \ref{sec_nPoint} little is known 
about possible solutions to their corresponding DSEs. For the 3-gluon vertex,
independent of the presence of quarks, a general procedure to construct a
solution to its Slavnov--Taylor identity (\ref{glSTI}) is in principle possible
\cite{Bar80,Bal80,Kim80}. Since this procedure involves unknown contributions
from the ghost-gluon scattering kernel, it cannot be readily applied to
express the vertex functions entirely in terms of the ghost and gluon
renormalisation functions $G$ and
$Z$. To achieve this one has to make additional  assumptions on the
scattering kernel. Since this kernel is related to the ghost-gluon vertex
the construction of the latter via the identity (\ref{eq:ghost-WTI}) should be
done before explicitely solving the Slavnov--Taylor identity (\ref{glSTI})
for the 3-gluon vertex. Noting that a simple solution to Eq.\ (\ref{glSTI})
is possible if ghosts are neglected completely  \cite{Bar80,Bal80,Kim80} we will
digress briefly in order to discuss the historically 
first and particularly drastic approximation to the gluon DSE.

\subsubsection{The Mandelstam Approximation}
\label{sub_Mand}  

The first approximation scheme for the gluon DSE in Landau gauge was 
originally proposed by Mandelstam \cite{Man79}. 
The essential truncating assumption is, even though working in Landau gauge,
to neglect all ghost contributions to the gluon DSE of pure QCD (without
quarks). As a justification for this, it was usually referred to perturbative
calculations which yield numerically small ghost contributions to the gluon
self-energy. Even though there was never any doubt about the importance of
ghosts for fundamental reasons such as transversality of the gluon propagator
and unitarity, it was asserted that their quantitative contributions to many
hadronic observables might remain negligible even beyond perturbation
theory. It will be seen later on in this section that presently available
solutions to the coupled system of gluon {\sl and} ghost DSEs yield
qualitatively quite different results as compared to the Mandelstam
approximation, and are thus counterexamples to this assertion. The exact
implications of the importance of ghosts, in particular, in the infrared, on
the effective interactions of quarks and on hadronic observables are more
subtle as will be discussed in detail in the appropriate context of later
sections.

\begin{figure}
  \centerline{
 \epsfig{file=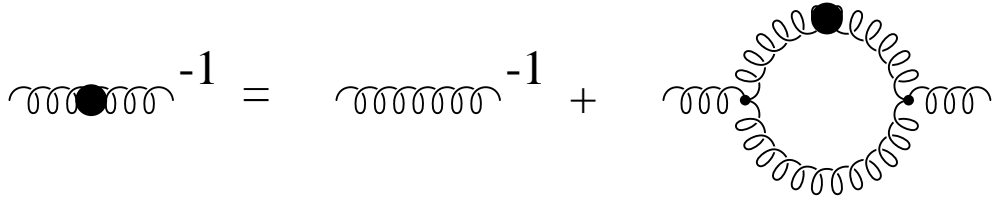,width=0.8\linewidth}}
  \caption{Diagrammatic representation of the gluon Dyson--Schwinger
           equation in Mandelstam's approximation.}
  \label{Mandelstam}
\end{figure}

As already stated, without ghosts the solution of Slavnov--Taylor identity
(\ref{glSTI}) assumes a particularly simple form \cite{Bar80,Bal80,Kim80}
(up to undetermined transverse terms):
\[  \Gamma_{\mu\nu\rho}(p,q,k)
       \, =\, - A_+(p^2,q^2)\,  \delta_{\mu\nu}\,  i(p-q)_\rho\,
          - \,  A_-(p^2,q^2)\,  \delta_{\mu\nu} i(p+q)_\rho  \]
\begin{equation}
   \hskip 6mm - \, 2\frac{A_-(p^2,q^2)}{p^2-q^2}
( \delta_{\mu\nu} pq \, -\,  p_\nu
q_\mu) \, i(p-q)_\rho\,  + \; \hbox{cyclic permutations} \; ,
  \label{3gv.MA}
\end{equation}
\[ \hskip -1cm\mbox{with} \quad    A_\pm (p^2,q^2)  =
      \frac{1}{2} \, \left( \frac{1}{Z(p^2)} \pm
      \frac{1}{Z(q^2)} \right) \quad .
\]
Assuming that the gluon renormalisation function $Z(p^2)$ is a slowly varying 
function one may then approximate this solution by
\begin{equation}
  \Gamma_{\mu\nu\rho}(p,q,k)
    = A_+(p^2,q^2) \Gamma^{\hbox{\tiny tl}}_{\mu\nu\rho}(p,q,k)  \;
    . \label{eq:MaAn}
\end{equation}
While this form for the full three-gluon vertex simplifies the 3-gluon loop
in the gluon DSE even more than the use of another bare vertex, which
corresponded to setting $\Gamma \equiv \Gamma^{\hbox{\tiny tl}}$,
the approximation (\ref{eq:MaAn}) was observed by Mandelstam to be
superior to the latter since it accounts for some of the dressing of the
vertex as it results from the corresponding Slavnov--Taylor identity. The
nature of this dressing is such that it cancels the dressing of one of the
gluon propagators in the 3-gluon loop, and without ghost contributions the
gluon DSE in the Mandelstam approximation assumes the comparatively simple form
\begin{eqnarray} \label{glMA}
  D^{-1}_{\mu\nu}(k)
    &=&  Z_3 \, {D^{\hbox{\tiny tl}}}^{-1}_{\mu\nu}(k) \\
&& \hskip -1cm +  g^2 N_c\, \frac{1}{2} \int \frac{d^4q}{(2\pi)^4}
    \; \Gamma^{\hbox{\tiny tl}}_{\mu\rho\alpha}(k,-p,q)
       \, D_{\alpha\beta}(q) D^{\hbox{\tiny tl}}_{\rho\sigma}(p) \,
                \Gamma^{\hbox{\tiny tl}}_{\beta\sigma\nu}(-q,p,-k)
  \; \nonumber
\end{eqnarray}
where $p=k+q$. This equation, the Mandelstam equation, is schematically
depicted in Fig.~\ref{Mandelstam}. It was already pointed out by Mandelstam
that in order to solve this equation self-consistently it is necessary to
implement an additional constraint: Without ghosts the
Slavnov--Taylor identity for the 3-gluon vertex and its
solution~(\ref{3gv.MA}) for $p^2 \to 0$ entail the masslessness condition
(\ref{massless.MA}) which may be written explicitely as
\begin{equation}
  \lim_{k^2 \to 0} \frac{k^2}{Z(k^2)} = 0 \, .
  \label{massless.Z}
\end{equation}
While imposing this additional condition seems consistent with the other
assumptions in the Mandelstam approximation, it has to be emphasised that
concluding (\ref{massless.Z}) as a result of (\ref{3gv.MA}) relies solely on
neglecting all ghost contributions in covariant gauges.

The Mandelstam equation in its original form is obtained from
Eq.~(\ref{glMA}) upon contraction with the transversal projector
${\mathcal P}_{\mu\nu}(k) = \delta_{\mu\nu} - k_\mu k_\nu /k^2$.
As stated in the last chapter in connection with the discussion of 
the photon DSE, this leads to an ambiguity in combination with an ultraviolet
momentum cutoff which violates gauge invariance.
The corresponding quadratic ultraviolet divergence  has in the early studies
been absorbed by a
suitably added counter term introduced in order to account for the masslessness
condition~(\ref{massless.Z}), see Refs.~\cite{Man79,Atk81}. The solution to
this equation proposed by Mandelstam's infrared analyses proceeds briefly as
follows: Assume $Z(k^2) \sim 1/k^2 $. 
This exclusively yields contributions which violate
the masslessness condition~(\ref{massless.Z}). Such terms have to be
subtracted. Since the kernel of Mandelstam's equation is
linear in $Z$, this is achieved by simply subtracting the corresponding
contribution from $Z$ in the integrand. Thus defining
\begin{equation} \label{MAsol}
        Z(k^2) =  \frac{b}{k^2} + C(k^2) \; , \quad  b = \hbox{const.},
\end{equation}
and retaining only the infrared subleading second term in the integrals,
one obtains a solution to this equation, which can be shown to
vanish in the infrared by some non-integer exponent of the
momentum \cite{Man79},
\begin{equation}
  C(k^2) \sim  (k^2)^{\gamma_0} \; , \quad
  \gamma_0 = \sqrt{\frac{31}{6} - 1} \simeq 1.273 \; ,
  \quad \hbox{for} \; k^2 \to 0 \; .
\end{equation}
Subsequently, an existence proof, a discussion of the singularity structure
and an asymptotic expansion of the solution generalising Mandelstam's
discussion of the leading behaviour of $C(k^2)$ in the infrared was given by
Atkinson et al. \cite{Atk81}.

\begin{figure}
 \centerline{ \epsfig{file=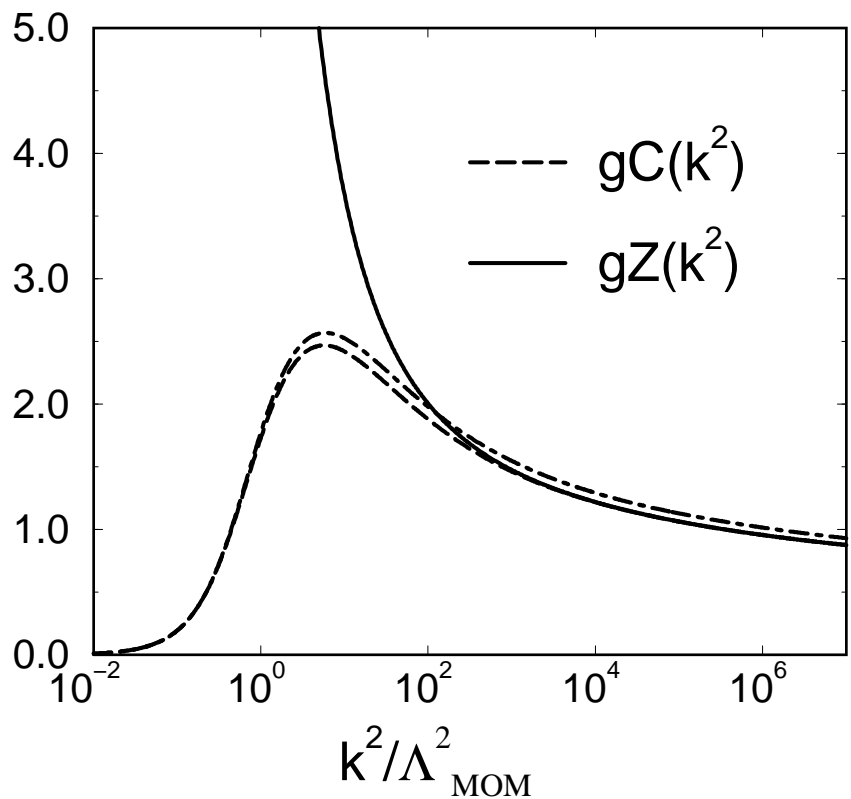,width=0.8\linewidth} }
 \caption[The gluon renormalisation function for Mandelstam's equation.]
         {The gluon renormalisation function $g Z(k^2) = 8\pi\sigma /k^2
          + gC(k^2)$ for Mandelstam's equation projected with ${\mathcal
	  R}_{\mu\nu}$. The dashed lines
          show the corresponding function $g C(k^2)$ for this equation and
	  Mandelstam's original equation (projected with ${\mathcal
	  P}_{\mu\nu}$).}
 \label{fig:gluon}
\end{figure}

Contracting instead of the transversal projector
${\mathcal P}_{\mu\nu}(k) = \delta_{\mu\nu} - k_\mu k_\nu /k^2$
with ${\mathcal R}_{\mu\nu}(k) = \delta_{\mu\nu}  - 4 k_\mu k_\nu /k^2$ leads
to a somewhat modified 
equation for the gluon renormalisation function, in particular, without
quadratically ultraviolet divergent terms~\cite{Bro89}.
The solution to this equation has an infrared behaviour quite similar to the 
solution of Mandelstam's original equation \cite{Hau96}.
In particular, explicitly separating
the leading infrared contribution according to (\ref{MAsol}) one obtains
a unique solution of the form
\begin{equation}
  C(k^2) \sim  (k^2)^{\gamma_0} \, , \;
  \gamma_0 = \frac{2}{9} \sqrt{229} \cos\left(\frac{1}{3}
  \arccos\left(-\frac{1099}{229\sqrt{229}}\right)\right) - \frac{13}{9}
  \simeq 1.271 \; ,
\end{equation}
for $k^2 \to 0$. Both equations can be solved using a combination of numerical 
and analytic methods~\cite{Hau98a}. In the infrared, the asymptotic expansion
technique of Ref.\ \cite{Atk81} is applied to calculate successive terms
recursively. The asymptotic expansions obtained this way were then matched to
the iterative numerical solution (see Ref.~\cite{Hau98a} for details).
As can be seen from Fig.~\ref{fig:gluon} the solutions to both versions
of Mandelstam's equation are very similar to each other.

A remark concerning the renormalisation of Mandelstam's equation is in order.
Logarithmic ultraviolet divergences are absorbed in the gluon
renormalisation constant $Z_3$ which can be shown to obey the identity
$Z_g Z_3 = 1$ in Mandelstam approximation \cite{Hau96}. This entails that the
product of the coupling and the gluon propagator, $g D_{\mu\nu}(k)$, does not
acquire multiplicative renormalisation in this approximation scheme. Using a
non-perturbative momentum subtraction scheme corresponding to the
renormalisation condition $Z(k^2 = \mu^2) = 1$
for some arbitrary renormalisation point $\mu^2 > 0 $, the resulting equation
can in both cases be cast in a renormalisation group invariant form
determining the renormalisation group invariant product $g Z(k^2)$ which
is equivalent to the running coupling $\bar g(t,g)$ of the scheme,
\begin{equation} \label{rc.MA}
  g Z(k^2) = \bar g(t_k,g) \; , \quad t_k = \frac{1}{2} \ln k^2/\mu^2 \; .
\end{equation}
The scaling behaviour of the solution near the
ultraviolet fixed point is determined by the coefficients $\beta_0 = 25/2$
and $\gamma_A^0 = 25/4 $ for Mandelstam's original equation or 
$\beta_0 = 14$ and $\gamma_A^0 = 7$ for the equation projected with 
${\mathcal R}_{\mu\nu}$.  These values are reasonably close to the
perturbative values for $N_f = 0$, {\it i.e.},  $\beta_0 = 11$ and
$\gamma_A^0 = 13/2$, the difference being attributed to neglected ghost
contributions.

The most important feature of the results concerning the Mandelstam
approximation is the infrared enhancement of the gluon propagator.
With ghosts completely neglected one obtains then an infrared enhanced 
quark-antiquark interaction 
\begin{equation}
g D_{\mu\nu}(k)  \, = \, {\mathcal P}_{\mu\nu}(k) \,\left( \,
 \frac{8\pi\sigma}{k^4}\,
 + \, \frac{gC(k^2)}{k^2} \, \right)  \label{ir_enh}
\end{equation}
which is renormalisation group invariant in the present approximation. It
especially allows to identify the string tension $\sigma $ and relate it to
the scale $\Lambda_{\hbox{\tiny MOM}}$ of the subtraction scheme \cite{Hau96}.

\subsubsection{Including Ghosts}
\label{sub_Ghost} 

\begin{figure}
  \centerline{ \epsfig{file=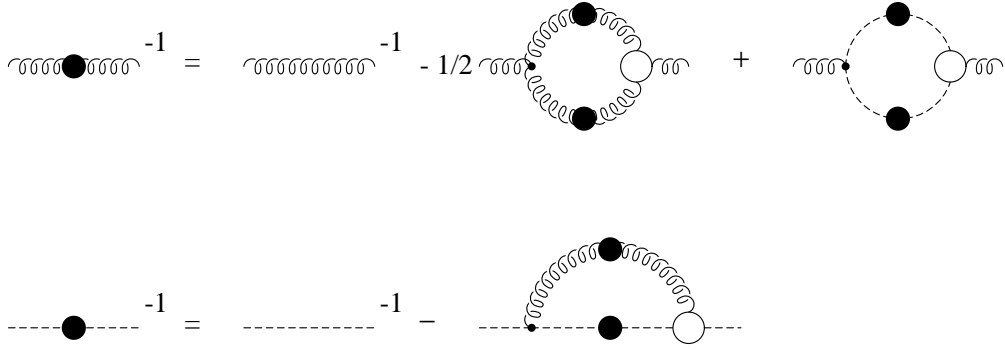,width=0.8\linewidth} }
  \caption{Diagrammatic representation of the gluon and ghost Dyson--Schwinger
           equations of QCD without quarks. In the gluon DSE terms with
           four-gluon vertices have been dismissed.}
  \label{GluonGhost}
\end{figure}

In this subsection it will be demonstrated that the inclusion of ghosts
change the solution for the gluon propagator drastically. 
This can be seen from 
the simultaneous solution of the coupled subsystem of gluon and ghost DSEs
(Eqs.\ (\ref{glDSEsim}) and (\ref{ghDSEmom})) without quarks, as depicted
in Fig.~\ref{GluonGhost}, see Refs.~\cite{Sme98,Hau98b}.
These solutions were obtained using the Landau gauge ($\xi = 0$) in
which one has $\widetilde Z_1 = 1$ \cite{Tay71}. Therefore, for the
ghost-gluon vertex the choice of its tree-level form in the DSEs might
seem justified as a first truncating assumption at least to the extend that
the asymptotic behaviour of the solutions in the ultraviolet does not
imply the necessity of renormalisation scale dependence by radiative
corrections of the ghost-gluon vertex in this gauge. This is in contrast to
the  other vertex functions in QCD which have to be dressed at least in such
a way as to account for their anomalous dimensions, if the solutions to the
DSEs in terms of propagators are expected to resemble their leading
perturbative behaviour at short distances. 
A good way to achieve this is in general is to construct vertex
functions from their respective Slavnov--Taylor identities which of course
respect the relations between the anomalous dimensions of vertex functions
and propagators, {\it i.e.}, their respective renormalisation scale
dependences.

As stated already in Chapter \ref{chap_Form},
the Slavnov--Taylor identities generally fail, however, to 
fully constrain the vertex functions. We have seen in the last chapter
that, {\it e.g.}, methods of refinement of such
constructions from implications of multiplicative renormalisability have been
developed for the fermion-photon vertex in QED. In the present context,
one can furthermore make use of the fact 
that the 3-gluon vertex has a considerably higher
symmetry than fermion vertices. This symmetry, namely its full Bose symmetry,
alleviates the problem of unconstraint terms considerably and the results are
not expected to be overly sensitive to such terms in the 3-gluon vertex.

At first sight it might seem appealing to retain the tree-level form of 
the ghost-gluon vertex  while taking the presence of the
ghost renormalisation function $G(k^2) $ into account in solving the
Slavnov--Taylor identity
(\ref{glSTI}) for the 3-gluon vertex. This can be shown to lead to a
contradiction: The STI for the 3-gluon vertex has
no fully Bose symmetric solution with this Ansatz~\cite{Sme98}.
One possibility to resolve this problem is to add structure to the
ghost-gluon scattering kernel. Its minimal modification necessary to admit a
symmetric solution to the Slavnov--Taylor identity for the 3-gluon vertex
while retaining the tree-level ghost-gluon vertex is (transverse in
$q_\nu$),
\begin{equation}
  \widetilde G_{\mu\nu} (q,p) = \delta_{\mu\nu} + \frac{1}{q^2}
  \frac{G(k^2) - G(q^2)}{G(p^2)}
  \Bigl( \delta_{\mu\nu} pq \, - \, p_\nu q_\mu \Bigr) \; . \label{glghkern}
\end{equation}
The requirement that the anomalous dimension of the ghost-gluon
scattering kernel is zero fixes its dependence on the renormalisation
functions of ghosts and gluons to a dependence on ratios $G(k^2)/G(p^2)$ and
$Z(k^2)/Z(p^2)$ with varying arguments. Interestingly, one further
simplifying Ansatz, namely that this dependence be at most linear in such
ratios, suffices to fully determine the kernel to be of the form
(\ref{glghkern}), see App.~B in~\cite{Sme98}.

This procedure is certainly not unambiguous, more complex structure to the
above scattering kernel cannot be excluded. It furthermore seems inconsistent
to neglect explicit one-particle irreducible four-gluon interactions while
resorting to some non-trivial but quite {\it ad hoc} assumption on the
ghost-gluon scattering kernel which also represents a sort of one-particle
irreducible four-point interactions.

Therefore a truncation based on using the tree-level form for the
ghost-gluon vertex function, $G_\mu (q,p) = iq_\mu $, while compatible
with the desired short distance behaviour of the solutions, at the same time
implies that some non-trivial assumptions on the otherwise unknown
ghost-gluon scattering kernel have to be made in order to solve the
Slavnov--Taylor identity for the 3-gluon vertex.

In light of this, a closer look at the ghost-gluon vertex function seems
necessary. A suitable source of information is provided by the
identity (\ref{eq:ghost-WTI}) involving the ghost-gluon vertex directly. 
Note that the derivation presented in Sec.\ \ref{sec_nPoint}  proceeds
not only via the usual BRS invariance but also requires neglecting irreducible
ghost-ghost scattering. This identity has the advantage that it allows to
express the ghost-gluon vertex in terms of the ghost renormalisation function
$G(k^2)$. The vertex constructed this way has the correct short
distance behaviour and therefore no anomalous dimension. It is compatible with 
the perturbative one-loop result, but some incompatibilities arise at
two-loop level \cite{Wat00}. Furthermore, the result suffices to
constrain $\widetilde G_{\mu\nu}(q,p) $ just enough to admit a simple
solution to the 3-gluon Slavnov--Taylor identity (\ref{glSTI}) without any
further assumptions \cite{Sme98}.

In Ref.\ \cite{Sme98} the ghost-gluon vertex $G_\mu(q,p)$ 
has been constructed explicitely
using the approximation described above. The corresponding form of
$G_\mu $ follows from $\widetilde
G_{\mu\nu}(q,p) $ as given by,
\begin{equation}
\widetilde G_{\mu\nu}(q,p) \, =\, \frac{G(k^2)}{G(q^2)} \, \delta_{\mu\nu} \,
+ \, \biggl( \frac{G(k^2)}{G(p^2) } \, - 1 \biggr) \, \frac{p_\mu q_\nu}{q^2}
\; , \;\; \mbox{and} \; \;  G_\mu(q,p) = i q_\nu  G_{\mu\nu}(q,p)  . \label{tG}
\end{equation}
Additional contributions to $\widetilde G_{\mu\nu}(q,p)$
which are purely transverse in $q_\nu$ can arise from the ghost-gluon
scattering kernel. Such terms cannot be constraint from the form of the
ghost-gluon vertex. In contrast to the truncation scheme based
on the tree-level ghost-gluon vertex as mentioned above, where precisely
such terms are necessary to solve the 3-gluon Slavnov--Taylor
identity,  unknown
contributions from the ghost-gluon scattering kernel can be neglected also
for the 3-gluon STI (\ref{glSTI}). Thus, using (\ref{tG}) in (\ref{glSTI})
one obtains,
\begin{equation}
 i k_\rho\Gamma_{\mu\nu\rho}(p,q,k) =
    G(k^2) \left( {\mathcal P}_{\mu\nu}(q) \frac{q^2 G(p^2)}{G(q^2) Z(q^2)}
    \, -\, {\mathcal P}_{\mu\nu}(p) \frac{p^2 G(q^2)}{G(p^2)Z(p^2) } \right) .
  \label{glWTI}
\end{equation}
Using the symmetry of the vertex the solution to (\ref{glWTI}) fixes the
vertex up to completely transverse parts. It can be derived straightforwardly
along the lines of the general procedure outlined in Refs.~\cite{Bal80,Kim80}. This
leads to the following solution to (\ref{glWTI}):
\[  \Gamma_{\mu\nu\rho}(p,q,k)
       \, =\, - A(p^2,q^2;k^2)\,  \delta_{\mu\nu}\,  i(p-q)_\rho\,
          - \,  B(p^2,q^2;k^2)\,  \delta_{\mu\nu} i(p+q)_\rho  \]
\begin{equation}
   \hskip 6mm - \, 2\frac{C(p^2,q^2;k^2)}{p^2-q^2}
( \delta_{\mu\nu} pq \, -\,  p_\nu
q_\mu) \, i(p-q)_\rho\,  + \; \hbox{cyclic permutations} \; ,
  \label{3gv}
\end{equation}
with  $ A = A_+ $, $ B = C = A_- $, and
\begin{equation}  \label{3gva}
 A_\pm (p^2,q^2;k^2)
    = G(k^2) \, \frac{1}{2} \, \left( \frac{G(q^2)}{G(p^2)Z(p^2)} \pm
      \frac{G(p^2)}{G(q^2)Z(q^2)} \right) \; .
\end{equation}
For comparison, the solution which results from using the non-trivial
ghost-gluon scattering kernel (\ref{glghkern}) together with 
the tree-level ghost-gluon vertex reads,  $A = A_+ $, $C = A_-
$, with
\begin{eqnarray}  \label{3gva_alt}
  A_\pm (p^2,q^2;k^2) =  G(k^2) \, \frac{1}{2} \, \left( \frac{1}{Z(p^2)} \pm
        \frac{1}{Z(q^2)} \right) \; , \;\; 
 B(p^2,q^2;k^2) =  \frac{1}{2} \, \left( \frac{G(q^2)}{Z(p^2)} -
        \frac{G(p^2)}{Z(q^2)} \right) \quad .
\end{eqnarray}
These two solutions differ by ratios of ghost renormalisation functions. It
has to be emphasised though that the tree-level ghost-gluon vertex
corresponding to the latter solution (\ref{3gva_alt}) only solves the
ghost-gluon Slavnov--Taylor identity (\ref{eq:ghost-WTI}) if at the same time a
tree-level ghost propagator is assumed. It is thus inconsistent
to insist on the tree-level ghost-gluon vertex (even by modifying
the ghost-gluon scattering kernel instead to solve the 3-gluon STI) as this
implies a trivial ghost propagator ($G(k^2)=1$). In fact, if ghost
contributions in Landau gauge are neglected completely, {\it e.g.}, in the
Mandelstam approximation, in which case this is consistent at the level of
Slavnov--Taylor identities, the solution for the 3-gluon vertex is obtained
from the above solutions by replacing all $G \to 1$. If ghost contributions
are to be taken into account, however, then also the ghost-gluon vertex
has to be dressed. Or else, Slavnov--Taylor identities are manifestly
violated.

As for undetermined transverse terms in the vertex functions of the 
ghost-gluon system, first note that the STI for the ghost-gluon
vertex leaves the following transverse contribution undetermined:
\begin{eqnarray}
   G_\mu(q,p) &=& iq_\nu  {\mathcal M}_{\mu\nu}(q,p) = \left\{i q_\mu (p^2 -pq) +
   ip_\mu (q^2 -pq)\right\}
  \, 2 f(k^2;p^2,q^2)   \nonumber \\
              &=& i(p+q)_\mu \, k^2  f(k^2;p^2,q^2) - i(p-q)_\mu \, (p^2-q^2)
   f(k^2;p^2,q^2)  \; ,
\end{eqnarray}
for some function $f$ (symmetric in its last two arguments). 
The different factors in this transverse contribution are chosen such that it
escapes the construction based on symmetry
arguments.\footnote{The possibility of such a term
was not noticed in \cite{Sme98},
where it was incorrectly asserted that, due to its symmetry, there would be
no such undetermined transverse terms in the ghost-gluon vertex.} In the
present truncation scheme such a contribution is dismissed as corresponding
to a non-trivial contribution to the neglected ghost-gluon scattering
kernel of the form
\begin{equation}
  {\mathcal M}_{\mu\nu}(q,p) = \left\{ \, \delta_{\mu\nu}\, (p^2 - q^2 ) \,-\,
   (p+ q)_\mu (p - q)_\nu \, \right\}  \, 2 f(k^2;p^2,q^2) \; .
\end{equation}
Furthermore, there are 4 purely transverse terms in the 3-gluon vertex
involving two more functions in the parameterisation of~\cite{Bal80} adopted
here,
\begin{eqnarray} \label{tr_3gl}
 \Gamma^{\hbox{\tiny tr}}_{\mu\nu\rho} (p,q,k) &=&  -i \, H(p^2,q^2,k^2) \,
 (p_\rho  q_\mu k_\nu - p_\nu q_\rho k_\mu ) \, +  \\
&& \hskip -1.2cm \Bigg( i \big(
 F(p^2,q^2;k^2)  ( \delta_{\mu\nu} pq  -  p_\nu q_\mu )
+  H(p^2,q^2,k^2)  \delta_{\mu\nu}  \big)
\; (p_\rho qk - q_\rho pk)
 \,   + \; \hbox{cycl. perm.}  \Bigg) \, . \nonumber
\end{eqnarray}
$H$ is fully symmetric in all its arguments, and $F$ is symmetric in the
first two. These 4 additional independent terms are the only undetermined
terms out of a total of 14 in the Lorentz structure of the 3-gluon vertex
(the others being fixed by the form~(\ref{3gv}) of the solution). In
contrast to fermion vertices in QED and QCD, in which the 8 transverse terms
out of 12 independent Lorentz tensor terms in an analogous
parameterisation~\cite{Bal80} are known to be important~\cite{Bro91}, the
Slavnov-Taylor identity for the 3-gluon vertex when combined with its full
Bose (exchange) symmetry puts much tighter constraints on the 3-gluon vertex
than the Ward--Takahashi/Slavnov--Taylor identities do on the
fermion-photon/gluon vertices. Furthermore, 
the infrared limit of the 3-gluon STI~(\ref{glWTI}), {\it i.e.}, the limit
$k \to 0$ for any of the three gluon momenta, puts additional constraints on
all Lorentz tensors. These constraints are saturated by the solution in the
form~(\ref{3gv}). As a result, the transverse terms (\ref{tr_3gl}) have to
vanish in this limit.

Another important difference between fermion vertices and the
3-gluon vertex function is that transverse terms in the vertices of
(electrons) quarks are of particular importance due to their coupling to
transverse (photons) gluons in the Landau gauge. In contrast, for the
3-gluon vertex function to be used in truncated gluon DSEs, it is well-known
that the relevant (unambiguous) contribution to the 3-gluon loop is
longitudinal in the external gluon momentum~\cite{Bro88a}. This implies that
even though the two 
gluons within the loop are transverse in Landau gauge, the third (external)
leg of the 3-gluon vertex {\sl must not} be connected to a transverse
projector.\footnote{It has to be contracted with $R_{\mu\nu}$ which projects
onto terms longitudinal in the external momentum, see Sec. \ref{sec_QED3}. 
This also removes the tadpole term as mentioned above.} 
The unconstraint terms of the 3-gluon vertex~(\ref{tr_3gl})
are, however, transverse with respect to {\sl all three} gluon momenta.

The DSEs (\ref{glDSEsim}) and (\ref{ghDSEmom}) with the vertex functions
given by (\ref{tG}) and (\ref{3gv}/\ref{3gva}) build a closed system of
equations for the renormalisation functions $G(k^2)$ and $Z(k^2) $ of ghosts
and gluons. Thereby explicit 4-gluon vertices (in the gluon DSE
(\ref{eq:Gluon-DSE})), irreducible 4-ghost correlations (in the identity for the
ghost-gluon vertex (\ref{eq:ghost-WTI})) and contributions from the ghost-gluon
scattering kernel (to the Slavnov--Taylor identity (\ref{glSTI}) as well as
to transverse parts of the ghost-gluon vertex) have been neglected. This is
the basic idea of the truncation scheme at the propagator level of the pure
gauge theory without quark. Its solution in an one-dimensional approximation
\cite{Sme98,Hau98b} will be presented below.

To extend this scheme and include the quark Dyson--Schwinger equation, the
quark-gluon vertex $\Gamma^a_\mu (p,q) $ has to be specified in
addition. Its the Slavnov--Taylor identity
reads~\cite{Eic74}, 
\begin{eqnarray}
  G^{-1} (k^2)  \,  ik_\mu \Gamma^a_\mu(p,q)    &=& \Bigl( g t^a -
  B^a(k,q)\Bigr) \, iS^{-1}(p) \label{qkSTI}\\  
  && \hskip 1.5cm - \, iS^{-1}(q) \, \Bigl( g t^a - B^a(k,q)\Bigr) \; , 
  \quad k = p -q \; . \nonumber 
\end{eqnarray}
Here,  $t^a$ is the $SU(3)$ generator in the fundamental representation, and
$B^a(k,q)$ is the ghost-quark scattering kernel which again represents 
connected 4-point correlations. In the present truncation scheme
in which such 4-point correlations are consistently neglected, the solution
to this Slavnov--Taylor identity is obtained from a particularly simple
extension to the construction of Ball and Chiu for the solution to the
analogous Ward--Takahashi identity of Abelian gauge theory \cite{Bal80}.
For $B^a(k,q) = 0$ the Slavnov--Taylor identity for the quark-gluon
vertex (\ref{qkSTI}) is solved by
\begin{displaymath}
  \Gamma^a_\mu(p,q) \, =\, - g t^a  \, G(k^2) \, \Bigg\{ \frac{1}{2} \left(
  A(p^2) + A(q^2)\right) i\gamma_\mu + \frac{p_\mu + q_\mu}{p^2-q^2}
  \biggl( \Bigl( A(p^2) 
\end{displaymath}
\begin{equation}
  \hskip 2cm - A(q^2)\Bigr)
  \frac{ip\hskip -5pt \slash  + iq\hskip -5pt \slash}{2} \,
- \, \left(B(p^2) - B(q^2)\right) \biggr) \Bigg\} \, + \;
  \hbox{transverse terms} \; . \label{qkSTIsol}
\end{equation}
This is furthermore justified for a study with emphasis on the infrared
behaviour of the propagators even beyond the present truncation scheme,
because it has been shown that $B^a(k,q) \to 0$ for $k \to 0$ in Landau
gauge \cite{Tay71}. Note that the only difference to the Abelian case
considered in~\cite{Bal80} at this point is the presence of the ghost
renormalisation function appearing in the quark-gluon vertex in Landau
gauge. This being a minor modification to the structure of the vertex, the
ghost renormalisation function in (\ref{qkSTIsol}) is, however, crucial for
the vertex to retain its correct anomalous dimension in the perturbative
limit. Furthermore, the presently available solutions demonstrate 
that the ghost
renormalisation function $G(k^2)$, being infrared enhanced as we will see
below, can give an essential contribution to the effective interaction of
quarks as part of the quark-gluon vertex also in the infrared.

At this point, additional unconstrained transverse terms in the
vertex can be significant. As discussed in the last chapter the
transverse part of the fermion-photon vertex in QED 
is crucial for multiplicative renormalisability~\cite{Bro91,Cur90}. 
Similar constructions to fix transverse pieces of the vertices in QCD is
still lacking. It is observed however, that an analogous transverse term in
the quark-gluon vertex is necessary in order to recover the correct
ultraviolet behaviour in the solution to the quark DSE in Landau gauge.  
First, in perturbation theory at 1-loop level the quark field
renormalisation constant $Z_2 = 1$ in Landau gauge. This means that there
should be no ultraviolet divergence at this level in the loop integral of the
quark DSE either. Obviously, this leading perturbative feature will be
manifest in the rainbow approximation, {\it i.e.}, the relativistic
Hartree--Fock 
approximation for quarks, which is obtained from replacing the quark-gluon
vertex by its tree-level form  $\Gamma_\mu(q,p) = \gamma_\mu $. At the level
of renormalisation group improvements at one-loop perturbation theory,
however, this is already inconsistent because of the non-vanishing anomalous
dimension of the vertex. One has $Z_{1F} = \widetilde Z_3^{-1} $ in Landau
gauge which implies that the renormalisation scale dependence of the
quark-gluon vertex is governed essentially by the one of the ghost
renormalisation function. This is explicitly verified by the solution to its
Slavnov--Taylor identity~(\ref{qkSTIsol}). It can, on the other hand, be
explicitly verified that the longitudinal pieces given in Eq.~(\ref{qkSTIsol})
change the divergence structure of the equation, in particular, leading to a
necessarily non-trivial $Z_2 \not= 1$. Secondly, a careful discussion of the 
leading
perturbative behaviour of the quark mass function reveals that with the
Ball--Chiu from for the vertex~(\ref{qkSTIsol}) alone the leading
perturbative behaviour of the running current quark mass cannot be
reproduced either. 

In the solution of the coupled gluon, ghost and quark DSE 
(or already in the quenched solution to the quark DSE) presented 
in Sec. \ref{sub_DCS} below the following transverse extension to the 
vertex~(\ref{qkSTIsol}),
which is analogous to the Curtis--Pennington vertex (\ref{CP})~\cite{Cur90}, 
will be seen to cure these problems, 
\begin{equation} \label{CP_ext}
\Gamma_\mu^{\mbox{\tiny tr}}(p,q) \, = \, G(k^2) \,\frac{1}{2}
 \Bigl( A(p^2) \, - \, A(q^2) \Bigr) \,
 \frac{(\gamma_\mu (p^2 - q^2) - (p+q)_\mu (p \hskip -5pt \slash - q \hskip
-5pt \slash ))  (p^2 + q^2)}{(M^2(p^2) + M^2(q^2))^2 + (p^2 - q^2 )^2}
\end{equation}
Here, $M(p^2) := B(p^2)/A(p^2)$. With this form there will be a) 
no quark field renormalisation necessary, {\it i.e.}, $ Z_2 = 1$ in Landau
gauge; b) the leading
contribution to the current mass in the ultraviolet will resemble
perturbation theory; and c) the correct scale dependence of the
renormalisation group improvement will be incorporated by the same explicit
overall factor $G(k^2)$, the ghost renormalisation function, as present in
the longitudinal pieces~(\ref{qkSTIsol}). To justify these observations on a
more sound basis for the quark vertex of Landau gauge QCD in analogy to the
available QED studies will be another important goal for future studies. 

With the form (\ref{qkSTIsol}) for the vertex function in the quark DSE,
improved by the additional transverse terms of~(\ref{CP_ext}), the present
truncation scheme is extended to a closed set of equations for all
propagators of Landau gauge QCD as parameterised by the four functions
$Z(k^2), G(k^2), A(k^2)$ and $B(k^2)$. Its simultaneous solution 
represents for the first time a systematic and complete solution to the DSEs 
of QCD at the level of propagators \cite{Ahl98}.

\begin{figure}
  \centering{\
        \epsfig{file=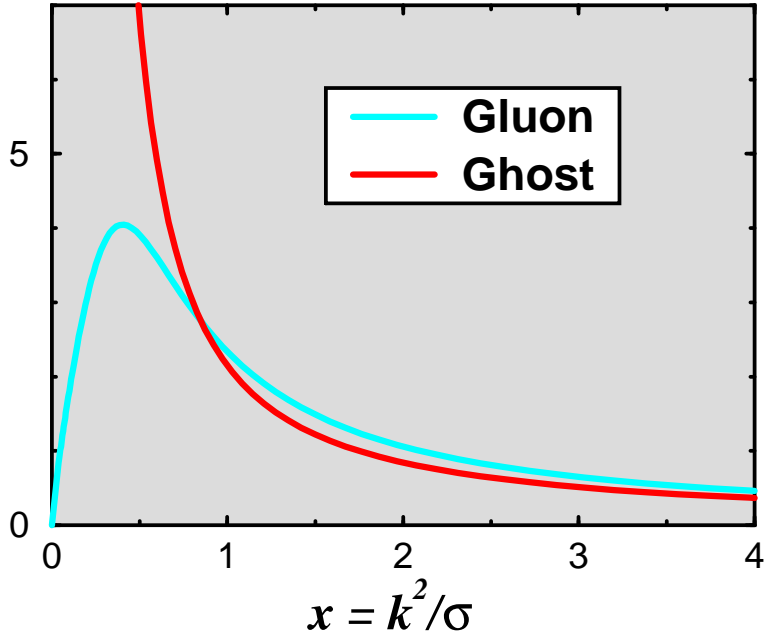,width=0.6\linewidth} }
  \caption{The renormalisation functions $Z(x)$ and $G(x)$.}
  \label{ZG}
\end{figure}

However, before turning to the very recent results including quarks it is
illustrative to discuss the solution of the coupled gluon-ghost system. The
non-linear integral equations for the renormalisation functions  $Z$ and $G$
of gluons and ghosts have been solved employing angle approximations
\cite{Sme98,Hau98b}, see Fig.\ \ref{ZG}. 
Such an approximation is as yet necessary in order to 
allow for an analytic infrared expansion of the solutions. This technique of
asymptotic infrared expansions which was originally developed for and applied
to the Mandelstam approximation~\cite{Atk81,Hau96} and subsequently extended to
the coupled system of gluon and ghost DSEs \cite{Sme98,Hau98b} proved to be a
necessary prerequisite to finding numerically stable solutions. An analogous
expansion will most likely have to be developed also for the four-dimensional
equations before the one-dimensional approximation can be abandoned. Some
preliminary progress towards solutions not relying on one-dimensional
approximations was obtained in a truncation scheme for gluons and ghosts with
bare vertices and considering in the gluon DSE the ghost loop only
in Ref.\ \cite{Atk98}. The modified angle approximation
approximation used in Refs.~\cite{Sme98,Hau98b} to arrive at the
one-dimensional system of equations is designed to preserve the leading order
of the integrands in the infrared limit of integration momenta. At the same
time, it accounts for the correct short distance behaviour of the solutions,
{\it i.e.}, the behaviour at large integration momenta. Nevertheless, due to
induced singularities some terms have to be dismissed, see Refs.\
\cite{Sme98,Atk97} for a discussion. The relative importance of such
a term  was assessed in~\cite{Sme98} where
the integral has been calculated without using an angle
approximation from the self-consistent solutions for $Z(k^2)$ and
$G(k^2)$ as obtained without this term. For all momenta the resulting
contribution was found to be negligible as compared to the other
terms retained in the solution to gluon DSE. Though even small terms can, in
principle, have a considerable effect on the non-linear self-consistency
problem, this supports the additional approximation to dismiss such terms.
Of course, progress on this issue is highly desirable.

The leading infrared behaviour of the gluon and ghost renormalisation function
can be extracted analytically. To this end one employs an Ansatz for these
functions of the form, 
\begin{equation}
Z(k^2) \sim (k^2)^\alpha \quad \hbox{and} \quad  G(k^2) \sim (k^2)^\beta \, .
\end{equation}
A complete analysis in Landau gauge based on this ansatz 
and keeping the vertex functions as general as possible can be found in Ref.\ 
\cite{Wat00}. The important point which makes such an analysis feasible
is the non-renormalisation of the ghost-gluon vertex, $\widetilde Z_1 =1$.
Then three possible solutions are found: $\beta =  0$, $\beta = -1$ and
$2\beta = -\alpha$ with $\beta > -1$. The first two solutions require, in 
addition, 
that the ultraviolet part of the integrals involved contributes to the infrared
power thereby spoiling the renormalisation group behaviour of, {\it e.g.}, the
running coupling, see the next section. The third solution $2\beta = -\alpha
> 2$ is, however, quite general.

As a matter of fact, such a solution has been obtained from the coupled gluon
system \cite{Sme97b} before the analysis in Ref.\ \cite{Wat00} had been
started. 
Defining $x := k^2/\sigma \to 0$ (for some scale $\sigma $)
and using $Z(x)G(x) \sim x^\kappa$ in the gluon DSE
immediately yields
\begin{equation}
  G(x) \sim x^{-\kappa} \quad \hbox{and} \quad  Z(x) \sim x^{2\kappa}
  \quad \hbox{as} \quad  x \to 0 \; .
  \label{irGZ}
\end{equation}
Furthermore, in order for a positive definite function $G(x)$ to result for
positive $x$ and assuming a positive definite $Z(x)$, as $x\to 0$, one
obtains the necessary condition $1/\kappa - 1/2 > 0$ which is equivalent to
\begin{equation}
  0 < \kappa < 2 \; . \label{0lkl2}
\end{equation}
The special case $\kappa = 0$ leads to a logarithmic singularity in the gluon
DSE for $x\to 0$. In particular, assuming that $ZG = c$
with some constant $c > 0$ and $x < x_0$ for a sufficiently small $x_0$,
one obtains $G^{-1}(x) \to c \,(3 N_c g^2/64\pi^2) \,\ln (x/x_0) +
\hbox{const}$ and thus $G(x) \to 0^-$ for $x \to 0$, showing that no positive
definite solution can be found in this case either.

From the gluon DSE one obtains
\begin{eqnarray}
  G(x) &\to & \left( g^2\gamma_0^G \left(\frac{1}{\kappa} - \frac{1}{2}
  \right) \right)^{-1}  c^{-1} x^{-\kappa} \; , \quad \gamma_0^G =
  \frac{1}{16\pi^2}  \; \frac{3 N_c}{4}  \; ,
  \label{loirG}\\
  Z(x) &\to &  \left( g^2\gamma_0^G \left(\frac{1}{\kappa} - \frac{1}{2}
  \right) \right) \,  c^{2} x^{2\kappa}   \; , \quad 0<\kappa < 2  \; ,
  \label{loirZ1}
\end{eqnarray}
where $\gamma_0^G$ is the leading order perturbative coefficient of the
anomalous dimension of the ghost field. Accordingly, the ghost-loop
gives infrared singular contributions $\sim x^{-2\kappa}$ to the gluon
equation while the 3-gluon loop yields terms proportional to
$ x^\kappa $ as $x\to 0$, which are thus subleading contributions to
the gluon equation in the infrared. With Eq.~(\ref{loirG}) the leading
asymptotic behaviour of the ghost equation for $x \to 0$ provides
\begin{equation}
Z(x) \,\to \,  g^2\gamma_0^G \, \frac{9}{4} \left(\frac{1}{\kappa} -
\frac{1}{2}   \right)^2 \left( \frac{3}{2}\, \frac{1}{2-\kappa} - \frac{1}{3}
+ \frac{1}{4\kappa} \right)^{-1} \, c^2 x^{2\kappa} 
\label{loirZ2}
\end{equation}
and mutual consistency between the gluon and ghost equation requires
\begin{equation}
  \left( \frac{3}{2}\, \frac{1}{2-\kappa} - \frac{1}{3} +
  \frac{1}{4\kappa} \right) \, \stackrel{!}{=} \, \frac{9}{4}
  \left(\frac{1}{\kappa} - \frac{1}{2} \right)  \; \Rightarrow \;
  \kappa \, = \, \frac{61 \stackrel{(+)}{-} \sqrt{1897}}{19} \,
         \simeq \, 0.92  \; .
  \label{kappa}
\end{equation}
Note that one of the roots is excluded by the range allowed to $\kappa
$.\footnote{We note here that the conversion of a tree-level pole
into an algebraic branchpoint with exponent larger than one has also been known
for the fermion propagator in QED, see, {\it e.g.}, supplement S4 in Ref.\ 
\cite{Jau76} and references therein. As can be seen from Eq.\ (S4-41) of Ref.\ 
\cite{Jau76} and the following paragraph the singularity
$(p^2+m^2)^{-1-\alpha /\pi}$ is related to the soft photon cloud.}

This leading behaviour of the gluon and ghost renormalisation functions and
thus their propagators is entirely due to ghost contributions. The details of
the treatment of the 3-gluon loop have no influence on above
considerations. This is in remarkable contrast to the Mandelstam
approximation, in which the 3-gluon loop alone determines the infrared
behaviour of the gluon propagator and the running coupling in Landau gauge
\cite{Man79,Atk81,Bro89,Hau96,Hau98a}. As a result of this, the running
coupling as obtained from the Mandelstam approximation is singular in the
infrared \cite{Hau96,Hau98a}. As will be shown in the next
section, the infrared behaviour derived from the present truncation scheme
implies an infrared stable fixed point. This is certainly a counter example
to the frequently quoted assertion that the presence of ghosts in Landau
gauge may have negligible influence on physical observables at hadronic
energy scales.

The qualitative infrared behaviour of ghosts and gluons reported in
Refs.~\cite{Sme97b,Sme98} and the conclusions on the dominance of the ghost
contributions in Landau gauge as described above were also confirmed in
Ref.~\cite{Atk97} where the same qualitative behaviour was obtained
neglecting the 3-gluon loop completely. A tree-level ghost-gluon vertex
together with a naive angle approximation in this case led to a numerical
value of $\kappa \simeq 0.77 $ for the same exponent determining the infrared
behaviour of gluons and ghosts~\cite{Atk97}. In the same approximation
but with no angle approximation employed a solution for $\kappa = 1^- $ is
reported in Ref.~\cite{Atk97}. ($\kappa = 1^- $ has to interpreted as
$\kappa = 1 - \epsilon$, $\epsilon \to 0^+$. Thus, this result already provides
a regularisation for the infrared singular ghost renormalisation function.)
Such forms (for $\kappa = 1$) have been
discussed in the literature before, independently in the two completely
different approaches of Refs.~\cite{Zwa90,Zwa92} and Refs.~\cite{Hae90,Sti95}
respectively. However, the evidence for the $\kappa = 1$ solutions has since
disappeared in both these approaches again~\cite{Zwa94,Dri98a}. For the
DSE study of Ref.~\cite{Atk98}, it cannot be excluded
that the $\kappa = 1$ behaviour obtained there might be an artifact of the {\sl
ghost-loop only approximation} and/or the use of a tree-level vertex.

For the infrared behaviour of gluon and ghost propagators as described
above, a comment on its implications for the vertex functions is in order.
Starting with the ghost-gluon vertex one realizes that the
limit of vanishing ghost momenta is regular:
\begin{eqnarray}
  G_\mu(q,p) \to & -\, ik_\mu & \qquad\mbox{for}\quad p \to 0  \\
  G_\mu(q,p) \to & 0     & \qquad\mbox{for}\quad q \to 0
  \quad ,  \label{glto0}
\end{eqnarray}
where it was used that $G(k^2) \sim (k^2)^{-\kappa}$ for $ k^2 \to 0$. On the
other hand, for vanishing gluon momentum $k \to 0$ the vertex diverges as
\begin{equation}
  G_\mu(q,p) \to 2 ip_\mu \, \frac{G(k^2)}{G(p^2)}\,
             \sim \, \frac{2 ip_\mu }{G(p^2)}\,  \frac{1}{ (k^2)^{\kappa}}
    \qquad\mbox{for}\quad k \to 0
  \quad .
\end{equation}
Since the exponent $\kappa$ is a (positive) irrational number, see
Eq.~(\ref{kappa}), the corresponding divergence cannot be easily interpreted. 
This divergence is in fact
weaker than a massless particle pole ($\kappa < 1$) and presumably lacking a
physical interpretation.

Similarly, the 3-gluon vertex as given in Eqs.~(\ref{3gv}),
(\ref{3gva}) shows analogous infrared divergences resulting from
the infrared enhanced ghost renormalisation function,
\begin{eqnarray}
\Gamma_{\mu\nu\rho}(p,q,k) &\to&   G(k^2) \, \Bigg\{ \Bigl( ip_\mu
\delta_{\nu\rho} + ip_\nu \delta_{\mu\rho} - 2 ip_\rho \delta_{\mu\nu} \Bigr)
\frac{1}{Z(p^2)} \label{diff3gl} \\
&&  +\, 2 ip_\rho \,  p^2  \, {\mathcal P}_{\mu\nu}(p)\, \biggl(
\frac{2G'(p^2)}{G(p^2)Z(p^2)} + \frac{Z'(p^2)}{Z^2(p^2)} \biggr) \Bigg\}
    \qquad\mbox{for}\quad k \to 0 \; .
 \nonumber
\end{eqnarray}
Note that such a limit usually implies some mild regularity
restrictions on the functions $A,B$ and $C$ of Eq.~(\ref{3gv}). In
particular, the form of~(\ref{diff3gl}) above is obtained from
Eqs.~(\ref{3gv}), (\ref{3gva}) provided that in addition,
\begin{equation}
  q^2 \, C(p^2,q^2;k^2) \, \to \, 0 \; , \quad\mbox{for} \quad q^2
  \, \to \,   0 \; .   \label{orgn_massless}
\end{equation}
This result, Eq.~(\ref{diff3gl}), then agrees with the differential
Slavnov--Taylor identity as obtained directly from Eq.~(\ref{glWTI}) in the
limit $k\to 0$. In the previous studies of the gluon DSE in the Mandelstam
approximation \cite{Man79,Atk81,Bro89,Hau96}, the analogous requirement for the
3-gluon vertex function to obey the differential Slavnov--Taylor identity
led to the so-called masslessness condition, Eq.~(\ref{massless.MA}),
$p^2/Z(p^2) \to 0$ for $p^2 \to 0$. In absence of ghost contributions, the
gluon DSE had to be supplemented by this as an additional constraint. This
original condition is violated by the infrared behaviour of the gluon propagator
found in~\cite{Sme97b}, {\it i.e.}, $Z(x) \to x^{2\kappa}$. The correct
replacement of this condition for the present case results from
Eq.\ (\ref{orgn_massless}). With $C = A_-$ from Eq.~(\ref{3gva}) one thus
obtains,
\begin{equation}
p^2 G(p^2) \to 0 \qquad \hbox{and} \quad \frac{p^2}{G(p^2)Z(p^2)} \to 0
\qquad \hbox{for} \quad p^2 \to 0 \; .
\label{massless}
\end{equation}
These two necessary conditions are obeyed by the infrared behaviour presented
above without further adjustments, {\it i.e.}, $G(x) \sim
1/(G(x)Z(x)) \sim x^{-\kappa}$ for $ x \to 0$ (with $\kappa \simeq 0.92$).
One realizes, however, that the original bounds on $\kappa $ obtained
from the consistency of the ghost DSE, when combined with conditions
(\ref{massless}), have to be restricted further to $ 0 < \kappa < 1$.
The possibility that the masslessness condition in its original form
might not hold for non-Abelian gauge theories has been pointed out
in the literature before~\cite{Kug95,Hat83}. The present results support this
conjecture.

Finally, since quarks cannot screen the divergence in the ghost
renormalisation function, see below, its presence in the quark-gluon vertex
function, see Eq.~(\ref{qkSTIsol}), has a similar effect there, too. It
will be demonstrated on the example of the running coupling in the next
section, how these apparently unphysical divergences in the elementary
correlation functions of gluons, ghosts and quarks can, in principle,
nevertheless cancel in physical quantities. The lack of a physical
interpretation of these divergences (other than maybe reflecting confinement)
should, however, not be too surprising for a Euclidean field theory violating
reflection positivity. Note also that in all skeletons of the kernels in
relativistic bound state equations for hadrons, the combination of dressed
quark-gluon vertices with the non-perturbative gluon propagator will give
contributions of the form 
\begin{equation}
\sim   g^2 \,   G^2(k^2)\, D_{\mu\nu} (k)  \;  \Gamma_\nu^{\hbox{\tiny
CP}}(p,p-k)  \otimes \Gamma_\mu^{\hbox{\tiny CP}}(q-k,q)  \; ,
\label{dressed-ladder-skeletton}
\end{equation}
where $\Gamma_\mu^{\hbox{\tiny CP}}$ is the vertex function
given in Eq. (\ref{CP}) for QED, here corresponding the sum of the terms in
Eqs. (\ref{qkSTIsol}) and (\ref{CP_ext}) without ghost
renormalisation function, coupling and colour factors.  
The resulting combination of the ghost
and gluon renormalisation functions, $g^2 G^2(k^2) Z(k)$, to this skeleton of
the quark interaction kernel is free of unphysical infrared divergences for
$k \to 0$. It is furthermore independent of the renormalisation scale in
Landau gauge and suited to define a non-perturbative running coupling as
will be discussed in the next section.

Generally, on external gluon lines the divergent contributions to scattering
amplitudes induced by the ghost renormalisation function in the vertex
functions will be over-compensated by the attached gluon propagators.
These divergences thus do not give rise to massless asymptotic gluon states.
In internal gluon lines from one-particle reducible contributions, the
divergence of the two vertex functions is exactly compensated by the gluon
propagator leaving one single ``massless'' pole, since with some dimensionless
constant $a>0$, $g^2 G^2(k^2) D(k) \to a/k^2$ for $k^2 \to 0$. If this
behaviour corresponding to a finite infrared fixed point should persist
 such a massless pole contribution to
quark and gluon scattering amplitudes in the Mandelstam variables could
nevertheless be compensated by an analogous pole from 1-PI contributions in
a way analogous to the mechanism described in Sec.~\ref{Sec2.5} in the
context of the non-perturbative approximation scheme of Ref.~\cite{Sti95}.
For this scheme it was demonstrated quite generally in
Refs.~\cite{Dri98a,Dri98b} how these unphysical but necessarily arising
{\sl shadow-poles} in Mandelstam variables that appear in 1-PI
correlations as well as in one-particle reducible contributions mutually
cancel in scattering amplitudes (introducing the notion of {\sl extended
irreducibility}).

The present situation is different, however, as far as the ghost fields are
concerned. The infrared enhanced ghost propagator will result in an
infrared divergence of external ghost legs which is too strong for a particle
interpretation. Note that a particle interpretation for ghost degrees of
freedom is, of course, anyhow precluded as they lead to negative
norm states. Furthermore, the infrared enhanced ghost propagator results in
non-decoupling of ghost clusters at large separations. For example, already
in the disconnected contributions to 4-point ghost correlations, for no pair
of (space-time) arguments with comparatively small separation will the
2-point correlations connecting this pair decouple from those connected to
the other largely (spacelike) separated pair.
At that point it is interesting to note that lattice calculations provide
evidence that in Coulomb gauge not only the ghost propagator but also the 
time-time component of the gluon propagator is infrared enhanced \cite{Cuc00}.
(The space-space components of the gluon propagator are infrared suppressed.)
Therefore, it is quite plausible that in Coulomb gauge this non-decoupling 
of clusters also applies to the Coulomb gluons (being also negative norm states).

Another example for the implications of the infrared divergences induced by
the ghosts which requires a more careful analysis is provided by ghost-box
contributions to 4-point gluon correlations. In particular, in colour singlet
channels it is expected that no long range correlations in contradiction with
cluster separation occur in these correlations. This will have to be studied
in more detail. There is no obvious conflict at least, as the Mandelstam
variables in such a case do not coincide with the momenta of the infrared
enhanced ghost 2-point correlations. Further damping of these enhanced
2-point correlations in such contributions is provided by the ghost-gluon
vertices which yield vanishing transverse contributions for vanishing ghost
leg momenta, as is seen from~(\ref{glto0}) above. However, this example
demonstrates that, with loosing the cluster decomposition in general, the
important question arises how to recover it for colour singlet clusters.

\subsubsection{Subtraction Scheme and Running Coupling}
\label{sub_Sub} 

Before the non-perturbative running coupling can be discussed, 
some introductory
remarks  on the choice of the non-perturbative subtraction scheme and its
relation to the definition of the running coupling are necessary. In
particular, it will become clear that the preliminary infrared discussion of
last section already yields an important first result: it implies the
existence of an infrared fixed point.

The starting point is the following identity for the renormalisation
constants
\begin{equation}
  \widetilde{Z}_1 \, = \, Z_g Z_3^{1/2} \widetilde{Z}_3 \, = \, 1 \; ,
  \label{wtZ1}
\end{equation}
which holds in the Landau gauge \cite{Tay71}.
It follows that the product
$g^2 Z(k^2) G^2(k^2)$ is renormalisation group invariant. In absence of any
dimensionful parameter this (dimensionless) product is therefore a function
of the running coupling $\bar g$,
\begin{equation}
  g^2 Z(k^2) G^2(k^2) = f( \bar{g}^2(t_k, g)) \; , \quad
  t_k = \frac{1}{2} \ln k^2/\mu^2 \; .
  \label{gbar}
\end{equation}
Here, the running coupling $\bar g(t,g)$ is the solution of
$d/dt \, \bar g(t,g) = \beta(\bar g) $ with $\bar g(0,g) = g$ and the
Callan--Symanzik $\beta$-function $\beta (g) = - \beta_0 g^3 + {\mathcal
O}(g^5)$. The perturbative momentum subtraction scheme is asymptotically
defined by $f(x) \to x$ for $ x\to 0$. This is realized by independently
setting
\begin{equation}
  Z(\mu^2) = 1 \quad \mbox{and} \quad G(\mu^2) = 1
  \label{persub}
\end{equation}
for some asymptotically large subtraction point $k^2 = \mu^2$.
If the renormalisation group invariant product 
$g^2 Z(k^2) G^2(k^2)$ is to have a physical
meaning, {\it e.g.}, in terms of a potential between static colour sources, it
should be independent under changes $(g,\mu) \to (g',\mu')$ according to
the renormalisation group for arbitrary scales $\mu'$. Therefore,
\begin{equation}
  g^2 Z({\mu'}^2) G^2({\mu'}^2) \, \stackrel{!}{=} \,  {g'}^2 = \bar g^2(\ln
(\mu'/\mu) , g) \; , \label{gbar'}
\end{equation}
and, $f(x) \equiv x $, $ \forall x$. This can thus be adopted as a physically
sensible definition of a non-perturbative running coupling in the Landau
gauge.

Note that this definition is an extension to the one used in the Mandelstam
approximation, Eq.~(\ref{rc.MA}). In Ref.~\cite{Hau96} the identity $Z_g Z_3
= 1$ has been obtained for this approximation (without ghosts) implying that $g
Z(k^2)$ would be the renormalisation group invariant product in this case. Its
according identification with the running coupling $ g Z(k^2)
= \bar g(t_k,g)$ is equivalent to the non-perturbative renormalisation
condition $Z(\mu^2) = 1 (\forall \mu$) which turns out to be possible in the
Mandelstam approximation (in which there is no ghost renormalisation
function).

In the present case, it is not possible to realize $f(x) \equiv x$ by simply
extending the perturbative subtraction scheme (\ref{persub}) to arbitrary
values of the scale $\mu$, as this would imply a relation between the functions
$Z$ and $G$ which is inconsistent with the leading infrared behaviour
of the solutions discussed in the last section. For two independent
functions the condition (\ref{persub}) is in general too restrictive to be
used for arbitrary subtraction points. Rather, in extending the perturbative
subtraction scheme, one is allowed to introduce functions of the coupling
such that
\begin{equation}
  Z(\mu^2) \, = \, f_A(g) \quad \hbox{and} \quad G(\mu^2) \, = \, f_G(g)
  \quad \hbox{with} \quad f_G^2 f_A \, = \, 1 \; ,
  \label{npsub}
\end{equation}
and the limits $ f_{A,\, G} \to 1 \, , \; g \to 0 $. Using this it is
straightforward to see that for $k^2 \not= \mu^2$ one has ($t_k = (\ln
k^2/\mu^2)/2$),
\begin{eqnarray}
  Z(k^2) &=& \exp\bigg\{ -2 \int_g^{\bar g(t_k, g)} dl \,
  \frac{\gamma_A(l)}{\beta(l)} \bigg\} \, f_A(\bar g(t_k, g)) \; ,
  \label{RGsol} \\
   G(k^2) &=& \exp\bigg\{ -2 \int_g^{\bar g(t_k, g)} dl \,
  \frac{\gamma_G(l)}{\beta(l)} \bigg\} \, f_G(\bar g(t_k, g)) \; .
  \nonumber
\end{eqnarray}
Here $\gamma_A(g)$ and $\gamma_G(g)$ are the anomalous dimensions
of gluons and ghosts respectively, and $\beta(g)$ is the Callan--Symanzik
$\beta$-function. Eq. (\ref{wtZ1}) corresponds to the following identity
for these scaling functions in Landau gauge:
\begin{equation}
  2 \gamma_G(g) \, +\, \gamma_A(g)  \, = \, -\frac{1}{g} \, \beta(g)  \; .
  \label{andim}
\end{equation}
One thus verifies that the product $g^2 Z G^2$ indeed gives the running
coupling ({\it i.e.}, Eq. (\ref{gbar}) with $f(x) \equiv x$). Perturbatively,
at one-loop level Eq. (\ref{andim}) is realized separately, {\it i.e.},
$\gamma_G(g) = - \delta \, \beta(g) /g$ and $ \gamma_A(g) = - (1-2\delta)\,
\beta(g)/g$ with $\delta = 9/44$ for $N_f=0$ and arbitrary
$N_c$. Non-perturbatively one can still separate these contributions from the
anomalous dimensions by introducing an unknown function $\epsilon(g)$,
\begin{equation}
  \gamma_G(g) \, =:\, - (\delta + \epsilon (g) ) \,
  \frac{\beta(g)}{g}
  \; \Rightarrow \quad  \gamma_A(g) \, =\, - (1 - 2\delta - 2\epsilon (g)
  ) \, \frac{\beta(g)}{g}  \;  . \label{RGgam}
\end{equation}
This allows to rewrite Eqs.~(\ref{RGsol}) as follows:
\begin{eqnarray}
  Z(k^2) &=& \biggl( \frac{\bar g^2(t_k, g)}{g^2} \biggr)^{1-2\delta}
  \exp\bigg\{ - 4\int_g^{\bar g(t_k, g)} dl \, \frac{\epsilon(l)}{l} \bigg\}
  \, f_A(\bar g(t_k, g)) \; , \label{RGsol1} \\
  G(k^2) &=& \biggl( \frac{\bar g^2(t_k, g)}{g^2}  \biggr)^{\delta}  \,
  \exp\bigg\{ 2 \int_g^{\bar g(t_k, g)} dl \, \frac{\epsilon(l)}{l} \bigg\}
  \, f_G(\bar g(t_k, g)) \; .
  \nonumber
\end{eqnarray}
This is generally possible, {\it i.e.}, also in the presence of quarks. In 
this case one has
$\delta = \gamma_0^G/\beta_0 = 9N_c/(44 N_c - 8 N_f)$ for $N_f$ flavours in
Landau gauge. The above representation of the renormalisation functions
expresses clearly that regardless of possible contributions from the unknown
function $\epsilon(g)$, the resulting exponentials cancel in the product $G^2
Z$. For a parameterisation of the renormalisation functions, these
exponentials can of course be absorbed by a redefinition of the functions
$f_{A,\, G}$. The only effect of such a redefinition is that the originally
scale independent functions $f_{A,\, G} (\bar g(t_k, g))$ will acquire a
scale dependence by this, if $\epsilon \not= 0$.

\begin{figure}[t]
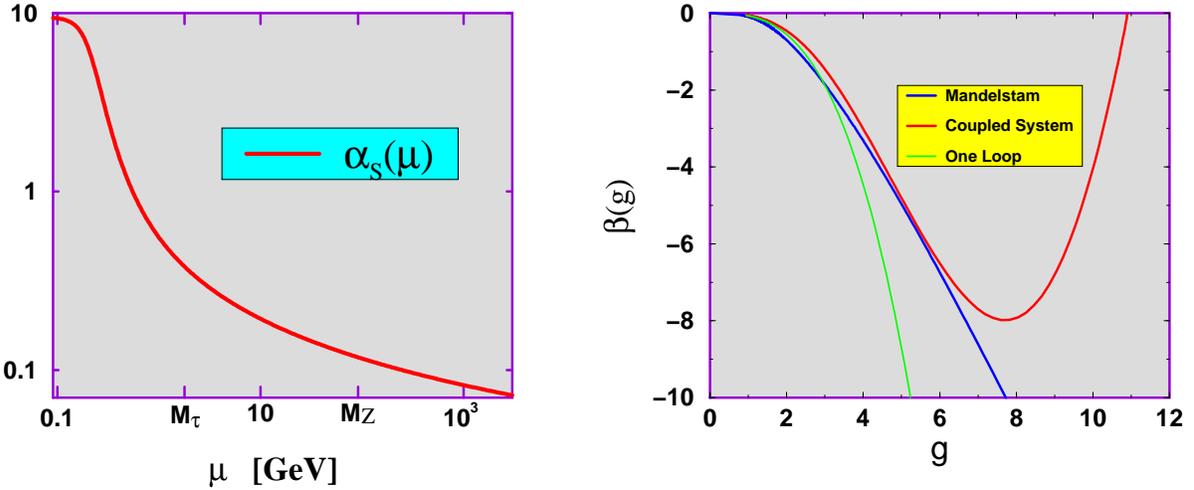

\hskip -.2cm
\parbox{6cm}{
  \epsfig{file=alpha.epsc,width=8.2cm}}
\hskip 2.6cm
\parbox{6cm}{
  \epsfig{file=beta4.epsc,width=8.2cm}}
  \caption[The running coupling $\alpha_S(\mu)$ for $t = 0$  and the
           corresponding $\beta$-function.]
          {The running coupling $\alpha_S(\mu)$ for $t = 0$ (left), and the
           corresponding $\beta$-function (right) in comparison with its
           leading perturbative form (one-loop) and the $\beta$-function
           as a result of the Mandelstam approximation from
           Ref.~\protect\cite{Hau96}.}
  \label{alpha}
\end{figure}

For the present truncation scheme it is possible, however, to obtain
explicitly scale independent equations thus showing that the solutions for the
renormalisation functions $G$ and $Z$ obey one-loop scaling at all scales
\cite{Sme98}. In particular, this implies that the products $g^{2\delta} G$
and $g^{2(1-2\delta)} 
Z$ are separately renormalisation group invariants that scheme (as
they are at one-loop level). As for the renormalisation scale dependence, the
non-perturbative nature of the result is therefore buried entirely in the
result for the running coupling.

The implications of the preliminary results for the infrared behaviour of the
solutions $G$ and $Z$ as given in the last section can now be discussed
without actually solving the gluon and ghost DSEs. From
Eqs.~(\ref{loirG}) and (\ref{loirZ1}) for $k^2 \to 0$ one finds,
\begin{equation}
   g^2 Z(k^2) G^2(k^2) \, = \, \bar g^2(t_k,g) \, \stackrel{t_k \to
  -\infty}{\longrightarrow} \, \left(
  \gamma_0^G \left( \frac{1}{\kappa} - \frac{1}{2} \right) \right)^{-1}
  =:\, g_c^2  \; .
  \label{gcrit}
\end{equation}
The critical coupling scales with the number of colours as $g_c^2 \sim 1/N_c$
implying that $g_c^2 N_c$ is constant in the present approach. This agrees
with the general considerations of the large $N_c$ limit in which $g^2 N_c$
is kept fixed as $N_c$ becomes large \cite{tHo74}. For $N_c=3$ and with
Eq.~(\ref{kappa}) for $\kappa$ one obtains $g_c^2 \simeq 119.1$ which
corresponds to a critical coupling $\alpha_c = g^2_c/(4\pi) \simeq
9.48$. This is a remarkable result in its own, if compared to the running
coupling as it was analogously obtained from the Mandelstam approximation
\cite{Hau96}. The dynamical inclusion of ghosts changes the infrared singular
coupling of the Mandelstam approximation to an infrared finite one implying
the existence of an infrared stable fixed point.

\begin{figure}[t]
\vskip 1.2cm
  \centerline{ \epsfig{file=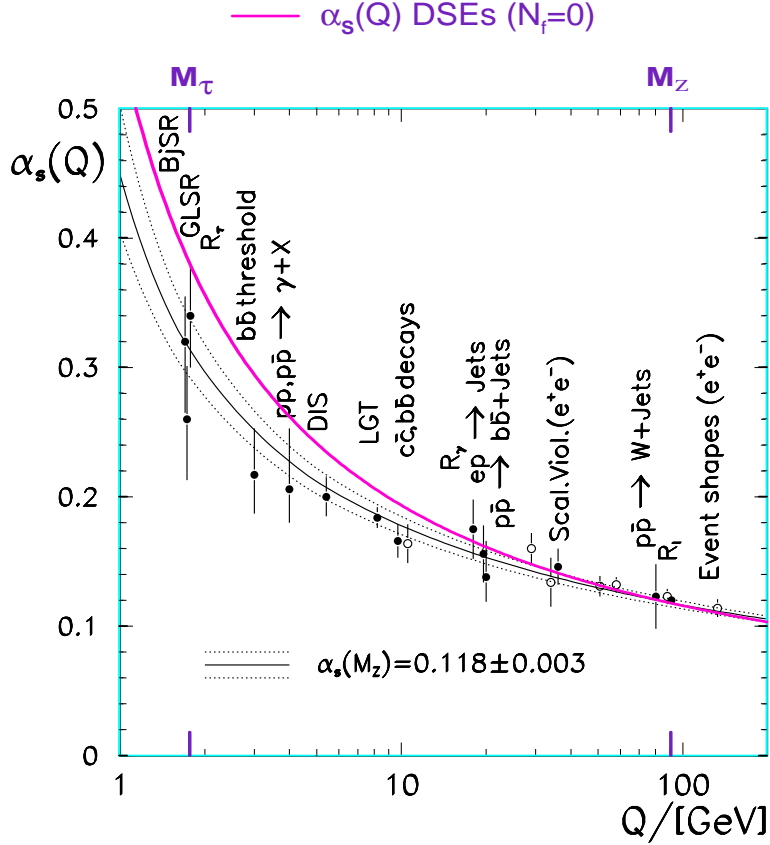,width=.6\textwidth} }
  \caption[The running coupling $\alpha_S$ from the Dyson--Schwinger
  equations of pure QCD compared to the collection of the {\it world's}
  experimental data.]
          {The running coupling $\alpha_S$ from the Dyson--Schwinger
  equations of pure QCD compared to the collection of the {\it world's}
  experimental data assembled by M.~Schmelling in Ref.~\protect\cite{Sch97}.
  The open dots result from global event shape variables.}
  \label{Fig:alpha_s}
\end{figure}

The running coupling as calculated from the coupled gluon and ghost DSEs,
see Fig.\ \ref{alpha},
involves an arbitrary renormalisation group invariant
scale $\sigma $ related to the infrared behaviour of
the propagators. Note that this is a non-perturbative realization of 
dimensional transmutation. The
phenomenologically well-known value of the running coupling $\alpha_S(M_Z) =
0.118$ at the mass of the $Z$-boson, $M_Z = 91.2$ GeV \cite{PDG98} can be
used to fix the overall momentum scale $\sigma$ directly which yields 
$\sigma \simeq (350 \hbox{MeV})^2$. For completeness, from the rough estimate
of the leading logarithmic behaviour of the renormalisation functions $Z$ and
$G$ at high momenta, this corresponds to a perturbative scale
$\Lambda_{\hbox{\tiny QCD}}$ in $250 \sim 300$ MeV.
This is not a very significant result, however, since the value for
$\Lambda_{\hbox{\tiny QCD}} $ depends on the number of quarks $N_f$, and the
order of the perturbative expansion, the scheme etc.. In particular, in the
present framework one expects it to change when quarks are included. 

Once the momentum scale is fixed, it is of course possible to extract the
corresponding value of the running coupling at other values of the momentum
scale. Phenomenologically, the running coupling is remarkably well known from
a variety of experiments in the range from around the $Z$-masses down to the
mass $M_\tau$ of the $\tau$-Lepton (at about 1.78 GeV).  A recent collection 
of the available measurements is shown in Fig.~\ref{Fig:alpha_s}. 
These range from deep-inelastic sum rules
(BjSR, GLSR), scaling violations in deep-inelastic scattering (DIS),
hadronic $\tau$ decays ($ R_\tau$), and $\Upsilon $ ($b\bar b$) spectroscopy
and decays in the energy range $Q\sim 1.6-10$ GeV, to measurements of total
cross sections, jet rates and global event shapes in $e^+e^-$,
$p\bar p$ and $ep$ collisions determining $\alpha_S$ at higher energies
typically in $Q \sim 30-130$GeV (see Ref.~\cite{Sch97} for details). Several
of the points in this figure represent hundreds of independent measurements,
and the agreement with the next-to-next-to-leading order evolution is a
truly remarkable success for perturbative QCD.

The running coupling from the Dyson-Schwinger equations in this same energy
range agrees far less with the experimental data, remarkably only that it
does much better than it actually should. From the ratio of the $Z$-- to the
$\tau$-mass, $M_Z/M_\tau \simeq 51.5$ (and $\alpha_S(M_Z) = 0.118$), one
obtains the value $\alpha_S(M_\tau) = 0.38$ which within errors might still
be acceptable experimentally~\cite{PDG98}. 
For $N_f=0$ and with using $\Lambda_{\hbox{\tiny
QCD}}$ in $250 \sim 300$ MeV, as a result of the estimate from the anomalous
dimensions given in the last section, such good agreement with the data is
actually not expected for a perturbative QCD coupling at one, two or three
loops. Even though the estimate of $\Lambda_{\hbox{\tiny QCD}}$ used here is
certainly not reliable, to parameterise the running coupling of the present
scheme by the $N_f =0$ perturbative (two-- or three-loop) form would require
$\Lambda_{\hbox{\tiny QCD}}$ to rather be in $0.8-1.2$GeV. This mismatch of
estimates might demonstrate the limitations of the present calculation better
than the apparent agreement with the experimental data.

Maybe more significantly, the running coupling over the full momentum range,
including the occurrence of the fixed point for $\mu \to 0$, is shown in
Fig.~\ref{alpha} with the momentum range of Fig.~\ref{Fig:alpha_s} marked by
$M_\tau$ and $M_Z$. The right panel of Fig.~\ref{alpha} shows the
corresponding $\beta$-function with its two fixed points in the ultraviolet
at $g = 0$ and in the infrared at $g = g_c \simeq 10.9$. For comparison
the leading perturbative form for $g \to 0$ (and $N_f = 0$), $\beta(g) \to -
\beta_0 \, g^3  = - 11/(16\pi^2) \, g^3$, is added to the figure as well as
the infrared singular non-perturbative result obtained from the analogous
subtraction scheme in the Mandelstam approximation according to
Sec.~\ref{sub_Mand} (see also Refs.~\cite{Hau96,Hau98a}). The solution to
the coupled system of gluon and ghost Dyson--Schwinger equations yields
better agreement with the leading perturbative form at small $g$ than the
Mandelstam approximation, since for the latter, due to neglecting ghosts, the
leading coefficient  $\beta_0 = 14/(16 \pi^2)$ differs from its corresponding
perturbative value. While this might still be regarded as small quantitative
discrepancy, the significant difference between the present result and the
Mandelstam approximation occurs for $g > 6$, once more demonstrating the
importance of ghosts in Landau gauge, in particular, in the infrared.

\subsubsection{Confined Gluons and Positivity}
 \label{sub_Pos} 

It has been mentioned in Sec.~\ref{Sec2.3} already that contradictions to 
positive (semi)definiteness of the norm of gluonic states (in the ghost free
subspace) arise from the superconvergence relations in
the asymptotically free, BRS invariant theory for not too many flavours ($N_f
< 10$). While this might still be a reincarnation of Haag's theorem occurring
already at the level of perturbative interactions, full non-perturbative
results as obtained from DSEs or lattice simulations
allow to assess their implications on positivity more directly.   

As one is dealing with Euclidean Green's functions, or Schwinger functions, for
the elementary correlations (of gluons, ghost and quarks) in QCD, the most 
fundamental way to investigate a possible particle interpretation for these
degrees of freedom is to resort to the Osterwalder--\-Schrader
reconstruction theorem which states that a G\aa rding--Wightman relativistic
quantum field theory can be constructed from a set of Schwinger functions if
those Euclidean correlation functions obey certain conditions, the
Osterwalder--\-Schrader axioms \cite{Haa96}. In particular, the axiom of
{\sl reflection positivity} for Euclidean Green's functions is thereby the
Euclidean counterpart to the positive definiteness of the norm in the Hilbert
space of the corresponding G\aa rding--Wightman quantum field theory.
This general Euclidean positivity condition involves arbitrary partial sums
of $n$-point correlation functions. To prove positivity of a theory based on
these conditions is a mathematically very hard task which has generally been 
possible only in very few special cases \cite{Fer91}. Here, the converse is
used. One counter example suffices to demonstrate the violation of positivity. 

For the special case of a single propagator $D(x-y)$, {\it i.e.}, for the
lowest partial sum, reflection positivity reads,  
\begin{equation}
  \int d^4\!x\,  d^4\!y  \;  \bar f(-x_0,\hbox{\bf x}) \,  D(x-y)\,
  f(y_0,\hbox{\bf y}) \, \ge \, 0 
  \label{OS2}
\end{equation}
where $f \in {\mathcal S}_+(\RR^4)$ is a complex valued test (Schwartz) function
with support in $\{(x_0,\hbox{\bf x}) : x_0 > 0 \}$. This special case of
reflection positivity can be shown to be a necessary {\sl and} sufficient
condition for the existence of a K\"all\'en--Lehmann
representation, {\it i.e.} a spectral representation of the
propagator as given in Eq.~(\ref{KL_rep}) with positive spectral function. 
Therefore, the construction of a counter example to this condition (by a
suitable choice of $f$) is a possibility to demonstrate that, as one
manifestation of confinement, the particular Euclidean correlations can have
no interpretation in terms of stable particle states.   

\begin{figure}[t]
\vskip 1cm
  \centering{\
        \epsfig{file=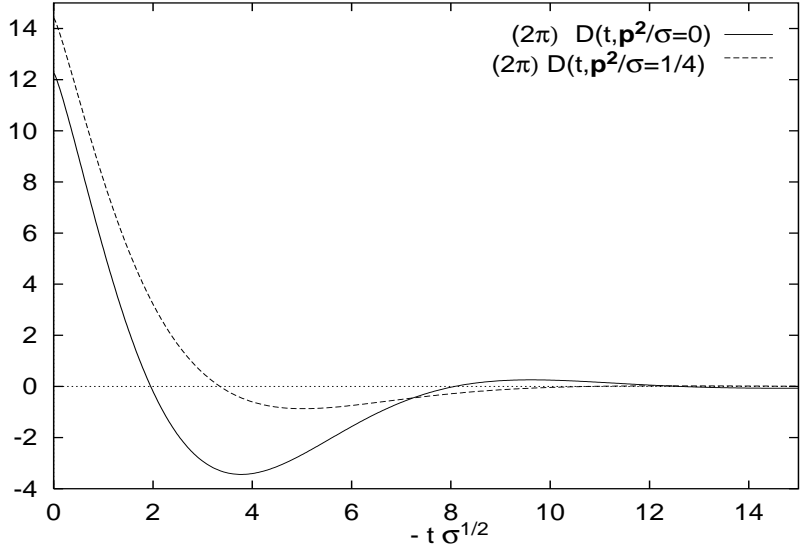,height=0.4\linewidth,width=0.5\linewidth}}
  \caption[The one-dimensional Fourier transform of the gluon propagator.]
          {The one-dimensional Fourier transform of the gluon propagator,
$D(t,\hbox{\bf p}^2)$.} 
  \label{gluon_ft}
\end{figure}

Heuristically, after three dimensional
Fourier transformation condition (\ref{OS2}) can be written, 
\begin{equation}
  \int_0^\infty \,   dt' \,  dt \; \bar f(t') \,  D(- (t'+t), \hbox{\bf p}) \,
  f(t) \, \ge \, 0 \; ,
\end{equation}
where $D(x_0,\hbox{\bf p}) := \int d^3\!x \, D(x_0,\hbox{\bf x}) \, e^{i
\,\hbox{\small\bf px}}$, and a separated momentum dependence of the analogous
Fourier transform of the test function  $f$ has to provide for a suitable
smearing around the three-momentum {\bf p}. 

In Fig.~\ref{gluon_ft} this Fourier transform of (essentially the
trace of) the gluon propagator, 
\begin{equation}
  D(t,\hbox{\bf p}^2) :=  \int \frac{dp_0}{2\pi}
    \frac{Z(p_0^2 + \hbox{\bf p}^2)}{p_0^2 + \hbox{\bf p}^2} \; e^{i p_0 t}
   \; , \label{eq:gluon_FT}
\end{equation}
is plotted for $\hbox{\bf p}^2 = 0$ and $\hbox{\bf p}^2 = \sigma
/4$. The fact that the DSE solution for sufficiently small $\hbox{\bf p}$ 
is negative over finite intervals in the Euclidean time in this
case shows trivially that reflection positivity is violated for the gluon
propagator. 

Another way to understand this connection between positivity and the
Euclidean correlation functions is based on an argument which was
first given by Mandula and Ogilvie \cite{Man87} and is phrased most clearly in
Ref.~\cite{Ais97}. The domain of analyticity of propagators in a 
G\aa rding--Wightman quantum field theory allows to write the spectral
representation for the (Euclidean) propagator $D(p)$ in the form,  
\begin{equation}
D(p) \, = \, \int_0^\infty \,  dm^2 \, \frac{\rho(m^2)}{p^2+m^2}  \; , 
\end{equation}
which gives rise to the Fourier transform,
\begin{eqnarray}
  D(t,\hbox{\bf p}^2) &=&  \int_0^\infty  \, dm^2 \, \rho(m^2) \;
   \frac{1}{2\omega} \;  \exp\{- \omega t\} \; , \quad \mbox{with} \; \omega
   \, = \, \sqrt{\mbox{\bf p}^2 + m^2} \\
 &=& \int_{\sqrt{\mbox{\bf p}^2}}^\infty  \, d\omega \; \rho(\omega^2-\mbox{\bf
   p}^2) \;  \exp\{- \omega t\} 
   \; . \label{gen_FT}
\end{eqnarray}
For a positive spectral function, $\rho(m^2) \ge 0$, it is therefore clear
that $\ln D(t,\hbox{\bf p}^2)$ has to be a convex function of the Euclidean
time,\footnote{Just as the free energy is a convex function of the external
fields in statistical mechanics.}  
\begin{equation} 
\frac{d^2}{dt^2} \, \ln D(t,\hbox{\bf p}^2)  \, \equiv \, \langle \omega^2
\rangle_{t,\mbox{\tiny \bf p$^2$}}  \, - \,\langle \omega
\rangle^2_{t,\mbox{\tiny\bf p$^2$}}   \, \ge 0 \; ,
\end{equation}   
because $\langle \omega \rangle_{t,\mbox{\tiny\bf p$^2$}}$ and $\langle
\omega^2 \rangle_{t,\mbox{\tiny\bf p$^2$}}$ are the moments of a positive
measure for all {\bf p}$^2$. 

Even though no negative $D(t, \hbox{\bf p}^2)$ have been reported in the
lattice calculations yet, the available results,
Refs.~\cite{Ber94,Mar95a,Nak95}, agree in indicating that the gluon
propagator is not a convex function of the Euclidean time and thus that
positivity is indeed violated for gluonic correlations. The corresponding
results will be discussed in Sec.\ \ref{sec_lattice}, a summary of lattice
results on the gluon propagator can be found on Ref.\ \cite{Man99}.

The ghost propagator in Euclidean momentum space, being negative for positive
renormalisation functions $G(k^2) > 0$, of course, is not reflection positive
either. This is due to the wrong spin-statistics of the ghost field and is the 
case at tree-level already. 

Besides this somewhat trivial minus sign, two general remarks on an infrared
enhanced propagator (as the one obtained for the ghost field here) might be
interesting to add. First, it is important to note that a propagator with an
infrared divergence $\propto 1/(k^2)^{1+\kappa}$ as found for the ghost
solution here (with $\kappa \simeq 0.93$), does not violate temperedness, as
long as $\kappa < 1$. Secondly, the Fourier transform $I_\kappa(t,\omega)$ of
correlations of a form $\propto 1/(p^2 + m^2)^{1+\kappa}$ (with $\kappa >
-1$, $\omega = \sqrt{\mbox{\bf p}^2 + m^2}$, and not necessarily $m = 0$) in
Euclidean momentum space generically reads,     
\begin{eqnarray}
I_\kappa (t,\omega) &:=&   \int \, \frac{dp_0}{2\pi} \,
  \frac{\omega^{2\kappa} }{(p_0^2 + \omega^2)}  {}_{ 1+\kappa} \,
 \; e^{i p_0 t} \label{gen_pl_ft} \\
&& \hskip -2cm =  \, \frac{(\omega t)^\kappa }{2^\kappa\Gamma(1+\kappa)}\,
    \frac{1}{\omega} \; \sqrt{\frac{\omega t}{2\pi}} \;
    \mbox{K}_{\kappa+\frac{1}{2}}(\omega t) \,
     \stackrel{\omega t \gg |\kappa|}{\longrightarrow} 
    \; \frac{(\omega t)^\kappa}{2^\kappa\Gamma(1+\kappa)} \,
  \frac{1}{2\omega} \; \exp\{-\omega t\} \; . \nonumber     
\end{eqnarray}
The limit of large $\omega t$ shows that $\ln I_\kappa (t,\omega)$ 
is not a convex function of $t$, if $\kappa > 0$.\footnote{For $\kappa =
0$ the limit $\omega t \gg |k|$ (indicated by the arrow in (\ref{gen_pl_ft}))
becomes an identity, and thus $\ln I_0 (t,\omega) $ linear in $t$ (with slope
$ -\omega $).} More generally, from the Bessel function (of the second kind)
$K_n(x)$, one indeed verifies that  $\ln I_\kappa (t,\omega)$ is a convex
function of $t$, if and only if $\kappa \le 0$. 

Euclidean correlations of a form~(\ref{gen_pl_ft}), {\it i.e.},
modifications of free particle propagators by an additional exponent $0 <
\kappa < 1$, in particular, infrared enhanced correlations (for $m=0$)
are thus an example of a manifest violation of positivity. However, they can in
principle have all the other necessary and important properties of a
local quantum field theory. 
These are for example temperedness which is necessary to define
Fourier transforms, or also, importantly, the expected domain of holomorphy
of the Green's functions which implies in particular that propagators are
analytic functions in the cut complex $k^2$-plane with singularities occurring
on the timelike real axis only. This is crucial for the incorporation of
timelike momenta in the analytically continued Euclidean formulation, {\it
e.g.}, to describe bound states. The general analyticity properties of the
Green's functions should therefore better be unaffected by possible
violations of positivity. The nature of these violations for propagators is
thus restricted to allowing the discontinuity at the cut, $\rho(k^2)$, to be
of indefinite sign. This is in complete agreement with the local description
of covariant gauge theories on indefinite metric spaces, see Sec.~\ref{Sec2.4}.
    
From all the evidence presented here as well as in Sec.~\ref{Sec2.3}, it seems
quite settled that the requirement of positivity on the level of the
elementary QCD correlations has to be abandoned anyway. Technically, this 
is fortunately the one property to give up with the least consequences of all.
This minimal relaxation of the axioms of quantum field theory seems
necessary in describing confining theories in which the elementary fields as
the carriers of charges may serve to implement locality but are not directly
associated with the observed particles.  

Concluding this section, a last remark concerns the occurrence of zeros in the
Fourier transform of the gluon propagator as a function of $t$,
Eq.~(\ref{eq:gluon_FT}), for sufficiently small $\hbox{\bf p}^2$. 
This implies that the gluon propagator in coordinate space cannot have a well
defined inverse for all $(t,\hbox{\bf x})$. While the corresponding
singularities in $G^{-1}(t,\hbox{\bf x})$ will cause problems in a
Hubbard--Stratonovich transformation to bosonise effective non-local quark
theories as the Global Colour Model, see Ref.~\cite{Tan97}, the same comment
that was given at the end of the last subsection applies here again. 
The infrared
structure of the vertex functions, renormalisation group invariance and the
existence of zeros in the gluon propagator, all lead to the same conclusion:
The combination of ghosts and gluons, $g^2 G^2(k^2) Z(k^2)/k^2 =
4\pi \alpha_S(k^2)/k^2 $ is the physically important quantity which  determines
the interactions of quarks in Landau gauge (and the current-current
interaction of the Global Colour Model, see the next section). 
It can be verified explicitly that
the Fourier transform of $\alpha_S(k^2)/k^2$ is free of such zeros and
positive.

\subsubsection{Dynamical Chiral Symmetry Breaking and Quark
Confinement}
\label{sub_DCS} 

The studies of DB$\chi$S via the study of the quark propagator DSE are
numerous, see Refs.\ \cite{Rob94} and \cite{Mir93} for a corresponding survey.
Almost all of these studies used a highly infrared singular kernel  for the
quark DSE in order to implement confinement. On the other hand, until recently
the ghost contribution in the quark-gluon vertex as discussed at the beginning
of this section have not been taken  into account. In this section we will
shortly review a few selected calculations used later in studies of meson
properties. It has to be noted, however, that the most recent of these
calculations do not employ an infrared singular kernel. In addition,  we will
also comment on results for the coupled gluon-ghost-quark DSEs in Landau gauge.
Please note that these are preliminary, and some of them are yet to be
published \cite{Ahl98}.

One model of quark confinement is based on the idea that
thresholds associated with quark production can be avoided in hadronic
processes by assuming that the momentum space quark propagator has no
singularities on (or near) the real axis in the complex $k^2$-plane which
precludes a K\"all\'en--Lehmann representation. The most simple realization
of this, being given by entire functions of $k^2$,
is often adopted in phenomenological
applications. In complex directions, singularities with timelike real
part might technically seem acceptable, if they are sufficiently far away
from the origin so as to not interfere with the hadronic processes
considered. This introduces an upper bound on the momentum range for the
applicability of such a model. Such a qualitative behaviour can in fact be
obtained from the quark DSE, {\it e.g.}, in the very simple model of gluonic
interactions mimicked by a 4-dimensional $\delta$-function in Euclidean 
momentum space~\cite{Mun83}.\footnote{This quite useful toy-model has been
used, {\it e.g.}, in a study of diquark confinement which will be discussed
in Sec.\ \ref{sec_Diq}.}

In phenomenological applications of Bethe--Salpeter equations for mesonic bound
states similar entire function forms of the quark propagators are frequently
also parameterised with quite remarkable success, see Chapter \ref{chap_Meson}.
Typical forms of such parameterisations of the quark structure functions
involve entire functions in the $k^2$-plane (with essential singularities at
infinity) and 3 -- 7 parameters for the light quark sector which are then
fitted to static meson properties, see {\it e.g.}
\cite{Rob00,Iva99,Rob96,Bur96} and references therein.

With respect to  phenomenological applications it is thereby interesting to
compare the different quark DSE models in quenched approximation to collect
further evidence on the origin, interpretation and justification of the various
models to be used in meson and baryon physics, see the next two chapters. The
probably most obvious question at this point is the role of ghosts. In this
chapter it became evident that
their contributions cannot be neglected in Landau gauge. The open question
can thus be phrased as to whether some of the more successful models might
nevertheless be able to model their influence without taking these
contributions into account explicitly.

To clarify this, consider the quark-current interaction of the
Global--Colour--Model which is motivated as being the result of an
expansion in the colour-octet quark currents $j^a_\mu(x)$  of the effective
quark theory given by the Feynman--Schwinger functional $Z_{\mbox{\tiny
YM}}[j]$ of the pure gauge theory with these currents as the external
fields.\footnote{As the measure in this gluonic functional of the
quark-currents is not Gaussian, of course, this can at best be understood in
the sense of Gell-Mann and Low's ``magic formula of perturbation theory''
\cite{Gel51}.}
The term quadratic in the quark-currents of this functional is inevitably
given by,
\begin{equation}
\ln Z_{\mbox{\tiny YM}}^{(2)}[j] \, := \,-  \frac{1}{2} \, \int \, d^4x d^4 y
\;  j^a_\mu(x)  \, {G^{(2)}_{\mu\nu}}^{ab}(x - y) \,  j^b_\nu(y) \; ,
\label{GCM_int}
\end{equation}
which {\sl defines} the interaction of the GCM. Obviously, the resulting
current interaction cannot be gauge invariant. It is only invariant under 
global colour rotations which is reflected in the model's name.  The
interpretation of this model interaction, also called the Abelian approximation, 
is motivated from analogy to what would occur in QED.
The central idea behind this
approximation can be traced to the assumption that, as follows from gauge
invariance for QED, the renormalisation constant of the gauge field is
related to the inverse coupling renormalisation by $Z_g^2 Z_3 = 1$. This
implies in particular for the gauge field propagator that
\begin{equation}
 {G^{(2)}_{\mu\nu}}^{ab}(x - y) \, = \, \delta^{ab}  \, g^2D_{\mu\nu}(x-y)
 \; ,  \label{GCM_AA}
\end{equation}
with $D$ being the gauge boson propagator,
\begin{equation}
  g^2D_{\mu\nu}(k) \, =\,  \left(\delta_{\mu\nu} - \frac{k_\mu k_\nu}{k^2}
  \right) \,   \frac{g^2Z(k^2)}{k^2} \, + \, g^2 \xi \, \frac{k_\mu
  k_\nu}{k^4} \; ,  \label{ab_glp}
\end{equation}
and thus $g^2Z(k^2)$, is a renormalisation group invariant which can be
identified with the running coupling in QED (as compared to $gZ(k^2)$ in the
Mandelstam approximation for QCD, see Sec.~\ref{sub_Mand}).  Perturbative
QCD does not show this behaviour. But again, neglecting the ghosts
of the non-Abelian theory in the covariant gauge, the Abelian approximation 
at this point effectively  might include some of
their contributions. Within this approximation, asymptotic
freedom is put in by replacing the infrared stable running coupling of QED
with the asymptotically free one of QCD (with $N_f $ quark flavours and
$N_c$ colours),
\begin{equation}
 g^2Z(k^2) \, \equiv \, \bar g^2(t_k,g^2 ) \, \to \, \frac{48\pi^2}{11N_c -
 2 N_f} \, \frac{1}{\ln(k^2/\Lambda^2)} \; , \quad \mbox{for}\; 
 k^2 \gg \Lambda^2 \; ,
\end{equation}
where $\Lambda$ denotes the QCD scale parameter. In the infrared the interaction
is then chosen such that phenomenologically acceptable models arise.
Especially, a certain strength of the interaction is needed to imply 
DB$\chi$S.

In the non-Abelian theory, the situation is of course different.
 In particular, in the
covariant gauges ghost contributions enter in this effective
interaction. Even though this might seem to be evident from its definition,
consider perturbation theory to demonstrate this fact explicitly.
The symmetry currents $j^a_\mu(x)$  not being renormalised, their interaction
$G^{(2)}$ has to be renormalisation group invariant as is the case for
the expression
(\ref{GCM_AA}) in QED, of course. In perturbative QCD in Landau gauge,
however, the product $g^2D$ has an anomalous dimension $\gamma $ which is
different from unity for all numbers of colours and all numbers of
flavours, and it is thus not renormalisation group invariant. Near the
ultraviolet fixed point, this anomalous dimension is given by
\begin{equation}
\gamma(g) \,  \stackrel{g\to 0}{\longrightarrow} \, \frac{\gamma^0_A}{\beta_0} \,
= \, \frac{(13/2) N_c - 2 N_f}{11 N_c - 2 N_f} \, \stackrel{!}{=}  \, \left\{
\begin{array}{l} 1 \; , \quad \mbox{AA} \; ,\quad {\rm not\; possible\; for}
\; N_c, N_f \;
\in \, \NN \\
\frac{1}{2}\; , \quad \mbox{MA} \; ,\quad \rightarrow \; N_f = N_c
\end{array} \right.
\end{equation}
with $\gamma_A^0$ and $\beta_0$ being the leading coefficients of the
gluon anomalous dimension and the $\beta$-function. Note that
in contrast to the Abelian approximation (AA) the result of the Mandelstam
approximation (MA), {\it c.f.} Sec.~\ref{sub_Mand}, that $gD$ is renormalisation
group invariant without ghosts, happens to resemble the perturbative result
for $N_f = N_c$. The effect of neglecting ghosts seems to have cancelled with
neglecting $N_f=N_c$ quark flavour at this point. Of course, the
quark loops as well as the ghost loop both contribute with a negative sign to
the perturbative divergences. The different strengths of these contributions
nevertheless make their compensation in the ratio $\gamma^0_A/\beta_0 $
possible.

While it thus seems impossible to neglect ghosts consistently,  ghosts do
implicitly enter in quark-gluon vertices, and that the interaction of the
GCM might be justified from a dressed-ladder exchange, {\it e.g.} as given
in Eq.~(\ref{dressed-ladder-skeletton}). For the GCM the quantity $G^{(2)}$
may in turn serve to define a non-perturbative running coupling, {\it e.g.},
from its part proportional to the metric $\delta^{ab} \delta_{\mu\nu}$ by,
\begin{equation}
\frac{1}{3(N_c^2-1)} \, \delta^{ab} \delta_{\mu\nu} \,
{G^{(2)}_{\mu\nu}}^{ab}(k) \, =: \,-  \frac{4\pi \alpha(k^2)}{k^2} \; .
 \label{GCM_rc}
\end{equation}
Perturbatively (at next-to-leading order) this is equivalent to the
definition for the running coupling adopted in Sec.~\ref{sub_Sub}. 
Non-perturbatively,
the definition (\ref{GCM_rc}) should, in principle, be obtainable from the term
of the quark-gluon vertex $\propto \gamma_\mu$ in a symmetric momentum
subtraction scheme in a quenched lattice approximation. It would be
interesting to assess, from such a lattice calculation, possible indications
towards the phenomenologically successful infrared enhancement for the GCM
coupling~(\ref{GCM_rc}). The simulation of Ref.~\cite{Sku98} seems to
indicate to the contrary. It is, however, obtained from an asymmetric
subtraction scheme and thus is very likely 
inconclusive, see Sec.~\ref{sec_lattice}. 
The effective interaction of the GCM in the covariant gauges, whatever it is, 
is {\sl not} the gluon propagator alone as suggested by
Eq.~(\ref{GCM_AA}). Ghosts contribute to the current-current interaction in
Eq.~(\ref{GCM_int}) as they are an unavoidable part of the (perturbative)
definition of $Z_{\mbox{\tiny YM}}[j]$.

The quark DSE~(\ref{quarkDSEmom}) (with $  S_{\mbox{\tiny
tl}}(k) = 1/(-i k \hskip -6pt \slash + Z_m m )$ denoting the tree-level
propagator), 
\begin{equation}
 S^{-1}(k)  \, = \, Z_2 S_{\mbox{\tiny tl}}^{-1}(k)  \, + \, g^2
 Z_{1F} \, \frac{4}{3}   \int \frac{d^4q}{(2\pi)^4} \;  \gamma_\mu \, S(q) \,
 \Gamma_\nu(q,k) \, D_{\mu\nu}(k-q) \; ,  \label{qkDSE_ch7}
\end{equation}
in the Abelian rainbow approximation effectively simplifies  to
\begin{equation}
 S^{-1}(k)  \, = \, Z_2 S_{\mbox{\tiny tl}}^{-1}(k)  \, + \, 
  \frac{16 \pi}{3}   \int \frac{d^4q}{(2\pi)^4} \; \alpha ((k-q)^2) 
  \gamma_\mu \, S(q) \,
 \gamma_\nu \, D_{\mu\nu}^{\rm free} (k-q) \; .  \label{qkDSE_AA}
\end{equation}
``Effective running couplings'' $\alpha ((k-q)^2)$ employed in the study of
meson properties can be found, {\it e.g.}, in Eqs.\
(\ref{alphaCraig},\ref{alphaPieter})
in the next chapter. Whereas the interaction in (\ref{alphaCraig}) contains a
$\delta$-function in momentum space, the form (\ref{alphaPieter}) is infrared
finite. Nevertheless, the results for the quark propagator are very similar.
This can be seen {\it e.g.} from the fact
that the quark condensate (renormalised at 1 GeV)
for both interaction types is very close to the phenomenological value,
\begin{equation}
\langle \bar q q (\mu^2 = 1 {\rm GeV}^2) \rangle =  - (0.240 {\rm GeV})^3 \, .
\end{equation} 
This condensate can be extracted either from the mass function or by integrating the
trace of the quark propagator. Parametrising the general form of the quark 
propagator in Euclidean momentum space,
\begin{equation}
  S(p) \, = \, \frac{1}{-ip\hskip -6pt \slash  A(p^2) + B(p^2)} \, = \,
\frac{Z_q(p^2)}{-ip\hskip -6pt \slash \,  + \, M(p^2)}  \; ,
\end{equation}
the mass function $M(p^2) = B(p^2)/A(p^2)$ is proportional to the quark
condensate at large $p^2$,
\begin{equation}
M(p^2)\simeq
   m -  \frac{4\pi \alpha (p^2)}{3p^2}
         \left( \frac{\alpha (p^2)}{\alpha (\mu ^2)}\right)^{-d_m}
   \langle \bar q q \,(\mu ^2) \rangle
\quad ,\label{ope}
\end{equation}
where $d_m=12/(11N_c-2N_f)$ is the anomalous dimension of the quark mass 
and $\mu$ is a large Euclidean renormalisation point. 

\begin{figure}
\centering{\
\epsfig{figure=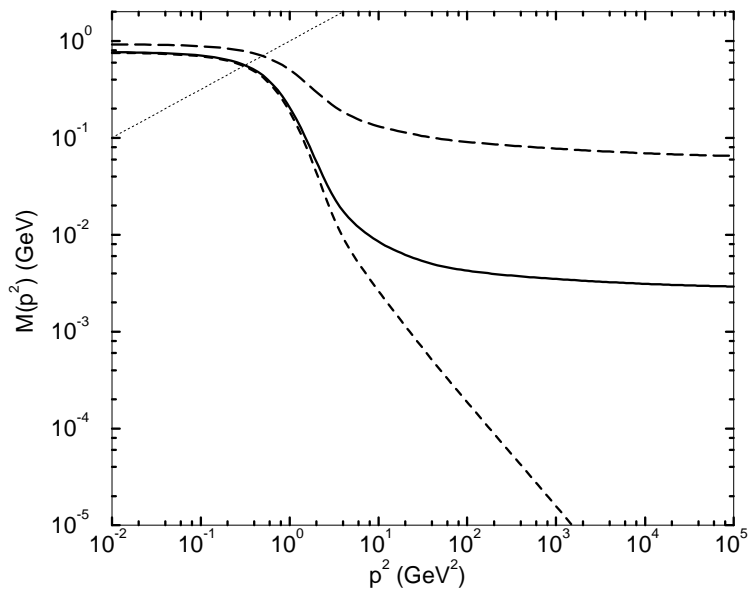,height=9.0cm}}
\caption[The renormalised dressed-quark mass function, $M(p^2)$, obtained by
solving the quark DSE with the interaction (\ref{alphaCraig}. (Adapted from
Ref.\ \cite{Mar97}.) ]
{The renormalised dressed-quark mass function, $M(p^2)$, obtained by
solving the quark DSE with the interaction (\ref{alphaCraig}): 
$u/d$-quark (solid line); $s$-quark (long-dashed
line); and chiral limit (dashed line).  The renormalisation point is
$\mu=19\,$GeV.  The intersection of the line $M^2(p)=p^2$ (dotted line)
with each curve defines the Euclidean constituent-quark mass, $M^E$.
(Adapted from Ref.\ \cite{Mar97}.)
\label{spplot}}
\end{figure}

The mass functions for the three light flavours and in the chiral limit obtained
in Ref.\ \cite{Mar97} are displayed in Fig.\ \ref{spplot}. Clearly one sees the
phenomenon of DB$\chi$S. The authors of Ref.\ \cite{Mar97} introduce as measure
the Euclidean mass $M^E$ defined by $(M^E)^2=M((M^E)^2)$. It has to be noted,
however, that the corresponding quark propagator does not possess poles for
time-like (real) $p^2$. Instead singularities for complex values at $p^2$ are
present.

\begin{figure}
\leftline{\hskip 2cm \epsfig{file=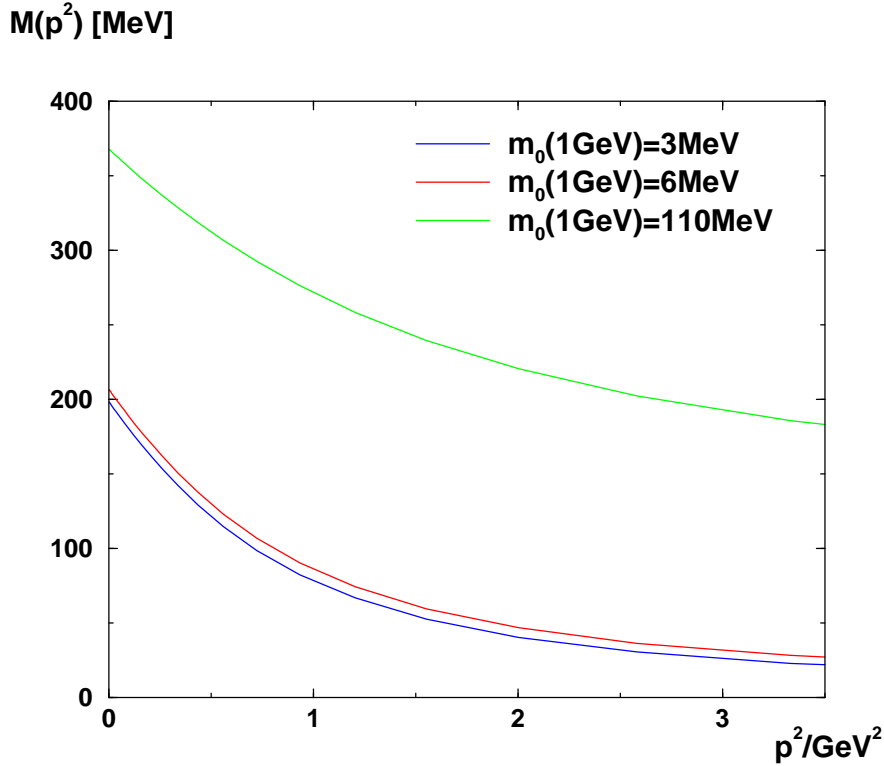,width=0.7\linewidth}}
\caption{Quark mass functions $M(p^2)$  for various
current masses from the quenched approximation (with $\bar\beta_0 = 11$).}
\label{qk_fig}
\end{figure}

Recently the coupled ghost, gluon and quark DSEs have been solved \cite{Ahl98}.
Before commenting on this solution it is illustrative to study the quark DSE in
quenched approximation using the ghost and gluon propagators of Ref.\
\cite{Sme98} discussed in the last sections together with its implied form 
for the quark-gluon vertex. The related truncation scheme is
based on the solution to the simplified Slavnov--Taylor identity obtained
from Eq.~(\ref{qkSTI}) by neglecting the quark-ghost scattering kernel,
\begin{eqnarray}
  G^{-1} (k^2)  \,  k_\mu \Gamma_\mu(p,q)  \, = \,   iS^{-1}(p) \,
  - \, iS^{-1}(q) \; ,  \quad k = p -q \; . \label{STI_ch7}
\end{eqnarray}
This identity being of analogous form as the Abelian Ward--Takahashi
identity but with the important additional ghost factor $G(k^2)$. 
The corresponding solution given in Eqs.\ (\ref{qkSTIsol},\ref{CP_ext})
is then employed in the quark DSE. The resulting effective quark interaction 
is only slightly infrared singular and even integrable. This allows, {\it
e.g.}, to use the Landshoff--Nachtmann model for the pomeron in this
approach. From the solution  a pomeron intercept of approximately
$ 2.7/$GeV \cite{Alk99a,Alk99b} has been estimated 
as compared to the typical phenomenological value of 2.0/GeV \cite{Cud89}.
The resulting quark mass functions are shown in Fig.\ \ref{qk_fig}.
It is reassuring that in this approach, where the only parameter, the scale
generated by dimensional transmutation, is fixed from the behaviour of the
running coupling at large momenta, DB$\chi$S occurs. It has to be emphasised
that except for lattice calculations this phenomenon has been found in
an {\it ab initio} calculation for the first time, {\it i.e.} without any
explicit
assumptions or modelling of the effective infrared quark interaction.
However, Fig.\ \ref{qk_fig} also makes clear that the calculated mass functions
are smaller than phenomenologically required. This is especially pronounced
in the quark condensate. At a renormalisation scale of 1 GeV the calculated
value is $\langle \bar q q \rangle  (\mu = 1 {\rm GeV}) = - (0.1 {\rm GeV})^3$
and thus an order of magnitude smaller than the empirical value.

This problem becomes even worse beyond the quenched approximation. First, it has
to be noted that only a solution for the coupled systems of propagator equations
has been found where the quark propagator resembles qualitatively the one in
quenched approximation \cite{Ahl98}. This does not exclude that  
other solutions exists, and as a matter of fact, such a solution with
non-trivial infrared behaviour of the quark renormalisation function $Z_q(p^2)$
seems possible. Concentrating on the solution found so far, one notes that
the quark loop
in the gluon DSE is screening and thus the effective quark-quark interaction
becomes smaller. This effect is somewhat weakened due to the back-reaction
via the quark DSE. But nevertheless the quark mass function and correspondingly
the quark condensate become even smaller. There is, however, an interesting
feature not present in the quenched approximation
(where $Z_q(p^2)$ is only slightly suppressed in the infrared): 
The calculated quark 
renormalisation function $Z_q(p^2)$ displays a very pronounced dip in the
infrared dropping to $Z_q(0) \approx 0.7$ slightly dependent on the number of
active flavours. This is for two reasons interesting: First, as can be seen from
Fig.\ \ref{latQ} such a behaviour is found in the most recent lattice
calculations. Secondly, this behaviour might have a profound influence on
the analytic properties of the quark propagator.
Therefore, the hope is that
despite the numerical mismatch for the amount of DB$\chi$S, this and related
investigations will shed light on the question of quark confinement in linear
covariant gauges.

\subsection{A Non-Perturbative Expansion Scheme in Landau
Gauge QCD}
\label{sec_Stingl} 

The results on the infrared behaviour of the gluon propagator in Landau gauge
with or without ghosts differ as much as possible. Certainly, including ghosts
provides DSEs which are consistent at the level  of
propagators. Nevertheless, it is of utter importance to have some other
non-perturbative scheme to compare with. In the next section we will contrast
the DSE solutions with lattice results. As already mentioned in Sec.\
\ref{Sec2.5} there is an alternative approach to the DSEs in Landau gauge QCD 
\cite{Hae90,Sti95}. A brief description of this approach in this section will
supplement the study of the infrared behaviour of the Green's functions in
Landau gauge QCD as inferred from the already presented truncation schemes. As
such it adds some new aspects while some others are given less emphasis in this
somewhat complementary approach to the Dyson--Schwinger hierarchy.

This approach is based on the observation that an exact vertex function and
its perturbative estimate differ by expressions which, in general, contain
essential singularities in the coupling of the same structure as in the
spontaneous mass scale, see Eq.\ (\ref{Lambda}). The idea therefore is to
improve the perturbative expansion in approximating the Euclidean proper
vertices $\Gamma$ by a double series in the coupling as well as in the degree
of a rational approximation  with respect to the spontaneous mass scale
$\Lambda $ being non-analytic in $g^2$. The proper
vertices are thereby approximated by a truncated double series in functions
$\Gamma ^{(r,p)}$ where $r$ denotes the degree of the rational
approximation while $p$ represents the order of perturbative corrections
in $g^2$ calculated from $\Gamma ^{(r,0)}$ instead of $\Gamma^{(0)}_{pert.}$,
see Ref.~\cite{Sti95}. The requirement that these non-perturbative terms
reproduce themselves in the Dyson--Schwinger equations is then used to
extract constraints on the finite set of dimensionless coefficients that are
used to parameterise the rational approximants.

To be specific, consider the 1-PI gluon 2-point function, {\it i.e.} the
inverse gluon propagator. It is transverse in Landau gauge and its functional
dependence on the gluon momentum and the spontaneous mass scale
$\Lambda$ is parameterised by a rational function with denominator of degree
$r$, introducing $2r+1$ dimensionless coefficients $\{\zeta_i \}$, $
\{\eta_i \}$,
\begin{equation}
\Gamma^{(r,0)}(k^2) \, = \, \frac{(k^2)^{r+1} + \zeta_{1} \Lambda^2 (k^2)^{r}
+ \dots \zeta_{r+1} (\Lambda^2)^{r+1}}{(k^2)^{r} + \eta_{1} \Lambda^2
(k^2)^{r-1} + \dots \eta_{r} (\Lambda^2)^{r}} \; .
\end{equation}
The idea here is to account for the polynomial dependence on $\Lambda$ and
thus on $k^2$ to parameterise non-perturbative effects such as the condensate
contributions of the operator product expansion. The logarithmic $\Lambda$
dependences have to be generated from radiative corrections, {\it i.e.}, from
higher perturbative orders ($p \geq 1$).

The lowest rational approximation $r = 0$ describes the dynamical
generation of a Schwinger mass~\cite{Sch62}, $m^2 = \zeta_1 \Lambda^2 $, from
the gluon DSE. Generally, for even orders $r$ there
will be at least one ``physical particle'' pole on the (negative) real
$k^2$-axis in the propagator. In a situation like this one would expect a
lowest isolated pole to stabilise as $r$ is increased, and a sequence of real
zeros and poles to develop and approximate the branch cut describing
multi-particle production for a propagator of physical degrees of freedom,
{\it i.e.}, one with a K\"all\'en--Lehmann representation. As this is not
anticipated for the elementary degrees of freedom in QCD, it was further
concentrated on the odd sequence of rational approximations, {\it i.e} an
even number of generally complex poles\footnote{This means that poles occur
only as complex conjugate pairs (for real coefficients). 
The occurrence of an even number of real
poles is rejected as being exotic. In particular, those closest to the origin
in the complex $k^2$-plane are assumed to come as complex pair.}
which have been interpreted as describing unstable short lived
excitations~\cite{Sti95}. For asymptotically large $k^2 \gg \Lambda^2 $ one
imposes the boundary condition that the rational approximants reduce to the
tree-level vertices, {\it e.g.}, for the gluon, $ \Gamma^{(r,0)}(k^2) \to
k^2 $. The power corrections to this limit can furthermore be compared to
condensate contributions in a corresponding operator product expansion. The
perturbative logarithms of the coefficient functions could in principle be
obtained analogously from $ \Gamma^{(r,p)}(k^2)$ for successively higher $p$.

The practical calculations in this approach are thus far restricted to the
lowest non-trivial degree $ r =1 $ in the odd rational approximation. The
general idea of the approach will be sketched for this most simple order in
the following. The 2-point gluon vertex function then has the
structure,
\begin{equation}
\Gamma^{(1,0)}(k^2) \, = \, k^2 \, +\, u_{1} \Lambda^2 \, +\, \frac{u_3
\Lambda^4}{k^2 + u_{2}\Lambda^2 } \; ,
\end{equation}
with one simple pole, and some linear combinations $u_i$ , $i=1...3 $, of the
$r=1$ coefficients introduced above. One can then infer from the hierarchical
coupling to the higher vertex functions in DSEs
together with the exchange symmetry of these higher vertices that analogous
simple poles have to occur also in the momentum variables corresponding to
each of their external legs~\cite{Sti95}. For example, the part of the
3-gluon vertex that replaces the tree-level vertex is found to have
the following structure ($A(x,y;z)  = A(y,x;z) $, and $A = 1$ at tree-level),
\begin{equation}
 \Gamma_{\mu\nu\rho}(p,q,k)
       \, =\, - A(p^2,q^2;k^2)\,  \delta_{\mu\nu}\,  i(p-q)_\rho\,
    + \; \hbox{cyclic permutations} \; ,
\end{equation}
containing poles and a polynomial numerator $N$ in the momenta and
$\Lambda^2$ of the form,
\begin{equation}  \label{3gvtlf}
 A(p^2,q^2;k^2)  =  \frac{ N(p^2,q^2;k^2) }{(p^2 + u_2 \Lambda^2) (q^2 +
 u_2 \Lambda^2)(k^2 + u_2 \Lambda^2)}  \quad .
\end{equation}
The additional requirement (besides the symmetry of $N$) that these vertex
functions, when replacing the tree-level vertices, should not change the
superficial degree of divergence of any diagram, is then used to constrain
the numerator polynomials further.\footnote{One might think that this
requirement is contained already in the asymptotic condition for the vertex to
reduce to tree-level at high momenta, in general however, some additional
restrictions are necessary~\protect\cite{Sti95}.}

These rational approximants for the 1-PI vertices at lowest perturbative
order $p=0$, {\it i.e.} order zero in the coupling $g^2$, replace the
tree-level vertices in a modified perturbative expansion. Combined with the
operator product expansion, this scheme corresponds to a Pad\'e
approximation with respect to the high momentum power corrections by
condensates~\cite{Ahl92}. Its actual use goes far beyond this, however.
Dyson--Schwinger equations are used to determine these approximants
self-consistently. The inhomogeneities of the DSEs, being
the tree-level vertices, lead to the following question: How can the
difference between these and the rational approximants at zeroth perturbative
order, {\it e.g.} for inverse the gluon propagator,
\begin{equation}
\Delta\Gamma (k^2) \, :=\,  \Gamma^{(1,0)}(k^2) \, -  \Gamma^{(0)}_{pert.}
(k^2) \, = \, \, u_{1} \Lambda^2 \, +\, \frac{u_3\Lambda^4}{k^2 + u_{2}  \Lambda^2
} \; ,
\end{equation}
be generated from loop-integrals containing explicit orders of the (bare)
coupling $g_0^2$? This mechanism can be understood qualitatively as follows:
Due to the presence of the mass scale $\Lambda$ there will be new types of
logarithmic ultraviolet divergences generated in the loops. These contain
factors
\begin{equation}
\propto \, \gamma_0 \, g_0^2 \, \ln \left( \Lambda_{UV}^2/\Lambda^2 \right)
\; ,
\end{equation}
where the renormalisation group invariant (since physical) mass scale
$\Lambda$ replaces the momentum variables that occur in analogous perturbative
divergences in which the constant $\gamma_0$ would be identified with the
leading coefficient of the corresponding anomalous dimension. For
perturbative divergences, the cutoff $\Lambda_{UV}$ dependence is traded for a
renormalisation scale $\mu$ dependence by subtracting  $ \gamma_0 \,
g^2 \, \ln (\Lambda_{UV}^2/\mu^2) $ with a suitable counter
term which is done successively order by order in the coupling. Here on the
other hand, it is used that, heuristically, the bare coupling is the running
coupling at the cutoff scale, $\bar g^2(\Lambda_{UV}) \to
g_0^2$, which tends to zero in the scaling limit ($\Lambda_{UV}\to \infty$),
{\it c.f.}, Eq.~(\ref{Lambda}),
\begin{equation}
 \bar g^2(\mu) = \frac{1}{\beta_0 \ln \left(
\mu^2/\Lambda^2 \right)} \; ,\quad \mbox{hence:} \quad
 \gamma_0 \, g_0^2 \, \ln \left( \Lambda_{UV}^2/\Lambda^2 \right) \to
 \frac{\gamma_0}{\beta_0} \; .
\end{equation}
In this way the contributions giving rise to the non-analyticities in $g_0^2
= 0$, and which thus cannot be taken care of by an asymptotic expansion in the
coupling, have to reproduce themselves in the equations at every order in
this expansion. This argument can be made more rigorous using dimensional
regularisation and the particular subtraction scheme of Ref.~\cite{Sti95}.
At one-loop this mechanism has been demonstrated to generate the rational
structure $\Delta\Gamma$ for the (inverse) gluon propagator as well as the
corresponding $r=1$ structure of the remaining six superficially divergent
vertex functions of Landau gauge QCD explicitly in
Refs.~\cite{Hae90,Sti95,Dri98a,Dri98b}.\footnote{It is intuitively clear from
the above that this generation of rational structure at lowest order
$\Gamma^{(r,0)}$ is restricted to the seven primitively divergent
vertices.}

Furthermore, the non-linear structure of the DSEs is
then used to determine the coefficients of the approximants by comparing their
poles and residues on one side of the equations with those generated from the
one-loop divergence structure on the other side as outlined above. One
conclusion from the resulting conditions is for example that the pole positions
in identical types of external legs of different proper vertex functions have
to be identical,  {\it e.g.}, $p_2 = - u_2 \Lambda^2 $ in the gluonic
2-point vertex as well as in the 3-point vertex as given above. Together
with additional constraints from the exchange symmetries as well as from the
form of the perturbative divergences the systems of equations that determine
the coefficients of the approximants are highly
over-determined. Some violations of exchange symmetries and some additional
contributions to the perturbative divergences have thus to be taken into
account and minimised for a given set of solutions that can be obtained in
absence of these additional constraints. The remaining violations can then
serve as an indicator for the quality of the particular level of
approximation. The present results suggest that the $r=1$ level describes the
system of gluon and ghost 2-point and 3-point functions quite
consistently~\cite{Sti95,Dri98a}.\footnote{The ghost propagator 
and the ghost-gluon vertex were in all presently available
solutions to this scheme taken to be the tree-level ones. 
While in the first studies~\cite{Hae90,Sti95} this was still an additional
assumption, the more complete solutions seemed to verify this assumption 
at a more complete level of the approximation~\cite{Dri98a,Dri98b}. This
together with an infrared finite or vanishing gluon propagator contradicts
the masslessness condition (\ref{massless.MA}) which  results
for the gluon propagator in Landau gauge if a trivial ghost sector is
assumed~\cite{Man79}.} And the results obtained for this
subsystem of the superficially divergent vertices in which
the influence of the 4-gluon vertex function is parameterised by additional
free coefficients, are furthermore confirmed reasonably well from a decoupled
study of the Dyson--Schwinger equation for the 4-gluon
vertex~\cite{Dri98b}. The violations mentioned above become more drastic,
however, when quarks are included. Also, the study of the 4-gluon equation
so far yields results which seems to be difficult to reconcile with those of
the 2-point and 3-point subsystem in presence of quarks. So the question
whether the approach is a viable possibility beyond a quenched approximation
remains somewhat unclear. There is certainly the possibility that the
situation improves considerably at much higher orders in $r$, given the
complexity of the approach at the presently employed order, however, this
conclusion would nevertheless cast some doubt on its practical use.
Beyond the question of practicability this approach has added something
conceptually new, namely a mechanism of compensating singularities, see
Sec.\ \ref{Sec2.5}. Due to its importance we will shortly reiterate on this
point below.

One previously raised criticism \cite{Haw98} concerning the apparently 
unphysical singularities in the proper vertex functions which are induced by
the zeros in the propagators, and which are a feature inherent in the approach,
has now been clarified \cite{Dri98a}. As seen from the structure of the gluon
3-point vertex function given above, the hierarchical coupling of the
different vertex functions implies common poles in external momenta. Therefore
these poles in the 2-point vertices which are zeros in the propagators
reappear in all higher vertex functions. It follows that poles in the external
lines are compensated by the corresponding zeros of the propagators attached to
these lines when computing scattering amplitudes and will not give rise to
unphysical asymptotic particles. As discussed in Sec.\ \ref{Sec2.5} the related
poles on internal lines are also automatically cancelled in this approximation
scheme \cite{Dri98a,Dri98b}: The necessary pole contributions reproduce
themselves in the hierarchy.
What still needs to be shown is that colourless hadrons do appear as asymptotic 
states.

\subsection{Lattice Results}
\label{sec_lattice} 

In this section we will discuss the results for QCD Green's functions as
obtained from lattice calculations. As we will see results from the lattice 
and from DSEs are complementary.
DSEs have to rely on truncation schemes. Therefore, to be at least able to 
judge their validity it is mandatory to confront them with some other 
non-perturbative methods. It is especially helpful to compare DSE results  
with the one of lattice calculations. Whereas the 
latter cannot investigate the deep infrared region they nevertheless allow
to distinguish between infrared suppressed and infrared enhanced functional
forms for propagators. Also the conclusion, that the gluon propagator in Landau
gauge violates positivity, can be drawn quite reliably from lattice data 
\cite{Man99}. 

As we will see lattice results provide evidence for an infrared suppressed gluon
and an infrared enhanced ghost propagator. Thus it is natural to
compare the DSE results of Ref.\ \cite{Sme98} with 
lattice results available for the gluon
\cite{Ber94,Mar95a,Lei98} as well as for the ghost propagator
\cite{Sum96}. These calculations use lattice versions to implement the Landau
gauge condition supplemented by different procedures to eliminate Gribov
copies (at least approximately). Recently, for the pure $SU(2)$ lattice gauge
theory, the influence of such copies of gauge equivalent configurations
present in the conventional Landau gauge, has been systematically
investigated for gluons and ghosts in \cite{Cuc97}.

\begin{figure}[t]
  \vskip 11cm
  \centering{ \epsfig{file=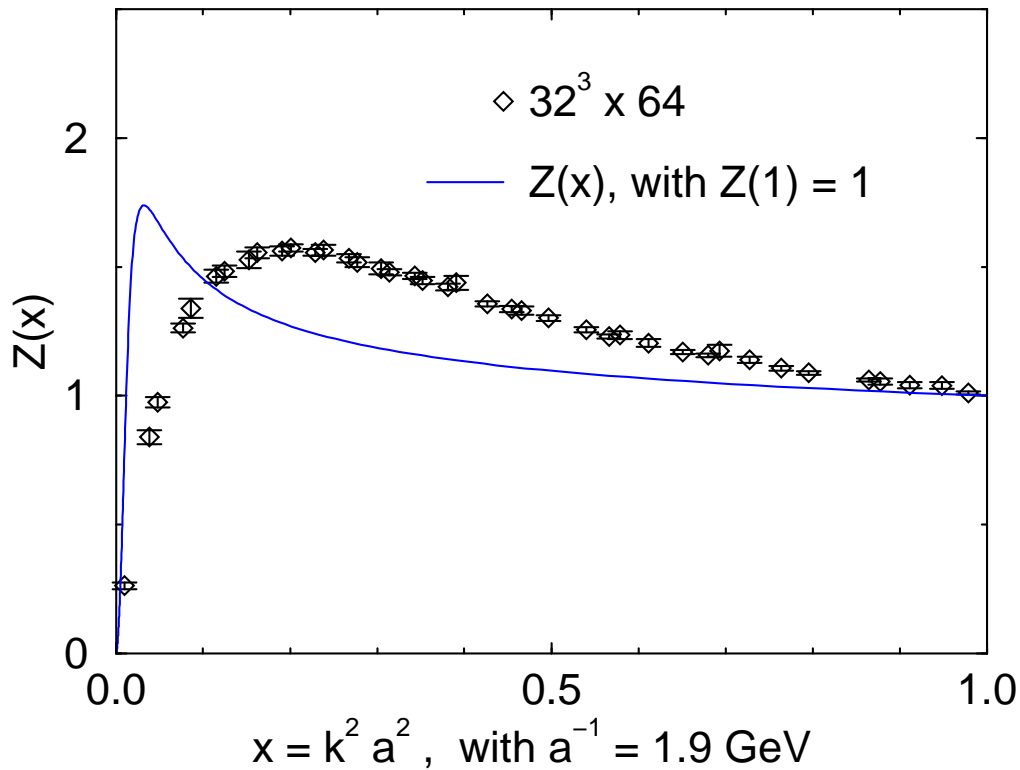,width=0.8\linewidth} }
  \vskip -10.5cm
  \caption{The gluon renormalisation function from Dyson--Schwinger
           equations compared to the
  corresponding lattice data from \protect\cite{Lei98}.}
  \label{Derek}
\end{figure}

Due to different normalisations, lattice sizes and, in general,
differing values of the lattice coupling $\beta$, it is not quite obvious
that all presently available lattice data sets for the gluon propagator in
Landau gauge from different groups are indeed consistent {\it e.g.}, compare
Refs.~\cite{Ber94} and \cite{Mar95a}.  Since it is not possible to compare
these different data sets without manipulating their scales, we will present
results only for a very recent and accurate lattice calculation for the gluon 
propagator in Landau gauge. 
In Fig.\ \ref{Derek} the lattice data of Ref.~\cite{Lei98} are shown.
For comparison the DSE solution (with normalisation $Z(x = 1) = 1$)
of Ref.\ \cite{Sme98} is also included.
The momentum scale in the
results, chosen to yield $\alpha_S(M_Z) = 0.118$ as described above, is not
used as a free parameter and is thus not adjusted. Rather, the gluon
propagator for this fixed momentum scale is plotted as a function of the
invariant momentum $x = k^2 a^2$ in units of the inverse lattice spacing with
$a^{-1} = 1.9$GeV from lattice phenomenology corresponding to the value $\beta
= 6.0$ for $SU(3)$ which was used in this particular simulation. 
Note that these lattice results obtained on an anisotropic $32^3 \times 64$ 
lattice with $\beta =6.0$ are without significant statistical
errors for momenta down to as low as $0.4$ GeV. A very recent investigation
\cite{Bon00} using an improved action and an improved gauge fixing unambiguously
demonstrates that the infrared suppression of the gluon propagator is not a
finite size effect. This at least indicates towards
a common trend for the gluon propagator to be very weak in the infrared.
The influence of Gribov copies on these lattice calculation is as yet not
entirely clear (nor is it on the presently available DSE solutions, of course).
However, the same qualitative behaviour and a very similar quantitative
behaviour, in particular in the infrared, was reported for the gluon
propagator from a lattice calculation employing the Laplacian
gauge\footnote{The Laplacian gauge also reduces to the Landau gauge in the
continuum, it was proposed as an effective procedure of finding
absolute minima, however \cite{Vin92,Vin95}.}
in~\cite{Ale00}.

\begin{figure}
  \vskip 10cm
  \centering{ \epsfig{file=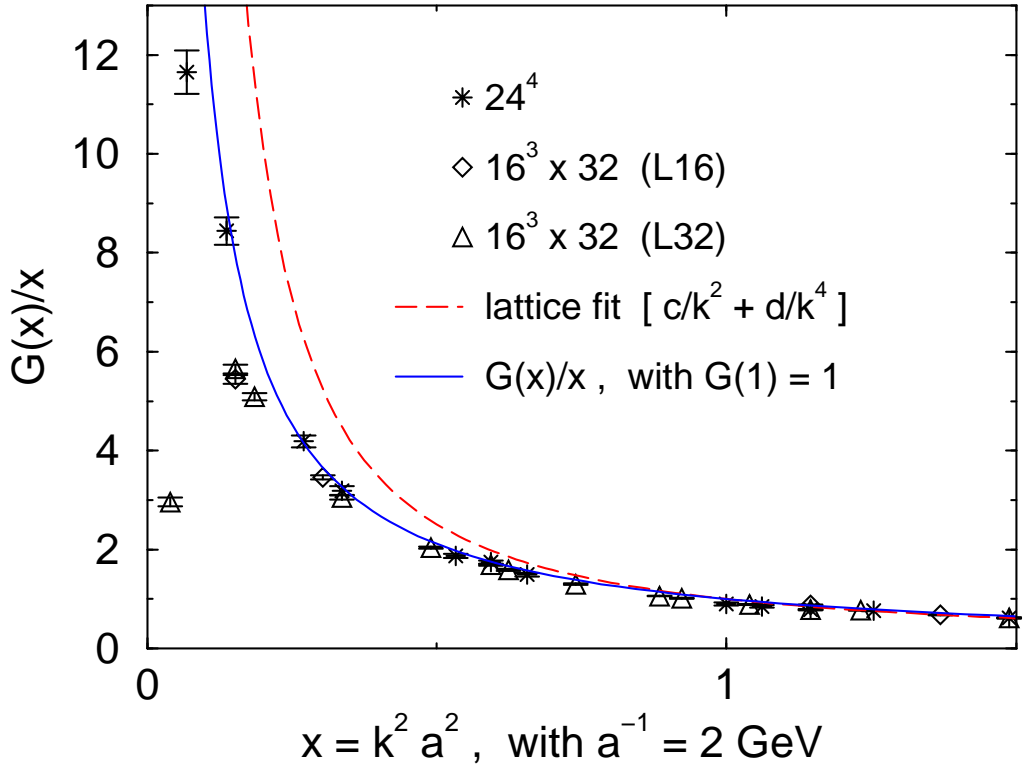,width=0.8\linewidth} }
  \vskip -9.5cm
  \caption[The ghost propagator from Dyson--Schwinger
           equations compared to lattice data.]
          {The ghost propagator from Dyson--Schwinger
           equations (solid line) compared to data from Fig.~1 in
           Ref.\ \protect\cite{Sum96} for $24^4$ and $16^3\times 32$
           lattices, and a fit of the form $c/x+d/x^2$ (with $c=0.744 , \;
           d=0.256 $) as obtained in Ref.\ \protect\cite{Sum96} for
           $x \ge 2$ (dashed line).}
  \label{Glat}
\end{figure}

In Fig. \ref{Glat} the infrared enhanced ghost propagator with
normalisation such that $G(x=1) = 1$ is compared to the results of
Ref.~\cite{Sum96} obtained on a symmetric $24^4$ and a non-symmetric $16^3
\times 32$ lattice for $SU(3)$ at $\beta = 6.0$ from Fig.~1 in
Ref.~\cite{Sum96} up to $x = 1.5$. Identical results modulo finite size
effects were obtained for an $8^4$ lattice (see Ref.~\cite{Sum96}). Again, $x
= k^2 a^2 $ with $a^{-1} = 2$ GeV and the momentum scale in the DSE result,
fixed from the $Z$-mass, is {\it not} adjusted.

The data extracted from the long direction of the $16^3 \times 32$
lattice might indicate the existence of a finite maximum in the ghost
propagator at very low momenta. The fact that the two lowest data
points in this set lie significantly below their neighbours of the $24^4$
lattice was interpreted by the authors of Ref.~\cite{Sum96} as a genuine
signal rather than a finite size effect. The reason for this being that, on
the small $8^4$ lattice, an enhancement was observed for the
lowest point (and attributed to the finite size) in contrast to the shift
downwards of the two points from the 32 direction of the non-symmetric
lattice. However, no such maximum was observed on any of the smaller
lattices.  The present DSE results do not confirm the existence of such a
maximum in the ghost propagator but coincide nicely with all those data
points of the differently sized lattices that lie on a universal curve. In
addition, note that the $24^4$ and $16^3 \times 32$ lattices are of roughly
the same size as the asymmetric $16^3 \times 48$ and $24^3 \times 48$
lattices used for the gluon propagator in Ref. \cite{Mar95a}. Considering
their estimate of finite size effects based on deviations between different
components of the gluon propagator at small momenta, one might be led to
question the significance of the maximum in the ghost propagator observed for
the one particular data set of Ref.~\cite{Sum96} at momenta too low to yield
finite size independent results for the gluon propagator, {\it c.f.},
Ref.~\cite{Mar95a}, on even larger lattices.

It is quite amazing to observe that the numerical DSE solution fits the lattice
data at low momenta ($x \le 1$) significantly better than the fit to an
infrared singular form with integer exponents, $D_G(k^2) = c/k^2  + d/k^4$,
as given in Ref.~\cite{Sum96}. Clearly, low momenta ($x<2$) were not included
in this fit, but the authors conclude that no reasonable fit of such a form
is possible if the lower momentum data is to be included. Therefore, apart
from the question about a possible maximum at the very lowest momenta, the
lattice calculation seems to confirm the existence of an infrared
enhanced ghost propagator of the form $D_G \sim 1/(k^2)^{1+\kappa}$ with
a non-integer exponent $0 < \kappa < 1$. The same qualitative conclusion has
in fact been obtained in a more recent lattice calculation of the ghost
propagator in $SU(2)$~\cite{Cuc97}, where its infrared dominant part was
fitted best by $D_G \sim 1/(k^2)^{1+\kappa}$ for an exponent $\kappa $ of
roughly $ 0.3$ (for $\beta = 2.7$). This also is in qualitative agreement
with the $SU(2)$ calculations of Ref.~\cite{Sum96}, again, with the exception
of one single data point for the smallest lattice momentum possible.

Furthermore, in Refs.~\cite{Sum96,Cuc97} the Landau gauge condition was
supplemented by algorithms to select gauge field configurations from the
fundamental modular region which is done to eliminate systematic errors that
might occur due to the presence of Gribov copies.\footnote{An investigation
of the effectiveness of the stochastic overrelaxation method to achieve this is
given in Ref.~\protect\cite{Cuc97}} Thus, the good agreement of the DSE result
with these lattice calculations suggests that the existence of such copies of
gauge configurations might have little effect on the solutions to Landau
gauge Dyson--Schwinger equations. This would also help to understand the
similarity of these solutions to the qualitative behaviour obtained by
Zwanziger for gluon and ghost propagators from implications of complete gauge
fixings \cite{Zwa92,Zwa94}.

\begin{figure}[t]
\hskip +1.5cm
\parbox{6cm}{
  \epsfig{file=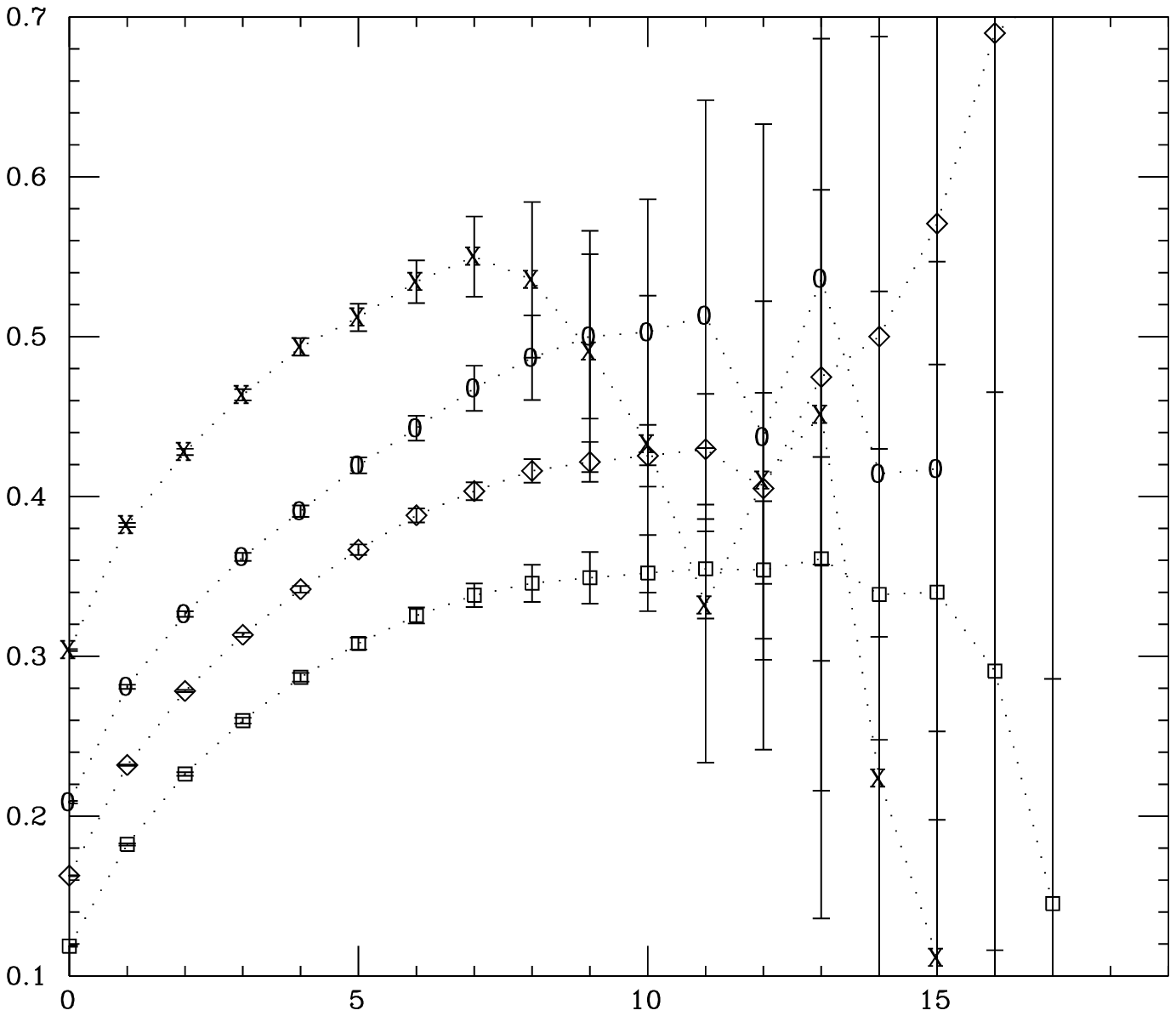,width=8cm}}
\hskip +.6cm
\parbox{6cm}{
  \epsfig{file=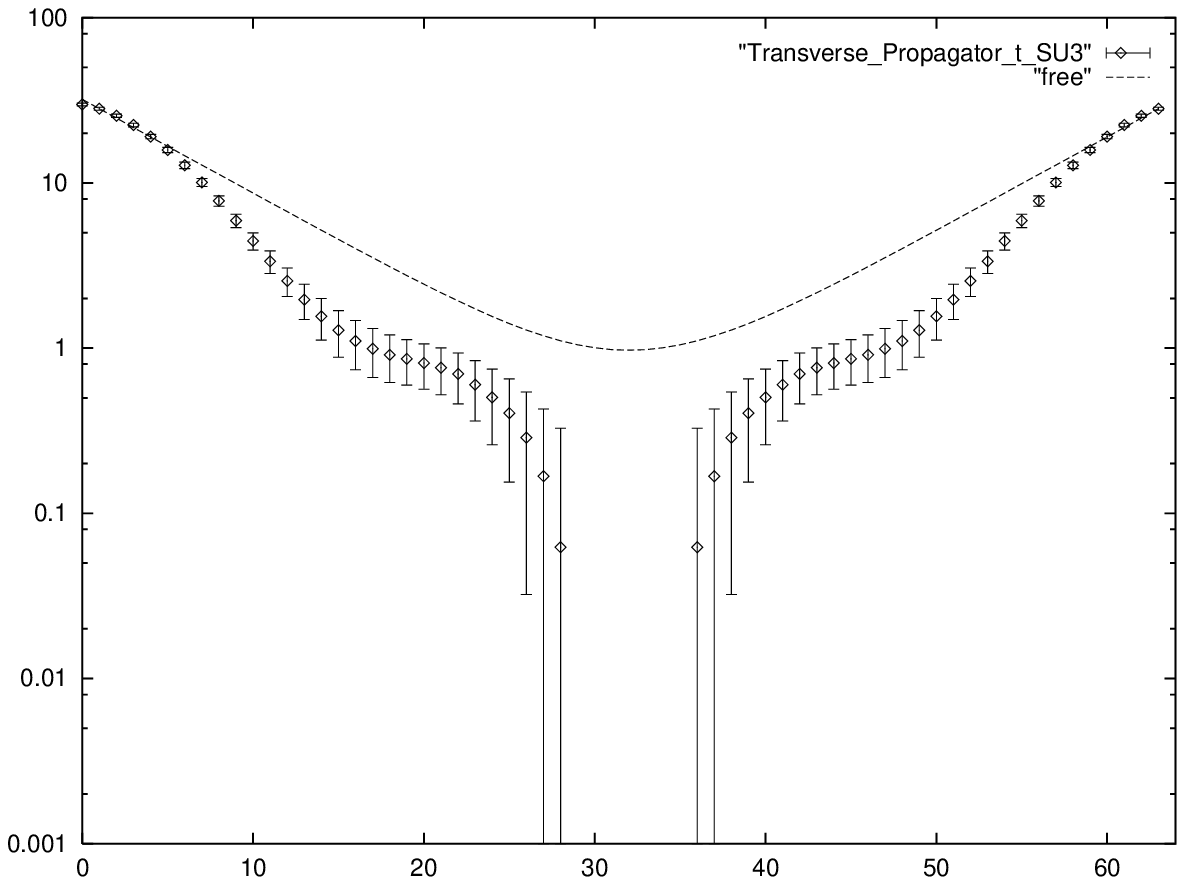,width=8.4cm}}
\vskip -.6cm
  \caption[Lattice results of the effective energy $\omega_{\mbox{\small
  eff}}(t,\mbox{\bf p})$.]
         {Lattice results of the effective energy $\omega_{\mbox{\small
  eff}}(t,\mbox{\bf p})$ at $(a\mbox{\bf p}) = 0, (2\pi/24,0,0)$,
  $(2\pi/24,2\pi/24,0)$, $(4\pi/24,0,0)$ on a $24^3\times 48$ lattice at $\beta
  = 6.0$  (left, Fig.~1 of Ref.~\cite{Mar95a}), and $D(t,\mbox{\bf p})$ at
  $(a\mbox{\bf p}) = (2\pi/48,0,0)$ for $\beta = 6.8$ on a $48^3 
  \times 64$ lattice (right, Fig.~2 of Ref.~\cite{Nak95}).}   
  \label{lat_glp_ft}
\end{figure}

Having seen that lattice data provide evidence for an infrared enhanced ghost
propagator and an infrared suppressed gluon propagator the question arises
whether lattice calculations support also the idea that the gluon propagator
does violate reflection positivity.
Even though no negative $D(t, \hbox{\bf p}^2)$ have been reported in the
lattice calculations yet, the available results,
Refs.~\cite{Ber94,Mar95a,Nak95}, agree in indicating that the gluon
propagator is not a convex function of the Euclidean time and thus that
positivity is indeed violated for gluonic correlations. In the left graph of
Fig.~\ref{lat_glp_ft}, the result of Ref.~\cite{Mar95a} for the effective
energy, 
\begin{equation} 
\omega_{\mbox{\small eff}}(t,\mbox{\bf p}) \, = \, \langle \omega
\rangle_{t,\mbox{\tiny\bf p$^2$}} \,:= \, - \, \frac{d}{dt} \ln 
D(t,\hbox{\bf p}^2)  \; ,  
\end{equation} 
is plotted which is clearly not the monotonically decreasing function of $t$
it would be for a positive gluonic spectral function. Similarly, the right
graph of Fig.~\ref{lat_glp_ft} shows the result of Refs.~\cite{Nak95,Ais97}
for $D(t,\hbox{\bf p}^2)$ together with the free propagator $\propto
\exp(-\omega t) $ which for {\bf p}$= (2\pi/(48 a),0,0)$ and with periodic
boundary conditions has the from
\begin{equation} 
D(t) \, \propto  \, \cosh(2\pi(t - N_t/2)/48) \; . 
\end{equation} 
The logarithmic plot again indicates quite clearly that this result is not a
convex function of $t$. Note that for a scale given by roughly $a^{-1} \simeq
6$GeV at $\beta = 6.8 $ the lowest momentum $2\pi/(48a)$ corresponds to
about $780$MeV. Using for the scale $\sigma \simeq (300\mbox{MeV})^2$ as 
in previous sections, one can estimate that the lattice result therefore
compares to a spatial momentum {\bf p}$^2/\sigma \gg 1$ for which the Fourier
transform of the DSE solution, Eq.~(\ref{eq:gluon_FT}), is found to no longer
be negative as a function of the Euclidean times $t$ either. The
qualitative behaviour as shown in Fig.~\ref{gluon_ft} is obtained for spatial
momenta {\bf p}$^2/\sigma < 1$.  The results of the lattice simulation are
thus not incompatible with the DSE solution. A closer comparison might
therefore be interesting. Good fits to the gluon data of Ref.~\cite{Nak95}
have been obtained in Ref.~\cite{Ais97} from parameterisations based on the
two lowest order rational approximants of the scheme described in
Sec.~\ref{sec_Stingl}. 

As noted already in Sec. \ref{Sec.2.4.2} the Kugo--Ojima Green's function has
been calculated on the lattice \cite{Nak00,Nak00a}. It is interesting to note 
that in these calculations also an infrared suppressed gluon and an infrared
enhanced ghost propagator have been found.

\begin{figure}
\centering{\ \epsfig{figure=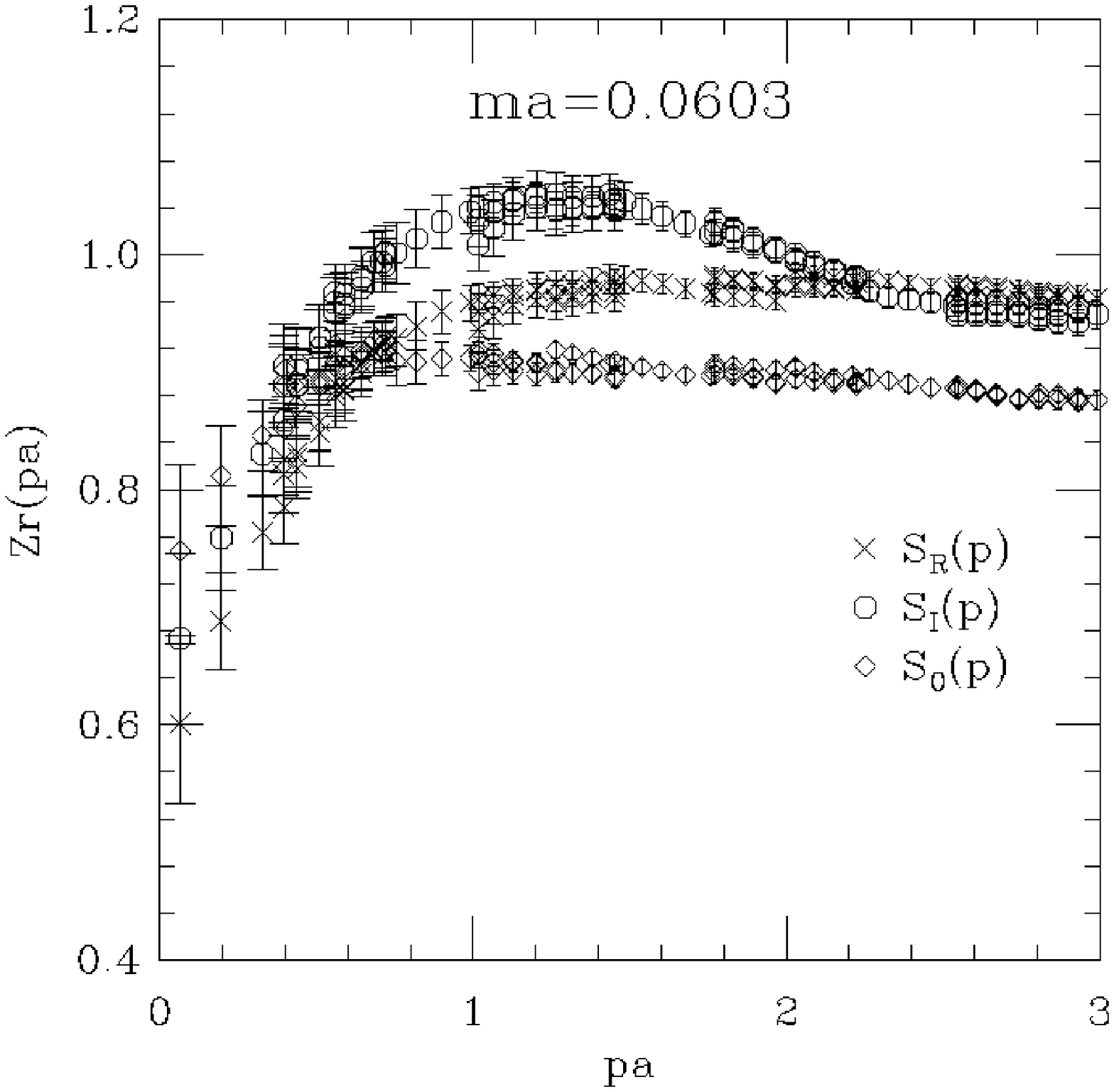,height=6.5cm} 
             \epsfig{figure=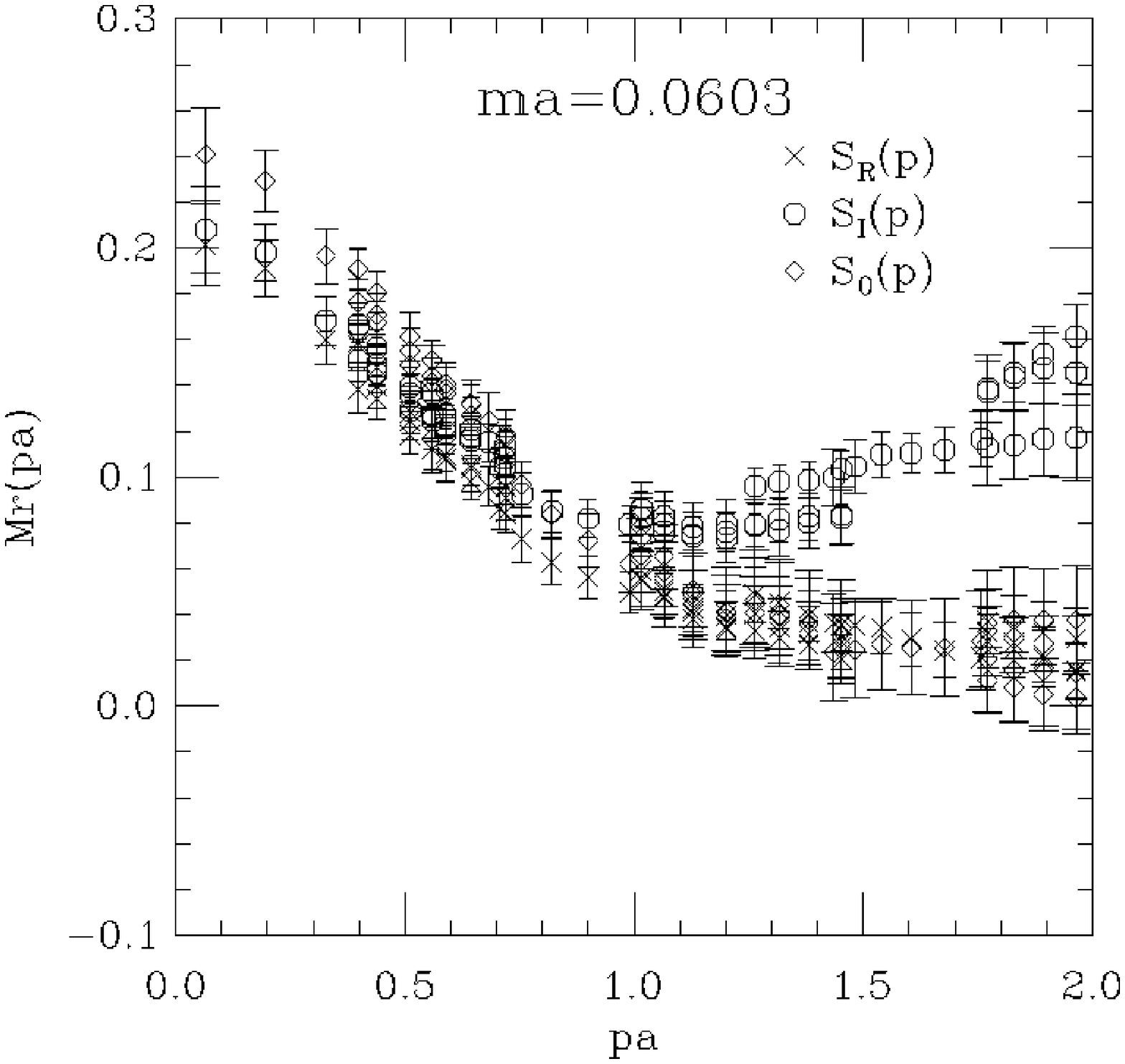,height=6.5cm} }

\caption{\label{latQ} {\it Left panel}: Lattice result for $Z_q(pa)$,
where $a\simeq 2.0\,\mbox{GeV}^{-1}$ is the lattice spacing, calculated with
$ma=0.0603$.
{\it Right panel}: Analogous plot for the mass function.  
(Adapted from Ref.\  \cite{Sku00a}.)}
\end{figure}

Lattice calculations of the infrared behaviour of the quark propagator are 
still in a very preliminary stage \cite{Sku00,Cud00,Bec00,Sku00a}. 
This first data provide evidence
that the quark renormalisation function $Z_q(p^2)$ is infrared suppressed, {\it
i.e.} the vector self-energy function $A(p^2)$ is infrared enhanced, and 
the mass function $M(p^2) = B(p^2)/A(p^2)$ approaches a value of $300\pm 30$MeV
at $p^2=0$. The mass function
$M(p^2)$ is a monotonically decreasing function of $p^2$ and the
only surprise, if any, is given by the fact that $M(p^2)$ decreases more slowly
than expected.  The results of the most recent calculations are shown in 
Fig.\ \ref{latQ}. 

\begin{figure}[t]
\vskip 1.2cm
\hskip -.8cm
\parbox{6cm}{
  \epsfig{file=Bou98_fig1.ps,width=8.8cm}}
\hskip 3.4cm
\parbox{6cm}{
  \epsfig{file=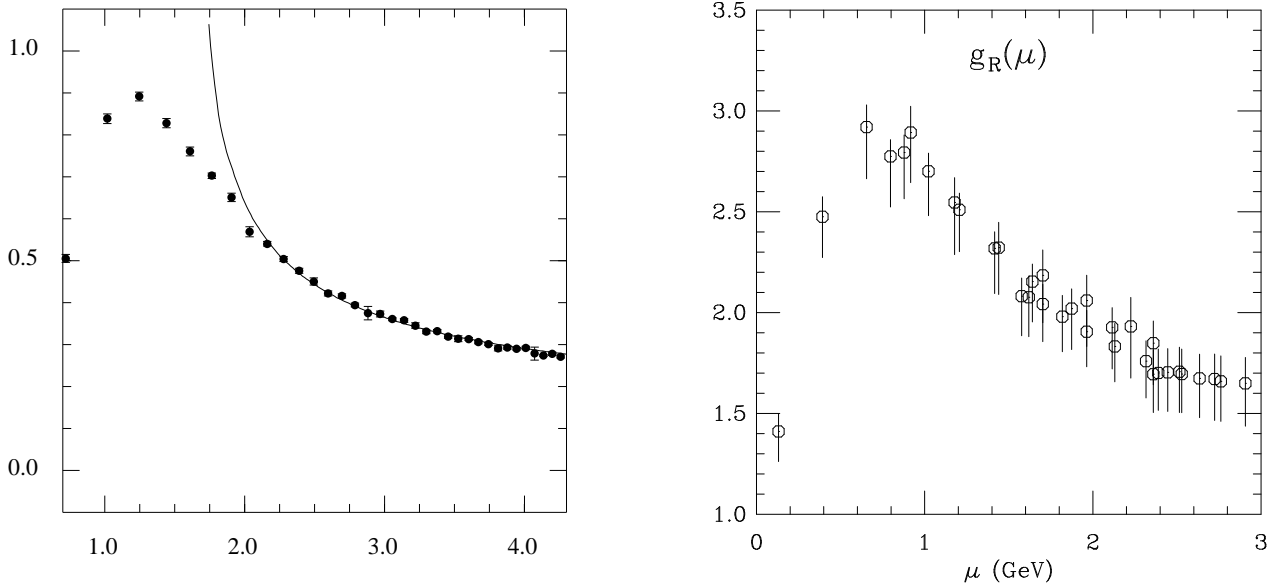,width=8cm}}
\vskip -.6cm
  \caption[Lattice results of the running coupling from the 3-gluon vertex.]
          {Lattice results of the running coupling from the 3-gluon vertex
  (left, together with a 3-loop fit, Fig.~1 of Ref.~\cite{Bou98}), and from
  the quark-gluon vertex for $\beta = 6.0$ on a $16^3 \times 48$ lattice
  (right, Fig.~2 of Ref.~\cite{Sku98}).}
  \label{lat_alpha}
\end{figure}

Recent lattice calculations of the running coupling are reported in
Refs.~\cite {All97,Bou98} based on the 3-gluon vertex, and Ref.~\cite{Sku98}
on the quark-gluon vertex. The non-perturbative definitions of these
couplings are related but manifestly different from the one adopted in Sec.\
\ref{sub_Sub}.
The most recent result from the 3-gluon vertex shown in the left graph of
Fig.~\ref{lat_alpha}, is obtained from an asymmetric momentum
subtraction scheme according to non-perturbatively requiring that the
transverse part of the 3-gluon vertex at the renormalisation point be given
by,
\begin{eqnarray}
&&\Gamma_{\mu\nu\rho}^T(p,-p,0) \, :=\,
{\mathcal P}_{\mu\kappa}(p) \, \Gamma_{\kappa\sigma\rho}(p,-p,0)  \, {\mathcal
 P}_{\sigma\nu}(p) \\
&&\Gamma_{\mu\nu\rho}^T(p,-p,0) \big|_{p^2 = \mu^2}  \, = \, \frac{1}{Z^{3/2}
(\mu^2)} \,\, 2 ip_\rho \,  \, {\mathcal P}_{\mu\nu}(p) \, .
\end{eqnarray}
Comparing to the according $k\to 0$ limit of the vertex as given in
Sec.~\ref{sub_Ghost}, one finds that this corresponds to a definition of the
running coupling $\bar g^2_{3GVas}$ which can explicitly be related to the
 one of Sec.~\ref{sub_Sub} ($\bar g^2(t,g)$ with $t = \ln \mu'/\mu$ and 
 $g := g(\mu)$),
\begin{equation}
\bar g^2(t,g^2)_{3GVas} \, = \, \bar g^2(t,g^2) \, \lim_{s \to 0} \,
\frac{G^2(s)}{G^2({\mu'}^2)} \, \left( 1 \, - \, \frac{\beta(\bar
g(t,g))}{\bar g(t,g)}
\, \right)^2 \; , \label{comp_rc}
\end{equation}
where according to Eq.~(\ref{gbar'}) in Sec.~\ref{sub_Sub} it was used that,
\begin{equation}
 \bar g^2(t,g^2) \, = \, g^2G^2({\mu'}^2) Z({\mu'}^2) \;,
 \quad
{\mu'}^2 \frac{d}{d{\mu'}^2} \ \left( \frac{2G'({\mu'}^2)}{G({\mu'}^2)}
 +  \frac{Z'({\mu'}^2)}{Z({\mu'}^2)} \right) \, = \, \frac{d}{dt} \ln \bar
 g(t,g^2) \; .
\end{equation}
An inessential difference in these two definitions of the running coupling
is the last factor in brackets in Eq.~(\ref{comp_rc}) which is what arises
typically by choosing the 3-gluon vertex instead of the ghost-gluon vertex as
is done here. This could of course be accounted for in comparing the results
of the different schemes.\footnote{Differences in these definitions should
actually occur at order $g^6$, {\it i.e.}, at 3-loop level. Here they seem
to occur at two-loop level. This is likely to be an artifact of the
truncation scheme in the DSEs.} However, the crucial difference is the ratio
of ghost renormalisation functions $G(s\to 0)/G({\mu'}^2)$. Of course, the
lowest lattice momentum used in the calculation is not strictly zero but
corresponds to some $s_{min}$. However, the considerations above show that
the asymmetric scheme can be extremely dangerous if infrared divergences
occur in vertex functions as the coupled system of DSEs indicates due to the
infrared enhancement of the ghost propagator in Landau gauge. In particular,
a large numerical factor arising from a large $G(s_{min})$ should thereby
result in a quite different scale $\Lambda_{\widetilde MOM}$ for the
asymmetric scheme. The additional $\mu'$-dependence from the ratio
$G(s_{min})/G({\mu'}^2)$ is harder to understand.\footnote{Note that the
perturbative limit, strictly speaking being {\sl all} momenta spacelike and
large, is never really reached if one momentum is put to zero as in the
asymmetric schemes. Therefore, one cannot expect to recover the perturbative
anomalous dimension for the vertex in the asymmetric limit $\mu \to \infty$.}
Clearly, from the infrared enhanced ghost renormalisation function this scale
dependence could account for the infrared suppressed couplings which seem to
be found in the asymmetric schemes.

Similarly, the results from the quenched calculation of the quark-gluon
vertex of Ref.~\cite{Sku98} given in the right graph of Fig.~\ref{lat_alpha}
are obtained from an analogous asymmetric scheme (with the gluon momentum set
to zero). It is thus expected to have the same problems in taking
the possible infrared divergences of the vertices into account which arise
in both, the 3-gluon and the quark-gluon vertex, as a result of the
infrared enhancement of the ghost propagator, {\it c.f.} Sec.\
\ref{sub_Ghost}.

Furthermore, definitions of the
coupling which lead to extremas at finite values of the scale correspond to
double valued $\beta$-functions with artificial zeros. If the maxima in the
couplings of the asymmetric schemes at finite scales $\mu  = \mu_0$, as shown
in Fig.~\ref{lat_alpha}, are no lattice artifacts, these results seem to
imply that the asymmetric schemes are less suited for a non-perturbative
extension of the renormalisation group to all scales $\mu \ge 0 $. Indeed,
the results for the running coupling from the 3-gluon vertex obtained for
the symmetric momentum subtraction scheme in Ref.~\cite{Bou98} differ from
those of the asymmetric scheme, in particular, in the infrared. The
suppression of the coupling seems to be much weaker if not absent in the
symmetric scheme. These results would be better to compare to the DSE
solution, however, they unfortunately seem to be much noisier thus far (see
Ref.~\cite{Bou98}).

The ultimate lattice calculation to compare to the present DSE coupling would
be obtained from a pure QCD calculation of the ghost-gluon vertex in Landau
gauge with a symmetric momentum subtraction scheme. This is unfortunately not
available yet.

With regard to a later chapter, ({\it c.f.}, Sec.\ \ref{sec_Diq}),  it is
interesting to note that there exist a lattice calculation of  diquark masses
\cite{Hes98}. Hereby the nucleon, delta, quark and diquark  correlation
functions in Landau gauge are analysed in order to extract  information on the
spin dependence of the quark-quark interaction. Evidence was found that the
nucleon-delta mass splitting can be attributed to the spin dependence of the
interaction between quarks in a colour anti-triplet state with spin 0 and 1,
respectively. The lightest diquark excitations are observed in the S=0 channel,
$m_{0^+}\approx 620$ MeV in the chiral limit. The mass of the spin 1 diquark as
been estimated to be 730 MeV. In view of a quark constituent mass of
approximately 310 MeV as found in the same calculation,
however, no evidence for a deeply bound diquark state has been found.

\section{Mesons as Quark-Antiquark Bound States}
\label{chap_Meson} 

Mesons are bound states of quarks and antiquarks. Hereby the pions as
(would-be) Goldstone bosons of the dynamically broken chiral symmetry play a
special role.  
Resolving this dichotomous nature of the pions provides a strong constraint
on an unified description of mesons. In the chiral limit quarks acquire a
dynamical (constituent) mass. Then a method is required which respects chiral
symmetry such that the pions become massless bound states. Phrased otherwise,
the binding energy between quarks and antiquarks in the pion channel has to
match exactly two times the constituent quark mass. Therefore, only a fully
Lorentz covariant scheme can yield a satisfactory description of pions. We
will see in this chapter that this is indeed possible.

\subsection{Bethe--Salpeter Equation for Mesons}
\label{sec_BSE} 

Applications of the QCD propagators in Bethe--Salpeter (BS) equations
\cite{Sal51} have
progressed during the last few years considerably. One caveat has to be
mentioned, however. Almost all of these investigations rely on the
rainbow-ladder approximation of the systems of DSEs and BS equations.
One important exception will be discussed in detail in Sec.\ \ref{sec_Diq}
where it is exemplified how the rainbow-ladder scheme provides for an
approximation that can far more reliably be used in pseudoscalar and vector
meson calculations than can in diquark calculations. Nevertheless, including a
non-trivial quark-gluon vertex function in a way that preserves
constraints from both, gauge invariance {\em and} chiral symmetry
is highly desirable but nobody succeeded yet in such an undertaking. 
In Ref.\ \cite{Mun95} a general
scheme for a consistent treatment of DSEs and BS equations has been formulated,
however, it has not been realized with a realistic quark-gluon vertex
constructed from the Slavnov--Taylor identities. Therefore, in the following
subsections we discuss the generalised ladder BS equation, {\it i.e.},
employing dressed propagators, for mesons.

\subsubsection{\label{sub_Der}Derivation of the Ladder Bethe--Salpeter
Equation}

The important step in the derivation of the homogeneous BS equation from the
inhomogeneous one is to realize that two-particle bound states can be
identified through the occurrence of poles in the corresponding four-point
Green's function. From the homogeneous BS equation which assumes the form  of
an eigenvalue problem in the ladder approximation one can determine the bound
state masses and covariant wave functions. Hereby the bound state mass has to 
be tuned such that the BS eigenvalue equals the given value of the coupling
constant. The covariant wave functions are then determined as the
eigenfunctions of the system.

Restricting ourselves to the rainbow-ladder approximation the
fermion-antifermion BS equations for QED and QCD are identical up to 
straightforwardly computable colour algebra  factors. Thus the derivation
presented in this section is equally valid for QED and QCD. Hereby the starting
point is the generating functional as given in Sec.\ \ref{Sec2.1}.  In Sec.\
\ref{sec_nPoint} we have already discussed an inhomogeneous BS equation.
The derivative of the effective action $\Gamma_{\mbox{\tiny QED}}$ with
respect to the 
gauge boson, the fermion and the antifermion fields defines the fermion-photon
vertex (\ref{photonfermionvertex}) whose DSE (\ref{fermion--photon}) is the
inhomogeneous BS equation:
$$
\Gamma_\mu (q,p) = Z_2 \gamma_\mu + \int  \frac{d^{\rm 4}l}{(2\pi)^{\rm 4}}
 S(q+l) \Gamma_\mu (q+l,p+l) S(p+l) K(p+l,q+l,l) \, . \qquad \qquad 
 (\protect{\ref{fermion--photon}})
$$
The kernel $K$ is defined
as the sum of all (amputated) contributions  which are two-particle
irreducible in the $s$-channel. Its perturbative expansion is
 pictorially represented in Fig. \ref{K1}. As
described in Sec.\ \ref{sec_nPoint} the transverse part of the quark-photon  
vertex
has a pole at momenta corresponding to the vector meson mass. Of course, one
can derive  inhomogeneous BS equations for the scalar, pseudoscalar,
axialvector and tensor channel of the quark-antiquark bound state. These
vertex functions will have poles related to the respective meson masses.
This property can be exploited to derive the homogeneous BS equation.

In order to keep the following discussion as transparent as possible we will
demonstrate the underlying principle using scalar fields. We assume to deal 
with three types of scalar fields. The two constituents are
supposed to have masses $m_1$ and $m_2$ and self-energies $\Sigma_1$ and
$\Sigma_2$. The four-point function $G^{\left( 4\right) }\left(
x_{1},x_{2},y_{1},y_{2}\right)$ describing the scattering of these two
constituents fulfils the inhomogeneous Bethe--Salpeter equation
\begin{equation}
\left( (\partial_\mu\partial^\mu) _{x_{2}}+m_{2}^{2}-\Sigma _{2}\right) 
\left( (\partial_\mu\partial^\mu) _{x_{1}}+m_{1}^{2}-\Sigma _{1}\right) 
G^{\left( 4\right) }\left(
x_{1},x_{2},y_{1},y_{2}\right) =\quad \quad \quad \quad \quad \quad
\label{X1.16}
\end{equation}
\vspace{-0.8cm}
\begin{eqnarray*}
&&\quad \quad \quad \quad \delta ^{\left( 4\right) }\left( x_{1}-y_{1}\right)
\delta ^{\left( 4\right) }\left( x_{2}-y_{2}\right) +\delta ^{\left( 4\right)
}\left( x_{1}-y_{2}\right) \delta ^{\left( 4\right) }\left( y_{1}-x_{2}\right)
\\ &&\quad \quad \quad \quad \quad \quad \quad
\quad \quad \quad \quad +\int d^{4}z_{1}d^{4}z_{2}K\left(
x_{1},y_{1},z_{1},z_{2}\right) G^{\left( 4\right) }\left(
z_{1},z_{2},y_{1},y_{2}\right) .
\end{eqnarray*}
Note that the introduction of the relative coordinate $x=x_1-x_2$ allows an
arbitrary parameter $\eta_P \in [0,1]$ 
in defining the coordinate $X=\eta_P  x_1 + (1-\eta_P  )x_2$ which, after
transforming to momentum space, results in the corresponding momenta
\begin{equation}
p_1=\eta_P  P +p \quad {\rm and} \quad p_2= (1-\eta_P  )P-p
\label{p12}
\end{equation}
for the constituents expressed in terms of the total and relative momenta, $P$
and $p$, respectively. Fourier transforming Eq.\ (\ref{X1.16}) leads to
\begin{equation}
\int \frac{d^{4}p^{\prime }}{\left( 2\pi \right) ^{4}}\left[ {D}\left(
p,p^{\prime },P\right) +{K}\left( p,p^{\prime },P\right) \right] {G}^{\left(
4\right) }\left( p^{\prime },p^{\prime \prime },P\right)
=\delta ^{\left( 4\right) }\left( p-p^{\prime \prime }\right) ,
\label{inhBSE}
\end{equation}
where
\begin{equation}
{D}\left( p,p^{\prime },P\right) :=\left( 2\pi \right) ^{4}\delta
^{\left( 4\right) }\left( p-p^{\prime }\right) \left( {G}_{1}^{\left(
2\right) }\right) ^{-1}\left( p_{1}\right) \left( {G}_{2}^{\left( 2\right)
}\right) ^{-1}\left( -p_{2}\right) \label{X1.21}
\end{equation}
is defined in terms of the inverse two-point Green's functions of the
constituents.

\begin{figure}
\begin{center}
\begin{minipage}{20mm}
  \epsfig{file=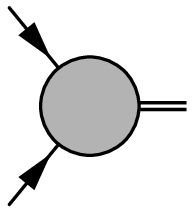,width=20mm}
\end{minipage}
\hspace{5mm}
=
\begin{minipage}{35mm}
  \epsfig{file=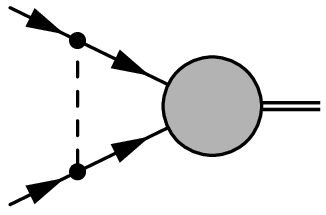,width=35mm}
\end{minipage}
\end{center}
\caption{Pictorial representation of the homogeneous Bethe--Salpeter equation
in the ladder approximation.
\label{BSE1}}
\end{figure}

The crucial step from the inhomogeneous to the homogeneous Bethe--Salpeter
equation consists in assuming the existence of a bound state of mass $M$ 
which is represented by a pole contribution in the total momentum of the
four-point Green's function at on-shell momenta  $P_{os}$, with
$P_{os}^{2} = - M^2$. For positive energies $P^0 > 0$ this contribution is of
the form, 
\begin{equation}
{G}^{\left( 4\right) }\left( p,p^{\prime },P_{os}\right)
=\frac{-i}{
\left( 2\pi \right) ^{4}}\frac{{\chi }\left( p,P_{os}\right)
{\bar{\chi}}\left( p^{\prime },P_{os}\right) }{2\omega \left( P^{0}-\omega
+i\epsilon \right) }+\rm{reg. \,terms},\quad
\omega := \sqrt{ {\bf P}^{2} + M^2 }\, .
\label{4point-pole}
\end{equation}
Here we introduced the definition of the Bethe--Salpeter amplitudes
\begin{eqnarray}
\chi \left( x_{1},x_{2},P\right) &:=& \left\langle 0\left| T\left\{ \Phi
\left( x_{1}\right) \Phi \left( x_{2}\right) \right\} \right| P\right\rangle
\; , \nonumber \\
\bar{\chi}\left( x_{1},x_{2},P\right) &:=& \left\langle 0\left| T \left\{ \Phi
^{\dagger }\left( x_{1}\right) \Phi ^{\dagger }\left( x_{2}\right) \right\}
\right| P\right\rangle \; , \label{X1.30}
\end{eqnarray}
together with their Fourier transforms
\begin{eqnarray}
\chi \left( x_{1},x_{2},P\right) &=:&e^{-iPX}\int \frac{d^{4}p}{\left( 2\pi
\right) ^{4}}e^{-ipx}{\chi }\left( p,P\right) \; ,  \nonumber \\
\bar{\chi}\left( x_{1},x_{2},P\right) &=:&e^{+iPX}\int \frac{d^{4}p}{\left(
2\pi \right) ^{4}}e^{-ipx}{\bar{\chi}}\left( p,P\right).
\label{X1.31}
\end{eqnarray}
Hereby $\left| 0\right\rangle$ denotes the ground state (vacuum) and $\left|
P\right\rangle$ the bound state. Close to the pole the regular terms can 
safely be neglected and the dependence of the four-point function on the
relative momenta $p$ and $p^\prime$ can be separated. 
Expanding ${G}^{\left(4\right) }$
and $\left[ {D}-{K}\right] $ in the inhomogeneous Bethe--Salpeter equation in
powers of $\left( P^{0}-\omega \right) $ yields the homogeneous Bethe--Salpeter
equation and the normalisation condition for the amplitude. The order $\left(
P^{0}-\omega \right) ^{-1}$ provides
\begin{equation}
\int \frac{d^{4}p^{\prime }}{\left( 2\pi \right) ^{4}}\left[ {D}\left(
p,p^{\prime },P_{os}\right) +{K}\left( p,p^{\prime },P_{os}\right)
\right] {\chi }\left( p,P_{os}\right) =0,  \label{X1.35}
\end{equation}
whereas to ${\mathcal O}\left( \left(P^{0}-\omega \right)
^{0}\right) $ one obtains
\begin{eqnarray}
\int \frac{d^{4}p}{\left( 2\pi \right) ^{4}}\frac{d^{4}p^{\prime }}{\left(
2\pi \right) ^{4}}{\rm tr} \left({\bar{\chi}}\left( p,P_{os}\right)
\left. \frac{\partial }{\partial P^{0}}\left( D\left( p,p^{\prime },P\right)
+ K\left( p,p^{\prime },P\right) \right) \right| _{P^{0}=\omega }
\!\!\! {
\chi }\left( p^{\prime },P_{os}\right) \right) \nonumber \\
 = 2i\omega .
\label{X1.39}
\end{eqnarray}
This ensures the residue to be equal to $1$ at the bound state pole.

The homogeneous Bethe--Salpeter equation (\ref{X1.35}) is a linear integral
equation for the amplitude ${\chi}$ whose overall normalisation is fixed by
(\ref{X1.39}). Approximating the kernel by the one-boson-exchange depicted in
the first diagram of Fig.\ \ref{K1}, Eq.\ (\ref{X1.35}) can be cast into an
eigenvalue problem for the coupling constant by using the vertex function
$\Gamma \left( p_1,p_2\right)$ instead of the amplitude:
\begin{equation}
{\chi}\left( p,P\right) =:G_{1}\left( p_{1}\right) G_{2}\left( p_{2}\right)
\Gamma \left( p_1,p_2\right) \; . \label{X1.37}
\end{equation}
In the context of BS equations it is advantageous
to use the total and relative momenta, see Eq.\ (\ref{p12}), as arguments of the
vertex functions, $\Gamma \left( p_1,p_2\right) \to \Gamma \left(p;P\right)$.
The homogeneous BS equation (\ref{X1.35}) in terms of the vertex function
then reads
\begin{equation}
\Gamma \left( p;P_{os}\right) = - \int \frac{d^{4}p^{\prime }}{\left( 2\pi
\right) ^{4}}K\left( p,p^{\prime },P_{os}\right) G_{1}\left( p_{1}^{\prime
}\right) G_{2}\left( p_{2}^{\prime }\right) \Gamma \left( p^{\prime
};P_{os}\right) .
\label{BSEvertex}
\end{equation}

In the ladder approximation the kernel $K$ is essentially given by 
the propagator of the exchanged particle  multiplied by one coupling constant
$g$ for each vertex, {\it i.e.}, by $g^{2}$. On inspection one finds that
(\ref{BSEvertex}) 
is an eigenvalue problem for $g^{2}$, if the $G_{1}$ and $G_{2}$ are the bare
propagators of the constituents. The ladder approximation to the BS
equation is pictorially represented in Fig.~\ref{BSE1}. If a parameter pair
$(g^2, P^0=M)$ exists the pole assumption is {\it a posteriori} justified and
$M$ is the bound state mass with $\chi$ being the corresponding amplitude (wave
function) as can be inferred from Eq.\ (\ref{4point-pole}) which reflects, of
course, nothing else than the Lehmann representation of the four-point
function.

The description of mesons, and especially the one of pions, requires to use a
generalised ladder approximation:  The intrinsically non-perturbative nature
of bound-state problems, and the complex structure of the QCD vacuum,
necessitates that one employs non-perturbative  gluon and quark propagators in
the BS equation kernel. There have been many studies of meson spectroscopy
using this 
framework; summaries can be found in Refs.~\cite{Tan97,Rob94,Mir93}. 
Typically, such studies employ an Ansatz for a dressed gluon propagator in 
solving a rainbow-approximate quark DSE, and then pair the input gluon
propagator with the calculated quark propagator to construct the
non-perturbatively dressed kernel for the meson BS equation in ladder
approximation. The resulting BS equation is then solved to obtain the
spectrum.  This rainbow-ladder truncation scheme has the feature that
Goldstone's theorem is manifest; {\it i.e.}, in the chiral limit, when the
current quark mass $m_q=0$, the pion is a zero-mass bound state in strongly
dressed quark-antiquark correlations~\cite{Del79}.  
As we will see in the following, with few-parameter models for the gluon
propagator, this can be used to provide fair descriptions of the light-light,
light-heavy and heavy-heavy meson spectra and decays.

In order to simplify notations we introduce ``multiple indices''
$E=\{i_c,i_f,i_D\}$ associated with the colour, the flavour and the 
Dirac structure of an amplitude. The homogenous BS equation for the vertex
function $\Gamma_M$, the subscript $M$ denoting meson, then reads: 
\begin{eqnarray} \label{bsemeson}
\lefteqn{\Gamma^{EF}_M(p;P) = }\\ & & \nonumber \int\,\frac{d^4k}{(2\pi)^4}
\,K_M^{EF;GH}(k,p;P)\,
\left(S(k+\frac{1}{2}P)\Gamma_M(k;P)S(k-\frac{1}{2}P)\right)^{GH}~,
\end{eqnarray} 
where we have symmetrised the momenta of the quark legs for reasons which will
become clear soon. The rainbow approximation is obtained from inserting 
a bare quark gluon vertex ($ \Gamma_\mu(k,p) \equiv \gamma_\mu $) 
into the quark DSE. Corresponding to this, the generalised ladder
approximation consists of employing
\begin{eqnarray}
\lefteqn{K_M^{EF;GH}(k,p;P)\,
\left(S(k+\frac{1}{2}P)\Gamma_M(k;P)S(k-\frac{1}{2}P)\right)^{GH}}\\
&& \nonumber
\equiv -g^2\,D_{\mu\nu}(p-k)\,
\left(\gamma_\mu\,\frac{\lambda^a}{2}\,
S(k+\frac{1}{2}P)\Gamma_M(k;P)S(k-\frac{1}{2}P)
\gamma_\nu\,\frac{\lambda^a}{2}\,\right)^{EF}
\end{eqnarray}
for the kernel in Eq.~(\ref{bsemeson}). This form of the kernel, the
dressed-ladder gluon exchange combined with the solution to the  
rainbow quark DSE, preserves the Goldstone boson character of the pion.
One observes that, in the chiral limit $m_q=0$, the meson 
BS equation~(\ref{bsemeson}) is obtained from the
rainbow DSE for the quark self-energy 
(defined by $S^{-1}(p) =:  i \gamma \cdot{p}  +  \Sigma(p)$),
\begin{eqnarray}
\label{rainbowDSE}
\Sigma(p) &= & m_q + g^2 \frac{4}{3} \int
     \frac{d^4k}{(2\pi)^4}\, \gamma_\mu \, S(k)\, \gamma_\nu  \,D_{\mu
     \nu}(p-k),
\end{eqnarray}
upon replacing 
\begin{equation}
\label{DSEtoBSE}
\gamma_\mu\,S(k)\,\gamma_\nu \to
\gamma_\mu\,S(k+P/2)\,\Gamma_M(k,P)\,S(k-P/2)\,\gamma_\nu~.
\end{equation}
Straightforward algebraic manipulations then reveal that 
the BS equation in the pseudoscalar channel for $P=0$ is identical to
the equation for the scalar quark self-energy, {\it i.e.}, to its  
chirally non-invariant dynamical contribution~\cite{Del79}.\footnote{An
extension of this constructive way to preserve the Goldstone boson character
of the pion will be presented in Sec. \ref{sub_Beyond}.} 
The derivation of this can equivalently be based on the observation 
that in the chiral limit the vertex function of a pion, with flavour index $a$ 
and vanishing momentum $P=0$, must be proportional to the result of an
infinitesimal chiral rotation of the quark self-energy,  
\begin{equation}
\label{chiRot}
\Gamma^a (p,P=0)  \propto  -i \frac d{d\alpha^a} 
\left( e^{i \alpha^b \frac 1 2 \tau^b \gamma_5} \Sigma (p)
e^{i \alpha^c \frac 1 2 \tau^c \gamma_5} \right)_{\alpha^a  =0} =
\frac 1 2 \tau^a \{ \gamma_5 , \Sigma (p) \} \, .
\end{equation}
In the chiral limit the dynamical breaking of chiral symmetry
is therefore always accompanied by a massless bound state in the
pion channel within the rainbow-ladder scheme. 

Before we are going to discuss the solutions of the ladder BS equations 
for the ground state mesons two comments on general properties are, however, in
order.

\subsubsection{Solutions of the Ladder
Bethe--Salpeter Equation in Minkowski Space}
\label{sub_Min} 

\begin{figure}
        \centerline{\epsfig{file=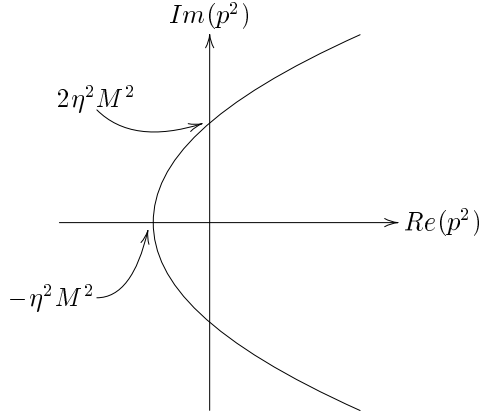} }
        \caption{Plot of the complex $p^2$ plane. The interior of the
          parabola shows the subset of the complex plane where the
          BS equation 'probes' the propagators of the
          constituents. $M$ is the mass of the bound state and $\eta$
          is the momentum partitioning parameter.}
        \label{BS_momregime}
  \end{figure}

An Euclidean formulation is used throughout this review which, as  described in
the beginning of Chapter \ref{chap_Basic}, has its justification in the fact
that the domain of holomorphy for the  Green's function of a quantum field
theory (fulfilling the usual axioms) allows a complex extension of the
Euclidean space. The on-shell momenta of bound states introduced in the last
section  are time-like and have to be represented as complex four-vectors in
an analytically continued Euclidean formulation. Being justified formally, at
least for quantum field theories without complications like confinement, there
are inherent practical difficulties, however. Looking at Eq.\ (\ref{bsemeson})
one realizes that the quark propagators have to be known in a parabolic region
of the complex $p^2$-plane. This region contains the positive real 
(half-)axis of space-like $p^2$, and it extends to $p^2=-M^2/4$ on the
negative real axis with its boundary intersecting the imaginary axis 
at $p^2=\pm iM^2/2$, see Fig.\ \ref{BS_momregime}. (This does, of 
course, depend on the momentum partitioning parameter $\eta_P$, {\it e.g.},
its extend into the time-like region is $p^2=-M^2\eta_P^2$ for one and
$p^2=-M^2(1-\eta_P)^2$ for the other constituent, such that
$\eta_P=1/2$ is the best choice to minimise this complex domain for both
quark propagators.) If the kernel of the quark DSE, {\it
i.e.}, in rainbow approximation the gluon propagator, is a known analytic
function in this parabolic region it is possible to solve the BS equation
without further approximation, see Sec.\  \ref{sec_GS} for a discussion of such
calculations. If the kernel of the quark DSE is known only
numerically at certain momenta, {\it e.g.}, on the space-like
$p^2$-axis, however, or if singularities occur in this parabolic region (note
that it always  contains the point $p^2=0$ for example), or if both is the
case, a reliable numerical evaluation of the integration kernel in the BS
equation is virtually impossible.

Thus a solution of the DSE directly in Minkowski space might provide some new
insight. Rewriting the integrals in the fermion DSE in rainbow approximation
with the help of dispersion relations it is possible to solve this equation
directly in Minkowski space \cite{Blu97}. However, for the extraction of the
imaginary part it has been a necessary prerequisite that the analytic structure
of the kernel is known explicitely. In Ref.\  \cite{Blu97} either a bare gauge
boson propagator or a bare gauge boson propagator multiplied with a
logarithmically running coupling has been used. Of course, the obtained results
are in accordance with the one obtained from the Euclidean DSE. Another
Minkowski space study of DSEs is employing the perturbation theory integral
representation (see below) in scalar $\phi^3$ theory \cite{Sau99}. This work
is, however, still unpublished.

In Refs.\ \cite{Kus95,Kus97a} the ladder BS equation for a scalar-scalar bound
state with scalar exchange has been solved in Minkowski space using the
perturbation theory integral representation \cite{Nak71}. This representation
is an extension of the spectral  representation for two-point Green's
functions. Hereby, for the relatively simple kernels tested so far, the results
for bare and dressed ladder kernels are in complete agreement with the results
obtained from an Euclidean approach and Wick rotation. The main advantage of
this method would, however, become vital when using non-ladder kernels. Note
that beyond ladder approximation the naive Wick rotation is not possible, {\it
i.e.}, one has to choose much more complicated integration contours. 
Furthermore, the method is quite general and it is also applicable if a Wick
rotation is not possible at all. 
Also in these cases it could provide the vertex function and the
amplitude for the entire range of allowed momenta. Therefore, progress in this
direction would be highly desirable.

\subsubsection{(In--)Consistency of the Relativistic
Description of Excited States}
\label{sub_Incon} 

At first sight the BS equation seems very suited to describe excited states.
Interpreting the spectrum of the homogeneous BS equation is, however, far from
being trivial \cite{Ahl99}. Common to the analytical\footnote{The only
analytically solvable example of a BS equation is the one for two (massive)
scalar particles bound by the ladder approximation to the exchange of a
massless scalar field \cite{Wic54,Cut54,Nak69}. Despite its relative simplicity
as compared to realistic systems this model, the Wick--Cutkosky model, displays
already the advantages ({\it e.g.}, full covariance) as well as the
shortcomings ({\it e.g.}, the existence of abnormal states) inherent to almost
all Bethe--Salpeter based approaches used until today.} and the numerical
solutions is the existence of abnormal states which have led to controversial
discussions regarding their physical interpretation \cite{Wic54,Cut54,Nak69}.
Even worse, for the case of constituents with unequal masses some eigenvalues
of the homogeneous Bethe--Salpeter equation become complex \cite{Kau69,Fuk93}.
Clearly, such a behaviour is unexpected and has to be understood. It has been
usually attributed to the use of the ladder approximation \cite{Wic54} which
destroys crossing symmetry from the very beginning. This conjecture in its
strict form is, however, refuted: Going beyond ladder approximation and
employing also crossed ladder exchanges  the abnormal states still exists
\cite{The99}. In this section we will provide evidence that the use of a
dressed ladder kernel is absolutely required if one  wants to interpret the
spectrum of the BS equation \cite{Ahl99}.

The abnormal solutions are ``excitations in relative time''. They will
obviously not appear in a purely non-relativistic treatment where the
constituents are considered for equal times only.\footnote{However, not all
abnormal solutions necessarily vanish in a three-dimensional reduction of the
BS equation \cite{Bij97}. On the contrary, the spectrum of a
three-dimensionally reduced equation will contain remnants of these abnormal
states.} In the Wick--Cutkosky model \cite{Wic54,Cut54} with constituents of
equal masses $m_1=m_2=m$ these abnormal states are easily identified: They only
exist for certain values of the coupling constant  ($\lambda :=g^{2}/16\pi ^{2}
m^2 >\lambda _{c}=1/4$). If the binding energy becomes very small the
corresponding coupling constant $\lambda$ vanishes for the normal solutions,
i.e.\ $\lambda \to 0$, whereas $\lambda \to \lambda _{c}=1/4$ for the abnormal
states. The latter behaviour is completely unexpected, a vanishing binding
energy should be related to no coupling at all. It can be shown that in this
model the abnormal solutions possess nodes when plotted as functions of the
relative time. For all normal solutions, including the ground state, there are
no nodes in relative time. It has to be noted, however, that for a general BS
equation there is no known method to identify abnormal states.

\begin{table}[tbp]
\begin{center}
\begin{tabular}{|c|c|c|}
\hline\hline
$~~~~~~~~~~~~~~M ,\mu ~~~~~~~~~~~~~~$ & symmetry group & BS
amplitude \\ \hline\hline $M =0,\mu =0$ & $O\left( 5\right) $ & $\chi \propto
Z_{nklm}\left(
\Omega _{5}\right) $ \\ \hline
$M =0,\mu \neq 0$ & $O\left( 4\right) $ & $\chi \propto Z_{klm}\left(
\Omega _{4}\right) $ \\ \hline
$M \neq 0,\mu =0$ & $O\left( 4\right) $ & $\chi \propto Z_{klm}\left(
\Omega _{4}\right) $ \\ \hline
$M \neq 0,\mu \neq 0$ & $O\left( 3\right) $ & $\chi \propto Y_{lm}\left(
\Omega _{3}\right) $ \\ \hline $\mu \rightarrow \infty $ & $O\left( 4\right) $ &
$\chi \propto Z_{klm}\left( \Omega
_{4}\right) $ \\ \hline
\end{tabular}
\end{center}
\caption[Summary of the symmetries of the scalar BS equation in
ladder approximation. (Adopted from Ref.\ \cite{Ahl99}.)]
{Summary of the symmetries of the scalar BS equation in
ladder approximation. $\mu$ denotes the mass of the exchange particle and $M$
is the Mass of the bound state. The functions $Z$ (or $Y$) denote the
spherical harmonics for the corresponding $n$-sphere $\Omega_n$.
(Adopted from Ref.\ \cite{Ahl99}.)
\label{t1}}
\end{table}

The appearance of complex eigenvalues is related to a crossing of an abnormal 
with a normal (or abnormal) state \cite{Ahl99}. Therefore, this problem is
related to the existence of abnormal states. It occurs for a wide range of
parameters. Increasing the mass of the exchange particle the higher lying
eigenvalues tend to become real again. This can be understood from the fact
that for an infinitely heavy exchange particle the BS equation assumes an
$O(4)$ 
symmetric form as in the case of a massless exchange particle, see also Table
\ref{t1} which summarises the symmetries of the scalar ladder BS equation. It
is interesting to note that the ladder BS spectrum of  QED also shows such a
phenomenon \cite{Ahl99}. First, one has to note that for QED in Feynman gauge
there exists a  continuum of solutions for $\alpha = e^{2}/4\pi > \pi/4 :=
\alpha_c$
\cite{Gol53}. Using a finite mesh for the numerical calculation this continuum
of eigenvalues\footnote{This continuum of eigenvalues sould be not confused
with a physical continuum, {\it i.e.}, the existence of scattering solutions.
Here we are discussing the formal spectrum of the BS equation. The Goldstein
continuum discussed here would imply (if taken at face value) that for all
couplings $\alpha > \pi/4$ normalizable solutions with any bound state mass
between zero and two times the constituent mass should exist.} is represented
by discrete eigenvalues which become more dense for a finer mesh. In this
continuum there are abnormal states and there occur crossings between these
states, {\it i.e.}, complex eigenvalues.\footnote{The existence of complex
eigenvalues has also been observed for an axialvector state in the ladder BS
equation of 2+1-dimensional QED with one four-component flavour, {\it i.e.},
in the confining phase \cite{All97a}. The poles of the fermion propagator
(calculated in the corresponding rainbow approximation) come in complex
conjugate pairs thereby signaling violation of positivity. It is suggestive
that the complex eigenvalues of the BS equation appear for values of the
fermion bare mass $m$ for which $-{\rm Re} (p^2_{\rm pole}) > {\rm Im}
(p^2_{\rm pole})$ as can be infered by comparing Table I in Ref.\ \cite{All96}
with Table II in Ref.\ \cite{All97a}. On the other hand, in the
non-relativistic limit $m\gg e^2$ this state exists again and becomes
degenerate with the positive parity scalar state. Furthermore, in Ref.\
\cite{All97a} unnatural parity states have been observed. It is quite likely
that these states are abnormal states, {\it i.e.}, excitations in relative
time.}

The important point to notice is that the scalar model possesses a critical
coupling beyond which the one-particle propagators are not renormalizable
\cite{Ros96,Ahl99}. For acceptable values of the coupling no
abnormal solutions to the BS equation exists. As for quenched QED one has to
notice that the critical value of the coupling for DB$\chi$S is $\alpha_c=
\pi/4$ in
Feynman gauge. Furthermore, from the discussion in Sec.\ \ref{sec_QED4} it is
clear that in quenched QED in rainbow approximation no finite mass solution for
the fermion DSE exists for values of the coupling exceeding the critical one.
On the other hand, for couplings below the critical coupling no abnormal states
appear in the spectrum of the BS equation. Therefore, these two examples
provide 
evidence that there is no problem with abnormal states as long as one applies
the BS equation only for values of the coupling constant where the underlying
theory is renormalizable and the full renormalized two-point functions are
well-defined \cite{Ahl99}. The important warning to keep in mind, however, is:
Calculating excited states from the BS equation (or from a three-dimensional
reduction of it) one first has to clarify the issue of abnormal states.
On the other hand, the use of the BS equation for the corresponding ground
states is quite safe.

\subsection{Ground State Mesons}
\label{sec_GS} 

\subsubsection{The Goldstone Boson Sector}
\label{sub_pion} 

Pions (and kaons) are very special. First of all, they are the lightest hadrons
and therefore play a significant role in nuclear physics as they provide the
long-range part of the nucleon-nucleon interaction. Furthermore, they are
produced in almost all reactions involving hadrons ranging from intermediate
energy electron-nucleon to ultrarelativistic heavy ion collisions. As stated
already in the last section their low mass can be understood from the
approximate chiral symmetry of QCD. Furthermore, symmetry considerations
provide the BS amplitude in the chiral limit, see Eq.\ (\ref{chiRot}). However,
in the real world with explicit current quark masses the situation is much more
complex as revealed by the up to now most complete study of the pion and kaon
BS amplitudes \cite{Mar97}.

The considerations in the last section demonstrated that the quark self-energy
has to be determined first. Identical kernels of the quark rainbow DSE and the
meson ladder BS equation have hereby to be chosen. The authors of Ref.\ \cite{Mar97}
employed a model kernel of the form $4\pi \alpha (k^2)/k^2$ with
\begin{equation}
\label{alphaCraig}
\alpha (k^2) = 2\pi^3 D k^2 \delta^{(4)} (k) +
\pi D \frac{k^4}{\omega^6} {\rm e}^{-k^2/\omega^2} + \frac{ \pi \gamma_m 
         (1 - \exp(-k^2/4 m_t^2)}       {\frac{1}{2}
        \ln\left[\tau + \left(1 + k^2/\Lambda_{\rm QCD}^2\right)^2\right]}
\end{equation}
where $\gamma_m = \frac{12}{33-2N_f} = \frac{12}{25}$ is the anomalous
dimension of the quark mass ($N_f=4$ has been used in Ref.\ \cite{Mar97}), and
a value of  $\Lambda_{\rm QCD}^{N_f=4}= 0.234\,{\rm GeV}$ has been assigned.
The parameter $\tau={\rm e}^2-1$ is chosen such that the perturbative Landau
pole is screened and the denominator of the last term in Eq.\
(\ref{alphaCraig}) becomes identical to one as $k^2$ vanishes. Obviously, this
last term is proportional to $k^2$ for low momenta and identical to the known
1-loop behaviour at large $k^2$. The parameter $m_t$ has hereby been  fixed to
be $m_t=0.5$GeV. The second term is modelled to provide interaction strength at
intermediate momenta of a few hundred MeV. To achieve this, 
values $\omega=0.3$GeV and
$D=0.781$GeV$^2$ have been used in  Ref.\ \cite{Mar97}. The first term, being a
delta function (adopted from  Ref.\ \cite{Mun83}), leads to a quark propagator
with complex conjugate poles, {\it i.e.}, to a positivity violating one. 
Choosing at a scale of 1 GeV the current masses to be $m_{u/d}^{1\,{\rm GeV}} =
5.5\,{\rm MeV}\,,\; m_{s}^{1\,{\rm GeV}} = 130\,{\rm MeV}$ ``Euclidean'' 
constituent masses of 560 MeV for up/down and 700 MeV for strange quarks are 
obtained.

The ladder BS equation in the pseudoscalar channel is an eigenvalue equation
for the vertex function. This vertex function does not only involve a
pseudoscalar term as in Eq.\ (\ref{chiRot}) but has the general form
\cite{Lle69} 
\begin{eqnarray}
\label{genpibsa}
\Gamma_H(k;P) & = &  T^H \gamma_5 \left[ i E_H(k;P) +
\gamma\cdot P F_H(k;P) \rule{0mm}{5mm}\right. \\
\nonumber & &
\left. \rule{0mm}{5mm}+ \gamma\cdot k \,k \cdot P\, G_H(k;P)
+ \sigma_{\mu\nu}\,k_\mu P_\nu \,H_H(k;P)
\right]\,,
\end{eqnarray}
where $T^H$ is the flavour Gell--Mann matrix for the meson $H$, {\it e.g.},  
$T^{K^+} = \frac{1}{2}\left(\lambda^4 + i \lambda^5\right)$.
For bound states with equal mass constituents the 
scalar functions $E_H$, $F_H$, $G_H$ and $H_H$ are even under 
$k\cdot P \to - k\cdot P$.
In general, the subleading Dirac components of $\Gamma_H(k;P)$ and therefore the
functions $F_H(k;P)$, $G_H(k;P)$ and $H_H(k;P)$ are nonzero.

To understand the relation of these functions to the quark propagator
one considers the renormalised axial-vector Ward--Takahashi in the chiral
limit
\begin{equation}
\label{avwti0}
-i P_\mu \Gamma_{5\mu}^H(k;P)  = {S}^{-1}(k+P/2)\gamma_5\frac{T^H}{2}
+  \gamma_5\frac{T^H}{2} {S}^{-1}(k-P/2)\,.
\end{equation}
In the chiral limit the axial-vector vertex has the form
\begin{eqnarray}
\label{truavv}
\Gamma_{5 \mu}^H(k;P) & = &
\frac{T^H}{2} \gamma_5
\left[ \rule{0mm}{5mm}\gamma_\mu F_R(k;P) + \gamma\cdot k k_\mu G_R(k;P)
- \sigma_{\mu\nu} \,k_\nu\, H_R(k;P) \right]\\
&+ & \nonumber
 \tilde\Gamma_{5\mu}^{H}(k;P)
+\,f_H\,  \frac{P_\mu}{P^2 } \,\Gamma_H(k;P)\,,
\end{eqnarray}
where $F_R$, $G_R$, $H_R$ and $\tilde\Gamma_{5\mu}^{H}$ are regular as
$P^2\to 0$. Note that $P_\mu \tilde\Gamma_{5\mu}^{H}(k;P) \sim {\mathcal O }
(P^2)$, and
$\Gamma_H(k;P)$ is the pseudoscalar BS amplitude of Eq.~(\ref{genpibsa}).
The residue of the pseudoscalar pole in the axial-vector vertex is $f_H$, the
leptonic decay constant of the meson $H$. The chiral limit axial-vector
Ward--Takahashi identity (\ref{avwti0}) implies,
\begin{eqnarray}
\label{bwti}
f_H E_H(k;0)  &= &  B(k^2)\,, \\
\label{fwti}
 F_R(k;0) +  2 \, f_H F_H(k;0)  & = & A(k^2)\,, \\
\label{rgwti}
G_R(k;0) +  2 \,f_H G_H(k;0)    & = & 2 A^\prime(k^2)\,,\\
\label{gwti}
H_R(k;0) +  2 \,f_H H_H(k;0)    & = & 0\,,
\end{eqnarray}
where $A(k^2)$ and $B(k^2)$ are the quark propagator functions. Note that 
Eq.\ (\ref{bwti}) is identical to Eq.\ (\ref{chiRot}). Surprisingly, these
equations, however, imply that the subleading Dirac structures are
non-vanishing in the chiral limit.

\begin{figure}
\centering{\
\epsfig{figure=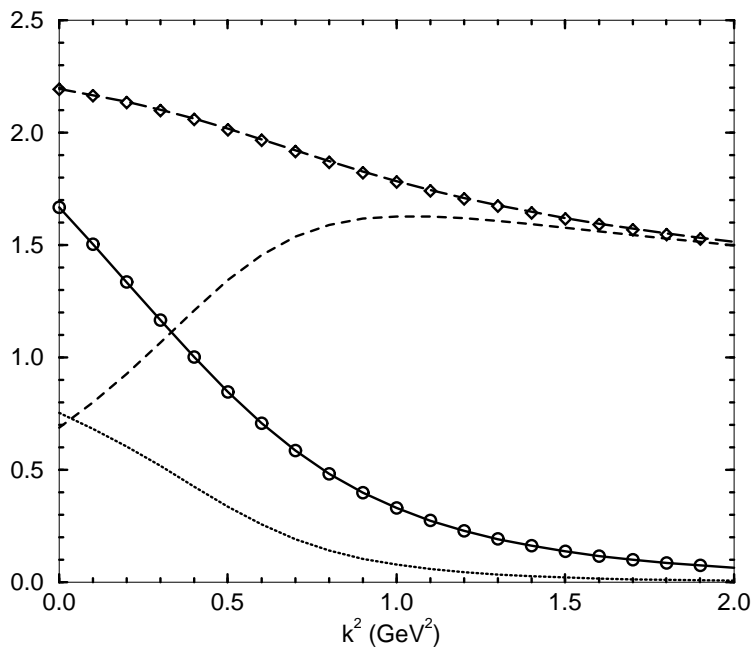,height=10.0cm}}
\caption[An illustration of the realisation  of the axial-vector
Ward--Takahashi identity for pion amplitudes in the chiral limit.
(Adapted from Ref.\ \protect\cite{Mar97}.)]
{An illustration of the realisation  of the identities
Eqs.~(\protect\ref{bwti}) and (\protect\ref{fwti}), which are a necessary
consequence of preserving the axial vector Ward-Takahashi identity. Shown  is
$f^0 E_H(k;0)$ (GeV, solid line); $F_R(k;0)$ (dimensionless, dashed line);
$f^0 F_H(k;0)$ (dimensionless, dotted line); and $ F_R(k;0) + 2 \, f_H
F_H(k;0) $ (long-dashed line).  In each curve the plotted points represent
the right-hand-side of these equations as obtained in the solution of the
chiral-limit quark DSE: $B(k^2)$ (GeV, $\circ$); $A(k^2)$ (dimensionless,
$\diamond$). (Adapted from Ref.\ \protect\cite{Mar97}.)
\label{wtiplot}}
\end{figure}

The BS equation has been solved in  Ref.\ \cite{Mar97} by a direct numerical
solution of the multidimensional integral equation and by employing an
expansion of the functions $E$, $F$, $G$ and $H$ in Chebyshev polynomials in
$k\cdot P /\sqrt{k^2P^2}$. This expansion converges quickly, and two moments
are sufficient to obtain an accurate solution. This is also seen from the fact
that physical observables calculated with theses solutions 
are independent of the arbitrary momentum
partitioning parameter $\eta_P$, see Eq.\ (\ref{p12}). Note that this has been
considered problematic in older studies of the BS equation \cite{Jai93} where
only the leading order Chebyshev moments have been kept in the numerical
solution.

The Goldstone boson character of the flavour-nonsinglet pseudoscalar has been
made explicit in the BS description as relativistic QCD bound states. Hereby the
axial-vector Ward--Takahashi identity is the key to realize the importance of
the non-leading Dirac invariants in the pseudoscalar BS amplitudes
\cite{Mar97,Gov84,Lan89a}. This axial-vector Ward--Takahashi identity is in 
turn then seen in the numerical solution, see Fig.\ \ref{wtiplot}.
Thus we summarise this subsection by stating that reliable BS amplitudes for
pions and kaons are available, {\it c.f.}, also the 3rd column in Table 2
below for the corresponding masses and decay constants. As we will see in the
following this is very helpful for calculating observables in hadronic
processes.

\subsubsection{A Dynamical $\eta^\prime$ Mass}
\label{sub_eta}  

In the preceding subsection the flavour-singlet pseudoscalar meson, the
$\eta^\prime$, has not been considered. Of course, due the anomalous
breaking of the $U_A(1)$ chiral symmetry the $\eta^\prime$ is not a
Goldstone boson.

More than twenty years ago Kogut and Susskind pointed out that for
dimensional reasons a non-vanishing contribution to the mass of the
pseudo-scalar flavour-singlet meson in the chiral limit can result from
its mixing with two non-perturbatively infrared enhanced gluons, 
corresponding to a momentum space propagator $D(k) \sim \sigma/k^4$ for $k^2
\to 0$ \cite{Kog74}. The identification of the string tension $\sigma$
shows that effects due to infrared enhanced gluons can be expected to be
complementary to instanton models. In particular, a description of the
$\eta$--$\eta'$ mixing driven by the string tension \cite{Mec97,Sme97a},
provides an interesting alternative to the standard solution of the $U_A(1)$
problem by instantons.

Phenomenologically, this mixing is described by the  $\eta_8 - \eta_0$  mass
matrix \cite{Wit79,Ven79},
\begin{eqnarray}
 \frac{1}{2}\ (\eta_8\ \ \  \eta_0)
 \left(
 \begin{array}{cc}
  \frac{4}{3}m_K^2-\frac{1}{3}m_\pi^2&\frac{2}{3}\sqrt{2}(m_\pi^2-m_K^2)\\
   & \\
  \frac{2}{3}\sqrt{2}(m_\pi^2-m_K^2) &\frac{2}{3}m_K^2+\frac{1}{3}m_\pi^2
                                       + \frac{2N_f}{f_0^2} \chi^2
 \end{array}
 \right)
 \left(
 \begin{array}{c}
  \eta_8\\
   \\
  \eta_0
 \end{array}
 \right)
\end{eqnarray}
where the screening mass in the flavour-singlet component, $m_0^2 := 2N_f
\chi^2/f_0^2 $, is given by a non-vanishing topological susceptibility,
\begin{equation}
\chi^2 := \frac{g^2}{(32\pi^2)^2} \int d^4x \, \langle \widetilde GG(x) \,
\widetilde GG(0) \rangle \quad \hbox{with}
\end{equation}
\begin{displaymath}
 \widetilde GG  =  \epsilon^{\mu\nu\rho\sigma} 2\partial_\mu \, {\rm tr}
(A_\nu  \partial_\rho A_\sigma -ig \frac 2 3 A_\nu A_\rho A_\sigma ) \; .
\end{displaymath}
In the Instanton Liquid Model the topological susceptibility, given by the
density of instantons, is $\chi^2 \approx 1 \hbox{fm}^{-4}$, and the mass
eigenvalues are $m_\eta \approx 530$MeV, $m_{\eta'} \approx 1170$MeV
together with a mixing angle of $\theta \approx -11.5^\circ$ \cite{Alk89a}.
This has to be compared with the experimental values $m_\eta =547$MeV and
$m_{\eta'} = 959$MeV.

Recently, indirect $SU(3)$ flavour breaking in the pseudoscalar meson mass
matrix has  been considered \cite{Kek00}, {\it i.e.}, the assumption that the
topological susceptibility leads only to a flavour-singlet component of the
mass matrix has been relaxed. Hereby, the corresponding ``weakening'' parameter
in the strange sector, $X=0.663$, has been determined from the two-photon
amplitudes $\pi^0,\eta, \eta^\prime \to \gamma  \gamma$ calculated in a DSE
based approach (see also Sect.\ \ref{sub_Int}  below). The corresponding
results are $m_\eta =588$MeV, $m_{\eta'} = 933$MeV and $\theta = -13.4^\circ$. 
One sees that this indirect flavour symmetry breaking improves the agreement
with the phenomenological $\eta$  and $\eta^\prime$ masses considerably. On
the other hand, as this effect is of quantitative importance only, we will
ignore it in the rest of this section and discuss the qualitative features.

\begin{figure}[t]
\centerline{\epsfig{file=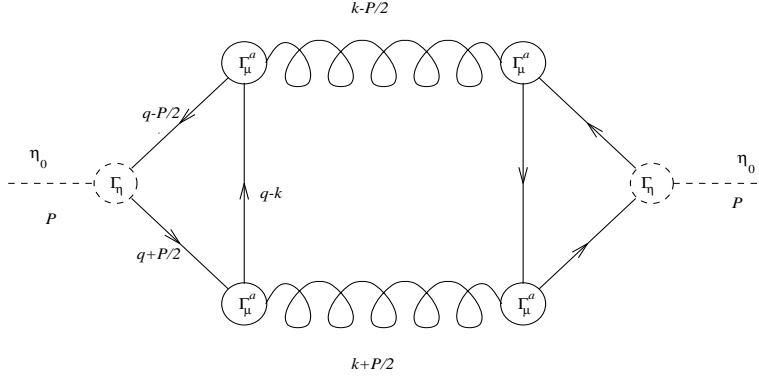,height=5cm,width=10cm}}
\vspace{5pt}
\caption{The diamond diagram $\Pi (P^2)$. (A factor 2 arises from crossed
gluon exchange.) }
\label{diamond}
\end{figure}

To explore the effect of infrared enhanced quark interactions, one concentrates
on the mixing of the flavour-singlet pseudo-scalar with two uncorrelated
gluons. According to the Kogut--Susskind argument, for infrared enhanced gluons
$\sim \sigma/k^4$, the corresponding diagram, see Fig.~\ref{diamond}, can
contribute to the topological susceptibility for the meson momentum $P \to 0$.
Of course, this diagram is not the only one capable of providing such a
contribution \cite{Fra98}, however, the consideration of this leading term
should be sufficient to demonstrate the effect qualitatively and to provide a
semi-quantitative estimate. To explore the Kogut--Susskind conjecture, 
the following model interaction for quarks in the Landau gauge
has been used in Refs.\ \cite{Mec97,Sme97a},
\begin{equation}
g^2 D_{\mu\nu}(k) = P_{\mu\nu} (k) \, \left( \frac{8\pi \sigma}{k^4} \, +\,
\frac{16\pi^2/9}{k^2 \ln(e+k^2/\Lambda^2)} \right) \; .
\label{int}\end{equation}
The second term, subdominant in the infrared, was added to simulate the
effect of the leading logarithmic contribution of perturbative QCD for $N_f =
3$. As mentioned at the beginning of this chapter, a quark interaction of the
form (\ref{int}) can, strictly speaking, not arise from gluons alone in
Landau gauge, since the product $g^2 D_{\mu\nu}$ is not renormalisation group
invariant for any finite number of flavours or colours. Even though this is
assumed in the {Abelian} approximation, ghost contributions do implicitly
enter in the RG invariant interaction by the dressing of the quark-gluon
vertex function as has been discussed in previous chapters.

From the axial anomaly, the quark triangle $\Gamma^{ab}_{\mu\nu}$ in
Fig. \ref{diamond} has the limit,\\
\begin{equation}
 P \to 0 \, , k^2 = 0 \; : \qquad  \Gamma^{ab}_{\mu\nu} \, \to \, \delta^{ab}
\epsilon_{\mu\nu\rho\sigma} \,
k^\rho P^\sigma \,
\frac{\sqrt{N_f} g^2}{f_0 8\pi^2}  \; ,
\end{equation}
with $f_0$ being the flavour-singlet decay constant.
This model independent form, determining the coupling of two gluons to the
pseudoscalar flavour-singlet bound state in the infrared, is particularly
suited for the present calculation, since the contribution to $\chi^2 $ is
obtained from $P\to 0$, and since the gluon interaction (\ref{int}) weights
the integrand so strongly in the infrared ($\sim \sigma/(k\pm P/2)^4$). With
this, all contributions containing ultraviolet dominant terms of the
interaction (\ref{int}) vanish for $P\to 0$, and one obtains
\cite{Mec97,Sme97a},
\begin{equation}
m_0^2 \, = \, \lim_{P^2\to 0} \Pi(P^2) \, =\, \frac{2 N_f}{f_0^2} \, \chi^2 \, =\, \frac{3N_f}{f_0^2} \,
\frac{\sigma^2}{\pi^4} \; .
\end{equation}
The phenomenological string tension $\sigma = 0.18$GeV$^2$ and $f_0\approx
f_\pi =93$MeV thus yield $m_0^2\approx 0.346$GeV$^2$, and the physical mass
eigenstates are, $m_{\eta^\prime} \approx 810$MeV and $m_{\eta} \approx
430$MeV, with a corresponding mixing angle $\theta \approx -30^\circ$.
Furthermore,
using $f_0^2 \simeq  f_\pi^2 ( 1 + \Pi'(P^2)
|_{P^2 \to 0} )$ with $\Lambda\approx 500 $MeV in (\ref{int}), one obtains an
additional contribution to the decay constant of the flavour-singlet of about
30\% as compared to the pion \cite{Mec97}.

As these values are reasonably close to experiment, one might conclude that the
$U_A(1)$-anomaly can be encoded in the infrared behaviour of QCD Green's
functions. Whether the Kogut--Susskind mechanism or the instanton based
solution to the $U_A(1)$ problem is realized in nature, can be assessed from
their respective temperature dependences. If the origin of the $\eta^\prime$
mass is predominantly due to instantons, the $\eta - \eta^\prime$ mixing
angle is expected to vary exponentially with temperature, leading to a
significant change of $\eta$ and $\eta^\prime$ production rates in
relativistic heavy ion collisions \cite{Alk89b}. On the other hand, lattice
calculations indicate that the string tension is almost temperature
independent up to the deconfinement transition. This offers the possibility
to study the physics of the $U_A(1)$ anomaly experimentally.

\subsubsection{An Unified Description of Light
and Heavy Mesons}
\label{sub_LH}

In Sec.\ \ref{sec_nPoint} we have described a solution for the DSE of the
quark-photon vertex function and remarked that its transverse part possesses 
a pole  at the vector meson mass \cite{Mar00}. Following the derivation of the
homogenous BS equation it is quite obvious then that the transverse part of
the quark-photon vertex fulfils Eq.\ (\ref{bsemeson}). As stated in Sec.\
\ref{sec_nPoint}, in general, twelve independent Dirac tensors are needed to
describe a vector vertex, four of them can be chosen to be longitudinal with
respect to the photon and/or vector meson momentum leaving eight independent
transverse terms. For a physical 
on-shell vector meson the corresponding BS vertex function is transverse,
\begin{equation}
 \left( P_\mu \Gamma ^V_\mu (p;P) \right) _{P^2=-m_V^2} = 0.
\end{equation}
Therefore, eight independent functions have to be determined in the solution of
the vector meson BS amplitudes \cite{Lle69}:
\begin{equation}
\Gamma ^V_\mu (p;P) = \sum_{i=1}^8 F_i(p^2,p\cdot P; P^2) T_\mu^i(p;P)
= \gamma_\mu^T F_1(p^2,p\cdot P; P^2) +\ldots \; ,
\end{equation} 
where $\gamma_\mu^T = \gamma_\mu - P_\mu \left(P\cdot \gamma \right) /P^2$
is the transverse projection of the four-vector $\gamma_\mu$.
Not surprisingly, the amplitude $F_1$ is quantitatively the most important one
\cite{Mar99}. 

The study in Ref.\ \cite{Mar99} uses a kernel very similar to the one used 
Ref.\ \cite{Mar97}, only the delta-function has been omitted, {\it i.e.},
\begin{equation}
\label{alphaPieter}
\alpha (k^2) = 
\pi D \frac{k^4}{\omega^6} {\rm e}^{-k^2/\omega^2} + \frac{ \pi \gamma_m 
         (1 - \exp(-k^2/4 m_t^2)}       {\frac{1}{2}
        \ln\left[\tau + \left(1 + k^2/\Lambda_{\rm QCD}^2\right)^2\right]} \,
        ,
\end{equation}
and $\omega$ and $D$ are treated as free parameters. Please note that the
coupling (\ref{alphaPieter}) is vanishing in the extreme infrared. On the other
hand, it is very strongly enhanced for momenta $p^2 \approx \omega ^2$. 
Table \ref{respseudo} (adapted from Ref.\ \cite{Mar99}) shows that pion and
kaon masses are described also very well.

\begin{table}
\begin{tabular}{l|ccccc}
& experiment  &         &$\omega=0.3\,{\rm GeV}$
                                &$\omega=0.4\,{\rm GeV}$
                                        &$\omega=0.5\,{\rm GeV}$\\
& (estimates) & Ref.~\cite{Mar97}& $D=1.25\,{\rm GeV}^2$
                                & $D=0.93\,{\rm GeV}^2$
                                        & $D=0.79\,{\rm GeV}^2$ \\ \hline
- $\langle \bar q q \rangle^0_{\mu=1 {\rm GeV}}$
                & $(0.236 \rm GeV)^3$ & $(0.241 \rm GeV)^3$
                                        & 0.242 & 0.241 & 0.243  \\
$m^{u=d}_{\mu=1 {\rm GeV}}$
                & 5 - 10 MeV  & 5.5 MeV & 5.54  &  5.54 &  5.35  \\
$m^{s}_{\mu=1 {\rm GeV}}$
                &100 - 300 MeV& 130 MeV & 124   &  125  &  123   \\
$m_\pi$         &  0.1385 GeV & 0.1385  & 0.139 & 0.138 & 0.138  \\
$f_\pi$         &  0.1307 GeV & 0.1307  & 0.130 & 0.131 & 0.131  \\
$m_K$           &  0.496 GeV  & 0.497   & 0.496 & 0.497 & 0.497  \\
$f_K$           &  0.160 GeV  & 0.154   & 0.154 & 0.155 & 0.157
\end{tabular}
\caption[Calculated values of the properties of light pseudoscalar mesons.
(Adapted from Ref.\ \protect\cite{Mar99}.)]
{\label{respseudo}
Calculated values of the properties of light pseudoscalar mesons for
the parametrisation of the effective coupling (\protect\ref{alphaPieter}),
using three different parameter sets, and also for the parametrisation
of Ref.~\protect\cite{Mar97}. (Adapted from Ref.\ \protect\cite{Mar99}.)}
\end{table}

\begin{table}
\begin{center}
\begin{tabular}{l|cccccc}
 & \multicolumn{2}{c}{$\rho$}      &
   \multicolumn{2}{c}{$K^\star$}   &
   \multicolumn{2}{c}{$\phi$}      \\
                & $m_\rho$ & $f_\rho$
                & $m_{K^\star}$ & $f_{K^\star}$ & $m_\phi$ & $f_\phi$ \\ \hline
experiment      & 0.770 & 0.216 & 0.892 & 0.225 & 1.020 & 0.236 \\ \hline
All amplitudes $F_1$-$F_8$
                &       &       &       &       &       &       \\ \hline
$\omega=0.3\,{\rm GeV}$, $D=1.20\,{\rm GeV}^2$
                & 0.747 & 0.197 & 0.956 & 0.246 & 1.088 & 0.255 \\
$\omega=0.4\,{\rm GeV}$, $D=0.93\,{\rm GeV}^2$
                & 0.742 & 0.207 & 0.936 & 0.241 & 1.072 & 0.259 \\
$\omega=0.5\,{\rm GeV}$, $D=0.79\,{\rm GeV}^2$
                & 0.74  & 0.215 & 0.94  & 0.25  & 1.08  & 0.266 \\  \hline
{amplitudes $F_1 \ldots F_5$ only}
                &       &       &       &       &       &       \\ \hline
$\omega=0.3\,{\rm GeV}$, $D=1.20\,{\rm GeV}^2$
                & 0.737 & 0.192 & 0.942 & 0.235 & 1.080 & 0.247 \\
$\omega=0.4\,{\rm GeV}$, $D=0.93\,{\rm GeV}^2$
                & 0.729 & 0.199 & 0.919 & 0.229 & 1.062 & 0.250 \\
$\omega=0.5\,{\rm GeV}$, $D=0.79\,{\rm GeV}^2$
                & 0.731 & 0.207 & 0.926 & 0.237 & 1.072 & 0.259

\end{tabular}
\end{center}
\caption[Results for the vector mesons. (Adapted from Ref.\
\protect\cite{Mar99}.) ]
{\label{resvecvarious}
Comparison of the results for the vector mesons for the three different
parameter sets for the effective interaction, using all eight BS
amplitudes (top) or the five leading ones (bottom). 
(Adapted from Ref.\ \protect\cite{Mar99}.)}
\end{table}

In Ref.\ \cite{Mar99} it has been found that only five of the eight covariants
are important for the vector meson masses and decay constants, see Table
\ref{resvecvarious}. (Two warnings are here in order. First, the relative
importance of amplitudes depends in general on the observable under
consideration. Second, in an other basis for the eight independent covariants
the number of relevant covariants might be different.)
Furthermore, for the $\rho$ and the $\phi$ meson the truncation to the leading
Chebyshev moment leads to very similar results. 
It is also interesting to note that the leading amplitudes for the 
$\rho$ are surprisingly akin to the ones for the
pion ({\it i.e.}, $E_\pi$ which is related to 
the scalar part of the quark self-energy in the chiral limit).
Up to now it is unknown whether this is a more general relation or whether it
is  valid only for the interaction (\ref{alphaPieter}) with the chosen 
parameters. Note also that in this calculation the $\rho$ and the $\omega$ are
degenerate. It is usually assumed that this degeneracy will be lifted by the
coupling to pions, see Sec.\ \ref{sub_Int}.

As we have seen in the sector with light, {\it i.e.}, up and down, and strange 
quarks the pseudoscalar and vector mesons can be understood quite well based on
solutions of the ladder BS equation. An investigation of scalar mesons with the
same quality as for pseudoscalar and vector mesons is not available, see,
however, Ref.\ \cite{Jai93} for a study which considers also scalar mesons.
There are indications that the ladder truncation is not a good approximation
for scalar mesons. The coupling of scalar mesons to two pseudoscalar mesons is
very strong complicating this issue enormously. There is evidence from lattice
calculations that scalar mesons are more like $\bar q^2q^2$ than $\bar qq$
states \cite{Alf00}. Interpretations of scalar mesons range accordingly from 
them being unitarised $\bar qq$ states to being `molecules' of two
pseudoscalars, {\it e.g.}, see Refs.\ \cite{Tor99,Pen99,Bla00,Jan94}. 
For the scalar-isoscalar mesons there is in addition the
issue of mixing with glueballs and/or annihilation into time-like gluons.
Due to this a straightforward application of the ladder BS equation to the
scalar mesons is certainly not adequate.

The rainbow DSE and ladder BS formalism is straightforwardly applicable to
mesons with heavy quarks \cite{Iva99}. First of all, the constituent quark mass
functions for charm and bottom quarks is almost trivial, at large momenta  they
obey the perturbative RG behaviour, and in the infrared they display a slight
enhancement. As the function $A(p^2)$ does also not deviate much from its
perturbative value the corresponding propagators can be approximated very well
by the free ones with a constant mass. Due to this the consequences of
heavy-quark symmetry are reproduced in this approach, {\it e.g.}, in the
heavy-quark limit pseudoscalar meson masses grow linearly with the mass of
their heaviest constituent, $m_P \propto \hat m_Q$.\footnote{For a review on
heavy-quark symmetry see, {\it e.g.}, Ref.\ \cite{Neu94}.} A note of warning
is, however, in order: For the solutions of the BS equation it is not only
required that the binding energy is much smaller than the quark mass but also
the momentum space widths of the amplitudes have to be significantly less
than the quark mass. Note also that in Ref.\ \cite{Iva99} parametrisations
have been employed for these amplitudes.
(Also the propagators of the light quarks have been
parametrised as entire functions.) These were then used to calculate
heavy-meson leptonic decays, semileptonic heavy-to-heavy and
heavy-to-light transitions ($B \to D^\ast$, $D$, $\rho$, $\pi$; $D \to
K^\ast$, $K$, $\pi$), radiative and strong decays ($B_{(s)}^\ast \to
B_{(s)}\gamma$; $D_{(s)}^*\to D_{(s)}\gamma$, $D \pi$), and the rare $B\to
K^\ast \gamma$ flavour-changing neutral-current process.

The leptonic decay constants are given by
\begin{eqnarray}
\nonumber f_P& = &f_V= \frac{\kappa_f}{\surd m_H}\,\frac{N_c}{2\sqrt{2}
\pi^2}\, \int\limits_0^\infty du \left(\sqrt{ u} - E_H\right)
\varphi_H(z)\left\{ \sigma_S^f(z) +
\frac{1}{2}\sqrt{u}\,\sigma_V^f(z)\right\}\,,
\label{lept}\\
\frac{1}{\kappa_f^2} &=&
\frac{N_c}{4\pi^2}\,
\int\limits_0^\infty du \varphi^2_H(z)
\left\{ \sigma_S^f(z) + \sqrt{u}\,\sigma_V^f(z)\right\}\,,
\nonumber
\end{eqnarray}
where $z=u-2 E_H \sqrt{ u}$, $f$ labels the lighter of the quarks inside the
meson, and $\varphi_H(z)$ is the scalar function characterising the dominant
Dirac-covariant in the heavy-meson BS amplitude, {\it e.g.}, the
term proportional to $\gamma_5$ for pseudoscalar mesons or the one
proportional to $\gamma_\mu$ for vector mesons. $\sigma_{S(V)}^f(z)$ is the
scalar (vector) part of the quark propagator with flavour $f$.
The semileptonic heavy-to-heavy pseudoscalar transition
form factors ($P_1$ $\to$ $P_2 \ell \nu$) acquire a particularly simple form in
the heavy-quark symmetry limit:\footnote{The transition form factors $f_\pm(t)$ 
with $t_\pm = (m_{P_2} \pm m_{P_1})^2$ contain all the information about
strong interaction effects in these processes. Their calculation is necessary
for a determination of the CKM matrix elements from a measurement of the
corresponding decay widths.}
\begin{eqnarray}
\label{fxi}
f_\pm(t)& := & \frac{m_{P_2} \pm
m_{P_1}}{2\sqrt{m_{P_2}m_{P_1}}}\,\xi_f(w),\\
\label{xif}
\xi_f(w)  & = & \kappa_f^2\,\frac{N_c}{4\pi^2}
\int\limits_0^1 \frac{d\tau}{W}
\int\limits_0^\infty du \, \varphi^2_H(z_W) 
\left[\sigma_S^f(z_W) + \sqrt{\frac{u}{W}} \sigma_V^f(z_W)\right]\,,
\end{eqnarray}
with $W= 1 + 2 \tau (1-\tau)(w-1)$, $z_W= u - 2 E_H \sqrt{u/W}$ and
\begin{equation}
w = \frac{m_{P_1}^2 + m_{P_2}^2 - t}{2 m_{P_1} m_{P_2}} = - v_{P_1} \cdot
v_{P_2}\,.
\end{equation}
The normalisation of the BS amplitude automatically ensures that the Isgur-Wise
function $\xi (w)$ \cite{Isg89} fulfils 
\begin{equation}
\label{xione}
\xi_f(w=1) = 1
\end{equation}
in the heavy-quark limit. Furthermore, one obtains from Eq.~(\ref{xif}) that
\begin{equation}
\rho^2 := -\left.\frac{d\xi_f}{dw}\right|_{w=1} \geq \frac{1}{3}\,.
\end{equation}
A similar analysis for the heavy-to-heavy transitions with vector mesons in the
final state and for heavy-to-light transitions yields relations between the
form factors that coincide with those observed in Ref.~\cite{Isg89}: In
the heavy-quark limit also these form factors are expressible solely in terms
of $\xi_f(w)$.

\begin{table}
\caption[The quantities used in $\chi^2$-fitting the model parameters of Ref.\
\protect\cite{Iva99}.]
{The 26 dimensionless quantities used in $\chi^2$-fitting the model
parameters. The light-meson
electromagnetic form factors are calculated in impulse
approximation and $\xi(w)$ is obtained
from $f_+^{B\to D}(t)$ via Eq.~(\protect\ref{fxi}).  (Table adapted from
Ref.~\protect\cite{Iva99}.)
\label{tableD} }
\begin{center}
\begin{tabular}{lll|lll}
        & Obs.  & Calc. & & Obs.  & Calc. \\\hline
$f_+^{B\to D}(0)$ & 0.73 & 0.58  &
        $f_\pi r_\pi$ & 0.44 $\pm$ 0.004 & 0.44   \\
$F_{\pi}(3.3\,{\rm GeV}^2)$ & 0.097 $\pm$  0.019  & 0.077 &
        B$(B\to D^\ast)$ & 0.0453 $\pm$ 0.0032 & 0.052\\
$\rho^2$ &  1.53 $\pm$ 0.36 & 1.84 &
        $\alpha^{B\to D^\ast}$ & 1.25 $\pm$ 0.26 & 0.94 \\
$\xi(1.1)$  & 0.86 $\pm$ 0.03& 0.84 &
        $A_{\rm FB}^{B\to D^\ast}$ & 0.19 $\pm$ 0.031 & 0.24 \\
$\xi(1.2)$  & 0.75 $\pm$ 0.05& 0.72 &
        B$(B\to \pi)$ & (1.8 $\pm$ 0.6)$ {\times 10^{-4}}$  & 2.2 \\
$\xi(1.3)$  & 0.66 $\pm$ 0.06& 0.63 &
        $f^{B\to \pi}_{+}(14.9\,{\rm GeV}^2)$ & 0.82 $\pm$ 0.17 & 0.82 \\
$\xi(1.4) $ & 0.59 $\pm$ 0.07& 0.56 &
        $f^{B\to \pi}_{+}(17.9\,{\rm GeV}^2)$ & 1.19 $\pm$ 0.28 & 1.00 \\
$\xi(1.5) $ & 0.53 $\pm$ 0.08& 0.50 &
        $f^{B\to \pi}_{+}(20.9\,{\rm GeV}^2)$ & 1.89 $\pm$ 0.53 & 1.28 \\
B$(B\to D)$ & 0.020 $\pm$ 0.007 & 0.013 &
        B$(B\to \rho)$ & (2.5 $\pm$ 0.9)$ {\times 10^{-4}}$ & 4.8 \\
B$(D\to K^\ast)$ & 0.047 $\pm$ 0.004  & 0.049 &
        $f_+^{D\to K}(0)$ & 0.73 &  0.61 \\
$\frac{V(0)}{A_1(0)} (D \to K^\ast)$ & 1.89 $\pm$ 0.25 & 1.74 &
        $f_+^{D\to \pi}(0)$ & 0.73 &  0.67 \\
$\frac{\Gamma_L}{\Gamma_T} (D \to K^\ast)$ & 1.23 $\pm$ 0.13 & 1.17 &
        $g_{B^\ast B\pi}$ & 23.0 $\pm$ 5.0 & 23.2 \\
$\frac{A_2(0)}{A_1(0)}(D \to K^\ast)$ & 0.73 $\pm$ 0.15 & 0.87 &
        $g_{D^\ast D\pi}$ & 10.0 $\pm$ 1.3 & 11.0 \\\hline
\end{tabular}
\end{center}
\end{table}

In Ref.\  \cite{Iva99} the model parameters were fitted to a total of 42 light
and heavy meson observables, see Table \ref{tableD} for the corresponding 
heavy meson quantities. The fitted constituent-heavy-quark masses are
\begin{equation}
\label{HQmass}
\hat M_c = 1.32\,\mbox{GeV~and}\; \hat M_b=4.65\,\mbox{GeV}\, .
\end{equation}
This entails that the heavy-meson binding energy, defined as the difference
between an averaged heavy-light meson ``mass'' and the heavy quark mass, is 
large. Using the averaged values for $D$- and $B$-meson masses, 
$m_D=1.99$GeV and $m_B=5.35$GeV, one obtains,
\begin{equation}
\label{EB}
\begin{array}{l}
E_D:= m_D - \hat M_c = 0.67\,{\rm GeV}\,, \\
E_B:= m_B - \hat M_b = 0.70\,{\rm GeV}\,,
\end{array}
\end{equation}
{\it i.e.}, $E_D/\hat M_c= 0.51$ and $E_B/\hat M_b= 0.15$. This provides
an indication that while an heavy-quark expansion is applicable
for the $b$-quark it will provide a poor approximation for the
$c$-quark.  The constituent-heavy-quark-masses in Eq.~(\ref{HQmass}) 
obtained in the Poincar\'e covariant approach~\cite{Iva99} are, 
respectively, $\sim 25$\% and $\sim 10$\% smaller than the values used 
in nonrelativistic models.

\begin{table}
\caption{Calculated values of a range of observables not included in fitting
the model's parameters.    (Table adapted from Ref.~\protect\cite{Iva99}.)
\label{tableE} }
\begin{center}
\begin{tabular}{lll|lll}
        & Obs.  & Calc. & & Obs.  & Calc. \\\hline
$f_K r_K$       &   0.472 $\pm$ 0.038 & 0.46 &
        $-f_K^2 r_{K^0}^2$ &  (0.19 $\pm$ 0.05)$^2$ & (0.10)$^2$   \\
$g_{\rho\pi\pi}$ & 6.05 $\pm$ 0.02 & 5.27  &
        $\Gamma_{D^{\ast 0} }$ (MeV)& $ < 2.1 $  & 0.020   \\
$g_{K^\ast K \pi^0}$ & 6.41 $\pm$ 0.06 & 5.96  &
        $\Gamma_{D^{\ast +}}$ (keV) &  $< 131$ & 37.9 \\
$g_\rho$ & 5.03 $\pm$ 0.012 & 5.27  &
        $\Gamma_{D_s^{\ast} D_s \gamma}$ (MeV)& $< 1.9$  & 0.001    \\
$f_{D^\ast}$ (GeV) &   & 0.290   &
        $\Gamma_{B^{\ast +} B^+ \gamma}$ (keV)&   & 0.030    \\
$f_{D^\ast_s}$ (GeV)&   & 0.298   &
        $\Gamma_{B^{\ast 0} B^0 \gamma}$ (keV)&  &  0.015 \\
$f_{B_s}$ (GeV) & 0.195 $\pm$ 0.035 & 0.194  &
        $\Gamma_{B_s^{\ast} B_s \gamma}$ (keV)&  &  0.011 \\
$f_{B^\ast}$ (GeV)&   & 0.200   &
        B($D^{\ast +}\!\to D^+ \pi^0$) & 0.306 $\pm$ 0.025 & 0.316 \\
$f_{B^\ast_s}$ (GeV)&   & 0.209   &
        B($D^{\ast +}\!\to D^0 \pi^+$) & 0.683 $\pm$ 0.014 & 0.683 \\
$f_{D_s}/f_D$ & 1.10 $\pm$ 0.06 &  1.10  &
        B($D^{\ast +}\!\to D^+ \gamma$) & 0.011~$^{+ 0.021}_{-0.007}$ & 0.001 \\
$f_{B_s}/f_B$  & 1.14 $\pm$ 0.08  & 1.07   &
        B($D^{\ast 0}\!\to D^0 \pi^0$) &  0.619 $\pm$ 0.029 & 0.826 \\
$f_{D^\ast}/f_D$ &       &  1.36  &
        B($D^{\ast 0}\!\to D^0 \gamma$) &  0.381 $\pm$ 0.029 &  0.174 \\
$f_{B^\ast}/f_B$  &       & 1.10   &
        B($B \to K^\ast \gamma$) & (5.7 $\pm$ 3.3)${\times 10^{-5}}$ & 11.4 \\\hline
\end{tabular}
\end{center}
\end{table}

With the model parameters fixed it is possible to calculate a wide range
of other light- and heavy-meson observables.  Some of the results are
summarised in Table \ref{tableE}, more results may be found in Ref.\
\cite{Iva99}. It is also  possible to check the fidelity of
heavy-quark symmetry limits.  The universal function characterising
semileptonic transitions in the heavy-quark symmetry limit, $\xi(w)$, can be
obtained with least uncertainties from $B\to D,D^\ast$ transitions.  
Using Eq.~(\ref{fxi}) to extract it from $f_+^{B\to D}(t)$
one obtains
\begin{equation}
\label{xifp}
\xi^{f_+}(1)= 1.08\,,
\end{equation}
which is a measurable deviation from Eq.~(\ref{xione}). Note that corrections
to the heavy-quark symmetry limit of the order of 30\% are encountered in 
$b\to c$
transitions and that these corrections can be as large as a factor of two in
$c\to d$ transitions. The investigation carried out in Ref.\ \cite{Iva99}
indicates that heavy and light mesons are both simply finite-size bound states
of dressed quarks and antiquarks.   

To summarise this subsection: The elements of the Poincar\'e-covariant DSE
framework are rich enough to account for the qualitative structure of almost all
mesons. Of course, possibilities for improvements are numerous. The most wanted
one, however, is in the treatment of the BS equation. The construction of a
BS kernel that respects Slavnov--Taylor and Ward--Takahashi identities 
would be a significant improvement. Especially, it would allow one consistent
route from the underlying properties of QCD Green's functions to meson
structure. Nevertheless, the level of understanding achieved so far justifies
studies of dynamical properties of mesons in this framework.

\subsubsection{Electromagnetic Coupling: Form  Factors and Decay}
\label{sub_EMFF}

As described in Subsection \ref{sub_pion}, the static properties of pions and
kaons such as the mass and decay constants have been studied at a fairly
fundamental level. Dynamic properties and scattering observables are much less
understood even though they are not less important. In this respect the 
elastic electromagnetic form factors of pions and kaons are a very
interesting next step. First of all, there are accurate data for $F_\pi$ at low
$Q^2$ to confront theoretical calculations with, and the charge radii
$r_\pi^2$, $r_{K^+}^2$, and $r_{K^0}^2$ are experimentally known.  Currently,
there are several experiments to determine both, the pion and the kaon form
factor in the range up to $3\,{\rm GeV}^2$ to better accuracy, {\it e.g.}, 
at the Jlab experiments E93-018 (Spokesperson O.~K.~Baker) and E93-021
(Spokesperson 
D.~Mack). Also the pion polarisabilities, {\it i.e.}, the second moment of
the form factor at $Q^2=0$, will be measured to a higher precision, {\it
e.g.}, in experiments at MAMI in Mainz (Spokesperson R.~Beck).

In Refs.\ \cite{Mar00,Mar00a} the electromagnetic form factors of pions and 
kaons have been calculated based on the solutions for the meson BS amplitudes
and the quark-photon vertex  of the homogeneous and inhomogeneous ladder BS
equations, respectively. As in the study of light vector mesons the required
dressed quark propagators are obtained from solutions of the quark DSE in
rainbow truncation using the effective interaction (\ref{alphaPieter}). Hereby
the model parameters are all fixed \cite{Mar99} and constrained only by
$m_\pi$, $m_K$, $f_\pi$ and $\langle\bar q q\rangle$. As can be inferred from
Secs.\ \ref{sec_nPoint} and \ref{sub_LH}, non-analytic effects from vector
mesons are automatically taken into account: the vector $\bar qq$ bound states
appear as poles in the quark-photon vertex solution~\cite{Mar00}. As a matter
of fact, in the domain $-m_V^2< Q^2 < 0.2 {\rm GeV}^2$  the quark-photon
vertex is well described by the sum of the Ball--Chiu vertex plus a second term
containing an explicit pole at the vector meson mass $p^2=-m_V^2$ \cite{Mar00}. 
This splitting yields (via the impulse approximation to be discussed below) 
a charge form factor for the pion that can be expressed as
\begin{equation}
 F_\pi(Q^2) \approx F^{\rm BC}_\pi(Q^2) -
         \frac{ g_{\rho\pi\pi} \; F_{V \pi\pi}(Q^2) \; Q^2}
         {g_\rho (Q^2 + m_\rho^2) } \,,
\label{piffbc}
\end{equation}
where $F^{\rm BC}_\pi(Q^2)$ is the result from the Ball--Chiu vertex.
The combination \mbox{$g_{\rho\pi\pi} F_{V \pi\pi}$} represents the $\pi\pi$
coupling to the vector $\bar q q$ correlation as present in the resonant term 
in the quark-photon vertex. The second term on the r.h.s.\ of
Eq.~(\ref{piffbc}) is written such that \mbox{$F_{V \pi\pi}(-m_\rho^2)=1$}.
As off-shell mesons are unphysical, $F_{V \pi\pi}(Q^2\not = - m_\rho^2)$
does not correspond to a physical process. Nevertheless, it is an interesting
quantity when discussing the relation of the approach presented here to Vector
Meson Dominance (VMD). To this end we note that  the departure of
$F_{V\pi\pi}(Q^2)$ from unity is a measure of the difference in
$\pi\pi$ coupling experienced by the effective vector $\bar q q$
correlation away from the $\rho$ mass-shell compared to the physical
$\rho\pi\pi$ coupling. On the other hand, using VMD (in the form as reviewed,
{\it e.g.}, in Ref.\ \cite{OCo95}), where the $\rho-\gamma$
coupling is described by the contraction  $\rho^{\mu \nu}F_{\mu \nu}$ of
their two field strength tensors, the pion form factor is given by
\begin{equation}
 F_\pi(Q^2) \approx  1 - \frac{ g_{\rho\pi\pi} \; Q^2}
            {g_\rho \, (Q^2 + m_\rho^2 - im_\rho \, \Gamma_\rho(Q^2))} \;.
 \label{piffvmd}
\end{equation}
The non-resonant constant term ``1'' arises from the photon coupling to the
charge of a point-like pion.  The resonant, $Q^2$-dependent term is due to 
the \mbox{$\rho-\gamma$} coupling 
and vanishes at \mbox{$Q^2=0$} in agreement with gauge invariance.
The width $\Gamma_\rho$ is non-vanishing due to $\pi\,\pi$
production only, and thus the form factor is real for $Q^2 > -4\,m_\pi^2$.
(In the model of Refs.\ \cite{Mar00,Mar00a} this width is neglected which is,
however, only of minor importance for the argument presented here.)
Employing VMD the charge radius \mbox{$r_\pi^2 = -6 F^\prime_\pi(0)$}
comes entirely from the resonant term and is \mbox{$6
g_{\rho\pi\pi}/(m_\rho^2 g_\rho) = 0.48~{\rm fm}^2$}, which compares
favourably with the experimental value $0.44~{\rm fm}^2$.

As the pion is a $\bar q q$ bound state, Eq.~(\ref{piffvmd}) cannot 
be the full truth for $F_\pi(Q^2)$ at space-like $Q^2$.  The photon
couples only to quark currents in the pion, and the vector meson
bound state is not a well-defined concept away from its pole. 
Addressing this question within the two contributions in the dressed
quark-photon vertex, Eq.\ (\ref{piffbc}) is employed to split also the 
pion charge radius into a  Ball--Chiu and a resonant contribution. 
The latter one is given by
 \begin{equation}
                \frac{6 \, g_{\rho \pi\pi}\, F_{V\pi\pi}(0)}
                {m_\rho^2 g_\rho}\; ,
\end{equation}
and a comparison to Eq.~(\ref{piffvmd}) reveals that  $F_{V\pi\pi}(0)$
characterises the necessary weakening of the VMD mechanism for $r_\pi^2$
to account for the distributed $\bar q q$ substructure. The numerical value
obtained in Ref.\ \cite{Mar00}, which is based on the solution of the
quark-photon vertex DSE,
is 0.58 or, phrased otherwise, approximately half of the pion radius is due
to the vector meson contribution and the other half is due to the longitudinal
part of the quark-photon vertex reflecting the electromagnetic Ward identity at
the level of quarks. Note that this implies that $g_{\rho \pi\pi}$ has to
decrease also by approximately 50\% when extrapolated from the physical value
at the $\rho$ meson mass-shell to the soft point $Q^2=0$. Of course, away
from the mass-shell $g_{\rho \pi\pi}$ does not have the meaning of a physical
coupling constant. 
At $Q^2=0$  the photon does not couple to a physical meson but to a 
distributed, interacting $\bar q q$ correlation, and a large portion of 
this coupling is already accounted for by the Ward-identity-preserving 
Ball--Chiu part of the vertex. The results presented in Ref.\ \cite{Mar00}
thus nicely reconcile the VMD picture with one based on the quark
substructure of pions.

\begin{figure}
\centering{\
\epsfig{figure=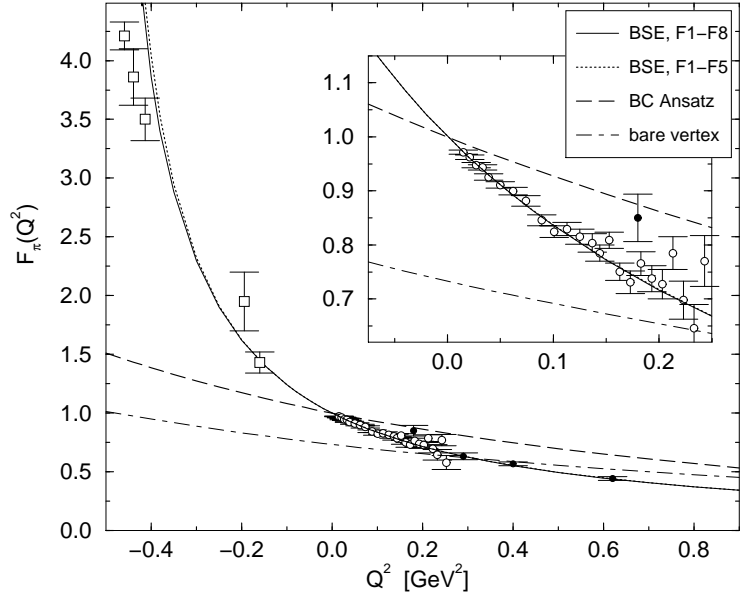,height=8.0cm} }
\caption{
The pion charge form factor $F_\pi(Q^2)$ as obtained from different
treatments of the quark-photon vertex.  The inset shows the $Q^2$ region
relevant for the charge radius.  (Adapted from  Ref.\ \protect\cite{Mar00}.)  
\label{fig:piFF} }
\end{figure}

The main result of Ref.\ \cite{Mar00} is represented in Fig.\ \ref{fig:piFF}.
Given the fact that the parameters have been fixed in the study of vector mesons
\cite{Mar99} the result is quite impressive. In Ref.\ \cite{Mar00a} the
calculation of the pion form factor has been extended up to 4 GeV$^2$. Given
the uncertainty of the experimental data this calculation should be
considered as a prediction. 
It is interesting to note that above 3.5 GeV$^2$ the result for the
pion form factor starts to deviate from the monopole fit. In Ref.\
\cite{Mar00a} also the kaon charge form factors have been calculated. The
$K^+$ form factor at several GeV$^2$ looks similar to the one of the pion,
at low momenta it is in agreement
with the experimental data which contains large errors, however. The
respective charge radii are given in Table \ref{resrK}.

\begin{table}
\begin{center}
\caption{\label{resrK}
Calculated charge radii compared with the experimental values.
(Adapted from Ref.\ \protect\cite{Mar00a}.)}
\begin{tabular}{l|rrr}
charge radii & experiment              & calculated &\\ \hline
$r_\pi^2$    & $ 0.44 \pm 0.01 $ fm$^2$&  0.45  fm$^2$ & \\
$r_{K^+}^2$  & $ 0.34 \pm 0.05 $ fm$^2$&  0.38  fm$^2$ & \\
$r_{K^0}^2$  & $-0.054\pm 0.026$ fm$^2$& $-$0.086 fm$^2$ &
\end{tabular}
\end{center}
\end{table}

In Ref.\ \cite{Haw99} the electromagnetic form factors of light vector mesons,
$G_E(Q^2)$, $G_M(Q^2)$ 
and $G_Q(Q^2)$, have been calculated in a DSE based approach. Hereby an algebraic
model for the quark propagator has been used. For the vector meson BS amplitudes
only the leading Dirac-covariants have been taken into account, and the 
corresponding
scalar function has been chosen to be of the same functional form as the
leading pion BS amplitude. The parameters have been adjusted by a fit to some
observables. The calculated static properties include the charge radii
$\langle r^2_{\rho ^+} \rangle ^{1/2} = 0.61$fm,
$\langle r^2_{K^{*+}}  \rangle ^{1/2} = 0.54$fm and
$\langle r^2_{K^{*0}}  \rangle    = -0.048$fm$^2$; the magnetic moments
$\mu_{\rho ^+} = 2.69 e/2m_\rho$, $\mu_{K^{*+}} = 2.37 e/2m_{K^{*+}}$ and
$\mu_{K^{*0}} = -0.40 e/2m_{K^{*0}}$; and the quadrupole moment
$\bar Q_\rho = 0.055$fm$^2$. 

We close this section by noting that in Ref.\ \cite{Hec98} the dipole moments
of the $\rho$ meson have been calculated with assuming electric dipole 
moments for quarks, {\it i.e.}, physics beyond the standard model. The
interesting observation made in Ref.\ \cite{Hec98} is that the use of
non-perturbative quark propagators (reflecting confinement) significantly
enhances the calculated dipole moments of the $\rho$ meson.

\subsubsection{Meson Interactions}
\label{sub_Int} 

The approach presented here is in principle capable of treating the interaction
of mesons with other mesons and photons. However, for an increasing number of
particles in the initial and the final states the practical difficulties will
rise enormously. Due to this, such investigations will be restricted to cases
which are of special interest. 
In this subsection we will review investigations of 
the coupling of the $\rho$ meson to two pions in order to demonstrate how a
finite width for a hadronic resonance is generated in this approach. The
calculation of the $\gamma \pi \rho$ form factor serves as an example that the 
approach presented here is able to provide information needed in nuclear
physics. 
The considerations of the $\pi \gamma \gamma $ and $\gamma \pi\pi \pi $ form
factors, on the other hand, provide some insight into a fundamental phenomenon
in quantum field theory, the anomaly.

Restricting oneself to the leading Dirac covariant for the $\rho$ meson BS
amplitude the $\rho \pi \pi $ vertex function can in principle be obtained
directly from the $\gamma \pi \pi $ vertex function, {\it i.e.}, the pion
charge form factor, discussed in  the last subsection.  
Since only the isovector part of the photon-quark vertex contributes to the
pion form factor, the $\rho^0 \pi
\pi $ vertex can be obtained by replacing $\Gamma_\mu /2$ in the expression
for the $\gamma \pi \pi $ vertex with $\gamma_\mu^T  V_\rho (q;Q)$ for the 
$\rho^0$ in the $\rho \pi \pi $ vertex. This has been the calculational
scheme in Ref.\  \cite{Mit97}. 
However, as this most recent  study of the $\rho \pi \pi $
coupling in the DSE approach  predates the solution of the photon-quark DSE,
the corresponding calculation is not as sophisticated as the corresponding one
for the pion form factor in the last subsection. For the pion only the leading
Dirac covariant in the chiral limit has been used, \mbox{$E_\pi(q;P) =
B(q^2,m)/f_\pi$}, and the $\rho$ amplitude has been parametrised as a
Gaussian, \mbox{$\Gamma^\rho(p^2)\propto e^{-p^2/a^2}$}. The width $a$ of this
Gaussian has been fixed by fitting the empirical value of $\rho \pi \pi $ at
the $\rho$ meson mass-shell, $g_{\rho\pi\pi}^{\rm expt}=6.05$. This
phenomenologically successful Ansatz has also been employed in studies of other
processes such as $\rho-\omega$ mixing \cite{Mit94} and the decay \mbox{$\rho
\rightarrow e^+ \; e^- $} \cite{Pic96}.  The resulting
$\rho\rightarrow\pi\pi$ decay width, given by
\begin{equation}
\Gamma_{\rho\rightarrow\pi\pi} = \frac{g_{\rho\pi\pi}^2}{4\pi}
\frac{m_\rho}{12}\left[1-\frac{4m_\pi^2}{m_\rho^2}\right]^{3/2},
\label{rhowidth}
\end{equation}
is $151~{\rm MeV}$.  The calculated form factor $F_{\rho\pi\pi}(Q^2)$
decreases rapidly with increasing $Q^2$ \cite{Mit97}. Its value at the soft
point is approximately 3, {\it i.e.}, half the value at the $\rho$ meson
pole. Therefore, the main result of Ref.\ \cite{Mit97} is a clear warning:
an approximation in which the $\rho \pi \pi $ vertex function is evaluated at
the soft point, and not at the pole, can induce an error by 100\% . Of course,
these investigations are exploratory in the sense that an independently
calculated  $\rho$ BS amplitude including subleading covariants will change 
the results. On the other hand, the strong suppression as $Q^2$ increases from
the mass-shell point to the space-like region indicates that the effective
coupling strength $g_{\rho\pi\pi}(Q^2)$ appropriate for the $\rho$ contribution
to the pion charge form factor and radius is significantly smaller than
what is typically assumed in the standard VMD approach. This is consistent 
with the findings that have been discussed in the last subsection based on a
more sophisticated calculation.

The parametrisation invoked in Ref.\ \cite{Mit97} enables a prediction for the
$\gamma\pi\rho$ vertex. This provides a check on the internal consistency of
this approach to meson physics beyond  phenomena that are dictated by chiral
symmetry.  Within nuclear physics the associated isoscalar $\gamma^*\pi\rho$
meson-exchange current contributes significantly to electron scattering off
light nuclei and is thus of phenomenological interest. Furthermore, it is an
anomalous process, {\it i.e.}, it can only occur due to the chiral anomaly of
QCD. 
As will be detailed below, this implies that independent of the model form of
the quark propagator the corresponding coupling constant at the soft point is
$g_{\gamma \pi \rho}=0.5$. Comparing to the empirical value $g_{\gamma \pi
\rho}^{\rm expt}=0.54\pm 0.03$ obtained from the experimental
$\rho^+\rightarrow\pi^+\gamma$ partial width ($67 \pm 7~{\rm keV}$), this 
indicates that the effects due to the momentum dependence of the effective
coupling are much weaker for this process. The calculated $\gamma\pi\rho$
form factor obtained with on-mass-shell $\pi$ and $\rho$ is much softer than
the one obtained in a VMD approach \cite{Tan97}.  

Hadronic processes involving an odd number of pseudoscalar mesons are of
particular interest because they are intimately connected to the anomaly
structure of QCD. The decay $\pi^0\to\gamma\gamma$ is the primary example of
such an anomalous process.  That such processes occur in the chiral limit
($m_\pi^2=0$) is a fundamental consequence of the quantisation of QCD; 
{\it i.e.}, of
the non-invariance of the QCD measure under chiral transformations even in the
absence of current quark masses \cite{Itz80}. The $\pi^0\to\gamma\gamma$ decay
rate can be calculated from a quark triangle diagram and agreement with the
observed rate requires that the number of colours $N_c$ equals three.  The
transition form factor for the related process $\gamma^\ast \pi^0 \to\gamma$
can be measured experimentally and has attracted a lot of theoretical interest
({\it e.g.}, see \cite{Kro96,Fra95a,Mar98,Kek99,Kek00,Ani00} and the references
therein), because it involves only one hadronic bound state and provides a good
test of QCD-based models and their interpolation between the soft and hard
domains. Another anomalous form factor, accessible to experiment, is the one 
that describes the transition $\gamma \pi ^* \to \pi \pi $. This provides
additional constraints on DSE based models not only because three hadronic
bound states are involved but also the pion BS amplitudes are tested at
time-like momenta \cite{Alk96a,Bis99,Bis00}. 

An important point in the treatment of anomalous processes in a DSE based
approach is the following: Using a quark-photon vertex that obeys the Ward
identity ({\it e.g.}, the Ball--Chiu vertex), and a pion BS amplitude with the
correct chiral limit, the integral appearing in the amplitude for the
anomalous   
process in the chiral limit at the soft point can be solved analytically.
First of all, as the longitudinal part of the quark-photon vertex and the
chiral limit pion BS amplitude can be expressed in terms of the functions
appearing in the quark propagator, $A(p^2)$ and $B(p^2)$, the corresponding
integrand contains exactly these functions. Introducing 
$C(s) = B^2(s)/(s\,A^2(s)) = M^2(s)/s$, the effective $\pi^0 \gamma \gamma$
coupling in the chiral limit can be written as
\begin{equation}
g^{(0)}_{\pi^0 \gamma \gamma} (0) =  - \int_0^\infty\,ds\, 
\frac{C'(s)}{(1+C(s))^3} = \int_0^\infty\,dC\,\frac{1}{(1+C)^3} = \frac 12 
\; ,
\end{equation} 
since $C(s)$ is a monotonic function for $s\ge 0$ with $C(s=0)=\infty$ and
$C(s=\infty)=0$.  Hence, the chiral limit value \cite{Itz80} is reproduced
{\it independent} of the details of the quark propagator. The same trick can
be applied to the $\gamma \to 3\pi$ amplitude in the analogous 
limits (soft point and massless pion):
 \begin{equation}
F^{3\pi}(0,0,0) = -\,\frac{eN_c}{2\pi^2}\int_0^\infty\,ds\,
        \frac{C'(s)C(s)}{(1+C(s))^4} =  \frac{eN_c}{2\pi^2f_\pi^3}
\int_0^\infty\,dC\,\frac{C}{(1+C)^4} = \frac{eN_c}{12\pi^2f_\pi^3}~.
\end{equation}
In order to obtain these results it is essential that the electromagnetic and
chiral Ward identities are satisfied by the vertex functions.  The subtle
cancellations that are required to obtain these results also make clear that
they cannot be obtained in model calculations where an arbitrary cutoff function
is introduced into each integral.  The fact that the pion BS
amplitude is proportional to the scalar part of the quark self-energy in the
chiral limit is crucial. Of course, these results remain valid with the
subleading Dirac covariants in the pion BS amplitude taken into account
\cite{Mar98}. Note also that the independence on the detailed form of the 
quark propagator also has to be seen in the calculation of the Wess--Zumino 
five-pseudoscalar term \cite{Pra88a}. It is a generic feature of the DSE based
approach which respects the underlying symmetry structure of QCD 
as closely as possible. 

Away from the soft point and for a realistic pion mass the corresponding 
form factors are, of course, model dependent. Especially at large momentum
transfers they are predictions which will be tested experimentally in the
near future. For more details we refer to the literature  
\cite{Fra95a,Kek00,Alk96a,Bis00}.

\subsubsection{Pion Loops}
\label{sub_loops} 

The meson fields have been treated so far as ladder $\bar q q$ bound states
whose structure and interactions are determined  by dressed quarks. Of course,
these mesons interact at a purely hadronic level.  To describe this type of
interactions meson loops have to be taken into account. As we will see in this
section meson loop effects are in most cases of interest numerically small. The
main reason to review some work in this direction is to provide a qualitative
demonstration that it is in principle possible to bridge the gap from QCD
Green's functions to hadronic interactions.

To describe the finite lifetime of the $\rho$ meson it is mandatory to couple
it to the two-pion channel \cite{Mit97,Tan97}. Here we discuss  the most
recent study in this direction, Ref.~\cite{Pic99}, which uses an algebraic
parametrisation of the quark propagator and of the light pseudoscalar and
vector meson BS amplitudes. 
Note that such parametrisations have also been used in
Ref.\ \cite{Haw99} to calculate the electromagnetic form factors of the light
vector mesons. As the on-shell $\rho\pi\pi$ coupling is fitted to the
experimental value, $g_{\rho\pi\pi}=6.03$, the correct $\rho$ width of 150 MeV
is reproduced. The corresponding polarisation operator displays the emergence
of an imaginary part for $q^2<-m_\rho^2$ with the expected non-analytic
behaviour. The real part of this polarisation operator leads to a mass shift
for the vector mesons. 
These contributions to the self-energies of the $\rho$ and
the $\omega$ mesons,  due to several pseudoscalar-pseudoscalar and
pseudoscalar-vector loops, are less than 10\% of the ``bare'' mass generated by
the quark core. A mass splitting $m_\omega = m_\rho \approx 25$ MeV is found
from the $\pi\pi$, $K\bar K$, $\omega\pi$, $\rho\pi$, $\omega\eta$, $\rho\eta$
and $K^*K$ channel. Despite the fact that this value is twice as large as the
experimental value  (12 MeV) this nevertheless demonstrates the effectiveness
of this approach which has neglected direct isospin breaking effects due to
different up and down quark masses. 
Thus, the inclusion of meson loops is on the one hand capable of describing
qualitative effects like the $\rho$ width and the $\omega -
\rho$ mass splitting, and on the other hand it only yields a small
correction to the predominant valence quark-antiquark structure of the vector
meson. This is emphasised by the fact that the two-pion loop provides a modest
increase in the $\rho$ charge-radius from 0.61 fm, calculated form the quark 
core only, to 0.67 fm when the pion loop is included in the calculation.

Despite this robustness of the ladder $\bar q q$ bound state picture when using
a physical pion mass, there is a question of fundamental interest concerning
the chiral limit. To this end one has to note that the non-analytic behaviour
underlying chiral perturbation theory is generated solely by pion loops
\cite{Ebe81,Gas83}. Therefore the question arises whether the approach
presented here is capable of producing this chiral-limit divergencies. This
question has been exemplified in the case of the pion charge radius
\cite{Alk95}. Chiral Perturbation Theory can be understood as the study of the
necessary consequences of the chiral Ward Identities via the construction of an
effective action, using field variables with pionic quantum numbers, in such a
way as to ensure that these identities are realised. It should be noted that in
this approach the pseudoscalar Lagrangian-field has no physical significance:
it is merely an
auxiliary field and should not be identified with the physical pion. At first
non-leading order in Chiral Perturbation Theory, O(E$^4$), the effective action
is only completely determined once the effect of one-pion loops, generated by
the O(E$^2$) part of the action, is included.  The regularisation of the
divergence of each of these loops introduces ten arbitrary parameters at this
level. The action at O(E$^4$) is completely specified once the effect of the
O(E$^2$) loops is taken into account and the ten parameters are fixed by comparison
with experimental data. These pseudoscalar loops are characteristic of the approach
and, indeed, are sometimes regarded as being the dominant feature.  The
expression for every physical observable receives a contribution from such
loops which depends on the mass of the particle in the loop and which
diverges in the chiral limit.  For example, in the case of the
electromagnetic pion charge radius, \mbox{$\langle r_\pi^2\rangle \sim \ln
m_\pi$}. In Ref.\ \cite{Alk95} the importance of these loop
contributions, evaluated at the real pion mass, relative to that of the
core of dressed-quarks, which is the dominant contribution at large
space-like $q^2$, has been estimated for the pion charge radius.

Hereby an Ansatz for the quark propagator and the leading Dirac-covariant for
the pion BS amplitude has been used. The photon coupling has been described
using the Ball--Chiu form of the quark-photon vertex. Neglecting pion
off-mass-shell effects the electromagnetic pion form factor can be written
in the form
\begin{equation}
F_\pi (Q^2) = {F}_\pi^{\rm quark \; core}(Q^2)
\left\{ 1 + I(Q^2,m_\pi^2) / f_\pi^2 \right\}
\end{equation}
where $I(Q^2,m_\pi^2)$ is a lengthy integral stemming from the pion loop. The
factorisation of ${F}_\pi^{\rm quark \; core}(Q^2)$ can be understood from the
fact that the pion loop contribution to the pion-photon coupling is nothing
else than the convolution of the quark-core pion-photon coupling with the 
$\pi\pi$ scattering amplitude. It is obvious that the pion loop describes some
additional structure of the pion in addition to the quark core.  However, the
one-pion-loop corrected value of the pion decay constant differs at most by 2\%
(slightly depending on the parameters for the quark propagator) from its quark
core value. The one-pion-loop corrected value of the pion charge radius has
been found between 6 and 14\% larger than its meson-tree-level value. To
discuss the chiral divergencies one may set the mass of all the external pions
to zero so that the contribution of the loop-pions in the chiral limit is 
easily identified when their mass $m_\pi^{\rm L}$ vanishes: 
\mbox{$m_\pi^{\rm L}\rightarrow 0$}. The quark core contribution, $r_\pi^{\rm
quark \; core}$, is regular in the chiral limit and only
weakly dependent on the current quark mass. At 
\mbox{$m_\pi^{\rm L} \approx 0.14$ GeV} the dominant contribution to the pion
charge radius is provided by the dressed-quark core. It is not until 
$m_\pi^{\rm L}$ becomes very small, $\sim
10$~MeV, that the pion cloud contribution becomes as important as the
quark core, and this contribution is well described by the form
\begin{equation}
\label{lnfit}
(r_\pi^2)^{\rm div} = (r_\pi^2)^{\rm quark \; core} \left[
  0.73 - 0.082\,\ln\left(\frac{(m_\pi^{\rm L})^2}{m_\rho^2}\right)\right]~,
\end{equation}
for \mbox{$m_\pi^{\rm L} < 0.14$ GeV}.

In Chiral Perturbation Theory the corresponding result reads \cite{Gas85}
\begin{equation}
\label{CHPT}
\langle r^2\rangle_\pi = \frac{12\,L_9^r}{F_0^2}
        - \frac{1}{32\pi^2\,F_0^2}
        \left( 2\,\ln\left[\frac{m_\pi^2}{\mu^2}\right] +
                \ln\left[\frac{m_K^2}{\mu^2}\right] + 3\right)
\end{equation}
where $\mu^2$ is the loop regularisation scale and $L_9^r$ is one of the ten
standard low-energy constants in the effective action of  Chiral Perturbation
Theory  into which an infinity from the divergence of the pseudoscalar loop has
been absorbed. Using accepted values for $L_9^r$ the first term in Eq.\
(\ref{CHPT}) provides 84 to 90 \% of the total value. One sees that at the 
physical value
of the pion mass, and with an accepted value of the renormalisation scale,
the main contribution to the charge radius is hidden in the parameter
\mbox{$L^r_9$}; it is not provided by the ``chiral logarithm''.  This can be
interpreted as a strong indication in favour of the importance of the
underlying quark-gluon degrees of freedom.

However, some warning is here in order. The parameter \mbox{$L^r_9$} cannot be
directly related to the dressed-quark core.  The flaw in such an
identification is obvious if one considers the scalar radius of the pion.  In
this case the analogous parameter in ChPT is \mbox{$L^r_4$}, which is inferred,
from fits to data, to be smaller than the ``chiral logarithm''. Nevertheless 
the scalar radius of the pion also receives its main
contribution from the dressed-quark core.

In the DSE based approach the charge radius of the pion receives a contribution
from its dressed-quark core, from pion loops, from $\rho_\mu$-$A_\nu$ mixing,
etc., each of which can be identified and calculated in a systematic manner.
The quark core contribution is finite in the chiral limit and, at
$m_\pi=0.14$~GeV, it is the dominant determining characteristic of the pion,
with the pion-loop contribution being a small, finite, additive correction of
less than 15\%.  The fact that the pion loop contribution is ultraviolet
finite  is a general property of the DSE approach: it is due to the internal
quark core structure which provides a natural cutoff in all integrals that
arise. Nevertheless the origin of the chiral divergence in the charge radius of
the pion is identified as arising from the pion loop. This feature  is present
in the DSE approach as an higher order correction, as are all meson loop
effects.

\bigskip

In this chapter we have reviewed the progress made in the last years in
describing meson as quark-antiquark 
bound states within the DSE approach. The main missing link
to the direct use of QCD Green's functions is the fact that one is still 
unable to include non-trivial quark-gluon vertex functions, obeying the
Slavnov--Taylor identities of QCD, in the BS equation. As we will see in the next
section this is not only a matter of principles but also a main obstacle in this
approach to describe baryons as bound states of confined quarks and confined
two-quark-correlations, the ``diquarks''.

\section{Baryons as Diquark-Quark Bound States}
\label{chap_Bary}

Many baryon models, too numerous to mention them all, exist. These range from
various sorts of bag \cite{Has78}, and skyrmion/soliton models
\cite{Adk83,Hol93,Wei96,Alk95a,Alk96,Chr96}, to non-relativistic 
\cite{Gel64,Kar68,Fai68}
as well as relativistic potential models \cite{Fey71}. In addition, hybrid
models exist which combine complementary aspects of the previous models.
Examples are the chiral bag \cite{Cho75,Rho94}, or also a recent hybrid model
combining the NJL soliton picture of baryons with the quark-diquark BS bound
state description within the NJL model \cite{Zue97}.

The naturally embedded framework in the present context is the description
of hadrons as bound states in relativistic Bethe--Salpeter/Faddeev equations
for particle poles in quark correlation functions (in the colour singlet
channel), the 4-point quark-antiquark Green's function for mesons, and the
6-quark Green's function for baryons.

The aim is, of course, to use the results from DSEs, in particular the ones for
the quark correlations, in the relativistic bound state equations for baryons
in a similar manner as for mesons. However, for baryons, in the simplest case
the nucleons, the relativistic bound state description is considerably less
understood, even on the phenomenological level. Their description as
relativistic quark-diquark BS bound states has been studied in the NJL model
in Refs.~\cite{Rei90,Buc92,Ish93a,Ish93b,Hua94,Buc95,Ish95,Asa95,Min99} and in
the Global Colour Model in Refs.~\cite{Bur88,Pra89}.  Other studies of nucleon
properties have simply parametrised the Faddeev amplitudes of the nucleons
\cite{Blo99,Blo00}. Some recent  developments in the quark-diquark bound state
picture based on more realistic models of quark interactions can be found in 
Refs.~\cite{Kus97,Hel97b,Oet98,Oet99,Oet00a,Oet00b,Oet00c}. While these latter
studies still entirely rely on simple model assumptions for  the quark
correlations, the mechanism modelled is sufficiently general to accommodate a
more detailed knowledge of the underlying dynamics of quarks and gluons as this
emerges. The hope is also here, of course, that the present gap between the
studies of quark and gluon correlations and the description of baryonic bound
states from these correlations will be closed in future. Until then, it will be
important to improve the understanding of the relativistic 3-quark bound state
problem from simple model assumptions, not for technical reasons alone. These
assumptions can then be replaced by more realistic ones successively.

While the non-relativistic Faddeev equation can be solved numerically in
potential models, see, {\it e.g.}, Ref.~\cite{Glo97}, the bound state problem
of 3-quark correlations in quantum field theory has to be truncated. At
present, the widely employed assumption is that the diquark correlations are
separable which allows to reduce the Faddeev problem to an effective
Bethe--Salpeter equation for the quark-diquark system. Separable diquark
correlations are obtained from the lowest lying poles in the particular
channel. The assumption therefore is that these pole contributions might
dominate the diquark correlations in the kinematic regime relevant within
baryons at not too high energies. However, as we will see in the following
section repulsive contributions to the quark-quark scattering kernel beyond the
ladder approximation are likely to prevent poles on the time-like $P^2$-axis
for diquark states.

\subsection{Goldstone Theorem and Diquark Confinement}
\label{sec_Diq}  

Besides the knowledge of the analytic structure of quark and gluon propagators
and an understanding of the absence of quark and transverse gluon
production thresholds in the $S$-matrix of hadronic colour-singlet processes, a
complete description of confinement also has to explain the absence of coloured
composite states. In particular, the existence of bound states from BS
equations  for coloured channels gives rise to the question, why such states
cannot be produced either.

A natural explanation might be obtained from a rigorous connection between
the representations of the gauge group and the sign of the BS
norm. It might for instance be established that colour non-singlet bound
states necessarily have to have negative norm and thus correspond to abnormal
solutions.

In this section it is demonstrated that there are important differences 
between otherwise analogous colour-singlet and non-singlet channels
beyond ladder approximation~\cite{Ben96,Hel97a}. In a
simple toy model of confinement and chiral symmetry breaking \cite{Mun83} 
it can be
explicitly verified that these differences suffice to remove the coloured
partner of physical mesons from the spectrum that would otherwise be bound 
 in the ladder approximation \cite{Ben96}. This same
mechanism has been verified to have the same effect in a diametrically
different model of chiral symmetry breaking, the Nambu--Jona--Lasinio (NJL)
model with an additional infrared cutoff to remove the quark thresholds
\cite{Hel97a}. This suggests, that the general 
mechanism as described below, indeed has the expected effect beyond the simple
model used for its demonstration at work.

In the last chapter it has become clear that the homogeneous BS equation for
relativistic bound states is derived under the assumption that the associated
two-body $T$-matrix has a pole in a given channel.  The absence of a solution
contradicts this assumption and establishes that no bound state exists with
the quantum numbers of the channel under consideration.

As noted in the last chapter a description of mesons in the rainbow-ladder
DS-BS equations approach is extremely rich and surprisingly successful.
However, it has the defect that it admits $(\bar 3)_c$-diquark bound
states~\cite{Pra89}. Clearly, such coloured states have not been observed.
Recently, a truncation scheme has been developed \cite{Ben96} that allows for
a systematic improvement in the kernels of the quark DSE and the meson as well
as the diquark BS equation. In this procedure the pion remains a Goldstone
boson at every order. The first correction to the rainbow-ladder approximation
is found to introduce a repulsive term in the BS kernel. Using a model gluon
propagator, singular in the infrared and with no Lehmann representation, the
quark DSE yields a quark propagator that has no Lehmann representation either.
Pairing these gluon and quark propagators in the meson channel, the repulsive
term is almost cancelled by attractive terms of the same order and therefore the
higher order terms are verified to only lead to very small changes in the meson
masses. However, due to the colour algebra of $SU(3)_c$, the
repulsive term is considerably stronger in the diquark channel and is not
cancelled by the attractive terms. This suffices to establish that the colour
anti-triplet 4-point Green's
function does not have a spectral representation with single
particle bound state poles, and that there are no corresponding diquark bound
states in this model. These arguments can be generalised to an arbitrary 
number of colours \cite{Hel97a}. The case $N_c=2$ is special: The diquarks are
then the (bosonic) baryons. Accordingly, the repulsive term does not
overwhelm the attractive one. On the other hand, 
for $N_c \to \infty$ the attractive interaction is
completely negligible, and there are no diquark correlations. Note that such a
picture is in complete agreement with the general arguments about the $1/N_c$
expansion \cite{tHo74,Wit79}.

\subsubsection{Beyond Rainbow-Ladder Approximation}
\label{sub_Beyond} 

First, we will have a look at the ladder diquark BS equation.
The analogue of the meson BS equation Eq.~(\ref{bsemeson}) for the quark-quark 
systems is
\begin{eqnarray}
\label{bsediquark}
\lefteqn{\Gamma^{EF}_D(p;P) = }\\
&&\nonumber
\int\,\frac{d^4k}{(2\pi)^4} \,K_D^{EF;GH}(k,p;P)\,
\left(S(k+\frac{1}{2}P)\Gamma_D(k;P)S^T(-k+\frac{1}{2}P)\right)^{GH}~,
\end{eqnarray}
where ``$T$'' denotes matrix-transpose and the subscript $D$ denotes diquark.
$K_D(k,p;P)$ is the kernel, $P$ is the total momentum of the system,
$k,p$ are the internal and external relative quark-quark momenta, and the
superscripts are associated with the colour, flavour and Dirac structure of
the amplitude; {\it i.e.}, $E=\{i_c,i_f,i_D\}$.
The absence of a solution to this equation entails that diquarks do not
appear in the strong-interaction spectrum.

In the iso-vector channel, each contribution to $K_M^{EF;GH}(k,p;P)$ is
matched by a direct analogue in $K_D^{EF;GH}(k,p;P)$ that can be obtained via
the replacement
\begin{equation}
\label{mesontodiquark}
S(k)\,\gamma_\mu\,\frac{\lambda^a}{2} \to
\left[\gamma_\mu\,\frac{\lambda^a}{2}\,S(-k)\right]^T~,
\end{equation}
for each antiquark in the meson kernel.
This means that a general one-to-one correspondence between every
contribution to the interaction kernels in the (iso-scalar) diquark channels
and the (iso-vector) meson channels  (with opposite
parity) can be established. In particular, for two flavours the corresponding
pairs of $(\bar 3)_c$ diquarks versus $(\ID )_c$ mesons thereby are:
the iso-scalar scalar(pseudo-scalar) diquarks versus the iso-triplet
pseudo-scalar(scalar) mesons,
and the iso-vector  axial-vector(vector) diquark versus the iso-vector
vector (axial-vector) mesons. This procedure breaks down only for the
iso-scalar meson channels which can mix with purely gluonic correlations, see
Sec.\ \ref{sub_eta}, and which thus have no partner in the diquark
correlations.

The ladder approximation to the diquark BS equation is defined by
\begin{eqnarray}
 \lefteqn{K_D^{EF;GH}(k,p;P)\,
 \left(S(k+\frac{1}{2}P)\Gamma_D(k;P)S^T(-k+\frac{1}{2}P)\right)^{GH}}\\
 && \nonumber
 \equiv g^2\,D_{\mu\nu}(p-k)\,
 \left(\gamma_\mu\,\frac{\lambda^a}{2}\,
 S(k+\frac{1}{2}P)\Gamma_D(k;P)\left[\gamma_\nu\,\frac{\lambda^a}{2}\,
 S(-k+\frac{1}{2}P)\right]^T\,\right)^{EF}
 \end{eqnarray}
in Eq.~(\ref{bsediquark}), which illustrates the application of
Eq.~(\ref{mesontodiquark}).

As discussed in the last chapter, in the chiral limit the
rainbow-approximate DSE and the pseudoscalar-meson BS equation in
ladder-approximation are identical when $P^2=0$; {\it i.e.}, one necessarily
has a massless pseudoscalar bound state when chiral symmetry is dynamically
broken \cite{Del79}. Goldstone's theorem is manifest. In any truncation, such
an identity between the quark DSE and the BS equation in the iso-vector
pseudoscalar meson channel is sufficient to ensure that this meson is a
Goldstone boson \cite{Mun95}. 

For the purpose of illustration and clarity the simple
model gluon propagator of Ref.~\cite{Mun83} is employed here,
\begin{equation}
g^2 D_{\mu\nu}(k^2) \equiv  16\pi^4\,G\,  \left(\delta_{\mu\nu} \, -\,
\frac{k_\mu k_\nu}{k^2} \right) \; \delta^4(k)~, \label{modelD}
\end{equation}
where $G=\eta^2/4$ with $\eta$ a mass scale. This one-parameter model is
appropriate in a study whose focus is confinement and DB$\chi$S. 
It allows for an
algebraic solution of the quark DSE and BS equations. The qualitative
features of the results presented here are not sensitive to this choice. 
In particular,  the same conclusions have been verified to hold also 
in the extremely different special case, the constant momentum space
interaction of the NJL model~\cite{Hel97a}.

As the next level of truncation for the quark DSE the quark-gluon vertex
function
\begin{eqnarray}
\label{vertexcorrect}
\Gamma_\nu^g(k,p) =
\gamma_\nu + \frac{1}{6}\int\frac{d^4 l}{(2\pi)^4}\,
g^2 D_{\rho\sigma}(p-l)\,
\gamma_\rho\,S(l+k-p)\gamma_\nu\,S(l)\gamma_\sigma
\end{eqnarray}
has been employed \cite{Ben96,Hel97a}. 
This is the first-order correction of the vertex by the dressed-gluon
propagator and the dressed-quark propagator, which is obtained as the
solution of this DSE. Here the explicit $3-$gluon vertex that could
contribute at this order is omitted. As a consequence one explores $3$- and
$4$-body interactions only to the extent that they are incorporated via the
non-perturbative dressing of the gluon propagator.

Using Eqs.~(\ref{vertexcorrect}) and (\ref{modelD}) in the quark DSE
one obtains
\begin{equation}
\label{newDSE}
 S^{-1}(p) = i \gamma\cdot p + m_q +
G\,\gamma_\mu S(p) \gamma_\mu + \frac{1}{8}\, G^2\,
\gamma_\mu S(p) \gamma_\nu S(p) \gamma_\mu S(p)\gamma_\nu~.
\end{equation}
Neglecting the O$(G^2)$ term yields the rainbow-approximation quark DSE of
this algebraic model.

\begin{figure}
  \centering{\
     \epsfig{file=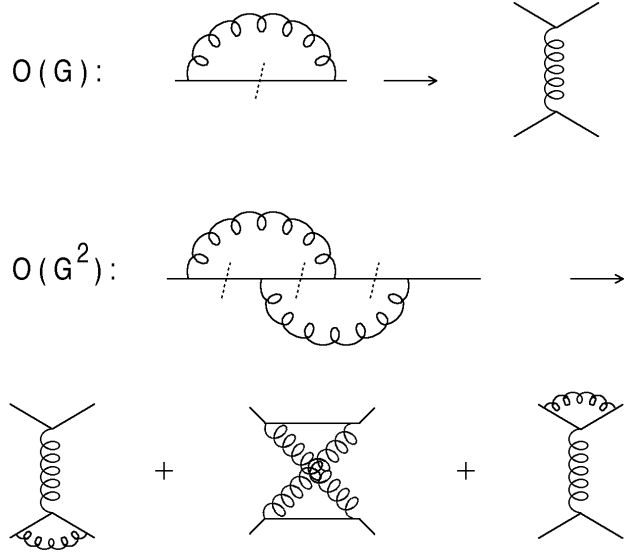,width=0.5\linewidth}  }
\caption[The kernel of the meson and diquark BS equations at next-to-leading
order. (Adapted from Ref.\ \cite{Ben96}.)]
{\label{OGtwo} Eq.~(\protect\ref{DSEtoBSEa}) illustrated to
O$(G^2)$, which provides the kernel of the meson and diquark BS equations.  
The internal solid lines represent dressed quark and gluon propagators.  The
internal quark lines of a given DSE contribution are cut sequentially to
introduce the meson total-momentum; {\it e.g.}, the 3 diagrams in the last
line are obtained by the sequential application of the cuts illustrated in
the line above. (Adapted from Ref.\ \cite{Ben96}.)}
\end{figure}

In extending the kernel of the BS equation one must preserve the Goldstone
boson character of the pion.  To this end one observes that the
ladder-approximate kernel in the meson BS equation can be obtained from the
expression for the quark self-energy in the rainbow-approximation quark DSE
via the replacement
\begin{equation}
\label{DSEtoBSEa}
\gamma_\mu\,S(k)\,\gamma_\nu \to
\gamma_\mu\,S(k+P/2)\,\Gamma_M(k,P)\,S(k-P/2)\,\gamma_\nu~,
\end{equation}
which is illustrated in the top diagram of Fig.~\ref{OGtwo}.

In the iso-vector channel this procedure can be implemented at every order;
{\it i.e.}, in every term of the quark DSE one may sequentially replace each
internal (non-perturbative) quark propagator in this way. This generates all
contributions of a given order to the kernel and ensures that Goldstone's
theorem is preserved at that order, as will become clear below. This is in
fact quite analogous to the Baym--Kadanoff procedure in many-body
physics~\cite{Kad62}. Its application to vacuum polarisation insertions does
not generate additional terms in the iso-vector kernel; a fact that very much
simplifies the study of iso-vector systems, allowing one to employ a model
gluon propagator and maintain Goldstone's theorem.  These arguments are
independent of the explicit form of the model gluon propagator.

The iso-scalar channel receives additional contributions from quark
annihilation diagrams as discussed in Sec.\ \ref{sub_eta}.
In the systematic generation described above these diagrams are
obtained from the quark DSE by applying Eq.~(\ref{DSEtoBSEa}) to the
dressed-quark loops in the gluon vacuum polarisation, which is implicit in
the dressed model gluon propagator.\footnote{Additional quark loops
arise in vertex corrections at orders beyond those employed here.}
While for a given model gluon propagator the corresponding diagrams thus
have to be added to the BS kernel at a specified truncation
explicitly, there is no need to simultaneously modify the quark DSE
in this case, since the iso-scalar meson channel is not protected by
Goldstone's theorem. 

The full O$(G^2)$ kernel for the iso-vector meson BS equation is illustrated 
in Fig.~\ref{OGtwo}. Using Eq.~(\ref{modelD}), the BS equation 
for colour singlet, iso-vector mesons is
\begin{eqnarray}
\label{mesonOGG}
\lefteqn{\Gamma_M(p;P) =
- G\,\gamma_\mu\,\chi_M\,\gamma_\mu}\\
&& \nonumber
- \frac{1}{8}\,G^2\,\gamma_\mu\,\left(
S_+\,\gamma_\nu\,S_+\,\gamma_\mu\,\chi_M
+ \underline{S_+\,\gamma_\nu\,\chi_M\,\gamma_\mu\,S_-}
+ \chi_M\,\gamma_\nu\,S_-\,\gamma_\mu\,S_-
\right)\,\gamma_\nu
\end{eqnarray}
where $S_\pm \equiv S(p\pm P/2)$ and $\chi_M\equiv S_+\,\Gamma_M(p;P)\,S_-$.
Using Eq.~(\ref{mesontodiquark}) the BS equation for $(\bar
3)_c$, iso-vector diquarks can be written
\begin{eqnarray}
\label{diquarkOGG}
\lefteqn{\Gamma_{D^{\bar 3}}^C(p;P) =
- \frac{1}{2}G\,\gamma_\mu\,\chi_{D^{\bar 3}}^C\,\gamma_\mu
- \frac{1}{16}\,G^2\,\gamma_\mu\,\left(
S_+\,\gamma_\nu\,S_+\,\gamma_\mu\,\chi_{D^{\bar 3}}^C
\right.}\\
&& \nonumber
\left.
+ 5\,\underline{
S_+\,\gamma_\nu\,\chi_{D^{\bar 3}}^C\,\gamma_\mu\,S_-}
+ \chi_{D^{\bar 3}}^C\,\gamma_\nu\,S_-\,\gamma_\mu\,S_-
\right)\,\gamma_\nu~,
\end{eqnarray}
with $\Gamma_{D^{\bar 3}}\equiv\Gamma_{D^{\bar 3}}^C C$ and
$\chi_{D^{\bar3}}^C\equiv S_+\,\Gamma_{D^{\bar 3}}^C(p;P)\,S_-$, where
$C=\gamma_2\gamma_4$ is the charge conjugation matrix.  In these equations
the underlined term is the ``crossed box'' contribution of Fig.~\ref{OGtwo}.
The differences in the numerical factors in both equations are entirely 
determined by relative weights which arise from the algebra of the $SU(3)_c$
generators for the various contributions.   

At O$(G)$ the only difference between
Eqs.~(\ref{mesonOGG}) and (\ref{diquarkOGG}) thus is that the coupling is
twice as strong in the meson equation. This would entail the existence of
scalar and axialvector diquark bound states with $m_{qq}^{\rm 0^+}>m_{\bar q
q}^{0^-}$ and $m_{qq}^{\rm 1^+}>m_{\bar q q}^{1^-}$~\cite{Cah89a}.

Only the term underlined in each of Eqs.~(\ref{mesonOGG}) and
(\ref{diquarkOGG}) is repulsive in the $0^-_{\bar q q}$, $1^-_{\bar q q}$,
$0^+_{q q}$, $1^+_{q q}$ channels.  Relative to its O$(G^2)$-companions
this term is five times as strong in the diquark equation than it is in
Eq.~(\ref{mesonOGG}), which provides for the 
possibility that it eliminates diquark bound states.
It might be interesting to see, how this result for the quite different
 weights of the repulsive to the attractive contributions in the
$(\ID )_c$ mesonic versus the $(\bar 3)_c$ diquark channel generalises for
arbitrary numbers of colours $N_c$. In Table~\ref{cweights} these weights are
given for general $SU(N_c)$. Negative signs indicate attractive
contributions. The crossed box contributions are the only repulsive
ones at this order. One can see that for all numbers of colours $N_c > 2$ the
diquark kernel is suppressed by a common overall factor $(N_c+1)/(N_c^2 - 1)$
in all contributions. For $N_c = 2$ this factor is equal to unity and no such
suppression occurs. The partial cancellation of the next order contributions
in the meson channel is also found to be independent of $N_c$, as the
repulsive crossed box term generally has the same strength as the
vertex correction contributions to the kernel in this channel.

The relevant factor for removing the diquark poles of the
ladder approximation from the spectrum is the factor $(N_c^2 - 1) - N_c$
which is unique to the repulsive crossed box term of the diquark channel.
Again, for the $SU(2)$ gauge group this factor is unity and a complete
degeneration between meson and diquark channels is obtained in this case.
The colour singlet diquarks for $N_c = 2$ can be interpreted as baryons which
are degenerate with mesons in the gauge group $SU(2)$ due to the
Pauli--G\"ursey symmetry.

\begin{table}
\caption[Weights of the different kernel contributions of Fig.~\ref{OGtwo}
corresponding to the meson and diquark BS equations.]
{Weights of the different kernel contributions of Fig.~\ref{OGtwo}
corresponding to the meson and diquark BS equations, Eqs.~(\ref{mesonOGG}) and
(\ref{diquarkOGG}) respectively, for general $SU(N_c)$ (minus signs indicate
attractive contributions).}
 \label{cweights}
\begin{center}
 \begin{tabular}{llll}
kernel   &  O$(G)$ ladder  & O$(G^2)$  vertex corr.  &  O$(G^2)$ crossed box
  \\ \hline \\[-8pt]
meson
& $ \displaystyle - \frac{3}{4} \frac{N_c^2-1}{2N_c} $  & $  \displaystyle -
\left(\frac{3}{4}\right)^2 \frac{N_c^2-1}{4N_c^2}  $ & $  \displaystyle
\left(\frac{3}{4}\right)^2 \frac{N_c^2-1}{4N_c^2} $  \\
\smallskip
diquark
& $ \displaystyle -  \frac{3}{4} \frac{N_c + 1}{2N_c} $ & $ \displaystyle -
\left(\frac{3}{4}\right)^2 \frac{N_c + 1}{4N_c^2}  $ & $  \displaystyle
\left(\frac{3}{4}\right)^2 \frac{N_c + 1}{4N_c^2} (N_c^2 - 1 - N_c)  $
\end{tabular}
\end{center}
\end{table}

In the large $N_c$ limit, on the other hand, the leading term, of ${\mathcal
O}(N_c)$, in the meson BS equation is the ladder kernel. The higher orders are
suppressed by $1/N_c$, and the ladder approximation for mesons thus gets better at
larger $N_c$. In the diquark BS equation this leading order ${\mathcal O}(N_c)$ is given
by the repulsive term in the kernel. This is in agreement with the general
large $N_c$ argument that mesons are the only relevant effective degrees of
freedom in this limit~\cite{tHo74,Wit79}.

\subsubsection{Disappearance of the Diquark Bound States}
\label{sub_Dis} 

In order to study the bound state spectrum one must solve the quark DSE. The
O$(G^2)$ corrections become noticeable for $p^2 < \eta^2/2 = 2G$, which is
important because it is the domain sampled in the BS equation.  It is also
important to note that one has dynamical chiral-symmetry breaking for any
$G>0$. Furthermore, the quark propagator possesses 
no pole on the real $p^2$-axis.
Therefore the quark propagator has certainly no Lehmann representation.

The model gluon propagator of Eq.~(\ref{modelD}) entails that the bound state
constituents have zero relative momentum; {\it i.e.}, $p=0$. In this case the
most general form of the $0^-_{\bar q q}$ BS amplitude is
\begin{equation}
\label{pibsaform}
\Gamma_{M}^{0^-}(P)=
\left[\theta_1^{0^-}(P^2)
+ i\frac{{P \kern-.5em\slash}}{\eta}  \, 
\theta_2^{0^-}(P^2)\right] \, \gamma_5 \; .
\end{equation}
Substituting Eq.~(\ref{pibsaform}) into Eq.~(\ref{mesonOGG}), with $p=0$, one
obtains a $2\times 2$ matrix eigenvalue problem of the form \mbox{$H
\,\Theta = \Theta$} where the elements of $H$ depend on $P^2$, and
$\Theta^T=(\theta_1,\theta_2$).  The eigenvalue problem is solved for 
det$\{ H(P^2)- \ID \} =0$.  The same procedure is applied to each of the other
channels.  The vector meson BS amplitude has the form
\begin{eqnarray}
\label{rhobsaform}
\Gamma_{M}^{1^-}(P)& = & \, \gamma_\mu \,  \epsilon^\lambda_\mu(P)  \,
\theta_1^{1^-}(P^2)  \, + \,
\frac{i}{\eta}\sigma_{\mu\nu}\, \epsilon_\mu^\lambda(P) \,   P_\nu \,
\theta_2^{1^-}(P^2)~,
\end{eqnarray}
where $\epsilon_\mu^\lambda(P)$, $\lambda=0,\,\pm 1$, is the polarisation vector:
$\epsilon^\lambda\cdot P=0$. $\Gamma_{D^{\bar 3}}^{0^+\,C}$ has the same form
as the pseudoscalar meson amplitude in Eq.~(\ref{pibsaform}); and
$\Gamma_{D^{\bar 3}}^{1^+\,C}$ the same form as the vector meson amplitude in
Eq.~(\ref{rhobsaform}).

The chiral limit results are presented in
Table~\ref{masstable}, the eigenvectors are:
\begin{equation}
 \label{mesoneigen}
 \begin{array}{lclcl}
  & & {\rm O}(G){\rm :O}(G) & & {\rm O}(G^2){\rm :O}(G^2) \\ \hline
 (\theta_1^{0^-},\theta_2^{0^-})& &  (0.83,0.55) && (0.87,0.49) \\
 (\theta_1^{1^-},\theta_2^{1^-}) & & (1,0) & & (0.99,-0.12)
 \end{array}
\end{equation}

\begin{table}
\caption[Calculated meson and diquark masses. (Adapted from Ref.\ 
\cite{Ben96}.)]
{\label{masstable} Calculated meson and diquark masses (in GeV). The
labels ``O$(G^n)$: O$(G^m)$'' mean that the solution of the O$(G^n)$ quark
Dyson--Schwinger equation was used in the O$(G^m)$ Bethe-Salpeter equation.
The mass scale $\eta=1.06$~GeV ($G=\eta^2/4$) and the non-zero quark mass
were chosen to reproduce the experimental ratio $m_\pi/m_\rho$ at
O$(G^2)$:O$(G^2)$.  ``Unbound'' means that there is no solution of the
associated homogeneous Bethe--Salpeter equation for real $P^2$. (Adapted 
from Ref.\ \cite{Ben96}.)}
\begin{center}
\begin{tabular}{llll}
$m_q=0$                   & O$(G)$: O$(G)$
                                & O$(G)$: O$(G^2)$
                                                & O$(G^2)$: O$(G^2)$ \\ \hline
$m_{\bar q q}^{0^-}$    & 0     & 0.30          & 0                     \\
$m_{qq}^{0^+} $         & 1.19  & Unbound       & Unbound               \\
$m_{\bar q q}^{1^-}$    & 0.750 & 0.745         & 0.823                 \\
$m_{qq}^{1^+} $         & 1.30  & Unbound       & Unbound               \\\hline
$m_q=0.012$             &       &               &                    \\ \hline
$m_{\bar q q}^{0^-}$    & 0.140 & 0.328         & 0.136                 \\
$m_{qq}^{0^+} $         & 1.21  & Unbound       & Unbound               \\
$m_{\bar q q}^{1^-}$    & 0.767 & 0.760         & 0.770                 \\
$m_{qq}^{1^+} $         & 1.32  & Unbound       & Unbound               \\\hline
\end{tabular}
\end{center}
\end{table}
The Goldstone theorem is manifest when the quark DSE and pseudoscalar meson
BS equation are truncated consistently; {\it i.e.}, at O$(G)$:O$(G)$ and
O$(G^2)$:O$(G^2)$. This can be shown analytically via a straightforward
generalisation of the arguments of Ref.~\cite{Del79}. If the dressing is
inconsistent, {\it e.g.}, O$(G)$:O$(G^2)$, the pseudoscalar is half as
massive as the vector meson. The O$(G^2)$ corrections only provide for a
small (10\%) mass increase in the vector meson channel, as one would expect
of a weak, net-repulsive interaction. It is weak because of the cancellation
between the vertex correction and crossed box contributions, which is a
necessary consequence of the preservation of Goldstone's theorem.

In the diquark channel, however, where the coefficient of the repulsive term
is larger and the cancellation incomplete, the O$(G^2)$ corrections have the
significant effect of eliminating the (bound state) pole on the real
$P^2$-axis. This means that the $(\bar 3)_c$ 4-quark correlations
do not have a spectral representation with asymptotic diquark-state
contribution.

\begin{figure}
\centerline{
  \epsfig{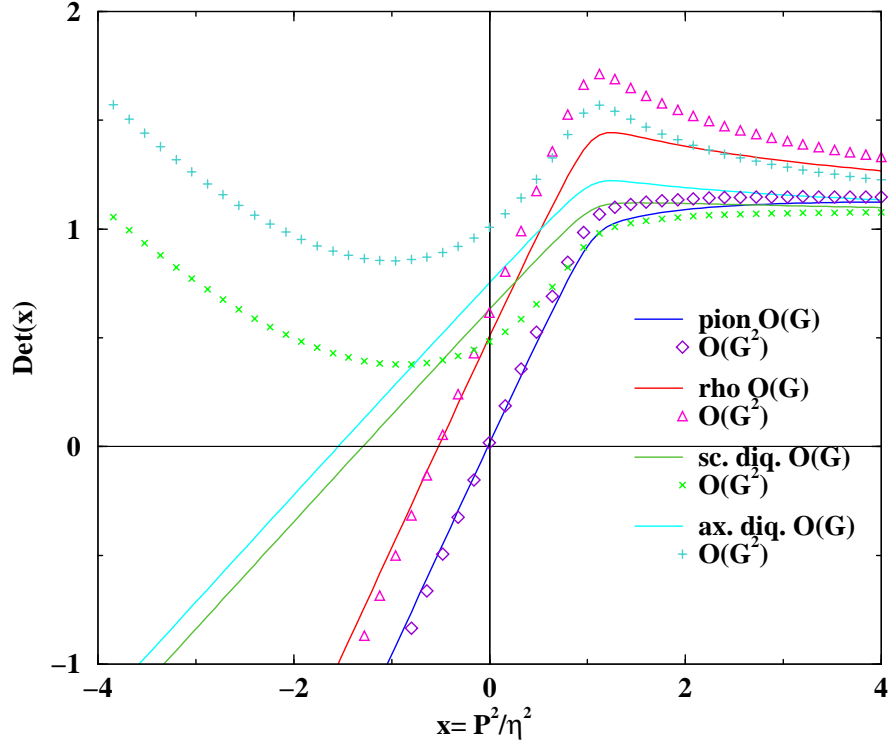}}
 \caption{det$\{ H(P^2) - \mbox{{\sf 1}\hskip -0.14em \vrule width .04em
 height 1.55ex} \, \} $  plotted as a function of $P^2$. This function
 vanishes at the square of the bound state mass in the channel under
 consideration. (Adapted from Ref.\ \cite{Ben96}.)}
\label{detHmI}
\end{figure}

The results for $m_q \neq 0$ are also presented in Table~\ref{masstable}.
The Goldstone boson character of the pseudoscalar is clear; {\it i.e.}, at a
consistent level of truncation its mass increases rapidly from zero as $m_q$
is increased.  
For $m_q=0.012$~GeV,  det$\{H(P^2)-\ID\} $ is plotted in Fig.~\ref{detHmI}.
It illustrates the effect of the O$(G^2)$ repulsive term in the
BS kernel, which shifts the zero in the meson channel very little
but completely eliminates it in the diquark channel.  

While the absence of a Lehmann representation for gluon and
quark correlations suffices to ensure there are no gluons and quarks in the
strong interaction spectrum, it neither entails nor precludes the existence of
bound states, whether coloured or not.  The rainbow-ladder truncation is
peculiar in the sense that it alone contains either purely attractive or
purely repulsive terms in the BS kernel in a given channel. In every
truncation beyond this there are both attractive and repulsive terms. The
systematic procedure for extending the rainbow-ladder kernel presented here
ensures the preservation of Goldstone's theorem at every step.

To illustrate these points the simple model gluon
propagator, Eq.~(\ref{modelD}), was employed. In a consistent truncation
there is almost complete cancellation between the O$(G^2)$ attractive and
repulsive terms in the meson channel.  This is an indication of why the
studies of meson spectroscopy and decays using a model gluon propagator in
rainbow-ladder approximation have been successful; {\it i.e.}, that the
truncation scheme may converge rapidly in the meson channel. The situation is
quite different in the diquark channel, however. The $SU(3)_c$ algebra
ensures that the coefficient of the O$(G^2)$ repulsive term is considerably
larger. This entails that the repulsive effect survives to ensure
that there is no diquark bound state. This effect cannot reasonably be
reproduced in rainbow-ladder approximation and indicates that this 
truncation is inadequate for the study of 4-quark correlations.

\subsection{Modelling of Diquarks}
\label{sec_diq} 

In the simple algebraic gluon model employed in the last section, the diquark 
poles are actually moved into the complex $P^2$-plane by the repulsive
contributions to the quark interactions. This affects the analytic properties
of the 4-quark Green's function, of course. As is seen already on the level
of the elementary quark correlations, the analytic structure of the
amplitudes in this model is quite different from the  
general analyticity properties of amplitudes in local 
quantum field theory. It is then, however,
important to study the domain of holomorphy of these correlations in detail
to verify the justification of an Euclidean formulation and its
limitations. Standard analyticity arguments cannot be used anymore.
Nevertheless, complex poles in diquark correlations
can in principle also serve to parametrise the {\sl relevant} part of
diquark correlations in baryons in a separable fashion. 

In a local quantum field theory, on the other hand, coloured
asymptotic states do in general exist, for the elementary fields as well as
possible coloured composites such as the diquarks, but not in the physical
subspace of the indefinite metric space of covariant gauge theories.   
Another realization of diquark confinement might therefore be due to a
negative norm for the corresponding amplitudes. 
The difference between physical, {\it e.g.}, meson
states and unphysical diquark states would then have to be due to their
respective norms. Very much like quarks and transverse gluons, a spectral
function for the discontinuity (at the cut along the time-like total
$P^2$-axis) in diquark correlations which is not positive would imply that
the corresponding states are in the unphysical part of the indefinite metric
space of gauge theories. It is a peculiarity of indefinite metric spaces that
possible components of such states in a properly defined positive subspace can
always be removed by an equivalence transformation, {\it i.e.}, one which
leaves physical matrix elements (in the positive subspace)
unchanged~\cite{Oeh94}. As described in the beginning of
Sec.~\ref{chap_Basic},  
given the standard analyticity properties of correlation functions, there is
then nothing to worry about the implementation of 
time-like ({\it e.g.}, bound state) momenta by analytically continuing the
Euclidean formulation. The absence of anomalous thresholds, easily
established fundamentally, is technically harder to realize in actual
calculations, however.

Leaving the question aside as to whether singularities in diquark
correlations occur at complex or time-like $P^2$ (with negative norm),
separable contributions due to isolated poles in the possibly complex
total momentum of diquark correlations are in either case the
ones of interest in the present description of baryons. The general model
building presented in the following thus assumes separable pole contributions
to parametrise diquark correlations. Quark correlations are parametrised
in the simplest case by free constituent fermion propagators. Model
correlations mimicking confinement by the absence of quark poles can
technically be implemented straightforwardly with minor modifications,
see Refs.~\cite{Hel97b,Oet98}. The notations used in this section resort to
Minkowski space conventions with formal Wick rotations performed in the last
step before the numerical calculations. Given there are no subtleties by
non-standard analyticity properties of the amplitudes, the procedure is
completely equivalent to a complex Euclidean formulation from the beginning.
It might seem more intuitive, though, to start the description of bound
states in Minkowski space.

The 4-point quark Green's function,
\begin{equation}
G_{\alpha\beta\gamma\delta}(x_1,x_2,x_3,x_4) \, =\,  \langle
T(q_\gamma(x_3) q_\alpha(x_1) \bar q_\beta(x_2) \bar q_\delta(x_4)) \rangle
\; .
\end{equation} 
is assumed to have a diquark pole in the total momentum $P$ which gives rise to
a contribution 
\begin{eqnarray}
G_{\alpha\gamma , \beta\delta}^{\hbox{\tiny pole}}(p,q,P) \, &:=&
e^{-iP\bar Y}  \, \int d^4\!X\, d^4\!y\, d^4\!z \,\,  e^{iqz} e^{-ipy} e^{iPX}
G_{\alpha\beta\gamma\delta}^{\hbox{\tiny pole}} (x_1,x_2,x_3,x_4) \nonumber \\
&=& \frac{i}{P^2 - m_d^2 +
i \epsilon} \; \chi_{\gamma\alpha}(p,P) \bar \chi_{\beta\delta}(q,P) \;
\; , \label{dq_pole_ms}
\end{eqnarray}
where the definitions $X = \eta_1 x_1 + \eta_3 x_3 $, with $ \eta_1 + \eta_3 = 1
$, and $ \bar Y = \eta_4 x_2 + \eta_2 x_4 $, with $ \eta_2 + \eta_4 = 1 $;
and $y = x_1 - x_3$, $z = x_2 - x_4$ have been used. $m_d$ denotes the diquark
mass.
The BS wave functions $\chi(p,P)$ of the diquark bound state are defined by
the matrix elements,
\begin{eqnarray}
\chi_{\alpha\beta}(x,y;\vec P) &:=& \langle q_\alpha(x) q_\beta(y) |P_+
\rangle \\
\bar \chi_{\alpha\beta}(x,y;\vec P) &:=& \langle P_+ | \bar q_\alpha(x) \bar
q_\beta(y) \rangle  
= \left(\gamma_0 \chi^\dagger(y,x;\vec P) \gamma_0 \right)_{\alpha\beta}
\label{conj_amp}\end{eqnarray}
Note that there is no need for time ordering here in contrast to
quark-antiquark bound states. The following normalisation of the states is
used,
\begin{equation}
\langle P'_\pm | P_\pm \rangle \, = \, 2 \omega_P \, (2\pi)^3 \delta^3 (\vec P'
- \vec P) \; , \quad \omega_P^2 = \vec P^2 + m_d^2 \; ,
\end{equation}
and the charge conjugate bound state is defined as $ |P_- \rangle \, =\, C | P_+
\rangle $. The contribution of the charge conjugate bound state is included in
(\ref{dq_pole_ms}) for $P_0 = - \omega_P $.
From invariance under space-time translations, the BS wave function has the
the general form,
\begin{equation}
\chi_{\alpha\beta}(x,y;\vec P) \, = \, e^{-iPX} \int
\frac{d^4p}{(2\pi)^4} \, e^{-ip(x-y)} \,    \chi_{\alpha\beta}(p,P) \; ,
\end{equation}
with  $X = \eta_x x + \eta_y y $,  $p := \eta_y p_\alpha  - \eta_x p_\beta $
, and $ \, P = p_\alpha  + p_\beta  $, where $p_\alpha$, $p_\beta$ denote the
momenta of the outgoing quarks in the Fourier transform  $
\chi_{\alpha\beta}(p_\alpha ,p_\beta;\vec P) $ of  $
\chi_{\alpha\beta}(x,y;\vec P) $. One thus has the
relation, $\chi_{\alpha\beta}(p,P) := \chi_{\alpha\beta}(p+ \eta_x P
,-p+\eta_y P ;\vec P)\vert_{P_0 = \omega} $. In the definition of
the conjugate amplitude, the convention
\begin{equation}
\bar\chi_{\alpha\beta}(x,y;\vec P) \, = \, e^{iP\bar X} \int
\frac{d^4p}{(2\pi)^4} \, e^{-ip(x-y)} \,    \bar\chi_{\alpha\beta}(p,P) \; ,
\end{equation}
with $ \bar X =  \eta_x y + \eta_y x $, ensures that hermitian conjugation
from Eq.~(\ref{conj_amp}) yields,
\begin{equation}
\bar \chi_{\alpha\beta} (p,P)\, = \, \left( \gamma_0 \chi^\dagger(p,P)
\gamma_0 \right)_{\alpha\beta} \; . \label{hcra}
\end{equation}
In the conjugate amplitude $\bar\chi_{\alpha\beta}
(p,P) $, the definition of relative and total momenta corresponds to $p =
\eta_x p'_\alpha  - \eta_y p'_\beta $ and $P = - p'_\alpha  -
p'_\beta $ for the outgoing quark momenta $p'_\alpha , p'_\beta$ in
\begin{equation}
   \bar\chi_{\alpha\beta}(p'_\alpha ,p'_\beta ;\vec P) \, =\,
\left(\gamma_0  \chi^\dagger (-p'_\beta , -p'_\alpha ;\vec P) \gamma_0
\right)_{\alpha\beta} \; .
\end{equation}
Note here that hermitian conjugation implies
for the momenta of the two respective quark legs, $ p_\alpha \to p'_\alpha =
- p_\beta $, and $ p_\beta \to p'_\beta = - p_\alpha $, which is equivalent
to $ \eta_x \leftrightarrow \eta_y$ and $P\to - P$.  Besides the hermitian
conjugation of Eq.~(\ref{hcra}), one has from the antisymmetry of the wave
function, $\chi_{\alpha\beta}(x,y;\vec P) = -\chi_{\beta\alpha}(y,x;\vec P)$.
For the corresponding functions of the relative coordinates/momenta, this
entails that $\eta_x$ and $\eta_y$ have to be interchanged in exchanging the
quark fields,
\begin{equation}
\chi(x;\vec P) \, = \, \left. - \chi^T(-x;\vec P)\right|_{\eta_x
\leftrightarrow \eta_y} \; , \quad  \chi(p, P) \, = \, \left. - \chi^T(-p,
P)\right|_{\eta_x \leftrightarrow \eta_y} \; . \label{dq_asym}
\end{equation}
This interchange of the momentum partitioning can be undone by a charge
conjugation, from which the following identity is obtained,
\begin{equation}
\chi^T(p,P) = - C \bar \chi(-p,-P) C^{-1} \; . \end{equation}
This last identity will become useful to relate $\bar\chi $ to $ \chi$ in
Euclidean space, which is a highly non-trivial task. 
In particular, this avoids the somewhat ambiguous definition
of the conjugation following from Eq.~(\ref{conj_amp}) in Euclidean space with
complex bound state momenta.

One last definition for diquark amplitudes concerns the truncation of the
quark legs, defining the amputated amplitudes $\widetilde\chi$, $\widetilde{
\bar\chi}$,
\begin{eqnarray}
\chi_{\alpha\beta}(p,P) &=& \left( S(p_\alpha) \widetilde
\chi(p,P) S^T(p_\beta)\right)_{\alpha\beta} \; , \\
\bar\chi_{\alpha\beta}(p,P) &=& \left( S^T(-p'_\alpha) \widetilde{
\bar\chi}(p,P) S(-p'_\beta)\right)_{\alpha\beta} \; . \end{eqnarray}
With the definitions above, the same relations hold for the amputated
amplitudes, in particular,
\begin{equation}
\widetilde{\bar\chi}(p,P)_{\alpha\beta} \, = \, \left( \gamma_0
\widetilde\chi^\dagger(p,P) \gamma_0 \right)_{\alpha\beta} \; .
\end{equation}

Once a gluonic interaction kernel between the quarks is specified, the diquark
amplitudes could in principle be obtained from homogeneous BS equations very
much like mesons are obtained in the quark-antiquark bound state problem.
However, as we have seen that correlations beyond the ladder approximation are
important, and because no symmetry-preserving way of including a non-trivial
quark-gluon vertex function is known, 
as a first step towards this more complete
calculation, parametrisations for scalar  and axialvector diquark amplitudes
are explored in the Bethe--Salpeter/Faddeev nucleon bound state equation. No
quark-quark BS equation for the diquark amplitudes is solved but the general
aspects anticipated for such solutions are studied in baryonic bound states.
For a given parametrisation the standard normalisation integrals are
calculated to fix the normalisation of the amplitudes. These integrals are
obtained from the inhomogeneous quark-quark BS equation under the usual
assumption that the gluonic interaction kernel be independent on the total
diquark momentum $P$. 
Given the exchange symmetry of the quark-quark BS equation such an assumption
seems reasonable also for general kernels.
For identical quarks with propagator $S(p)$ the resulting normalisation
condition reads:
\begin{eqnarray}
1 &=&  \frac{-i}{4 m_d^2} \int \frac{d^4p}{(2\pi)^4} \, \left\{ \hbox{tr}
\left( S^T(p_\beta) \widetilde{\bar\chi}(p,P) \left(
P\frac{\partial}{\partial P} S(p_\alpha)\right) \widetilde\chi(p,P) \right)
\right. \nonumber\\
&&\hskip 3cm + \,  \left.
\left(\widetilde{\bar\chi}(p,P) S(p_\alpha)  \widetilde\chi(p,P) \left(
P\frac{\partial}{\partial P} S^T(p_\beta)\right) \right)
\right\} \; , \label{dq_norm}
\end{eqnarray}
with $p_\alpha = p + \eta_1 P$ and $ p_\beta = -p + \eta_2 P$.

Contrary to the quark-antiquark and quark-quark case
the kernel of the quark-diquark BS equation necessarily depends on the total
momentum of the baryonic bound state. To demonstrate this it is sufficient
to consider only the leading Dirac-covariant of the scalar isoscalar diquark,
\begin{equation}
\chi(p,P) =  \gamma_5C \,  \frac{1}{N_s} \tilde P(p^2,pP) \; ,
\label{sdq_amp_def}
\end{equation}
where $C$ is the charge conjugation matrix. The normalisation constant $N_s$
is explicitly separated from the invariant function $\tilde P(p,P)$. This
constant is then fixed from the normalisation condition in Eq.~(\ref{dq_norm})
for a given choice of $\tilde P$. For further simplicity, it might seem
reasonable to neglect the dependence of this invariant function on the scalar
$pP$ in addition. This simplification, yielding the leading moment of an
expansion of the angular dependence in terms of orthogonal polynomials, is
known to give the dominant contribution to bound state amplitudes in many
circumstances.  However, in the present case the antisymmetry of the diquark
wave function, {\it c.f.}, Eqs. (\ref{dq_asym}), for identical quarks,
entails that
\begin{equation}
\tilde P(p^2, pP) \, =\, \left. \tilde P(p^2, - p P) \right|_{\eta_1
\leftrightarrow \eta_2} \; .
\end{equation}
For general $ \eta_1 = 1 - \eta_2 \not= 1/2 $ and thus for $\bar p := p
\big|_{\eta_1 \leftrightarrow \eta_2} \not= p$, it is therefore not possible
to neglect the $pP$ dependence in the amplitude completely without violating
the quark-exchange symmetry. To correct this, one may assume instead that the
amplitude depends on both scalars, $p^2$ and $pP$ in a specific way as to
guaranty this symmetry, namely via the scalar  $x :=
p_\alpha p_\beta - \eta_1 \eta_2 \, m_d^2 = (\eta_2 - \eta_1) pP - p^2  = -
(\eta_2 - \eta_1) \bar pP - \bar p^2 $  with $\bar p  =   \eta_1 p_\alpha -
\eta_2 p_\beta $, and with the definitions of $p_{\{\alpha , \beta\}}$ given
above. For $\eta_1 = \eta_2 = 1/2$ this coincides with the usual relative
momentum ($x = p^2$). Note further that $P^2 = m_d^2 $ is not a free
variable of the amplitude. For the two remaining scalars built out of the two
momenta $p$ and $P$ the particular choice with definite exchange symmetry is
given by the two independent combinations $p_\alpha p_\beta$, which is
essentially the same $x$ as above, and $ p_a^2 - p_\beta^2 $. The latter has
to appear in odd powers, {\it i.e.}, in higher moments. These are neglected
thus setting,
\begin{equation}
\tilde P(p^2, pP) = P(x)  =  P( (\eta_2 - \eta_1) pP - p^2 )  = P( (\eta_1 -
\eta_2) \bar pP - \bar p^2 )\; . \label{dq_rel_mom}
\end{equation}

The parametrisations of diquark amplitudes used in the baryon models of
Refs.~\cite{Kus97,Hel97b,Oet98} actually neglect possible sub-structure
completely, corresponding to $P(p^2, pP) \equiv 1$. In this case, the
normalisation constant $N_s$ is ill-defined. An additional free parameter is
introduced as the strength of the diquark-quark coupling which in a more
realistic description corresponds to exactly the normalisation of the diquark
amplitude as obtained from Eq.~(\ref{dq_norm}). More recent studies
\cite{Oet99,Oet00b} have improved on
this model assumption and included a sub-structure of diquarks. As the
quark-diquark coupling is then not at ones
disposal, it is assessed this way whether in such a calculation the
quark-diquark approximation can lead to a viable description of baryons for
coupling strengths as obtained from BS diquark amplitudes rather
than adjusted as free parameter.

The discussion of this section demonstrates some general aspects in such
a parametrisation, most importantly, due to the exchange symmetry of
the amplitude. The precise shape of the function $P(x)$ can be conveniently 
fixed from the calculated electromagnetic form factor of the nucleon, see
Sec.\ \ref{sec_EM} below. It will turn out that a dipole form 
$P_d(x) \, = \, \lambda^4/(x + \lambda^2)^2 $ for the diquark amplitude
will lead to good results for form factors. The
respective width $\lambda$  indirectly determines the coupling strengths 
to the quarks. These widths are adjusted in the calculation so as to yield the
diquark normalisation $1/N_s$ necessary from fixing the nucleon mass. From
solutions to diquark BS equations in a next step, the question of the necessary
diquark-quark coupling strengths to yield sufficiently strong binding to the
3rd quark in baryons by the mechanism described below will be shifted
further to the strength of the gluonic quark-quark interaction kernel in the
diquark channels. The finite width of the diquark correlations
implies a further technical improvement. While some ultraviolet
regularisation is necessary with the point-like diquark-quark couplings of
Refs.~\cite{Kus97,Hel97b,Oet98}, in a more realistic calculation sufficient
ultraviolet convergence is naturally provided by the damping from the diquark
amplitude.

\subsection{Quark-Diquark Bethe--Salpeter Equations}
\label{sec_qdq} 

\subsubsection{\label{sub_Gen} General Structure of Bethe--Salpeter
 Amplitudes of Octet and Decuplet Baryons}

\begin{figure}
 \begin{center}
   \epsfig{file=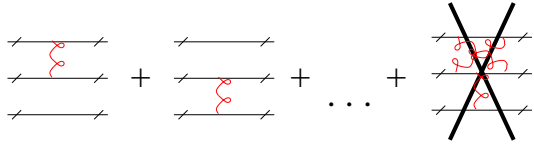,width=7cm}
 \end{center}
 \caption{Examples for admitted and excluded graphs in the 3-quark
          interaction kernel $K$. (Adapted from Ref.\ \cite{Oet00c}.)}
 \label{gluekern}
\end{figure}

\begin{figure}
 \begin{center}
   \epsfig{file=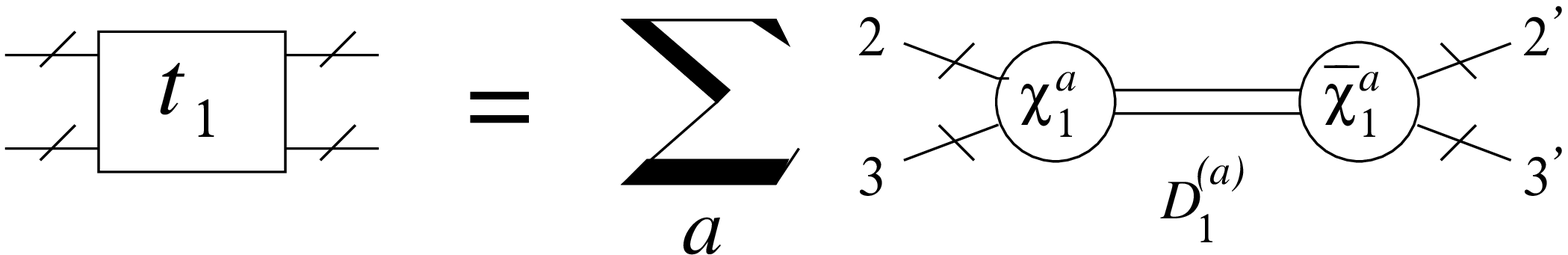,width=7cm}
 \end{center}
 \caption{The separable $t$-matrix. (Adapted from Ref.\ \cite{Oet00c}.)}
 \label{tsep_fig}
\end{figure}

Without a genuine 3-quark scattering kernel, see Fig.\ \ref{gluekern}, and
assuming separability of the quark-quark $t$-matrix, see Fig.\ \ref{tsep_fig},
the Dyson series for the full 3-quark propagator reduces to a coupled set of
Bethe--Salpeter/Faddeev equations for baryons as bound states of quark and
diquark correlations, see, {\it e.g.}, Ref.~\cite{Hua94}. Hereby, neither for
quark nor for diquark correlations a particle interpretation has to be assumed.
In particular, the separability assumption for the quark-quark $t$-matrix may
or may not be realized by a sum over particle pole contributions. 
To be concrete we will choose the following ansatz for the quark-quark
scattering matrix (returning thereby to Euclidean notations)
\begin{eqnarray} \label{tsep}
 t(k_\alpha,k_\beta;p_\alpha,p_\beta) \equiv t(k,p,P) &=&
 \chi^5_{\alpha\beta}(k,P) \;D(P)\;\bar
  \chi^5_{\gamma\delta}(p,P) \; +\; \\
  & &
\chi_{\alpha\beta}^\mu(k,P) \;D^{\mu\nu}(P)\;
    \bar  \chi_{\gamma\delta}^\nu(p,P)  \; . \nonumber
\end{eqnarray}
The relative momenta are defined as
\begin{equation}
 k[p]= \sigma\, k_\alpha[p_\alpha]- (1-\sigma)\, k_\beta[p_\beta],
    \qquad \sigma \in [0,1]\; .
\end{equation}

The diquark propagators used in the scalar and the axialvector channel
are taken to be
\begin{eqnarray}
D(P) &=& -\frac{1}{P^2+m_{sc}^2}\, C(P^2,m=m_{sc}) \; ,
 \label{Ds} \\
D^{\mu\nu}(P) &=& -\frac{1}
   {P^2+m_{ax}^2} \left( \delta^{\mu\nu}+  (1-\xi) \frac{P^\mu P^\nu}{m_{ax}^2}
\right)\, C(P^2,m=m_{ax})  \; .
 \label{Da}
\end{eqnarray}
With the choice $C(P^2,m)=1$ and $\xi=0$ they correspond to free propagators of
a spin-0 and  a spin-1 particle. This will be the form to be employed in this
and the next section. Later on we will review investigations using
nontrivial forms for $C$ which remove the
free-particle poles from the real axis and thus may mimic confinement.

Equipped with the separable form of the two-quark correlations,
Eq.~(\ref{tsep}), we will now derive the BS equation for the nucleon and the
Delta baryon. The generalisation to octet and decuplet baryons (using three
flavours in the isospin limit) is given in Ref.\ \cite{Oet98}, and we will
shortly comment on it at the end of this section. To complete the model
definition, we have to specify the functional form of the quark propagator. We
take it to be a free fermion propagator with a constituent mass $m_q$,
\begin{equation}
 S(p)= \frac{i{p\hspace{-.5em}/\hspace{.11em}} -m_q}{p^2+m_q^2}
 i\, C(p^2,m=m_q) \;
  \qquad  {\rm with}\;\; C(p^2,m=m_q)=1.
 \label{qprop}
\end{equation}
Of course, when using modified diquark propagators, $C\not = 1$, the quark
propagator will be altered accordingly. 

\begin{figure}
 \begin{center}
   \epsfig{file=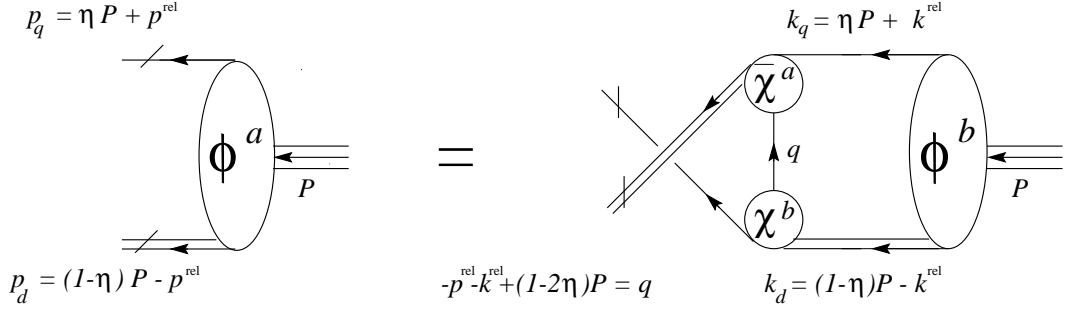,width=14cm}
 \end{center}
 \caption{The coupled set of BS equations
          for the vertex functions $\Phi$. (Adapted from Ref.\
          \cite{Oet00c}.)}
 \label{bse_fig}
\end{figure}

The nucleon BS amplitudes (or wave functions) can be described by
an effective multi-spinor characterising the
scalar and axialvector correlations,
\begin{equation}
 \Psi (p,P) u (P,s) \equiv
    \pmatrix {\Psi^5 (p,P) \cr \Psi^\mu  (p,P)} u(P,s).
\end{equation}
$u(P,s)$ is a positive-energy Dirac spinor of spin $s$, $p$ and $P$ are the
relative and total momenta of the quark-diquark pair, respectively.
The vertex functions are defined by truncation of the legs,
\begin{equation}
 \pmatrix{ \Phi^5  \cr \Phi^\mu } =
    S^{-1}  \pmatrix{  D^{-1} & 0 \cr 0  & (D^{\mu\nu})^{-1} }
 \pmatrix{ \Psi^5  \cr \Psi^\nu } .
 \label{amp}
\end{equation}
The diquark propagators $D$ and $D^{\mu\nu}$ are given in
Eqs.\ (\ref{Ds},\ref{Da}), and the quark propagator $S$ in Eq.\ (\ref{qprop}).
The coupled system of BS equations for the nucleon amplitudes or their vertex
functions can be written in the following compact form,
\begin{equation}
  \int \frac{d^4p'}{(2\pi )^4} G^{-1}(p,p',P)
  \pmatrix{\Psi^5 \cr \Psi^{\mu'}}(p',P) =0 \;,
  \label{bse_nuc}
\end{equation}
in which $G^{-1}(p,p',P)$ is the inverse of the full quark-diquark 4-point
function. It is the sum of the disconnected part and the interaction kernel.
The latter results from the reduction of the Faddeev equation for separable
quark-quark correlations. It describes the exchange of the quark with one of
those in the diquark which is necessary to implement Pauli's principle in the
baryon, {\it i.e.}, it describes the minimal dynamical coupling necessary to
account for the full exchange symmetry in the quark-diquark
model~\cite{Rei90}. Due to the overall colour antisymmetry of the baryon
(being a colour singlet) the other quantum numbers have to be symmetrised
leading to Pauli attraction instead of Pauli repulsion familiar from most
ordinary fermionic many-body systems. Taking into account the coupled channel
nature of scalar and axialvector diquark contributions within the nucleon
one obtains
\begin{eqnarray}
 G^{-1} (p,p',P) &=&
    (2\pi)^4 \;\delta^4(p-p')\; S^{-1}(p_q)\;
      \pmatrix{ D^{-1}\!(p_d) &  0 \cr 0 & (D^{\mu'\mu})^{-1}\! (p_d) }
          \nonumber\\
        &&  \nonumber\\
 & - &  \frac{1}{2}
  \pmatrix{ - \chi{ (p_2^2) } \; S^T{ (q) }\; \bar\chi{
  (p_1^2) } &
     \sqrt{3}\; \chi^{\mu'}{ (p_2^2) }\; S^T{ (q) }\;\bar\chi {
  (p_1^2) } \cr
    \sqrt{3}\;\chi{ (p_2^2) }\; S^T{ (q) }\;\bar\chi^{\mu}{ (p_1^2) } &
      \chi^{\mu'}{ (p_2^2) }\; S^T{ (q) }\;\bar\chi^{\mu}{ (p_1^2) }
     } \; . 
 \label{Kdef}
\end{eqnarray}
The flavour and colour factors have been taken into account explicitly,
and $\chi, \, \chi^{\mu}$ stand for the Dirac structures of the
diquark-quark vertices.
The freedom to partition the total momentum between quark and diquark
introduces the parameter $\eta \in [0,1]$ with $p_q=\eta P+p$ and
$p_d=(1-\eta)P - p$ as usual. The momentum of the exchanged quark is then 
given by
$q=-p-p'+(1-2\eta)P$. The relative momenta of the quarks in the diquark
vertices  $\chi$ and  $\bar\chi$ are $p_2=p+p'/2-(1-3\eta)P/2$ and
$p_1=p/2+p'-(1-3\eta)P/2$, respectively.
Invariance under 4-dimensional translations implies that for
every solution  $\Phi(p,P;\eta_1)$ of the BS equation there exists
a family of solutions of the form $\Phi(p+(\eta_2-\eta_1)P,P;\eta_2)$.
The corresponding BS equations are pictorially represented in Fig.\ 
\ref{bse_fig}.
The necessary presence of the total momentum $P$ of the baryonic bound
state in the exchange kernel for $\eta \not= 1/2$ was apparently not taken
into account in the studies of Refs.~\cite{Kus97,Hel97b}.

As stated in the last section the quark exchange kernel
of the reduced Bethe--Salpeter/Faddeev problem for baryons, for
$\eta \not= 1/2$, necessarily depends on the total momentum of the baryonic
bound state $P$. Since this has important implications on the normalisations
and charges of the bound state amplitudes, it is preferable to use the
residual freedom in choosing the momentum partitionings in the relativistic
bound state problem such as to keep the bound state momentum dependence of
the exchange kernel to a necessary minimum. While the exchange quark momentum
is found to be $P$-independent for $\eta = 1/2 $, this choice, however,
necessarily introduces $P$-dependence in the diquark amplitudes:
The dominant momentum dependence of the diquark amplitudes are given by the 
scalars $x_1$ and $x_2$,
\begin{eqnarray}
x_1 \, &=& \, - p_1^2 - (1-2\sigma) ((1-\eta) p_1P - p_1k) \; , \label{x_1}\\
x_2 \, &=& \, - p_2^2 + (1-2\sigma') ((1-\eta) p_2P - p_2p) \; . \label{x_2}
\end{eqnarray}
These coincide with $p_{1,2}^2$ only for symmetric quark momentum
partitionings, {\it i.e.}, $\sigma = \sigma' = 1/2$, {\it c.f.}, the
discussion at the end of the last section. 
These symmetrised arguments of the diquark amplitudes $x_{1,2}$ are
independent of the total nucleon momentum, only if $ \sigma = \sigma' =
\frac{1}{2} $ and thus $\eta = \frac{1}{3}$.
This conclusion can be generalised~\cite{Oet99}: The exchange symmetry
of the diquark amplitudes suffices to show that these can generally be
independent of $P$ only if $\eta = 1/3$ and $\sigma = \sigma' = 1/2 $.
This is the only choice leading to diquark amplitudes independent of
the total nucleon bound state momentum $P$, and this follows from the exchange
symmetry alone and is {\sl not} a result of the particular parametrisations
employed in the model calculation.

In actual calculations the variable $\eta$ is varied around the value 
$\eta = 1/3$ \cite{Oet00b}. 
The diquark momentum partitionings are fixed  to $\sigma =
\sigma ' = (1-2\eta )/(1-\eta )$  for given $\eta$.  
While $P$-independent
diquark amplitudes can be obtained only for the value $\eta = 1/3$ with this
choice the exchange quark carries total momentum whenever $\eta \not= 1/2$.
This entails that the exchange kernel of the reduced Bethe--Salpeter/Faddeev
equation for baryons unavoidably depends on the total momentum of the baryonic
bound state. This implies some considerable extensions to the calculations,
{\it e.g.}, of electromagnetic form factors, which become necessary with the
inclusion of diquark sub-structure~\cite{Oet99,Oet00b}.

Using the positive energy projector with nucleon bound state mass $M_n$,
\begin{equation}
 \Lambda^+= \frac{1}{2}\left( 1+ \frac{P\hspace{-.5em}/\hspace{.15em}}{iM_n}
 \right),
\end{equation}
the vertex functions can be decomposed into their most general Dirac
structures,
\begin{eqnarray}
 \Phi^5(p,P)&=& (S_1 +\frac{i}{M_n}{p\hspace{-.5em}/\hspace{.15em}} S_2) 
 \Lambda^+, \qquad
   \label{phi5deco}\\
 \Phi^\mu(p,P)&=& \frac{P^\mu}{iM_n} (A_1 +\frac{i}{M_n}
 {p\hspace{-.5em}/\hspace{.15em}} A_2) \gamma_5
      \Lambda^+  + 
      \gamma^\mu (A_3 +\frac{i}{M_n}
              {p\hspace{-.5em}/\hspace{.15em}}A_4)
                   \gamma_5\Lambda^+ + \label{phimudeco} \\ & & 
                   \frac{p^\mu}{iM_n}( A_5 + \frac{i}{M_n} 
              {p\hspace{-.5em}/\hspace{.15em}}A_6)
                  \gamma_5\Lambda^+ \; . \nonumber 
\end{eqnarray}
In the rest frame of the nucleon, $P = ( \vec 0 , iM_n)$,
the unknown scalar functions $S_i$ and $A_i$ are functions of $p^2=p^\mu
p^\mu$ and of $P \cdot p$.
Certain linear combinations of these eight covariant components then
lead to a full partial wave decomposition, see the next section.

The BS solutions are normalised by the canonical condition
\begin{eqnarray}
  M_n \Lambda^+ \;& \stackrel{!}{=}&
  -\int \frac{d^4\,p}{(2\pi)^4}
  \int \frac{d^4\,p'}{(2\pi)^4} \label{normnuc} 
   \bar \Psi(p',P_n) \left[ P^\mu \frac{\partial}{\partial P^\mu}
    G^{-1} (p',p,P) \right]_{P=P_n} \hskip -.5cm  \Psi(p,P_n) \; .
 \end{eqnarray}

The effective multi-spinor for the delta baryon representing
the BS wave function can be characterised as $\Psi_\Delta^{\mu\nu}(p,P)
u^\nu(P)$ where $u^\nu(P)$ is a Rarita-Schwinger spinor.
The momenta are defined analogous to the nucleon case.
As the delta state is flavour symmetric, only the axialvector
diquark contributes  and,  accordingly, the corresponding BS equation reads,
\begin{eqnarray}
  \int \frac{d^4p'}{(2\pi )^4} G^{-1}_\Delta (p,p',P)
  \Psi_\Delta^{\mu'\nu}(p',P) =0 \; ,
  \label{bse_del}
\end{eqnarray}
where the inverse quark-diquark propagator $G^{-1}_\Delta$ in the
$\Delta$-channel is given by
\begin{eqnarray}
  G^{-1}_\Delta(p,p',P) &=&  (2\pi)^4 \delta^4(p-p')\; S^{-1} (p_q) \;
        (D^{\mu\mu'})^{-1} (p_d) + 
      \chi^{\mu'}(p_2^2)\; S^T(q)\;\bar\chi^\mu(p_1^2).
\end{eqnarray}
The general decomposition of the corresponding vertex function
$\Phi^{\mu\nu}_\Delta$, obtained as in Eq.\ (\ref{amp})
by truncating the quark and diquark legs of the BS wave function
$\Psi_\Delta^{\mu\nu}$, reads
\begin{eqnarray}
 \Phi^{\mu\nu}_\Delta (p,P) &=& (D_1 + \frac{i}{M_\Delta} 
                  {p\hspace{-.5em}/\hspace{.15em}}   D_2) \
                        \Lambda^{\mu\nu} + 
        \frac{P^\mu}{iM_\Delta} (E_1 + \frac{i}{M_\Delta} 
                 {p\hspace{-.5em}/\hspace{.15em}}   E_2)
        \frac{p^{\lambda T}}{iM_\Delta} \Lambda^{\lambda\nu} +
        \label{Deldec}\\
   & &  \gamma^\mu (E_3 + \frac{i}{M_\Delta} 
             {p\hspace{-.5em}/\hspace{.15em}}   E_4 )
        \frac{p^{\lambda T}}{iM_\Delta} \Lambda^{\lambda\nu} + 
        \frac{p^\mu}{iM_\Delta} ( E_5 + \frac{i}{M_\Delta} 
                {p\hspace{-.5em}/\hspace{.15em}}  E_6)
        \frac{p^{\lambda T}}{iM_\Delta} \Lambda^{\lambda\nu} \; . \nonumber
\end{eqnarray}
Here, $\Lambda^{\mu\nu}$ is the Rarita-Schwinger projector,
\begin{eqnarray}
  \Lambda^{\mu\nu} \!= \Lambda^+
             \left(
             \delta^{\mu\nu}-\frac{1}{3}\gamma^\mu\gamma^\nu
             +\frac{2}{3} \frac{P^\mu P^\nu}{M_\Delta^2} -
             \frac{i}{3} \frac{P^\mu\gamma^\nu-P^\nu\gamma^\mu}{M_\Delta}
              \right)
             \end{eqnarray}
 which obeys the
constraints
\begin{equation}
  P^\mu \Lambda^{\mu\nu} = \gamma^\mu \Lambda^{\mu\nu} =0.
\end{equation}
Therefore, the only  non-zero components arise from the
contraction with the transverse relative momentum
$p^{\mu T}=p^\mu - P^\mu (p\cdot P)/P^2$.
The invariant functions $D_i$ and $E_i$ in Eq.~(\ref{Deldec}) again depend
on $p^2$ and $ p\cdot P$.
The partial wave decomposition in the rest frame is made explicit below.

Finally, we want to comment on an extension to three flavours. In the isospin
limit the strange quark constituent mass is the only source of flavour symmetry
breaking. The equations describing octet and decuplet baryons have been derived
under the premises of flavour and spin conservation, {\it i.e.}, only those
wave function components with the same spin and flavour content couple to
each other \cite{Oet98}. The
flavour structure of the eight equations describing $N, \Lambda, \Sigma, \Xi,
\Delta, \Sigma^*, \Xi^*$ and $\Omega$ can be found in Appendix A of Ref.\
\cite{Oet98}. The $\Lambda$-hyperon is hereby of special interest. First, its
measured polarisation asymmetry in the process $p\gamma \rightarrow K^+\Lambda$
will provide  a stringent test for the diquark-quark model for time-like
momenta, see below. As discussed in \cite{Kro97}, there are only scalar
diquarks involved in this process. Secondly, broken $SU(3)$-flavour symmetry 
induces a component of the total antisymmetric flavour singlet into wave and
vertex function. In non-relativistic quark models with $SU(6)$ symmetry such a
component is forbidden by the Pauli principle. As the flavour singlet is  
composed of scalar diquarks and quarks only, this generates two additional
scalar amplitudes besides the usual two from the octet $\Lambda$ state. The
axialvector part of the vertex function remains unchanged in flavour space.
In Ref.\ \cite{Oet98} it has been found that the flavour singlet amplitudes
are numerically small. Thus, one can safely
regard the $\Lambda$-hyperon as an almost pure octet state in flavour space.

\subsubsection{\label{sub_Par}Partial wave decomposition }

In a relativistic system only the total angular momentum, 1/2 for
the nucleon and 3/2 for the $\Delta$, is a good quantum number.
Nevertheless it is instructive to decompose the BS amplitudes into 
partial waves in the rest frame. However, it has to be noted that 
these partial waves start to mix when the covariant amplitude is boosted.

In the rest frame the Pauli--Lubanski operator for an
arbitrary multi-spinor $\psi$ is given by
\begin{equation}
 W^i=\frac{1}{2}\epsilon_{ijk}{\mathcal L}^{jk} \; .
\end{equation}
Its eigenvalues are the total angular momentum
\begin{equation}
 {\mathbf W}^2{\psi}=J(J+1) \psi  \; .
\end{equation}
The tensor ${\mathcal L}^{jk}$ is the sum of an orbital part, $L^{jk}$, and
a spin part, $S^{jk}$. For a three-particle system they are given by
\begin{eqnarray}
 L^{jk}&=&\sum_{a=1}^3 (-i)\left(p_a^j\frac{\partial}{\partial p_a^k}-
         p_a^k\frac{\partial}{\partial p_a^j}\right) \; , \\
 2(S^{jk})_{\alpha\alpha',\beta\beta',\gamma\gamma'}&=&
 (\sigma^{jk})_{\alpha\alpha'}\otimes{\delta}_{\beta\beta'}\otimes{\delta}_{\gamma\gamma'}+
 {\delta}_{\alpha\alpha'}\otimes (\sigma^{jk})_{\beta\beta'}
 \otimes{\delta}_{\gamma\gamma'}+ \nonumber \\
  & & {\delta}_{\alpha\alpha'}\otimes{\delta}_{\beta\beta'}\otimes(\sigma^{jk})_{\gamma\gamma'} \; ,
\end{eqnarray}
such that ${\mathcal L}^{jk}=L^{jk}+\frac{1}{2}S^{jk}$.
The tensor $L^{jk}$ is proportional to the unit matrix in Dirac space.
The definition of $\sigma^{\mu\nu}:=-\frac{i}{2}[\gamma^\mu,\gamma^\nu]$ differs
by a minus sign from its Minkowski counterpart.
The tensors $L$ and $S$ are written
as a sum over the respective tensors for each of the three constituent quarks
which are labelled $a=1,2,3$ and with respective Dirac indices
${ \alpha\alpha', \beta\beta',\gamma\gamma'}$.

Defining the spin matrix $\Sigma^i=\frac{1}{2}\epsilon_{ijk}\sigma^{jk}$ the
Pauli--Lubanski operator can be written as
\begin{eqnarray}
(W^i)_{\alpha\alpha',\beta\beta',\gamma\gamma'} &=& L^i\,
  {\delta}_{\alpha\alpha'}\otimes{\delta}_{\beta\beta'}\otimes\delta_{\gamma\gamma'}
  +(S^i)_{\alpha\alpha',\beta\beta',\gamma\gamma'}\; ,  \\
L^i    &=&(-i)\epsilon_{ijk} \left[ p^j\frac{\partial}{\partial p^k}
                   + q^j \frac{\partial}{\partial q^k} \right] \; , \label{Ldef} \\
(S^i)_{\alpha\alpha',\beta\beta',\gamma\gamma'}   &=& \frac{1}{2} \left(
   (\Sigma^{i})_{\alpha\alpha'}\otimes{\delta}_{\beta\beta'}\otimes
    {\delta}_{\gamma\gamma'}+
 {\delta}_{\alpha\alpha'}\otimes (\Sigma^{i})_{\beta\beta'}\otimes
 {\delta}_{\gamma\gamma'}+ \right. \nonumber \\
    & & \left.{\delta}_{\alpha\alpha'}\otimes{\delta}_{\beta\beta'}\otimes
 (\Sigma^{i})_{\gamma\gamma'}
 \right) \, .
\end{eqnarray}
Hereby the relative momentum $p$ between quark and diquark
and the relative momentum $q$ within the diquark has been introduced
via a canonical transformation:
\begin{equation}
P=p^1+p^2+p^3,\quad p=\eta (p^1+p^2)-(1-\eta)p^3, 
\quad q=\frac{1}{2}(p^1-p^2) \; .
\end{equation}
Taking into account only the leading Dirac covariant in the diquark amplitudes
no orbital angular momentum is carried by the diquarks,
\begin{equation}
 {\mathbf L}^2 {\chi}(q)= {\mathbf L}^2 {\chi^\mu}(q) =0 \; .
\end{equation}
The Pauli--Lubanski operator then simplifies and can be calculated by
straightforward but tedious algebra \cite{Oet98,Oet00c}.

\begin{table}
\caption[Components of the octet baryon wave function with their
respective spin and orbital angular momentum.
(Adapted from Ref.\ \cite{Alk99}.)]
{Components of the octet baryon wave function with their
respective spin and orbital angular momentum.
$(\gamma_5 C)$ corresponds to
scalar and $(\gamma^\mu C),\,\mu=1 \dots 4,$ to axialvector
diquark correlations. Note that the partial waves in the first row
possess a non-relativistic limit. (Adapted from Ref.\ \cite{Alk99}.)}
\begin{center}
\begin{tabular}{p{0.1cm}p{1.8cm}cccc}\hline
 &&&&& \\
\multicolumn{2}{l}{\parbox{1.9cm}{ { \mbox{``non-relat.''} \mbox{partial waves}} }  } &
  $\pmatrix{ \scriptstyle \chi \cr \scriptstyle 0 } \scriptstyle{(\gamma_5 C)}$    &
  $\scriptstyle\hat P^4{\pmatrix{ \scriptstyle 0\cr \scriptstyle \chi}} (\gamma^4 C)$    &
  ${\pmatrix{\scriptstyle i\sigma^i\chi \cr \scriptstyle 0}} \scriptstyle (\gamma^i C)$    &
  ${\pmatrix{\scriptstyle i\left(\hat p^i(\vec{\sigma}\hat{\vec{p}})-\frac{\sigma^i}{3}\right) \chi\cr \scriptstyle 0}} \scriptstyle (\gamma^i C)$ \\
 &  {spin} & {1/2} & {1/2} & {1/2} & {3/2} \\
 &  {orb.ang.mom.} & { $s$} & {$s$} & {$s$} & {$d$} \\
 &                  &        &       &       &      \\
\multicolumn{2}{l}{\parbox{1.9cm}{ { \mbox{``relat.''} \mbox{partial waves} }  }} &
  $\pmatrix{ \scriptstyle 0 \cr \scriptstyle \vec \sigma \vec p \chi } \scriptstyle (\gamma_5 C)$    &
  $\scriptstyle \hat P^4{\pmatrix{ \scriptstyle (\vec{\sigma}\vec{p})\chi\cr \scriptstyle 0}}(\gamma^4 C)$    &
  ${\pmatrix{\scriptstyle 0\cr \scriptstyle i\sigma^i(\vec{\sigma}\vec{p})\chi}}\scriptstyle (\gamma^i C)$    &
  ${\pmatrix{\scriptstyle 0\cr \scriptstyle i\left(p^i-\frac{\sigma^i(\vec{\sigma}\vec{p})}{3}\right)\chi}} \scriptstyle (\gamma^i C)$ \\
 &  {spin} & \centering{1/2} & \centering{1/2} & \centering{1/2} & {3/2} \\
 & {orb.ang.mom.} & \centering{ $p$} & \centering{$p$} & \centering{$p$} & {$p$}\\ \hline
\end{tabular}
\end{center}
\label{wave}
\end{table}

In the nucleon (or generally in octet baryons) there  is one $s$-wave
associated with the scalar diquark and two $s$-waves associated with the
axialvector diquark, one of them connected with its  virtual time component,
see Table \ref{wave}. In the non-relativistic limit only two $s$-waves out of
the eight components would survive. It is remarkable that the relativistic
description leads to four accompanying $p$-waves, the ``lower components'',
and a $d$-wave which are expected to give substantial contributions to the
fraction of the nucleon spin carried by orbital angular momentum. At least,
these $p$-waves would not be present in a non-relativistic model.

In the delta (or generally in decuplet baryons) only one $s$-wave is
found by the method described above. 
Two $d$-waves that could in principle survive the non-relativistic
limit are present and one $d$-wave can be attributed to the virtual
time-component of the axialvector diquark. All even partial waves are
accompanied by relativistic ``lower'' components that could be even more
important as in the nucleon case.

The relativistic decomposition of nucleon and $\Delta$ quark-diquark wave
functions yields a rich structure in terms of partial waves, for more details
see Refs.\ \cite{Oet98,Oet00c}. Well-known problems from certain
non-relativistic quark model descriptions are avoided from the beginning in a
relativistic treatment: First, photo-induced $N-\Delta-$transitions that are
impossible in spherically symmetric non-relativistic nucleon ground states 
will occur in this model through overlaps in the axialvector part of the
respective wave functions. Additionally, photo-induced transitions from scalar
to axialvector diquarks can take place, thus creating an overlap of the nucleon
scalar diquark correlations with the $\Delta$ axialvector diquark correlations.
Secondly, the total baryon spin will mainly be due to the quark spin in the
$s$-waves and the orbital angular momentum of the relativistic $p$ waves (that
are absent in a non-relativistic description). Which fraction of, {\em e.g.},
the nucleon spin, is carried by the quark spins is related to the matrix
element of the flavour singlet axial current and is subject of an on-going
investigation.

\subsubsection{\label{sub_Num} Numerical Results for Ground State Baryons}

The quark-diquark BS equations have been solved in the corresponding bound
state rest frame using an expansion in Chebyshev moments, see Refs.\
\cite{Kus97,Oet98,Oet99,Oet00c} for details. This method exploits the
approximate O(4) symmetry of the BS equation and proves to be very efficient
for obtaining numerically accurate solutions of the full 4-dimensional
equations. Actually, there is not much difference in the requirements for
computational resources in solving the 4-dimensional equation this way or
solving a reduced three-dimensional approximation. Given the non-covariance
of the reduced equation (and the related shortcomings when calculating
observables, {\it e.g.}, see Ref.\ \cite{Oet00a}) further use of
three-dimensional reductions seems questionable.

In Ref.\ \cite{Oet00b} the nucleon and delta amplitudes have been calculated
using two different parameter sets. For one set a constituent quark mass of 
$m_q=0.36$ GeV has been employed. Due to the
free-particle poles in the quark and diquark propagators used in Ref.\
\cite{Oet00b} the axialvector diquark mass is below 0.72 GeV and the delta mass
below 1.08 GeV. On the other hand, nucleon and delta masses are fitted by
second set and the parameter space is constrained by these two
masses accordingly. In particular, this implies $m_q>0.41$ GeV. 
The first parameter set (in which the delta is too light) results in much
better nucleon properties, see below. Especially for the nucleon magnetic 
moments this is easily understood: For weak binding, {\it i.e.}, close to the
non-relativistic limit, the magnetic moment is roughly proportional to
$M_n/m_q$. Thus a standard value for the constituent mass around 0.33 GeV would
be highly desirable. However, one then faces the question how to describe
confinement and thereby avoid unphysical thresholds.

In Ref.\ \cite{Oet98} the quark and diquark propagators have been modified by
choosing in Eqs.\ (\ref{Da},\ref{Ds},\ref{qprop})
\begin{equation} 
C(P^2,m) = 1- e^{-d(P^2+m^2)/m^2}. 
\label{conff}
\end{equation} 
This cancels the pole of the propagators at the expense of
introducing  an essential singularity at time-like infinity. Such a form is
sufficient to prevent any unphysical threshold in the BS equation. Furthermore,
point-like diquarks have been used. The nucleon and the delta masses have been
used as  to fix the coupling strengths in the scalar and axialvector channel.
The calculated hyperon masses are in good agreement with the experimental ones,
see Table \ref{masses}. The wave functions for baryons with distinct
strangeness content but same spin differ mostly due to flavour Clebsch-Gordan
coefficients, the respective invariant functions being very similar.

\begin{table}
\caption[Octet and decuplet masses obtained with two different parameter sets.
(Adapted from Ref.\ \cite{Alk99}.)]
{Octet and decuplet masses obtained with two different parameter sets.
Set I represents a calculation with weakly confining propagators ($d=10$ in Eq.\
(\protect\ref{conff})), Set II with strongly confining propagators ($d=1$ in Eq.\
(\protect\ref{conff})). All masses are given in GeV.
(Adapted from Ref.\ \cite{Alk99}.)}
\begin{center}
\begin{tabular}{lllllllll}\hline
 & $m_u$ & $m_s$ & $M_\Lambda$ & $M_\Sigma$ & $M_\Xi$ & $M_{\Sigma^*}$ & $M_{\Xi^*}$ & $M_\Omega$ \\ \hline
Set I & 0.5 & 0.65 & 1.123 & 1.134 & 1.307 & 1.373 & 1.545 & 1.692 \\
Set II& 0.5 & 0.63 & 1.133 & 1.140 & 1.319 & 1.380 & 1.516 & 1.665 \\
Exp.  &     &      & 1.116 & 1.193 & 1.315 & 1.384 & 1.530 & 1.672 \\
\hline
\end{tabular}
\end{center}
\label{masses}
\end{table}

\subsection{Electromagnetic Form Factors}
\label{sec_EM} 

\begin{figure}
\begin{center}
 \epsfig{file=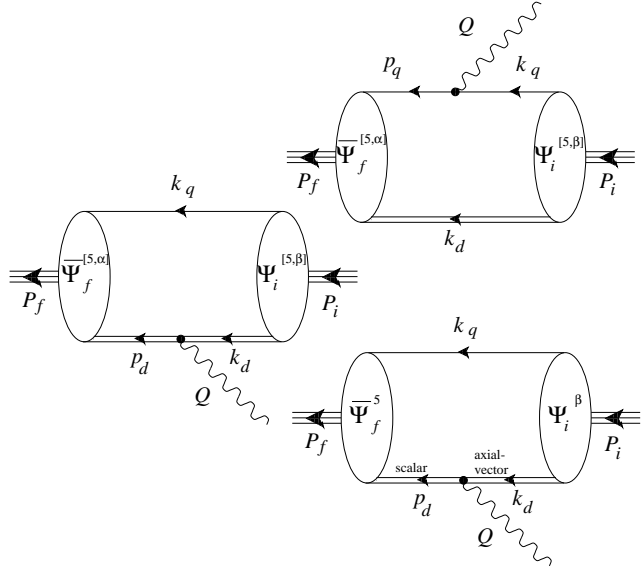,width=0.5\linewidth}
\end{center}
\caption{Impulse approximate contributions to the electromagnetic
current. (Adapted from Ref.\ \cite{Oet00b}.)}
\label{impulse}
\end{figure}

The Sachs form factors $G_E$ and $G_M$ can be extracted from the solutions
of the BS equations using the relations
\begin{equation}
 G_E =\frac{M_n}{2P^2} \; \mbox{Tr} \langle J^\mu \rangle P^\mu, \quad
 G_M =\frac{iM_n^2}{Q^2}  \; \mbox{Tr}\langle J^\mu \rangle \gamma^\mu_T\;,
\end{equation}
where $P=(P_i+P_f)/2$, $\gamma^\mu_T=\gamma^\mu-
 P^\mu  {P\hspace{-.5em}/\hspace{.15em}} /P^2$, and the spin-summed matrix 
element $ \langle J^\mu \rangle $ is given by
\begin{eqnarray}
 \langle J^\mu \rangle &\equiv&  \langle P_f,s_f | J^\mu | P_i,s_i \rangle \,
 \sum_{s_f,s_i} u(P_f,s_f) \bar{u}(P_i,s_i)  \nonumber\\
   &=& \int \frac {d^4p_f}{(2\pi)^4} \int \frac {d^4p_i}{(2\pi)^4}
       \bar \Psi (P_f,p_f) \; J^\mu \; \Psi (P_i,p_i).  \label{curmelt}
\end{eqnarray}
The current $J^\mu$ herein is obtained as in Ref.~\cite{Oet99,Oet00b},
see, however, also Refs.\ \cite{Hel97b,Blo99,Blo00} where electromagnetic
nucleon form factors have been calculated within a BS quark-diquark model.
This current represents a sum of all possible couplings of the photon to the 
inverse quark-diquark propagator $G^{-1}$ given in Eq.\ (\ref{Kdef}). This
construction which ensures current conservation can be systematically derived
from the general ``gauging technique'' employed in Refs. \cite{Kvi99,Ish00}.

\begin{figure}
\begin{center}
 \epsfig{file=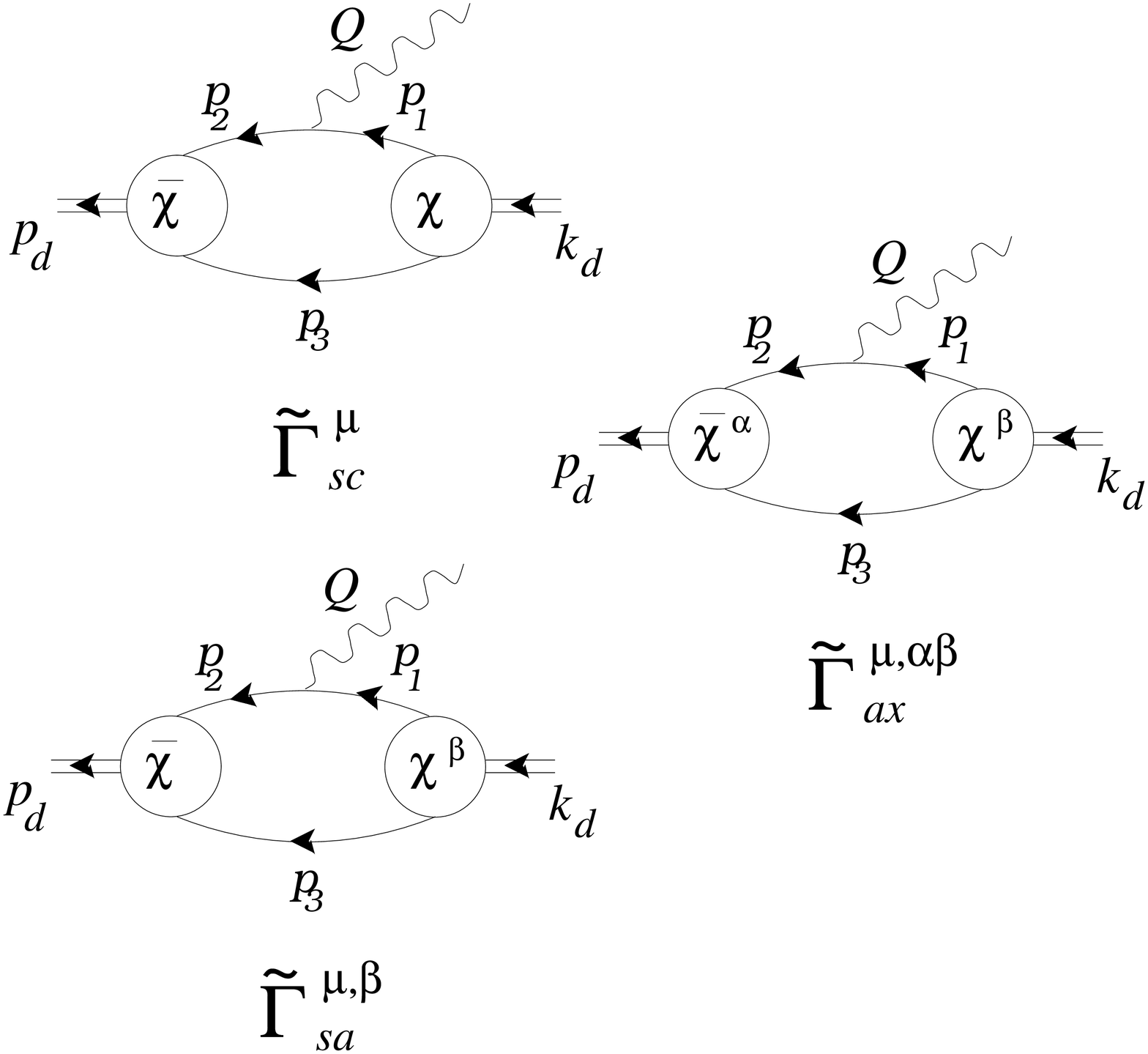, width=0.5\linewidth}
\end{center}
\caption{Resolved vertices: photon-scalar diquark, photon-axialvector diquark
  and anomalous scalar-axialvector diquark transition. (Adopted from Ref.\
  \cite{Oet00b}.)}
\label{emresolve}
\end{figure}

The two contributions to the current that arise from
coupling the photon to the disconnected part of $G^{-1}$, the
first term in Eq.\ (\ref{Kdef}), yield the couplings to the quark and the 
diquark in impulse approximation. They are graphically represented
by the middle and the upper diagram
in Fig.~\ref{impulse}. The corresponding kernels, to be
multiplied by the charge of the respective
quark or diquark upon insertion into the r.h.s. of Eq.~(\ref{curmelt}), read,
\begin{eqnarray}
 J^\mu_q &=& (2\pi)^4\, \delta^4(p_f-p_i-\eta Q) 
              \Gamma^\mu_q \,
              \tilde D^{-1} (k_d), \label{iaqv}\\
 J^{\mu}_{sc[ax]} &=& (2\pi)^4\, \delta^4(p_f-p_i+(1-\eta) Q) 
              \Gamma^{\mu,[\alpha\beta]}_{sc[ax]}\,
              S^{-1} (k_q). \label{iadqv}
\end{eqnarray}
Here, the inverse diquark propagator $\tilde D^{-1}$ comprises
both, scalar and axialvector components.
The vertices in Eqs.~(\ref{iaqv}) and
(\ref{iadqv}) are the ones for a free quark, a spin-0 and a spin-1 particle,
respectively,
\begin{eqnarray}
 \Gamma^\mu_q \!&=&\! -i \gamma^\mu, \qquad  \Gamma^{\mu}_{sc} \; =\;
 -(p_d+k_d)^\mu, \qquad \mbox{and}  \label{qandsdqv}\\
 \Gamma^{\mu,\alpha\beta}_{ax} \!&=&\! -(p_d+k_d)^\mu\; \delta^{\alpha\beta} +
       p_d^\alpha \;\delta^{\mu\beta} + k_d^\beta \;\delta^{\mu\alpha}
       + 
        \kappa\; (Q^\beta \; \delta^{\mu\alpha} -Q^\alpha\; \delta^{\mu\beta}).
      \label{spin1vert}
\end{eqnarray}
The Dirac indices $\alpha,\beta$ in (\ref{spin1vert}) refer to the
vector indices of the final and the initial state wave function, respectively.
The axialvector diquark can have an anomalous magnetic moment $\kappa$.
In Ref.\ \cite{Oet00b} its value has been obtained from a calculation for 
vanishing momentum transfer
in which the quark substructure of the diquarks is resolved.
The corresponding contributions are represented by the upper and
the right diagram in Fig.~\ref{emresolve}.
It turns out that $\kappa = 1$ is a reasonable value independent of the 
parameters used. This is intuitively understandable:
The magnetic moments of two
quarks with charges $q_1$ and $q_2$ add up to $(q_1+q_2)/m_q$, the magnetic
moment of the axialvector diquark is $(1+\kappa)(q_1+q_2)/m_{ax}$, and if the
axialvector diquark is weakly bound, $m_{ax}\approx 2 m_q$, then $\kappa
\approx 1$.

The vertices in Eqs. (\ref{qandsdqv}) and (\ref{spin1vert}) satisfy
their respective Ward-Takahashi identities, {\it i.e.} those for free
quark and diquark propagators ({\it c.f.}, Eqs. (\ref{Ds},\ref{Da}) and
(\ref{qprop})),
and thus describe the minimal coupling of the photon to quark and diquark.

Important additional contributions arise due to photon-induced transitions
between scalar and axialvector
diquarks as represented by the lower diagram in Fig.~\ref{impulse}.
These yield purely transverse currents and do thus not affect current
conservation. The tensor structure of these contributions resembles that of
the triangle anomaly. In particular, the structure of the
vertex describing the transition from axialvector  to
scalar diquark is given by
\begin{equation}
 \Gamma^{\mu\beta}_{sa}=-i\frac{\kappa_{sa}}{2M_n}\, \epsilon^{\mu\beta\rho\lambda}
                 (p_d+k_d)^\rho Q^\lambda ,
 \label{sa_vert}
\end{equation}
and the analogous expression for the reverse transition from an scalar 
to axialvector. 
The tensor structure of these anomalous diagrams is in the limit $Q\to 0$ 
determined by the quark loop in a way as
represented by the lower diagram in Fig.~\ref{emresolve}. The  constant 
$\kappa_{sa}$ turns out to be approximately 2 \cite{Oet00b}.

Upon performing the flavour algebra for the current matrix elements of the
impulse approximation, one obtains the following explicit forms for
proton and neutron,
\begin{eqnarray}
 \langle J^\mu \rangle^{\rm imp}_{p} &=&
   \frac{2}{3} \langle J^\mu_q \rangle^{\rm sc-sc} +
   \frac{1}{3} \langle J^\mu_{sc} \rangle^{\rm sc-sc} +
   \langle J^\mu_{ax} \rangle^{\rm ax-ax} + 
   \frac{\sqrt{3}}{3} \left( \langle J^\mu_{sa} \rangle^{\rm sc-ax} +
         \langle J^\mu_{as} \rangle^{\rm ax-sc} \right) \; , \label{jimp}\\
 \langle J^\mu \rangle^{\rm imp}_{n} &=&
   -\frac{1}{3}\left( \langle J^\mu_q \rangle^{\rm sc-sc} -
    \langle J^\mu_q \rangle^{\rm ax-ax} -
    \langle J^\mu_{sc} \rangle^{\rm sc-sc} +  
       \langle J^\mu_{ax} \rangle^{\rm ax-ax} \right)- \nonumber \\ &&
   \frac{\sqrt{3}}{3} \left( \langle J^\mu_{sa} \rangle^{\rm sc-ax} +
         \langle J^\mu_{as} \rangle^{\rm ax-sc} \right) \; .\label{jimn}
\end{eqnarray}
The superscript `sc-sc' indicates that
the current operator is to be sandwiched between scalar
nucleon amplitudes for both the final and the initial state in
Eq.~(\ref{curmelt}).
Likewise `sc-ax' denotes current operators that are sandwiched
between scalar amplitudes in the final and axialvector amplitudes in
the initial state, {\em etc.}. Note that the
axialvector amplitudes contribute to the proton current
only in combination with diquark current couplings.

\begin{figure}
\begin{center}
 \epsfig{file=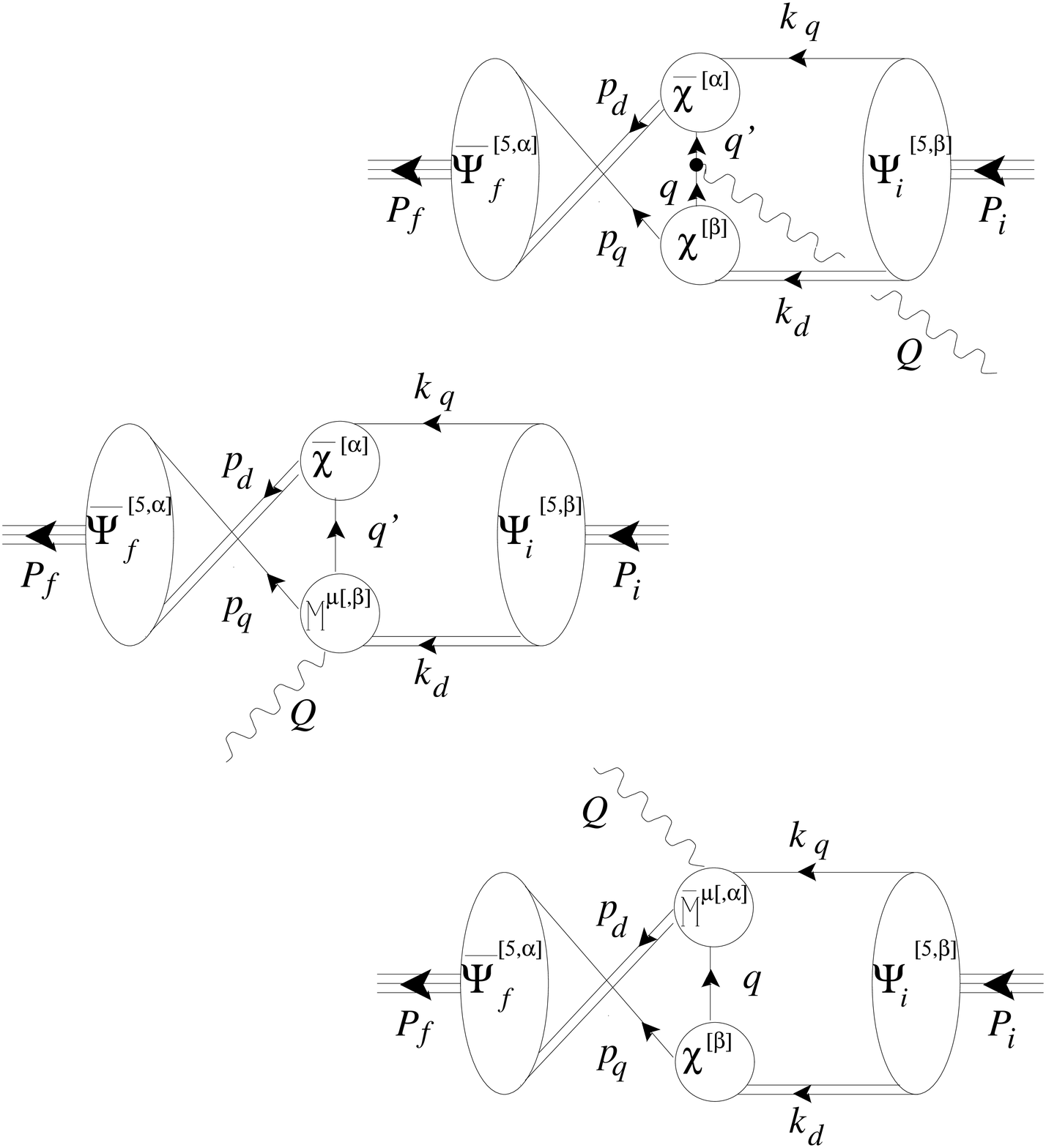,width=0.5\linewidth}
\end{center}
\caption{Exchange quark and seagull diagrams. (Adopted from Ref.\
  \cite{Oet00b}.)}
\label{7dim}
\end{figure}

Current conservation requires that the photon also has to be coupled to
the interaction kernel in the BS equation, {\it i.e.}, to the second term in
the inverse quark-diquark propagator $G^{-1}$ of Eq.~(\ref{Kdef}). The
corresponding contributions were derived in \cite{Oet99} and are
represented by the diagrams in Fig.\ \ref{7dim}. In particular, 
in addition to the photon coupling with the
exchange-quark (with vertex $\Gamma^\mu_q$), irreducible  ``seagull''
interactions of the photon with the diquark substructure have to be taken
into account. These diquark-quark-photon
vertices are constrained by Ward identities. The explicit construction of
Ref.~\cite{Oet99} yields the following seagull couplings:
\begin{eqnarray}
  M^{\mu[,\beta]} & = & q_q \frac{(4p_1-Q )^\mu}
                                 {4 p_1\cdot Q -Q^2}
                 \left[ \chi^{[\beta]}( p_1-Q/2 ) -\chi^{[\beta]}( p_1) \right]
                        + \nonumber \\ &&
               q_{ex} \frac{(4p_1+Q )^\mu}
                                 {4 p_1\cdot Q +Q^2}
                 \left[ \chi^{[\beta]}( p_1+Q/2 ) -\chi^{[\beta]}( p_1)
                                 \right] \; . \label{SeagullsM}
\end{eqnarray}
Here, $q_q$ denotes the charge of the quark with momentum $p_q$, $q_{ex}$ the
charge
of the exchanged quark with momentum $q'$, and $p_1$ is the relative momentum
of the two, $p_1=(p_q-q')/2$ (see Fig.~\ref{7dim}). The conjugate vertices
$\bar M^{\mu[,\alpha]}$ are
obtained from the conjugation of the diquark amplitudes $ \chi^{[\beta]}$ in
Eq.~(\ref{SeagullsM}) together with the replacement $p_1 \to p_2 = (q-k_q)/2$.

The diagrams of Figs.~\ref{impulse} and \ref{7dim} have been evaluated in Ref.\
\cite{Oet00b} using the numerical solutions for the BS amplitudes.
The continuation of these from the nucleon rest frame to the Breit frame is
described in detail in Refs.\ \cite{Hel97b,Oet99}. A warning is here, however,
in order: For finite momentum transfer,
care is needed in treating the singularities of the
quark and diquark propagators that appear in the single
terms of Eq.~(\ref{curmelt}). In Ref.~\cite{Oet99} it was shown that
for some kinematical situations explicit residues have
to be taken into account in the calculation of the impulse
approximation diagrams.

The results of Ref.~\cite{Oet00b} for the proton electric form factor are in
excellent agreement with the phenomenological dipole behaviour. Also the
calculated neutron electric form factor agrees very well with data. The
magnetic moments come out somewhat too small showing that  stronger axialvector
diquark correlations would be favourable for larger values of the magnetic
moments. 

\begin{figure}
 \begin{center}
 \epsfig{file=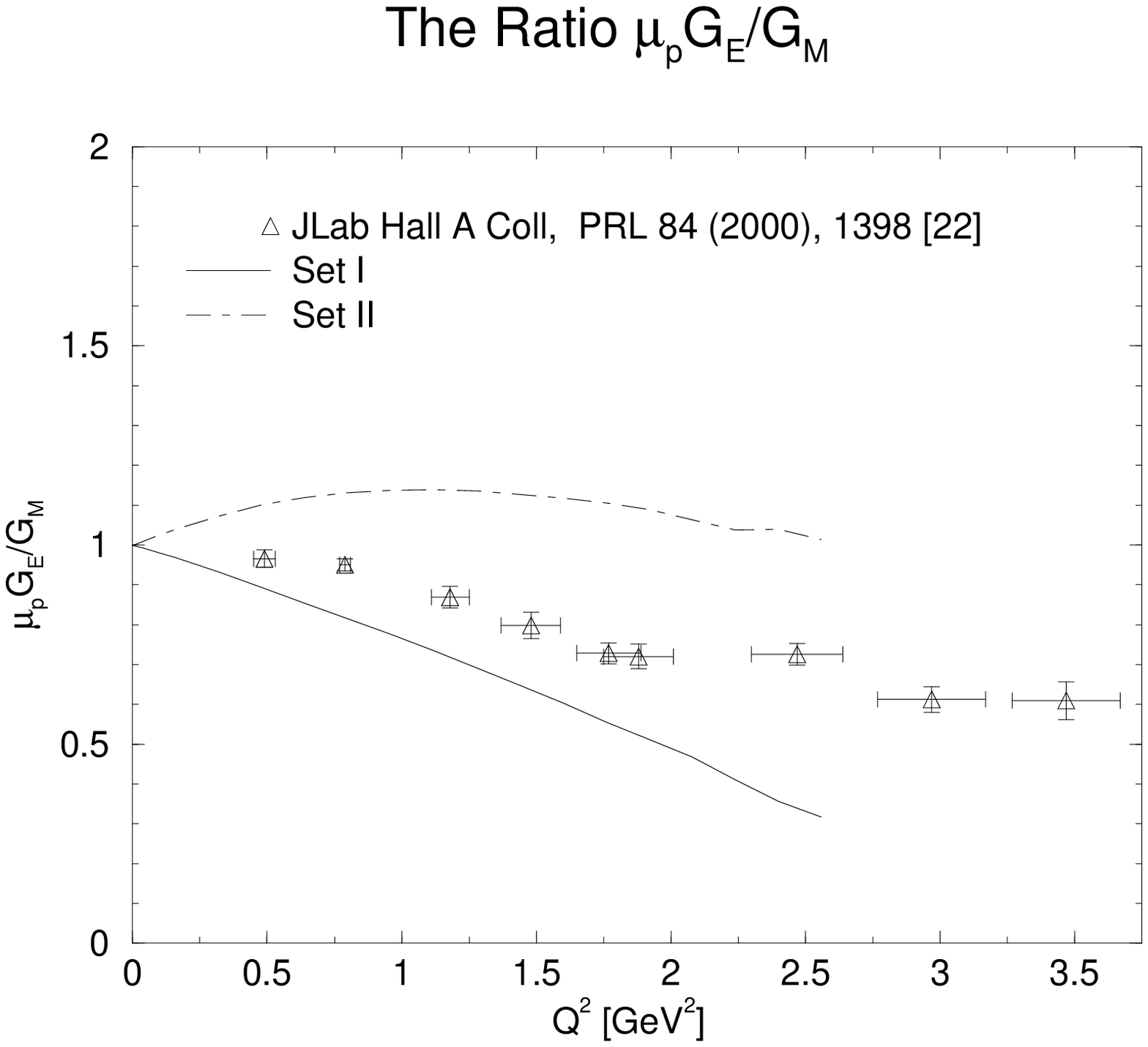,width=10cm}
 \caption{The ratio $(\mu_p\;G_E)/G_M$ compared to the data
          from Ref. \cite{Jon99}.}
 \label{gemfig}
 \end{center}
\end{figure}

Recent data from Jefferson Lab, see Ref.~\cite{Jon99},  for the ratio $\mu_p\,
G_E/G_M$ are compared to the results of Ref.\ \cite{Oet00b} in Fig.\
\ref{gemfig}. The ratio obtained from parameters with weak axialvector
correlations lies above the experimental data, and that  from parameters with
strong axialvector correlations below. Thus, the experimental observation that
this ratio decreases significantly with increasing $Q^2$ (about 40\% from $Q^2
= 0 $ to $3.5$ GeV$^2$), can be well reproduced with axialvector diquark
correlations of a certain strength included. The reason for this is the
following: The impulse approximate photon-diquark couplings yield
contributions that tend to fall off slower with increasing $Q^2$ than those of
the quark. This is the case for both, the electric and the magnetic form
factor. If no axialvector diquark correlations inside the nucleon are
maintained, however, the only diquark contribution to the electromagnetic
current arises from $\langle J^\mu_{sc} \rangle^{\rm sc-sc}$, see
Eqs.~(\ref{jimp},\ref{jimn}). Although this term does provide for a substantial
contribution to $G_E$, its respective contribution to $G_M$ is of the order of
10$^{-3}$. This reflects the fact that an on-shell scalar diquark would have
no magnetic moment at all, and the small contribution to $G_M$ may be
interpreted as an off-shell effect. Consequently, too large a ratio $\mu_p\,
G_E/G_M$ results, if only scalar diquarks are maintained \cite{Oet99}. For
the parameters leading to the correct $\Delta$ mass but quite weak axialvector
correlations this effect is still visible, although already with these weak
axialvector correlations the scalar-to-axialvector transitions 
have the qualitatively new effect of yielding a  
ratio $\mu_p\, G_E/G_M$ that, for sufficiently large $Q^2$, decreases with
increasing $Q^2$.  These transitions almost
exclusively contribute to $G_M$, and it thus follows that the stronger
axialvector correlations enhance this effect. The ratio $\mu_p
\, G_E/G_M$ imposes an upper limit on the relative importance of the
axialvector correlations of estimated 30\% (to the BS norm of the nucleons).

The nucleon electromagnetic form factors have also been calculated in Refs.\
\cite{Blo99,Blo00} using a scalar diquark only. In these investigations a
different approach has been  taken: Not only the quark and diquark propagators
have been modelled by entire functions but also the nucleon Faddeev amplitude
(or more precisely, a BS-like quark-diquark amplitude)  has been
parametrised by a  one-parameter entire function. The parameters of the quark
propagator have been fixed by a fit to meson observables, the remaining three
parameters (width of the nucleon amplitude, width of the  diquark amplitude and
diquark correlation length) have been determined by optimising a fit to the
proton electric form factor. The corresponding diagrams have been calculated in
a generalised  impulse approximation. This takes care of the internal structure
of the diquark. Due to the use of a dressed quark-photon vertex current
conservation is maintained for the constituents. 
For the baryonic bound states, however, it has been demonstrated in Ref.\
\cite{Bla99} that the approach of Refs.\ \cite{Blo99,Blo00} suffers from an
overcounting problem. This is substantiated by the fact that the impulse
approximation of Refs.\ \cite{Blo99,Blo00} cannot be obtained from the general
treatment of a relativistic three-body problem presented in Ref.\ \cite{Kvi99}
in the limit of separable quark-quark correlations. (Note that such a
derivation is possible
in the approach of Refs.\ \cite{Oet99,Oet00b}.) As the particular diagrams
under dispute provide small contributions to observables only, the results of
Refs.\ \cite{Blo99,Blo00} are nevertheless interesting.

The neutron electric form factor obtained in Ref.\ \cite{Blo99} is much larger
than the experimental data at all (space-like) momenta. This is very likely a
defect due to neglecting the axialvector diquark as comparison with Refs.\
\cite{Hel97b,Oet99} reveals: Also in these studies only the scalar diquarks have
been taken into account and the neutron electric form factor has been
overestimated. On the other hand, the result for the proton magnetic moment 
obtained in Ref.\ \cite{Blo99} is close is to its phenomenological value, the
(absolute value of the) neutron magnetic moment is somewhat too small. 

Summarising this section: The electromagnetic properties of the nucleon can be
described reasonably well within a Faddeev/BS equation approach. It would be
interesting to see whether a more sophisticated study using 
``confined'' quarks and diquarks, a
calculated nucleon amplitude (from the quark-diquark BS equation), and a
reasonable amount of axialvector diquark correlations could provide a more
accurate description.   

\subsection{Strong and Weak Form Factors}
\label{sec_SWFF} 

In Ref.\ \cite{Blo00} (using the Ans\"atze of Ref.\ \cite{Blo99} and taking
into account only a scalar diquark)
the pseudoscalar, isoscalar- and isovector-vector, axial-vector and scalar
nucleon form factors have been calculated. The pion-nucleon and the axial
coupling based on a solution of the BS equation has been calculated with
``confined'' constituents and only a scalar diquark in Ref.\ \cite{Hel97b}. 
For ``free'' constituents and with the axialvector diquark included
corresponding results have been reported in Ref.\ \cite{Oet00b}. 

The coupling of the pion to the nucleon, described by a pseudoscalar operator,
and the pseudovector currents of weak processes such as the neutron
$\beta$-decay are connected to each other in the soft limit by the
Goldberger-Treiman relation.

The matrix element of the pseudoscalar density $J_5^a$ can be
parametrised as
\begin{equation}
  \langle J^a_5 \rangle = \Lambda^+(P_f)\,\tau^a \gamma_5 g_{\pi NN}(Q^2)\,
  \Lambda^+(P_i) \; ,    \label{psdens}
\end{equation}
which yields
\begin{equation}
 g_{\pi NN}(Q^2) = - \frac{2M_n^2}{Q^2} {\rm Tr} \langle J^a_5 \rangle
          \gamma_5   \frac{\tau^a}{2}. \label{gpNNtr}
\end{equation}
As discussed in the last chapter the chiral Ward identity provides for the 
leading  Dirac covariant  of the pion-quark vertex
\begin{equation}
 \Gamma_5^a (P,p)
         = \gamma_5 \frac{B(p^2)}{f_\pi} \tau^a ,
 \label{vertpion}
\end{equation}
where $f_\pi$ is the pion decay constant and $B(p^2)$ is the scalar part of the
quark self-energy in the chiral limit. The three additionally possible Dirac 
structures have been neglected in the calculations of Refs.\
\cite{Blo00,Oet00b}. (Note that with free constituent propagators as in Ref.\
\cite{Oet00b} one simply has  $B(p^2)=m_q$, the constituent quark mass.)

The matrix elements of the pseudovector current are para\-me\-trised
by the form factor $g_A(Q^2)$ and the induced pseudoscalar form factor
$g_P(Q^2)$,
\begin{equation}
 \langle J^{a,\mu}_5 \rangle = \Lambda^+(P_f)\,\frac{\tau^a}{2}
  \left[ i\gamma^\mu \gamma_5 g_A(Q^2) +Q^\mu \gamma_5 g_P(Q^2) \right]\,\Lambda^+(P_i).
  \label{nucax}
\end{equation}
For $Q^2 \to 0$ the Goldberger-Treiman relation,
\begin{equation}
 g_A(0) =  f_\pi  g_{\pi NN}(0)/{M_n} \; ,
\end{equation}
then follows from current conservation and the observation
that only the induced pseudoscalar form factor $g_P(Q^2)$ has a pole on the
pion mass-shell.

By definition, $g_A$ describes the regular part of the pseudovector current
and $g_P$ the induced pseudoscalar form factor.
They can be extracted from Eq.~(\ref{nucax}) as follows:
\begin{eqnarray}
 g_A (Q^2) & = &- \frac{i}{4\left( 1+\frac{Q^2}{4 M_n^2}\right) }
             {\rm Tr} \langle J^{a,\mu}_5 \rangle \left( \gamma_5 \gamma^\mu -
             i \gamma_5 \frac{2M_n}{Q^2} Q^\mu \right) \tau^a \; ,
 \label{gatrace} \nonumber \\
           & & \\
 g_P (Q^2) & = & \frac{2M_n}{Q^2} \left( g_A (Q^2) - \frac{M_n}{Q^2}
              {\rm Tr} \langle J^{a,\mu}_5 \rangle Q^\mu \gamma_5 \;\tau^a
               \right) \; .
 \label{gptrace}
\end{eqnarray}
Chiral symmetry constraints may be used to construct the axialvector-quark 
vertex.
In the chiral limit, the Ward-Ta\-ka\-ha\-shi identity for this vertex reads,
\begin{equation}
 Q^\mu \Gamma^{\mu,a}_5 = \frac{\tau^a}{2}
      \left( S^{-1}(k) \gamma_5 + \gamma_5 S^{-1}(p) \right)\; ,  \quad (Q=k-p).
\label{pvWTI}
\end{equation}
This constraint is satisfied by the form of the vertex proposed in
Ref.~\cite{Del79},
\begin{equation}
 \Gamma^{\mu,a}_5 =
       -i \gamma^\mu \gamma_5 \frac{\tau^a}{2} +  \frac{Q^\mu}{Q^2}
       f_\pi \Gamma_5^a .
 \label{vertpv}
\end{equation}
The second term which contains the massless pion pole
does not contribute to $g_A$ as can be seen from Eq.\ (\ref{gatrace}).
From these quark contributions to the pion coupling
and the pseudovector current alone, Eqs. (\ref{vertpv}) and (\ref{gpNNtr})
would thus yield,
\begin{equation}
 \lim_{Q^2 \rightarrow 0} \frac{Q^2}{2M_n} g_P(Q^2) =
   \frac{f_\pi}{M_n} g_{\pi NN} (0).
   \label{gpgpNN}
\end{equation}
Here, the Goldberger-Treiman relation follows if the pseudovector
current was conserved or, off the chiral limit, from PCAC.

Current conservation is a non-trivial requirement
in the relativistic bound state description of nucleons, however.
First, we ignored the pion and pseudovector couplings to the
diquarks in the simple argument above. For scalar diquarks alone which
themselves do not couple to either of the two, as can be inferred
from parity and covariance,
pseudovector current conservation could in principle
be maintained by including the couplings to the interaction kernel of the
nucleon BS equation in much the same way as was done for the electromagnetic
current.

Unfortunately, when axialvector diquarks are included, even this will not
suffice to maintain current conservation. As observed recently in
Ref.~\cite{Ish00}, a doublet of axialvector {\em and} vector diquarks has to be
introduced, in order to comply with chiral Ward identities in general. The
reason essentially is that vector and axialvector diquarks mix under a chiral
transformation whereas this is not the case for scalar and pseudoscalar
diquarks. Since vector diquarks on the one hand introduce six additional
components to the nucleon wave function, but are on the other hand not expected
to influence the binding strongly, here vector diquark correlations have been
neglected so far. 

In Ref.\ \cite{Oet00b} the relevant couplings of the currents to the diquarks
have been estimated from the quark loops at $Q^2=0$. In Ref.\ \cite{Blo00}
the impulse approximation discussed already in the last section has been 
employed.

In Ref.\ \cite{Oet00b} it was found that large contributions to $g_{\pi NN}(0)$
and $g_A(0)$ arise from the scalar-axialvector transitions, and  
both quantities are overestimated,
$g_A(0) = 1.35 - 1.49$ instead of 1.27 and $g_{\pi NN}(0) = 16 - 17$
instead of $\approx 13.2$.  As mentioned, the axialvector diquark 
contributions violate the Goldberger--Treiman relation.
Some compensations occur between the small contributions from
the axialvector diquark impulse-coupling and the comparatively large
ones from scalar-axialvector transitions which provide for the dominant effect
to yield $g_A(0) > 1$. 
Summing all contributions the Goldberger--Treiman relation is violated by 14 --
18\%.

In Ref.\ \cite{Blo00} $g_{\pi NN}(0)$ is slightly overestimated, too. On the
other hand, $g_A(0)$ results generally to be smaller than one. This might
simply reflect the importance of scalar-axialvector transitions for this
quantity. It is interesting to note that a non-vanishing $f_{\omega NN}$ has
been obtained. As this quantity might be important for meson-exchange models
of nuclear physics (where it has been neglected so far) this issue certainly
deserves to be studied further. The analysis of the $\sigma$-term in Ref.\
\cite{Blo00} illustrates the only method known up to now that allows an
unambiguous off-shell extrapolation in the estimation of meson-nucleon form
factors. An important element in the calculation of the scalar form factor
presented in Ref.\ \cite{Blo00} is the dressed-quark scalar vertex obtained
from the  solution of the inhomogeneous BS equation. As anticipated, it
possesses a pole at the scalar mass, and the residue of this pole provides the
$\sigma$-meson nucleon coupling. The solution of the inhomogeneous BS equation
allows to extract this coupling at every possible value of $Q^2$, especially
also at the soft point $Q^2=0$ where this coupling is directly related with the
$\sigma$-term. In this way one obtains $\sigma = 14$MeV whereas at the
$\sigma$-meson pole a several times larger value would have been extracted.

\begin{figure}
\centerline{\epsfxsize 8cm \epsfbox{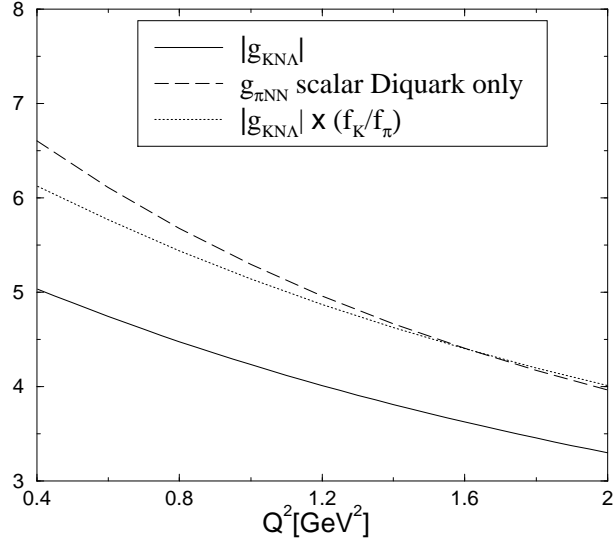}}
\caption{Comparison of the $g_{KN\Lambda} (Q^2)$ form factor with 
a ``projected'' $g_{\pi NN} (Q^2)$, 
only the scalar diquark component of the nucleon
amplitude has been taken into account in the calculation. (Adopted from Ref.\
\cite{Fis99}.)}
\label{gklnxmg}
\end{figure}

As already stated, transitions from nucleons to the $\Lambda$-hyperon project
out the scalar diquark component of the nucleon amplitude \cite{Kro97}.
Therefore, corresponding processes provide a test for the assumption of
a separable quark-quark $t$-matrix, {\it i.e.}, the diquark hypothesis. A first
and illustrative step in this direction is the $g_{KN\Lambda} (Q^2)$ form
factor. Its within this model calculated (absolute) value at all space-like 
$Q^2$ is much smaller then $g_{\pi NN} (Q^2)$ \cite{Fis99}. Hereby the direct
contribution to flavour symmetry breaking, $m_s>m_u$, is of minor importance.
Some 20\% are due to the larger kaon decay constant, $f_K>f_\pi$, but the most
important effect is the projection onto the scalar diquark component of the
nucleon amplitude, see Fig.\ \ref{gklnxmg}.

\subsection{Hadronic Reactions}
\label{sec_Prod} 

Production processes where the quark propagator is tested for time-like
momenta could be very interesting with respect to the employed
parametrisations of confinement.  On the other hand, such processes are fairly
complicated and the issue might be obscured, if too numerous diagrams for
subprocesses contribute. For this reason processes like kaon photoproduction
with a $\Lambda$-hyperon in the final channel, $p\gamma \to K \Lambda$, are
well suited for such investigations. As stated already, flavour algebra leads
to the restriction of scalar diquarks. Furthermore, the kaon does not couple
to the scalar diquark. Therefore, restricting to the impulse approximation 
leads to the justification of a spectator model. 

It has to be emphasised that the investigations reported in this section all 
have quite an exploratory character. They more or less serve to demonstrate the
feasibility of such calculations.

\subsubsection{\label{sec_Deep} Deep Inelastic Structure Functions}

Before going to processes with strangeness in the final channel the results of
the chronologically first application of the diquark-quark BS solution will
be briefly 
reviewed here. In Ref.\ \cite{Kus97} the nucleon structure function $F_1(x)$
has been calculated. Hereby the nucleon was modelled as consisting of a valence
quark and a scalar diquark of equal mass. The photon coupled only to the quark
whereas the diquark was treated as a spectator. Despite the still exploratory
character of this application, for BS equation based diquark
models the first of its kind, it already produced interesting results. 
The spin structure of the nucleon has 
been found to contribute non-trivially to the structure function $F_1(x)$: Its
valence-quark contribution is governed by the ``non-relativistic'' components
only in the case of very weak binding. The shape of the unpolarised
valence-quark distribution has been found to be mainly determined by
relativistic kinematics and does not depend on the details of the quark-diquark
dynamics. 

Furthermore, it is interesting to note that due to the full covariance of the
model the structure function has the correct support  ($x\in [0,1]$) from the
very beginning. No projection technique is needed. The valence-quark
distribution shows a clear peak at approximately $x=1-m_{sc}/M$. Assuming that 
the scalar diquark mass $m_{sc}$ is approximately 2/3 of the nucleon bound state
mass $M$, the empirically observed peak at $x\approx 1/3$ would be reproduced.

The investigations reported in Ref.\ \cite{Kus97} are certainly a clear
motivation to study all experimentally observable nucleon structure functions
in a more sophisticated version of the diquark-quark BS model in the future.

\subsubsection{\label{sec_Photo} Kaon Photoproduction off the Proton}

\begin{figure}[b]
\begin{center}
 \epsfig{file=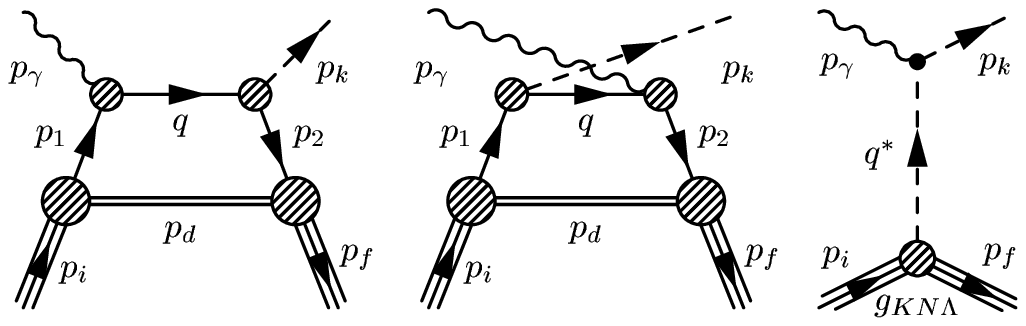,width=12cm}
\end{center}
\caption{Impulse approximation diagrams contributing
  to $\gamma p \rightarrow K\Lambda$.}
\label{gp>kl}
\end{figure}

As stated above a spectator picture emerges for the reaction $\gamma p
\rightarrow K\Lambda$ in the impulse approximation, and this process is
described within a diquark model by the diagrams
shown in Fig.\ \ref{gp>kl}.
The model specific input that goes into the calculation of the cross section
and the asymmetries for kaon photoproduction are the wave functions for the
baryons (which are solutions of their BS equations) and the propagators of
quarks 
and diquarks. Hereby it is important to note that the quark propagator in
between the incoming photon and the outgoing kaon in the left diagram of Fig.\
\ref{gp>kl} is tested in a parabolic region of complex momenta $q^2$ such that
the lowest real value of $q^2$ is given by $q^2_{\rm min} = - (\eta_P M +E)^2$
where $\eta_P$ is the momentum partitioning parameter, $M$ is the nucleon mass,
and $E$ is the photon energy in the nucleon rest frame. Obviously, for large
photon energies the quark propagator at time-like momenta becomes important.

\begin{figure}
\begin{center}
 \epsfig{file=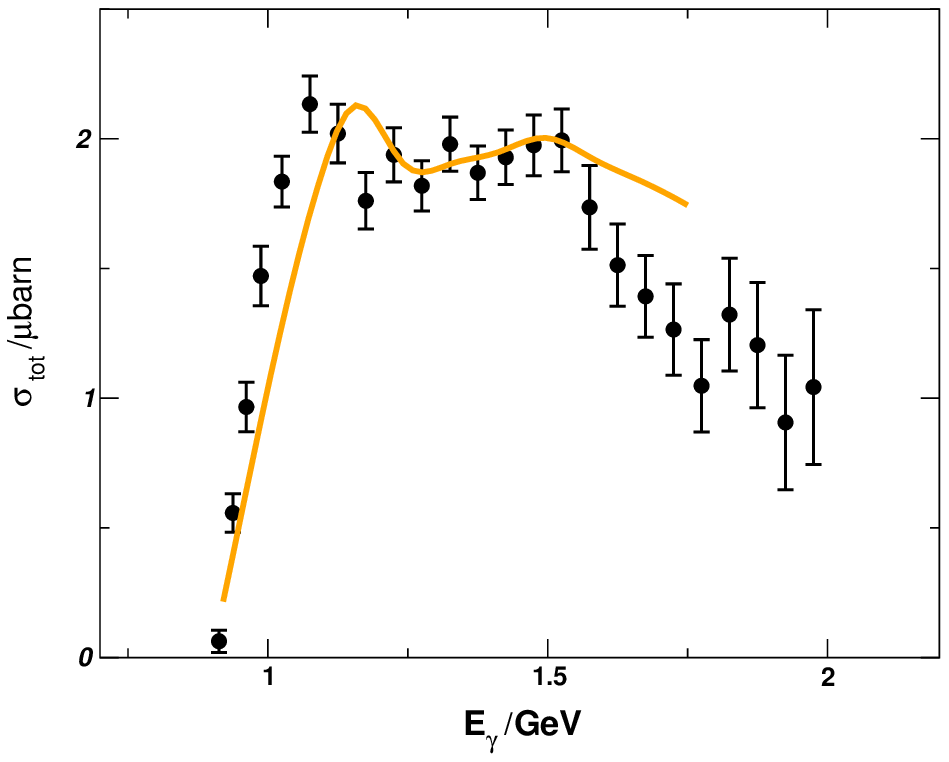,width=6cm}
 \hspace{1.5cm}
 \epsfig{file=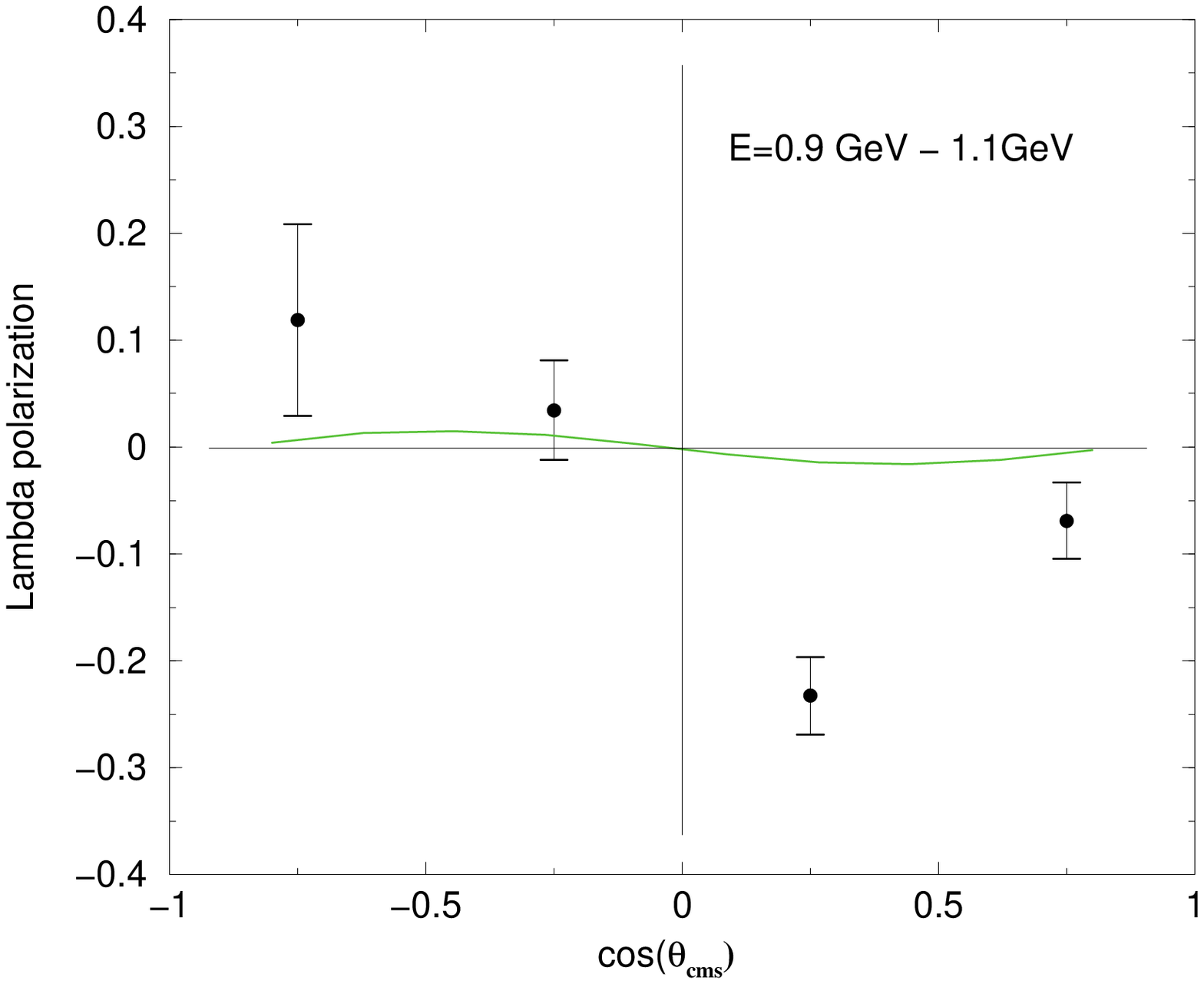,width=6cm}
\end{center}
\caption{Kaon photoproduction. The left panel
 shows the total cross section compared to the data of the
 SAPHIR collaboration, the right panel the
 asymmetry. (Adapted from Ref.\ \cite{Alk99}.)}
\label{photoprod}
\end{figure}

The results of Ref.\ \cite{Alk99} for $p\gamma \rightarrow K\Lambda$ using
pointlike diquarks are shown in Fig.\ \ref{photoprod}. As anticipated it has
been found that the total cross section strongly depends on the behaviour of
the quark propagator in the time-like region. To obtain the results shown  in
Fig.\ \ref{photoprod} the function $C(p^2,m)$ in Eqs.\ (\ref{Ds},\ref{qprop})
has been chosen to be $\exp(-0.25 | 1+p^2/m^2 |)$.\footnote{A discussion of 
production processes within the diquark model using qualitatively different 
Ans\"atze for propagators can be found in Ref.~\cite{Ahl00a}.}  Using instead
$C(p^2,m)=\exp(-(1+p^2/m^2))$ as done in previous calculations leads
to enormously large cross sections. For the form chosen, however, the total
cross section matches the data quite nicely and it even follows the peculiar
structure of the measured data in the region above 1~GeV. This is due to
interference with the kaon exchange diagram shown in Fig.\ \ref{gp>kl}. The
$\Lambda$ polarisation is shown in the right panel of Fig.\ \ref{photoprod},
and the comparison reveals that it falls short of the measured data by a
substantial factor. However, the characteristic change of sign as required by
the data is present.

\subsubsection{\label{sec_Stran} Associated Strangeness Production in $pp$ 
Collisions}

The application of the diquark-quark BS model to the associated strangeness
production $pp\to pK\Lambda$ \cite{Fis99} is based on the assumption that for 
intermediate reaction energies the interaction mechanism is dominated by the
exchange of pseudoscalar mesons but that nevertheless the subhadronic quark
structure is tested.  The calculated total cross sections have been found
significantly too small, however, this effect might be attributed to the strong
initial and final state interactions. The form of the calculated differential
cross sections agree, nevertheless, nicely with experiments. 

Calculating the depolarisation tensor, which is a measure of the relation 
between the spin of the incoming proton and the outgoing $\Lambda$-hyperon, 
and which is only marginally influenced
by initial and final state interactions, one finds a pronounced dependence on 
the parametrisations of the quark and diquark propagator. Hereby it is again
the behaviour at large time-like momenta which, in this case, is amplified by
the fact that even the signs one obtains for the pion and kaon exchange
diagrams can differ in different models for the propagators. As mostly
parallel spins  
of the proton and the $\Lambda$-hyperon contribute to the pion diagrams and
antiparallel spins to the kaon diagrams a completely different interference
pattern is obtained in the results for the  depolarisation tensor.

All the results mentioned for hadronic reactions are based on the assumption for
point-like diquarks (see, however, Ref.~\cite{Ahl00a}), 
and because a reasonable description of form factors
requires extended diquarks, all these calculations can only be considered as
exploratory. But they have nevertheless proven that a fully Lorentz covariant
approach based on the substructure of hadrons can be applied even to quite
complicated processes.


\section{Concluding Remarks}
\label{chap_concl}  

In this review we have attempted to demonstrate the long way from the dynamics
of confined quark and glue to a description of hadrons and their processes in
one coherent approach. The results obtained so far encourage to pursue it. Of
course, improvements are possible and highly desirable at practically every
step. The hope is that future work will bridge more of the gaps remaining 
between them. 

First of all, a consistent qualitative picture for all QCD propagators and
vertex functions will hopefully emerge in the near future from lattice and
Dyson--Schwinger calculations. These may then be employed in the rich meson
phenomenology based on the Bethe--Salpeter equation. We have seen that many
aspects of meson physics can be understood quite well on the basis of these
investigations. Diquark correlations have also been studied from simple models
of quark and gluon propagators. In a coherent calculation, these could of
course be obtained as solutions to  Bethe--Salpeter equations  constructed from
Dyson--Schwinger equations. Coupling the diquark correlations to a further
quark propagator, the way to baryon wave functions from results of
Dyson--Schwinger equations for quarks and gluons is straight ahead. The
complications for physical observables such as form factors in a
Bethe--Salpeter/Faddeev calculation of baryons from the results for the
structure of diquark and quark correlations, are under control. The numerical
machinery to calculate these observables in this extended diquark-quark
approach is currently being developed.

The ideal calculations of this kind have the minimal parameter set of QCD. This
allows to critically assess the necessary truncating assumptions on the
infinite hierarchy of equations of motion for the QCD Green's functions. No
abundant parameters are present to tune, and thus to cover up possible
insufficiencies of these calculations. The program so far yielded reliable
results for the coupled system of equations for the gluon and ghost propagators
in Landau gauge. In the momentum range overlapping with available lattice
results good agreement is obtained. Confinement of gluons can be attributed to
the violation of positivity found to be manifest in the solution. This feature
is also seen in lattice results. The quenched solution to the quark
Dyson--Schwinger equation from these results compare nicely to lattice results
obtained more recently. The necessary non-perturbative mass scale is produced
by the dynamics encoded in the Dyson--Schwinger equations of gluons, ghosts and
quarks. The positivity and the analyticity properties of the quark propagator
are interesting open questions. These can be addressed with the full unquenched
solutions which became available very recently. 

We hope to have demonstrated that this approach is worth to be further pursued,
and that a coherent description of hadronic states and processes based on the
dynamics of confined correlations of quark, glue and ghosts can be realized in
the near future. Hereby the method based on the QCD Green's functions has to be
understood as being complementary to other techniques. Lattice calculations
provide evidence that truncations of the infinite hierarchy of Dyson--Schwinger
equations have been done reasonably. On the other hand, results of these
truncated equations help to understand the lattice data. Solutions of
Bethe--Salpeter equations provide explicit examples of different realization of
symmetries in the meson and baryon spectrum. Meson and baryon properties as
well as hadronic reactions may in the end serve as experimental tests of our
understanding of confinement. And this connects to what might be the most
challenging goal in hadron physics: Not only to understand the mysterious
phenomenon of confinement but also to find a way to verify it experimentally.

\section*{Acknowledgements}

We have benefited from many discussions with colleagues and friends. Our warm
thanks to all of them. A probably incomplete list of people we would like to
mention in particular, includes   
F.~Coester, A.~Cucchieri, H.~Gies,  N.~Ishii,
O.~Jahn, A.~C.~Kalloniatis, K.~Langfeld, F.~Lenz, P.~Maris,
S. Nedelko, J. Negele, K.~Nishijima, M.~Oettel, M.~A.~Pichowsky,
M.~Pennington,  L.~O'Raifeartaigh, H.~Reinhardt,  C.~D.~Roberts, M.~Schaden,
S.~Schmidt, A.~Schreiber, P.~Tandy, M.~Thies, K.~Yazaki,
P.~Watson, H.~Weigel and A.~G.~Williams.   

We thank S.~Ahlig for a critical reading of the manuscript and for providing
Figs. \ref{fig:MQED} and \ref{gp>kl}.

This work has been supported in part by DFG (Al 279/3-3) and COSY (contract 
no.\ 41376610).

\newpage

\appendix

\section{Real vs. Complex Ghost Fields: Ghost-Antighost
Symmetry and $SL(2,\mbox{\sf R})$} 
\label{App.BRS}

In this appendix we summarise the relation between real and complex ghost
fields, {\it i.e.}, between the hermiticity assignments 
$c^\dagger(x)  = c(x) $ and $(i\bar c(x))^\dagger = i \bar c(x)$ versus 
$c(x)^\dagger = \bar c(x)$ in the operator formalism. The latter assignment
is used in Sec.~\ref{chap_Basic}. It has been stressed in the literature,
see, {\it e.g.} Ref.~\cite{Nak90}, that the correct hermiticity assignment
should be the former, however. This is true for the standard
Faddeev-Popov gauges with $\xi\not=0 $. In Landau gauge ($\xi =0$) we can
make use of the additional ghost-antighost symmetry to establish the
equivalence of both formulations, employing independent real or complex ghost
fields (corresponding to the ghost and antighost degrees of freedom). The
ghost-antighost symmetry and thus the complex formulation with  $c(x)^\dagger
= \bar c(x)$ can be maintained for $\xi \not= 0$ at the expense
of quartic ghost interactions in (a special case of)        
the so-called Curci--Ferrari gauge(s) \cite{Cur76a,Cur76b}.
     
Let us first consider the hermiticity assignment $c^\dagger(x)  = c(x) $ and
$(i\bar c(x))^\dagger = i \bar c(x)$ corresponding to two independent and
real Euclidean (Grassmann) ghost $c(x) \to u(x)$ and antighost fields $\bar
c(x)  \to i v(x) $. The gauge fixing part $\mathcal{L_{G\!F}}$ of the
effective Lagrangean $\mathcal{L}_{\mbox{\tiny eff}}$ in Eq.~(\ref{Leff}) of
Sec.~\ref{chap_Basic} then reads,
\begin{equation}
\mathcal{L_{G\!F}} = iB^a \partial_\mu A_\mu^a + \frac{\xi}{2} B^a B^a 
   + i v^a \partial_\mu D_\mu^{ab} u^b \; ,     \label{Lgf}
\end{equation}
where we introduced the Nakanishi-Lautrup auxiliary field $B$ as a real
Euclidean field. Its integration yields the usual gauge fixing term $(\partial
A)^2/2\xi $ introduced in Sec.~\ref{chap_Basic}.  The last term in
Eq.~(\ref{Lgf}) specifies the ghost part $\mathcal{L}_{\mbox{\tiny
ghost}}$,  which is hermitean:  
\begin{equation} 
\mathcal{L}_{\mbox{\tiny ghost}}  = \frac{1}{2} \big(
\mathcal{L}_{\mbox{\tiny ghost}} + \mathcal{L}_{\mbox{\tiny ghost}}^\dagger
\big) =  - \frac{i}{2} \, ( u, v ) \left( \begin{array}{cc} 
                                  0 &  D\partial \\
                                 -\partial D  & 0 \end{array} \right) 
         \left( {u \atop v} \right) 
                                 \; . 
\end{equation}
In Landau gauge we have $\partial_\mu D_\mu  = D_\mu \partial_\mu$ and
thus,\footnote{This holds {\em on-shell}, {\it i.e.}, after the constraint
$\partial A=0$ is implemented. On the level of the Lagrangean
it is slightly inconsistent. To be precise, we would
have to symmetrise the Faddeev-Popov operator first which can be achieved in
the Landau gauge by shifting the $B$-field. We will do that anyway for
the more general Curci--Ferrari gauges below. For the present argument 
we ignore this subtlety.}        
\begin{equation} 
\mathcal{L}_{\mbox{\tiny ghost}}^{\xi=0} = - \frac{i}{2} \, ( u, v )\, \partial
D \, \varepsilon  \, \left( {u \atop v} \right) \quad \mbox{with} \qquad 
                    \varepsilon = \left( \begin{array}{cc} 
                                  \; 0   & \;  1 \\
                                 -1 \;  &  \; 0 \end{array} \right)
\label{LGghost}  
\end{equation}
which is the metric in a 2-dimensional spinor space. Since
\begin{equation} 
A^T \varepsilon A = \varepsilon \quad \mbox{for $2\times 2$-matrices $A$ with
det$A = 1$,} \; i.e., \; A \in SL(2, \CC) \; ,    \label{sl2c}
\end{equation}
the ghost Lagrangean for $\xi = 0$ has a global $SL(2,\RR)$ symmetry, the
subgroup of $SL(2,\CC)$ that preserves the above hermiticity assignment,
\begin{equation}   
 \left( {u \atop v} \right)  \, \mapsto \, A \,  \left( {u \atop v} \right)
 \; , \quad  A \in SL(2,\RR) \; .  
\end{equation}
Note that both, the ghost number symmetry and the Faddeev-Popov(FP)
conjugation, are contained in this global symmetry. The former is generated
by the (hermitean) ghost charge $Q_c$, {\it c.f.}, Sec. 3.4.2 in 
Ref.~\cite{Nak90},
\begin{equation}  
           [i Q_c , u^a(x) ] = u^a(x) \; , \qquad [i Q_c , v^a(x) ] = - v^a(x)
           \; ,
\end{equation}
or, by exponentiation,
\begin{equation}  
          e^{i Q_c \theta} \, : \quad  \left( {u \atop v} \right) \,  \mapsto 
    \, \left(  \begin{array}{cc} e^\theta \, &\,  0 \\
                      0 \, &\,   e^{-\theta} \end{array} \right) 
         \left( {u \atop v} \right)   \; .
\end{equation}
This corresponds to a (non-compact) Abelian subgroup of
$SL(2,\RR)$. FP conjugation $\mathcal{C}_{\mbox{\tiny FP}}$, {\it i.e.}, the
ghost-antighost symmetry of the Landau gauge, can be represented by
\begin{equation}  
       \mathcal{C}_{\mbox{\tiny FP}} \, : \quad  \left( {u \atop v} \right)
       \, \mapsto \, \left( 
       \begin{array}{cc} \; 0  &\,  1\\
                      -1 \, &   \, 0 \end{array} \right)   \left( {u
       \atop v} \right)   \; ,     \label{FPgr}
\end{equation}
which corresponds to a rotation by $\pi/2$ along the compact direction in
$SL(2,\RR)$. 

For completeness, we give an explicit basis for the $sl(2,\RR)$ algebra as
follows, 
\begin{equation} 
\sigma^0 = \Bigg(\begin{array}{cc} 1 \, & \; 0 \\
                                   0 \, & -1 \, \end{array}\Bigg) \; , \quad
\sigma^+ = \Bigg(\begin{array}{cc} 0 \, & \; 1 \\
                                   0 \, & \; 0  \end{array}\Bigg) \; , \quad
\sigma^- = \Bigg(\begin{array}{cc} 0 \, & \; 0 \\
                                   1 \, & \; 0 \end{array}\Bigg) \; ,
\label{sl2r}
\end{equation}
with $[\sigma^0,\sigma^\pm ] = \pm 2\sigma^\pm $ and $[\sigma^+,\sigma^-] =
\sigma_0$. The Noether charges generating the global $SL(2,\RR)$ with this
Lie algebra can be identified with $iQ_c$, $iQ_{\bar c\bar c}/2$ and
$-iQ_{cc}/2$  corresponding to $\sigma^0$, $\sigma^+$ and
$\sigma^-$ , respectively. Their explicit forms are given in Secs. 3.4.2 and
3.5.1 of Ref.~\cite{Nak90}. 
 
The connection with the complex formulation is now possible analogously to the
Cayley map,
\begin{equation} 
       z \, \mapsto \, h(z) = \frac{z-i}{z+i} \; , 
\end{equation}
which maps the upper half of the complex plane biholomorphically
onto the unit disc. Just as $SL(2,\RR)$ is (locally) isomorphic to the
automorphisms of the upper half-plane, with a two-to-one homomorphism
provided by
\begin{equation}
              SL(2,\RR) \ni A = \left( \begin{array}{cc} \alpha \,&\, \beta \\
                                \gamma \,  &\, \delta \end{array}   \right) \,
                                \mapsto  h_A(z) = \frac{\alpha z + \beta}{
                                \gamma z + \delta } \; ,
\end{equation}
so is $SU(1,1)$ two-to-one with the automorphisms of the unit disk in the
complex plane. We therefore introduce complex ghost fields $\eta^a(x)$ with 
$\bar\eta = \eta^\dagger$ by
\begin{equation} 
               \left({\eta \atop \bar\eta}\right) \, := \, S  
               \left( {u \atop v} \right)   \; , \quad \mbox{with} \; S = 
 \frac{1}{\sqrt{2}} \left(\begin{array}{cc}  1  \, & -i\,  \\
                                             1 \, &  \;\, i  \end{array}
               \right) \; ,
\end{equation}
with conventions such that the Cayley map $h(z) =
h_{({\scriptscriptstyle\sqrt{2}}S)}(z)$. Since 
${\scriptstyle\sqrt{-i}} S\in SL(2,\CC)$, from (\ref{sl2c}) we then find
immediately that the Landau gauge ghost Lagrangean of Eq.~(\ref{LGghost}) 
reads,  
\begin{equation} 
\mathcal{L}_{\mbox{\tiny ghost}}^{\xi=0} = -  \frac{1}{2} \, (\eta , \bar\eta
)\, \partial D   \left( \begin{array}{cc} 
                                  \; 0   & \;  1 \\
                                 -1 \;  &  \; 0 \end{array} \right)   
 \left( {\eta \atop \bar\eta} \right) \, =  \,  \bar\eta^a \, 
                    \partial_\mu \!D_\mu^{ab}  \, \eta^b  \; .
                                  \label{LGghc} 
\end{equation}
The map $S$ provides for an isomorphism $SL(2,\RR) \simeq SU(1,1)$,
\begin{equation}
SL(2,\RR) \ni  A \mapsto  SAS^{-1} \in SU(1,1) = \bigg\{ M :=
        \Bigg(\begin{array}{cc}   a \, & \, b \\ 
              \bar b \, & \, \bar a \end{array} \Bigg)  : \; 
        a,b \in \CC  \, , \; \mbox{det}B = 1 \bigg\} \; .
\end{equation}
Explicitly we have 
\begin{equation}
  M = SAS^{-1} = \frac{1}{2} \Bigg( \begin{array}{cc} 
   \alpha + \delta + i(\beta -\gamma) \, & \,  
                     \alpha - \delta  - i(\beta +\gamma) \\
   \alpha - \delta + i(\beta +\gamma) \, & \,  
                     \alpha + \delta  - i(\beta +\gamma) \end{array} \Bigg)
   \; ,
\end{equation}
or $\alpha = \mbox{Re}(a+b)$,  $\beta = \mbox{Im}(b-a)$,  $\gamma =
\mbox{Im}(a+b)$,  $\delta = \mbox{Re}(a-b)$. The global symmetry of
$\mathcal{L}_{\mbox{\tiny ghost}}^{\xi=0}$ in the complex formulation 
is now  $SU(1,1)$, the 3-parameter subgroup of $SL(2,\CC)$ that preserves the
hermiticity assignment $\bar\eta = \eta^\dagger $ under
\begin{equation}  
\Bigg({\eta \atop \bar\eta} \Bigg) \, \mapsto \, M \Bigg({\eta \atop \bar\eta}
\Bigg) \; .
\end{equation}
One verifies furthermore that 
\begin{equation}  
      M^\dagger \Bigg( \begin{array}{cc}  1 \, & \;0 \\
                            0\, & -1\, \end{array} \Bigg) M  \, = \, 
            \Bigg( \begin{array}{cc}  1 \, & \;0 \\
                            0\, & -1\, \end{array} \Bigg) \,
\end{equation}
and with det$M = 1$ thus $M\in SU(1,1)$. It is now simple to transcribe the
action of the ghost number and FP conjugation symmetries under the map $S$:
\begin{equation}
e^{iQ_c \theta} \; : \quad  \eta \mapsto  (\cosh\theta)  \, \eta  +
(\sinh\theta)  \, \bar\eta \; ; \quad  \mbox{and} \quad
\mathcal{C}_{\mbox{\tiny FP}}   \; : \quad   \eta \mapsto i \eta \; .
\end{equation}
The original ghost number symmetry is no-longer diagonal, 
\begin{equation}
           [i Q_c , \eta^a(x) ] = \bar\eta^a(x) \; , \qquad 
                         [i Q_c , \bar\eta^a(x) ] = \eta^a(x)
           \; .
\end{equation}
Now, in the complex basis, the FP conjugation is diagonal which arose from a
rotation by $\lambda = \pi/2$ along the compact direction of $SL(2,\RR)$. This
$U(1)$-subgroup gives rise to a conserved global ``$U(1)$ ghost number''
symmetry of the complex fields corresponding to 
          $ \displaystyle \eta \mapsto e^{i\lambda } \eta $.

Before we continue to discuss the BRS and anti-BRS symmetries, note that
the full symmetry of the Landau gauge can be maintained for $\xi\not=0 $ in a
slightly more general setting for the covariant gauge
fixing~\cite{Cur76a,Bau82,Thi85,Nak90}. As observed in Ref.~\cite{Bau82},
renormalisability, global gauge, BRS and Lorentz invariance allow, as the
most general form in four dimensions, the addition of another independent term
to $\mathcal{L_{GF}}$ in Eq.~(\ref{Lgf}) which can be expressed in terms of
real or complex ghost fields as follows, 
\begin{equation}   
\frac{\zeta}{2} (B^a + g f^{abc} v^b u^c )^2 =  \frac{\zeta}{2} 
\Big(B^a + i \frac{g}{2} 
       f^{abc} (\eta^b \eta^c - \bar\eta^b\bar\eta^c ) \Big)^2  \; ,
\end{equation}
and which introduces a second gauge parameter $\zeta$. This parameter
controls the hermiticity of the Lagrangean. In particular, for $\zeta = 0$,
in the standard Faddeev-Popov gauges with $\xi \not=0$, the FP conjugation
symmetry is broken and only the real formulation thus leads to a hermitean
Lagrangean. The full global $SL(2,\RR) \simeq SU(1,1)$ which establishes the
equivalence of both formulations on the other hand  is maintained for $\zeta
= \xi$. For $\xi \not=0$, one can therefore introduce,
\begin{eqnarray}
 \mathcal{L'_{GF}}  &=&  
i B^a \partial_\mu A_\mu^a + \frac{\xi}{4} \Big( B^a B^a + (B^a + g f^{abc}
v^b u^c )^2 \Big) + i v^a \partial_\mu D_\mu^{ab} u^b \; ,     
 \label{Lgfp}  \\
  &=& i B^a \partial_\mu A_\mu^a + \frac{\xi}{4} \bigg\{ B^a B^a + \Big(B^a + 
 i \frac{g}{2} f^{abc} (\eta^b \eta^c - \bar\eta^b\bar\eta^c )\Big)^2 \bigg\} 
    \, + \frac{1}{2}\,(\eta + \bar\eta)^a \, \partial_\mu \!D_\mu^{ab}  \, (
 \eta - \bar\eta)^b   \; , \nonumber
\end{eqnarray}
to generalise the symmetry of the Landau gauge. The effect of this
modification is most easily seen from shifting the $B$-field,
\begin{equation} 
B' :=  B + \frac{g}{2} \, (v \times u) =
B + i\, \frac{g}{4}  \bigg( \eta \times \eta - \bar\eta \times \bar \eta 
     \bigg) \; ,
\quad \mbox{with} \; \;  
( v \times u )^a \equiv f^{abc} v^b u^c  \; .  \label{defBp}
\end{equation}
In terms of the $B'$-field, it is straightforward to rewrite the gauge
fixing Lagrangean $\mathcal{L'_{GF}}$ of Eq.~(\ref{Lgfp}),
\begin{eqnarray}
 \mathcal{L'_{GF}} &=& iB' \partial A + \frac{\xi}{2} B'B'  + \frac{\xi}{2} 
      \Big( \frac{g}{2} ( v \times u )
     \Big)^2  +  \, \frac{i}{2} \,  v \, \Big(  \partial D \,+\, 
      D\partial  \Big)\, u \; ,  \label{Lgfpr} \\
 &=&  i B' \partial A + \frac{\xi}{2} B'B'  - \frac{\xi}{2} 
      \bigg( \frac{g}{2} ( \bar\eta \times \eta )
     \bigg)^2  +  \, \frac{1}{2} \,  \bar\eta \, \Big(  \partial D \,+\, 
      D\partial  \Big)\, \eta \; .  \label{Lgfpc}
\end{eqnarray}
For the quartic ghost-interaction in the complex version, the 2nd
Eq.~(\ref{Lgfpc}) above, we have made use of the Jacobi identity  to rewrite
$ ( \eta \times \eta - \bar\eta \times \bar \eta )^2 = 4 (\bar\eta \times\eta
)^2 $. At the expense of these quartic ghost self-interactions, the essential
effect of the additional term in (\ref{Lgfp}) is to ``symmetrise'' the
Faddeev-Popov operator of the conventional covariant gauges, $\partial D  \to
( \partial D \,+\,  D\partial)/2 $. The $SL(2,\RR)$ or the $SU(1,1)$ invariance
of ghost self-interactions in the real or the complex version, respectively,  
is most easily seen from,
\begin{eqnarray}
 f^{abc} (\bar\eta^b \eta^c) \,  f^{ade} (\bar\eta^d \eta^e) &=&
           \frac{2}{N_c} \,  (\bar\eta^a\eta^a) (\bar\eta^b\eta^b) \,+\,
    d^{abc} (\bar\eta^b \eta^c) \,  d^{ade} (\bar\eta^d \eta^e) \nonumber\\
        &=&  \frac{1}{2N_c} \Big( (\eta^a,\bar\eta^a) \,\varepsilon\,
           \bigg({\eta^a\atop\bar\eta^a}\bigg) \Big)^2 + \frac{1}{4} \Big( 
              d^{abc}   (\eta^b,\bar\eta^b) \,\varepsilon\,
           \bigg({\eta^c\atop\bar\eta^c}\bigg) \Big)^2 \; . 
\end{eqnarray}
One therefore verifies that
$\mathcal{L'_{GF}}$ is invariant under the global $SL(2,\RR) \simeq SU(1,1)$
just as $\mathcal{L}_{\mbox{\tiny ghost}}^{\xi=0}$. 
In absence of the (anti-)BRS transformations discussed below, $B'$, $A$, and
the quark fields $q$, could be taken to belong to the trivial (singlet)
representation. Note, however, that this is not true for the original $B$
field introduced in (\ref{Lgf}) which transforms non-trivially. Under FP
conjugation for instance, 
\begin{equation} 
      \mathcal{C}_{\mbox{\tiny FP}} 
     \, : \quad B \mapsto  B + g (v \times u) \; ,
      \quad \mbox{while}  \; B' \mapsto B' \; . \label{FPB}
\end{equation}
For $\xi = 0$ we recover the Landau gauge Lagrangean.    
Its symmetrised ghost part, when the $SL(2,\RR) \simeq SU(1,1)$
invariant $B'$ field is employed as in Eqs.~(\ref{Lgfpr}) or (\ref{Lgfpc}), 
is to replace $L_{\mbox{\tiny ghost}}^{\xi=0}$ for an {\em off-shell}
extension of the discussion of the global ghost symmetries and the relation
between the real and the complex formulation, {\it i.e.}, one which proceeds 
analogously to that leading from Eqs.~(\ref{LGghost}) to (\ref{LGghc}) 
without need, however, to employ the constraint $\partial A = 0$.   

Under the map $S$, we furthermore obtain from Eqs.~(\ref{sl2r})
(a basis for $su(1,1)$, of course), 
\begin{equation}
\sigma_0 \mapsto 
  \Bigg(\begin{array}{cc} 0 \, & \; 1 \\
                                   1 \, & \; 0  \end{array}\Bigg) \; , \quad
\sigma^1 =  \sigma^+ + \sigma^- \mapsto 
    \Bigg(\begin{array}{cc} 0 \, & -i \, \\
                                   i \, & \; 0  \end{array}\Bigg) \; , \quad
\sigma^2 =  \sigma^+ - \sigma^- \mapsto 
    \Bigg(\begin{array}{cc} i \; &  \, 0 \\
                                  0 \; & -i \,  \end{array}\Bigg) \; .
\end{equation}
And the corresponding Noether currents, in terms of the complex ghost fields,
are given by,
\begin{eqnarray}  
      J^0_\mu &=& -\frac{1}{2} \bigg( \eta \Big(\partial_\mu + D_\mu \Big)
      \eta - \bar\eta \Big(\partial_\mu + D_\mu \Big) \bar\eta \bigg) \; ,
      \qquad Q^0 = \int \!\! d^3x  \, J^0_0 \, = \, Q_c  \; ,\\
      J^1_\mu &=& -\frac{i}{2} \bigg( \eta \Big(\partial_\mu + D_\mu \Big)
      \eta + \bar\eta \Big(\partial_\mu + D_\mu \Big) \bar\eta \bigg) \; ,
      \qquad Q^1 = \int \!\! d^3x \, J^1_0 \,  = \, \frac{1}{2} \Big( Q_{\bar
      c\bar c} - Q_{cc} \Big)  \; , \nonumber \\
      J^2_\mu &=& \phantom{-} \frac{i}{2} \bigg( \bar\eta \Big(\partial_\mu +
      D_\mu \Big)  \eta + \eta \Big(\partial_\mu + D_\mu \Big) \bar\eta
      \bigg) \; , \qquad Q^2 = \int \!\! d^3x \, J^2_0 \, = \, 
       \frac{1}{2} \Big( Q_{\bar c\bar c} + Q_{cc} \Big)   \; .\nonumber
\end{eqnarray}

We now introduce BRS variations $\delta \Phi \equiv \lambda \delta_{\bf
B}\Phi$ of generic fields $\Phi $
with global Grassmann parameter $\lambda $ and the BRS charge $Q_B$ such that
in the operator formulation,
\begin{equation} 
                  \delta_{\bf B} \Phi =  i \{ Q_B , \Phi \} \; ,
\end{equation}
where $\{,\}$ denotes the ghost-number graded commutator, {\it i.e.}, the
anti-commutator if both operators have odd ghost number and the
commutator otherwise. In presence of the full ghost-antighost symmetry one
also has anti-BRS invariance, defined by
\begin{equation}
\bar\delta_{\bf B} \Phi  =  i \{ \overline{Q}_B , \Phi \}   :=
\mathcal{C}_{\mbox{\tiny FP}} \, \delta_{\bf B} \, \mathcal{C}_{\mbox{\tiny
FP}}^{-1} \; . 
\end{equation}
With the Euclidean conventions introduced in Sec.~\ref{chap_Basic}
for the real fields in Landau gauge we have ({\it c.f.}, Eqs.~(\ref{FPgr}) and
(\ref{FPB}) for the action of $\mathcal{C}_{\mbox{\tiny FP}}$),
\begin{equation}
\begin{array}{l@{\hskip 2cm}l} 
 \delta_{\bf B} A = -D u &  \bar\delta_{\bf B} A = -D v \\
\displaystyle \delta_{\bf B} u = -\frac{g}{2} \, (u \times u ) &
  \bar\delta_{\bf B} u = - B - g \, (v\times u)  \\
 \delta_{\bf B} v = B  & \displaystyle
  \bar\delta_{\bf B} v = - \frac{g}{2} \, (v\times v)  \\
   \delta_{\bf B} B = 0  &  \bar\delta_{\bf B} B = - g \, (v \times
 B) \\
    \delta_{\bf B} q = i g \, (t^a u^a) \, q &  \bar\delta_{\bf B} q =
 i g \, (t^a v^a) \, q  \; .  
\end{array} \label{realBRS}
\end{equation}
As usual, one furthermore has $\delta_{\bf B} \delta_{\bf B}  = 
\bar\delta_{\bf B} \bar\delta_{\bf B} = \delta_{\bf B} \bar\delta_{\bf B}  + 
\bar\delta_{\bf B} \delta_{\bf B} = 0 $.

For the complex formulation, employing the gauge fixing
corresponding to $\mathcal{L'_{GF}}$ of Eq.~(\ref{Lgfpc}) which generalises
the Landau gauge in an $SU(1,1)$ symmetric way, it is convenient to introduce
{\em complex} BRS variations as follows,
\begin{equation}
         \begin{array}{l@{\hskip 2cm}l} 
 \displaystyle \delta_{\bf B}' := \frac{1}{\sqrt{2}} \Big( \delta_{\bf B} - i
        \bar\delta_{\bf B} \Big) &
 \displaystyle \bar\delta_{\bf B}' :=
        \frac{1}{\sqrt{2}} \Big( \delta_{\bf B} + i \bar\delta_{\bf B} \Big) 
           \; . \end{array}
\end{equation}
It is then straightforward to verify that for these, one obtains,
\begin{equation}
\begin{array}{@{\hskip -1cm}l@{\hskip 1cm}l} 
  \delta_{\bf B}' A = -D \eta &  \bar\delta_{\bf B}' A = -D
  \bar\eta  \\
 \displaystyle \delta_{\bf B}' \eta = -\frac{g}{2} \, (\eta \times \eta ) & 
 \displaystyle \bar\delta_{\bf B}' \eta = - iB' - \frac{g}{2} \, 
    (\bar\eta\times \eta)   \\
 \displaystyle  \delta_{\bf B}' \bar\eta = iB' - \frac{g}{2} \, 
    (\bar\eta\times \eta)  & 
 \displaystyle \bar\delta_{\bf B}' \bar\eta = - \frac{g}{2} \, 
    (\bar\eta\times \bar\eta )  \\
 \displaystyle  \delta_{\bf B}' B' =  - \frac{g}{2} \, (\eta \times B')  - i
  \frac{g^2}{8} \, \Big(  \bar\eta \times ( \eta \times \eta ) \Big) &
 \displaystyle \bar\delta_{\bf B} B' = - \frac{g}{2} \, (\bar\eta \times B' )
   + i \frac{g^2}{8} \, \Big( \eta \times ( \bar\eta \times \bar\eta )\Big) \\
  \delta_{\bf B}' q = i g \, (t^a \eta^a)\, q  &  \bar\delta_{\bf B}'
 q =  i g \, (t^a \bar\eta^a) \, q   \; ,
\end{array} 
\hskip -1.5cm  \label{compBRS}
\end{equation}
and again, of course,  $\delta_{\bf B}' \delta_{\bf B}'  = 
\bar\delta_{\bf B}' \bar\delta_{\bf B}' = \delta_{\bf B}' \bar\delta_{\bf B}'
+  \bar\delta_{\bf B}' \delta_{\bf B}' = 0 $.

Clearly, originally real fields will in general no-longer remain real under
the complex BRS transformations just as the real (anti-)BRS transformations
lead to $B',A$ and quark fields that transform non-trivially under
$SL(2,\RR)$. This is because the invariance of the symmetrically covariant
gauge fixed theory is a {\em semi-direct} product of the global $SL(2,\RR)
\simeq SU(1,1)$ with the BRS symmetry \cite{Cur76a,Cur76b,Thi85}. Though 
the former is an invariant subgroup, the latter is not and 
the transformations of both do not commute with each other.

For the complex BRS transformations given above one readily verifies that the
gauge fixing Lagrangean in Eq.~(\ref{Lgfpc}) can be represented by
\begin{equation} 
\mathcal{L'_{GF}} \,=\, \delta_{\bf B}' \bigg[ \bar\eta \Big( \partial A - i
\frac{\xi}{2} \, B' \Big) \bigg]  \; .
\end{equation}
The complex formulation can also be cast in a form which is less symmetric
with respect to the complex BRS transformations which resembles the familiar
real formulation, Eqs.~(\ref{Lgfpr}) and (\ref{realBRS}), more closely,
however. This is possible with a second shift of the $B$-field, {\it c.f.},
(\ref{defBp}), 
\begin{equation}
 B'' := B' + i \frac{g}{2} \, (\bar\eta \times \eta) = 
B + i\, \frac{g}{4}  \bigg( \eta \times \eta - \bar\eta \times \bar \eta 
    + 2 \,  \bar\eta \times \eta   \bigg) \; .
\end{equation}
The complex BRS transformations of Eqs.~(\ref{compBRS}) can then be written,
\begin{equation}
\begin{array}{@{\hskip -1cm}l@{\hskip 1cm}l} 
 \displaystyle  \delta_{\bf B}' \bar\eta = iB'' & 
 \displaystyle \bar\delta_{\bf B}' \eta = - iB'' -  g \, 
    (\bar\eta\times \eta)   \\
 \displaystyle  \delta_{\bf B}' B'' =  0  &
 \displaystyle \bar\delta_{\bf B} B'' = - g \, (\bar\eta \times B'' ) 
\end{array} 
\end{equation}
with the other transformations in Eqs.~(\ref{compBRS}) remaining unchanged.
In terms of this field $B''$, Eq.~(\ref{Lgfpc}) yields,
\begin{eqnarray} 
\mathcal{L'_{GF}} &=&  i B'' \Big(\partial A - \xi \frac{g}{2} \, (\bar\eta
\times \eta ) \Big) 
+ \frac{\xi}{2} B''B''  - \xi \bigg( \frac{g}{2} ( \bar\eta \times \eta )
     \bigg)^2  +  \, \bar\eta \, \partial D  \, \eta \; , \label{Lgfpp} \\
  & = & \delta_{\bf B}' \bigg[ \bar\eta \bigg( F(A)  - i
\frac{\xi}{2} \, \Big( B'' + i \frac{g}{4} \, (\bar\eta \times\eta)  \Big) 
   \bigg)  \bigg]  \; , \quad\mbox{with} \quad 
      F(A) =  \partial A - \xi \frac{g}{2} \, (\bar\eta \times \eta )  \; ,  
\end{eqnarray}
and with the constraint  $F(A) = i \xi B''$ 
for the $B''$-field in (\ref{Lgfpp}) being equivalent to $ \partial A = i \xi
B' $ from (\ref{Lgfpc}).  The form for the gauge fixing Lagrangean given in
Eq.~(\ref{Lgfpc}) should be viewed as the correct extension beyond the Landau
gauge of the complex formulation introduced in Sect.~\ref{chap_Basic} in
Eqs.~(\ref{Leff}) and (\ref{BRS_2}) with the identification $\eta \to c $,
$\lambda\delta_{\bf B}' \to \delta$  and the hermiticity assignment 
$\bar c = c^\dagger$. In particular, integrating the $B'$-field in
Eq.~(\ref{Lgfpc}) (or the $B''$-field in (\ref{Lgfpp}) above), the (unique) 
result is,  
\begin{equation} 
    \mathcal{L'_{GF}}  \, = \, \frac{1}{2\xi}\Big(\partial_\mu A^a_\mu \Big)^2
    \,-  \, \frac{\xi}{2} \bigg( \frac{g}{2} f^{abc} \bar c^b c^c 
     \bigg)^2  + \frac{1}{2} \, \bar c^a \, \Big(\partial_\mu D_\mu^{ab}  +
    D^{ab}_\mu \partial_\mu  \Big) c^b \; . 
\end{equation}
And the {\em on-shell} BRS transformation for $\bar c$ then reads,
\begin{equation} 
     \delta \bar c^a =  \frac{1}{\xi} \, F^a(A) \, \lambda \; , \quad 
          \mbox{with}  
        \quad      F^a(A) = \partial_\mu A^a_\mu - \xi \, 
      \frac{g}{2} \, f^{abc}  \bar c^b c^c \; .
\end{equation}
The Landau gauge limit $\xi \to 0$ is smooth for the $SL(2,\RR) \simeq
SU(1,1)$ symmetric gauge fixing, the renormalisability is maintained
\cite{Bau82}, and the important observation that $\widetilde Z_1 = 1$ in
Landau gauge has also been verified at one-loop level in Ref.~\cite{Bau82}. 

The relation between the BRS-algebra and de Rham cohomology is quite well
established. Pedagogical accounts of this can be found in, {\it e.g.},
Refs.~\cite{Nak90,Bir91,Nis96}. The semi-direct product of
the $SL(2,\RR)$ with the double BRS-algebra generated by the charges $Q_B$
and $\overline{Q}_B$ can be obtained by a Inonu--Wigner contraction of the
simple $OSp(1,2)$ superalgebra in which the Curci--Ferrari mass
term~\cite{Cur76a,Cur76b} is sent to zero, see Refs.~\cite{Thi85,Nak90}.
For completeness we furthermore mention that a relation of the double
BRS-algebra to the theory of double complexes (the Dolbeault cohomology for
complex manifolds) was realized in
Refs.~\cite{Thi79,Qui81,Hoy82,Hoy83,Thi85}. The $SL(2,\RR)$ ghost-antighost
symmetry was combined with the Lorentz symmetry to arrive at a superspace
formulation in Refs.~\cite{Bon81,Hir81,Del82}.

\section{Conventions for Fourier Transformations} 
\label{app_FT}

In this appendix we summarise our conventions for Fourier transforming Green's
functions from Euclidean space to the corresponding momentum space. For all 
propagators $D(x,y)$ which depend in a translationally invariant background only
on $\xi = x-y$ we use
\begin{equation}
  D(p)    := \int\! d^4 \xi \e^{-i p\xi } D(x,y) \, .
\end{equation}

For the fermion-photon, quark-gluon and gluon-ghost vertex functions we use
notations adopted to having in- and out-going fermion (ghost) legs. The
momentum  of the photon is defined as outgoing whereas the two fermion momenta
are chosen differently, one incoming and one outgoing: 
\begin{equation}
  \Gamma_\mu(k,q,p)
    = \int d^4x \, d^4y \, d^4z
      \e^{i kx} \e^{i qy} \e^{-i pz} \Gamma^a_\mu(x;y,z) \, .
\end{equation}
Momentum conservation allows to define a reduced vertex function:
\begin{equation}   \label{fpv_def_a}
     \Gamma_\mu(k,q,p) =   - ie  \, (2\pi)^4 \delta^4(k+q-p)
     \Gamma_\mu(q,p)  \; ,
\quad \hbox{\hskip .8cm 
\begin{minipage}[c]{0.25\linewidth}
  \epsfig{file=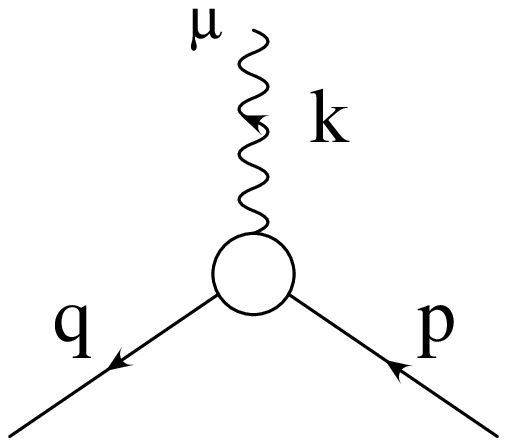,width=0.7\linewidth}
\end{minipage}}
\end{equation}
The momentum dependence of the quark-gluon and the ghost-gluon vertex is
defined completely analogous, see Sect.\
\ref{sec_nPoint}. 

For the three-gluon vertex function, on the other hand, we choose a momentum
dependence emphasing its complete Bose symmetry:
\begin{equation}
  \Gamma^{abc}_{\mu\nu\rho}(x,y,z)
    = \int \frac{d^4p}{(2\pi)^4} \frac{d^4q}{(2\pi)^4} \frac{d^4k}{(2\pi)^4}
        \e^{i px} \e^{i qy} \e^{i kz} \Gamma^{abc}_{\mu\nu\rho}(p,q,k) \; .
\end{equation}
The corresponding reduced vertex is then defined as
\begin{equation}
    \Gamma^{abc}_{\mu\nu\rho}(k,p,q) =:
                                   g f^{abc} (2\pi)^4 \delta^4(k+p+q)
\Gamma_{\mu\nu\rho}(k,p,q)     \; .
\quad \hbox{
\begin{minipage}[c]{0.25\linewidth}
  \epsfig{file=3GluonVertex.eps,width=0.8\linewidth}
\end{minipage}}
\end{equation}
Hereby
the indices are given in counterclockwise order starting at the dot.

\section{Dyson--Schwinger Equation for the Gluon Propagator}
\label{app_gluon-DSE}

\begin{figure}[t]
  \centering\epsfig{file=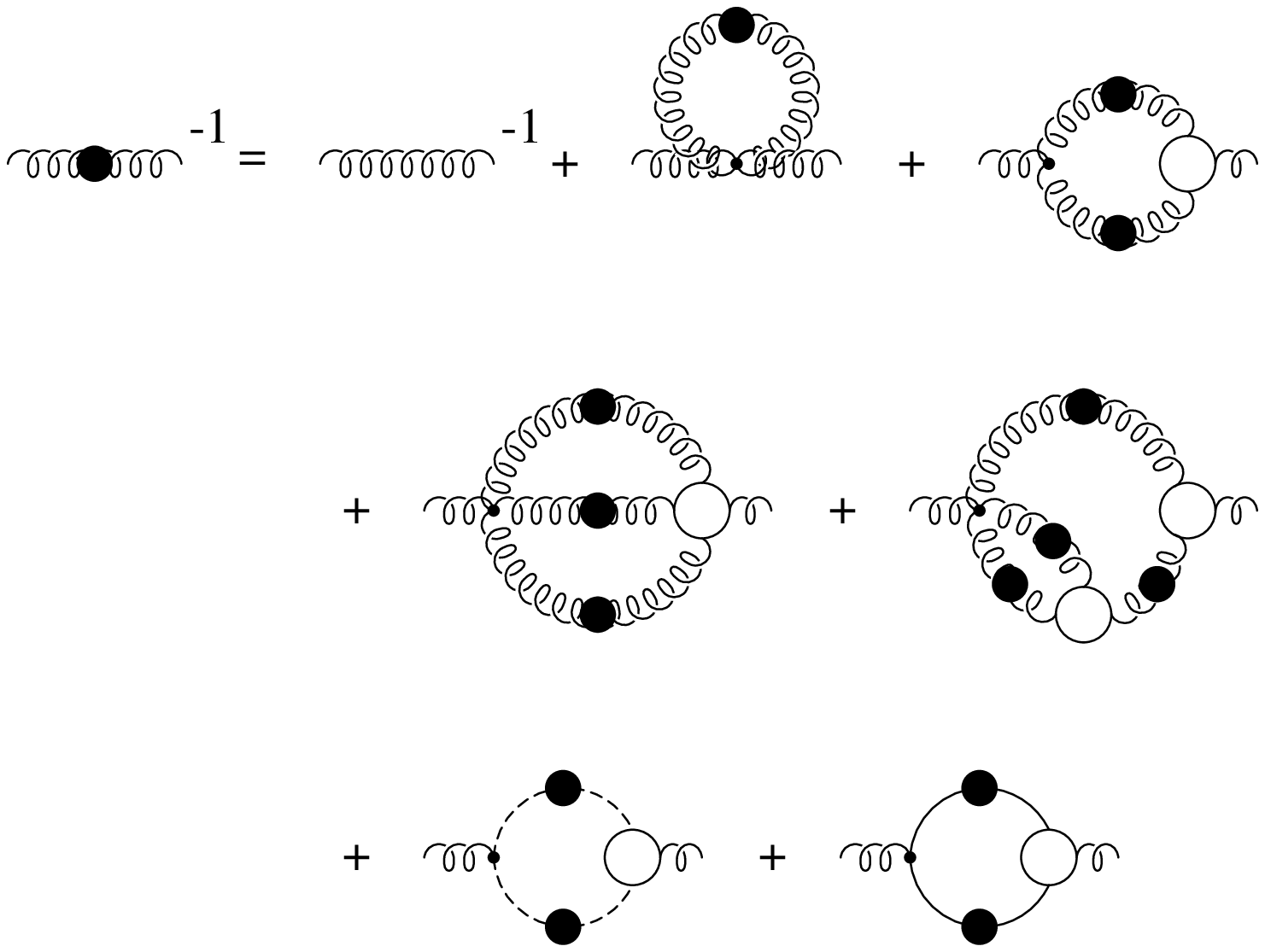,width=.8\linewidth}
  \caption[Pictorial representation of the gluon DSE.]
    {Pictorial representation of the gluon DSE.}
   \label{fig:Gluon-DSE-A}
\end{figure}

In this appendix we will give the explicit form of the DSE for the 
gluon propagator. One obtains
\begin{eqnarray}
  -\frac{\delta^2 \Gamma_{\mbox{\tiny QCD}}}
  {\delta A^a_\mu(x) \,  \delta A^b_\nu(y)} &=&
    - Z_3 \left[ \delta_{\mu\nu} \,  \partial ^2
        - \left( 1 - \frac{1}{Z_3 \,  \xi} \right) \partial _\mu \partial _\nu \right]
          \delta^{ab} \,  \delta(x-y)  \nonumber \\
    &+& Z_1 \,  g f^{ade} \int d^4z \, 
        \langle  A^b_\nu(y) A^c_\rho(z) \rangle ^{-1} \biggl\{
          \langle  A^c_\rho(z) A^d_\sigma(x) \,  \partial _\mu A^e_\sigma(x) \rangle  
	  \nonumber \\
       && \qquad - \langle  A^c_\rho(z) A^d_\sigma(x) \,  \partial _\sigma A^e_\mu(x) \rangle 
        - \langle  A^c_\rho(z) \,  \partial _\sigma A^d_\sigma(x) A^e_\mu(x) \rangle  \biggr\}  
	\nonumber\\
    &+& Z_4 \,  g^2 f^{afg} f^{gde} \Bigl\{
          \delta^{bf} \langle  A^d_\mu(x) A^e_\nu(x) \rangle 
        + \delta^{be} \langle  A^f_\nu(x) A^d_\mu(x) \rangle   \nonumber \\
       && \qquad + \delta^{bd} \delta_{\mu\nu} \langle  A^f_\rho(x) A^e_\rho(x) \rangle 
              \Bigr\} \,  \delta(x-y) \nonumber\\
    &+& Z_4 \,  g^2 f^{afg} f^{gde} \int d^4z \, 
        \langle  A^b_\nu(y) A^c_\rho(z) \rangle ^{-1}
        \langle  A^c_\rho(z) A^f_\sigma(x) A^d_\mu(x) A^e_\sigma(x) \rangle _c  \nonumber\\
    &-& Z_{1F} i g \,  t^a \,  \gamma_\mu \int d^4z \, 
        \langle  A^b_\nu(y) A^c_\rho(z) \rangle ^{-1}
        \langle  A^c_\rho(z) \,  \bar{\psi}(x) \,  \psi(x) \rangle _c  \nonumber\\
    &+& \widetilde Z_1 \,  g f^{ade} \int d^4z \, 
        \langle  A^b_\nu(y) A^c_\rho(z) \rangle ^{-1}
       \langle  A^c_\rho(z) \bigl( \partial _\mu \,  \bar{c}^d(x) \bigr) \, 
       c^e(x) \rangle _c \; 
\end{eqnarray}
where the functional identity
\begin{eqnarray}
  \frac{\delta}{\delta A^a_\mu(x)}
    &=&   \int d^4z \,
         \frac{\delta J^b_\nu(z)}{\delta A^a_\mu(x)} \,
         \frac{\delta}{\delta J^b_\nu(z)}  \\
    &=& - \int d^4z \,
         \frac{\delta^2 \Gamma}{\delta A^a_\mu(x) \, \delta A^b_\nu(z)} \,
         \frac{\delta}{\delta J^b_\nu(z)} \nonumber
\end{eqnarray}
has been used.
The resulting gluon DSE reads
\begin{eqnarray}
  {D^{-1}}^{ab}_{\mu\nu}(x-y) &=&
      Z_3 \, {D^{-1}_{(0)}}^{ab}_{\mu\nu}(x-y)  \nonumber \\
    &+& Z_{1} \, \frac{1}{2}
      \int d^4x_1 \, d^4x_2 \, d^4y_1 \, d^4y_2  \nonumber \\
     &&   \qquad \times {\Gamma^{(0)}}^{acd}_{\mu\alpha\beta}(x,x_1,x_2) \,
        D^{de}_{\beta\gamma}(x_2-y_1) \, D^{cf}_{\alpha\delta}(x_1-y_2) \,
        \Gamma^{bef}_{\nu\gamma\delta}(y,y_1,y_2)  \nonumber\\
    &+& Z_{4} \, \frac{1}{2} \, \int d^4x_1 \, d^4x_2 \,
        {\Gamma^{(0)}}^{abcd}_{\mu\nu\alpha\beta}(x,y,x_1,x_2) \,
        D^{cd}_{\alpha\beta}(x_1-x_2)  \nonumber \\
    &+& Z_{4} \, \frac{1}{6} \,
        \int d^4x_1 \, d^4x_2 \, d^4x_3 \,
             d^4y_1 \, d^4y_2 \, d^4y_3 \,
        {\Gamma^{(0)}}^{acde}_{\mu\alpha\beta\gamma}(x,x_1,x_2,x_3) \nonumber \\
       && \qquad\times
        D^{cm}_{\alpha\lambda}(x_1-y_3) \,
        D^{dl}_{\beta\sigma}(x_2-y_2)   \,
        D^{ek}_{\gamma\rho}(x_3-y_1)    \,
        \Gamma^{bklm}_{\nu\rho\sigma\lambda}(y,y_1,y_2,y_3)  \nonumber\\
    &+& Z_{4} \, \frac{1}{2} \,
        \int d^4x_1 \, d^4x_2 \, d^4x_3 \,
             d^4y_1 \, d^4y_2 \, d^4y_3 \,
             d^4z_1 \, d^4z_2 \\
      &&  \qquad\times
        {\Gamma^{(0)}}^{acde}_{\mu\alpha\beta\gamma}(x,x_1,x_2,x_3) \,
        D^{ck}_{\alpha\rho}(x_1-y_1)   \,
        D^{dm}_{\beta\lambda}(x_2-y_3) \,
        D^{ep}_{\gamma\delta}(x_3-z_1) \nonumber \\
     &&   \qquad\times
        \Gamma^{klm}_{\rho\sigma\lambda}(y_1,y_2,y_3) \,
        D^{lq}_{\sigma\kappa}(y_2-z_2) \,
        \Gamma^{bpq}_{\nu\delta\kappa}(y,z_1,z_2) \nonumber\\
    &+& Z_{1F} \int d^4x_1 \, d^4x_2 \, d^4y_1 \, d^4y_2  \nonumber\\
     &&   \qquad \times {\Gamma^{(0)}}^a_\mu(x,x_1,x_2) \,
        S(x_2-y_1) \, S(y_2-x_1) \, \Gamma^b_\nu(y,y_1,y_2)  \nonumber\\
    &+& \widetilde Z_{1} \int d^4x_1 \, d^4x_2 \, d^4y_1 \, d^4y_2  \nonumber \\
     &&   \times {\Gamma^{(0)}}^{acd}_\mu(x,x_1,x_2) \,
        G^{de}(x_2-y_1) \, G^{fc}(y_2-x_1) \, \Gamma^{bef}_\nu(y,y_1,y_2)
	\label{glDSE}
\end{eqnarray}
Finally, after Fourier transforming into momentum space this equation becomes
\begin{eqnarray}
 {D^{-1}}^{ab}_{\mu\nu}(p) &=&
      Z_3 \,  {D^{-1}_{(0)}}^{ab}_{\mu\nu}(p)  \nonumber\\
    &+& Z_{1} \,  \frac{1}{2}
      \int d^4q \,  d^4k \, 
        {\Gamma^{(0)}}^{acd}_{\mu\alpha\beta}(p,-q,-k) \, 
        D^{de}_{\beta\gamma}(k) \,  D^{cf}_{\alpha\delta}(q) \, 
        \Gamma^{bef}_{\nu\gamma\delta}(-p,k,q)  \nonumber\\
    &+& Z_{4} \,  \frac{1}{2} \,  \int d^4q_1 \,  d^4q_2 \, 
        {\Gamma^{(0)}}^{abcd}_{\mu\nu\alpha\beta}(p,-p,-q_1,q_2) \, 
        D^{cd}_{\alpha\beta}(q_1) \nonumber \\
    &+& Z_{4} \,  \frac{1}{6} \, 
        \int d^4k_1 \,  d^4k_2 \,  d^4k_3 \, 
        {\Gamma^{(0)}}^{acde}_{\mu\alpha\beta\gamma}(p,-k_1,-k_2,-k_3) \nonumber\\
     &&   \qquad\times
        D^{cm}_{\alpha\lambda}(k_1) \, 
        D^{dl}_{\beta\sigma}(k_2)   \, 
        D^{ek}_{\gamma\rho}(k_3)    \, 
        \Gamma^{bklm}_{\nu\rho\sigma\lambda}(p,k_3,k_2,k_1)  \nonumber\\
    &+& Z_{4} \,  \frac{1}{2} \, 
        \int d^4k_1 \,  d^4k_2 \,  d^4k_3 \,  d^4k_4  \nonumber\\
     &&   \qquad\times
        {\Gamma^{(0)}}^{acde}_{\mu\alpha\beta\gamma}(p,-k_1,-k_2,-k_3) \, 
        D^{ck}_{\alpha\rho}(k_1)   \, 
       D^{dm}_{\beta\lambda}(k_2) \, 
        D^{ep}_{\gamma\delta}(k_3) \nonumber\\
     &&   \qquad\times
        \Gamma^{klm}_{\rho\sigma\lambda}(k_1,-k_4,k_2) \, 
        D^{lq}_{\sigma\kappa}(k_4) \, 
        \Gamma^{bpq}_{\nu\delta\kappa}(-p,k_3,k_4) \nonumber\\
    &+& Z_{1F} \int d^4q \,  d^4k \, 
        {\Gamma^{(0)}}^a_\mu(p,q,k) \, 
        S(-q) \,  S(k) \,  \Gamma^b_\nu(-p,k,q)   \nonumber\\\
    &+& \widetilde Z_{1} \int d^4q \,  d^4k \, 
        {\Gamma^{(0)}}^{acd}_\mu(p,q,k) \, 
        G^{de}(-q) \,  G^{fc}(k) \,  \Gamma^{bef}_\nu(-p,k,q)
  \label{eq:Gluon-DSE}
\end{eqnarray}
which is the Dyson-Schwinger equation for the (inverse) gluon propagator, 
derived in Refs.\ \cite{Eic74,Bak77} and schematically depicted in Fig.\
\ref{fig:Gluon-DSE-A}.

In linear covariant gauges the gluon propagator has a general structure 
described  by one scalar function $Z(k^2)$, see Eq.\ (\ref{eq:Gluon-Prop}).
On the other hand,
in the axial gauge the gluon propagator involves in general two scalar functions
$g$ and $f$:\footnote{Colour indices are suppressed in the following formulas.
As usual, we assume that the gluon propagator is colour diagonal.}
\begin{eqnarray}
D_{\mu\nu}(p) &=& \Delta(p^2,(pt)^2) \, \left\{ {\mathcal M}_{\mu\nu}(p) \,
g(p^2,(pt)^2) \,- \, t^2 {\mathcal P}_{\mu\nu}(t) \, f(p^2,(pt)^2) \right\} 
\label{glp_axg}\\
\Delta(p^2,(pt)^2) &=& 
 \left(g(p^2,(pt)^2) - t^2 f(p^2,(pt)^2)\right)^{-1} \,
\left(p^2 g(p^2,(pt)^2) - (pt)^2 f(p^2,(pt)^2)\right)^{-1} \; . \nonumber
\end{eqnarray}
Here, the transverse projector ${\mathcal P}_{\mu\nu}(t) =
\delta_{\mu\nu} \, - \, t_\mu t_\nu /t^2 $ appears in addition to the tensor
\begin{equation}
{\mathcal M}_{\mu\nu}(k) =  \delta_{\mu\nu} \, - \, \frac{k_\mu t_\nu + k_\nu
t_\mu}{kt} \, +\, t^2 \frac{k_\mu k_\nu}{(kt)^2 } \; .
\end{equation}
The tree-level propagator is obtained for $g = 1$ and $f=0$.  With the above
parameterisation  of the gluon propagator
(\ref{glp_axg}), the vacuum polarisation tensor follows to be of the form,
\begin{equation}
\Pi_{\mu\nu}(p) =  p^2 \, {\mathcal P}_{\mu\nu}(p) \, g(p^2,(pt)^2) \,- \, (pt)^2
\, {\mathcal N}_{\mu\nu}(p) \, f(p^2,(pt)^2) \; .
\end{equation}
Due to the presence of $f$  another tensor is involved,
\begin{equation}
 {\mathcal N}_{\mu\nu}(k) =  \delta_{\mu\nu} \, - \, \frac{k_\mu t_\nu + k_\nu
t_\mu}{kt} \, +\, k^2 \frac{t_\mu t_\nu}{(kt)^2 } \; ,
\end{equation}
and one has,
\begin{equation}
   D_{\mu\rho}(p) \, \Pi_{\rho\nu}(p) \,= \,  \delta_{\mu\nu} \, - \,
\frac{p_\mu t_\nu}{pt} \; , \quad \Pi_{\mu\rho}(p) \, D_{\rho\nu}(p)
   \, = \, \delta_{\mu\nu} \, - \,  \frac{t_\mu p_\nu}{pt}   \; .
\end{equation}
Thus, despite the fact the gluon propagator DSE contains in axial gauge one term
less than in linear covariant gauges (the ghost loop) its explicit form is much
more complicated. 

\section{3-Gluon Vertex in Axial Gauge}
\label{app_3-gluon}

As mentioned in Sect.\  \ref{sec_nPoint} the Slavnov--Taylor
identity for the 3-gluon vertex $\Gamma_{\mu\nu\rho}$
in axial gauge has the comparatively simple form,
\begin{equation}  
ik_\rho \Gamma_{\mu\nu\rho} (p,q,k) \, =\, \Pi_{\mu\nu}(q) \, -  \,
\Pi_{\mu\nu} (p) \; .
\end{equation}
Nevertheless allowing for the most
general tensor structure, however, its solution becomes quite
involved. With the additional requirement that it is free from
kinematic singularities is implemented the solution has been obtained in
Ref.~\cite{Kim80}, and can be cast in the form,
\begin{eqnarray}  \label{axg_3gv}
i\Gamma_{\mu\nu\rho} (p,q,k) &=& \left( T_1(p,k)_{\mu\rho\nu} -
T_1(q,k)_{\nu\rho\mu}\right) \, g(k^2,(kt)^2) \\
  && \hskip -2cm + \,  \left( T_2(p,k)_{\mu\rho\nu} -
T_2(q,k)_{\nu\rho\mu}\right) \, f(k^2,(kt)^2)  
 \, + \, T_3(p,q)_{\mu\nu\rho} \, \frac{1}{2}\,  \left( g(p^2,(qt)^2) -
g(q^2,(pt)^2) \right)  \nonumber\\
  && \hskip -2cm  + \, T_4(p,q)_{\mu\nu\rho} \, \frac{1}{2}\,  \left(
f(p^2,(qt)^2) - f(q^2,(pt)^2) \right) \, +\quad \hbox{cycl. permutations} \;
. \nonumber
\end{eqnarray}
This compact form hides the complexity of this solution which, however,
can be seen quite clearly from the explicit expressions for the tensors
$T_1, T_2, T_3$ and $T_4$:
\begin{eqnarray}
T_1(p,q)_{\mu\nu\rho} &=& \delta_{\mu\nu} q_\rho \, - \, \frac{1}{2}
\left(\delta_{\mu\nu} pq - q_\mu p_\nu \right) \, \left(
\frac{(p-q)_\rho}{p^2 - q^2} \, + \, \frac{t_\rho}{pt + qt} \right)
\nonumber\\
T_2(p,q)_{\mu\nu\rho} &=& - \, \delta_{\mu\nu} t_\rho qt \, - \, t_\mu t_\nu
q_\rho  \nonumber \\
&& - \, \frac{1}{2} \left(\delta_{\mu\nu} pt qt - t_\mu p_\nu qt - t_\nu
q_\mu pt + t_\mu t_\nu pq \right) \, \left( \frac{(p-q)_\rho}{p^2 - q^2} \, +
\, \frac{t_\rho}{pt + qt} \right) \nonumber\\
T_3(p,q)_{\mu\nu\rho} &=& - \, \left(\delta_{\mu\nu} pq
- q_\mu p_\nu \right) \, \left( \frac{(p-q)_\rho}{p^2 - q^2} \, - \,
\frac{t_\rho}{pt + qt} \right) \nonumber \\
T_4(p,q)_{\mu\nu\rho} &=& \left(\delta_{\mu\nu} pt qt - t_\mu p_\nu qt - t_\nu
q_\mu pt + t_\mu t_\nu pq \right) \, \left( \frac{(p-q)_\rho}{p^2 - q^2} \, -
\, \frac{t_\rho}{pt + qt} \right) \, . \nonumber
\end{eqnarray}
This solution for the 3-gluon vertex is not
only free from singularities of the type $1/(p^2 -q^2)$ but also from the
typical axial gauge singularities of the form $1/(pt)$~\cite{Kim80}.

\bibliographystyle{nuclph}

\bibliography{dseref}

\begin{thebibliography}{100}

\bibitem{Leh55}
H.~Lehmann, K.~Symanzik, and W.~Zimmermann,
\newblock Nuov. Cim. {\bf 1} (1955) 205.

\bibitem{Itz80}
C.~Itzykson and J.-B. Zuber,
\newblock {\em Quantum Field Theory},
\newblock McGraw-Hill, 1980.

\bibitem{Haa96}
R.~Haag,
\newblock {\em Local Quantum Physics},
\newblock Springer Verlag, 2nd edition, 1996.

\bibitem{Haa58}
R.~Haag,
\newblock Phys. Rev. {\bf 112} (1958) 669.

\bibitem{Nis58}
K.~Nishijima,
\newblock Phys. Rev. {\bf 111} (1958) 995.

\bibitem{Zim58}
W.~Zimmermann,
\newblock Nuov. Cim. {\bf 10} (1958) 597.

\bibitem{Kin62}
T.~Kinoshita,
\newblock J.~Math.~Phys. {\bf 3} (1962) 650,
\newblock T.~D.~Lee and M.~Nauenberg, Phys.~Rev.~{\bf 133}, 6B (1964) 1549.

\bibitem{Blo37}
F.~Bloch and A.~Nordsieck,
\newblock Phys. Rev. {\bf 52} (1937) 54.

\bibitem{Mar78}
W.~Marciano and H.~Pagels,
\newblock Phys. Rep. {\bf 36} (1978) 137.

\bibitem{Kin76}
T.~Kinoshita and A.~Ukawa,
\newblock Phys. Rev. {\bf D13} (1976) 1573.

\bibitem{Nel81}
C.~A. Nelson,
\newblock Nucl. Phys. {\bf B186} (1981) 187.

\bibitem{Pog76}
E.~C. Poggio and H.~R. Quinn,
\newblock Phys.~Rev. {\bf D14} (1976) 578.

\bibitem{Gro74}
D.~J. Gross and A.~Neveu,
\newblock Phys.~Rev. {\bf D10} (1974) 3235.

\bibitem{Dys49}
F.~J. Dyson,
\newblock Phys.~Rev. {\bf 75} (1949) 1736.

\bibitem{Sch51}
J.~S. Schwinger,
\newblock Proc.~Nat.~Acad.~Sc. {\bf 37} (1951) 452; {\it ibid.}, 455.

\bibitem{Tay71}
J.~C. Taylor,
\newblock Nucl.~Phys. {\bf B33} (1971) 436.

\bibitem{Sla72}
A.~Slavnov,
\newblock Theor.~Math.~Phys. {\bf 10} (1972) 99.

\bibitem{Man79}
S.~Mandelstam,
\newblock Phys.~Rev. {\bf D20} (1979) 3223.

\bibitem{Atk81}
D.~Atkinson, J.~K. Drohm, P.~W. Johnson, and K.~Stam,
\newblock J.~Math.~Phys. {\bf 22} (1981) 2704.

\bibitem{Bro89}
N.~Brown and M.~R. Pennington,
\newblock Phys.~Rev. {\bf D39} (1989) 2723.

\bibitem{Hau96}
A.~Hauck, L.~v.~Smekal, and R.~Alkofer,
\newblock in {\em Quark Confinement and the Hadron Spectrum II}, edited by
  N.~Brambilla and M.~Prosperi, p. 258, World Scientific, 1997,
\newblock preprint, ANL-PHY-8386-TH-96, UNITU-THEP-5/1996.

\bibitem{Bak81a}
M.~Baker, J.~S. Ball, and F.~Zachariasen,
\newblock Nucl.~Phys. {\bf B186} (1981) 531 ; {\it ibid.}, 560.

\bibitem{Bak81b}
M.~Baker, J.~S. Ball, and F.~Zachariasen,
\newblock Nucl.~Phys. {\bf B226} (1983) 4555.

\bibitem{Ale82}
A.~Alekseev, B.~A. Arbuzov, and V.~A. Baikov,
\newblock Teor. Mat. Fiz. {\bf 52} (1982) 187.

\bibitem{Sch82}
W.~J. Schoenmaker,
\newblock Nucl.~Phys. {\bf B194} (1982) 535.

\bibitem{Cud91}
J.~R. Cudell and D.~A. Ross,
\newblock Nucl.~Phys. {\bf B358} (1991) 247.

\bibitem{Bue95}
K.~B\"uttner and M.~R. Pennington,
\newblock Phys.~Rev. {\bf D52} (1995) 5220.

\bibitem{Tan97}
P.~C. Tandy,
\newblock Prog.~Part.~Nucl.~Phys. {\bf 39} (1997) 117.

\bibitem{Rob94}
C.~D. Roberts and A.~G. Williams,
\newblock Prog.~Part.~Nucl.~Phys. {\bf 33} (1994) 477.

\bibitem{Mir93}
V.~A. Miransky,
\newblock {\em Dynamical Symmetry Breaking in Quantum Field Theories},
\newblock World Scientific, 1993,
\newblock and references therein.

\bibitem{Rob00}
C.~D. Roberts and S.~M. Schmidt,
\newblock Prog. Part. Nucl. Phys. {\bf 45} (2000) S1.

\bibitem{Gup50}
S.~N. Gupta,
\newblock Proc.~Roy.~Soc. {\bf A63} (1950) 681.

\bibitem{Ble50}
K.~Bleuler,
\newblock Helv.~Phys.~Acta {\bf 23} (1950) 567.

\bibitem{Str76}
F.~Strocchi,
\newblock Phys.~Lett. {\bf B62} (1976) 60.

\bibitem{Oji80}
I.~Ojima,
\newblock Z.~Phys. {\bf C5} (1980) 227.

\bibitem{Nak90}
N.~Nakanishi and I.~Ojima,
\newblock {\em Covariant Operator Formalism of Gauge Theories and Quantum
  Gravity}, volume~27 of {\em Lecture Notes in Physics},
\newblock World Scientific, 1990.

\bibitem{Bec75}
C.~Becchi, A.~Rouet, and R.~Stora,
\newblock Comm.~Math.~Phys. {\bf 42} (1975) 127,
\newblock {\it see also:} I. V. Tyutin, Lebedev preprint FIAN No. 39 (1975).

\bibitem{Fer91}
R.~Fern\'andez, J.~Fr\"ohlich, and A.~D. Sokal,
\newblock {\em Random Walks, Critical Phenomena, and Triviality in Quantum
  Field Theory},
\newblock Springer Verlag, 1991.

\bibitem{Oeh80}
R.~Oehme and W.~Zimmermann,
\newblock Phys.~Rev. {\bf D21} (1980) 471; {\it ibid.}, 1661.

\bibitem{Oeh95}
R.~Oehme,
\newblock Int.~J.~Mod.~Phys. {\bf A10} (1995) 1995.

\bibitem{Nis96}
K.~Nishijima,
\newblock Czech.~J.~Phys. {\bf 46} (1996) 1.

\bibitem{Fad67}
L.~D. Faddeev and V.~N. Popov,
\newblock Phys.~Lett. {\bf B25} (1967) 29.

\bibitem{Baa92a}
P.~van Baal,
\newblock Nucl.~Phys. {\bf B369} (1992) 259.

\bibitem{Baa92b}
P.~van Baal,
\newblock Global Issues in Gauge Fixing,
\newblock in {\em Non-perturbative approaches to QCD}, edited by D.~Diakonov,
  Trento, 1995, Procs. of the ECT* Workshop,
\newblock hep-th/9511119.

\bibitem{Gre97}
O.~W. Greenberg,
\newblock Phys.~Lett. {\bf B416} (1998) 144.

\bibitem{Oeh90}
R.~Oehme,
\newblock Phys.~Rev. {\bf D42} (1990) 4209,
\newblock Phys.~Lett. {\bf B252} (1990) 641.

\bibitem{Oeh94}
R.~Oehme and W.~Xu,
\newblock Phys.~Lett. {\bf B333} (1994) 172.

\bibitem{Sme97b}
L.~von Smekal, A.~Hauck, and R.~Alkofer,
\newblock Phys. Rev. Lett. {\bf 79} (1997) 3591.

\bibitem{Sme98}
L.~von Smekal, A.~Hauck, and R.~Alkofer,
\newblock Ann. Phys. {\bf 267} (1998) 1.

\bibitem{Lei98}
D.~Leinweber, J.~I. Skullerud, and A.~G. Williams,
\newblock Phys.~Rev. {\bf D58} (1998) 031501.

\bibitem{Bon00}
F.~Bonnet, P.~O. Bowman, D.~B. Leinweber, and A.~G. Williams,
\newblock Phys. Rev. {\bf D62} (2000) 051501.

\bibitem{Mar95a}
P.~Marenzoni, G.~Martinelli, and N.~Stella,
\newblock Nucl.~Phys. {\bf B455} (1995) 339.

\bibitem{Nak95}
A.~Nakamura, H.~Aiso, M.~Fukuda, T.~Iwamiya, T.~Nakamura, and M.~Yoshida,
\newblock Gluon Propagators and Confinement,
\newblock in {\em RCNP Confinement 1995}, pp. 90--95, 1995,
\newblock e-print hep-lat/9506024.

\bibitem{War50}
J.~C. Ward,
\newblock Phys. Rev. {\bf 78} (1950) 1824.

\bibitem{Tak57}
Y.~Takahasi,
\newblock Nuov. Cim. {\bf 6} (1957) 370,
\newblock {\it see also:} H. S. Green, Proc. Phys. Soc. {\bf A66} (1953) 873.

\bibitem{Lle80}
C.~H. Llewellyn-Smith,
\newblock in {\em Quantum Flavour Dynamics, Quantum Chromodynamics and Unified
  Theories}, edited by K.~T. Mahanthappa and J.~Randa, New York, 1980, Plenum
  Press.

\bibitem{Kug79}
T.~Kugo and I.~Ojima,
\newblock Prog. Theor. Phys. Suppl. {\bf 66} (1979) 1.

\bibitem{Jog00}
S.~D. Joglekar and A.~Misra,
\newblock Int. J. Mod. Phys. {\bf A10} (2000) 1453.

\bibitem{Nie75}
N.~K. Nielsen,
\newblock Nucl. Phys. {\bf B101} (1975) 173.

\bibitem{Gam99}
P.~Gambino and P.~A. Grassi,
\newblock Phys. Rev. {\bf D62} (1999) 076002.

\bibitem{Gli87}
J.~Glimm and A.~Jaffe,
\newblock {\em Quantum Physics: A Functional Integral Point of View},
\newblock Springer Verlag, 1987.

\bibitem{Pro99}
L.~V. Prokhorov,
\newblock Phys. Uspekhi {\bf 42} (1999) 1099.

\bibitem{Bag99}
E.~Bagan, M.~Lavelle, and D.~McMullan,
\newblock Ann. Phys. {\bf 282} (2000) 471; {\it ibid.}, 503.

\bibitem{Gri78}
V.~N. Gribov,
\newblock Nucl.~Phys. {\bf B139} (1978) 1.

\bibitem{Sin78}
I.~M. Singer,
\newblock Comm.~Math.~Phys. {\bf 60} (1978) 7.

\bibitem{Nas91}
C.~Nash,
\newblock {\em Differential Topology and Quantum Field Theory},
\newblock Academic Press, 1991.

\bibitem{Zhi87}
A.~R. Zhitnitskii,
\newblock Sov. J. Nucl. Phys. {\bf 46} (1987) 147; {\it ibid.}, 343.

\bibitem{Jac78}
R.~Jackiw, I.~Muzinich, and C.~Rebbi,
\newblock Phys.~Rev. {\bf D17} (1978) 1576.

\bibitem{Chr80}
N.~H. Christ and T.~D. Lee,
\newblock Phys.~Rev. {\bf D22} (1980) 939.

\bibitem{Har78}
B.~Harrington and H.~Shepard,
\newblock Phys. Rev {\bf D17} (1978) 2122.

\bibitem{Ros79}
P.~Rossi,
\newblock Nucl. Phys. {\bf B149} (1979) 170.

\bibitem{Gro81}
D.~J. Gross, R.~D. Pisarski, and L.~G. Yaffe,
\newblock Rev. Mod. Phys. {\bf 53} (1981) 43.

\bibitem{Rei97}
H.~Reinhardt,
\newblock Nucl. Phys. {\bf B503} (1997) 505.

\bibitem{Jah98}
O.~Jahn and F.~Lenz,
\newblock Phys. Rev. {\bf D58} (1998) 085006.

\bibitem{For98}
C.~Ford, U.~G. Mitreuter, T.~Tok, A.~Wipf, and J.~M. Pawlowski,
\newblock Ann.~Phys. {\bf 269} (1998) 26.

\bibitem{For99}
C.~Ford, T.~Tok, and A.~Wipf,
\newblock Phys.~Lett. {\bf B456} (1999) 155.

\bibitem{Kra98a}
T.~C. Kraan and P.~{van~Baal},
\newblock Nucl. Phys {\bf B533} (1998) 627.

\bibitem{Kra98b}
T.~C. Kraan and P.~{van~Baal},
\newblock Nucl. Phys {\bf A642} (1998) 299.

\bibitem{Jah99}
O.~Jahn,
\newblock J. Phys. {\bf A33} (2000) 2997.

\bibitem{Rei93}
H.~Reinhardt, K.~Langfeld, and L.~{v. Smekal},
\newblock Phys. Lett. {\bf B300} (1993) 111.

\bibitem{Qua98}
M.~Quandt and H.~Reinhardt,
\newblock Intern. J.~Mod. Phys. {\bf A13} (1998) 4049.

\bibitem{Alf76}
V.~{de Alfaro}, S.~Fubini, and G.~Furlan,
\newblock Phys. Lett. {\bf B65} (1976) 163.

\bibitem{Alf78}
V.~{de Alfaro}, S.~Fubini, and G.~Furlan,
\newblock Phys. Lett. {\bf B73} (1978) 463.

\bibitem{Act79}
A.~Actor,
\newblock Rev. Mod. Phys. {\bf 51} (1979) 461.

\bibitem{Wu68}
T.~T. Wu and C.~N. Yang,
\newblock in {\em Properties of Matter under Unusual Conditions}, edited by
  H.~Mark and S.~Fernbach, p. 349, Interscience Pub., J. Wiley \& Sons, New
  York, 1969.

\bibitem{Cal77}
C.~Callan, R.~Dashen, and D.~Gross,
\newblock Phys. Lett. {\bf B66} (1977) 375.

\bibitem{Cal78}
C.~G. Callan, R.~F. Dashen, and D.~J. Gross,
\newblock Phys.~Rev. {\bf D17} (1978) 2717.

\bibitem{Ste00}
J.~V. Steele and J.~W. Negele,
\newblock Phys. Rev. Lett. {\bf 85} (2000) 4207.

\bibitem{tHo75}
G.~t'Hooft,
\newblock in {\em High Energy Physics EPS Int. Conference}.

\bibitem{Man76}
S.~Mandelstam,
\newblock Phys.~Rep. {\bf 23} (1976) 245.

\bibitem{Del91}
G.~{Dell' Antonio} and D.~Zwanziger,
\newblock Comm.~Math.~Phys. {\bf 138} (1991) 291.

\bibitem{Riv87}
R.~J. Rivers,
\newblock {\em Path integral methods in quantum field theory},
\newblock Cambridge University Press, 1987.

\bibitem{Bau00a}
L.~Baulieu and D.~Zwanziger,
\newblock Nucl. Phys. {\bf B581} (2000) 604.

\bibitem{Bau00b}
L.~Baulieu, P.~A. Grassi, and D.~Zwanziger,
\newblock e-print  (2000),
\newblock hep-th/0006036.

\bibitem{Miz94}
M.~Mizutani and A.~Nakamura,
\newblock Nucl. Phys. B (Proc. Suppl.) {\bf 34} (1994) 253.

\bibitem{Ren99}
F.~DiRenzo and L.~Scorzato,
\newblock in {\em LATTICE 99},
\newblock Nucl. Phys. B (Proc. Suppl.) {\bf 83-84} (2000) 822.

\bibitem{Sho00}
F.~Shoji, T.~Suzuki, H.~Kodama, and A.~Nakamura,
\newblock Phys. Lett. {\bf B476} (2000) 199.

\bibitem{Zwa92}
D.~Zwanziger,
\newblock Nucl.~Phys. {\bf B378} (1992) 525.

\bibitem{Zwa94}
D.~Zwanziger,
\newblock Nucl.\ Phys. {\bf B412} (1994) 657.

\bibitem{Ber94}
C.~Bernard, C.~Parrinello, and A.~Soni,
\newblock Phys.~Rev. {\bf D49} (1994) 1585.

\bibitem{Sum96}
H.~Suman and K.~Schilling,
\newblock Phys.~Lett. {\bf B373} (1996) 314.

\bibitem{Cuc97}
A.~Cucchieri,
\newblock Nucl.~Phys. {\bf B508} (1997) 353.

\bibitem{Bir91}
D.~Birmingham, M.~Blau, M.~Rakowski, and G.~Thompson,
\newblock Phys.~Rep. {\bf 209} (1991) 129.

\bibitem{Fuj79}
K.~Fujikawa,
\newblock Prog. Theor. Phys. {\bf 61} (1979) 627.

\bibitem{Hir79}
P.~Hirschfeld,
\newblock Nucl. Phys. {\bf B157} (1979) 37.

\bibitem{Bau98}
L.~Baulieu and M.~Schaden,
\newblock Int. J. Mod. Phys. {\bf A13} (1998) 985.

\bibitem{Sha84}
B.~Sharpe,
\newblock J. Math. Phys. {\bf 25} (1984) 3324.

\bibitem{Bau96}
L.~Baulieu, A.~Rozenberg, and M.~Schaden,
\newblock Phys.~Rev. {\bf D54} (1996) 7852.

\bibitem{Sch98}
M.~Schaden and A.~Rozenberg,
\newblock Phys.~Rev. {\bf D57} (1998) 3670,
\newblock see also, M.~Schaden, {\sl Quantization in the Presence of Gribov
  Copies}, in Procs. of 4th Workshop on Quantum Chromodynamics, Paris, France,
  1-6 Jun 1998, hep-th/9810162.

\bibitem{Sch98b}
M.~Schaden,
\newblock Phys. Rev. {\bf D59} (1998) 014508.

\bibitem{Sch99}
M.~Schaden,
\newblock e-print  (1999),
\newblock hep-th/9909011.

\bibitem{Sch00}
M.~Schaden,
\newblock talk given at the 5th Workshop on QCD, Villefranche-sur-Mer, France,
  3-7 Jan. 2000, e-print, hep-th/0003030.

\bibitem{Wes83}
G.~B. West,
\newblock Phys.~Rev. {\bf D27} (1983) 1878.

\bibitem{Nis94}
K.~Nishijima,
\newblock Int.~J.~Mod.~Phys. {\bf A9} (1994) 3799,
\newblock {\it ibid.}, {\bf A10} (1995) 3155.

\bibitem{Nis96b}
K.~Nishijima and N.~Takase,
\newblock Int.~J.~Mod.~Phys. {\bf A11} (1996) 2281.

\bibitem{Hua98}
K.~Huang,
\newblock {\em Quantum Field Theory, from Operators to Path Integrals},
\newblock John Wiley Inc., 1998.

\bibitem{Goe98}
M.~G\"ockeler, R.~Horsley, V.~Linke, P.~Rakow, G.~Schierholz, and H.~St\"uben,
\newblock Phys.~Rev.~Lett. {\bf 80} (1998) 4119.

\bibitem{Sme91}
L.~v.~Smekal, P.~A. Amundsen, and R.~Alkofer,
\newblock Nucl.~Phys. {\bf A529} (1991) 633.

\bibitem{Xu96}
W.~J. Xu,
\newblock {\em Asymptotic Limits and Sum Rules for Propagators in Quantum
  Chromodynamics},
\newblock PhD thesis, University of Chicago, 1996,
\newblock hep-th/9607045; see also, R.~Oehme and W.~J.~Xu, Phys.~Lett. {\bf
  B384} (1996) 269.

\bibitem{Thi85}
J.~Thierry-Mieg,
\newblock Nucl. Phys. {\bf B261} (1985) 55.

\bibitem{Nak79}
N.~Nakanishi,
\newblock Prog. Theor. Phys. {\bf 62} (1979) 1396.

\bibitem{Oji78}
I.~Ojima,
\newblock Nucl. Phys. {\bf B143} (1978) 340.

\bibitem{Kug95}
T.~Kugo,
\newblock at Int. Symp. on BRS Symmetry, Kyoto, 18-22 Sep. 1995, e-print,
  hep-th/9511033.

\bibitem{Fer71}
R.~Ferrari and L.~E. Picasso,
\newblock Nucl. Phys. {\bf B31} (1971) 316.

\bibitem{Bra74}
R.~A. Brandt and W.-C. Ng,
\newblock Phys. Rev. {\bf D10} (1974) 4198.

\bibitem{Len94a}
F.~Lenz, H.~W.~L. Naus, K.~Otha, and M.~Thies,
\newblock Ann.~Phys. {\bf 233} (1994) 17; {\it ibid.}, 51.

\bibitem{Hat82}
H.~Hata,
\newblock Prog.~Theor.~Phys. {\bf 67} (1982) 1607.

\bibitem{Eli75}
S.~Elitzur,
\newblock Phys.~Rev. {\bf D12} (1975) 3978.

\bibitem{Wit81}
E.~Witten,
\newblock Nucl.~Phys. {\bf B188} (1981) 513.

\bibitem{Fro81}
J.~Fr\"ohlich, G.~Morchio, and F.~Strocchi,
\newblock Nucl.~Phys. {\bf B190} (1981) 553.

\bibitem{Sto95}
D.~Stoll and M.~Thies,
\newblock e-print  (1995),
\newblock hep-th/9504068.

\bibitem{Abb76}
L.~F. Abbot, J.~S. Kang, and H.~J. Schnitzer,
\newblock Phys.~Rev. {\bf D13} (1976) 2212.

\bibitem{Lan93a}
K.~Langfeld, L.~v.~Smekal, and H.~Reinhardt,
\newblock Phys.~Lett. {\bf B308} (1993) 279.

\bibitem{Sme94}
L.~v.~Smekal, K.~Langfeld, H.~Reinhardt, and R.~F. Langbein,
\newblock Phys.~Rev. {\bf D50} (1994) 6599.

\bibitem{Lan96}
R.~F. Langbein, K.~Langfeld, H.~Reinhardt, and L.~{v. Smekal},
\newblock Mod. Phys. Lett. {\bf A11} (1996) 631.

\bibitem{Gaw85}
K.~Gawedzki and A.~Kupiainen,
\newblock Nucl.~Phys. {\bf B257} (1985) 474.

\bibitem{Lan95}
K.~Langfeld, L.~v.~Smekal, and H.~Reinhardt,
\newblock Phys.~Lett. {\bf B362} (1995) 128.

\bibitem{Wit78}
E.~Witten,
\newblock Nucl.~Phys. {\bf B145} (1978) 110.

\bibitem{Col77}
J.~C. Collins, A.~Duncan, and S.~D. Joglekar,
\newblock Phys.~Rev. {\bf D16} (1977) 438.

\bibitem{Lan93b}
K.~Langfeld, L.~v.~Smekal, and H.~Reinhardt,
\newblock Phys.~Lett. {\bf B311} (1993) 207.

\bibitem{Nak99}
H.~Nakajima and S.~Furui,
\newblock in {\em LATTICE 99},
\newblock Nucl. Phys. B (Proc. Suppl.) {\bf 83-84} (2000) 521.

\bibitem{Nak00a}
H.~Nakajima and S.~Furui,
\newblock e-print  (2000),
\newblock hep-lat/0006002.

\bibitem{Ara62}
H.~Araki, K.~Hepp, and D.~Ruelle,
\newblock Helv. Phys. Acta {\bf 35} (1962) 164.

\bibitem{Rue62}
D.~Ruelle,
\newblock Helv. Phys. Acta {\bf 65} (1962) 147.

\bibitem{Str78}
F.~Strocchi,
\newblock Phys. Rev. {\bf D17} (1978) 2010.

\bibitem{Str77}
F.~Strocchi,
\newblock Comm. Math. Phys. {\bf 56} (1977) 57.

\bibitem{Iof69}
M.~Z. Iofa and V.~J. Fainberg,
\newblock JETPh {\bf 56} (1969) 1644,
\newblock Theor. Math. Phys. 1 (1969) 187.

\bibitem{Efi68}
G.~V. Efimov,
\newblock Comm. Math. Phys. {\bf 7} (1968) 138.

\bibitem{Hae90}
U.~H\"abel, R.~K\"onning, H.-G. Reusch, M.~Stingl, and S.~Wigard,
\newblock Z.~Phys. {\bf A336} (1990) 423; {\it ibid.}, 435.

\bibitem{Sti95}
M.~Stingl,
\newblock Z.~Phys. {\bf A353} (1996) 423.

\bibitem{Leu80}
H.~Leutwyler,
\newblock Phys. Lett. {\bf B96} (1980) 154.

\bibitem{Dit00}
W.~Dittrich and H.~Gies,
\newblock {\em Probing the Quantum Vacuum},
\newblock Springer Verlag, Berlin-Heidelberg, 2000.

\bibitem{Efi93}
G.~V. Efimov and M.~A. Ivanov,
\newblock {\em The Quark Confinement Model of Hadrons},
\newblock IOP Publishing, London, 1993.

\bibitem{Efi95}
G.~V. Efimov and S.~N. Nedelko,
\newblock Phys. Rev. {\bf D51} (1995) 176.

\bibitem{Efi96}
J.~V. Burdanov, G.~V. Efimov, S.~N. Nedelko, and S.~A. Solunin,
\newblock Phys. Rev. {\bf D54} (1996) 4483.

\bibitem{Bur96}
C.~J. Burden, C.~D. Roberts, and M.~J. Thomson,
\newblock Phys.~Lett. {\bf B371} (1996) 163.

\bibitem{Rob96}
C.~D. Roberts,
\newblock Nucl.~Phys. {\bf A605} (1996) 475.

\bibitem{Bub92}
M.~Buballa and S.~Krewald,
\newblock Phys. Lett. {\bf B294} (1992) 19.

\bibitem{Hab95}
H.~Haberzettl,
\newblock Nucl. Phys. {\bf A582} (1995) 603.

\bibitem{Ahl92}
J.~Ahlbach, A.~Streibl, and M.~Schaden,
\newblock Non-perturbative Solutions to the Dyson-Schwinger Equations of Pure
  QCD,
\newblock in {\em Proceedings of the Workshop on QCD Vacuum Structure}, edited
  by B.~M\"uller and H.~Fried, Paris, 1992, World Scientific.

\bibitem{Dri98a}
L.~Driesen, J.~Fromm, J.~Kuhrs, and M.~Stingl,
\newblock Eur.~Phys.~J. {\bf A4} (1999) 381.

\bibitem{Dri98b}
L.~Driesen and M.~Stingl,
\newblock Eur.~Phys.~J. {\bf A4} (1999) 401.

\bibitem{Gur99}
P.~Emirdag, R.~Easther, G.~Guralnik, and S.~Hahn,
\newblock in {\em LATTICE 99},
\newblock Nucl. Phys. B (Proc. Suppl.) {\bf 83-84} (2000) 938; S.C. Hahn and G.
  S. Guralnik, e-print (1999) hep-th/9901019, talk given at the 4th Workshop on
  Quantum Chromodynamics, Paris, France, 1-6 June 1998.

\bibitem{Gar96}
S.~Garcia, Z.~Guralnik, and G.~S. Guralnik,
\newblock e-print  (1996),
\newblock hep-th/9612079.

\bibitem{Fri72}
H.~M. Fried,
\newblock {\em Functional Methods and Models in Quantum Field Theory},
\newblock MIT Press, Cambridge, 1972.

\bibitem{Bau99}
L.~Baulieu and D.~Zwanziger,
\newblock Nucl. Phys. {\bf B548} (1999) 527.

\bibitem{Gei99}
K.~Geiger,
\newblock Phys. Rev. {\bf D60} (1999) 034012.

\bibitem{Cuc00}
A.~Cucchieri,
\newblock private communication; A. Cucchieri and D. Zwanziger, e-print (2000),
  hep-lat/0008026.

\bibitem{Fin82}
J.~R. Finger and J.~E. Mandula,
\newblock Nucl.~Phys. {\bf B199} (1982) 168.

\bibitem{Adl84}
S.~L. Adler and A.~C. Davis,
\newblock Nucl.~Phys. {\bf B244} (1984) 469.

\bibitem{LeY84}
A.~LeYaouanc, L.~Oliver, O.~Pene, and J.~C. Raynal,
\newblock Phys. Rev. {\bf D29} (1984) 1233.

\bibitem{Koc86}
A.~Kocic,
\newblock Phys. Rev. {\bf D33} (1986) 1785.

\bibitem{Alk87}
R.~Alkofer and P.~A. Amundsen,
\newblock Phys.~Lett. {\bf B187} (1987) 395.

\bibitem{Alk88}
R.~Alkofer and P.~A. Amundsen,
\newblock Nucl. Phys. {\bf B306} (1988) 305.

\bibitem{Alk89}
R.~Alkofer, P.~A. Amundsen, and K.~Langfeld,
\newblock Z. Phys. {\bf C42} (1989) 199.

\bibitem{Kle88}
S.~P. Klevansky and R.~H. Lemmer,
\newblock Phys. Rev. {\bf D38} (1988) 3559.

\bibitem{Wil97}
T.~Wilke and S.~Klevansky,
\newblock Ann. Phys. {\bf 258} (1997) 81.

\bibitem{Elw95}
U.~Ellwanger, M.~Hirsch, and A.~Weber,
\newblock Z.~Phys. {\bf C69} (1996) 687.

\bibitem{Elw96}
U.~Ellwanger, M.~Hirsch, and A.~Weber,
\newblock Eur.~Phys.~J. {\bf C1} (1998) 563.

\bibitem{Eic74}
E.~J. Eichten and F.~L. Feinberg,
\newblock Phys.~Rev. {\bf D10} (1974) 3254.

\bibitem{Bou98}
P.~Boucaud, J.~P. Leroy, J.~Micheli, O.~Pene, and C.~Roiesnel,
\newblock JHEP {\bf 12} (1998) 004,
\newblock JHEP {\bf 10} (1998) 017.

\bibitem{Bar80}
U.~Bar-Gadda,
\newblock Nucl.~Phys. {\bf 163} (1980) 312.

\bibitem{Bal80}
J.~S. Ball and T.~W. Chiu,
\newblock Phys.~Rev. {\bf D22} (1980) 2542; {\it ibid.}, 2550.

\bibitem{Kim80}
S.~K. Kim and M.~Baker,
\newblock Nucl.~Phys. {\bf B164} (1980) 152.

\bibitem{Mar00}
P.~Maris and P.~C. Tandy,
\newblock Phys.~Rev. {\bf C61} (2000) 045202.

\bibitem{Fra95}
M.~R. Frank,
\newblock Phys.~Rev. {\bf C51} (1995) 987.

\bibitem{Dom64}
C.~deDominicis and P.~C. Martin,
\newblock J. Math. Phys. {\bf 5} (1964) 14; {\it ibid.}, 31.

\bibitem{Vas72}
A.~N. Vasiliev and A.~K. Kazanskii,
\newblock Theor. Math. Phys. {\bf 12} (1972) 875; {\it ibid.}, {\bf 14} (1973)
  215.

\bibitem{Cor89}
J.~M. Cornwall and J.~Papavassiliou,
\newblock Phys.~Rev. {\bf D40} (1989) 3474.

\bibitem{Cor82}
J.~M. Cornwall,
\newblock Phys.~Rev. {\bf D26} (1982) 1453.

\bibitem{Pap00}
J.~Papavassiliou,
\newblock Phys.~Rev.~Lett. {\bf 84} (2000) 2782,
\newblock Phys. Rev. {\bf D62} (2000) 045006.

\bibitem{Wat99}
N.~J. Watson,
\newblock Nucl. Phys. {\bf B552} (1999) 461,
\newblock e-print (1999), hep-ph/9912303, to appear in the proceedings of the
  International Workshop on Physical Variables in Gauge Theories, Dubna,
  Russia, 21 - 25 Sep 1999.

\bibitem{Sch62a}
J.~Schwinger,
\newblock Phys. Rev. {\bf 128} (1962) 2425.

\bibitem{Ada97}
C.~Adam,
\newblock Ann. Phys. {\bf 259} (1997) 1.

\bibitem{Ada96}
C.~Adam,
\newblock Czech. J. Phys. {\bf 46} (1996) 893,
\newblock (e-print hep-ph/9501273).

\bibitem{Rad99}
T.~Radozycki,
\newblock Phys. Rev. {\bf D60} (1999) 105027.

\bibitem{Che99}
M.~N. Chernodub, M.~I. Polikarpov, and V.~I. Zakharov,
\newblock Phys. Lett. {\bf B457} (1999) 147.

\bibitem{Cam98}
A.~Campbell-Smith and N.~E. Mavromatos,
\newblock Acta Phys. Polon. {\bf B29} (1998) 3819,
\newblock (e-print cond-mat/9810324).

\bibitem{Tre95}
S.~B. Treiman, R.~Jackiw, B.~Zumino, and E.~Witten,
\newblock {\em Current Algebras and Anomalies},
\newblock Word Scientific, 1995.

\bibitem{Pol87}
A.~M. Polyakov,
\newblock {\em Gauge Fields and Strings},
\newblock Harwood Academic Publishers, 1987.

\bibitem{Mat99}
T.~Matsumaya, H.~Nagahiro, and S.~Uchida,
\newblock Phys. Rev. {\bf D60} (1999) 105020,
\newblock T. Matsumaya and H. Nagahiro, e-print (2000) hep-th/0004069.

\bibitem{Hos89}
Y.~Hoshino and T.~Matsumaya,
\newblock Phys. Lett. {\bf B222} (1989) 493.

\bibitem{Don94}
Z.~Dong, H.~J. Munzcek, and C.~D. Roberts,
\newblock Phys. Lett. {\bf B333} (1994) 536.

\bibitem{Bur98}
C.~J. Burden and P.~C. Tijiang,
\newblock Phys. Rev. {\bf D58} (1998) 085019.

\bibitem{Bas99}
A.~Bashir, A.~Kizilers{\"u}, and M.~R. Pennigton,
\newblock e-print  (1999),
\newblock hep-ph/9907418.

\bibitem{Mar96}
P.~Maris,
\newblock Phys. Rev. {\bf D54} (1996) 4650.

\bibitem{Bro88a}
N.~Brown and M.~R. Pennington,
\newblock Phys.~Rev. {\bf D38} (1988) 2266.

\bibitem{Sch98a}
A.~W. Schreiber, T.~Sizer, and A.~G. Williams,
\newblock Phys. Rev. {\bf D58} (1998) 125014.

\bibitem{Gus99}
V.~P. Gusynin, A.~W. Schreiber, T.~Sizer, and A.~G. Williams,
\newblock Phys. Rev. {\bf D60} (1999) 065007.

\bibitem{Kiz00}
A.~Kizilers{\"u}, T.~Sizer, and A.~G. Williams,
\newblock e-print  (2000),
\newblock hep-ph/0001147.

\bibitem{App86}
T.~Appelquist, M.~Bowick, D.~Karabali, and L.~Wijewardhana,
\newblock Phys. Rev. {\bf D33} (1986) 3704.

\bibitem{App88}
T.~Appelquist, D.~Nash, and L.~Wijewardhana,
\newblock Phys. Rev. Lett. {\bf 60} (1988) 2575.

\bibitem{Gus98}
V.~P. Gusynin, V.~A. Miransky, and A.~V. Shpagin,
\newblock Phys. Rev. {\bf D58} (1998) 085023.

\bibitem{Kon95}
K.~I. Kondo and P.~Maris,
\newblock Phys. Rev. Lett. {\bf 74} (1995) 18.

\bibitem{Ait97}
I.~J.~R. Aitchison, N.~E. Mavromatos, and D.~McNeill,
\newblock Phys. Lett. {\bf B402} (1997) 154.

\bibitem{Ahl00}
S.~Ahlig, R.~Alkofer, P.~Maris, and L.~von Smekal,
\newblock work in progress .

\bibitem{Mar93}
P.~Maris,
\newblock PhD thesis, University of Groningen, 1993.

\bibitem{Mar95}
P.~Maris,
\newblock Phys. Rev. {\bf D52} (1995) 6087.

\bibitem{Abd98}
E.~Abdalla and R.~Banerjee,
\newblock Phys. Rev. Lett. {\bf 80} (1998) 238.

\bibitem{Dal99}
D.~Dalmazi, A.~de~Souza~Datra, and M.~Hott,
\newblock Phys.~Rev. {\bf D61} (2000) 125018.

\bibitem{Blo95}
J.~C.~R. Bloch and M.~R. Pennington,
\newblock Mod. Phys. Lett {\bf A10} (1995) 1225.

\bibitem{Pen98}
M.~R. Pennington,
\newblock in {\em Proceedings of the Workshop on Nonperturbative Methods in
  Quantum Field Theory}, edited by A.~W. Schreiber, A.~G. Williams, and A.~W.
  Thomas, p.~49, Adelaide, 1998, World Scientific.

\bibitem{Rob87}
C.~D. Roberts and R.~T. Cahill,
\newblock Aust. J. Phys. {\bf 40} (1987) 499.

\bibitem{Cah98}
R.~T. Cahill and S.~M. Gunner,
\newblock Fizika {\bf B7} (1998) 171.

\bibitem{Fuk76}
R.~Fukuda and T.~Kugo,
\newblock Nucl. Phys. {\bf B117} (1976) 250.

\bibitem{Cor74}
J.~Cornwall, R.~Jackiw, and E.~Tomboulis,
\newblock Phys. Rev. {\bf D10} (1974) 2428.

\bibitem{Atk79}
D.~Atkinson and M.~Fry,
\newblock Nucl. Phys. {\bf B156} (1979) 301.

\bibitem{Cur90}
D.~C. Curtis and M.~Pennington,
\newblock Phys.~Rev. {\bf D42} (1990) 4165.

\bibitem{Kiz95}
A.~Kizilers{\"u}, M.~Reenders, and M.~R. Pennington,
\newblock Phys. Rev. {\bf D52} (1995) 1242.

\bibitem{Cur93}
D.~C. Curtis and M.~Pennington,
\newblock Phys.~Rev. {\bf D48} (1993) 4933.

\bibitem{Atk94}
D.~Atkinson, J.~C.~R. Bloch, V.~P. Gusynin, M.~R. Pennington, and M.~Reenders,
\newblock Phys. Lett. {\bf B329} (1994) 117.

\bibitem{Bur93}
C.~J. Burden and C.~D. Roberts,
\newblock Phys. Rev. {\bf D47} (1993) 5581.

\bibitem{Bas94}
A.~Bashir and M.~R. Pennigton,
\newblock Phys. Rev. {\bf D50} (1994) 7679.

\bibitem{Bas98}
A.~Bashir, A.~Kizilers\"u, and M.~R. Pennington,
\newblock Phys.~Rev. {\bf D57} (1998) 1242.

\bibitem{Haw96}
F.~T. Hawes, A.~G. Williams, and C.~D. Roberts,
\newblock Phys. Rev. {\bf D54} (1996) 5361.

\bibitem{Haw97}
F.~T. Hawes, T.~Sizer, and A.~G. Williams,
\newblock Phys. Rev. {\bf D55} (1997) 3866.

\bibitem{Bue96}
K.~B\"uttner,
\newblock {\em Confinement and the Infrared Behaviour of the Gluon Propagator},
\newblock PhD thesis, University of Durham, 1996.

\bibitem{Bla74}
S.~Blaha,
\newblock Phys.~Rev. {\bf D10} (1974) 4268.

\bibitem{Gro90}
F.~Gross and J.~Milana,
\newblock Phys.~Rev. {\bf D43} (1990) 2401.

\bibitem{Pag77}
H.~Pagels,
\newblock Phys.~Rev. {\bf D15} (1977) 2991.

\bibitem{Wes82}
G.~B. West,
\newblock Phys.~Lett. {\bf B115} (1982) 468.

\bibitem{Nat85}
K.~R. Natroshvili, A.~A. Khelashvili, and V.~Y. Khmaladze,
\newblock Teor.\ Mat.\ Fiz. {\bf 65} (1985) 360.

\bibitem{Vac89}
L.~G. Vachnadze, K.~R. Natroshvili, A.~A. Khelashvili, and V.~Y. Khmaladze,
\newblock Teor.\ Mat.\ Fiz. {\bf 80} (1989) 264.

\bibitem{Vac94}
L.~G. Vachnadze, N.~A. Kiknadze, K.~S. Turashvili, and A.~A. Khelashvili,
\newblock Theor.\ Math.\ Phys. {\bf 100} (1994) 811.

\bibitem{Gai90}
P.~Gaigg, W.~Kummer, and M.~Schweda, editors,
\newblock {\em Physical and Nonstandard Gauges}, volume 361 of {\em Lecture
  notes in physics}, Springer Verlag, 1990.

\bibitem{Nak00}
Y.~Nakawaki and G.~McCartor,
\newblock e-print  (2000),
\newblock hep-th/0004140.

\bibitem{Lit98}
D.~F. Litim and J.~M. Pawlowski,
\newblock Phys. Lett. {\bf B435} (1998) 181; Nucl. Phys. Proc. Suppl. {\bf 74}
  (1999) 329.

\bibitem{Wil74}
K.~G. Wilson and I.~G. Kogut,
\newblock Phys. Rep. {\bf 12} (1974) 75; F. Wegner and A. Houghton, Phys. Rev.
  {\bf A8} (1973) 401.

\bibitem{Lav89}
M.~Lavelle and M.~Schaden,
\newblock Phys.~Lett {\bf B217} (1989) 551.

\bibitem{Len94b}
F.~Lenz, H.~W.~L. Naus, and M.~Thies,
\newblock Ann.~Phys. {\bf 233} (1994) 317.

\bibitem{Fre76}
G.~Frenkel and J.~C. Taylor,
\newblock Nucl.~Phys. {\bf B109} (1976) 439.

\bibitem{Nak83}
N.~Nakanishi,
\newblock Phys.~Lett. {\bf B131} (1983) 381.

\bibitem{Bas89}
A.~Bassetto,
\newblock Good and Bad News Concerning Axial Gauges,
\newblock In Gaigg et~al. \cite{Gai90}.

\bibitem{Vac95}
L.~G. Vachnadze, N.~A. Kiknadze, and A.~A. Khelashvili,
\newblock Theor.\ Math.\ Phys. {\bf 102} (1995) 34.

\bibitem{Wat00}
P.~Watson,
\newblock PhD thesis, University of Durham, UK, 2000.

\bibitem{Hau98a}
A.~Hauck, L.~von Smekal, and R.~Alkofer,
\newblock Comput. Phys. Commun. {\bf 112} (1998) 149.

\bibitem{Hau98b}
A.~Hauck, L.~von Smekal, and R.~Alkofer,
\newblock Comput. Phys. Commun. {\bf 112} (1998) 166.

\bibitem{Bro91}
N.~Brown and N.~Dorey,
\newblock Mod.~Phys.~Lett. {\bf A6} (1991) 317.

\bibitem{Ahl98}
S.~Ahlig, L.~v.~Smekal, and R.~Alkofer,
\newblock work in progress .

\bibitem{Atk98}
D.~Atkinson and J.~C.~R. Bloch,
\newblock Mod.~Phys.~Lett. {\bf A13} (1998) 1055.

\bibitem{Atk97}
D.~Atkinson and J.~C.~R. Bloch,
\newblock Phys. Rev. {\bf D58} (1998) 094036.

\bibitem{Jau76}
J.~M. Jauch and F.~Rohrlich,
\newblock {\em The Theory of Photons and Electrons},
\newblock Springer Verlag, 2nd edition, 1976.

\bibitem{Zwa90}
D.~Zwanziger,
\newblock Nucl.~Phys. {\bf B345} (1990) 461.

\bibitem{Hat83}
H.~Hata and I.~Niigata,
\newblock Nucl.~Phys. {\bf B389} (1993) 133.

\bibitem{tHo74}
G.~t'Hooft,
\newblock Nucl.~Phys. {\bf B72} (1974) 461.

\bibitem{Sch97}
M.~Schmelling,
\newblock e-print  (1997),
\newblock hep-ex/9701002.

\bibitem{PDG98}
C.~Caso et~al.,
\newblock Eur.~Phys.~J. {\bf C3} (1998) 1.

\bibitem{Man87}
J.~E. Mandula and M.~Ogilvie,
\newblock Phys.~Lett. {\bf B185} (1987) 127.

\bibitem{Ais97}
H.~Aiso, J.~Fromm, M.~Fukuda, T.~Iwamiya, A.~Nakamura, T.~Nakamura, M.~Stingl,
  and M.~Yoshida,
\newblock Nucl.~Phys.~(Proc. Suppl.) {\bf B53} (1997) 570.

\bibitem{Man99}
J.~E. Mandula,
\newblock Phys. Rep. {\bf 315} (1999) 273.

\bibitem{Mun83}
H.~J. Munczek and A.~M. Nemirovsky,
\newblock Phys.~Rev. {\bf D28} (1983) 181.

\bibitem{Iva99}
M.~A. Ivanov, Y.~L. Kalinovsky, and C.~D. Roberts,
\newblock Phys. Rev. {\bf 60} (1999) 034018,
\newblock and references therein.

\bibitem{Gel51}
M.~Gell-Mann and F.~E. Low,
\newblock Phys.~Rev. {\bf 84} (1951) 350.

\bibitem{Sku98}
J.~I.Skullerud,
\newblock Nucl.~Phys.~Proc.~Suppl. {\bf 63} (1998) 242.

\bibitem{Mar97}
P.~Maris and C.~D. Roberts,
\newblock Phys. Rev. {\bf C56} (1997) 3369.

\bibitem{Alk99a}
R.~Alkofer, S.~Ahlig, and L.~von Smekal,
\newblock in {\em Nuclear and Particle physics with CEBAF at Jefferson Lab},
\newblock Fizika {\bf B8} (1999) 277; (e-print hep-ph/9901322).

\bibitem{Alk99b}
R.~Alkofer, S.~Ahlig, and L.~von Smekal,
\newblock in {\em Understanding Deconfinement in QCD}, edited by D.~Blaschke,
  F.~Karsch, and C.~D. Roberts, ECT* Trento, 1999, World Scientific,
\newblock (e-print hep-ph/9905324).

\bibitem{Cud89}
J.~Cudell, A.~Donnachie, and P.~Landshoff,
\newblock Nucl. Phys. {\bf B322} (1989) 55.

\bibitem{Sch62}
J.~Schwinger,
\newblock Phys.~Rev. {\bf 125} (1962) 397.

\bibitem{Haw98}
F.~T. Hawes, P.~Maris, and C.~D. Roberts,
\newblock Phys. Lett. {\bf B440} (1998) 353.

\bibitem{Vin92}
J.~C. Vink and U.-J. Wiese,
\newblock Phys. Lett. {\bf B289} (1992) 122.

\bibitem{Vin95}
J.~C. Vink,
\newblock Phys. Rev. {\bf D51} (1995) 1292.

\bibitem{Ale00}
C.~Alexandrou,
\newblock talk given at {\it Quark Confinement and the Hadron Spectrum IV},
  Vienna, July 3-8, 2000 .

\bibitem{Sku00a}
J.~I. Skullerud and A.~G. Williams,
\newblock e-print  (2000),
\newblock hep-lat/0007028.

\bibitem{Sku00}
J.~I. Skullerud and A.~G. Williams,
\newblock in {\em LATTICE 99},
\newblock Nucl. Phys. B (Proc. Suppl.) {\bf 83-84} (2000) 209.

\bibitem{Cud00}
J.~R. Cudell, A.~LeYaouanc, and C.~Pittori,
\newblock in {\em LATTICE 99},
\newblock Nucl. Phys. B (Proc. Suppl.) {\bf 83-84} (2000) 890.

\bibitem{Bec00}
D.~Becirevic, V.~Lubicz, G.~Martinelli, and M.~Testa,
\newblock in {\em LATTICE 99},
\newblock Nucl. Phys. B (Proc. Suppl.) {\bf 83-84} (2000) 893.

\bibitem{All97}
B.~All\'es, D.~Henty, H.~Panagopoulos, C.~Parinello, C.~Pittori, and D.~G.
  Richards,
\newblock Nucl.~Phys. {\bf B502} (1997) 325.

\bibitem{Hes98}
M.~He{\ss}, F.~Karsch, E.~Laermann, and I.~Wetzorke,
\newblock Phys. Rev. {\bf D58} (1998) 111502,
\newblock F. Karsch, M. He{\ss}, E. Laermann and I. Wetzorke, in {\it
  LATTICE98}, Nucl.Phys. (Proc.Suppl.) {\bf 73} (1999) 213.

\bibitem{Sal51}
E.~E. Salpeter and H.~A. Bethe,
\newblock Phys. Rev. {\bf 84} (1951) 1232.

\bibitem{Mun95}
H.~J. Munczek,
\newblock Phys.~Rev. {\bf D52} (1995) 4736.

\bibitem{Del79}
R.~Delbourgo and M.~D. Scadron,
\newblock J.~Phys. {\bf G5} (1979) 1621.

\bibitem{Blu97}
A.~Blumhofer and J.~Manus,
\newblock Nucl. Phys. {\bf B515} (1998) 522.

\bibitem{Sau99}
V.~Sauli,
\newblock private communication.

\bibitem{Kus95}
K.~Kusaka and A.~G. Williams,
\newblock Phys. Rev. {\bf D51} (1995) 7026.

\bibitem{Kus97a}
K.~Kusaka, K.~Simpson, and A.~G. Williams,
\newblock Phys. Rev. {\bf D56} (1997) 5071.

\bibitem{Nak71}
N.~Nakanishi,
\newblock {\em Graph Theory and Feynman Integrals},
\newblock Gordon Breach, New York, 1971.

\bibitem{Ahl99}
S.~Ahlig and R.~Alkofer,
\newblock Ann. Phys. {\bf 275} (1999) 113.

\bibitem{Wic54}
G.~C. Wick,
\newblock Phys. Rev. {\bf 96} (1954) 1124.

\bibitem{Cut54}
R.~E. Cutkosky,
\newblock Phys. Rev. {\bf 96} (1954) 1135.

\bibitem{Nak69}
N.~Nakanishi,
\newblock Suppl.~Prog.~Theor.~Phys. {\bf 43} (1969) 1.

\bibitem{Kau69}
W.~B. Kaufmann,
\newblock Phys. Rev. {\bf 187} (1969) 2951.

\bibitem{Fuk93}
I.~Fukui and N.~Set\^{o},
\newblock Progr. Theor. Phys. {\bf 89} (1993) 205,
\newblock {\it ibid.}, {\bf 95} (1996) 433.

\bibitem{The99}
L.~Theussl and B.~Desplanques,
\newblock e-print  (1999),
\newblock nucl-th/9908007.

\bibitem{Bij97}
J.~Bijtebier,
\newblock Nucl. Phys. {\bf A623} (1997) 498.

\bibitem{Gol53}
J.~S. Goldstein,
\newblock Phys. Rev. {\bf 91} (1953) 1516.

\bibitem{All97a}
T.~W. Allen and C.~J. Burden,
\newblock Phys. Rev. {\bf D55} (1997) 4954.

\bibitem{All96}
T.~W. Allen and C.~J. Burden,
\newblock Phys. Rev. {\bf D53} (1996) 5842,
\newblock erratum Phys. Rev. {\bf D54} (1996) 6567.

\bibitem{Ros96}
R.~Rosenfelder and A.~W. Schreiber,
\newblock Phys. Rev. {\bf D53} (1996) 3337; {\it ibid.}, 3354.

\bibitem{Lle69}
C.~H. Llewellyn-Smith,
\newblock Ann. Phys. {\bf 53} (1969) 521.

\bibitem{Jai93}
P.~Jain and H.~J. Munczek,
\newblock Phys.~Rev. {\bf D48} (1993) 5403,
\newblock and references therein.

\bibitem{Gov84}
J.~Govaerts, J.~E. Mandula, and J.~Weyers,
\newblock Nucl. Phys. {\bf B237} (1984) 59.

\bibitem{Lan89a}
K.~Langfeld, R.~Alkofer, and P.~A. Amundsen,
\newblock Z.\ Phys.\ {\bf C42} (1989) 159.

\bibitem{Kog74}
J.~Kogut and L.~Susskind,
\newblock Phys. Rev. {\bf D10} (1974) 3468.

\bibitem{Mec97}
A.~Mecke,
\newblock Diploma Thesis, University of T\"ubingen, 1997.

\bibitem{Sme97a}
L.~von Smekal, A.~Mecke, and R.~Alkofer,
\newblock in {\em Intersections between Particle and Nuclear Physics, 6th
  Conference}, edited by T.~W. Donelly, p. 746, World Scientific, 1997,
\newblock e-print hep-ph/9707210.

\bibitem{Wit79}
E.~Witten,
\newblock Nucl.~Phys. {\bf B156} (1979) 269.

\bibitem{Ven79}
G.~Veneziano,
\newblock Nucl.~Phys. {\bf B159} (1979) 461,
\newblock P.~Di~Veccia and G.~Veneziano, Nucl.~Phys. {\bf B171} (1980), 253.

\bibitem{Alk89a}
R.~Alkofer, M.~Nowak, J.~Verbaarschot, and I.~Zahed,
\newblock Phys.~Lett. {\bf B233} (1989) 205.

\bibitem{Kek00}
D.~Kekez, D.~Klabucar, and M.~D. Scadron,
\newblock J. Phys. {\bf G26} (2000) 1335,
\newblock (hep-ph/0003234).

\bibitem{Fra98}
M.~Frank and T.~Meissner,
\newblock Phys. Rev. {\bf C57} (1998) 345.

\bibitem{Alk89b}
R.~Alkofer, P.~A. Amundsen, and H.~Reinhardt,
\newblock Phys.~Lett. {\bf B218} (1989) 75.

\bibitem{Mar99}
P.~Maris and P.~C. Tandy,
\newblock Phys.~Rev. {\bf C60} (1999) 055214.

\bibitem{Alf00}
M.~Alford and R.~L. Jaffe,
\newblock Nucl. Phys. {\bf B578} (2000) 367.

\bibitem{Tor99}
N.~A. T{\"o}rnqvist,
\newblock Eur. Phys. J. {\bf C11} (1999) 359.

\bibitem{Pen99}
M.~R. Pennington,
\newblock e-print  (1999),
\newblock hep-ph/9905241, talk given at Workshop on Hadron Spectroscopy, Rome,
  Italy, 8-12 Mar 1999.

\bibitem{Bla00}
D.~Black, A.~H. Fariborz, and J.~Schechter,
\newblock Phys. Rev. {\bf D61} (2000) 074001; {\it ibid.}, 074030.

\bibitem{Jan94}
G.~Janssen, B.~C. Pearce, K.~Holinde, and J.~Speth,
\newblock Phys. Rev. {\bf D52} (1995) 2690.

\bibitem{Neu94}
M.~Neubert,
\newblock Phys. Rep. {\bf 245} (1994) 259.

\bibitem{Isg89}
N.~Isgur and M.~Wise,
\newblock Phys. Lett. {\bf B232} (1989) 113,
\newblock Phys. Lett. {\bf B237} (1990) 527.

\bibitem{Mar00a}
P.~Maris and P.~C. Tandy,
\newblock Phys.~Rev. {\bf C62} (2000) 055204.

\bibitem{OCo95}
H.~O'Connell, B.~Pearce, A.~Thomas, and A.~Williams,
\newblock Prog. Part. Nucl. Phys. {\bf 39} (1995) 201.

\bibitem{Haw99}
F.~T. Hawes and M.~A. Pichowsky,
\newblock Phys.~Rev. {\bf C59} (1999) 1743.

\bibitem{Hec98}
M.~B. Hecht and B.~H.~J. McKellar,
\newblock Phys. Rev. {\bf C57} (1998) 2638.

\bibitem{Mit97}
K.~L. Mitchell and P.~C. Tandy,
\newblock Phys. Rev. {\bf C55} (1997) 1477.

\bibitem{Mit94}
K.~L. Mitchell, P.~C. Tandy, C.~D. Roberts, and R.~T. Cahill,
\newblock Phys. Lett. {\bf B335} (1994) 282.

\bibitem{Pic96}
M.~A. Pichowsky and T.-S. Lee,
\newblock Phys. Lett. {\bf B379} (1996) 1.

\bibitem{Kro96}
P.~Kroll and M.~Raulfs,
\newblock Phys. Lett. {\bf B387} (1996) 848.

\bibitem{Fra95a}
M.~R. Frank, K.~L. Mitchell, C.~D. Roberts, and P.~C. Tandy,
\newblock Phys. Lett. {\bf B359} (1995) 17.

\bibitem{Mar98}
P.~Maris and C.~D. Roberts,
\newblock Phys. Rev. {\bf C58} (1998) 3659.

\bibitem{Kek99}
D.~Kekez and D.~Klabucar,
\newblock Phys. Lett. {\bf B457} (1999) 359.

\bibitem{Ani00}
I.~V. Anikin, A.~E. Dorokhov, and L.~Tomio,
\newblock Phys. Lett. {\bf B475} (2000) 361.

\bibitem{Alk96a}
R.~Alkofer and C.~D. Roberts,
\newblock Phys. Lett. {\bf B369} (1996) 101.

\bibitem{Bis99}
B.~Bistrovic and D.~Klabucar,
\newblock Phys. Rev. {\bf D61} (2000) 033006.

\bibitem{Bis00}
B.~Bistrovic and D.~Klabucar,
\newblock Phys. Lett. {\bf B478} (2000) 127.

\bibitem{Pra88a}
J.~Praschifka, C.~D. Roberts, and R.~T. Cahill,
\newblock Ann. Phys. {\bf 188} (1988) 20.

\bibitem{Pic99}
M.~A. Pichowsky, S.~Walawalkar, and S.~Capstick,
\newblock Phys. Rev. {\bf D60} (1999) 054030.

\bibitem{Ebe81}
D.~Ebert and M.~K. Volkov,
\newblock Fort. Phys. {\bf 29} (1981) 35.

\bibitem{Gas83}
J.~Gasser and H.~Leutwyler,
\newblock Ann. Phys. {\bf 158} (1983) 142.

\bibitem{Alk95}
R.~Alkofer, A.~Bender, and C.~D. Roberts,
\newblock Int. J. Mod. Phys. {\bf A10} (1995) 3319.

\bibitem{Gas85}
J.~Gasser and H.~Leutwyler,
\newblock Nucl. Phys. {\bf B250} (1985) 465; {\it ibid.}, 517.

\bibitem{Has78}
P.~Hasenfratz and J.~Kuti,
\newblock Phys.~Rep. {\bf 40} (1978) 75.

\bibitem{Adk83}
G.~S. Adkins, C.~R. Nappi, and E.~Witten,
\newblock Nucl.~Phys. {\bf B228} (1983) 552.

\bibitem{Hol93}
E.~G.~Holzwarth,
\newblock {\em Baryons as Skyrme Solitons},
\newblock World Scientific, 1993.

\bibitem{Wei96}
H.~Weigel,
\newblock Int. J. Mod. Phys. {\bf A11} (1996) 2419.

\bibitem{Alk95a}
R.~Alkofer and H.~Reinhardt,
\newblock {\em Chiral Quark Dynamics},
\newblock Springer Verlag, Berlin-Heidelberg, 1995,
\newblock Lecture Notes in Physics, Vol.\ m33.

\bibitem{Alk96}
R.~Alkofer, H.~Reinhardt, and H.~Weigel,
\newblock Phys.~Rep. {\bf 265} (1996) 139.

\bibitem{Chr96}
C.~V. Christov, A.~Blotz, H.~C. Kim, P.~Pobylitsa, T.~Watabe, T.~Meissner,
  E.~Ruiz-Arriola, and K.~Goeke,
\newblock Prog.~Part.~Nucl.~Phys. {\bf 37} (1996) 91.

\bibitem{Gel64}
M.~Gell-Mann,
\newblock Phys.~Rev. {\bf 8} (1964) 214.

\bibitem{Kar68}
G.~Karl and E.~Obryk,
\newblock Nucl.~Phys. {\bf B8} (1968) 609.

\bibitem{Fai68}
D.~Faiman and A.~W. Hendry,
\newblock Phys.~Rev. {\bf 173} (1968) 1720.

\bibitem{Fey71}
R.~P. Feynman, M.~Kislinger, and F.~Ravndal,
\newblock Phys.~Rev. {\bf D3} (1971) 2706.

\bibitem{Cho75}
A.~Chodos and C.~Thorn,
\newblock Phys.~Rev. {\bf D12} (1975) 2733.

\bibitem{Rho94}
M.~Rho,
\newblock Phys.~Rep. {\bf 240} (1994) 1.

\bibitem{Zue97}
U.~Z\"uckert, R.~Alkofer, H.~Weigel, and H.~Reinhardt,
\newblock Phys.~Rev. {\bf C55} (1997) 2030.

\bibitem{Rei90}
H.~Reinhardt,
\newblock Phys.~Lett. {\bf B244} (1990) 316.

\bibitem{Buc92}
A.~Buck, R.~Alkofer, and H.~Reinhardt,
\newblock Phys.~Lett. {\bf B286} (1992) 29.

\bibitem{Ish93a}
N.~Ishii, W.~Bentz, and K.~Yazaki,
\newblock Phys.~Lett. {\bf B301} (1993) 165.

\bibitem{Ish93b}
N.~Ishii, W.~Bentz, and K.~Yazaki,
\newblock Phys.~Lett. {\bf B318} (1993) 26.

\bibitem{Hua94}
S.~Huang and J.~Tjon,
\newblock Phys.~Rev. {\bf C49} (1994) 1702.

\bibitem{Buc95}
A.~Buck and H.~Reinhardt,
\newblock Phys.~Lett. {\bf B356} (1995) 168.

\bibitem{Ish95}
N.~Ishii, W.~Bentz, and K.~Yazaki,
\newblock Nucl.~Phys. {\bf A587} (1995) 617.

\bibitem{Asa95}
H.~Asami, N.~Ishii, W.~Bentz, and K.~Yazaki,
\newblock Phys.~Rev. {\bf C51} (1995) 3388.

\bibitem{Min99}
H.~Mineo, W.~Bentz, and K.~Yazaki,
\newblock Phys. Rev. {\bf C60} (1999) 065201.

\bibitem{Bur88}
C.~J. Burden, R.~T. Cahill, and J.~Praschifka,
\newblock Austral. J. Phys. {\bf 42} (1989) 147.

\bibitem{Pra89}
J.~Praschifka, R.~T. Cahill, and C.~D. Roberts,
\newblock Intern. J.~Mod. Phys. {\bf A4} (1989) 4929.

\bibitem{Blo99}
J.~C.~R. Bloch, C.~D. Roberts, S.~M. Schmidt, A.~Bender, and M.~R. Frank,
\newblock Phys. Rev. {\bf C60} (1999) 062201.

\bibitem{Blo00}
J.~C.~R. Bloch, C.~D. Roberts, and S.~M. Schmidt,
\newblock Phys. Rev. {\bf C61} (2000) 065207.

\bibitem{Kus97}
K.~Kusaka, G.~Piller, A.~W. Thomas, and A.~G. Williams,
\newblock Phys.~Rev. {\bf D55} (1997) 5299.

\bibitem{Hel97b}
G.~Hellstern, R.~Alkofer, M.~Oettel, and H.~Reinhardt,
\newblock Nucl.\ Phys. {\bf A627} (1997) 679.

\bibitem{Oet98}
M.~Oettel, G.~Hellstern, R.~Alkofer, and H.~Reinhardt,
\newblock Phys.~Rev. {\bf C58} (1998) 2459.

\bibitem{Oet99}
M.~Oettel, M.~A. Pichowsky, and L.~von Smekal,
\newblock Eur. Phys. J. {\bf A8} (2000) 251.

\bibitem{Oet00a}
M.~Oettel and R.~Alkofer,
\newblock Phys. Lett. {\bf B484} (2000) 243.

\bibitem{Oet00b}
M.~Oettel, R.~Alkofer, and L.~von Smekal,
\newblock Eur. Phys. J. {\bf A8} (2000) 553.

\bibitem{Oet00c}
M.~Oettel,
\newblock PhD thesis, University of T{\"u}bingen, 2000,
\newblock \\ http://w210.ub.uni-tuebingen.de/dbt/volltexte/2000/177.

\bibitem{Glo97}
L.~Y. Glozman, Z.~Papp, W.~Plessas, K.~Varga, and R.~F. Wagenbrunn,
\newblock Phys.~Rev. {\bf C57} (1998) 3406.

\bibitem{Ben96}
A.~Bender, C.~D. Roberts, and L.~v.~Smekal,
\newblock Phys.~Lett. {\bf B380} (1996) 7.

\bibitem{Hel97a}
G.~Hellstern, R.~Alkofer, and H.~Reinhardt,
\newblock Nucl.\ Phys. {\bf A625} (1997) 697.

\bibitem{Kad62}
L.~P. Kadanoff and G.~Baym,
\newblock {\em Quantum Statistical Mechanics},
\newblock Benjamin Inc., New York, 1962.

\bibitem{Cah89a}
R.~T. Cahill, C.~D. Roberts, and J.~Praschifka,
\newblock Phys. Rev. {\bf D36} (1989) 2804.

\bibitem{Kro97}
P.~Kroll, M.~Sch{\"u}rmann, K.~Passek, and W.~Schweiger,
\newblock Phys. Rev. {\bf D55} (1997) 4315.

\bibitem{Alk99}
R.~Alkofer, S.~Ahlig, C.~Fischer, and M.~Oettel,
\newblock in {\em MENU 99},
\newblock $\pi N$ Newsletter {\bf 15} (1999) 238; (e-print nucl-th/9911020).

\bibitem{Kvi99}
A.~N. Kvinikhidze and B.~Blankleider,
\newblock Phys. Rev. {\bf C60} (1999) 044003; {\it ibid.}, 044004.

\bibitem{Ish00}
N.~Ishii,
\newblock e-print  (2000),
\newblock nucl-th/0004063.

\bibitem{Jon99}
M.~K. Jones et~al.,
\newblock Phys. Rev. Lett. {\bf 84} (2000) 1398.

\bibitem{Bla99}
B.~Blankleider and A.~N. Kvinikhidze,
\newblock Phys. Rev. {\bf C62} (2000) 039801.

\bibitem{Fis99}
C.~Fischer,
\newblock Diploma Thesis, University of T\"ubingen, 1999.

\bibitem{Ahl00a}
S.~Ahlig, R.~Alkofer, C.~Fischer, M.~Oettel, H.~Reinhardt, and H.~Weigel,
\newblock to be published .

\bibitem{Cur76a}
G.~Curci and R.~Ferrari,
\newblock Phys. Lett. {\bf B63} (1976) 91.

\bibitem{Cur76b}
G.~Curci and R.~Ferrari,
\newblock Nuovo Cim. {\bf 32A} (1976) 151.

\bibitem{Bau82}
L.~Baulieu and J.~Thierry-Mieg,
\newblock Nucl. Phys. {\bf B197} (1982) 477.

\bibitem{Thi79}
J.~Thierry-Mieg and Y.~{Ne'eman},
\newblock Ann. Phys. {\bf 123} (1979) 247.

\bibitem{Qui81}
M.~{Quir\'os}, F.~J. {de Urr\'\i es}, J.~Hoyos, M.~L. {Maz\'on}, and
  E.~Rodriguez,
\newblock J. Math. Phys. {\bf 22} (1981) 1767.

\bibitem{Hoy82}
J.~Hoyos, M.~{Quir\'os}, J.~{Ram\'\i rez Mittelbrunn}, and F.~J. {de Urr\'\i
  es},
\newblock J. Math. Phys. {\bf 23} (1982) 1504.

\bibitem{Hoy83}
J.~Hoyos, M.~{Quir\'os}, J.~{Ram\'\i rez Mittelbrunn}, and F.~J. {de Urr\'\i
  es},
\newblock Nucl. Phys. {\bf B218} (1983) 159.

\bibitem{Bon81}
L.~Bonora and M.~Tonin,
\newblock Phys. Lett. {\bf B98} (1981) 48.

\bibitem{Hir81}
A.~C. Hirschfeld and H.~Leschke,
\newblock Phys. Lett. {\bf B101} (1981) 48.

\bibitem{Del82}
R.~Delbourgo and P.~D. Jarvis,
\newblock J. Phys. {\bf A15} (1982) 611.

\bibitem{Bak77}
M.~Baker and C.~Lee,
\newblock Phys.~Rev. {\bf D15} (1977) 2201.

\end{thebibliography}

\newpage

\addcontentsline{toc}{section}{{ List of Figures}}
\listoffigures


\addcontentsline{toc}{section}{{ List of Tables}}
\listoftables

\end{document}